\documentclass[12pt,english]{book}
\usepackage[T1]{fontenc}
\usepackage[latin1]{inputenc}
\usepackage{geometry}
\usepackage{fancyhdr}
\usepackage{rotating}
\usepackage{subfigure}
\usepackage{graphicx}
\usepackage{setspace}
\makeatletter

\newcommand{\noun}[1]{\textsc{#1}}

\providecommand{\tabularnewline}{\\}

 \newcommand{\lyxaddress}[1]{
   \par {\raggedright #1 
   \vspace{1.4em}
   \noindent\par}
 }
 \newcommand{\lyxrightaddress}[1]{
   \par {\raggedleft \begin{tabular}{l}\ignorespaces
   #1
   \end{tabular}
   \vspace{1.4em}
   \par}
 }
 \usepackage{verbatim}

\usepackage{framed}

\usepackage{babel}

\def    \be             {\begin{equation}}
\def    \ee             {\end{equation}}
\def    \ba             {\begin{eqnarray}}
\def    \ea             {\end{eqnarray}}

\def\beq{\begin{equation}}
\def\eeq{\end{equation}}
\def\beqn{\begin{eqnarray}}
\def\ba{\begin{eqnarray}}
\def\eeqn{\end{eqnarray}}
\def\ea{\end{eqnarray}}

\def\m{{\tt -}}

\def\Q{{\bf Q}}
\def\R{{\bf R}}
\def\z{{\zeta}}
\def\one {\bf  1}

\def\A{\mathcal{A}}

\newcommand{\beqa}{\begin{eqnarray}}
\newcommand{\eeqa}{\end{eqnarray}}

\newcommand{\n}{\hspace*{-2.5mm}}
\newcommand{\COMMA}{\hspace{0.1cm},}
\newcommand{\STOP}{\hspace{0.1cm}.}

\makeatother
\begin{document}
\begin{center}{\huge \thispagestyle{empty}}\end{center}{\huge \par}

\begin{center}{\large UNIVERSITÀ DEGLI STUDI DI LECCE}\end{center}{\large \par}

\begin{center}{\large DIPARTIMENTO DI FISICA}\end{center}{\large \par}

\begin{center}\vspace{1cm}\end{center}

\begin{center}\includegraphics[%
  width=4cm]{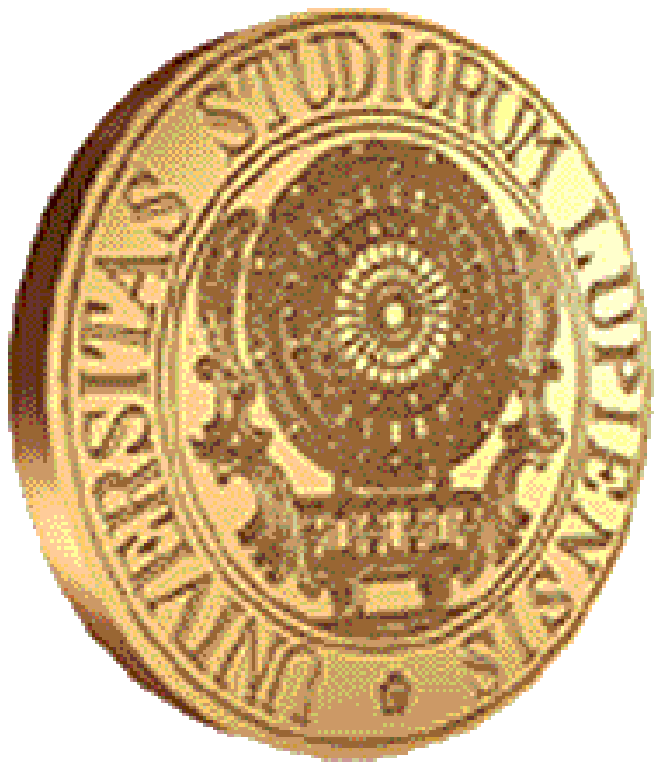}\vspace{2cm}\end{center}

\begin{center}\textbf{\huge QCD at Hadron Colliders and in}\\
\textbf{\huge Ultra High Energy Cosmic Rays}\vspace{3cm}\end{center}

\lyxaddress{\textbf{Advisor}\\
Claudio Corianò}

\lyxrightaddress{\textbf{Candidate}\\
Alessandro Cafarella}

\lyxrightaddress{\vspace{0.5cm}}

\begin{center}\noindent

\rule{16cm}{0.1mm}\renewcommand{\headrulewidth}{0pt}\end{center}

\begin{center}\noun{\large Tesi di Dottorato di Ricerca in Fisica
-- XVII Ciclo}\fancyhf{}

\renewcommand{\headrulewidth}{0pt}

\fancyhead[LE,RO]{\thepage}

\fancyhead[RE]{\nouppercase\rightmark}

\fancyhead[LO]{\nouppercase\leftmark}\end{center}

\tableofcontents{}

\chapter*{List of publications}

\addcontentsline{toc}{chapter}{List of publications}\markboth{}{List of publications}

\section*{Research papers}

The chapters presented in this thesis are based on the following research works.

\begin{enumerate}
\item A.~Cafarella, C.~Corianò and M.~Guzzi, \emph{An $x$-space analysis of
evolution equations: Soffer's inequality and the nonforward evolution},
J.High Energy Phys.\textbf{~11} (2003) 059
\item A.~Cafarella and C.~Corianò, \emph{Direct solution of renormalization
group equations of QCD in $x$-space: NLO implementations at leading
twist}, Comput.Phys.Commun.\textbf{~160} (2004) 213
\item A.~Cafarella, C.~Corianò and A.~Faraggi, \emph{Large scale air shower
simulations and the search for new Physics at Auger}, Int.J.Mod.Phys.A
\textbf{19, 22} (2004) 3729
\item A.~Cafarella, C.~Corianò and T.N.~Tomaras, \emph{Cosmic ray signals from
mini black holes in models with extra dimensions: an analytical/Monte
Carlo study}, J.High Energy Phys.\textbf{~06} (2005) 065
\item A.~Cafarella and C.~Corianò, \emph{The kinetic interpretation of the
DGLAP equation, its Kramers-Moyal expansion and positivity of helicity
distributions}, Int.J.Mod.Phys.A \textbf{20, 21} (2005) 4863
\item A.~Cafarella, C.~Corianò, M.~Guzzi and J.~Smith, \emph{On the scale variation of the total cross section for Higgs production at the LHC and at the Tevatron}, hep-ph/0510179, submitted to Eur.Phys.J.C
\item V.~Barone, A.~Cafarella, C.~Corianò, M.~Guzzi and P.~Ratcliffe, \emph{ Double transverse-spin asymmetries in Drell-Yan processes with antiprotons}, hep-ph/0512121, submitted to Phys.Lett.B
\end{enumerate}

\section*{Other works}

This paper has been completed during the doctoral course, but it is unrelated with the thesis topic.

\begin{enumerate}
\item A.~Cafarella, M.~Leo and R.A.~Leo, \emph{Numerical analysis of the
one-mode solutions in the Fermi-Pasta-Ulam system}, Phys.Rev.E \textbf{69}
(2004) 046604
\end{enumerate}

\section*{Conference proceedings}

\begin{enumerate}
\item A.~Cafarella, C.~Corianò, M.~Guzzi and D.~Martello, \emph{Superstring
relics, supersymmetric fragmentation and UHECR}, hep-ph/0208023,
published in the proceedings of the \emph{1st International Conference
on String Phenomenology , Oxford, England, 6-11 July 2002}, editors
S.~Abel, A.~Faraggi, A.~Ibarra and M.~Plumacher, World Scientific
(2003)
\item A.~Cafarella, C.~Corianò and M.~Guzzi, \emph{Solving renormalization
group equations by recursion relations}, hep-ph/0209149, published
in \emph{Proceedings of the Workshop: Nonlinear Physics: Theory and
Experiment. II}, editors M.J.~Ablowitz, M.~Boiti, F.~Pempinelli,
B.~Prinari, World Scientific (2003)
\item A.~Cafarella, C.~Corianò and A.~Faraggi, \emph{Ultrahigh-energy cosmic
rays and air shower simulations: a top-bottom view}, hep-ph/0306236,
published in the proceedings of \emph{XV edizione degli Incontri sulla
Fisica delle Alte Energie, Lecce, 23-26 Aprile 2003}, Società Italiana
di Fisica (2004)
\item A.~Cafarella and C.~Corianò, \emph{Identifying string relics at Auger?},
hep-ph/0309159, published in the proceedings of the conference
\emph{String Phenomenology 2003, Durham, England, 29 July - 4 August
2003}, editors V. Sanz, S. Abel, J. Santiago and A. Faraggi, World
Scientific (2004)
\item A.~Cafarella, C.~Corianò and T.N.~Tomaras, \emph{Searching for extra
dimensions in high energy cosmic rays}, hep-ph/0410190, to
be published in the proceedings of \emph{XIII International Symposium
on Very High Energy Cosmic Ray Interactions, Pylos (Greece), 2004,
September 6 -12}
\item A.~Cafarella, C.~Corianò and T.N.~Tomaras, \emph{QCD, supersymmetry
and low energy gravity}, hep-ph/0412037, to be published in
the proceedings of \emph{6th Conference on Quark Confinement and the
Hadron Spectrum, Villasimius, Italy, 21-25 September 2004}
\end{enumerate}

\chapter*{Introduction\addcontentsline{toc}{chapter}{Introduction}\markboth{}{Introduction}}
\fancyhead[LO]{\nouppercase{Introduction}}

\subsection*{QCD at Hadron Colliders }

Quantum Chromodynamics is the accepted theory of the strong interactions.
After decades spent to test this theory and after its successfull
description of the rich and complex phenomenology of the hadronic
world, nobody seriously doubts that the simple Lagrangian which is
at the core of its formulation is the correct one.

At the same time, the wider dynamics of what has come to be known
as the \emph{Standard Model of elementary particles} has been tested
with an incredible accuracy at LEP and at the Tevatron, confirming
the validity of the formulation of the model as a local gauge theory
based on the gauge group $SU(3)\times SU(2)\times U(1)_{Y}$, where
$U(1)_{Y}$ is the abelian group describing the hypercharge. After
the incorporation of spontaneous symmetry breaking of the $SU(2)\times U(1)$
symmetry to $U(1)_{em}$ via the Higgs mechanism, the model can describe
the electroweak interactions, leaving a broken gauge theory which
is exact only in the color sector $(SU(3))$ and in the electromagnetic
$U(1)_{em}$ abelian group. The application of these field theoretical 
methods to the analysis of specific processes is 
a difficult scientific enterprise and requires the identification
of special theoretical tools, the proof of  theorems
in quantum field theory and the development of sophisticated software
-- based on theory -- for accurate predictions of the experimental observables
of a large array of processes. In this effort one discovers that
there are energy regions of the interactions where a perturbative
picture of the theory can be slowly built, thanks to the property
of asymptotic freedom, and others where this is not possible due to
a strong coupling constant. In the perturbative region, being the
theory still characterized by the property of quark confinement, a
class of theorems known as \emph{factorization theorems} allow the
definition of a calculable scheme within which to compare the theory
with the experiment. The essential ingredient of this procedure is
the possibility to separate -- in inclusive processes -- the perturbative
part of a process -- what is known as the \emph{hard scattering} -- from
the non-perturbative parts, termed \emph{parton distribution functions}.
In exclusive processes, at large momentum transfers, a similar picture
holds, and allows to describe nucleon form factors and other elastic
processes by \emph{hadronic wave functions}, which nowadays are
special cases of some non-local matrix elements on the light cone
termed \emph{nonforward parton distributions}. 

As we have just mentioned, perturbative QCD appears naturally at large
energy and momentum transfers and is light-cone dominated. Parton
distributions are non-local correlators defined for light cone separations
of the field operators and their definition intrinsically introduces
a scale. This description goes under the name of \emph{parton model}.
The changes induced on this operators due to the changes of the scale
at which they are defined can be described by renormalization group
methods. In the case of ordinary parton distributions these equations
are called DGLAP equations and they have been studied in various contexts
(for polarized and unpolarized collision processes) and at various
orders in perturbation theory. One of the objective of this thesis
is to elaborate on the phenomenology of these equations in great detail,
to next-to-leading order (NLO) (chapter \ref{chap:NLOcode}) and 
up to next-to-next-to-leading order (NNLO) in $\alpha_{s}$, the QCD strong
coupling constant (chapter \ref{chap:NNLOcode}). This perturbative order is likely
to be the {}``final state'' for the study of the DGLAP equations
in perturbative QCD, since it provides an accuracy which clearly satisfies
the requirements of precisions needed at hadron colliders. The study
of these equations is rather comprehensive, involves all the leading-twist
distributions up to NLO and is further extended to NNLO for the unpolarized
distributions. Applications of these studies include a prediction
for the total cross section for Higgs detection at the LHC at NNLO (chapter \ref{chap:Higgs}),
where the dependence on the predictions on the factorization scale
is studied in depth and an NLO application for the study of the Drell
Yan process for the PAX experiment (chapter \ref{chap:NLOcode}) at few GeV's. These applications
will all be discussed in the first three chapters of this thesis work. 
In chapter \ref{chap:kinetic} and in chapter \ref{chap:Soffer} we elaborate on a kinetic approach to the DGLAP 
evolution by introducing a Kramers-Moyal expansion of this equation after testing 
both analytically and numerically the positivity of the kernels up to NLO. 
In chapter \ref{chap:Soffer} we continue and enlarge this analysis and apply it to the study of an 
inequality due to Soffer and to its behaviour under the action of the renormalization 
group. Incidentally, we also present an ansatz for the evolution 
of nonforward parton distributions and prove the existence of 
a kinetic form in the nonsinglet case.

\subsection*{QCD in the Astroparticle Domain}

In the second part of this thesis we will try to provide some applications
of standard perturbative QCD methods in the study of the extensive
air showers which are generated by proton and neutrino primaries in
high and ultra high energy cosmic rays (UHECR). QCD enters in the
modeling of the fragmentation region and our QCD predictions are appropriately
interfaced with an air shower simulator which is well known in the
cosmic rays community, CORSIKA. 

This second part of the thesis can be read quite independently from
the first part, but has been made possible by the developments of
various computational tools described in the first part. In this second
part we will define some observables of the air showers which can
help us discriminate between standard scenarios and {}``new physics''
scenarios in the context of UHECR (chapter \ref{chap:airshowers}), having in mind the ongoing experimental
efforts of various experimental collaborations, such as Auger, to
assess the existence (or the absence) of a cutoff in the upper part
of the cosmic rays spectrum. These issues have been addressed in two
papers. In a first paper we have introduced some specific observables
for the study of the extensive air showers and modified the hadronization
codes used by the various CORSIKA routines and studied their impact
on the lateral distributions and the multiplicity distributions, while
in a second paper, on which chapter \ref{chap:MBH} is based, we have analyzed in great detail the characteristic
patterns of the showers in the context of one specific scenarios for
new physics which goes under the name of theory of \emph{extra dimensions}.
In this last case an attempt has been made to understand the so called
\emph{Centauro events} in cosmic rays physics as evaporating mini
black holes, which are predicted to form in theories with extra dimensions.
The methodology that we develope is quite general and can be used
to discern between standard and new physics scenario also in the context
of other theories, since we propose and exemplify a general strategy. This second part of the thesis is heavily
computational and the author has developed software for the complex
analysis of the data generated by the simulator. In particular, the
modeling of the hadronization of the evaporating mini black hole has
been developed using a basic physics picture of the stages appearing
in its decay and a hadronization pattern of the partonic states which
is in concordance with the {}``democratic'' coupling of gravity
to all the states of the standard model.

\chapter{Direct solution of Renormalization Group Equations of QCD in $x$-space:\\NLO implementations at leading twist\label{chap:NLOcode}}
\fancyhead[LO]{\nouppercase{Chapter 1. Direct solution of RGE of QCD in $x$-space: NLO implementations}}
\section{Objectives of the chapter and Background} 
In this chapter we formulate and implement an algorithm for the solution of 
evolution equations in QCD which has some specific peculiarities 
compared to other methods based on the use of Mellin moments or of Laguerre 
polynomials. 
We present the motivation of this work and discuss the implementation 
of the method up to next to leading order (NLO) in $\alpha_s$, the QCD strong coupling constant. We also briefly discuss some of the technical aspects of the 
derivation of the recursion relations which allow to rewrite an operatorial 
solution of these equations in terms of some special functions determined 
recursively. The method is implemented numerically and a brief discussion 
of the structure of the program is included.
The content of this chapter is based on the original research article \cite{CPC}. 
In the final part of this chapter we provide an application of the results on the 
evolution of the transverse spin distributions to the case of Drell-Yan lepton pair 
production in the few GeV region, near the $J/\psi$ resonance, up to NLO in $\alpha_s$, 
where an enhancement is expected.  
Predictions for the transverse spin asymmetries of this process are presented.  

\subsection{Factorization and Evolution} 
Factorization theorems play a crucial role in the application of perturbation theory to hadronic 
reactions. The proof of these theorems and the actual implementation of their implications has spanned a 
long time and has allowed to put the parton model under a stringent experimental test. Prior to embark on 
the discussion of our contributions to the study of evolution algorithms in Bjorken $x$-space, 
we provide here a brief background on the topic in order to make our treatment self-contained.

In sufficiently inclusive cross sections, leading power factorization theorems allow to write down a 
hadronic cross section in terms of parton distributions and of some hard scatterings, the latter being 
calculable at a given order in perturbation theory using the fundamental QCD Lagrangian. 
Specifically, for a hadronic cross section, for instance  a proton-proton cross section $\sigma_{pp}$, 
the result of the calculation can be summarized by the formula 
\ba
\sigma_{pp} =\sum_{f} \int^{1}_{0} dx_{1} \int^{1}_{0} dx_{2} f_{h_{1}\rightarrow f}(x_{1},Q^{2}) f_{h_{2}\rightarrow f}(x_{2},Q^{2})\hat{\sigma }(x_{1},x_{2},\hat{s},\hat{t},Q^2)\,,
\label{pr}
\ea
where the integral is a function of some variables $x_1$ and $x_2$ which describe the QCD dynamics at parton level in the Deep Inelastic Scattering (DIS) limit or, equivalently, at large energy and momentum transfers. 
These variables are termed \emph{Bjorken variables} and are scale invariant. 
 This formula is a statement about the computability of a hadronic collisions in terms of 
some ``building blocks'' of easier definition.

The variable $Q^2$, in the equation above, can be identified with the factorization scale of the process. $\hat{\sigma}$ can be computed at a given order in perturbation theory in an expansion  in $\alpha_s$, 
while the $f_{h_i\to f}(x, Q^2)$ are the parton distributions. These describe the probability for a hadron $h$ to prepare for the scattering 
a parton $f$, which undergoes the collision. An equivalent interpretation of the functions 
$f_{h_i\to f}(x, Q^2)$ is to characterize the density of partons of type $f$ into a hadron of type $h$. 
A familiar notation, which simplifies the previous notations shown above, 
is to denote by $q_i(x,Q^2)$ the density of quarks in a hadron (a proton in this case) 
of flavour $i$ and by $g(x,Q^2)$ the corresponding density of gluons. For instance, the annihilation 
channel of a quark with an antiquark in a generic process is accounted for by the contribution 
\be
\int_0^1 dx_1 \int_0^1 dx_2 q(x_1,Q^2) \overline{q}(x_2,Q^2) \hat{\sigma}_{q \overline{q}}(x_1,x_2,Q^2) 
\ee
 and so on for the other contributions, such as the quark-gluon sector ($qg$) or the gluon-gluon sector 
($gg$) each of them characterized by a specific hard scattering cross section $\hat{\sigma}_{qg}$,  or 
$\hat{\sigma}_{qg}$, and so on. In this separation of the cross section into contributions of hard scatterings 
$\hat{\sigma}$ and parton distributions $f(x,Q^2)$ the scale at which the separation occurs is, in a way 
artificial, in the sense that the hadronic cross section $\sigma$ should not depend on $Q^2$ or 
\be
\frac{d \sigma}{d Q^2} =0.
 \ee
However, a perturbative computation performed by using the factorization formula, however, shows that this 
is not the case, since the perturbative expansion of $\hat{\sigma}$
\be
\hat{\sigma}= \hat{\sigma}^{(0)} +\alpha_s(Q^2)^2 \hat{\sigma}^{(1)}+ \alpha_s(Q^2)\hat{\sigma}^{(2)}\qquad
\ee 
naturally brings in the dependence on the factorization scale $Q$. This dependence is weaker if we are able 
to push our computation of the hard scattering to a sufficiently high order in $\alpha_s$. The order at which the perturbative expansion stops is also classified as a ``leading order'' (LO), 
``next-to-leading order'' (NLO) or, even better, a ``next-to-next-to-leading'' (NNLO) 
contribution if more and more terms in the expansion are included, 

At the same time, the parton distributions $f(x,Q^2)$ carry a similar dependence on the 
scale $Q$, which is summarized in a renormalization group equation (RGE) which resums the logarithmic 
violations to the scaling behaviour induced by the perturbative expansion.
Also in this case we need to quantify this effect and reduce its impact on the prediction 
of the cross section. The topic of this chapter, therefore, is about the quantification of this effect and the 
implementation of an algorithm which reorganizes the solutions of these RGE's in a suitable way to render 
the numerical implementation quick and accurate up to NLO. This strategy is fully worked out for all the leading twist distributions, with and without polarization. The results of these chapter can be applied for 
the precise determination of both polarized and polarized cross sections in pp collisions.      
In chapter \ref{chap:Higgs} this methodology will be pushed even further, to NNLO, which is the state of the art in QCD 
and will be applied to the case of the Higgs croiss section, which is important for the discovery of this 
particle at the LHC.

\section{Parton Dynamics at NLO: a short overview}
Our understanding of the QCD dynamics has improved steadily along the years. 
In fact we can claim that precision studies of the QCD background in a variety of 
energy ranges, from intermediate to high energy -- 
whenever perturbation theory and factorization theorems hold -- 
are now possible and the level of accuracy reached in this area is due both 
to theoretical improvements and to the flexible numerical implementations 
of many of the algorithms developed at theory level.  

Beside their importance in the determination of the QCD background in the search of new physics, 
these studies convey an understanding of the structure of the nucleon from first principles, 
an effort which is engaging both theoretically and experimentally.  

It should be clear, however, that perturbative methods 
are still bound to approximations, from the order of the perturbative 
expansion to the phenomenological description of the parton distribution functions, being them related to a parton model view of the QCD dynamics.

Within the framework of the parton model, evolution equations of DGLAP-type 
-- and the corresponding initial conditions on the parton distributions --
are among the most important 
blocks which characterize the description of
the quark-gluon interaction and, as such, deserve continuous attention.
Other parts of this description require the computation of hard scatterings 
with high accuracy and an understanding of the fragmentation region as well.
The huge success of the parton model justifies all the effort. 

In this consolidated program, we believe that any attempt to study the renormalization
group evolution describing the perturbative change of the 
distributions with energy, from a different -- in this case, numerical -- standpoint is of interest.

In this chapter we illustrate an algorithm based on the
use of recursion relations for the solution of evolution equations
of DGLAP type. We are going to discuss some of the salient features 
of the method and illustrate 
its detailed implementation as a computer code 
up to next-to-leading order in $\alpha_s$. 

In this context, it is worth to recall that the most common method 
implemented so far in the solution of the DGLAP equations is the one based
on the Mellin moments, with all its good points and limitations. 

The reason for such limitations are that, while it is rather straightforward to solve
the equations in the space of moments, 
their correct inversion is harder to perform, since this 
requires the computation of an integral 
in the complex plane and the search for an optimized path.

In this respect, several alternative implementations of the NLO 
evolution are available from the previous literature, either based on
the use of {}``brute force'' algorithms \cite{bruteforce} or 
on the Laguerre expansion \cite{FurmanskiPetronzio,CorianoSavkli}, all with their positive features and their limitations. 

Here we document an implementation to NLO of a method based on an
ansatz \cite{Rossi} which allows to rewrite the evolution
equations as a set of recursion relations for some scale invariant
functions, \( A_{n}(x) \) and \( B_{n}(x) \), which appear in the
expansion. The advantage, compared to others, 
of using these recursion relations is that just few iterates 
of these are necessary in order to obtain a stable solution. 
The number of iterates is determined at run time.
We also mention that 
our implementation can be extended to more complex cases, including 
the cases of nonforward parton distributions and of supersymmetry. Here 
we have implemented the recursion method in all the important cases of 
initial state scaling violations connected to the evolutions of both 
polarized and unpolarized parton densities, including the less known transverse spin distributions.

 Part of this chapter is supposed to illustrate the algorithm, having 
worked out in some detail the structure of the recursion relations rather explicitly, 
especially in the case of nonsinglet evolutions, such as for transverse 
spin distributions. 

One of the advantages of the method is its analytical base,  
since the recursion relations 
can be written down in explicit form and at the same time 
one can perform a simple analytical matching between various regions in the 
evolutions, which is a good feature of $x$-spaced methods. 
While this last aspect is not 
relevant for ordinary QCD, it is relevant in other theories, such as 
for supersymmetric extensions of the parton model, once one assumes 
that, as the evolution scale $Q$ raises, new anomalous dimensions are needed 
in order to describe the mixing between ordinary and supersymmetric partons.

\section{Definitions and Conventions}

In this section we briefly outline our definitions and conventions.

We will be using the running of the coupling constant up to two-loop level\begin{equation}
\label{eq:alpha_s}
\alpha _{s}(Q^{2})=\frac{4\pi }{\beta _{0}}\frac{1}{\log (Q^{2}/\Lambda _{\overline{MS}}^{2})}\left[ 1-\frac{\beta _{1}}{\beta ^{2}_{0}}\frac{\log \log (Q^{2}/\Lambda _{\overline{MS}}^{2})}{\log (Q^{2}/\Lambda _{\overline{MS}}^{2})}+O\left( \frac{1}{\log ^{2}(Q^{2}/\Lambda _{\overline{MS}}^{2})}\right) \right] ,
\end{equation}
where\begin{equation}
\beta _{0}=\frac{11}{3}N_{C}-\frac{4}{3}T_{f},\qquad \beta _{1}=\frac{34}{3}N^{2}_{C}-\frac{10}{3}N_{C}n_{f}-2C_{F}n_{f},
\end{equation}
and\begin{equation}
N_{C}=3,\qquad C_{F}=\frac{N_{C}^{2}-1}{2N_{C}}=\frac{4}{3},\qquad T_{f}=T_{R}n_{f}=\frac{1}{2}n_{f},
\end{equation}
where \( N_{C} \) is the number of colors, \( n_{f} \) is the number
of active flavors, which is fixed by the number of quarks with \( m_{q}\leq Q \).
We have taken for the quark masses 
$ m_{c}=1.5\, \textrm{GeV}, m_{b}=4.5\, \textrm{GeV}$ 
and $ m_{t}=175\, \textrm{GeV} $, these are necessary in order 
to identify the thresholds at which the number of flavours $n_f$ 
is raised as we increase the final evolution scale.

 In our conventions 
$\Lambda_{QCD}$ is denoted by $\Lambda _{\overline{MS}}^{(n_{f})}$
and is given by

\begin{equation}
\Lambda _{\overline{MS}}^{(3,4,5,6)}=0.248,\, 0.200,\, 0.131,\, 0.050\, \textrm{GeV}.
\end{equation}

We also define the distribution of a given helicity $(\pm)$,
\( f^{\pm }(x,Q^{2}) \), which is the probability 
of finding a parton of type \( f \) at a scale \( Q \), 
where \( f=q_{i},\overline{q}_{i},g \),
in a longitudinally polarized proton with the spin aligned ($+$) 
or anti-aligned ($-$) respect 
to the proton spin and carrying a fraction \( x \) of the proton's
momentum. 

As usual, we introduce the longitudinally polarized parton
distribution of the proton\begin{equation}
\label{eq:long_distr}
\Delta f(x,Q^{2})\equiv f^{+}(x,Q^{2})-f^{-}(x,Q^{2}).
\end{equation}
We also introduce another type of parton density, termed 
{\em transverse spin 
distribution}, which is defined as 
the probability of finding a parton of type \( f \) in a transversely
polarized proton with its spin parallel ($\uparrow$) minus the probability 
of finding it antiparallel ($\downarrow$) to the proton
spin
\begin{equation}
\Delta _{T}f(x,Q^{2})\equiv f^{\uparrow }(x,Q^{2})-f^{\downarrow }(x,Q^{2}).
\end{equation}

Similarly, the unpolarized (spin averaged) parton distribution of the proton
is given by
\begin{equation}
\label{eq:unp_distr}
f(x,Q^{2})\equiv f^{+}(x,Q^{2})+f^{-}(x,Q^{2})=f^{\uparrow }(x,Q^{2})+f^{\downarrow }(x,Q^{2}).
\end{equation}
We also recall that 
taking linear combinations of Equations (\ref{eq:unp_distr}) and
(\ref{eq:long_distr}), one recovers 
the parton distributions of a given helicity 
\begin{equation}
f^{\pm }(x,Q^{2})=\frac{f(x,Q^{2})\pm \Delta f(x,Q^{2})}{2}.
\end{equation}
In regard to the kernels, the notations \( P \), \( \Delta P \), \( \Delta _{T}P \),
\( P^{\pm} \),  will be used to denote the Altarelli-Parisi
kernels in the unpolarized, longitudinally polarized, transversely
polarized, and the positive (negative) helicity cases respectively.

The DGLAP equation is an integro-differential equation whose general
mathematical structure is

\begin{equation}
\frac{\textrm{d}}{\textrm{d}\log Q^{2}}f(x,Q^{2})=P(x,\alpha_{s}(Q^{2}))\otimes f(x,Q^{2}),\label{eq:DGLAP}\end{equation}
where the convolution product is defined by\begin{equation}
\left[a\otimes b\right](x)=\int_{x}^{1}\frac{\textrm{d}y}{y}a\left(\frac{x}{y}\right)b(y)=\int_{x}^{1}\frac{\textrm{d}y}{y}a(y)b\left(\frac{x}{y}\right).\end{equation}

Let us now turn to the evolution equations, starting from the unpolarized
case. Defining\begin{equation}
\label{eq:definizioni}
q_{i}^{(\pm )}=q_{i}\pm \overline{q}_{i},\qquad q^{(+)}=\sum _{i=1}^{n_{f}}q_{i}^{(+)},\qquad \chi _{i}=q_{i}^{(+)}-\frac{1}{n_{f}}q^{(+)},
\end{equation}
the evolution equations are  
\begin{equation}
\frac{\textrm{d}}{\textrm{d}\log Q^{2}}q_{i}^{(-)}(x,Q^{2})=P_{NS^{-}}(x,\alpha _{s}(Q^{2}))\otimes q_{i}^{(-)}(x,Q^{2}),
\end{equation}
\begin{equation}
\frac{\textrm{d}}{\textrm{d}\log Q^{2}}\chi _{i}(x,Q^{2})=P_{NS^{+}}(x,\alpha _{s}(Q^{2}))\otimes \chi _{i}(x,Q^{2}),
\end{equation}
for the nonsinglet sector and
\begin{equation}
\label{eq:singlet}
\frac{\textrm{d}}{\textrm{d}\log Q^{2}}\left( \begin{array}{c}
q^{(+)}(x,Q^{2})\\
g(x,Q^{2})
\end{array}\right) =\left( \begin{array}{cc}
P_{qq}(x,\alpha _{s}(Q^{2})) & P_{qg}(x,\alpha _{s}(Q^{2}))\\
P_{gq}(x,\alpha _{s}(Q^{2})) & P_{gg}(x,\alpha _{s}(Q^{2}))
\end{array}\right) \otimes \left( \begin{array}{c}
q^{(+)}(x,Q^{2})\\
g(x,Q^{2})
\end{array}\right) 
\end{equation}
for the singlet sector.

Equations analogous to (\ref{eq:definizioni} -- \ref{eq:singlet}),
with just a change of notation, are valid in the longitudinally polarized
case and, due to the linearity of the evolution equations, also for
the distributions in the helicity basis. In the transverse case instead,
there is no coupling between gluons and quarks, so the singlet sector
(\ref{eq:singlet}) is missing. In this case we will solve just the
nonsinglet equations \begin{equation}
\frac{\textrm{d}}{\textrm{d}\log Q^{2}}\Delta _{T}q_{i}^{(-)}(x,Q^{2})=\Delta _{T}P_{NS^{-}}(x,\alpha _{s}(Q^{2}))\otimes \Delta _{T}q_{i}^{(-)}(x,Q^{2}),
\end{equation}
\begin{equation}
\frac{\textrm{d}}{\textrm{d}\log Q^{2}}\Delta _{T}q_{i}^{(+)}(x,Q^{2})=\Delta _{T}P_{NS^{+}}(x,\alpha _{s}(Q^{2}))\otimes \Delta _{T}q_{i}^{(+)}(x,Q^{2}).
\end{equation}

We also recall that the perturbative expansion, up to next-to-leading order, of the kernels
is\begin{equation}
P(x,\alpha _{s})=\left( \frac{\alpha _{s}}{2\pi }\right) P^{(0)}(x)+\left( \frac{\alpha _{s}}{2\pi }\right) ^{2}P^{(1)}(x)+\ldots .
\end{equation}

\section{Mathematical structure of the kernel}

Altarelli-Parisi kernels are defined as distributions. The most general
form is the following\begin{equation}
P(x)=P_{1}(x)+\frac{P_{2}(x)}{(1-x)_{+}}+P_{3}\delta(1-x),\label{eq:general_kernel}\end{equation}
with a regular part $P_{1}(x)$, a {}``plus distribution'' part
$P_{2}(x)/(1-x)_{+}$ and a delta distribution part $P_{3}\delta(1-x)$.
For a generic function $\alpha(x)$ defined in the $[0,1)$ interval
and singular in $x=1$, the plus distribution $[\alpha(x)]_{+}$ is
defined by\begin{equation}
\int_{0}^{1}f(x)[\alpha(x)]_{+}\textrm{d}x=\int_{0}^{1}\left(f(x)-f(1)\right)\alpha(x)\textrm{d}x,\end{equation}
where $f(x)$ is a regular test function. Alternatively, an operative
definition (that assumes full mathematical meaning only when integrated)
is the following\begin{equation}
[\alpha(x)]_{+}=\alpha(x)-\delta(1-x)\int_{0}^{1}\alpha(y)\textrm{d}y.\label{eq:def_plus}\end{equation}
From (\ref{eq:def_plus}) it follows immediately that each plus distribution
integrate to zero in the $[0,1]$ interval\begin{equation}
\int_{0}^{1}[\alpha(x)]_{+}\textrm{d}x=0.\end{equation}

\begin{framed}

\textbf{\textit{Convolution of a generic kernel.}}

We want to make the convolution of the generic kernel (\ref{eq:general_kernel})
with a function $f(x)$. To improve numerical stability, in the code
we indeed multiply the functions to be convolved with the kernel (the
coefficients $A_{n}$ and $B_{n}$) by an $x$ factor. To
compute the formulas that we implement in the code, we introduce the
notation $\bar{f}(x)=xf(x)$. The treatment of the regular and the
delta-function parts is trivial\begin{equation}
P_{1}(x)\otimes\bar{f}(x)=xP_{1}(x)\otimes f(x)=x\int_{x}^{1}\frac{\textrm{d}y}{y}P_{1}(y)f\left(\frac{x}{y}\right)=\int_{x}^{1}\textrm{d}yP_{1}(y)\bar{f}\left(\frac{x}{y}\right)\label{eq:conv_1}\end{equation}
\begin{equation}
P_{3}\delta(1-x)\otimes\bar{f}(x)=\int_{x}^{1}\frac{\textrm{d}y}{y}P_{3}\delta(1-y)\bar{f}\left(\frac{x}{y}\right).\label{eq:conv_3}\end{equation}

Let us now treat the more involved case of the plus distribution part\begin{eqnarray}
\frac{P_{2}(x)}{(1-x)_{+}}\otimes f(x) & = & \frac{P_{2}(x)}{1-x}\otimes f(x)-\left(\int_{0}^{1}\frac{\textrm{d}y}{1-y}\right)P_{2}(x)\delta(1-x)\otimes f(x)\nonumber \\
 & = & \int_{x}^{1}\frac{\textrm{d}y}{y}\frac{P_{2}(y)}{1-y}f\left(\frac{x}{y}\right)-\int_{0}^{1}\frac{\textrm{d}y}{1-y}\int_{x}^{1}\frac{\textrm{d}y}{y}P_{2}(y)\delta(1-y)f\left(\frac{x}{y}\right)\nonumber \\
 & = & \int_{x}^{1}\frac{\textrm{d}y}{y}\frac{P_{2}(y)}{1-y}f\left(\frac{x}{y}\right)-P_{2}(1)f(x)\int_{0}^{1}\frac{\textrm{d}y}{1-y}\nonumber \\
 & = & \int_{x}^{1}\frac{\textrm{d}y}{y}\frac{P_{2}(y)}{1-y}f\left(\frac{x}{y}\right)-P_{2}(1)f(x)\int_{x}^{1}\frac{\textrm{d}y}{1-y}\nonumber \\
 &  & \qquad\qquad\qquad\qquad\,\,\,-P_{2}(1)f(x)\int_{0}^{x}\frac{\textrm{d}y}{1-y}\\
 & = & \int_{x}^{1}\frac{\textrm{d}y}{y}\frac{P_{2}(y)f(x/y)-yP_{2}(1)f(x)}{1-y}+f(x)\log(1-x),\end{eqnarray}
which yields\begin{equation}
\frac{P_{2}(x)}{(1-x)_{+}}\otimes\bar{f}(x)=\int_{x}^{1}\textrm{d}y\frac{P_{2}(y)\bar{f}(x/y)-P_{2}(1)\bar{f}(x)}{1-y}+\bar{f}(x)\log(1-x).\label{eq:conv_2}\end{equation}

\end{framed}

\begin{framed}
\textbf{\textit{Mellin Moments}}

The $n$-th Mellin moment of a function of the Bjorken variable $f(x)$
is defined by\begin{equation}
f_{n}=\int_{0}^{1}x^{n-1}f(x)\textrm{d}x.\end{equation}
An important property of Mellin moments is that the Mellin moment
of the convolution of two functions is equal to the product of the
individual Mellin moments\begin{equation}
\left[f\otimes g\right]_{n}=f_{n}g_{n}.\label{eq:Mellin_product}\end{equation}
Let us prove it.\begin{equation}
\left[f\otimes g\right]_{n}=\int_{0}^{1}\textrm{d}x\, x^{n-1}\left[f\otimes g\right](x)=\int_{0}^{1}\textrm{d}x\, x^{n-1}\int_{x}^{1}\frac{\textrm{d}y}{y}f(y)g\left(\frac{x}{y}\right).\end{equation}
Exchanging the $x$ and $y$ integrations \begin{equation}
\left[f\otimes g\right]_{n}=\int_{0}^{1}\textrm{d}y\, f(y)\int_{0}^{y}\frac{\textrm{d}x}{y}x^{n-1}g\left(\frac{x}{y}\right),\end{equation}
and introducing the new variable $z=x/y$\begin{equation}
\left[f\otimes g\right]_{n}=\int_{0}^{1}\textrm{d}y\, f(y)y^{n-1}\int_{0}^{1}\textrm{d}z\, z^{n-1}g(z)=f_{n}g_{n}.\end{equation}

This leads to an alternative formulation of DGLAP equation, that is
also the most widely used to numerically solve the evolution equations.
By taking the first Mellin moment of both sides of the integro-differential
equation (\ref{eq:DGLAP}) we are left with the differential equation\begin{equation}
\frac{\textrm{d}}{\textrm{d}\log Q^{2}}f_{1}(Q^{2})=P_{1}(Q^{2})f_{1}(Q^{2})\end{equation}
that can be easily solved to give\begin{equation}
f_{1}(Q^{2})=\int_{0}^{1}f(x,Q^{2})\textrm{d}x.\end{equation}
To get the desired solution $f(x,Q^{2})$ there is a last step, the
inverse Mellin transform of the first moment of the parton distributions,
involving a numerical integration on the complex plane. This is the
most difficult (and time-consuming) task that the algorithms of solution
of DGLAP equation based on Mellin transformation -- by far the most
widely used -- must accomplish. But the most severe restriction of
the flexibility of such evolution codes (for example QCD-Pegasus by
Vogt \cite{QCDPegasus}) as compared to $x$-space methods is that
the initial distribution are needed in a form which facilitates the
inversion of the moments, i.e.~a functional form; but very often
parton sets are given in discrete grids of $x$ and $Q$ values.
\end{framed}

From Feynman diagrams calculations one can get just the regular part
$P_{1}(x)$ of each kernel. The remaining distributional parts (plus
distribution and delta distribution) emerge from a procedure of regularization,
that introduce the plus distribution part to regularize the eventual
singularity in $x=1$ and the delta distribution to fulfill some physical
constraints, the \emph{sum rules}.

The first one is the \emph{baryon number sum rule} (BNSR), asserting
that the baryon number (number of quarks less number of antiquarks)
of the hadron must remain equal to its initial value (3 in the case
of the proton) throughout the evolution, i.e.~for each value of $Q^{2}$\begin{equation}
q_{1}^{(-)}(Q^{2})=\int_{0}^{1}q^{(-)}(x,Q^{2})\textrm{d}x=3.\label{eq:BNSR}\end{equation}
Deriving (\ref{eq:BNSR}) with respect to $\log Q^{2}$ and having
in mind that $q^{(-)}$ evolves with $P_{NS}^{V}$, we get\begin{equation}
\int_{0}^{1}\textrm{d}x\left[P_{NS}^{V}(Q^{2})\otimes q^{(-)}(Q^{2})\right](x)=0.\end{equation}
Making use of the property of the Mellin moment of a convolution (\ref{eq:Mellin_product})
this implies\begin{equation}
\left(\int_{0}^{1}P_{NS}^{V}(x,Q^{2})\textrm{d}x\right)\left(\int_{0}^{1}q^{(-)}(x,Q^{2})\textrm{d}x\right)=0,\end{equation}
from which, using (\ref{eq:BNSR}), we find the BNSR condition on
the kernel\begin{equation}
\int_{0}^{1}P_{NS}^{V}(x)\textrm{d}x=0.\label{eq:BNSR_kernel}\end{equation}

The other constraint is the \emph{momentum sum rule} (MSR), asserting
that the total momentum of the hadron is constant throughout the evolution.
Having in mind that $x$ is the fraction of momentum carried out by
each parton, this concept is translated by the relation\begin{equation}
\int_{0}^{1}\left(xq^{(+)}(x,Q^{2})+xg(x,Q^{2})\right)\textrm{d}x=1\label{eq:MSR}\end{equation}
that must hold for each value of $Q^{2}$. Deriving with respect to
$\log Q^{2}$ and using the singlet DGLAP equation\begin{eqnarray}
\int_{0}^{1}\textrm{d}x\, x\left\{ \left[P_{qq}(Q^{2})\otimes q^{(+)}(Q^{2})\right](x)+\left[P_{qg}(Q^{2})\otimes g(Q^{2})\right](x)\right.\nonumber \\
\left.+\left[P_{gq}(Q^{2})\otimes q^{(+)}(Q^{2})\right](x)+\left[P_{gg}(Q^{2})\otimes g(Q^{2})\right](x)\right\}  & = & 0.\end{eqnarray}
Using (\ref{eq:Mellin_product}) we get\begin{eqnarray}
\left[\int_{0}^{1}x\left(P_{qq}(x,Q^{2})+P_{gq}(x,Q^{2})\right)\textrm{d}x\right]\left[\int_{0}^{1}xq^{(+)}(x,Q^{2})\textrm{d}x\right]\nonumber \\
+\left[\int_{0}^{1}x\left(P_{qg}(x,Q^{2})+P_{gg}(x,Q^{2})\right)\textrm{d}x\right]\left[\int_{0}^{1}xg(x,Q^{2})\textrm{d}x\right] & = & 0,\end{eqnarray}
from which we find the MSR conditions on the singlet kernels\begin{equation}
\int_{0}^{1}x\left(P_{qq}(x,Q^{2})+P_{gq}(x,Q^{2})\right)\textrm{d}x=0,\label{eq:MSR1_kernel}\end{equation}
\begin{equation}
\int_{0}^{1}x\left(P_{qg}(x,Q^{2})+P_{gg}(x,Q^{2})\right)\textrm{d}x=0.\label{eq:MSR2_kernel}\end{equation}

\begin{framed}

\textbf{\textit{The kernels regularization procedure: an example.}}

We illustrate now an example of the regularization procedure of the
Altarelli-Parisi kernels through the sum rules. The LO kernels computed
by diagrammatic techniques for $x<1$ are\begin{equation}
P_{qq}^{(0)}(x)=P_{NS}^{(0)}(x)=C_{F}\left[\frac{1+x^{2}}{1-x}\right]=C_{F}\left[\frac{2}{1-x}-1-x\right]\end{equation}
\begin{equation}
P_{qg}^{(0)}(x)=2T_{f}\left[x^{2}+(1-x)^{2}\right]\end{equation}
\begin{equation}
P_{gq}^{(0)}(x)=C_{F}\left[\frac{1+(1-x)^{2}}{x}\right]\end{equation}
\begin{equation}
P_{gg}^{(0)}(x)=2N_{C}\left[\frac{1}{1-x}+\frac{1}{x}-2+x(1-x)\right].\end{equation}
We want to analytically continue this kernels to $x=1$ curing the
ultraviolet singularities in $P_{qq}^{(0)}(x)$ and $P_{gg}^{(0)}(x)$.
We start introducing the plus distribution prescription in $P_{qq}^{(0)}(x)$.
We make the replacement\begin{equation}
\frac{1}{1-x}\longrightarrow\frac{1}{(1-x)_{+}}\end{equation}
to avoid the singularity and we add a term $k\delta(1-x)$ (where
$k$ has to be determined) to fulfill the BNSR (\ref{eq:BNSR_kernel}).
So we have\begin{equation}
P_{qq}^{(0)}(x)\longrightarrow C_{F}\left[\frac{2}{(1-x)_{+}}-1-x+k\delta(1-x)\right].\end{equation}
Imposing by the BNSR that $P_{qq}^{(0)}(x)$ integrates to zero in
$[0,1]$ and remembering that the plus distribution integrates to
zero we get\begin{equation}
\int_{0}^{1}P_{qq}^{(0)}(x)\textrm{d}x=C_{F}\left[-1-\frac{1}{2}+k\right]=0,\end{equation}
hence $k=3/2$, and the regularized form of the kernel is\begin{equation}
P_{qq}^{(0)}(x)=C_{F}\left[\frac{2}{(1-x)_{+}}-1-x+\frac{3}{2}\delta(1-x)\right].\end{equation}
Noticing that\begin{equation}
\int_{0}^{1}\frac{x}{(1-x)_{+}}\textrm{d}x=\int_{0}^{1}\frac{x-1+1}{(1-x)_{+}}\textrm{d}x=\int_{0}^{1}\left(-1+\frac{1}{(1-x)_{+}}\right)\textrm{d}x=-1\end{equation}
it can be easily proved that the MSR (\ref{eq:MSR1_kernel}) is satisfied.
Let us now regularize $P_{gg}(x)$. We make the replacement\begin{equation}
P_{gg}^{(0)}(x)\longrightarrow2N_{C}\left[\frac{1}{(1-x)_{+}}+\frac{1}{x}-2+x(1-x)\right]+k\delta(1-x).\end{equation}
 Imposing the other MSR (\ref{eq:MSR2_kernel}) we get\begin{eqnarray}
\int_{0}^{1}\left\{ 2N_{C}\left[\frac{x}{(1-x)_{+}}+1-2x+x^{2}(1-x)\right]+kx\delta(1-x)\right.\nonumber \\
\left.+2T_{f}\left[x^{3}+x(1-x)^{2}\right]\right\} \textrm{d}x & = & 0,\end{eqnarray}
from which we find\begin{equation}
k=\frac{11}{6}N_{C}-\frac{2}{3}T_{f}=\frac{\beta_{0}}{2},\end{equation}
so the regularized form of the kernel is\begin{equation}
P_{gg}^{(0)}(x)=2N_{C}\left[\frac{1}{(1-x)_{+}}+\frac{1}{x}-2+x(1-x)\right]+\frac{\beta_{0}}{2}\delta(1-x).\end{equation}

\end{framed}

\section{The Ansatz and some Examples}

In order to solve the evolution equations directly in \( x \)-space,
we assume solutions of the form\begin{equation}
\label{eq:ansatz}
f(x,Q^{2})=\sum _{n=0}^{\infty }\frac{A_{n}(x)}{n!}\log ^{n}\frac{\alpha _{s}(Q^{2})}{\alpha _{s}(Q_{0}^{2})}+\alpha _{s}(Q^{2})\sum _{n=0}^{\infty }\frac{B_{n}(x)}{n!}\log ^{n}\frac{\alpha _{s}(Q^{2})}{\alpha _{s}(Q_{0}^{2})},
\label{rossis}
\end{equation}
for each parton distribution \( f \), where $Q_0$ defines the initial 
evolution scale. The justification of this ansatz can be found, 
at least in the case of the photon structure function, 
in the original work of Rossi \cite{Rossi}, and its connection 
to the ordinary solutions of the DGLAP equations is most easily 
worked out by taking moments of the scale invariant coefficient 
functions $A_n$ and $B_n$ and comparing them to 
the corresponding moments 
of the parton distributions, as we are going to illustrate 
in section \ref{sec:moments}. The link between Rossi's expansion 
and the solution of the evolution equations 
(which are ordinary differential equations) in the space 
of the moments up to NLO will be discussed in that section, from which 
it will be clear that 
Rossi's ansatz involves a resummation 
of the ordinary Mellin moments of the parton distributions.   

Setting \( Q=Q_{0} \) in (\ref{eq:ansatz})
we get\begin{equation}
\label{eq:boundary}
f(x,Q_{0}^{2})=A_{0}(x)+\alpha _{s}(Q_{0}^{2})B_{0}(x).
\end{equation}
Inserting (\ref{eq:ansatz}) in the evolution equations, we obtain
the following recursion relations for the coefficients \( A_{n} \)
and \( B_{n} \) 
\begin{equation}
\label{eq:An_recurrence}
A_{n+1}(x)=-\frac{2}{\beta _{0}}P^{(0)}(x)\otimes A_{n}(x),
\end{equation}

\begin{equation}
B_{n+1}(x)=-B_{n}(x)-\frac{\beta_{1}}{4\pi\beta_{0}}A_{n+1}(x)-\frac{2}{\beta_{0}}P^{(0)}(x)\otimes B_{n}(x)-\frac{1}{\pi\beta_{0}}P^{(1)}(x)\otimes A_{n}(x) 
\label{eq:Bn_recurrence}
\end{equation}
obtained by equating left-hand sides and right-hand-side of the equation 
of the same logarithmic power 
in $\log^n\alpha_s(Q^2)$ and $\alpha_s \log^n \alpha_s(Q^2)$. 
Any boundary condition satisfying (\ref{eq:boundary}) can be chosen at the lowest scale $Q_0$ and in our case we choose\begin{equation}
\label{eq:initial}
B_{0}(x)=0,\qquad f(x,Q_{0}^{2})=A_{0}(x).
\end{equation}

The actual implementation of the recursion relations is the main effort 
in the actual writing of the code. Obviously, 
this requires particular care in the handling of 
the singularities in $x$-space, being all the kernels defined as 
distributions. Since the distributions are integrated, 
there are various ways to render the integrals finite, 
as discussed in the previous literature on the method \cite{Storrow} 
in the case of the photon structure function. 
In these previous studies the 
edge-point contributions -- i.e.~the terms which multiply $\delta(1-x)$ 
in the kernels -- are approximated using a sequence of functions 
converging to the $\delta$ function in a distributional sense.

This technique is not very efficient. We think that 
the best way to proceed is to actually perform the integrals explicitly in the recursion relations and let the subtracting 
terms appear under the same integral together with the 
bulk contributions ($x<1$) (see also \cite{Gordon}). This procedure is best exemplified 
by the integral relation 
\begin{equation}
\int_{x}^{1}\frac{\textrm{d}y}{y}\frac{1}{(1-y)_{+}}f\left(\frac{x}{y}\right)=\int_{x}^{1}\frac{\textrm{d}y}{y}\frac{f(x/y)-yf(x)}{1-y}+f(x)\log(1-x)\end{equation}in which, on the right hand side, regularity of both the first 
and the second term is explicit. For instance, the singlet evolution equations become 
in the unpolarized 
case
\begin{eqnarray}
\frac{\textrm{d}q^{(+)}(x)}{\textrm{d}\log Q^{2}} & = & C_{F}\left[2\int_{x}^{1}\frac{\textrm{d}y}{y}\frac{q^{(+)}(x/y)-yq^{(+)}(x)}{1-y}\right.\nonumber \\
 &  & \left.\qquad+2q^{(+)}(x)\log(1-x)-\int_{x}^{1}\frac{\textrm{d}y}{y}(1+y)q^{(+)}\left(\frac{x}{y}\right)+\frac{3}{2}q^{(+)}(x)\right]\nonumber \\
 &  & +2T_{R}n_{f}\int_{x}^{1}\frac{\textrm{d}y}{y}\left[y^{2}+(1-y)^{2}\right]g\left(\frac{x}{y}\right)\end{eqnarray}
\begin{eqnarray}
\frac{\textrm{d}g(x)}{\textrm{d}\log Q^{2}} & = & C_{F}\int_{x}^{1}\frac{\textrm{d}y}{y}\frac{1+(1-y)^{2}}{y}q^{(+)}\left(\frac{x}{y}\right)\nonumber \\
 &  & +2N_{C}\left[\int_{x}^{1}\frac{\textrm{d}y}{y}\frac{g(x/y)-yg(x)}{1-y}+g(x)\log(1-x)\right.\nonumber \\
 &  & \left.\qquad\quad+\int_{x}^{1}\frac{\textrm{d}y}{y}\left(\frac{1}{y}-2+y(1-y)\right)g\left(\frac{x}{y}\right)\right]+\frac{\beta_{0}}{2}g(x).\end{eqnarray}

\section{An Example: The Evolution of the Transverse Spin Distributions}
LO and NLO recursion relations for the coefficients of the expansion 
can be worked out quite easily. We illustrate here in detail 
the implementation of a nonsinglet evolutions, such 
as those involving transverse spin distributions. 
For the first recursion relation (\ref{eq:An_recurrence}) in this case 
we have
\beqn
&&A^{\pm}_{n+1}(x)=-\frac{2}{\beta_{0}}\Delta_{T}P^{(0)}_{qq}(x)\otimes A^{\pm}_{n}(x)=\nonumber\\ 
&&C_{F}\left(-\frac{4}{\beta_{0}}\right)\left[\int^{1}_{x}\frac{dy}{y}\frac{y A^{\pm}_{n}(y) - x A^{\pm}_{n}(x)}{y-x} + A^{\pm}_{n}(x) \log(1-x)\right]+\nonumber\\
&&C_{F}\left(\frac{4}{\beta_{0}}\right) \left(\int_{x}^{1}\frac{dy}{y} A^{\pm}_{n}(y)\right) + C_{F}\left(-\frac{2}{\beta_{0}}\right)\frac{3}{2} A^{\pm}_{n}(x)\,.
\eeqn
As we move to NLO, it is convenient to summarize 
the structure of the transverse kernel $\Delta_{T}P^{\pm, (1)}_{qq}(x)$ as  

\beqn
&&\Delta_{T}P^{\pm, (1)}_{qq}(x)= K^{\pm}_{1}(x)\delta(1-x) + K^{\pm}_{2}(x)S_{2}(x) +K^{\pm}_{3}(x)\log(x)\nonumber\\
&&+ K^{\pm}_{4}(x)\log^{2}(x) +K^{\pm}_{5}(x)\log(x)\log(1-x) + K^{\pm}_{6}(x)\frac{1}{(1-x)_{+}} + K^{\pm}_{7}(x)\,.    
\eeqn

Hence, for the $(+)$ case we have 

\beqn
&&\Delta_{T}P^{+, (1)}_{qq}(x)\otimes A^{+}_{n}(x) = K^{+}_{1} A^{+}_{n}(x) + \int^{1}_{x}\frac{dy}{y}\left[K^{+}_{2}(z) S_{2}(z) + K^{+}_{3}(z)\log(z) \right.\nonumber\\
&& \left. + \log^{2}(z)K^{+}_{4}(z) + \log(z)\log(1-z)K^{+}_{5}(z)\right] A^{+}_{n}(y) +  \nonumber\\ 
&&K^{+}_{6}\left\{\int^{1}_{x}\frac{dy}{y} \frac{yA^{+}_{n}(y) - xA^{+}_{n}(x)}{y-x} + A^{+}_{n}(x)\log(1-x) \right\} + K^{+}_{7}\int^{1}_{x}\frac{dy}{y}A^{+}_{n}(y)\,, 
\eeqn

where $z={x}/{y}$. For the $(-)$ case we get a similar expression.
  
For the $B^{\pm}_{n+1}(x)$  we get (for the $(+)$ case) 

\ba
&&B^{+}_{n+1}(x) = - B^{+}_{n}(x) + \frac{\beta_{1}}{2\beta^{2}_{0}} \left\{2C_{F}\left[\int^{1}_{x}\frac{dy}{y}\frac{y A^{+}_{n}(y) - x A^{+}_{n}(x)}{y-x} + A^{+}_{n}(x) \log(1-x)\right]\right.+\nonumber\\
&&\left.-2C_{F}\left(\int_{x}^{1}\frac{dy}{y} A^{+}_{n}(y)\right) + C_{F}\frac{3}{2} A^{+}_{n}(x)\right\}-\frac{1}{4\pi\beta_{0}}K^{+}_{1} A^{+}_{n}(x)+ \int^{1}_{x}\frac{dy}{y}\left[ K^{+}_{2}(z) S_{2}(z) + \right.\nonumber\\
&&+ \left.K^{+}_{3}(z)\log(z)+\log^{2}(z)K^{+}_{4}(z) + \log(z)\log(1-z)K^{+}_{5}(z)\right]\left(-\frac{1}{4\pi\beta_{0}}\right)A^{+}_{n}(y)+\nonumber\\
&&K^{+}_{6}\left(-\frac{1}{4\pi\beta_{0}}\right)\left\{\left[\int^{1}_{x}\frac{dy}{y} \frac{yA^{+}_{n}(y) - xA^{+}_{n}(x)}{y-x} + A^{+}_{n}(x)\log(1-x) \right] + K^{+}_{7}\int^{1}_{x}\frac{dy}{y}A^{+}_{n}(y)\right\}-\nonumber\\
&&C_{F}\left(-\frac{4}{\beta_{0}}\right)\left[\int^{1}_{x}\frac{dy}{y}\frac{y B^{\pm}_{n}(y) - x B^{\pm}_{n}(x)}{y-x} + B^{\pm}_{n}(x) \log(1-x)\right]+\nonumber\\
&&C_{F}\left(\frac{4}{\beta_{0}}\right) \left(\int_{x}^{1}\frac{dy}{y} B^{\pm}_{n}(y)\right) + C_{F}\left(-\frac{2}{\beta_{0}}\right)\frac{3}{2} B^{\pm}_{n}(x)\,\nonumber
\ea
where in the $(+)$ case we have the expressions 

\ba
&&K_{1}^{+}(x)=\frac{1}{72}C_{F} (-2 n_{f} (3+4\pi^{2}) + N_{C}(51 + 44\pi^{2} - 216 \zeta(3))+ 9C_{F}(3-4\pi^{2}+48\zeta(3))\nonumber\\
&&K_{2}^{+}(x)= \frac{2 C_{F}(-2C_{F}+N_{C})x}{1+x}\nonumber \\
&&K_{3}^{+}(x)= \frac{C_{F}(9C_{F}-11 N_{C}+2n_{f})x}{3(x-1)}\nonumber \\
&&K_{4}^{+}(x)=\frac{C_{F}N_{C}x}{1-x}\nonumber \\ 
&&K_{5}^{+}(x)=\frac{4C_{F}^{2}x}{1-x}\nonumber \\
&&K_{6}^{+}(x)=-\frac{1}{9}C_{F}(10n_{f}+N_{C}(-67 + 3\pi^{2}))\nonumber\\
&&K_{7}^{+}(x)=\frac{1}{9}C_{F}(10n_{f}+N_{C}(-67 + 3\pi^{2})),\nonumber\\
\ea
and for the $(-)$ case
\ba
&&K_{1}^{-}(x)=\frac{1}{72}C_{F} (-2 n_{f} (3+4\pi^{2}) + N_{C}(51 + 44\pi^{2} - 216 \zeta(3))+ 9C_{F}(3-4\pi^{2}+48\zeta(3))\nonumber\\
&&K_{2}^{-}(x)= \frac{2 C_{F}(+2C_{F}-N_{C})x}{1+x}\nonumber \\
&&K_{3}^{-}(x)= \frac{C_{F}(9C_{F}-11 N_{C}+2n_{f})x}{3(x-1)}\nonumber\\
&&K_{4}^{-}(x)=\frac{C_{F}N_{C}x}{1-x}\nonumber\\
&&K_{5}^{-}(x)=\frac{4C_{F}^{2}x}{1-x}\nonumber\\
&&K_{6}^{-}(x)=-\frac{1}{9}C_{F}(10n_{f}+N_{C}(-67 + 3\pi^{2}))\nonumber\\
&&K_{7}^{-}(x)=-\frac{1}{9}C_{F}(10n_{f}-18 C_{F}(x-1)+N_{C}
(-76 +3\pi^{2}+9x)).\nonumber\\
\ea
The terms containing similar distribution (such as ``+'' distributions 
and $\delta$ functions) have been combined 
together in order to speed-up the computation of the recursion relations.

\section{Comparisons among Moments\label{sec:moments}}

It is particularly instructive to illustrate here briefly the relation 
between the Mellin moments of the parton distributions, which evolve 
with standard ordinary differential equations, and those of the 
arbitrary coefficient $A_n(x)$ and $B_n(x)$ which characterize 
Rossi's expansion up to next-to-leading order. This relation, as we are going to show, involves a 
resummation of the ordinary moments of the parton distributions. 

Specifically, here we will be dealing with the relation between 
the Mellin moments of the coefficients appearing in the expansion 
\beqa
A(N) &=& \int_0^1\,dx \, x^{N-1} A(x)\nonumber \\
B(N) &=&\int_0^1\,dx \, x^{N-1} B(x) \nonumber \\
\eeqa
and those of the distributions  
\beq
\Delta_T q^{(\pm)}(N,Q^2)=\int_0^1\,dx \, x^{N-1} \Delta_T q^{(\pm)}(x,Q^2)).
\eeq 
For this purpose we recall that the general (nonsinglet) solution to NLO for the latter moments is given by 
\begin{eqnarray} \label{evsol}
\nonumber
\Delta_T q_{\pm} (N,Q^2) &=& K(Q_0^2,Q^2,N)
\left( \frac{\alpha_s (Q^2)}{\alpha_s (Q_0^2)}\right)^{-2\Delta_T 
P_{qq}^{(0)}(N)/ \beta_0}\! \Delta_T q_{\pm}(N, Q_0^2)
\label{solution}
\end{eqnarray}
with the input distributions $\Delta_T q_{\pm}^n (Q_0^2)$ at the input scale 
$Q_0$.
We also have set 
\beq
K(Q_0^2,Q^2,N)= 1+\frac{\alpha_s (Q_0^2)-
\alpha_s (Q^2)}{\pi\beta_0}\!
\left[ \Delta_T P_{qq,\pm}^{(1)}(N)-\frac{\beta_1}{2\beta_0} \Delta_T 
P_{qq\pm}^{(0)}(N) \right]. 
\eeq
In the expressions above we have introduced the corresponding moments for the LO and NLO kernels 
($\Delta_T P_{qq}^{(0),N}$,
$ \Delta_T P_{qq,\pm}^{(1),N})$. 

The relation between the moments of the coefficients of the nonsinglet
$x$-space expansion and those of the parton distributions at any $Q$, as expressed by eq.~(\ref{solution}) can be easily written down
\beq
A_n(N) + \alpha_s B_n(N)=\Delta_T q_\pm(N,Q_0^2)K(Q_0,Q,N)\left(\frac{-2 \Delta_T P_{qq}(N)}{\beta_0}\right)^n.
\label{relation}
\eeq

As a check of this expression, notice that the initial condition is easily obtained from  
(\ref{relation}) setting $Q\to Q_0, n\to 0$, thereby obtaining 
\beq
A_0^{NS}(N) + \alpha_s B_0^{NS} (N)= \Delta_T q_\pm(N,Q_0^2),
\eeq
which can be solved with $A_0^{NS}(N)=\Delta_T q_\pm(N,Q_0^2)$ and 
$B_0^{NS} (N)=0$. 

It is then evident that the expansion (\ref{rossis}) involves a resummation of the logarithmic contributions, as shown in eq.~(\ref{relation}). 

In the singlet sector we can work out a similar relation both to LO

\beq
A_n(N) = e_1\left(\frac{-2 \lambda_1}{\beta_0}\right)^n 
+e_2 \left(\frac{-2 \lambda_2}{\beta_0}\right)^n 
\eeq

with 
\beqa
e_1 &=& \frac{1}{\lambda_1 - \lambda_2}\left( P^{(0)}(N)- \lambda_2 \one \right)
\nonumber \\
e_2 &=& \frac{1}{\lambda_2 - \lambda_1}\left( - P^{(0)}(N) + \lambda_1 \one\right)
\nonumber \\
\lambda_{1,2}&=& \frac{1}{2}\left( 
P^{(0)}_{qq}(N) + P^{(0)}_{gg}(N) \pm \sqrt{\left(P^{(0)}_{qq}(N)- P^{(0)}_{gg}(N)\right)^2  
+ 4 P^{(0)}_{qg}(N)P^{(0)}_{gq}(N)}\right),
\eeqa
and to NLO 
\beq
A_n(N) + \alpha_s B_n(N) = \chi_1\left(\frac{-2 \lambda_1}{\beta_0}\right)^n 
+\chi_2 \left(\frac{-2 \lambda_2}{\beta_0}\right)^n, 
\eeq

where 
\beqa
\chi_1 &=& e_1 + \frac{\alpha}{2 \pi}\left( \frac{-2}{\beta_0}e_1 \R e_1 
+\frac{ e_2 \R e_1}{\lambda_1 - \lambda_2 - \beta_0/2}\right) \nonumber \\
\chi_2 &=& e_2 + \frac{\alpha}{2 \pi}\left( \frac{-2}{\beta_0}e_2 \R e_2  
+\frac{ e_1 \R e_2}{\lambda_2 - \lambda_1 - \beta_0/2}\right)\nonumber \\
\eeqa
with 
\beq
\R= P^{(1)}(N) -\frac{\beta_1}{2 \beta_0}P^{(0)}(N).
\eeq
We remark that $A_n(N)$ and $B_n(N)$, $P^{(0)}(N)$, $P^{(1)}(N)$, in this case, 
are all 2-by-2 singlet matrices.

\section{Initial conditions\label{sec:initial}}

As input distributions in the unpolarized case, we have used the 
models of Ref.\cite{GRV}, valid to NLO in the \( \overline{\textrm{MS}} \)
scheme at a scale \( Q_{0}^{2}=0.40\, \textrm{GeV}^{2} \)
\begin{eqnarray}
x(u-\overline{u})(x,Q_{0}^{2}) & = & 0.632x^{0.43}(1-x)^{3.09}(1+18.2x)\nonumber \\
x(d-\overline{d})(x,Q_{0}^{2}) & = & 0.624(1-x)^{1.0}x(u-\overline{u})(x,Q_{0}^{2})\nonumber \\
x(\overline{d}-\overline{u})(x,Q_{0}^{2}) & = & 0.20x^{0.43}(1-x)^{12.4}(1-13.3\sqrt{x}+60.0x)\nonumber \\
x(\overline{u}+\overline{d})(x,Q_{0}^{2}) & = & 1.24x^{0.20}(1-x)^{8.5}(1-2.3\sqrt{x}+5.7x)\nonumber \\
xg(x,Q_{0}^{2}) & = & 20.80x^{1.6}(1-x)^{4.1}
\end{eqnarray}
and \( xq_{i}(x,Q_{0}^{2})=x\overline{q_{i}}(x,Q_{0}^{2})=0 \) for
\( q_{i}=s,c,b,t \).

Following \cite{GRSV}, we have related the unpolarized input distribution
to the longitudinally polarized ones 

\begin{eqnarray}
x\Delta u(x,Q_{0}^{2}) & = & 1.019x^{0.52}(1-x)^{0.12}xu(x,Q_{0}^{2})\nonumber \\
x\Delta d(x,Q_{0}^{2}) & = & -0.669x^{0.43}xd(x,Q_{0}^{2})\nonumber \\
x\Delta \overline{u}(x,Q_{0}^{2}) & = & -0.272x^{0.38}x\overline{u}(x,Q_{0}^{2})\nonumber \\
x\Delta \overline{d}(x,Q_{0}^{2}) & = & x\Delta \overline{u}(x,Q_{0}^{2})\nonumber \\
x\Delta g(x,Q_{0}^{2}) & = & 1.419x^{1.43}(1-x)^{0.15}xg(x,Q_{0}^{2})
\end{eqnarray}
and \( x\Delta q_{i}(x,Q_{0}^{2})=x\Delta \overline{q_{i}}(x,Q_{0}^{2})=0 \)
for \( q_{i}=s,c,b,t \).
Being the transversity
distribution experimentally unknown, following \cite{MSSV}, 
we assume the saturation of
Soffer's inequality
\begin{equation}
x\Delta _{T}q_{i}(x,Q_{0}^{2})=\frac{xq_{i}(x,Q_{0}^{2})+x\Delta q_{i}(x,Q_{0}^{2})}{2}.
\end{equation}

These input distributions will be used in the next section in the analysis of the 
transverse asymmetries for polarized pp collisions in the Drell-Yan process. 

\section{Transversity in the Drell-Yan process}
The Drell-Yan mechanism for lepton pair production is, at parton level, described by 
the annihilation of a quark-antiquark pair $(q \bar{q})$ into an s-channel (virtual) 
photon which decays into a lepton pair. The process is characterized by a distinct 
signature, since the two leptons $(l^+ l^-)$ of the final state can be more easily 
identified. As we have discussed in the previous sections of this chapter, 
the study of the transverse spin distributions of the proton is an ongoing process which requires more experimental data in the future in order to provide us with a clearer understanding of these functions. The natural question to ask is: what are the processes 
mediated at parton level by these particular matrix elements, which therefore 
carry information, over to the final state, 
on the distribution of transverse spin in the proton. 
In this section
we focalize our attention on the phenomenology of the transversely polarized
distributions $h_1^{a}(x,Q^2)$  (or $\Delta_T(x, Q^2)$ in other notations) which is the missing part in the QCD description
of the spin structure of the nucleon at leading twist.

Since $h_1^{a}(x,Q^2)$ is chirally-odd, as discussed by Jaffe, 
one of the possible way to access the measurement of
this observable is the Drell-Yan process. 

In fact, in deep inelastic scattering processes $h_1^{a}(x,Q^2)$ is severely suppressed
in the operator product expansion since it needs a mass insertion 
in the unitarity diagram to appear. So the Drell-Yan process remains up to now the theoretically cleanest way to observe $h_1^{a}(x,Q^2)$. 

In a polarized proton-antiproton collision one can construct an asymmetry $A_{TT}$ which is
proportional to the product of transversity of the proton's quark and the transversity of the
antiproton's anti-quark, which are equal due to the charge conjugation.
The most recent experimental proposal of antiproton-proton scattering 
with polarization has been presented by the  PAX collaboration.
In the PAX experiment polarized antiprotons are produced by spin filtering an internal
polarized gas target and scattered off protons at intermediate energy. The computation 
of the related transverse spin asymmetries in Drell-Yan, which we are going to discuss 
below, is performed using the factorization formula for the cross section    
${d}\delta\sigma\equiv\left({d}\sigma^{\uparrow\uparrow}-
{d}\sigma^{\uparrow\downarrow}\right)/2$. This is given as a double
convolution of transversity distributions with the corresponding
transversely polarized partonic cross section.
In the case of antiproton-proton collision with a di-muon production
the expression is

\ba
\label{sig}
\frac{d \delta\sigma}{dM dy d\phi} &=&
\sum_q \tilde{e}_q^2
\int_{x_1^0}^1 {d}x_1 \int_{x_2^0}^1 {d}x_2 \left[
\Delta_T q(x_1,\mu_F^2) \Delta_T q(x_2,\mu_F^2) +
\Delta_T\bar q(x_1,\mu_F^2) \Delta_T \bar q(x_2,\mu_F^2) \right] \nonumber \\
&& \times\frac{d\Delta_T\hat\sigma}{d M dy d\phi} \,
\ea

where $\mu_F^2$ is the factorization scale of the process, 
and the charge $\tilde{e}$ is quite general, since it may encompass both electromagnetic 
and electroweak effects. 
$S$ is the center of mass initial energy and $M$ represents the invariant mass
of the virtual photon.
The $y$ variable is called rapidity and it is connected with the variables
$x_1^0$, $x_2^0$ by a scaling parameter $\tau={M^2/S}$ as $x_1^0=\sqrt{\tau} e^y$,
$x_2^0=\sqrt{\tau} e^{-y}$.
$\phi$ is the azimuthal angle of one muon.

The next-to-leading order $\alpha_s$ expression \cite{vogelsang1} for the hard scattering term
is written in the modified minimal subtraction $\overline{MS}$ scheme as follows

\ba
\label{sigmanlo}
\frac{{d} \Delta_T \hat \sigma^{(1),
{\overline{{MS}}}}}{{d}
M{d}y{d}\phi} &=&
\frac{2\alpha^2}{9SM} C_F \frac{\alpha_s(\mu_R^2)}{2\pi}
\frac{4\tau(x_1x_2+\tau)}{x_1x_2(x_1+x_1^0)(x_2+x_2^0)}\cos (2\phi)
\nonumber \\
&\times& \left\{
\delta(x_1-x_1^0)\delta(x_2-x_2^0) \left[
\frac{1}{4}\ln^2\frac{(1-x_1^0)(1-x_2^0)}{\tau}
+\frac{\pi^2}{4}-2 \right]
\right. \nonumber \\ && \left.
+ \delta(x_1-x_1^0) \left[ \frac{1}{(x_2-x_2^0)_{+}}
\ln\frac{2x_2(1-x_1^0)}{\tau(x_2+x_2^0)}
+\left(\frac{\ln(x_2-x_2^0)}{x_2-x_2^0}\right)_{+}
+\frac{1}{x_2-x_2^0}\ln\frac{x_2^0}{x_2}\right]
\right. \nonumber \\ && \left.
+ \frac{1}{2[(x_1-x_1^0)(x_2-x_2^0)]_{+}}
+ \frac{(x_1+x_1^0)(x_2+x_2^0)}{(x_1 x_2^0+x_2x_1^0)^2}
- \frac{3\ln\left(\frac{x_1x_2+\tau}{x_1x_2^0+x_2x_1^0} \right)}{(x_1-x_1^0)
(x_2-x_2^0)}
\right\}+ \nonumber \\
&& [1 \leftrightarrow 2]\,.\nonumber\\
\ea
In this expression the renormalization scale $\mu_R$ has been set to coincide with 
the factorization scale $\mu_F=M$. We recall that a different choice for these two scales 
would be responsible for the generation of additional logs $(\log(\mu_F^2/\mu_R^2))$ both 
in the evolution and in the hard scatterings. This issue will be discussed in more 
detail in the computation of the Higgs total cross section, where it has more relevance.

The expression of the LO cross section is quite simple and it reads
\be
\label{LO}
\frac{d\Delta_T\hat\sigma^{(0)}}{d
Mdyd\phi}=\frac{2\alpha^2}{9SM}\cos (2\phi) \delta(x_1-x_1^0)
\delta(x_2-x_2^0) \, .
\ee
To lowest order (LO) $x_1^0$ and $x_2^0$ coincide
with the momentum fractions carried by the incident partons.

\begin{figure}[t]
{\centering \resizebox*{12cm}{!}{\rotatebox{-90}{\includegraphics{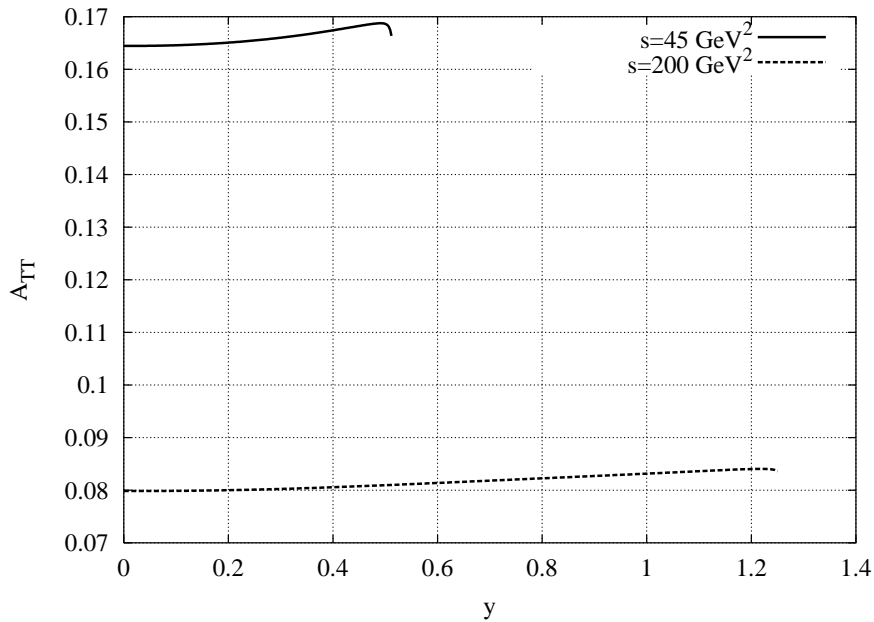}}} \par}
\caption{NLO asymmetry calculated at $M=4$ GeV with MRST inputs.}
\label{Asym1}
\end{figure}

\begin{figure}[t]
{\centering \resizebox*{12cm}{!}{\rotatebox{-90}{\includegraphics{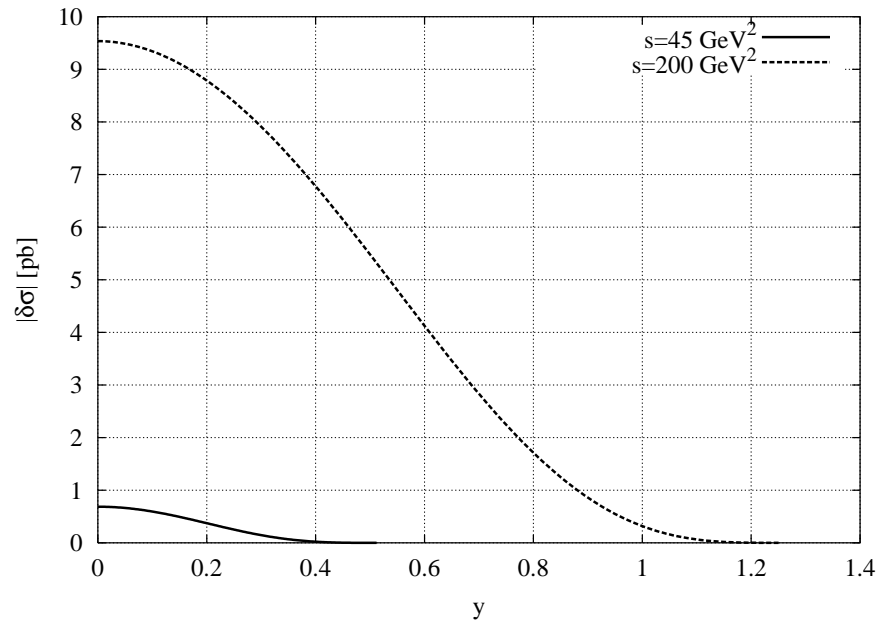}}} \par}
\caption{NLO cross section calculated at $M=4$ GeV with MSRT inputs.}
\label{cro1}
\end{figure}

\begin{figure}[t]
{\centering \resizebox*{12cm}{!}{\rotatebox{-90}{\includegraphics{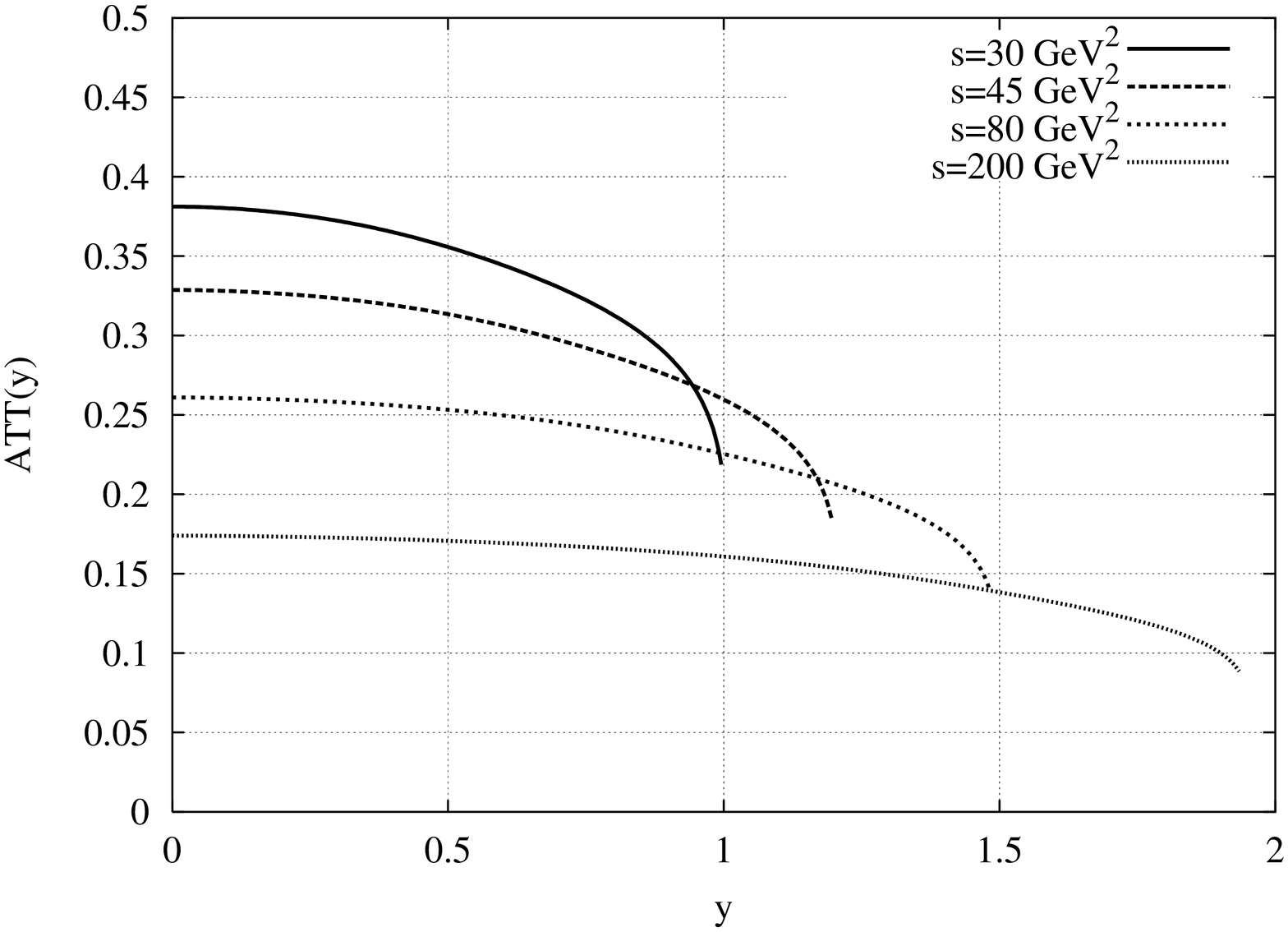}}} \par}
\caption{NLO Asymmetry calculated integrating $M$ between 2 and 3 GeV with GRV inputs.}
\label{Asym2}
\end{figure}

\begin{figure}[t]
{\centering \resizebox*{12cm}{!}{\rotatebox{-90}{\includegraphics{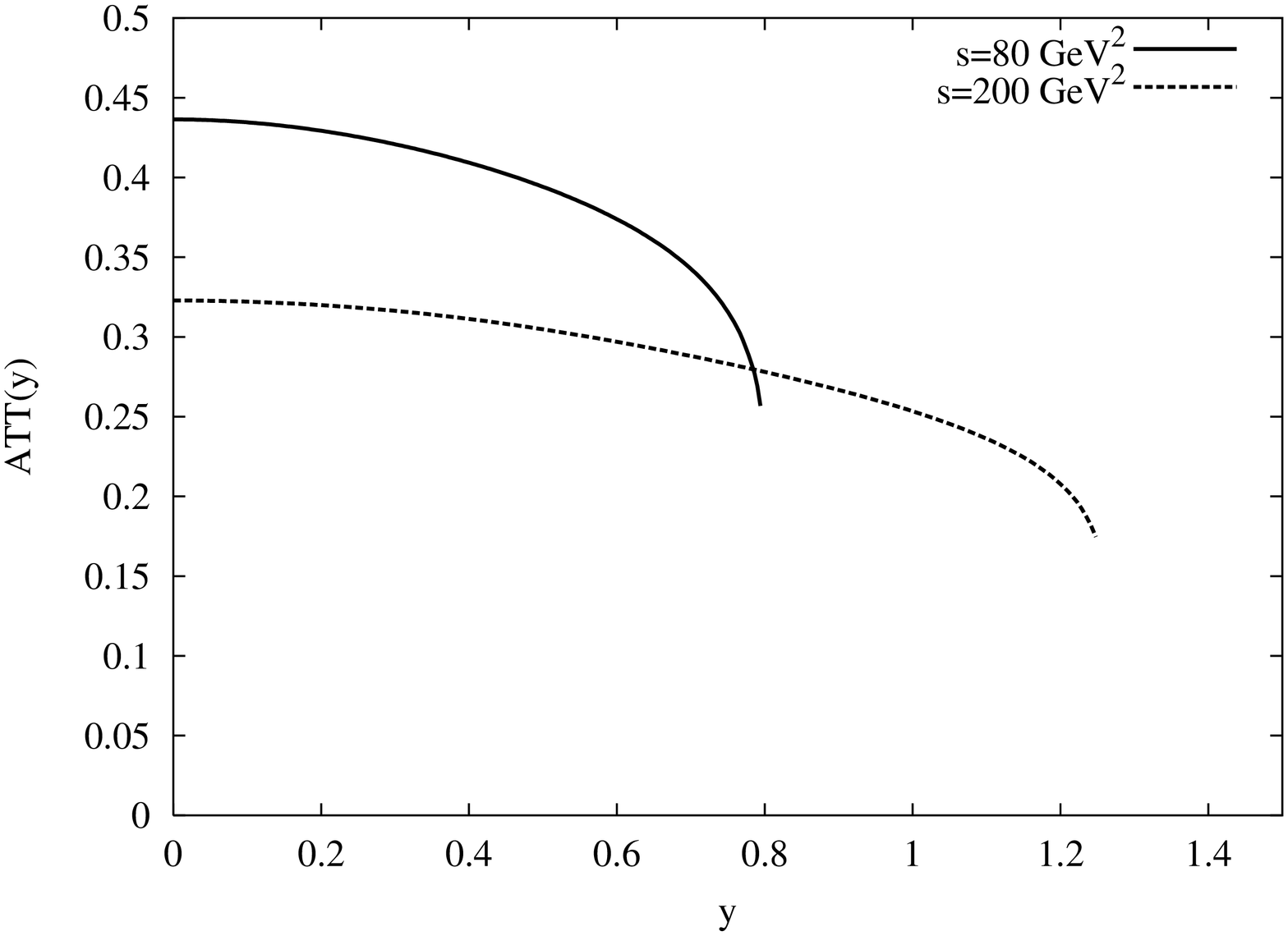}}} \par}
\caption{NLO Asymmetry calculated integrating $M$ between 4 and 7 GeV with GRV inputs.}
\label{Asym3}
\end{figure}

The asymmetry depending on $y$ variable can be constructed

\be
\label{asy1}
A_{TT}(y) \equiv
\frac{\int_{M_0}^{M_1}{d}M
\left(\int_{-\pi/4}^{\pi/4}-\int_{\pi/4}^{3\pi/4}+
\int_{3\pi/4}^{5\pi/4}-\int_{5\pi/4}^{7\pi/4} \right){d}\phi
\,\,d\delta\sigma/dM dy d\phi}
{\int_{M_0}^{M_1}{d}M\int_0^{2\pi}d\phi
\,\,d\sigma/dM dy d\phi}\,.
\ee
A measurement of the asymmetry with a sufficient accuracy can be done
in the PAX experiment in the dilepton mass region below the $J/\Psi$
threshold and a systematic study at leading order has be done in this last
two years \cite{efremov1,barone1}.
We have numerically analyzed \cite{barone2} a region very close to the $J/\Psi$
resonance $M\approx 4$ GeV$^2$ to NLO since could be crucial in order to enhance 
the cross section. In fact asymmetries would be very difficult to measure in a region characterized by a fast falling cross section. Near a resonance the possibility of a successfull 
measurement is sharply enhanced.  
The cross section for dilepton production increases by almost 2
orders of magnitude in going from $M=4$ to $M=3$ GeV, since this cross section
involves unknown quantities related to $q\bar{q}$-$J/\Psi$ coupling.
However, independently of these unknown quantities, the $q\bar{q}$-$J/\Psi$
coupling is a vector one, with the same spinor and Lorentz structure as
$q \bar{q}$-$\gamma^{*}$ coupling. These unknown quantities cancel in the ratio
giving $A_{TT}$, while the helicity structure remains, so that the asymmetry
is given by 
\ba
A_{TT}\simeq \hat{a}_{TT}\frac{h_1^{u}(x_1,M^2)h_1^{u}(x_2,M^2)}{u(x_1,M^2)u(x_2,M^2)}\,,
\ea
where $\hat{a}_{TT}$ is the double spin asymmetry of the elementary QED process
$q \bar{q}\rightarrow l^{+} l^{-}$
\ba
\hat{a}_{TT}=\frac{\sin^2{\theta}}{1+\cos^2{\theta}} \cos{2\phi}
\ea
This substantially enhances the sensitivity of the PAX experiment
to $A_{TT}$ and the amount of direct information achievable
on $h_1^{u}(x_1,M^2)h_1^{u}(x_2,M^2)$.
In order to evolve the transversity we have used the longitudinal bound
on the inputs
\ba
h_1^{q}(x,Q_{0}^{2})=\Delta q(x,Q_{0}^{2})
\ea
and we performed the evolution at LO and NLO starting by the same initial point
$Q_0^2=1$ GeV$^2$ for the GRV's and $Q_{0}^{2}=0.4$ GeV$^2$ for the set of distributions 
termed MRST's. We show below some plots of the asymmetry vs the rapidity
calculated at NLO using different input parton distribution functions
and at different values of $S$. In fig.~(\ref{Asym1}) we present NLO results for the 
transverse double spin asymmetries for an invariant mass of the lepton pair equal to 
4 GeV and in fig.~(\ref{cro1}) we plot the unpolarized cross section at the same energy.
The asymmetry is about 15 percent, considering a GRV input model for the transverse 
spin distributions, with a sizeable cross section. 
We also show results for the integrated asymmetries in figs.~(\ref{Asym2}) and 
(\ref{Asym3}), which grow up to 35-40 percent, and are therefore sizeable. 

\section{Documentation of the Code}
In this section we describe the variables, the parameters and the 
functions introduced in the numerical implementation of the program. The 
notation that we use is the standard one adopted by Computer Physics Communication for the 
documentation of the code. This section consists of a list of all the functions 
and  variables defined in the program, the output files, and some comments concerning the performance of the implementation.   

\subsection{Names of the input parameters, variables and of the output files}

\subsubsection{Notations \label{subsubsec:dist_ind}}

\begin{tabular}{lll}
\hline 
0&
gluons, \( g \)&
\texttt{g}\\
1-6&
quarks, \( q_{i} \), sorted by their mass values(\( u,d,s,c,b,t \))&
\texttt{u,d,s,c,b,t}\\
7-12&
antiquarks, \( \overline{q_{i}} \)&
\texttt{au,ad,as,ac,ab,at}\\
13-18&
\( q^{(-)}_{i} \)&
\texttt{um,dm,sm,cm,bm,tm}\\
19-24&
\( \chi _{i} \) (unpolarized and longitudinally polarized cases)&
\texttt{Cu,Cd,Cs,Cc,Cb,Ct}\\
&
\( q^{(+)}_{i} \) (transversely polarized case)&
\texttt{Cu,Cd,Cs,Cc,Cb,Ct}\\
25&
\( q^{(+)} \)&
\texttt{qp}\\
\hline
\end{tabular}

\subsubsection{Input parameters and variables}

\begin{tabular}{lll}
\hline 
\texttt{process}&
0&
unpolarized\\
&
1&
longitudinally polarized\\
&
2&
transversely polarized\\
\hline
\end{tabular}\\
\begin{tabular}{lll}
\hline 
\texttt{spacing}&
1&
linear\\
&
2&
logarithmic\\
\end{tabular}\\
\begin{tabular}{ll}
\hline 
\texttt{GRID\_PTS}&
Number of points in the grid\\
\texttt{NGP}&
Number of Gaussian points, \( n_{G} \)\\
\texttt{ITERATIONS}&
Number of terms in the sum (\ref{eq:ansatz})\\
\texttt{extension}&
Extension of the output files\\
\hline
\end{tabular}\\
\begin{tabular}{ll}
\hline 
\texttt{step}&
grid step (linear spacing case)\\
\texttt{lstep}&
step in \( \log _{10}x \) (logarithmic spacing case)\\
\texttt{X{[}i{]}}&
\( i \)-th grid point, \( x_{i} \)\\
\texttt{XG{[}i{]}{[}j{]}}&
\( j \)-th Gaussian abscissa in the range \( [x_{i},1] \), \( X_{ij} \)\\
\texttt{WG{[}i{]}{[}j{]}}&
\( j \)-th Gaussian weights in the range \( [x_{i},1] \), \( W_{ij} \)\\
\texttt{nf, Nf}&
number of active flavors, \( n_{f} \)\\
\texttt{n\_evol}&
progressive number of the evolution step is \( n_{f}-3 \)\\
\texttt{Q{[}i{]}}&
values of \( Q \)  in the corresponding grid \\
\texttt{lambda{[}i{]}}&
\( \Lambda _{\overline{MS}}^{(n_{f})} \), where \( i=n_{f}-3 \)\\
\texttt{A{[}i{]}{[}j{]}{[}k{]}}&
coefficient \( A_{j}(x_{k}) \) for the distribution with index \( i \)\\
\texttt{B{[}i{]}{[}j{]}{[}k{]}}&
coefficient \( B_{j}(x_{k}) \) for the distribution with index \( i \)\\
\texttt{beta0}&
\( \beta _{0} \)\\
\texttt{beta1}&
\( \beta _{1} \)\\
\texttt{alpha1}&
\( \alpha _{s}(Q_{in}) \), where \( Q_{in} \) is the lower \( Q \)
of the evolution step\\
\texttt{alpha2}&
\( \alpha _{s}(Q_{fin}) \), where \( Q_{fin} \) is the higher \( Q \)
of the evolution step\\
\hline
\end{tabular}

\subsubsection{Output files}

The generic name of an output file is: \texttt{\textbf{X}}\texttt{\textbf{\emph{n}}}\texttt{\textbf{i.ext}},
where

\begin{description}
\item [\texttt{X}]is U in the unpolarized case, L in the longitudinally
polarized case and T in the transversely polarized case;
\item [\texttt{\emph{n}}]is a progressive number that indicates the scale
\( Q^{2} \) at a given stage: \( n=0 \) refers to the initial
scale, the highest value of \( n \) refers to the final scale and
the intermediate values refer to the quarks production thresholds
(1 for charm, 2 for bottom and 3 for top);
\item [\texttt{i}]is the identifier of the distribution, reported in
the third column of the table in subsubsection \ref{subsubsec:dist_ind};
\item [\texttt{ext}]is an extension chosen by the user.
\end{description}

\subsection{Description of the program}

\subsubsection{Main program}

At run time, the program asks the user to select a linear or a logarithmic
spacing for the \( x \)-axis. The logarithmic spacing is useful in order 
to analyze the small-\( x \) behavior. Then the program stores as external
variables the grid points \( x_{i} \) and, for each of them, calls
the function \texttt{gauleg} which computes the Gaussian points
\( X_{ij} \) and weights \( W_{ij} \) corresponding to the integration
range \( [x_{i},1] \), with \( 0\leq j\leq n_{G}-1 \). After that,
the user is asked to enter the type of process, the final value
of \( Q \) and an extension for the names of the output files. At this point the
program computes the initial values of the parton distributions for
gluons, up, down, anti-up and anti-down (see Section \ref{sec:initial})
at the grid points and stores them in the arrays \texttt{A{[}i{]}{[}0{]}{[}k{]}}
(see (\ref{eq:initial})), setting to zero the initial distributions
of the heavier quarks.

The evolution is done in the various regions of the evolutions, all characterized by a specific flavour number. 
Each new flavour comes into play
only when the scale $Q$ reaches the corresponding quark mass. 
In that case $n_f$ is increased by 1 everywhere in the program.
The recurrence relations (\ref{eq:An_recurrence}) and (\ref{eq:Bn_recurrence})
are then solved iteratively for both the nonsinglet and the singlet
sector, and at the end of each energy step the evolved distributions
are reconstructed via the relation (\ref{eq:ansatz}). The distributions
computed in this way become the initial conditions for the subsequent
step. The numerical values of the distributions at the end of each
energy step are printed to files.

\subsubsection{Function \texttt{writefile}}

\texttt{void writefile(double {*}A,char {*}filename);}

This function creates a file, whose name is contained in the string
\texttt{{*}filename}, with an output characterized by 
two columns of data: the left column contains
all the values of the grid points \( x_{i} \) 
and the right one the corresponding values of the array \texttt{A{[}i{]}}.

\subsubsection{Function \texttt{alpha\_s}}

\texttt{double alpha\_s(double Q,double beta0,double beta1,double
lambda);}

Given the energy scale \texttt{Q}, the first two terms of the perturbative
expansion of the \( \beta  \)-function \texttt{beta0} and \texttt{beta1}
and the value \texttt{lambda} of \( \Lambda _{\overline{MS}}^{(n_{f})} \),
\texttt{alpha\_s} returns the two-loop running of the coupling constant,
using the formula (\ref{eq:alpha_s}).

\subsubsection{Function \texttt{gauleg}}

\texttt{void gauleg(double x1,double x2,double x{[}{]},double w{[}{]},int
n);}

This function is taken from \cite{NumRecipes} with just some minor changes.
Given the lower and upper limits of integration \texttt{x1} and \texttt{x2},
and given \texttt{n}, \texttt{gauleg} returns arrays \texttt{x{[}0,...,n-1{]}}
and \texttt{w{[}0,...,n-1{]}} of length \texttt{n}, containing the
abscissas and weights of the Gauss-Legendre \texttt{n}-point quadrature
formula.

\subsubsection{Function \texttt{interp}}

\texttt{double interp(double {*}A,double x);}

Given an array \texttt{A}, representing a function known only at the
grid points, and a number \texttt{x} in the interval \( [0,1] \),
\texttt{interp}
returns the linear interpolation of the function at
the point \texttt{x}.

\subsubsection{Function \texttt{IntGL}}

\texttt{double IntGL(int i,double kernel(double z),double {*}A);}

Given an integer \texttt{i} (corresponding to a grid point \( x_{i} \)),
a one variable function \texttt{kernel(z)} and an array \texttt{A},
representing a function \( g(x) \) known at the grid points, \texttt{IntGL}
returns the result of the integral
\begin{equation}
\int _{x_{i}}^{1}\frac{\textrm{d}y}{y}kernel\left( \frac{x_{i}}{y}\right) g(x_{i}),
\end{equation}
computed by the Gauss-Legendre technique.

\subsubsection{Function \texttt{IntPD}}

\texttt{double IntGL(int i,double {*}A);}

Given an integer \texttt{i}, to which it corresponds a grid point \( x_{i} \),
and an array \texttt{A}, representing a function \( f(x) \) known
at the grid points, \texttt{IntGL} returns the result of the convolution\begin{equation}
\frac{1}{(1-x_{i})_{+}}\otimes f(x_{i})=\int _{x_{i}}^{1}\frac{\textrm{d}y}{y}\frac{yf(y)-x_{i}f(x_{i})}{y-x_{i}}+f(x_{i})\log (1-x_{i}),
\end{equation}
computed by the Gauss-Legendre technique.

\subsubsection{Function \texttt{S2}}

\texttt{double S2(double z);}

This function evaluates the Spence function
\( S_{2}(z) \) using the expansion\begin{equation}
S_{2}(z)=\log z\log (1-z)-\frac{1}{4}\left( \log z\right) ^{2}+\frac{\pi ^{2}}{12}+\sum ^{\infty }_{n=1}\frac{(-z)^{n}}{n^{2}}
\end{equation}
arrested at the 50th order.

\subsubsection{Function \texttt{fact}}

\texttt{double fact(int n);}

This function returns the factorial \( n! \)

\subsubsection{Initial distributions}

\texttt{double xuv(double x);}~\\
\texttt{double xdv(double x);}~\\
\texttt{double xdbmub(double x);}~\\
\texttt{double xubpdb(double x);}~\\
\texttt{double xg(double x);}~\\
\texttt{double xu(double x);}~\\
\texttt{double xubar(double x);}~\\
\texttt{double xd(double x);}~\\
\texttt{double xdbar(double x);}~\\
\texttt{double xDg(double x);}~\\
\texttt{double xDu(double x);}~\\
\texttt{double xDubar(double x);}~\\
\texttt{double xDd(double x);}~\\
\texttt{double xDdbar(double x);}

Given the Bjorken variable \texttt{x}, these functions return the
initial distributions at the input scale (see Section \ref{sec:initial}).

\subsubsection{Regular part of the kernels}

\texttt{double P0NS(double z);}~\\
\texttt{double P0qq(double z);}~\\
\texttt{double P0qg(double z);}~\\
\texttt{double P0gq(double z);}~\\
\texttt{double P0gg(double z);}~\\
\texttt{double P1NSm(double z);}~\\
\texttt{double P1NSp(double z);}~\\
\texttt{double P1qq(double z);}~\\
\texttt{double P1qg(double z);}~\\
\texttt{double P1gq(double z);}~\\
\texttt{double P1gg(double z);}~\\
\texttt{double DP0NS(double z);}~\\
\texttt{double DP0qq(double z);}~\\
\texttt{double DP0qg(double z);}~\\
\texttt{double DP0gq(double z);}~\\
\texttt{double DP0gg(double z);}~\\
\texttt{double DP1NSm(double z);}~\\
\texttt{double DP1NSp(double z);}~\\
\texttt{double DP1qq(double z);}~\\
\texttt{double DP1qg(double z);}~\\
\texttt{double DP1gq(double z);}~\\
\texttt{double DP1gg(double z);}~\\
\texttt{double tP0(double z);}~\\
\texttt{double tP1m(double z);}~\\
\texttt{double tP1p(double z);}

Given the Bjorken variable \texttt{z}, these functions return the
part of the Altarelli-Parisi kernels that does not contain singularities.

\subsection{Running the code}

In the plots shown in this chapter we have divided the interval \( [0,1] \)
of the Bjorken variable \( x \) in 500 subintervals (\texttt{GRID\_PTS}=501),
30 Gaussian points (\texttt{NGP}=1), and we have retained 10 terms
in the sum (\ref{eq:ansatz}) (\texttt{ITERATIONS}=10). In the figures
\ref{fig:Lu_log} and \ref{fig:Lg_log} the flag \texttt{spacing}
has been set to 2, in order to have a logarithmically spaced grid. This feature
turns useful if one intends to analyze the small-\( x \) behavior. We have tested our 
implementation in a detailed study of Soffer's inequality up to NLO 
\cite{CafarellaCorianoGuzzi}.

\begin{figure}[tbh]
{\centering \resizebox*{8cm}{!}{\rotatebox{-90}{\includegraphics{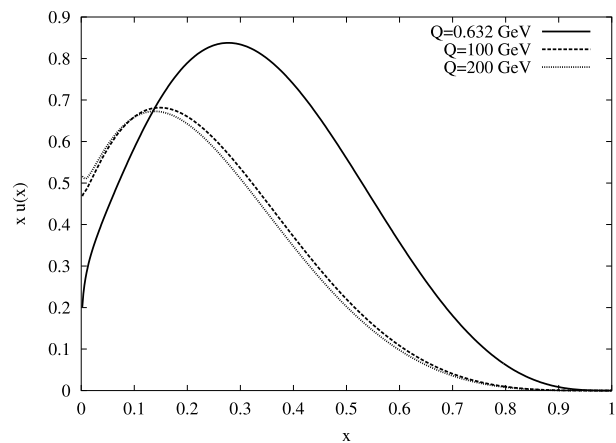}}} \par}

\caption{Evolution of the unpolarized quark up distribution \protect\( xu\protect \)
versus \protect\( x\protect \) at various \protect\( Q\protect \)
values.}
\end{figure}

\begin{figure}[tbh]
{\centering \resizebox*{8cm}{!}{\rotatebox{-90}{\includegraphics{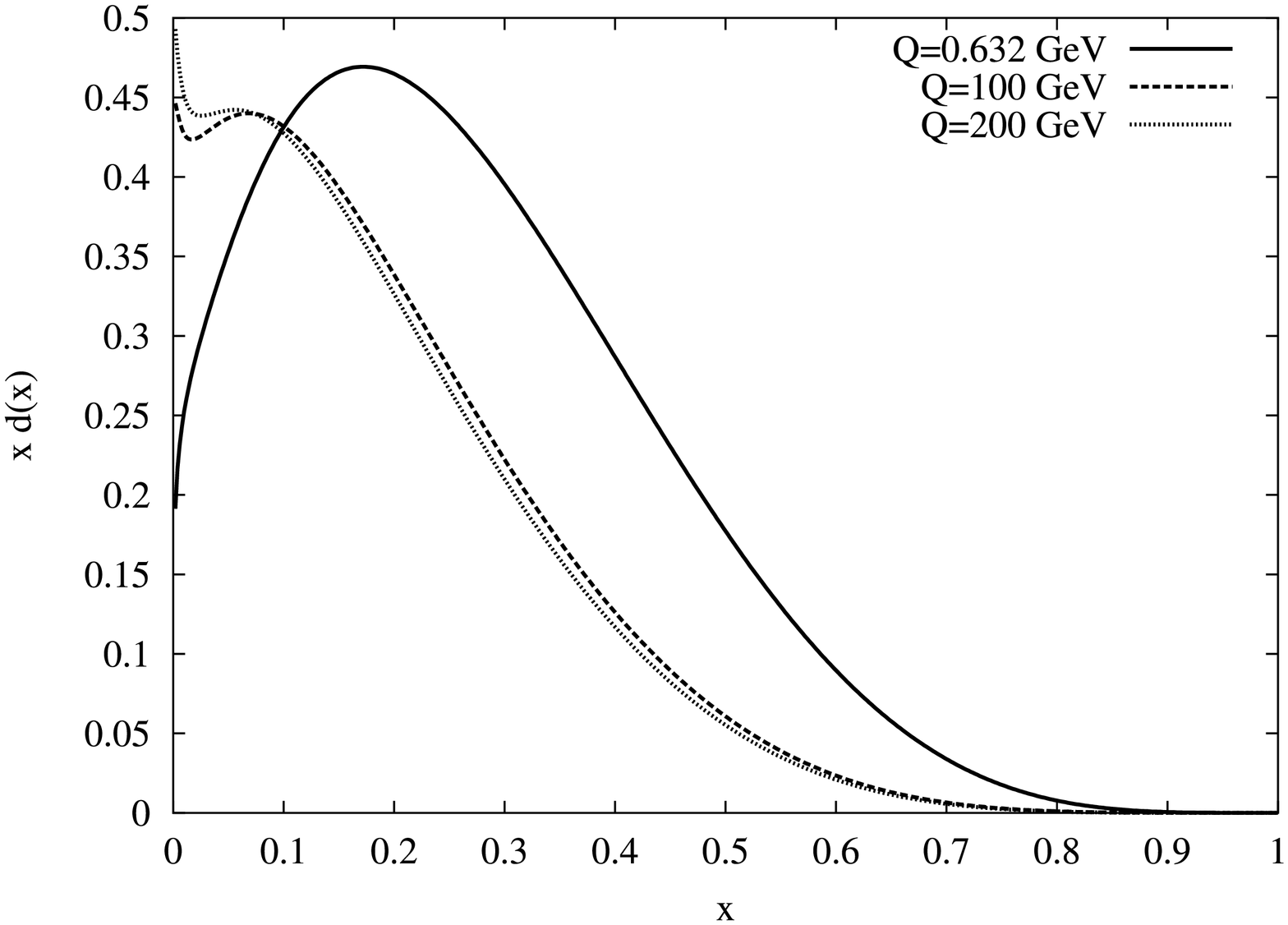}}} \par}

\caption{Evolution of \protect\( xd\protect \) versus \protect\( x\protect \)
at various \protect\( Q\protect \) values.}
\end{figure}

\begin{figure}[tbh]
{\centering \resizebox*{8cm}{!}{\rotatebox{-90}{\includegraphics{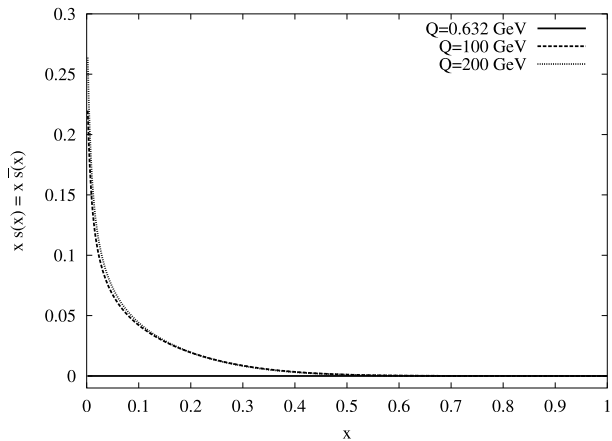}}} \par}

\caption{Evolution of \protect\( xs=x\overline{s}\protect \) versus \protect\( x\protect \)
at various \protect\( Q\protect \) values.}
\end{figure}

\begin{figure}[tbh]
{\centering \resizebox*{8cm}{!}{\rotatebox{-90}{\includegraphics{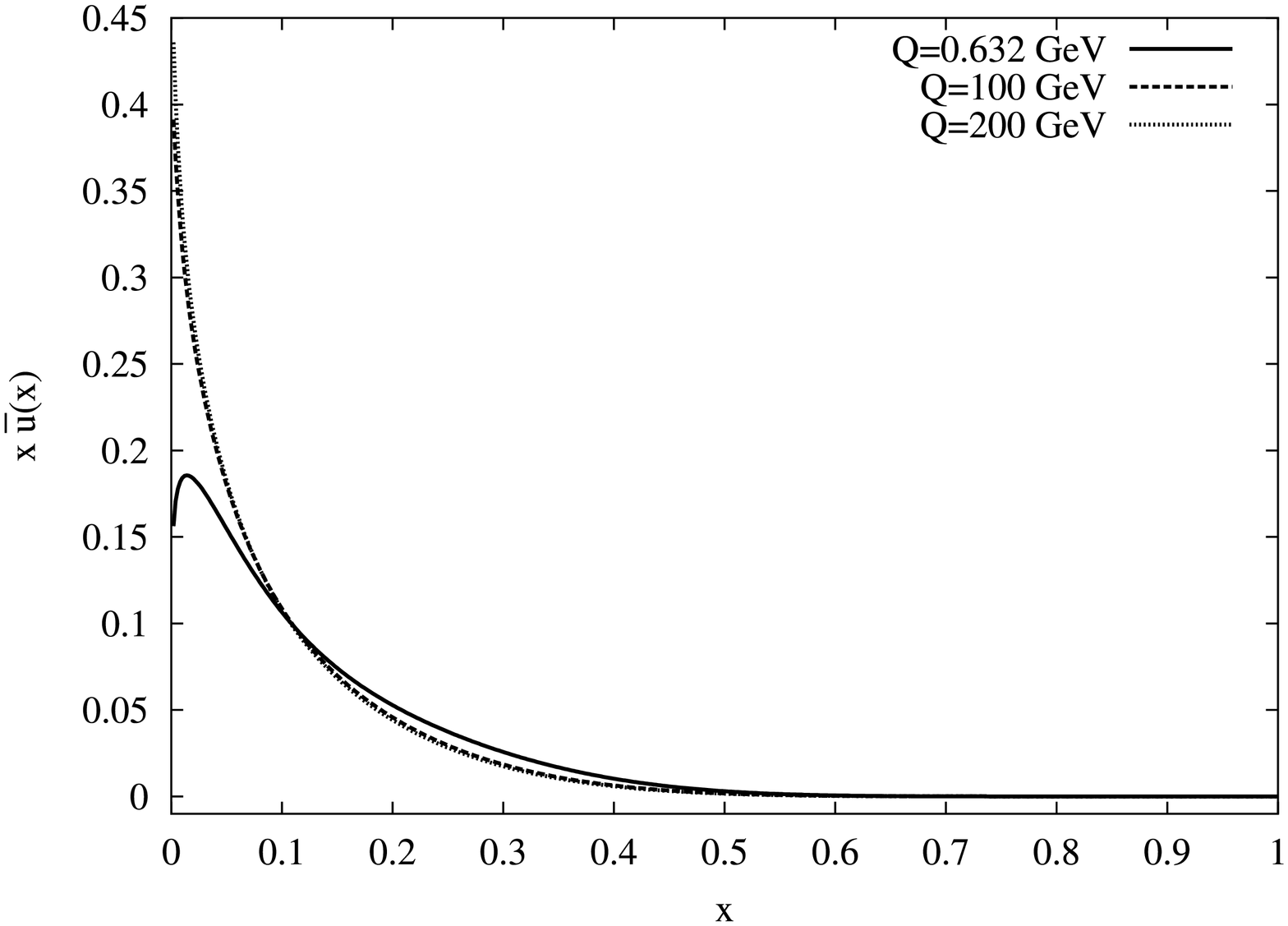}}} \par}

\caption{Evolution of the unpolarized antiquark up distribution \protect\( x\overline{u}\protect \)
versus \protect\( x\protect \) at various \protect\( Q\protect \)
values.}
\end{figure}

\begin{figure}[tbh]
{\centering \resizebox*{8cm}{!}{\rotatebox{-90}{\includegraphics{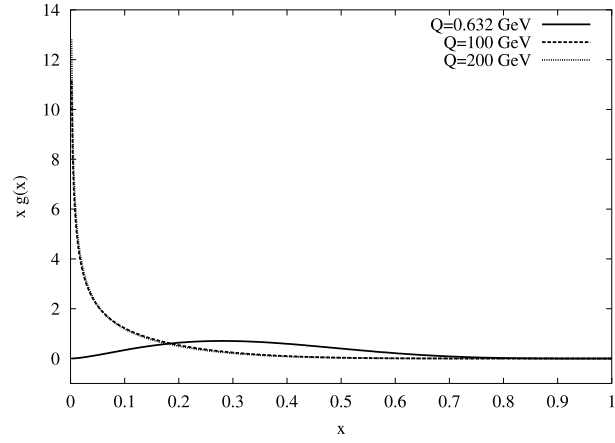}}} \par}

\caption{Evolution of the unpolarized gluon distribution \protect\( xg\protect \)
versus \protect\( x\protect \) at various \protect\( Q\protect \)
values.A}
\end{figure}

\begin{figure}[tbh]
{\centering \resizebox*{8cm}{!}{\rotatebox{-90}{\includegraphics{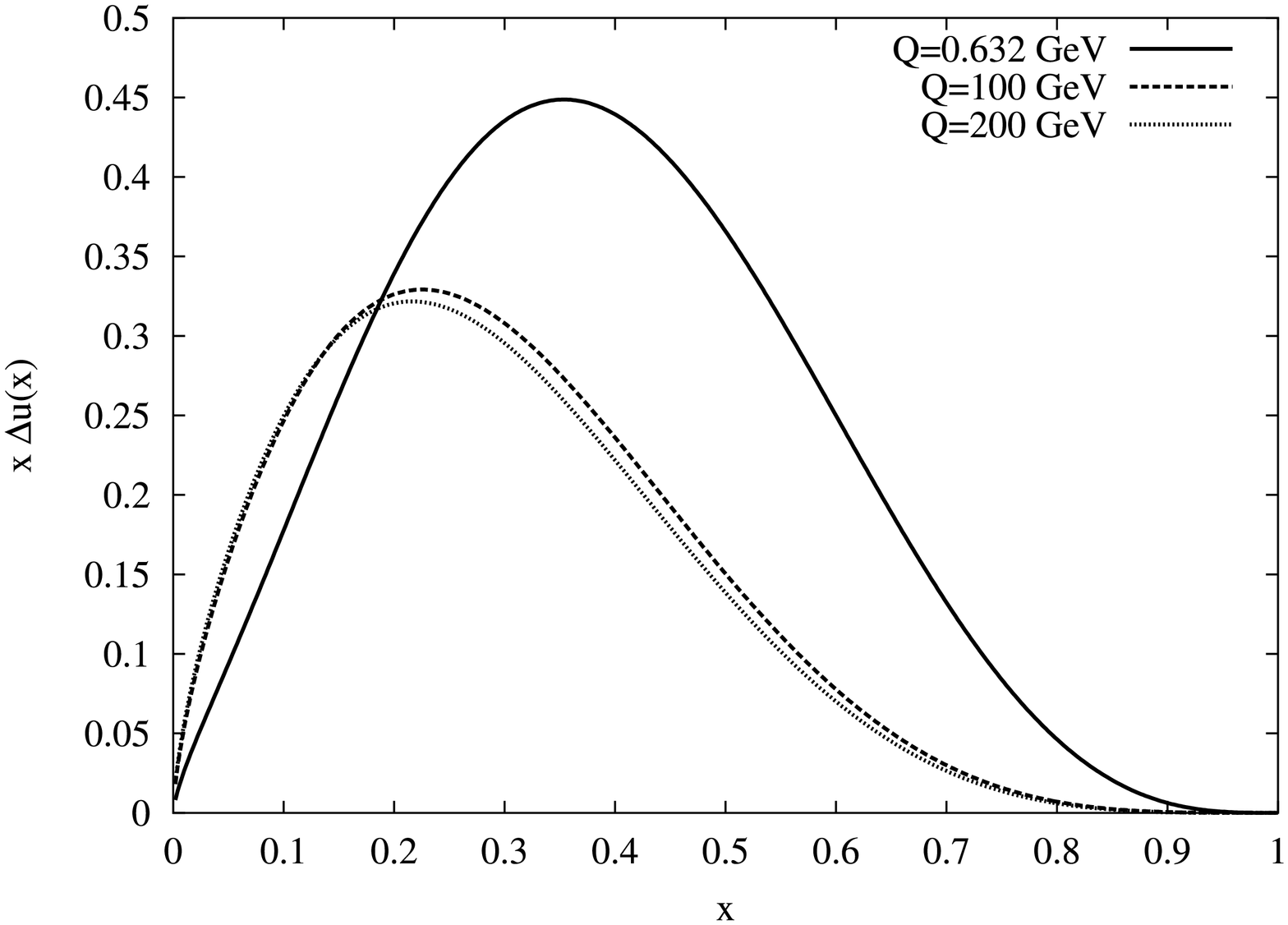}}} \par}

\caption{Evolution of the longitudinally polarized quark up distribution \protect\( x\Delta u\protect \)
versus \protect\( x\protect \) at various \protect\( Q\protect \)
values.\label{fig:Lu}}
\end{figure}

\begin{figure}[tbh]
{\centering \resizebox*{8cm}{!}{\rotatebox{-90}{\includegraphics{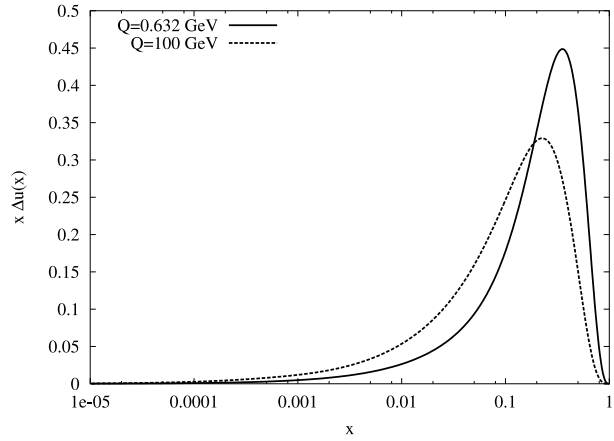}}} \par}

\caption{As in figure (\ref{fig:Lu}), but now the \protect\( x\protect \)-axis
is in logarithmic scale, to show the small-\protect\( x\protect \)
behavior.\label{fig:Lu_log}}
\end{figure}

\begin{figure}[tbh]
{\centering \resizebox*{8cm}{!}{\rotatebox{-90}{\includegraphics{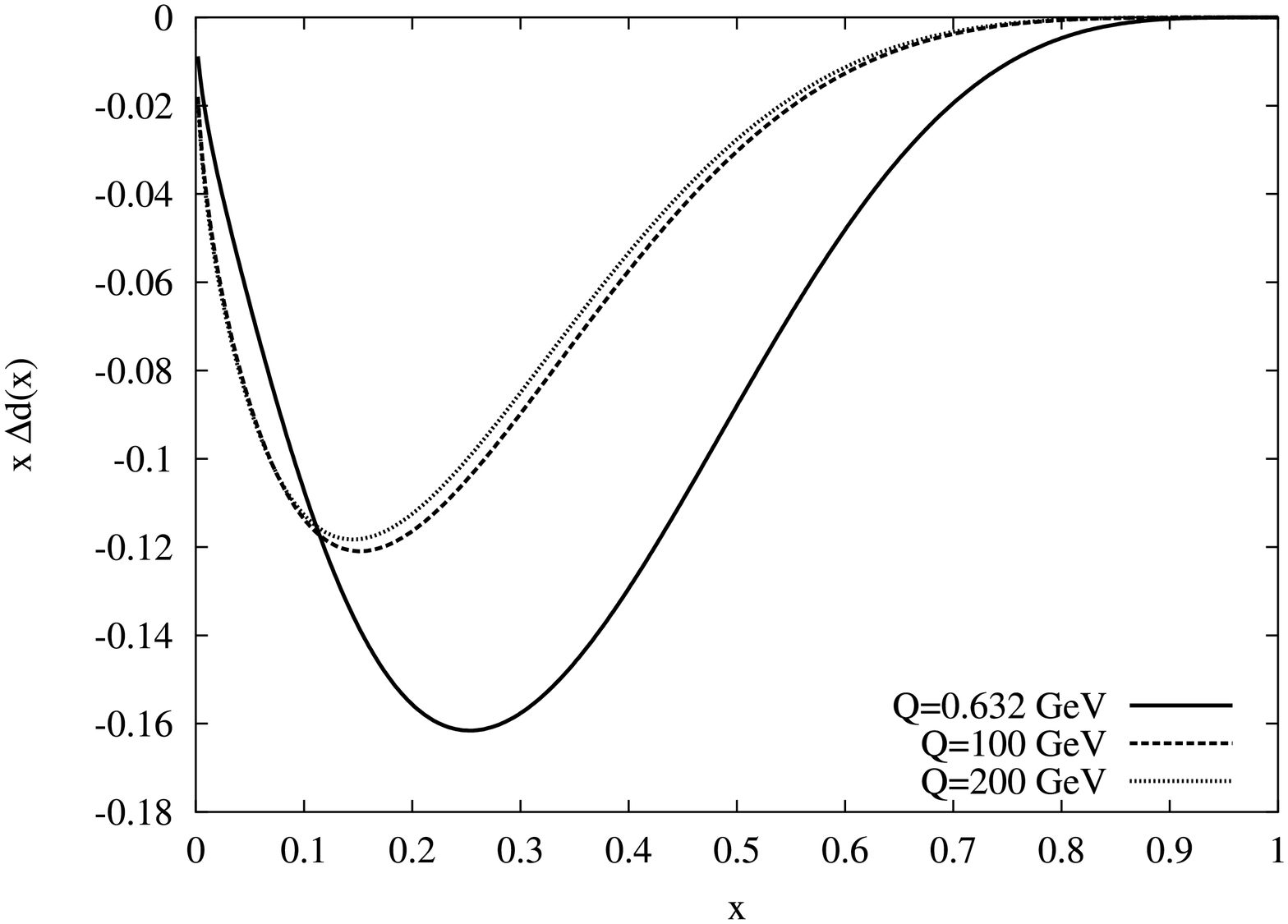}}} \par}

\caption{Evolution of \protect\( x\Delta d\protect \) versus \protect\( x\protect \)
at various \protect\( Q\protect \) values.}
\end{figure}

\begin{figure}[tbh]
{\centering \resizebox*{8cm}{!}{\rotatebox{-90}{\includegraphics{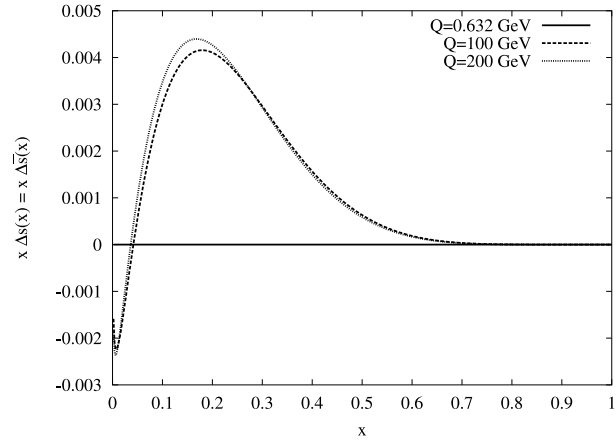}}} \par}

\caption{Evolution of \protect\( x\Delta s=x\Delta \overline{s}\protect \)
versus \protect\( x\protect \) at various \protect\( Q\protect \)
values.}
\end{figure}

\begin{figure}[tbh]
{\centering \resizebox*{8cm}{!}{\rotatebox{-90}{\includegraphics{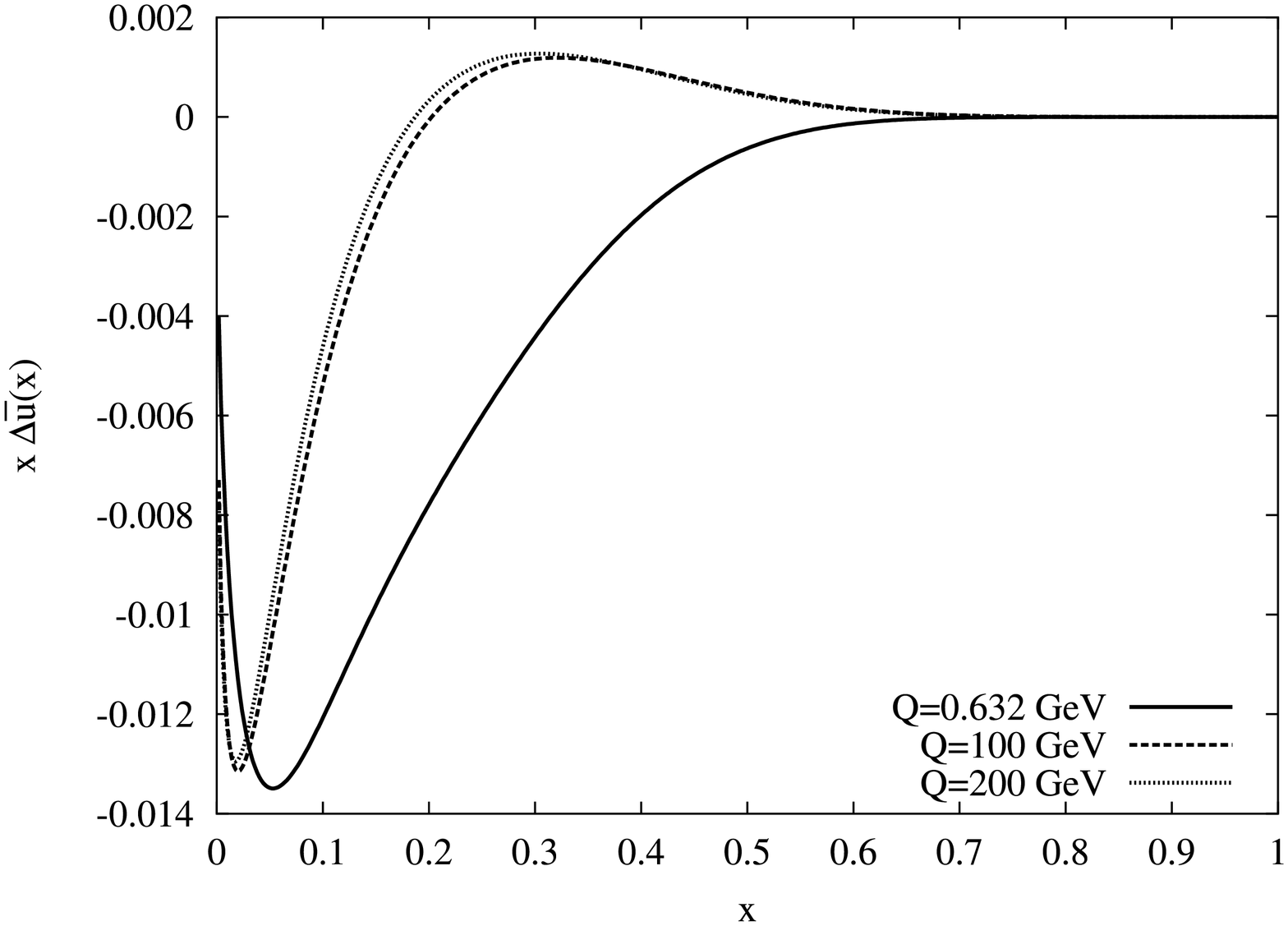}}} \par}

\caption{Evolution of the longitudinally polarized antiquark up distribution
\protect\( x\Delta \overline{u}\protect \) versus \protect\( x\protect \)
at various \protect\( Q\protect \) values.}
\end{figure}

\begin{figure}[tbh]
{\centering \resizebox*{8cm}{!}{\rotatebox{-90}{\includegraphics{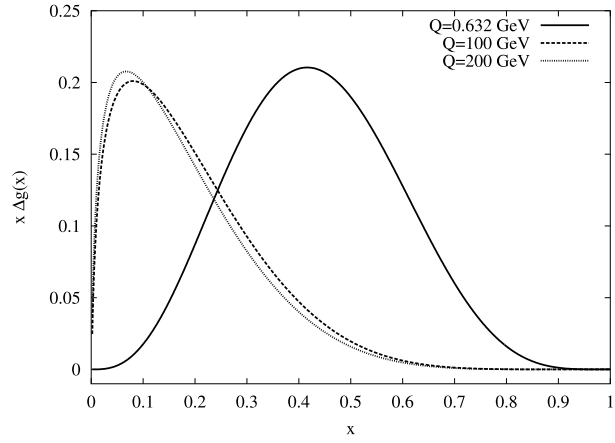}}} \par}

\caption{Evolution of the longitudinally polarized gluon distribution \protect\( x\Delta g\protect \)
versus \protect\( x\protect \) at various \protect\( Q\protect \)
values.\label{fig:Lg}}
\end{figure}

\begin{figure}[tbh]
{\centering \resizebox*{8cm}{!}{\rotatebox{-90}{\includegraphics{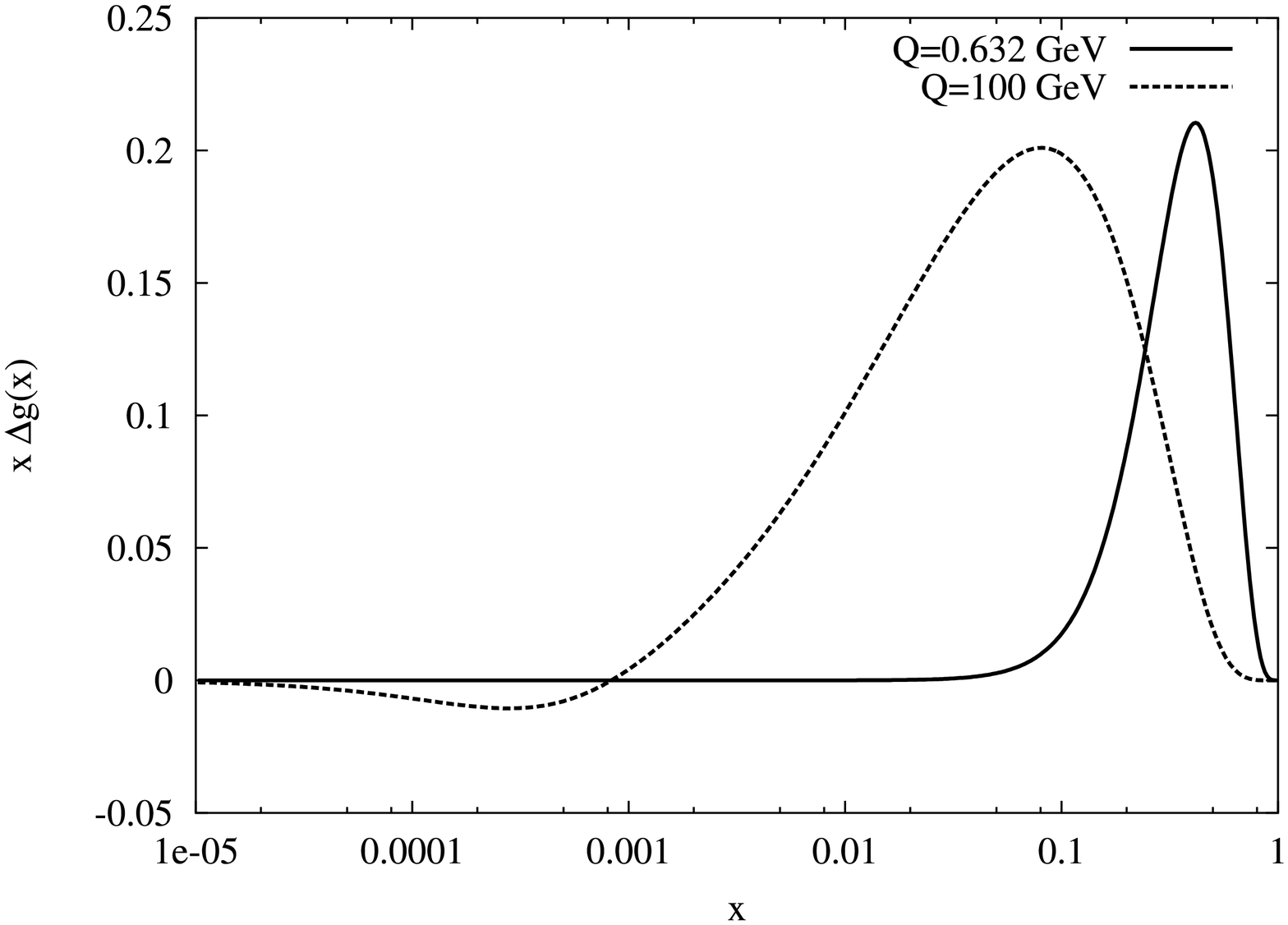}}} \par}

\caption{As in figure (\ref{fig:Lg}), but now the \protect\( x\protect \)-axis
is in logarithmic scale.\label{fig:Lg_log}}
\end{figure}

\begin{figure}[tbh]
{\centering \resizebox*{8cm}{!}{\rotatebox{-90}{\includegraphics{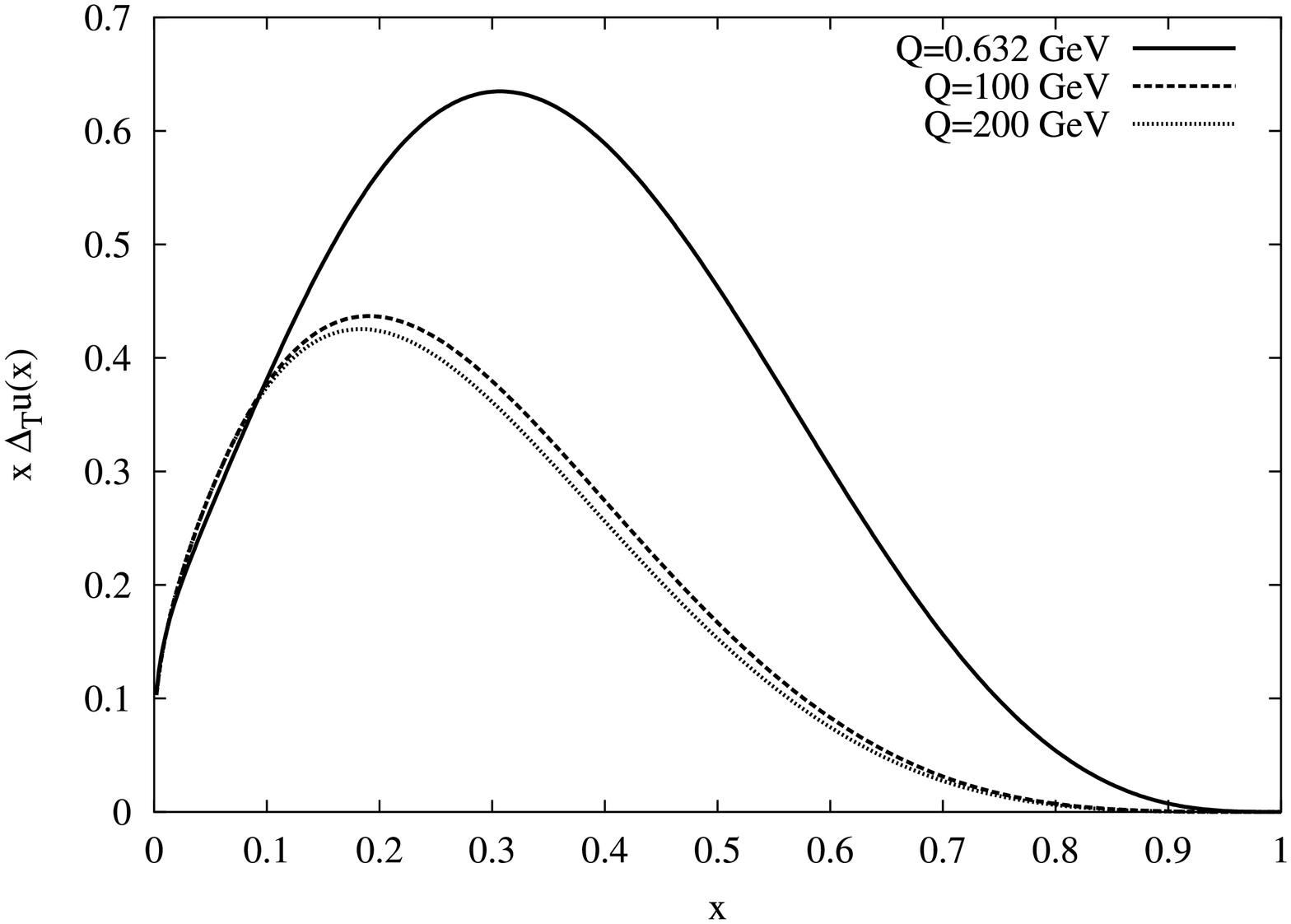}}} \par}

\caption{Evolution of the transversely polarized quark up distribution \protect\( x\Delta _{T}u\protect \)
versus \protect\( x\protect \) at various \protect\( Q\protect \)
values.}
\end{figure}

\begin{figure}[tbh]
{\centering \resizebox*{8cm}{!}{\rotatebox{-90}{\includegraphics{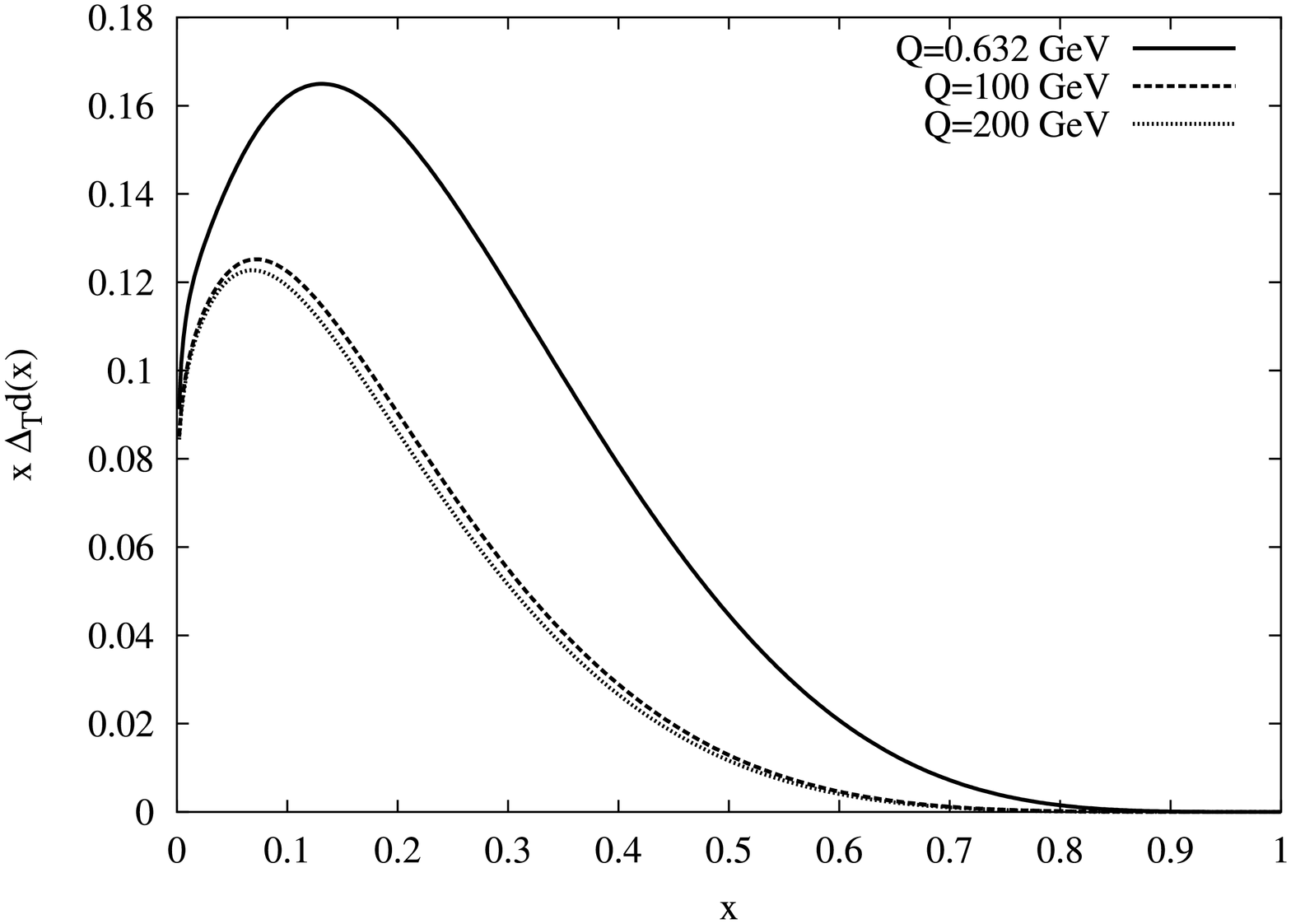}}} \par}

\caption{Evolution of \protect\( x\Delta _{T}d\protect \) versus \protect\( x\protect \)
at various \protect\( Q\protect \) values.}
\end{figure}

\begin{figure}[tbh]
{\centering \resizebox*{8cm}{!}{\rotatebox{-90}{\includegraphics{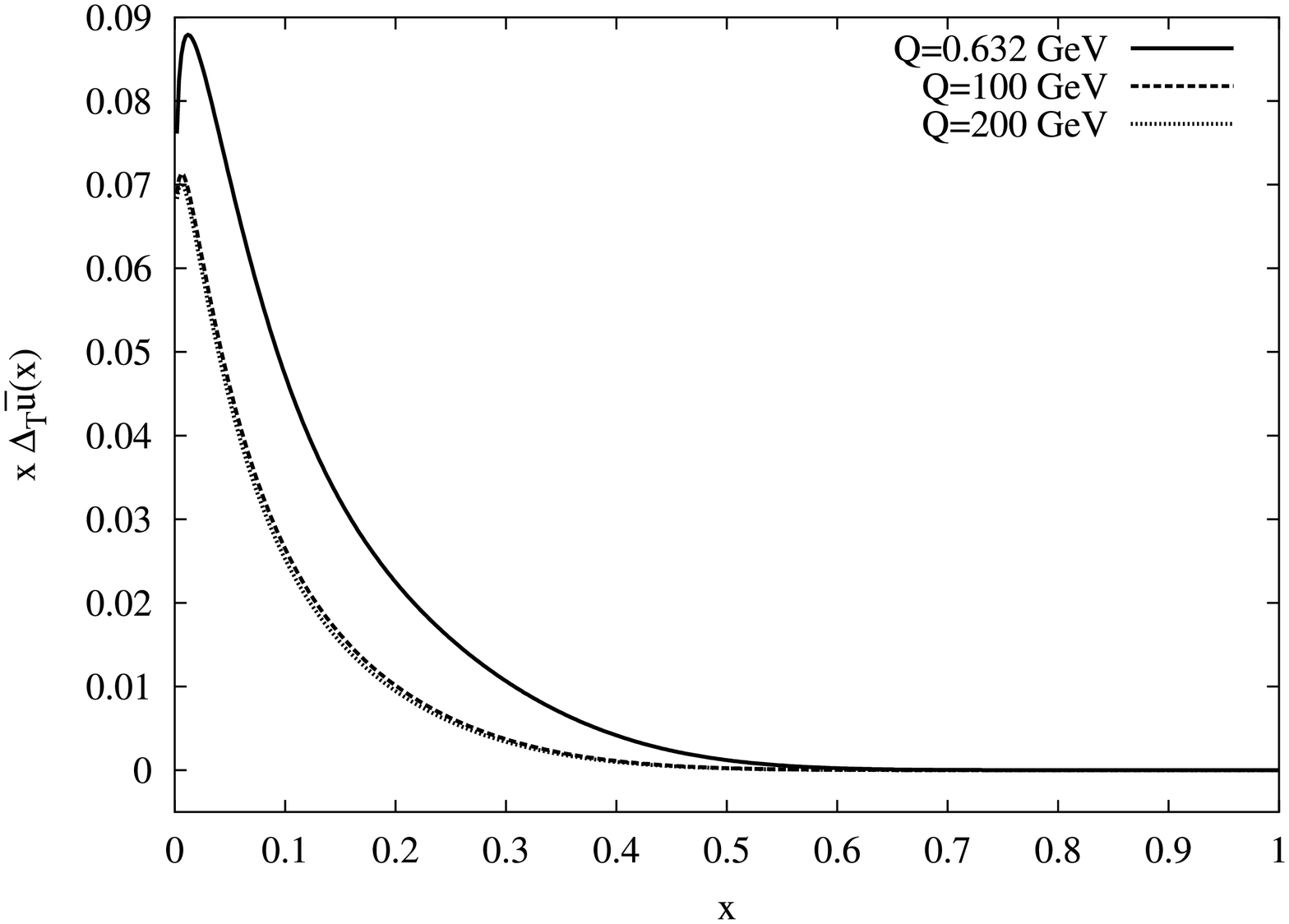}}} \par}

\caption{Evolution of the transversely polarized antiquark up distribution
\protect\( x\Delta _{T}\overline{u}\protect \) versus \protect\( x\protect \)
at various \protect\( Q\protect \) values.}
\end{figure}

\section{Unpolarized kernels}

The reference for the unpolarized kernels is \cite{FurmanskiPetronzio}, rearranged for our 
purposes. 
We remind that the plus distribution is defined by\begin{equation}
\int _{0}^{1}\textrm{d}x\frac{f(x)}{(1-x)_{+}}=\int _{0}^{1}\textrm{d}x\frac{f(x)-f(1)}{1-x}
\end{equation}
and the Spence function is\begin{equation}
S_{2}(x)=-2\textrm{Li}_{2}(-x)-2\log x\log (1+x)+\frac{1}{2}\log ^{2}x-\frac{\pi ^{2}}{6},
\end{equation}
where the dilogarithm is defined by\begin{equation}
\textrm{Li}_{2}(x)=\int _{x}^{0}\frac{\log (1-t)}{t}\textrm{d}t.
\end{equation}

\begin{equation}
P^{(0)}_{NS^{-}}(x)=P^{(0)}_{NS^{+}}(x)=P^{(0)}_{qq}(x)=C_{F}\left[ \frac{2}{(1-x)_{+}}-1-x+\frac{3}{2}\delta (1-x)\right] 
\end{equation}
\begin{equation}
P^{(0)}_{qg}(x)=2T_{f}\left[ x^{2}+(1-x)^{2}\right] 
\end{equation}
\begin{equation}
P^{(0)}_{gq}(x)=C_{F}\left[ \frac{1+(1-x)^{2}}{x}\right] 
\end{equation}
\begin{equation}
P^{(0)}_{gg}(x)=2N_{C}\left[ \frac{1}{(1-x)_{+}}+\frac{1}{x}-2+x(1-x)\right] +\frac{\beta (0)}{2}\delta (1-x)
\end{equation}

\begin{eqnarray}
P^{(1)}_{NS^{-}}(x) & = & \left\{ \frac{C_{F}}{18}\left[ 162C_{F}(x-1)+4T_{f}(11x-1)+N_{C}(89-223x+3\pi ^{2}(1+x))\right] \right\} \nonumber \\
 &  & +\left\{ \frac{C_{F}\left[ 30C_{F}-23N_{C}+4T_{f}+12C_{F}x+(N_{C}-24C_{F}+4T_{f})x^{2}\right] }{6(x-1)}\right\} \log x\nonumber \\
 &  & +\left\{ \frac{C_{F}\left[ C_{F}-N_{C}-(C_{F}+N_{C})x^{2}\right] }{2(x-1)}\right\} \log ^{2}x\nonumber \\
 &  & +\left\{ \frac{2C_{F}^{2}(1+x^{2})}{x-1}\right\} \log x\log (1-x)\nonumber \\
 &  & -\left\{ \frac{C_{F}(2C_{F}-N_{C})(1+x^{2})}{1+x}\right\} S_{2}(x)\nonumber \\
 &  & -\left\{ \frac{C_{F}}{9}\left[ N_{C}(3\pi ^{2}-67)+20T_{f}\right] \right\} \frac{1}{(1-x)_{+}}\nonumber \\
 &  & +\left\{ \frac{C_{F}}{72}\left[ N_{C}(51+44\pi ^{2}-216\zeta (3))-4T_{f}(3+4\pi ^{2})\right. \right. \nonumber \\
 &  & \qquad \qquad \left. \left. +9C_{F}(3-4\pi ^{2}+48\zeta (3))\right] \right\} \delta (1-x)
\end{eqnarray}

\begin{eqnarray}
P^{(1)}_{NS^{+}}(x) & = & \left\{ \frac{C_{F}}{18}\left[ 18C_{F}(x-1)+4T_{f}(11x-1)+N_{C}(17-151x+3\pi ^{2}(1+x))\right] \right\} \nonumber \\
 &  & +\left\{ \frac{C_{F}\left[ 6C_{F}(1+2x)-(11N_{C}-4T_{f})(1+x^{2})\right] }{6(x-1)}\right\} \log x\nonumber \\
 &  & +\left\{ \frac{C_{F}\left[ C_{F}-N_{C}-(C_{F}+N_{C})x^{2}\right] }{2(x-1)}\right\} \log ^{2}x\nonumber \\
 &  & +\left\{ \frac{2C_{F}^{2}(1+x^{2})}{x-1}\right\} \log x\log (1-x)\nonumber \\
 &  & +\left\{ \frac{C_{F}(2C_{F}-N_{C})(1+x^{2})}{1+x}\right\} S_{2}(x)\nonumber \\
 &  & -\left\{ \frac{C_{F}}{9}\left[ N_{C}(3\pi ^{2}-67)+20T_{f}\right] \right\} \frac{1}{(1-x)_{+}}\nonumber \\
 &  & +\left\{ \frac{C_{F}}{72}\left[ N_{C}(51+44\pi ^{2}-216\zeta (3))-4T_{f}(3+4\pi ^{2})\right. \right. \nonumber \\
 &  & \qquad \qquad \left. \left. +9C_{F}(3-4\pi ^{2}+48\zeta (3))\right] \right\} \delta (1-x)
\end{eqnarray}
\begin{eqnarray}
P^{(1)}_{qq}(x) & = & \frac{C_{F}}{18x}\left\{ x\left[ 18C_{F}(x-1)+N_{C}\left( 17-151x+3\pi ^{2}(1+x)\right) \right] \right. \nonumber \\
 &  & \qquad \quad \left. +4T_{f}\left[ 20-x\left( 19+x(56x-65)\right) \right] \right\} \nonumber \\
 &  & +\left\{ \frac{C_{F}\left[ 6C_{F}(1+2x)-11N_{C}(1+x^{2})+8T_{f}\left( 2x(2x(1+x)-3)-1\right) \right] }{6(x-1)}\right\} \log x\nonumber \\
 &  & +\left\{ \frac{C_{F}\left[ C_{F}-N_{C}+4T_{f}-(C_{F}+N_{C}+4T_{f})x^{2}\right] }{2(x-1)}\right\} \log ^{2}x\nonumber \\
 &  & +\left\{ \frac{2C_{F}^{2}(1+x^{2})}{x-1}\right\} \log x\log (1-x)\nonumber \\
 &  & +\left\{ \frac{C_{F}(2C_{F}-N_{C})(1+x^{2})}{1+x}\right\} S_{2}(x)\nonumber \\
 &  & -\left\{ \frac{C_{F}}{9}\left[ N_{C}(3\pi ^{2}-67)+20T_{f}\right] \right\} \frac{1}{(1-x)_{+}}\nonumber \\
 &  & +\left\{ \frac{C_{F}}{72}\left[ N_{C}(51+44\pi ^{2}-216\zeta (3))-4T_{f}(3+4\pi ^{2})\right. \right. \nonumber \\
 &  & \qquad \qquad \left. \left. +9C_{F}(3-4\pi ^{2}+48\zeta (3))\right] \right\} \delta (1-x)
\end{eqnarray}
\begin{eqnarray}
P^{(1)}_{qg}(x) & = & \left\{ \frac{1}{9x}\left[ T_{f}\left( 3C_{F}x(42-87x+60x^{2}-\pi ^{2}(2+4(x-1)x))\right. \right. \right. \nonumber \\
 &  & \qquad \qquad \left. \left. \left. +N_{C}(40+x(450x-36-436x^{2}+\pi ^{2}(3+6(x-1)x)))\right) \right] \right\} \nonumber \\
 &  & +\left\{ \frac{T_{f}}{3}\left[ 6N_{C}+8N_{C}x(6+11x)+3C_{F}(3-4x+8x^{2})\right] \right\} \log x\nonumber \\
 &  & +\left\{ 8(C_{F}-N_{C})T_{f}(1-x)x\right\} \log (1-x)\nonumber \\
 &  & +\left\{ T_{f}\left[ C_{F}(1-2x+4x^{2})-N_{C}(3+2x(3+x))\right] \right\} \log ^{2}x\nonumber \\
 &  & +\left\{ 2(C_{F}-N_{C})T_{f}\left[ 1+2(x-1)x\right] \right\} \log ^{2}(1-x)\nonumber \\
 &  & -\left\{ 4C_{F}T_{f}\left[ 1+2(x-1)x\right] \right\} \log x\log (1-x)\nonumber \\
 &  & +\left\{ 2N_{C}T_{f}\left[ 1+2x(1+x)\right] \right\} S_{2}(x)
\end{eqnarray}
\begin{eqnarray}
P^{(1)}_{gq}(x) & = & \left\{ \frac{1}{18x}\left[ C_{F}\left( N_{C}(18-3\pi ^{2}(2+(x-2)x)+2x(19+x(37+44x)))\right. \right. \right. \nonumber \\
 &  & \qquad \qquad \left. \left. \left. -9C_{F}x(5+7x)-16T_{f}(5+x(4x-5))\right) \right] \right\} \nonumber \\
 &  & +\left\{ \frac{C_{F}}{6}\left[ 3C_{F}(4+7x)-2N_{C}\left( 36+x(15+8x)\right) \right] \right\} \log x\nonumber \\
 &  & +\left\{ \frac{C_{F}}{3x}\left[ N_{C}\left( 22+x(17x-22)\right) -4T_{f}\left( 2+(x-2)x\right) \right. \right. \nonumber \\
 &  & \qquad \qquad \left. \left. -3C_{F}\left( 6+x(5x-6)\right) \right] \right\} \log (1-x)\nonumber \\
 &  & +\left\{ \frac{C_{F}}{2x}\left[ C_{F}(x-2)x+N_{C}\left( 2+3x(2+x)\right) \right] \right\} \log ^{2}x\nonumber \\
 &  & +\left\{ \frac{C_{F}(N_{C}-C_{F})\left[ 2+(x-2)x\right] }{x}\right\} \log ^{2}(1-x)\nonumber \\
 &  & -\left\{ \frac{2C_{F}N_{C}\left( 2+(x-2)x\right) }{x}\right\} \log x\log (1-x)\nonumber \\
 &  & -\left\{ \frac{C_{F}N_{C}\left( 2+x(2+x)\right) }{x}\right\} S_{2}(x)
\end{eqnarray}
\begin{eqnarray}
P^{(1)}_{gg}(x) & = & \left\{ \frac{1}{18x}\left[ 24C_{F}T_{f}(x-1)\left( x(11+5x)-1\right) +4N_{C}T_{f}\left( x(29+x(23x-19))-23\right) \right. \right. \nonumber \\
 &  & \qquad \quad \left. \left. +N_{C}^{2}\left( 6\pi ^{2}(x(2+(x-1)x)-1)-x(25+109x)\right) \right] \right\} \nonumber \\
 &  & +\left\{ \frac{N_{C}^{2}\left[ 11(1-4x)x-25\right] -4N_{C}T_{f}(1+x)-6C_{F}T_{f}(3+5x)}{3}\right\} \log x\nonumber \\
 &  & +\left\{ \frac{2C_{F}T_{f}x(x^{2}-1)+N_{C}^{2}\left[ 1+x\left( 2+x(3+(x-6)x)\right) \right] }{(1-x)x}\right\} \log ^{2}x\nonumber \\
 &  & +\left\{ \frac{4N_{C}^{2}\left[ 1+(x-1)x\right] ^{2}}{(x-1)x}\right\} \log x\log (1-x)\nonumber \\
 &  & -\left\{ \frac{2N_{C}^{2}\left( 1+x+x^{2}\right) ^{2}}{x(1+x)}\right\} S_{2}(x)\nonumber \\
 &  & -\left\{ \frac{N_{C}}{9}\left[ N_{C}(3\pi ^{2}-67)+20T_{f}\right] \right\} \frac{1}{(1-x)_{+}}\nonumber \\
 &  & +\left\{ \frac{N_{C}}{3}\left[ N_{C}(8+9\zeta (3))-4T_{f}\right] -C_{F}T_{f}\right\} \delta (1-x)
\end{eqnarray}

\section{Longitudinally polarized kernels}

The reference for the longitudinally polarized kernels is \cite{Vogelsang96}.

\begin{equation}
\Delta P^{(0)}_{NS^{-}}(x)=\Delta P^{(0)}_{NS^{+}}(x)=\Delta P^{(0)}_{qq}(x)=C_{F}\left[ \frac{2}{(1-x)_{+}}-1-x+\frac{3}{2}\delta (1-x)\right] 
\end{equation}
\begin{equation}
\Delta P_{qg}^{(0)}(x)=2T_{f}(2x-1)
\end{equation}
\begin{equation}
\Delta P_{gq}^{(0)}(x)=C_{F}(2-x)
\end{equation}
\begin{equation}
\Delta P_{gg}^{(0)}(x)=2N_{C}\left[ \frac{1}{(1-x)_{+}}-2x+1\right] +\frac{\beta (0)}{2}\delta (1-x)
\end{equation}

\begin{equation}
\Delta P^{(1)}_{NS^{-}}(x)=P^{(1)}_{NS^{+}}(x)
\end{equation}
\begin{equation}
\Delta P^{(1)}_{NS^{+}}(x)=P^{(1)}_{NS^{-}}(x)
\end{equation}
\begin{eqnarray}
\Delta P^{(1)}_{qq}(x) & = & \left\{ \frac{C_{F}}{18}\left[ 162C_{F}(x-1)+8T_{f}(4+z)+N_{C}\left( 89-223x+3\pi ^{2}(1+x)\right) \right] \right\} \nonumber \\
 &  & +\left\{ \frac{C_{F}}{6(x-1)}\left[ N_{C}(x^{2}-23)-6C_{F}(4x^{2}-2x-5)\right. \right. \nonumber \\
 &  & \qquad \qquad \qquad \left. \left. +8T_{f}\left( 2+x(5x-6)\right) \right] \right\} \log x\nonumber \\
 &  & +\left\{ \frac{C_{F}\left[ C_{F}-N_{C}+4T_{f}-(C_{F}+N_{C}+4T_{f})x^{2}\right] }{2(x-1)}\right\} \log ^{2}x\nonumber \\
 &  & +\left\{ \frac{2C_{F}^{2}(1+x^{2})}{x-1}\right\} \log x\log (1-x)\nonumber \\
 &  & -\left\{ \frac{C_{F}(2C_{F}-N_{C})(1+x^{2})}{1+x}\right\} S_{2}(x)\nonumber \\
 &  & -\left\{ \frac{C_{F}}{9}\left[ N_{C}(3\pi ^{2}-67)+20T_{f}\right] \right\} \frac{1}{(1-x)_{+}}\nonumber \\
 &  & +\left\{ \frac{C_{F}}{72}\left[ N_{C}(51+44\pi ^{2}-216\zeta (3))-4T_{f}(3+4\pi ^{2})\right. \right. \nonumber \\
 &  & \qquad \qquad \left. \left. +9C_{F}(3-4\pi ^{2}+48\zeta (3))\right] \right\} \delta (1-x)
\end{eqnarray}
\begin{eqnarray}
\Delta P^{(1)}_{qg}(x) & = & \left\{ \frac{T_{f}}{3}\left[ C_{F}\left( \pi ^{2}(2-4x)-66+81x\right) +N_{C}\left( 72-66x+\pi ^{2}(2x-1)\right) \right] \right\} \nonumber \\
 &  & +\left\{ T_{f}\left[ 2N_{C}(1+8x)-9C_{F}\right] \right\} \log x\nonumber \\
 &  & +\left\{ 8(N_{C}-C_{F})T_{f}(x-1)\right\} \log (1-x)\nonumber \\
 &  & +\left\{ T_{f}\left[ C_{F}(2x-1)-3N_{C}(1+2x)\right] \right\} \log ^{2}x\nonumber \\
 &  & +\left\{ 2(C_{F}-N_{C})T_{f}(2x-1)\right\} \log ^{2}(1-x)\nonumber \\
 &  & +\left\{ 4C_{F}T_{f}(1-2x)\right\} \log x\log (1-x)\nonumber \\
 &  & +\left\{ 2N_{C}T_{f}(1+2x)\right\} S_{2}(x)
\end{eqnarray}
\begin{eqnarray}
\Delta P^{(1)}_{gq}(x) & = & \left\{ \frac{C_{F}}{18}\left[ 9C_{F}(8x-17)-8T_{f}(4+x)+N_{C}\left( 82+3\pi ^{2}(x-2)+70x\right) \right] \right\} \nonumber \\
 &  & +\left\{ \frac{C_{F}}{2}\left[ N_{C}(8-26x)+C_{F}(x-4)\right] \right\} \log x\nonumber \\
 &  & +\left\{ \frac{C_{F}}{3}\left[ 4T_{f}(x-2)-3C_{F}(2+x)+N_{C}(10+x)\right] \right\} \log (1-x)\nonumber \\
 &  & +\left\{ \frac{C_{F}}{2}\left[ 3N_{C}(2+x)-C_{F}(x-2)\right] \right\} \log ^{2}x\nonumber \\
 &  & +\left\{ C_{F}(C_{F}-N_{C})(x-2)\right\} \log ^{2}(1-x)\nonumber \\
 &  & +\left\{ 2C_{F}N_{C}(x-2)\right\} \log x\log (1-x)\nonumber \\
 &  & -\left\{ C_{F}N_{C}(2+x)\right\} S_{2}(x)
\end{eqnarray}
\begin{eqnarray}
\Delta P^{(1)}_{gg}(x) & = & \left\{ \frac{1}{18}\left[ 180C_{F}T_{f}(x-1)+8N_{C}T_{f}(19x-4)+N_{C}^{2}\left( 6\pi ^{2}(1+2x)-305-97x\right) \right] \right\} \nonumber \\
 &  & +\left\{ \frac{1}{3}\left[ N_{C}^{2}(29-67x)+6C_{F}T_{f}(x-5)-4N_{C}T_{f}(1+x)\right] \right\} \log x\nonumber \\
 &  & +\left\{ \frac{N_{C}^{2}(2x^{2}+x-4)-2C_{F}T_{f}(x^{2}-1)}{x-1}\right\} \log ^{2}x\nonumber \\
 &  & +\left\{ \frac{4N_{C}^{2}x(2x-1)}{x-1}\right\} \log x\log (1-x)\nonumber \\
 &  & -\left\{ \frac{2N_{C}^{2}x(1+2x)}{1+x}\right\} S_{2}(x)\nonumber \\
 &  & -\left\{ \frac{N_{C}}{9}\left[ N_{C}(3\pi ^{2}-67)+20T_{f}\right] \right\} \frac{1}{(1-x)_{+}}\nonumber \\
 &  & +\left\{ \frac{1}{3}\left[ N_{C}^{2}(8+9\zeta (3))-3C_{F}T_{f}-4N_{C}T_{f}\right] \right\} \delta (1-x)
\end{eqnarray}

\section{Transversely polarized kernels}

The reference for the transversely polarized kernels is \cite{Vogelsang98}.

\begin{equation}
\Delta _{T}P^{(0)}_{NS^{-}}(x)=\Delta _{T}P^{(0)}_{NS^{+}}(x)=C_{F}\left[ \frac{2}{(1-x)_{+}}-2+\frac{3}{2}\delta (1-x)\right] 
\end{equation}

\begin{eqnarray}
\Delta _{T}P_{NS^{-}}^{(1)} & = & \left\{ \frac{C_{F}}{9}\left[ 20T_{f}-18C_{F}(x-1)+N_{C}(9x-76+3\pi ^{2})\right] \right\} \nonumber \\
 &  & +\left\{ \frac{C_{F}(9C_{F}-11N_{C}+4T_{f})x}{3(x-1)}\right\} \log x\nonumber \\
 &  & +\left\{ \frac{C_{F}N_{C}x}{1-x}\right\} \log ^{2}x\nonumber \\
 &  & +\left\{ \frac{4C_{F}^{2}x}{x-1}\right\} \log x\log (1-x)\nonumber \\
 &  & +\left\{ \frac{2C_{F}(2C_{F}-N_{C})x}{1+x}\right\} S_{2}(x)\nonumber \\
 &  & -\left\{ \frac{C_{F}}{9}\left[ N_{C}(3\pi ^{2}-67)+20T_{f}\right] \right\} \frac{1}{(1-x)_{+}}\nonumber \\
 &  & +\left\{ \frac{C_{F}}{72}\left[ N_{C}(51+44\pi ^{2}-216\zeta (3))-4T_{f}(3+4\pi ^{2})\right. \right. \nonumber \\
 &  & \qquad \qquad \left. \left. +9C_{F}(3-4\pi ^{2}+48\zeta (3))\right] \right\} \delta (1-x)
\end{eqnarray}
\begin{eqnarray}
\Delta _{T}P_{NS^{+}}^{(1)} & = & \left\{ \frac{C_{F}}{9}\left[ N_{C}(3\pi ^{2}-67)+20T_{f}\right] \right\} \nonumber \\
 &  & +\left\{ \frac{C_{F}(9C_{F}-11N_{C}+4T_{f})x}{3(x-1)}\right\} \log x\nonumber \\
 &  & +\left\{ \frac{C_{F}N_{C}x}{1-x}\right\} \log ^{2}x\nonumber \\
 &  & +\left\{ \frac{4C_{F}^{2}x}{x-1}\right\} \log x\log (1-x)\nonumber \\
 &  & +\left\{ \frac{2C_{F}(N_{C}-2C_{F})x}{1+x}\right\} S_{2}(x)\nonumber \\
 &  & -\left\{ \frac{C_{F}}{9}\left[ N_{C}(3\pi ^{2}-67)+20T_{f}\right] \right\} \frac{1}{(1-x)_{+}}\nonumber \\
 &  & +\left\{ \frac{C_{F}}{72}\left[ N_{C}(51+44\pi ^{2}-216\zeta (3))-4T_{f}(3+4\pi ^{2})\right. \right. \nonumber \\
 &  & \qquad \qquad \left. \left. +9C_{F}(3-4\pi ^{2}+48\zeta (3))\right] \right\} \delta (1-x)
\end{eqnarray}

\section{Conclusions}
We have illustrated and documented a method for solving 
in a rather fast way the NLO evolution equations 
for the parton distributions at leading twist. 
The advantages of the method compared to other implementations 
based on the inversion of the Mellin moments, 
as usually done in the case of QCD, 
are rather evident. 
We have also shown how Rossi's ansatz, originally formulated in the case 
of the photon structure function, relates to the solution of 
DGLAP equations formulated in terms of moments. 
The running time of the implementation is truly modest, even for a large 
number of iterations, and allows to get a very good accuracy.

\chapter{NNLO extension of the solution of the\\Renormalization Group Equations\label{chap:NNLOcode}}
\fancyhead[LO]{\nouppercase{Chapter 2. NNLO extension of the solution of the RGE}}

\section{Introduction to the Chapter} 

We have seen in Chapter 1 that renormalization group equations play a key role in the computation of cross 
sections in hadronic collisions. We have also seen that the dependence on the factorization scale of the 
factorization formula is largely reduced when higher order corrections in $\alpha_S$ are kept into account. 
As we have already discussed, the use of efficient algorithms for the solution of these equations is of 
remarkable relevance in order to efficiently compare the theory with the experiments. As we increase the 
collision energy in a scattering process, the factorization scale has a wider interval allowed 
for its variation. It is of some help to summarize this issue in a simple but rather clarifying way. 

Consider a proton proton collision at LHC energy (these can cover a range between few TeV's and 14 TeV) 
and let us assume that we use the parton model description of the process based on the factorization formula. 
As we have mentioned, $Q$, the factorization scale (sometimes also called ``$\mu_f$'') is a fraction of the total 
energy available in the center of mass frame of the two colliding beams. In order to decide on the most 
accurate value for $\mu_f$ so to come up with a prediction, let's say, for a multi jet cross section, 
we have to look at the average $p_T$ of the final state jet and use that energy value as an indication for the 
underlying value of $\mu_f$. Of course this procedure is approximate and requires a study of the behaviour of the 
final result for the cross section in terms of various choices of $\mu_f$. A drastic reduction on the 
dependence of the result on the value of $\mu_f$ is obtained if we include higher order corrections. As we have 
already stressed before, both the hard scatterings and the evolution of the parton distributions 
have to be accurate enough for this reduced sensitivity to emerge. In general the computation 
of hard scatterings, which are process dependent, are laborious enough to be limited to a specific set of 
``golden plated modes'', such as Drell-Yan lepton pair production and few more, and require difficult 
technical studies by various research groups to be performed. For instance, in the case of the Higgs total cross section and 
of the cross section for the production of a Higgs particle with an associated gauge boson at the LHC, considerable progress has been done in the last few years. On the other hand the computation of the kernels of the renormalization group equations 
for the parton distributions has also to be known at the same level of accuracy as the hard scatterings. 
Therefore it is imperative, especially for the detection of the Higgs boson at LHC energy, 
to be able to move from the NLO to the NNLO case in the study of some ``golden plated modes'' involving 
this particle. It is unlikely that in the near future most of the NLO hard scatterings will be extended to 
include their NNLO corrections, but, as we have mentioned, in the case of the Higgs this process is under 
intensive investigation. The computation of the NNLO evolution kernels, which are process independent, has been completed quite recently and has set a landmark on the applicability of perturbative QCD to hadron colliders. 
The numerical study of these kernels is a nontrivial task and the accurate solution of the associated RGE's is also nontrivial. 

In this second chapter our aim is to extend up to NNLO the results of chapter \ref{chap:NLOcode} using a similar methodology in order to solve these equations. At this time our numerical implementation 
is the second code which is able to evolve 
parton distributions up to this level of accuracy. It is obvious that most of the discussion in this chapter overlaps with the  methods of chapter \ref{chap:NLOcode}, but with some key differences that we will underline. 
The new implementation contains new features which are not present in our previous discussion since the new NNLO 
kernels are far more complex than those known at NLO. These issues are discussed in some detail in this and in the next chapter.

After the submission of this thesis, the preliminar work presented in this chapter has been completed and published in \cite{NNLOlogarithmic}.

\section{Definitions and Conventions}

In this section we present our definitions and conventions, which are similar to those of chapter 
\ref{chap:NLOcode} with some important modifications.
We introduce the 3loop evolution of the coupling 
\cite{alpha_s}\begin{equation}
\alpha_{s}(Q^{2})=\frac{4\pi}{\beta_{0}L}\left\{ 1-\frac{\beta_{1}}{\beta_{0}^{2}}\frac{\log L}{L}+\frac{1}{\beta_{0}^{3}L^{2}}\left[\frac{\beta_{1}^{2}}{\beta_{0}}\left(\log^{2}L-\log L-1\right)+\beta_{2}\right]+O\left(\frac{1}{L^{3}}\right)\right\} ,\label{eq:alpha_s_nnlo}\end{equation}
where\begin{equation}
L=\log\frac{Q^{2}}{\Lambda_{\overline{MS}}^{2}},\end{equation}
and the beta function is defined by\begin{equation}
\beta(\alpha_{s})=\frac{\textrm{d}\alpha_{s}(Q^{2})}{\textrm{d}\log Q^{2}},\label{eq:beta_def}\end{equation}
and its three-loop expansion \cite{betafunction} is\begin{equation}
\beta(\alpha_{s})=-\frac{\beta_{0}}{4\pi}\alpha_{s}^{2}-\frac{\beta_{1}}{16\pi^{2}}\alpha_{s}^{3}-\frac{\beta_{2}}{64\pi^{3}}\alpha_{s}^{4}+O(\alpha_{s}^{5}),\label{eq:beta_exp}\end{equation}
where\begin{equation}
\beta_{0}=\frac{11}{3}N_{C}-\frac{4}{3}T_{f},\end{equation}
\begin{equation}
\beta_{1}=\frac{34}{3}N_{C}^{2}-\frac{10}{3}N_{C}n_{f}-2C_{F}n_{f},\end{equation}
 \begin{equation}
\beta_{2}=\frac{2857}{54}N_{C}^{3}+2C_{F}^{2}T_{f}-\frac{205}{9}C_{F}N_{C}T_{f}-\frac{1415}{27}N_{C}^{2}T_{f}+\frac{44}{9}C_{F}T_{f}^{2}+\frac{158}{27}N_{C}T_{f}^{2}\end{equation}
and\begin{equation}
N_{C}=3,\qquad C_{F}=\frac{N_{C}^{2}-1}{2N_{C}}=\frac{4}{3},\qquad T_{f}=T_{R}n_{f}=\frac{1}{2}n_{f},\end{equation}
where $N_{C}$ is the number of colors, $n_{f}$ is the number of
active flavors, that is fixed by the number of quarks with $m_{q}\leq Q$;
$\Lambda_{\overline{MS}}^{(n_{f})}$ is calculated using the known
value of $\alpha_{s}(m_{Z})$ and imposing the continuity of $\alpha_{s}$
at the quark masses thresholds.
We recall that the perturbative expansion of the kernels now includes the NNLo contributions 
is\begin{equation}
P(x,\alpha_{s})=\left(\frac{\alpha_{s}}{2\pi}\right)P^{(0)}(x)+\left(\frac{\alpha_{s}}{2\pi}\right)^{2}P^{(1)}(x)+\left(\frac{\alpha_{s}}{2\pi}\right)^{3}P^{(2)}(x)+\ldots.\label{eq:kernel_expansion}\end{equation} 
whose specific form can be found in the original literature \cite{NNLO_nonsinglet,NNLO_singlet}. 

We solve Eq. (\ref{eq:DGLAP}) directly in $x$-space, 
assuming a solution of the form\begin{eqnarray}
f(x,Q^{2}) & = & \sum_{n=0}^{\infty}\frac{A_{n}(x)}{n!}\log^{n}\frac{\alpha_{s}(Q^{2})}{\alpha_{s}(Q_{0}^{2})}+\alpha_{s}(Q^{2})\sum_{n=0}^{\infty}\frac{B_{n}(x)}{n!}\log^{n}\frac{\alpha_{s}(Q^{2})}{\alpha_{s}(Q_{0}^{2})}\nonumber \\
 &  & +\left(\alpha_{s}(Q^{2})\right)^{2}\sum_{n=0}^{\infty}\frac{C_{n}(x)}{n!}\log^{n}\frac{\alpha_{s}(Q^{2})}{\alpha_{s}(Q_{0}^{2})}\label{eq:ansatz2}\end{eqnarray}
for each parton distribution $f$, where $Q_{0}$ defines the initial
evolution scale. 

As in chapter \ref{chap:NLOcode}, also in this case we derive the following recursion relations for the coefficients
$A_{n}$, $B_{n}$ and $C_{n}$ (see the frame below for details)\begin{equation}
A_{n+1}(x)=-\frac{2}{\beta_{0}}P^{(0)}(x)\otimes A_{n}(x),\label{eq:An_recurrence2}\end{equation}
\begin{equation}
B_{n+1}(x)=-B_{n}(x)-\frac{\beta_{1}}{4\pi\beta_{0}}A_{n+1}(x)-\frac{2}{\beta_{0}}P^{(0)}(x)\otimes B_{n}(x)-\frac{1}{\pi\beta_{0}}P^{(1)}(x)\otimes A_{n}(x),\label{eq:Bn_recurrence2}\end{equation}

\begin{eqnarray}
C_{n+1}(x) & = & -2C_{n}(x)-\frac{\beta_{1}}{4\pi\beta_{0}}B_{n}(x)-\frac{\beta_{1}}{4\pi\beta_{0}}B_{n+1}(x)-\frac{\beta_{2}}{16\pi^{2}\beta_{0}}A_{n+1}(x)\nonumber \\
 &  & -\frac{2}{\beta_{0}}P^{(0)}(x)\otimes C_{n}(x)-\frac{1}{\pi\beta_{0}}P^{(1)}(x)\otimes B_{n}(x)\nonumber \\
 &  & -\frac{1}{2\pi^{2}\beta_{0}}P^{(2)}(x)\otimes A_{n}(x).\label{eq:Cn_recurrence}\end{eqnarray}

\begin{framed}

\textbf{\textit{Derivation of the recursion relations and renormalization scale dependence.}}

We introduce the shortcut notation\begin{equation}
L(Q^{2})=\log\frac{\alpha_{s}(Q^{2})}{\alpha_{s}(Q_{0}^{2})}\end{equation}
and, making use of the beta function definition (\ref{eq:beta_def}),
we compute its derivative\begin{equation}
\frac{\textrm{d}L(Q^{2})}{\textrm{d}\log Q^{2}}=\frac{\alpha_{s}(Q_{0}^{2})}{\alpha_{s}(Q^{2})}\frac{\textrm{d}}{\textrm{d}\log Q^{2}}\frac{\alpha_{s}(Q^{2})}{\alpha_{s}(Q_{0}^{2})}=\frac{1}{\alpha_{s}(Q^{2})}\frac{\textrm{d}\alpha_{s}(Q^{2})}{\textrm{d}\log Q^{2}}=\frac{\beta(\alpha_{s})}{\alpha_{s}(Q^{2})}\end{equation}
Inserting our ansatz (\ref{eq:ansatz2}) for the solution into the
DGLAP equation (\ref{eq:DGLAP}) we get for the LHS\begin{eqnarray}
 &  & \sum_{n=1}^{\infty}\left\{ \frac{A_{n}(x)}{n!}nL^{n-1}\frac{\beta(\alpha_{s})}{\alpha_{s}}+\alpha_{s}\frac{B_{n}(x)}{n!}nL^{n-1}\frac{\beta(\alpha_{s})}{\alpha_{s}}\right.\nonumber \\
 &  & \left.\qquad+\alpha_{s}^{2}\frac{C_{n}(x)}{n!}nL^{n-1}\frac{\beta(\alpha_{s})}{\alpha_{s}}\right\} \nonumber \\
 &  & +\sum_{n=0}^{\infty}\left\{ \beta(\alpha_{s})\frac{B_{n}(x)}{n!}L^{n}+2\alpha_{s}\beta(\alpha_{s})\frac{C_{n}(x)}{n!}L^{n}\right\} .\end{eqnarray}
Note that the first sum starts at $n=1$, because the $n=0$ term
in (\ref{eq:ansatz2}) does not have $Q^{2}$ dependence. Tranforming
$n\rightarrow n-1$ in the first sum, using the three-loop expansion
of the beta function (\ref{eq:beta_exp}) and neglecting all terms
of order $\alpha_{s}^{4}$ or more, the previous formula becomes\begin{eqnarray}
 &  & \sum_{n=0}^{\infty}\left\{ \frac{A_{n+1}(x)}{n!}L^{n}\left(-\frac{\beta_{0}}{4\pi}\alpha_{s}-\frac{\beta_{1}}{16\pi^{2}}\alpha_{s}^{2}-\frac{\beta_{2}}{64\pi^{3}}\alpha_{s}^{3}\right)\right.\nonumber \\
 &  & +\frac{B_{n+1}(x)}{n!}L^{n}\left(-\frac{\beta_{0}}{4\pi}\alpha_{s}^{2}-\frac{\beta_{1}}{16\pi^{2}}\alpha_{s}^{3}\right)+\frac{C_{n+1}(x)}{n!}L^{n}\left(-\frac{\beta_{0}}{4\pi}\alpha_{s}^{3}\right)\nonumber \\
 &  & \left.+\frac{B_{n}(x)}{n!}L^{n}\left(-\frac{\beta_{0}}{4\pi}\alpha_{s}^{2}-\frac{\beta_{1}}{16\pi^{2}}\alpha_{s}^{3}\right)+2\frac{C_{n}(x)}{n!}L^{n}\left(-\frac{\beta_{0}}{4\pi}\alpha_{s}^{3}\right)\right\} .\label{eq:recrelLHS}\end{eqnarray}
Using the kernel expansion (\ref{eq:kernel_expansion}), we get for
the RHS\begin{eqnarray}
 &  & \sum_{n=0}^{\infty}\frac{L^{n}}{n!}\left\{ \frac{\alpha_{s}}{2\pi}\left[P^{(0)}\otimes A_{n}\right](x)+\frac{\alpha_{s}^{2}}{4\pi^{2}}\left[P^{(1)}\otimes A_{n}\right](x)\right.\nonumber \\
 &  & \quad\quad\quad+\frac{\alpha_{s}^{3}}{8\pi^{3}}\left[P^{(2)}\otimes A_{n}\right](x)+\frac{\alpha_{s}^{2}}{2\pi}\left[P^{(0)}\otimes B_{n}\right](x)\nonumber \\
 &  & \left.\quad\quad\quad+\frac{\alpha_{s}^{3}}{4\pi^{2}}\left[P^{(1)}\otimes B_{n}\right](x)+\frac{\alpha_{s}^{3}}{2\pi}\left[P^{(0)}\otimes C_{n}\right](x)\right\} .\label{eq:recrelRHS}\end{eqnarray}
Equating (\ref{eq:recrelLHS}) and (\ref{eq:recrelRHS}) term by term
and grouping the terms proportional respectively to $\alpha_{s}$,
$\alpha_{s}^{2}$ and $\alpha_{s}^{3}$ we get the three desired recursion
relations (\ref{eq:An_recurrence2}), (\ref{eq:Bn_recurrence2}) and
(\ref{eq:Cn_recurrence}).

\end{framed}

Setting $Q=Q_{0}$ in (\ref{eq:ansatz2}) we get\begin{equation}
f(x,Q_{0}^{2})=A_{0}(x)+\alpha_{s}(Q_{0}^{2})B_{0}(x)+\left(\alpha_{s}(Q^{2})\right)^{2}C_{0}(x).\label{eq:boundary2}\end{equation}
Any boundary condition satisfying (\ref{eq:boundary2}) can be chosen
at the lowest scale $Q_{0}$ and in our case we choose\begin{equation}
B_{0}(x)=C_{0}(x)=0,\qquad f(x,Q_{0}^{2})=A_{0}(x).\end{equation}
What the program does is starting with a parameterized form of the
parton distribution functions at a low energy scale $Q_{0}$ (typically
of the order of $1\,\textrm{GeV}$), computed by some specialized
groups by fitting experimental data, imposing the boundary condition
(\ref{eq:boundary2}) for the coefficients $A_{0}(x)$, $B_{0}(x)$
and $C_{0}(x)$, computing iteratively $A_{n}(x)$, $B_{n}(x)$ and
$C_{n}(x)$ up to a certain value of $n$ by the recursion relations
(\ref{eq:An_recurrence2} -- \ref{eq:Cn_recurrence}) and then computing
the sum (\ref{eq:ansatz2}). It should be stressed that the solution of the RGE 
found by this method is characterized by two scales, 
the final evolution scale $Q\def \mu_F$, 
which is taken to be the factorization scale of a given process and $\lambda_{QCD}$. 
However, the hard scatterings can be renormalized at a scale 
different from $\mu_F$ and this clearly introduces a new scale in the hadronic 
cross section which also is an artifact of the perturbative expansion. Keeping track 
of this new scale dependence requires a careful analysis of the perturbative 
expansion of the kernels, as we are going to illustrate below.   

In order to study the renormalization scale dependence of the NNLO kernel it is 
convenient to solve the RGE for the coupling constant thereby expressing the 
coupling at a scale $\mu_F$ in terms of the coupling at a different point, 
$\mu_R$, which is the renormalization point of the theory, which is also arbitrary

\ba
\label{implicit}
\frac{1}{a_s(\mu_F^2)}=\frac{1}{a_s(\mu_R^2)}
+\beta_0 \ln \left(\frac{\mu_F^2}{\mu_R^2} \right)
-b_1\ln\left\{\frac{a_s(\mu_F^2) \, [ 1 + b_1 a_s(\mu_R^2) ]}
{a_s(\mu_R^2) \, [ 1 + b_1 a_s(\mu_F^2) ]} \right\}
\ea
where $a_s(\mu^2)=\alpha_s(\mu^2)/(4\pi)$.
It is possible to introduce a renormalization scale dependence
in $\alpha_s$ through a Taylor expansion of $\alpha_s(\mu_F^2)$
in terms of $\alpha_s(\mu_R^2)$
\ba
\alpha_s(\mu_F^2)=\alpha_s(\mu_R^2)-\left[\frac{\alpha_s^2(\mu_R^2)}{4\pi}
+\frac{\alpha_s^3(\mu_R^2)}{(4\pi)^2}(-\beta_0^2 L^2+\beta_1 L)\right]
\ea
where the $\mu_F^2$ dependence is shifted into the factor $L=\ln(\mu_F^2/\mu_R^2)
$ 
Then, a more handy expression for the $\alpha_s$ coupling obtained
by an inverse power of $L_{\Lambda}=\ln(\mu_R^2/\Lambda^2_{QCD})$ expansion
can be used for $\alpha_s(\mu_R^2)$.
Up to NNLO this gives
\ba
\label{exp1}
&&\frac{\alpha_s(\mu_R^2)}{4\pi}=\frac{1}{\beta_0 L_{\Lambda}}-
\frac{1}{(\beta_0 L_{\Lambda})^2} b_1\ln L_{\Lambda}+
\frac{1}{(\beta_0 L_{\Lambda})^3} \left[b_1^2 \left(\ln^2 L_{\Lambda}
-\ln L_{\Lambda}-1\right) + b_2 \right] \nonumber\\
\ea
and

\ba
\label{kern2}
&&P_{ij}(x,\mu_F^2,\mu_R^2)=\frac{\alpha_s(\mu_R^2)}{4\pi}P_{ij}^{(0)}(x) \nonumber\\
&&\hspace{2.5cm}+\frac{\alpha_s^2(\mu_R^2)}{(4\pi)^2}\left(P_{ij}^{(1)}(x)
-\beta_0P_{ij}^{(0)}(x) L\right)
\nonumber\\
&&\hspace{2.5cm}+\frac{\alpha_s^3(\mu_R^2)}{(4\pi)^3}\left[P_{ij}^{(2)}(x)-
2\beta_0 L P_{ij}^{(1)}(x)
-\left(\beta_1 L - \beta_0^2 L^2 \right) P_{ij}^{(0)}(x)\right] \nonumber\\
&&\hspace{2.5cm}+\frac{\alpha_s^4(\mu_R^2)}{(4\pi)^4}\left[P_{ij}^{(3)}(x)
- 3\beta_0 L \, P_{ij}^{(2)}(x)
-\left( 2\beta_1 L - 3\beta_0^2 L^2 \right) P_{ij}^{(1)}(x)\right.\nonumber\\
&&\hspace{2.5cm}\left.-\left(\beta_2 L - 5/2\, \beta_1 \beta_0 L^2 + \beta_0^3 L^3
\right) P_{ij}^{(0)}(x)\right] \nonumber\\
\ea
These new linear combinations of kernels carry 
an explicit dependence on $\mu_F$ and are suitable for the study of the dependence of the 
hadronic cross section on $\mu_R$. As we have mentioned, for a complete picture to emerge, 
the NNLO corrections to the hard scatterings have to be computed in generality, with 
$\mu_R$ and $\mu_F$ held distinct.

\section{Nonsinglet and singlet structure of the kernels up to NNLO}

Let us first introduce the notations
\begin{equation}
q_{i}^{(\pm)}=q_{i}\pm\overline{q}_{i},\qquad q^{(\pm)}=\sum_{i=1}^{n_{f}}q_{i}^{(\pm)}.\end{equation}
The general structure of the nonsinglet splitting functions is given
by\begin{equation}
P_{q_{i}q_{k}}=P_{\overline{q}_{i}\overline{q}_{k}}=\delta_{ik}P_{qq}^{V}+P_{qq}^{S},\end{equation}
\begin{equation}
P_{q_{i}\overline{q}_{k}}=P_{\overline{q}_{i}q_{k}}=\delta_{ik}P_{q\bar{q}}^{V}+P_{q\bar{q}}^{S}.\end{equation}
This leads to three independently evolving types of nonsinglet distributions:
the evolution of the flavor asymmetries\begin{equation}
q_{NS,ik}^{(\pm)}=q_{i}^{(\pm)}-q_{k}^{(\pm)}\end{equation}
and of linear combinations thereof is governed by\begin{equation}
P_{NS}^{\pm}=P_{qq}^{V}\pm P_{q\bar{q}}^{V}.\end{equation}
The sum of the valence distributions of all flavors $q^{(-)}$ evolves
with\begin{equation}
P_{NS}^{V}=P_{qq}^{V}-P_{q\bar{q}}^{V}+n_{f}\left(P_{qq}^{S}-P_{q\bar{q}}^{S}\right)\equiv P_{NS}^{-}+P_{NS}^{S}.\label{eq:PNSv}\end{equation}

The quark-quark splitting function $P_{qq}$ can be expressed as\begin{equation}
P_{qq}=P_{NS}^{+}+n_{f}\left(P_{qq}^{S}+P_{q\bar{q}}^{S}\right)\equiv P_{NS}^{+}+P_{ps}.\label{eq:Pqq}\end{equation}
 The nonsinglet contribution dominates Eq.~(\ref{eq:Pqq}) at large
$x$, where the \emph{pure singlet} term $P_{ps}=P_{qq}^{S}+P_{q\bar{q}}^{S}$
is very small. At small $x$, on the other hand, the latter contribution
takes over, as $xP_{ps}$ does not vanish for $x\rightarrow0$, unlike
$xP_{NS}^{+}$. The gluon-quark and quark-gluon entries in Eq.~(\ref{eq:singlet})
are given by\begin{equation}
P_{qg}=n_{f}P_{q_{i}g},\end{equation}
\begin{equation}
P_{gq}=P_{gq_{i}}\end{equation}
in terms of the flavor-independent splitting functions $P_{q_{i}g}=P_{\bar{q}_{i}g}$
and $P_{gq_{i}}=P_{g\bar{q}_{i}}$. With the exception of the first
order part of $P_{qg}$, neither of the quantities $xP_{qg}$, $xP_{gq}$
and $xP_{gg}$ vanishes for $x\rightarrow0$.

In the expansion in powers of $\alpha_{s}$ (\ref{eq:kernel_expansion}),
the flavor-diagonal (valence) quantity $P_{qq}^{V}$ is of order $\alpha_{s}$,
while $P_{q\bar{q}}^{V}$ and the flavor-independent (sea) contributions
$P_{qq}^{S}$ and $P_{q\bar{q}}^{S}$ are of order $\alpha_{s}^{2}$.
A non-vanishing difference $P_{qq}^{S}-P_{q\bar{q}}^{S}$ occurs for
the first time at the third order.

Our next step is to choose a proper basis of nonsinglet distributions
that allows us to reconstruct, through linear combinations, the distribution
of each parton, i.e.~the gluon distribution $g$, the quark distributions
$q_{i}$ and the antiquark distributions $\bar{q}_{i}$, namely $2n_{f}+1$
relevant distributions. The singlet evolution gives us 2 distributions,
$g$ and $q^{(+)}$, so we need to evolve $2n_{f}-1$ independent
nonsinglet distributions. We choose

\begin{enumerate}
\item $q^{(-)}$, evolving with $P_{NS}^{V}$;
\item $q_{NS,1i}^{(-)}=q_{1}^{(-)}-q_{i}^{(-)}$ (for each $i$ such that
$2\leq i\leq n_{f}$), evolving with $P_{NS}^{-}$;
\item $q_{NS,1i}^{(+)}=q_{1}^{(+)}-q_{i}^{(+)}$ (for each $i$ such that
$2\leq i\leq n_{f}$), evolving with $P_{NS}^{+}$.
\end{enumerate}
We can easily prove that\begin{equation}
q_{i}^{(\pm)}=\frac{1}{n_{f}}\left(q^{(\pm)}+\sum_{k=1,k\neq i}^{n_{f}}q_{NS,ik}^{(\pm)}\right).\label{eq:comb_linNS}\end{equation}
Choosing $i=1$ in (\ref{eq:comb_linNS}), we compute $q_{1}^{(-)}$
from the evolved nonsinglets of type 1 and 2 and $q_{1}^{(+)}$ from
the evolved singlet $q^{(+)}$ and nonsinglet of type 3. Then from
the nonsinglets 2 and 3 we compute respectively $q_{i}^{(-)}$ and
$q_{i}^{(+)}$ for each $i$ such that $2\leq i\leq n_{f}$, and finally
$q_{i}$ and $\bar{q}_{i}$.

Life can be made easier going down from NNLO to NLO, as we have $P_{qq}^{S,(1)}=P_{q\bar{q}}^{S,(1)}$.
This implies (see Eq.~(\ref{eq:PNSv})) that $P_{NS}^{V,(1)}=P_{NS}^{-,(1)}$,
i.e.~the nonsinglets $q^{(-)}$ and $q_{NS,ik}^{(-)}$ evolve with
the same kernel, and the same does each linear combination thereof,
in particular $q_{i}^{(-)}$ for each flavor $i$. The basis of $2n_{f}-1$
nonsinglet distributions that we choose to evolve at NLO is

\begin{enumerate}
\item $q_{i}^{(-)}$ (for each $i\leq n_{f}$), evolving with $P_{NS}^{-,(1)}$;
\item $q_{NS,1i}^{(+)}=q_{1}^{(+)}-q_{i}^{(+)}$ (for each $i$ such that
$2\leq i\leq n_{f}$), evolving with $P_{NS}^{+,(1)}$,
\end{enumerate}
and the same we do at LO, where we have in addition $P_{NS}^{+,(0)}=P_{NS}^{-,(0)}$,
being $P_{q\bar{q}}^{V,(0)}=0$.

\begin{framed}
\textbf{\textit{Remark: the NLO versus the NNLO decomposition}}

Prior to moving toward the implementation of this algorithm, which will be discussed below, we 
pause for a moment in order to remark on the differences between the singlet/nonsinglet decomposition 
of the kernels implemented in chapter \ref{chap:NLOcode} and those discussed in this chapter.  


In the transverse case there is no coupling between gluons
and quarks, so the singlet sector (\ref{eq:singlet}) is missing.
This means that $\Delta_{T}q^{(+)}$ is a nonsinglet evolving with
$\Delta_{T}P_{NS}^{+}$; but the same kernel evolves also $\Delta_{T}q_{NS,ik}^{(-)}$,
so the linear combinations $\Delta_{T}q_{i}^{(+)}$ for each flavor
$i$. So the $2n_{f}$ relevant distributions that we evolve
at LO and NLO are

\begin{enumerate}
\item $\Delta_{T}q_{i}^{(-)}$ (for each $i\leq n_{f}$), evolving with
$\Delta_{T}P_{NS}^{-}$;
\item $\Delta_{T}q_{i}^{(+)}$ (for each $i\leq n_{f}$), evolving with
$\Delta_{T}P_{NS}^{+}$.
\end{enumerate}
\end{framed}

\section{Harmonic polylogarithms} 
We summarize here some key features of the kernel which are relevant for a better understanding of the implementation and refer to the original literature for more details. 

Higher order computations in QCD involve a special type of trascendental functions, 
termed harmonic polylogarithms (HPL). Although their definition is a straightforward extension of 
that of ordinary Spence functions, additional loop integrations in the Feynman 
diagram expansion generate multiple integrals of the logarithmic contributions which appear at leading order in $\alpha_s$. 
 The study of these functions in physics has been addressed in great generality 
in the last few years and has been quite useful in order to classify and study 
their behaviour. HPL are, in the case of parton distributions and 
of coefficient functions, studied in the standard domain $(0,1)$, They show branch 
cuts away from this region and can be analytically continued away from it. 
The NNLO kernel is expressed in terms of various polylogarithms. 
This computation  \cite{NNLO_nonsinglet,NNLO_singlet}
has been performed in the $N$-Mellin space, where the nonsinglet and the singlet
anomalous dimensions are related to the DGLAP kernel by the well known
Mellin transform
\ba
\gamma(N)^{(n)}=-\int_0^1 dx\,x^{N-1}P^{(n)}(x)\,.
\ea

Inverting the above relation, one obtains the expression for the $P^{(n)}(x)$
in the $x$-space that we will use in our implementation.
By this procedure the isomorphism between harmonic sums of $N$ and HPL is manifest and this can be demonstrated by algebraic
procedure.
Our notation for the harmonic polylogarithms $H_{m_1,...,m_w}(x)$,
$m_j = 0,\pm 1$ follows Ref.~\cite{RemiddiVermaseren}.
The lowest-weight ($k = 1$) functions $H_m(x)$ are
given by
\beq
\label{first}
  H_0(x)       \: = \: \ln x \:\: , \quad\quad
  H_{\pm 1}(x) \: = \: \mp \, \ln (1 \mp x) \:\: .
\eeq
The higher-weight ($k \geq 2$) functions are recursively defined as
\beq
\label{second}
H_{m_1,...,m_k}(x) \: = \:
\left\{ \begin{array}{cl}
\displaystyle{ \frac{1}{k!}\,\ln^k x \:\: ,}
   & \quad {\rm if} \:\:\: m^{}_1,...,m^{}_k = 0,\ldots ,0 \\[2ex]
\displaystyle{ \int_0^x \! dz\: f_{m_1}(z) \, H_{m_2,...,m_k}(z)
   \:\: , } & \quad {\rm otherwise}
\end{array} \right.
\eeq
with
\beq
\label{third}
  f_0(x)       \: = \: \frac{1}{x} \:\: , \quad\quad
  f_{\pm 1}(x) \: = \: \frac{1}{1 \mp x} \:\: .
\eeq
\beq
\label{fourth}
  H_{{\footnotesize \underbrace{0,\ldots ,0}_{\scriptstyle k} },\,
  \pm 1,\, {\footnotesize \underbrace{0,\ldots ,0}_{\scriptstyle l} },
  \, \pm 1,\, \ldots}(x) \: = \: H_{\pm (k+1),\,\pm (l+1),\, \ldots}(x)
  \:\: .
\eeq
these identities and many more have been used in the investigation 
of the kernel.

\section{Description of the program}
\subsection{Main program}

At run time, the program asks the user to choose linear the perturbative
order, the input model (currently implemented MRST and Alekhin, see
Table \ref{tab:initialconditions}), the final value of $Q$ and an
extension for the output files. Then the program stores as global
variables the grid points $x_{i}$ and, for each of them, calls the
function \texttt{gauleg} that computes the Gaussian abscissas $X_{ij}$
and weights $W_{ij}$ corresponding to the integration range $[x_{i},1]$,
with $0\leq j\leq n_{G}-1$. At this point the program stores the
initial values of the parton distributions at the grid points in the
arrays \texttt{A{[}i{]}{[}0{]}{[}k{]}}.

The evolution is done by energy steps. Each flavor comes into play
only when the energy in the center of mass frame reaches the corresponding
quark mass, so each step is determined by the number of flavors involved.
The recurrence relations (\ref{eq:An_recurrence2} -- \ref{eq:Cn_recurrence})
are then solved iteratively for both the nonsinglet and the singlet
sector, and at the end of each energy step the evoluted distributions
are reconstructed via relation (\ref{eq:ansatz2}). The distributions
computed in this way become the initial conditions for the subsequent
step. The numerical values of the distributions at the end of each
energy step are printed to files.

\subsection{External files}

Some external files are used by the program.

\begin{enumerate}
\item \texttt{xpns2e.f} and \texttt{xpij2e.f} are Fortran codes by Moch,
Vermaseren and Vogt \cite{NNLO_nonsinglet,NNLO_singlet} in which
are the NNLO kernels are defined. Very few modifications have been
done to make them compatible with our code. These files need \texttt{hplog.f}
to work.
\item \texttt{hplog.f} is a Fortran code by Gehrmann and Remiddi \cite{GehrmannRemiddi}
in which a subroutine that computes numerically harmonic polylogarithms
up to weight 4 is implemented. Harmonic polylogarithms are defined
in \cite{RemiddiVermaseren}.
\item \texttt{partonww.f} is just a merging of the three Fortran codes \texttt{mrst2001lo.f},
\texttt{mrst2001.f} and \texttt{mrstnnlo.f} by the MRST group \cite{MRST1,MRST2}
to access to their grids of LO, NLO and NNLO parton densities. Very
few modifications have been done.
\item \texttt{lo2002.dat}, \texttt{alf119.dat} and \texttt{vnvalf1155.dat}
are the MRST parton densities grids at LO, NLO and NNLO respectively.
\item \texttt{a02m.f} is the Fortan code by Alekhin \cite{Alekhin} to access
to his grids of LO, NLO and NNLO parton densities.
\item \texttt{a02m.pdfs\_1\_vfn}, \texttt{a02m.pdfs\_2\_vfn} and \texttt{a02m.pdfs\_3\_vfn}
are the Alekhin parton densities grids in the variable flavor number
scheme at LO, NLO and NNLO respectively.
\end{enumerate}

\subsection{Functions}

\subsubsection{Function \texttt{writefile}}

\texttt{void writefile(double {*}A,char {*}filename);}

This function creates a file, whose name is contained in the string
\texttt{{*}filename}, with two columns of data: the left one contains
all the grid points $x_{i}$ and the right one the corresponding values
\texttt{A{[}i{]}}.

\subsubsection{Function \texttt{alpha\_s}}

\texttt{double alpha\_s(int order,double Q,double lambda);}

Given the perturbative \texttt{order}, the energy scale \texttt{Q}
and the value \texttt{lambda} of $\Lambda_{\overline{MS}}^{(n_{f})}$,
and making use of the values of $\beta_{0}$, $\beta_{1}$ and $\beta_{2}$,
stored as global variables, \texttt{alpha\_s} returns the running
of the coupling constant, using the formula (\ref{eq:alpha_s_nnlo}).

\subsubsection{Function \texttt{gauleg}}

\texttt{void gauleg(double x1,double x2,double x{[}{]},double w{[}{]},int
n);}

This function is taken from \cite{NumRecipes} with just minor changes.
Given the lower and upper limits of integration \texttt{x1} and \texttt{x2},
and given \texttt{n}, \texttt{gauleg} returns arrays \texttt{x{[}0,...,n-1{]}}
and \texttt{w{[}0,...,n-1{]}} of lenght \texttt{n}, containing the
abscissas and weights of the Gauss-Legendre \texttt{n}-point quadrature
formula.

\subsubsection{Function \texttt{interp}}

\texttt{double interp(double {*}A,double x);}

Given an array \texttt{A}, representing a function known only at the
grid points, and a number \texttt{x} in the interval $[0,1]$, \texttt{interp}
returns the linear interpolation of the function at the point \texttt{x}.

\subsubsection{Kernels}

\texttt{double P0NS(int i,double x);}~\\
\texttt{double P0qq(int i,double x);}~\\
\texttt{double P0qg(int i,double x);}~\\
\texttt{double P0gq(int i,double x);}~\\
\texttt{double P0gg(int i,double x);}~\\
\texttt{double P1NSm(int i,double x);}~\\
\texttt{double P1NSp(int i,double x);}~\\
\texttt{double P1qq(int i,double x);}~\\
\texttt{double P1qg(int i,double x);}~\\
\texttt{double P1gq(int i,double x);}~\\
\texttt{double P1gg(int i,double x);}~\\
\texttt{double P2NSm(int i,double x);}~\\
\texttt{double P2NSp(int i,double x);}~\\
\texttt{double P2NSv(int i,double x);}~\\
\texttt{double P2qq(int i,double x);}~\\
\texttt{double P2qg(int i,double x);}~\\
\texttt{double P2gq(int i,double x);}~\\
\texttt{double P2gg(int i,double x); }

Given the Bjorken variable \texttt{x}, these functions return, depending
on the value of the index \texttt{i}

\begin{enumerate}
\item the regular part of the kernel $P_{1}(x)$;
\item the plus distribution part of the kernel $P_{2}(x)$;
\item the delta distribution part of the kernel $P_{3}(x)$,
\end{enumerate}
as in Eq.~(\ref{eq:general_kernel}).

\subsubsection{Function \texttt{convolution}}

\texttt{double convolution(int i,double kernel(int,double),double
{*}A);}

Given an integer \texttt{i}, to which corresponds a grid point $x_{i}$,
a two variable function \texttt{kernel(i,x)}, representing a kernel,
and an array \texttt{A}, representing a function $xf(x)=\bar{f}(x)$
known at the grid points, \texttt{convolution} returns the sum of
the three pieces (\ref{eq:conv_1}, \ref{eq:conv_2}, \ref{eq:conv_3})
computed by the Gauss-Legendre technique.

\subsubsection{Function \texttt{RecRel\_A}}

\texttt{double RecRel\_A(double {*}A,int k,double P0(int,double));}

Given an array \texttt{A}, representing the function $A_{n}(x)$ known
at the grid points, an integer \texttt{k} (to which corresponds a
grid point $x_{k}$) and a two variable function \texttt{P0(i,x)},
representing a leading order kernel, \texttt{RecRel\_A} returns the
RHS of Eq.~(\ref{eq:An_recurrence2}) for $x=x_{k}$.

\subsubsection{Function \texttt{RecRel\_B}}

\texttt{double RecRel\_B(double {*}A,double {*}B,int k,double P0(int,double),}~\\
 \texttt{double P1(int,double));}

Given two array \texttt{A} and \texttt{B}, representing the functions
$A_{n}(x)$ and $B_{n}(x)$ known at the grid points, an integer \texttt{k}
(to which corresponds a grid point $x_{k}$) and two functions \texttt{P0(i,x)}
and \texttt{P1(i,x)}, representing respectively the LO and NLO part
of a kernel, \texttt{RecRel\_B} returns the RHS of Eq.~(\ref{eq:Bn_recurrence2})
for $x=x_{k}$.

\subsubsection{Function \texttt{RecRel\_C}}

\texttt{double RecRel\_C(double {*}A,double {*}B,double {*}C,int k,double
P0(int,double),}~\\
 \texttt{double P1(int,double),double P2(int,double));}

Given three array \texttt{A}, \texttt{B} and \texttt{C} representing
the functions $A_{n}(x)$, $B_{n}(x)$ and $C_{n}(x)$ known at the
grid points, an integer \texttt{k} (to which corresponds a grid point
$x_{k}$) and three functions \texttt{P0(i,x)}, \texttt{P1(i,x)} and
\texttt{P2(i,x)}, representing respectively the LO, NLO and NNLO part
of a kernel, \texttt{RecRel\_C} returns the RHS of Eq.~(\ref{eq:Cn_recurrence})
for $x=x_{k}$.

\subsubsection{Function \texttt{Li2}}

\texttt{double Li2(double x);}

This function evaluates an approximated value of the dilogarithmic
function $Li_{2}(x)$ using the expansion\begin{equation}
Li_{2}(x)=\sum_{n=1}^{\infty}\frac{x{}^{n}}{n^{2}}\end{equation}
arrested at the 50th order.

\subsubsection{Function \texttt{fact}}

\texttt{double fact(int n);}

This function returns the factorial $n!$

\subsection{Distribution indices and identificatives\label{subsec:dist_ind}}

\begin{tabular}{lll}
\hline 
0&
gluons, $g$&
\texttt{g}\tabularnewline
1-6&
quarks, $q_{i}$, sorted by increasing mass ($u,d,s,c,b,t$)&
\texttt{u,d,s,c,b,t}\tabularnewline
7-12&
antiquarks, $\overline{q_{i}}$&
\texttt{au,ad,as,ac,ab,at}\tabularnewline
13-18&
$q_{i}^{(-)}$&
\texttt{um,dm,sm,cm,bm,tm}\tabularnewline
19-24&
$q_{i}^{(+)}$&
\texttt{up,dp,sp,cp,bp,tp}\tabularnewline
25&
$q^{(-)}$&
\texttt{qm}\tabularnewline
26-30&
$q_{NS,1i}^{(-)},\quad i\neq1$&
\texttt{dd,sd,cd,bd,td}\tabularnewline
31&
$q^{(+)}$&
\texttt{qp}\tabularnewline
32-36&
$q_{NS,1i}^{(+)},\quad i\neq1$&
\texttt{ds,ss,cs,bs,ts}\tabularnewline
\hline
\end{tabular}

\subsection{Input parameters and variables}

\begin{tabular}{ll}
\hline 
\texttt{GRID\_PTS}&
Number of points in the grid\tabularnewline
\texttt{NGP}&
Number of Gaussian points, $n_{G}$\tabularnewline
\texttt{ITERATIONS}&
Number of terms in the sum (\ref{eq:ansatz2})\tabularnewline
\texttt{extension}&
Extension of the output files\tabularnewline
\texttt{order}&
Perturbative order (0=LO, 1=NLO, 2=NNLO)\tabularnewline
\texttt{input}&
Input model (1=MRST, 2=Alekhin)\tabularnewline
\hline
\end{tabular}\\
\begin{tabular}{ll}
\hline 
\texttt{X{[}i{]}}&
$i$-th grid point, $x_{i}$\tabularnewline
\texttt{XG{[}i{]}{[}j{]}}&
$j$-th Gaussian abscissa in the range $[x_{i},1]$, $X_{ij}$\tabularnewline
\texttt{WG{[}i{]}{[}j{]}}&
$j$-th Gaussian weight in the range $[x_{i},1]$, $W_{ij}$\tabularnewline
\texttt{nf, Nf}&
number of active flavors, $n_{f}$\tabularnewline
\texttt{nfi}&
number of active flavors at the input scale\tabularnewline
\texttt{Q{[}i{]}}&
values of $Q$ between which an evolution step is performed\tabularnewline
\texttt{lambda{[}nf{]}}&
$\Lambda_{\overline{MS}}^{(n_{f})}$\tabularnewline
\texttt{A{[}i{]}{[}j{]}{[}k{]}}&
coefficient $A_{j}(x_{k})$ for the distribution with index $i$\tabularnewline
\texttt{B{[}i{]}{[}j{]}{[}k{]}}&
coefficient $B_{j}(x_{k})$ for the distribution with index $i$\tabularnewline
\texttt{C{[}i{]}{[}j{]}{[}k{]}}&
coefficient $C_{j}(x_{k})$ for the distribution with index $i$\tabularnewline
\texttt{beta0}&
$\beta_{0}$\tabularnewline
\texttt{beta1}&
$\beta_{1}$\tabularnewline
\texttt{beta2}&
$\beta_{2}$\tabularnewline
\texttt{alpha1}&
$\alpha_{s}(Q_{in})$, where $Q_{in}$ is the lower $Q$ of the evolution
step\tabularnewline
\texttt{alpha2}&
$\alpha_{s}(Q_{fin})$, where $Q_{fin}$ is the higher $Q$ of the
evolution step\tabularnewline
\hline
\end{tabular}

\subsection{Output files}

The generic name of an output file is: \texttt{\textbf{U}}\texttt{\textbf{\emph{n}}}\texttt{\textbf{\_i.ext}},
where:

\begin{description}
\item [\texttt{\emph{n}}]indicates the scale $Q^{2}$ to which data refers:
$n=i$ refers to the initial scale, the higher value of $n=f$ refers
to the final scale; if $n$ is a number, then it refers to the $n$-th
quark (ordered by increasing mass) production thresholds;
\item [\texttt{i}]is the identificative of the distribution, reported in
the third column of the table in subsection \ref{subsec:dist_ind};
\item [\texttt{ext}]is an extension chosen by the user.
\end{description}

\section{Running the code}

In this section we show some pdf plots obtained by running the code
in many different situations.

\begin{figure}[!tbh]
\begin{center}\subfigure[MRST LO]{\includegraphics[%
  width=5.5cm,
  angle=-90]{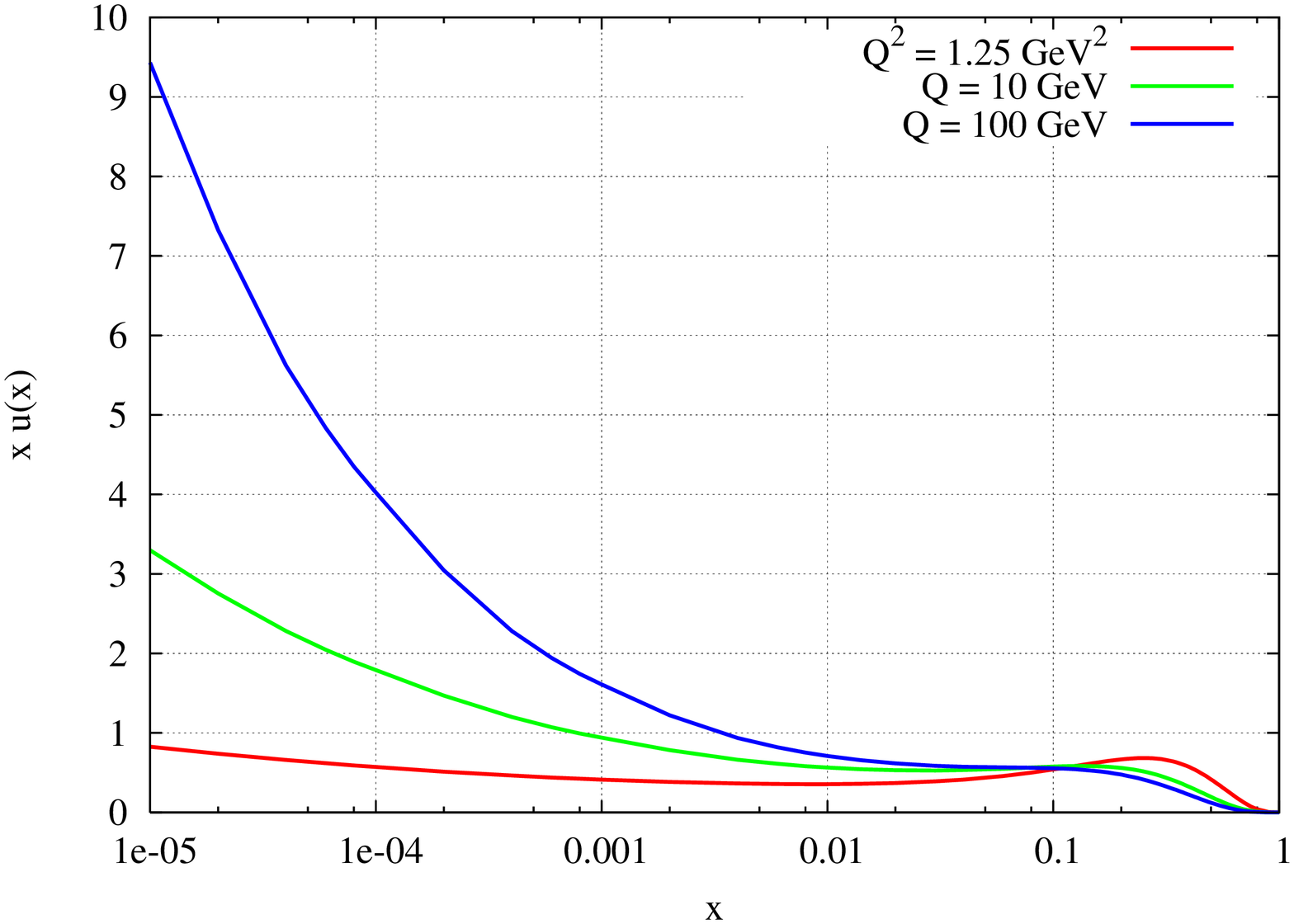}}\subfigure[Alekhin LO]{\includegraphics[%
  width=5.5cm,
  angle=-90]{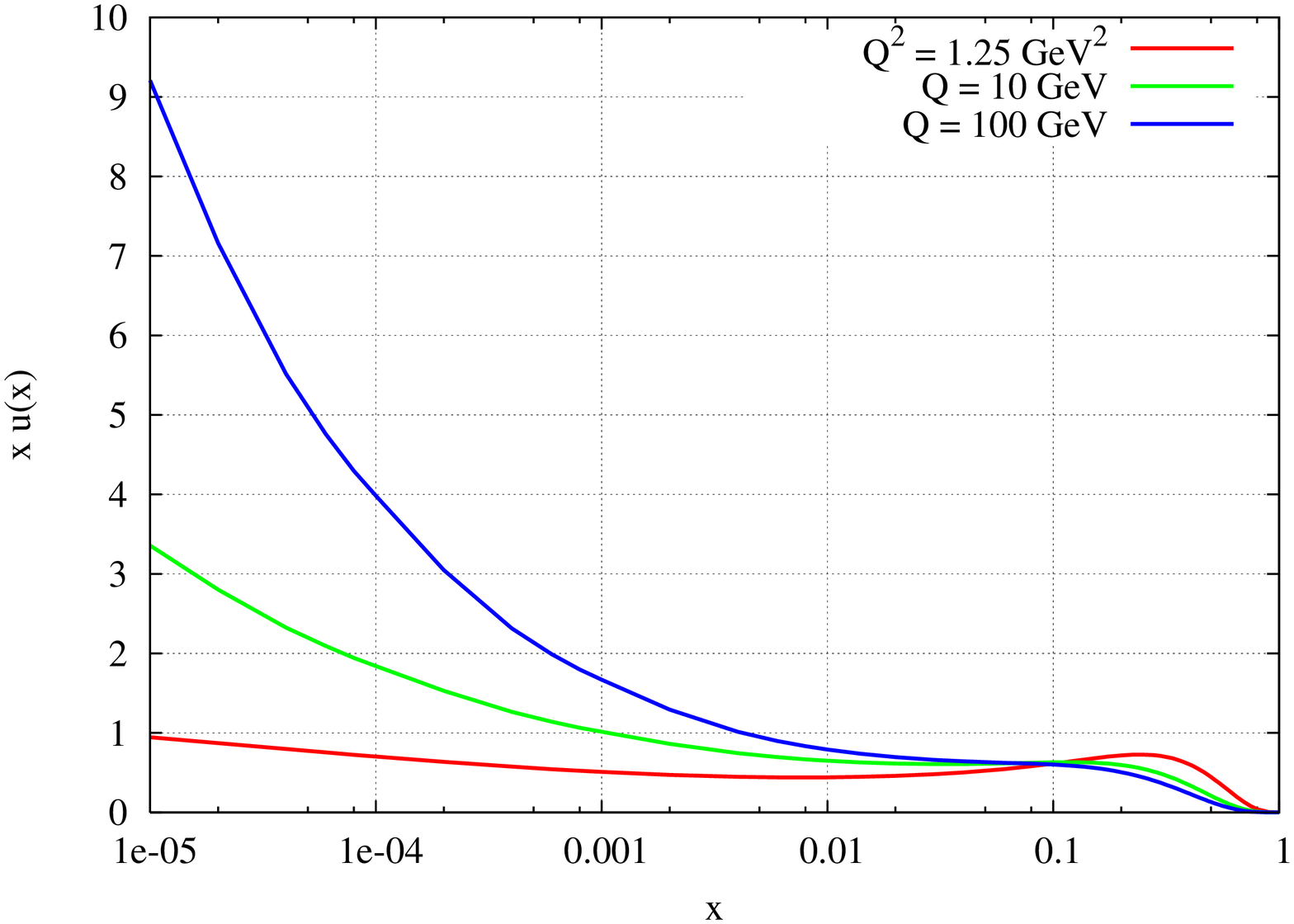}}\end{center}

\begin{center}\subfigure[MRST NLO]{\includegraphics[%
  width=5.5cm,
  angle=-90]{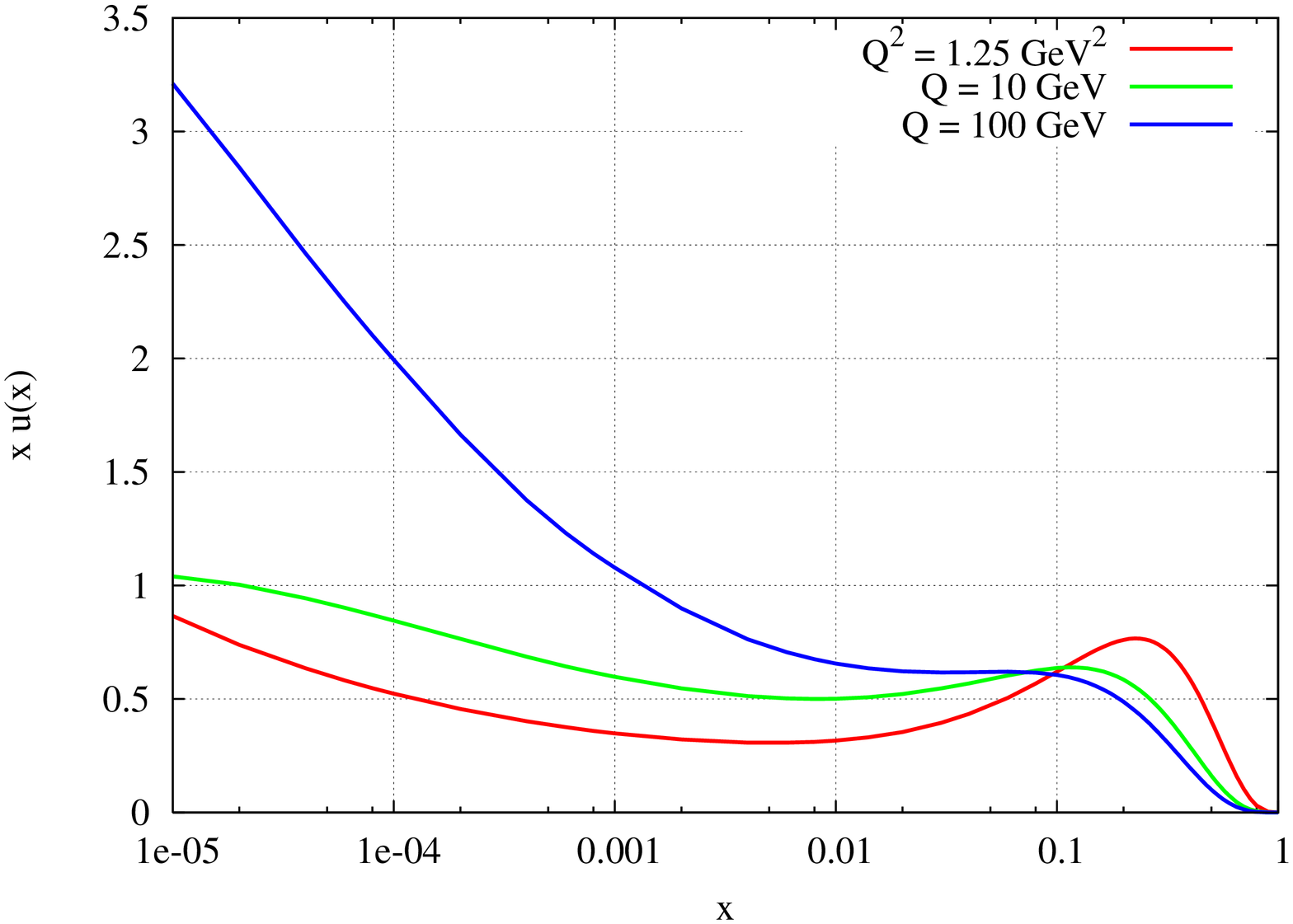}}\subfigure[Alekhin NLO]{\includegraphics[%
  width=5.5cm,
  angle=-90]{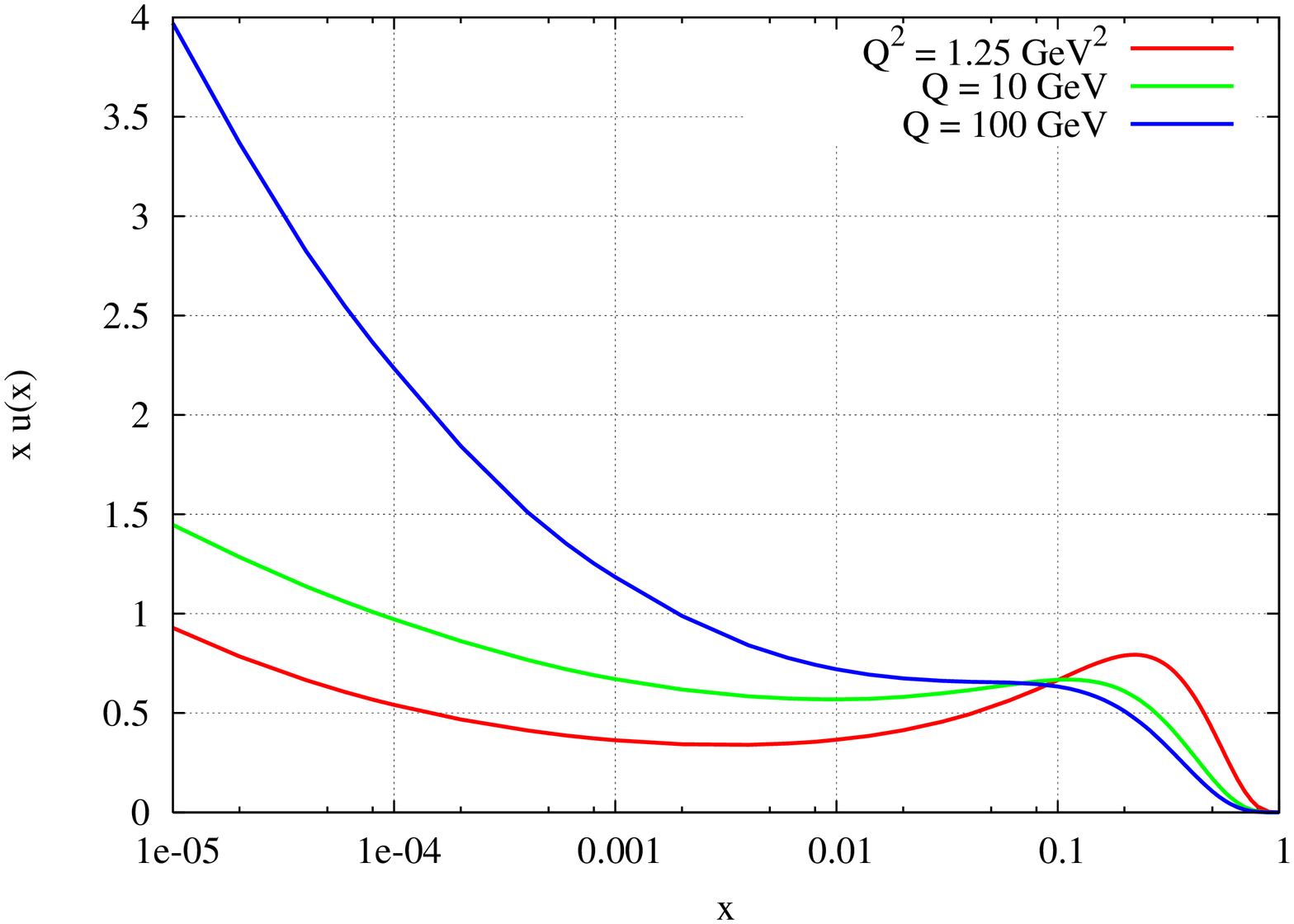}}\end{center}

\begin{center}\subfigure[MRST NNLO]{\includegraphics[%
  width=5.5cm,
  angle=-90]{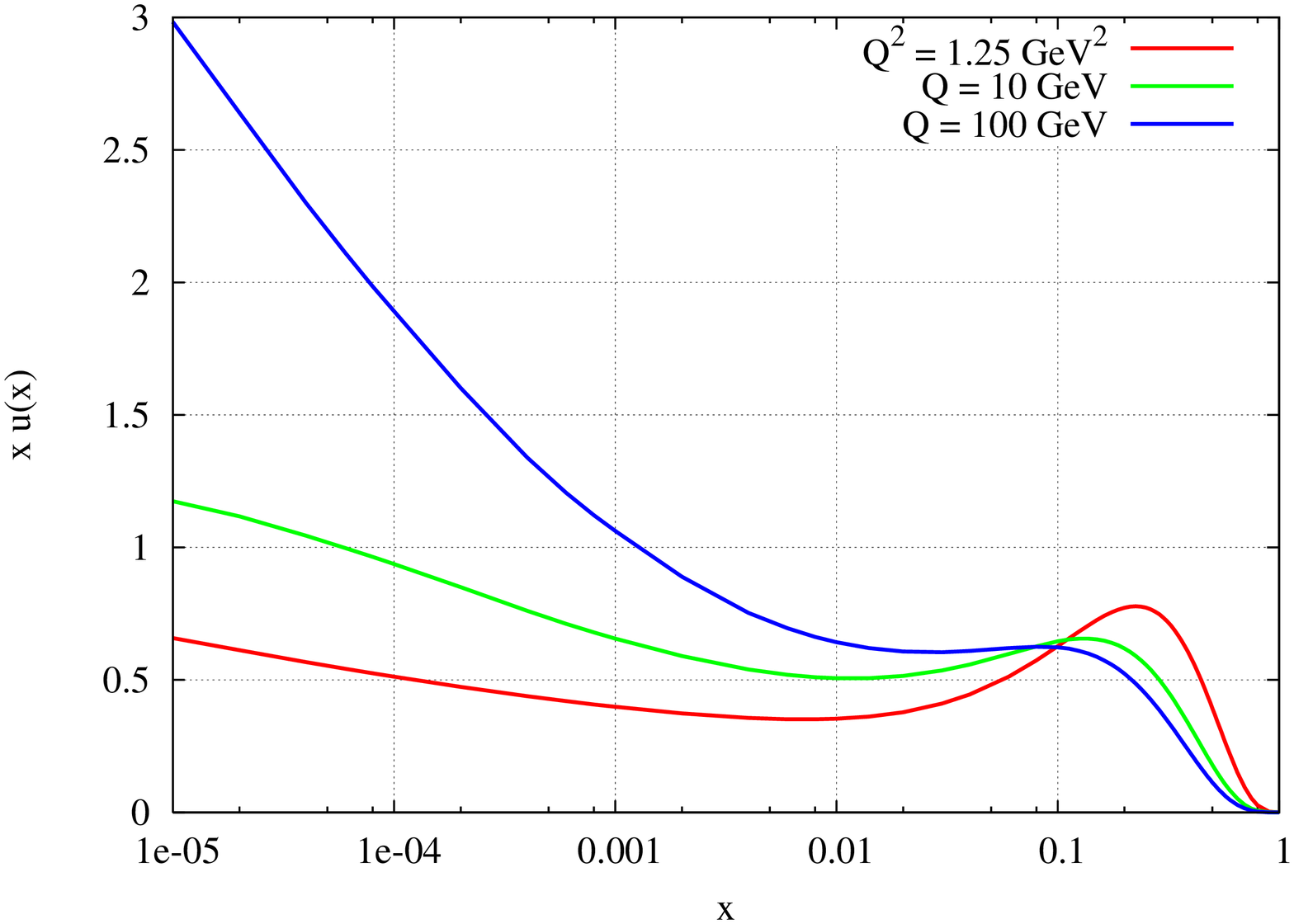}}\subfigure[Alekhin NNLO]{\includegraphics[%
  width=5.5cm,
  angle=-90]{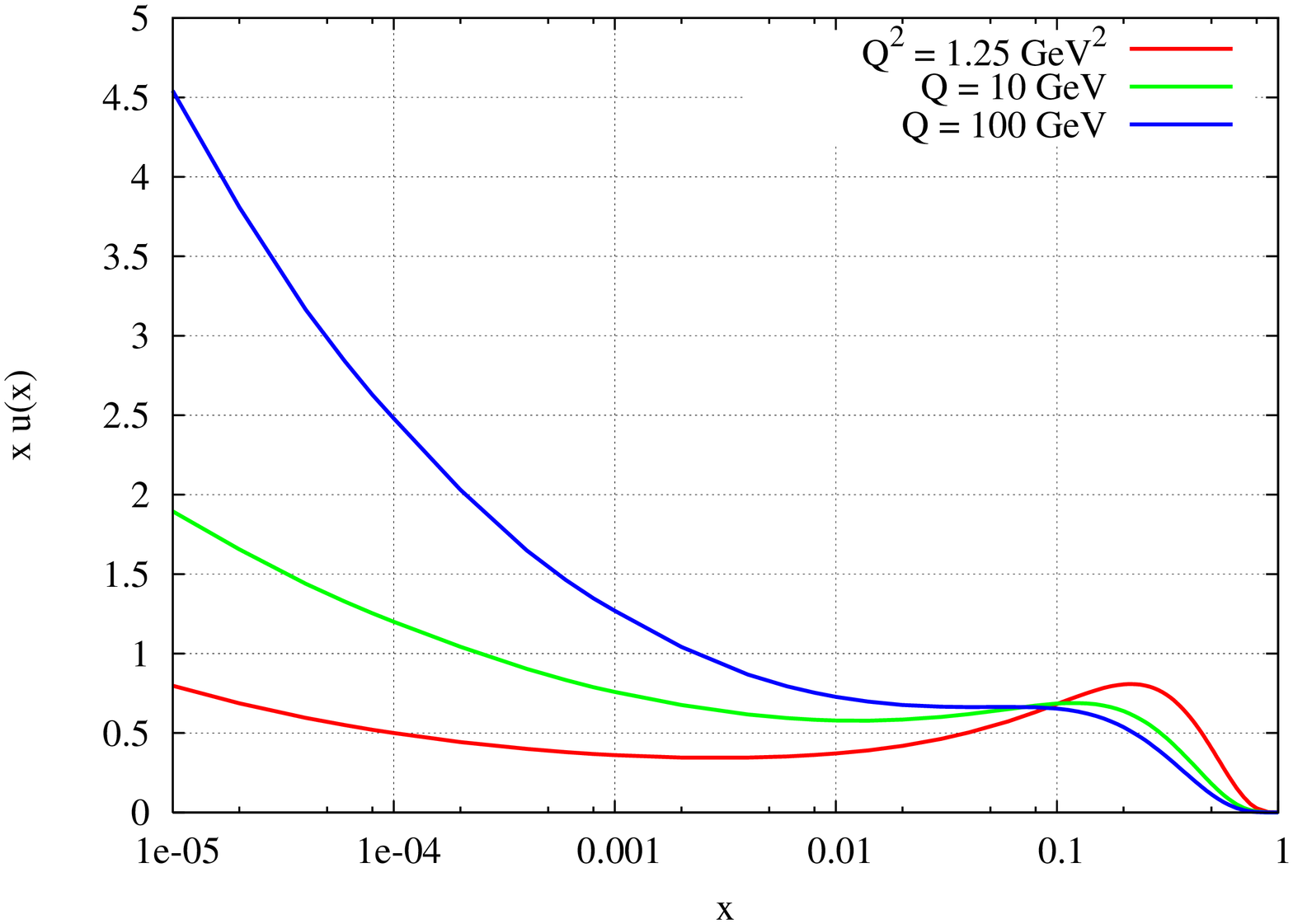}}\end{center}

\caption{Evolution of the quark up distribution $xu$ versus $x$ at different
$Q$ values. All perturbative orders and both input models are shown.\label{fig:Unp_up}}
\end{figure}

\begin{figure}[!tbh]
\begin{center}\subfigure[MRST LO]{\includegraphics[%
  width=5.5cm,
  angle=-90]{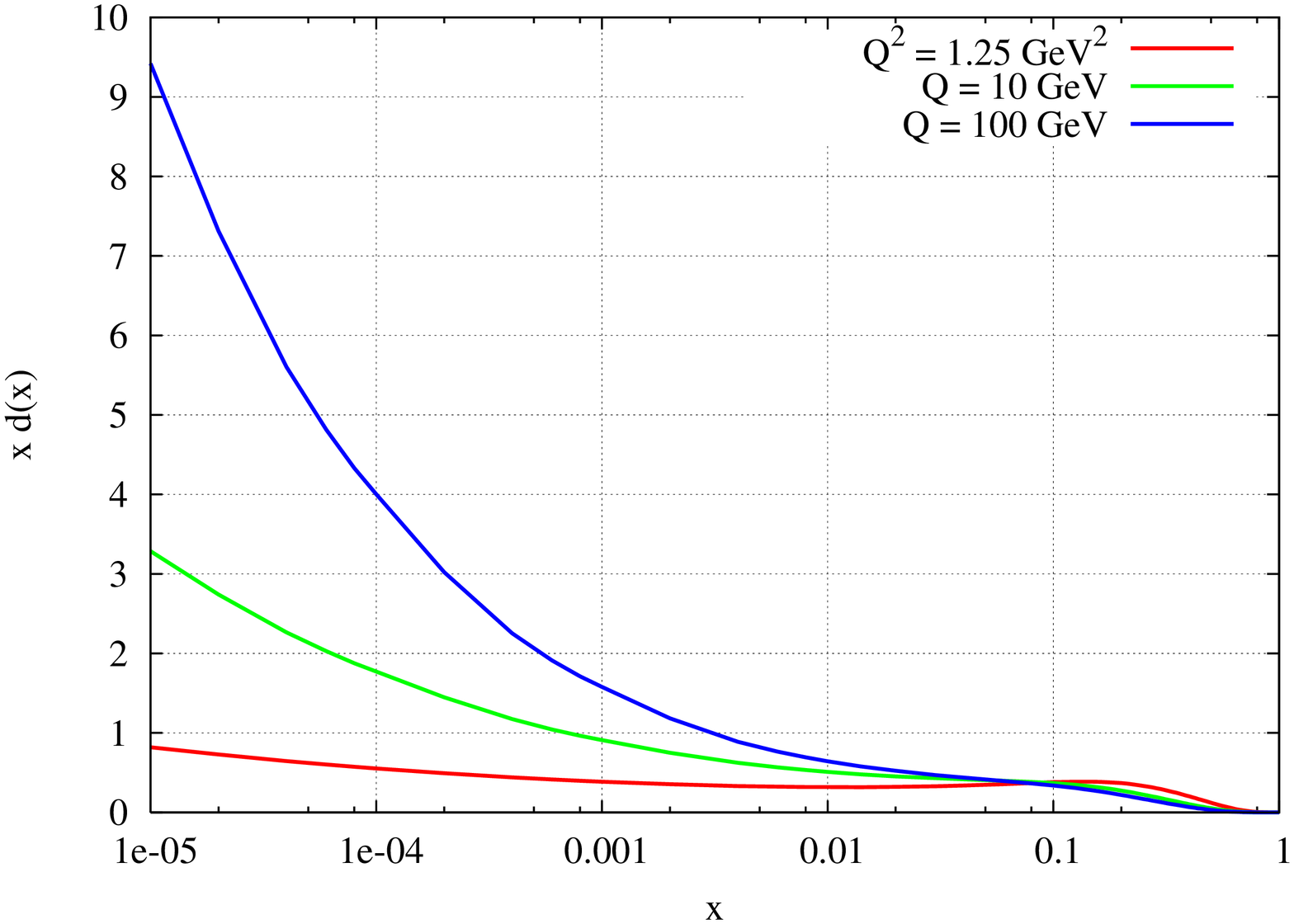}}\subfigure[Alekhin LO]{\includegraphics[%
  width=5.5cm,
  angle=-90]{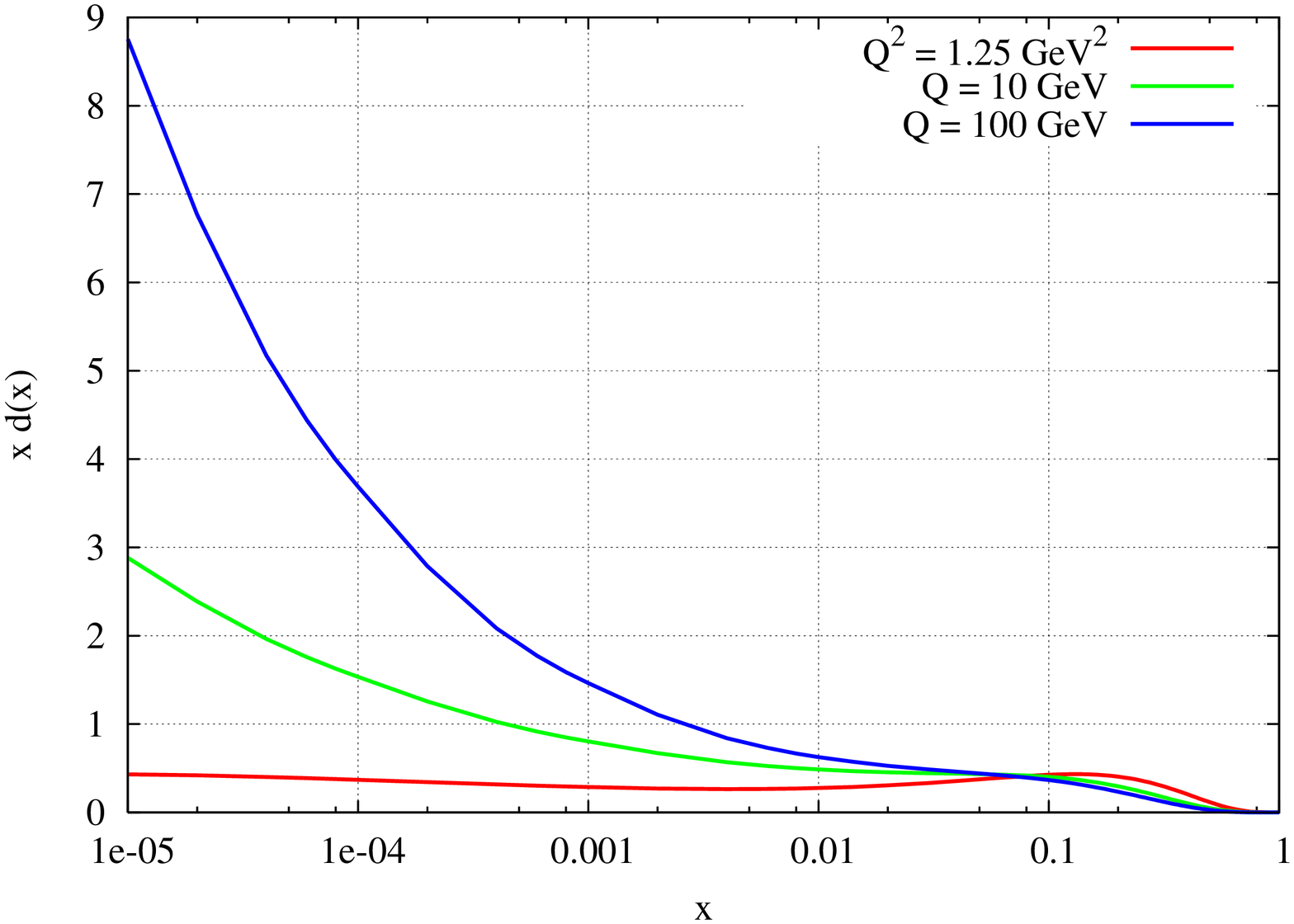}}\end{center}

\begin{center}\subfigure[MRST NLO]{\includegraphics[%
  width=5.5cm,
  angle=-90]{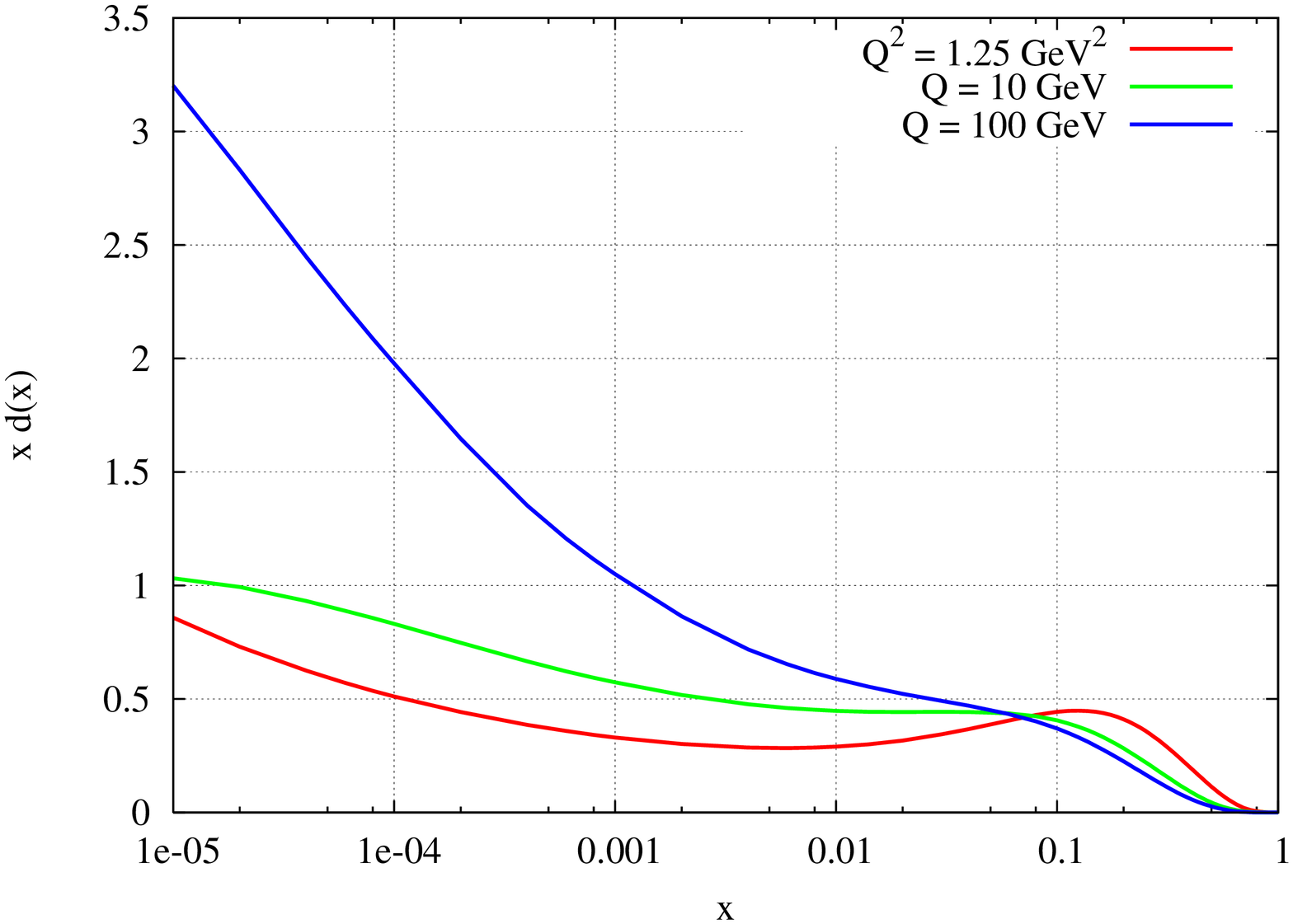}}\subfigure[Alekhin NLO]{\includegraphics[%
  width=5.5cm,
  angle=-90]{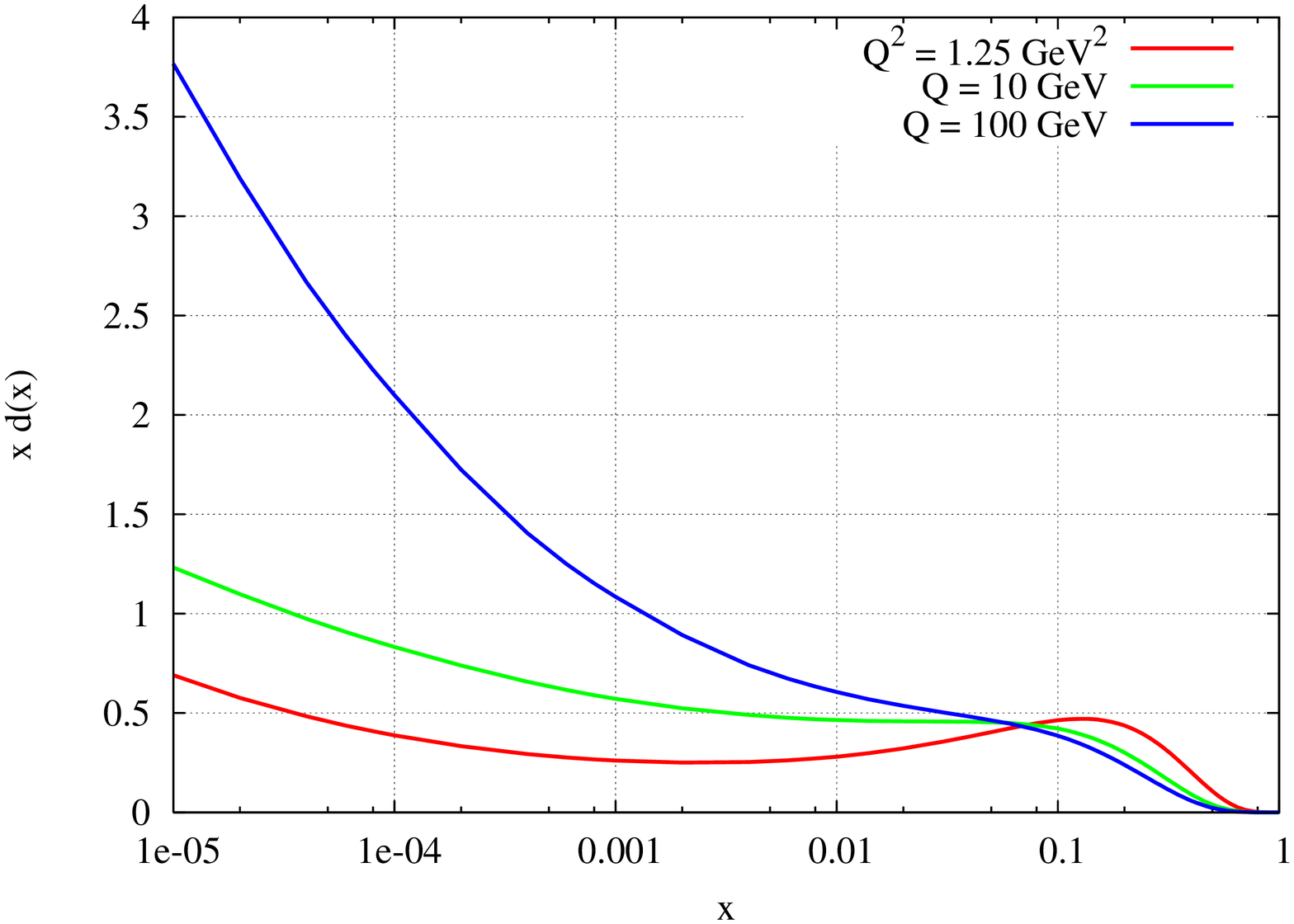}}\end{center}

\begin{center}\subfigure[MRST NNLO]{\includegraphics[%
  width=5.5cm,
  angle=-90]{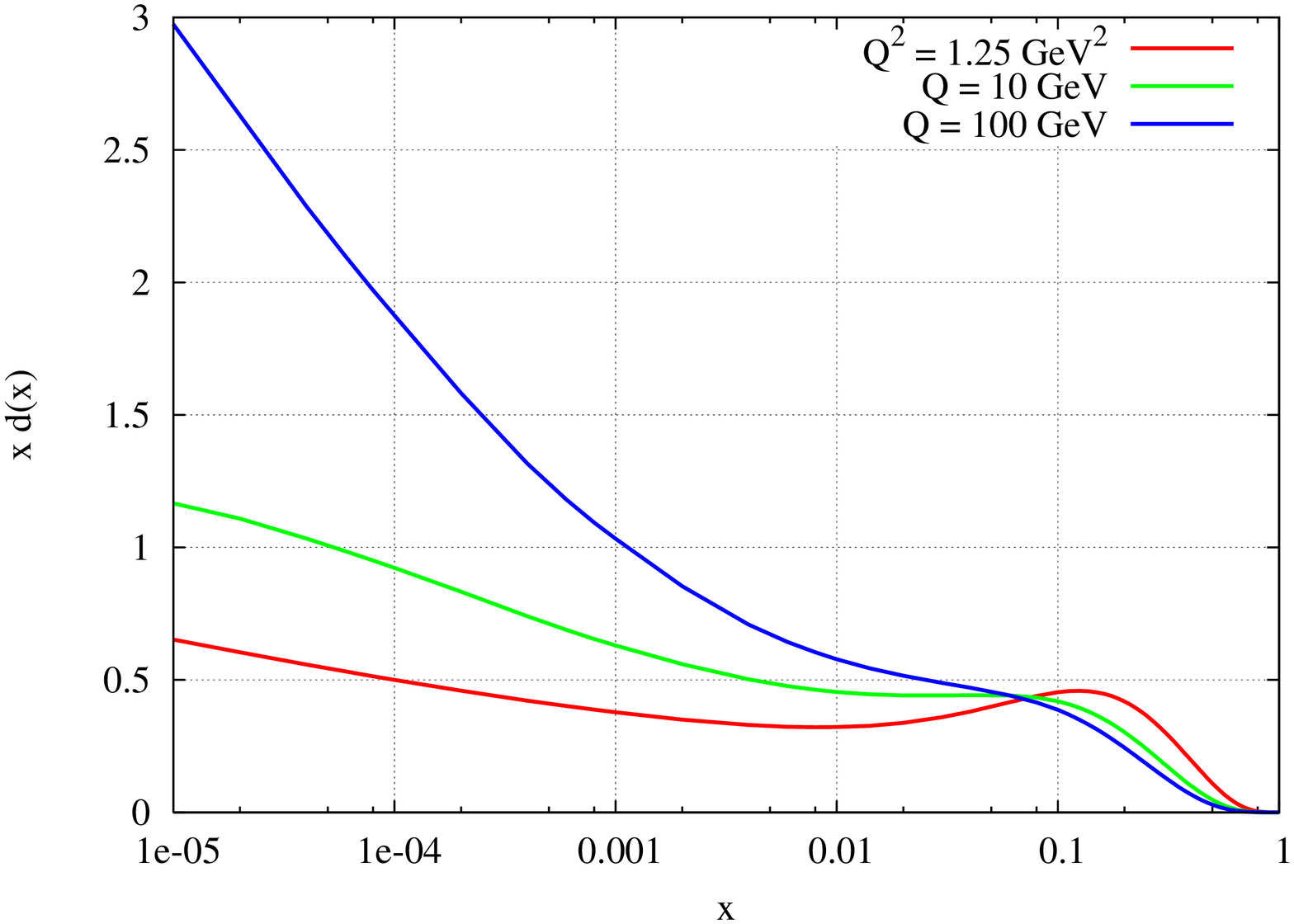}}\subfigure[Alekhin NNLO]{\includegraphics[%
  width=5.5cm,
  angle=-90]{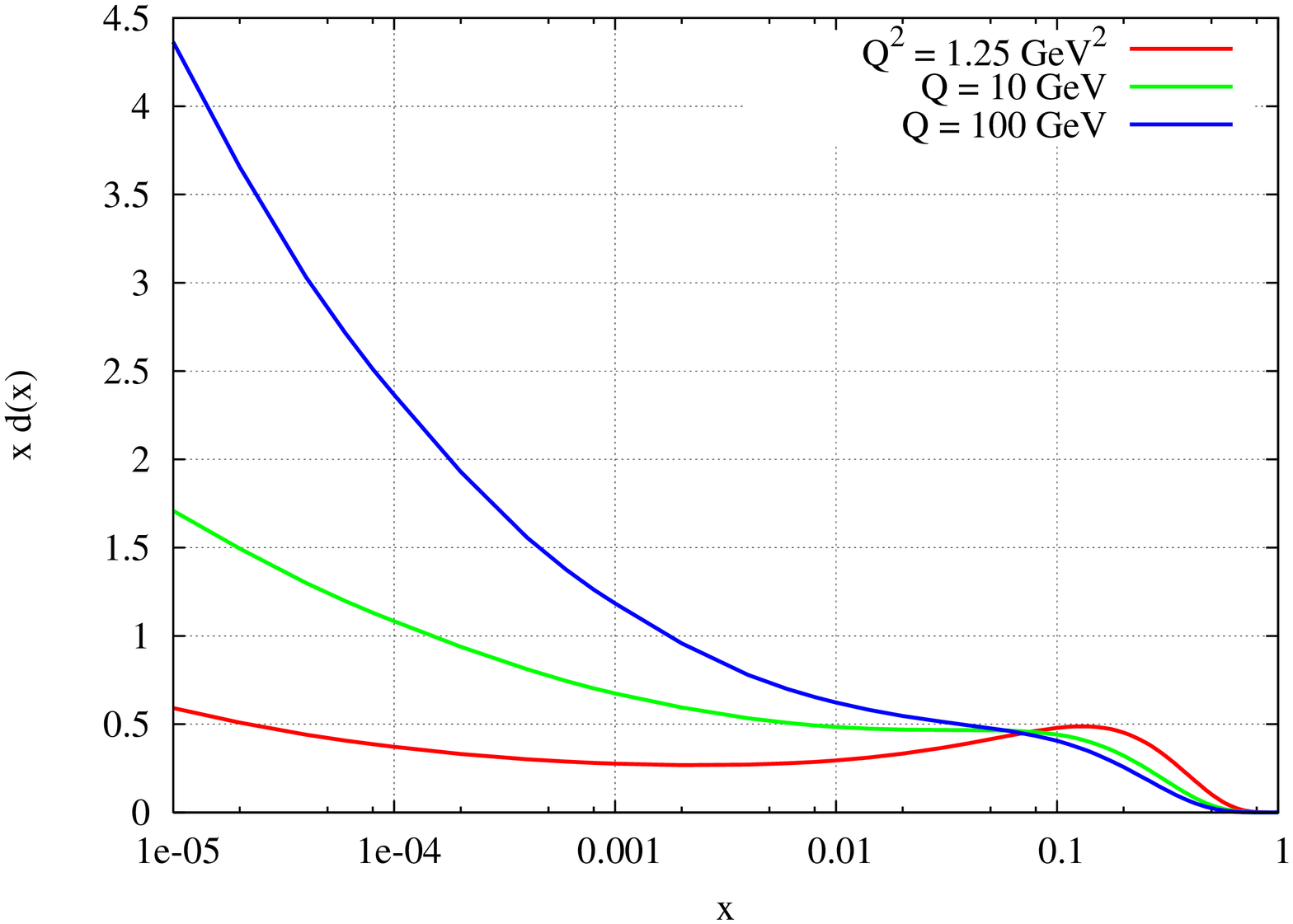}}\end{center}

\caption{Evolution of the quark down distribution $xd$ versus $x$ at different
$Q$ values. All perturbative orders and both input models are shown.\label{fig:Unp_down}}
\end{figure}

\begin{figure}[!tbh]
\begin{center}\subfigure[MRST LO]{\includegraphics[%
  width=5.5cm,
  angle=-90]{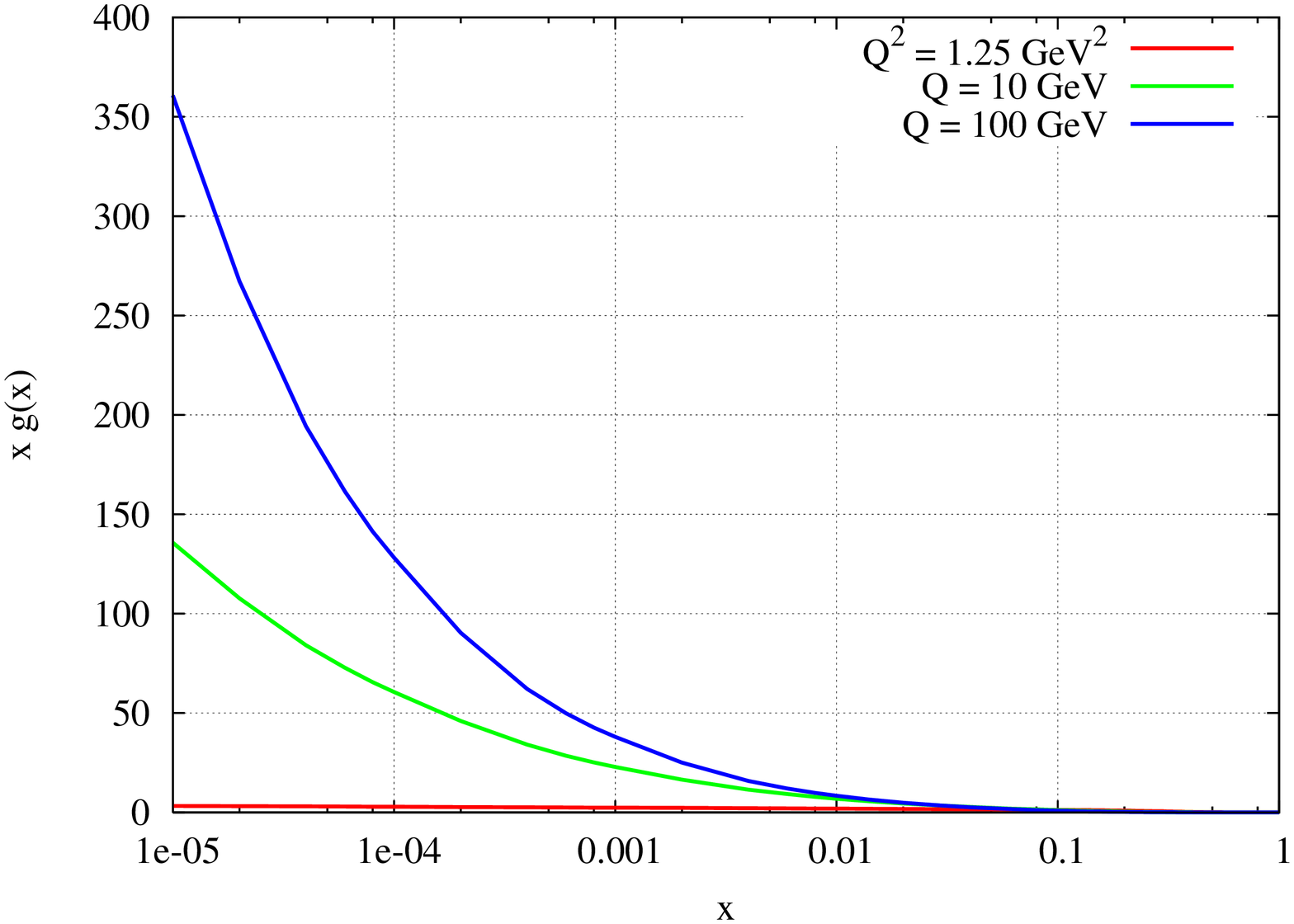}}\subfigure[Alekhin LO]{\includegraphics[%
  width=5.5cm,
  angle=-90]{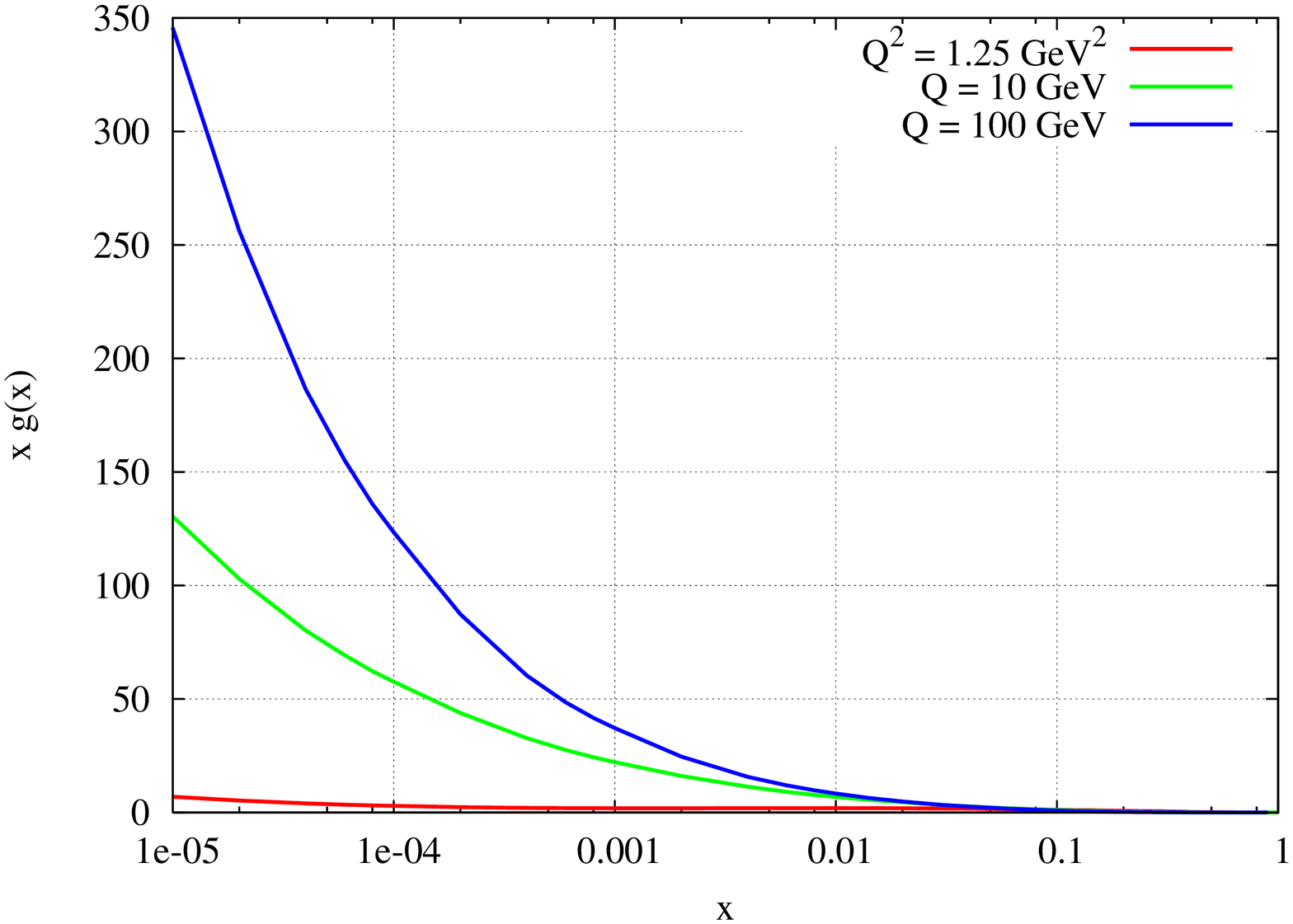}}\end{center}

\begin{center}\subfigure[MRST NLO]{\includegraphics[%
  width=5.5cm,
  angle=-90]{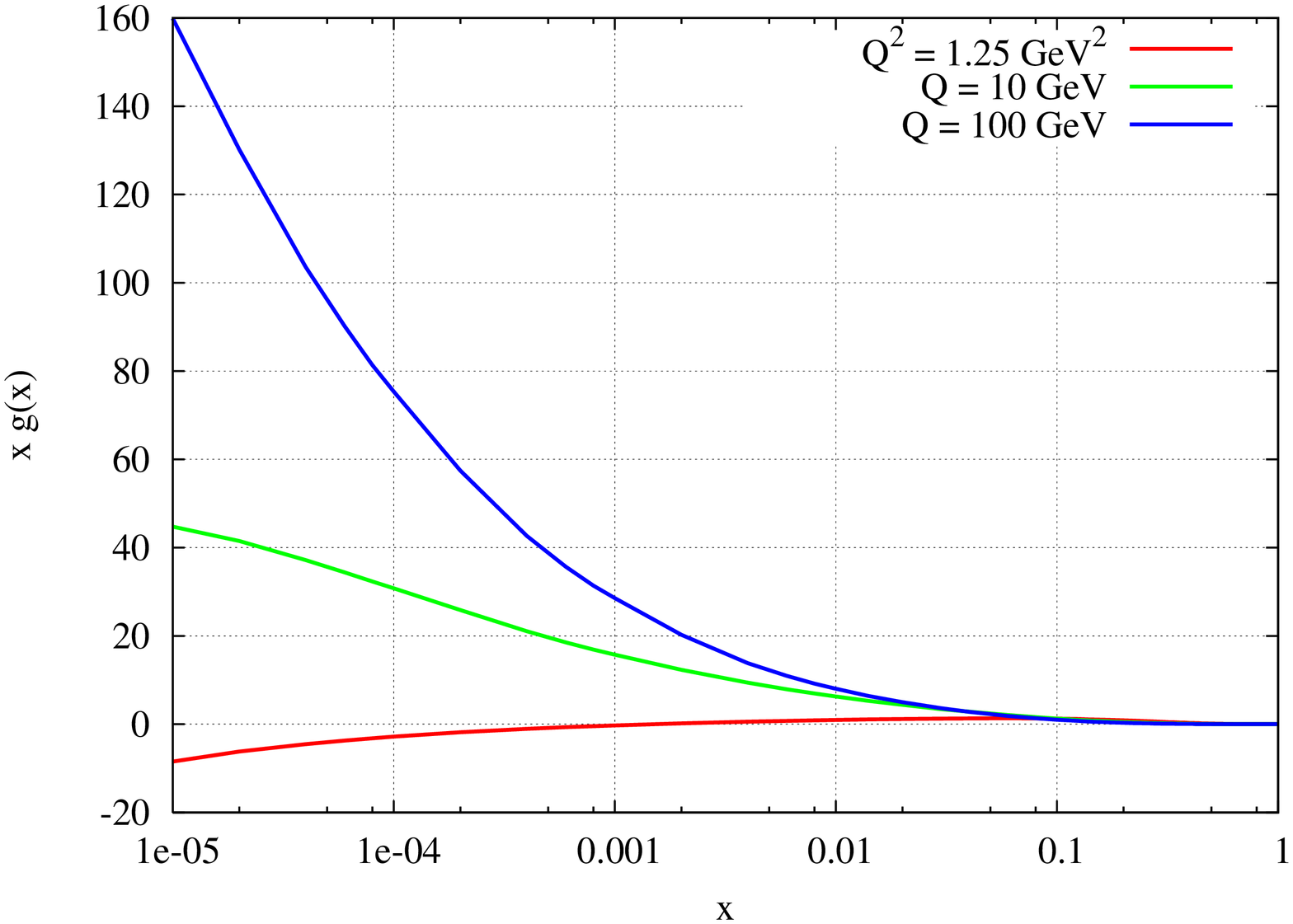}}\subfigure[Alekhin NLO]{\includegraphics[%
  width=5.5cm,
  angle=-90]{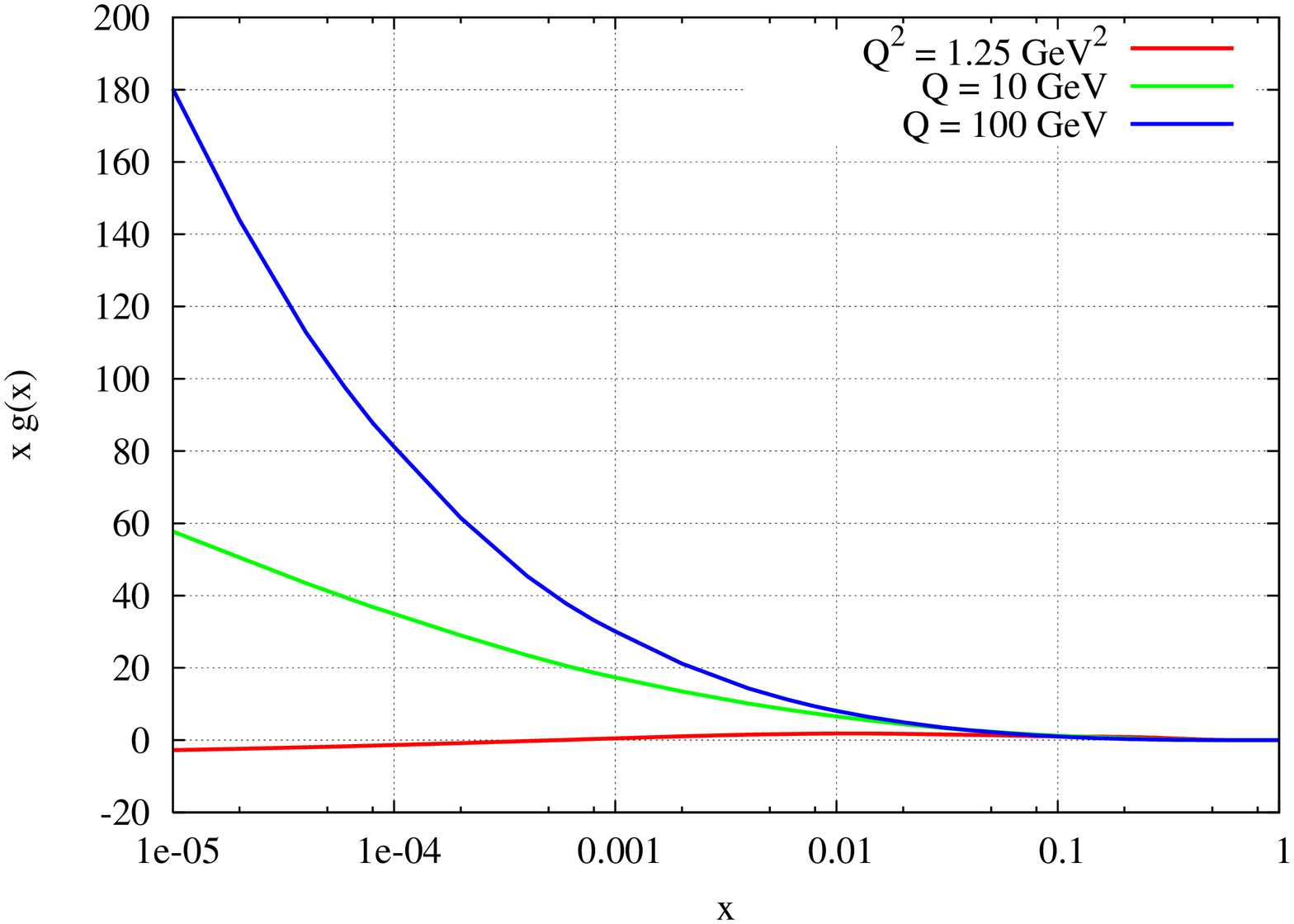}}\end{center}

\begin{center}\subfigure[MRST NNLO]{\includegraphics[%
  width=5.5cm,
  angle=-90]{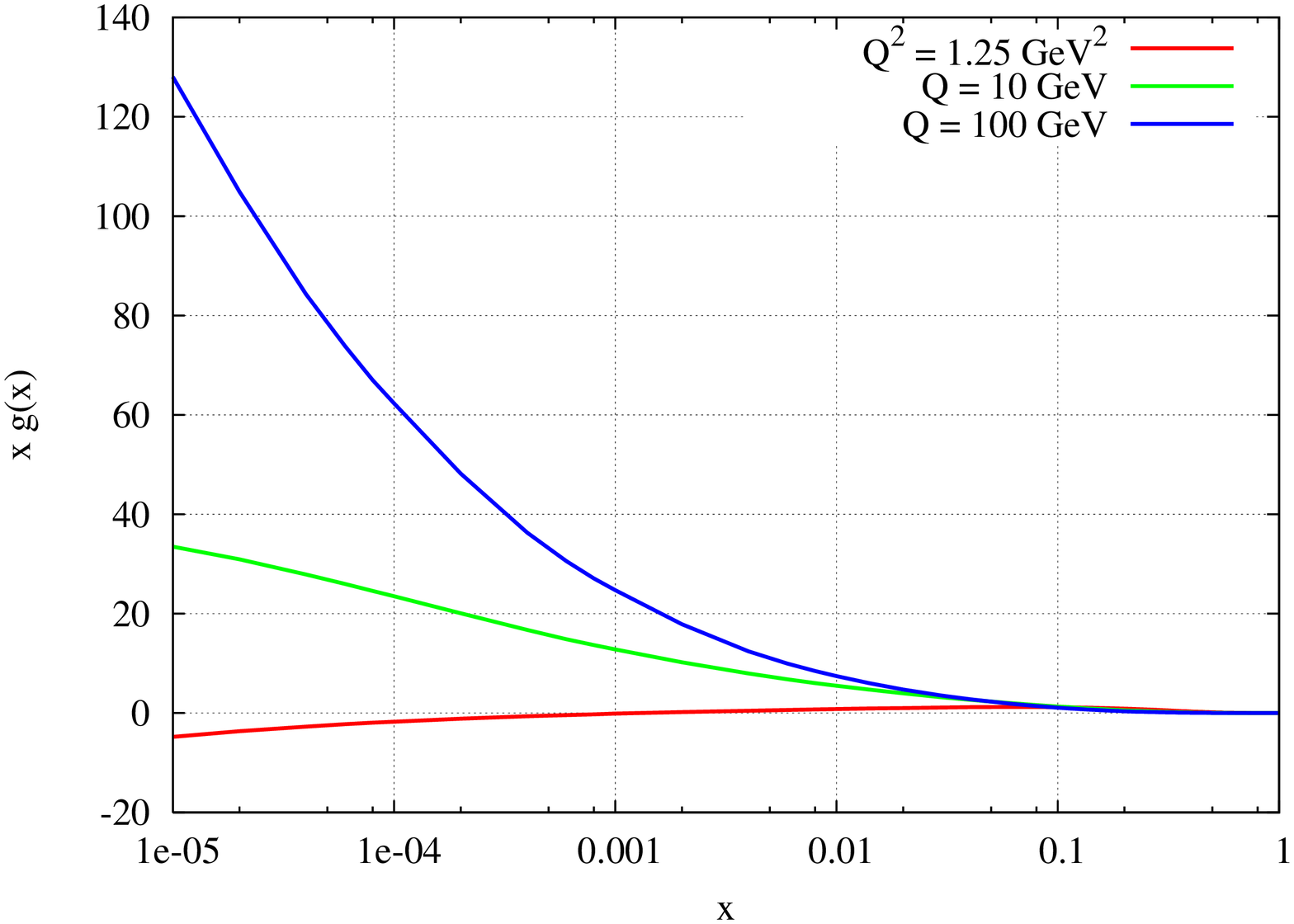}}\subfigure[Alekhin NNLO]{\includegraphics[%
  width=5.5cm,
  angle=-90]{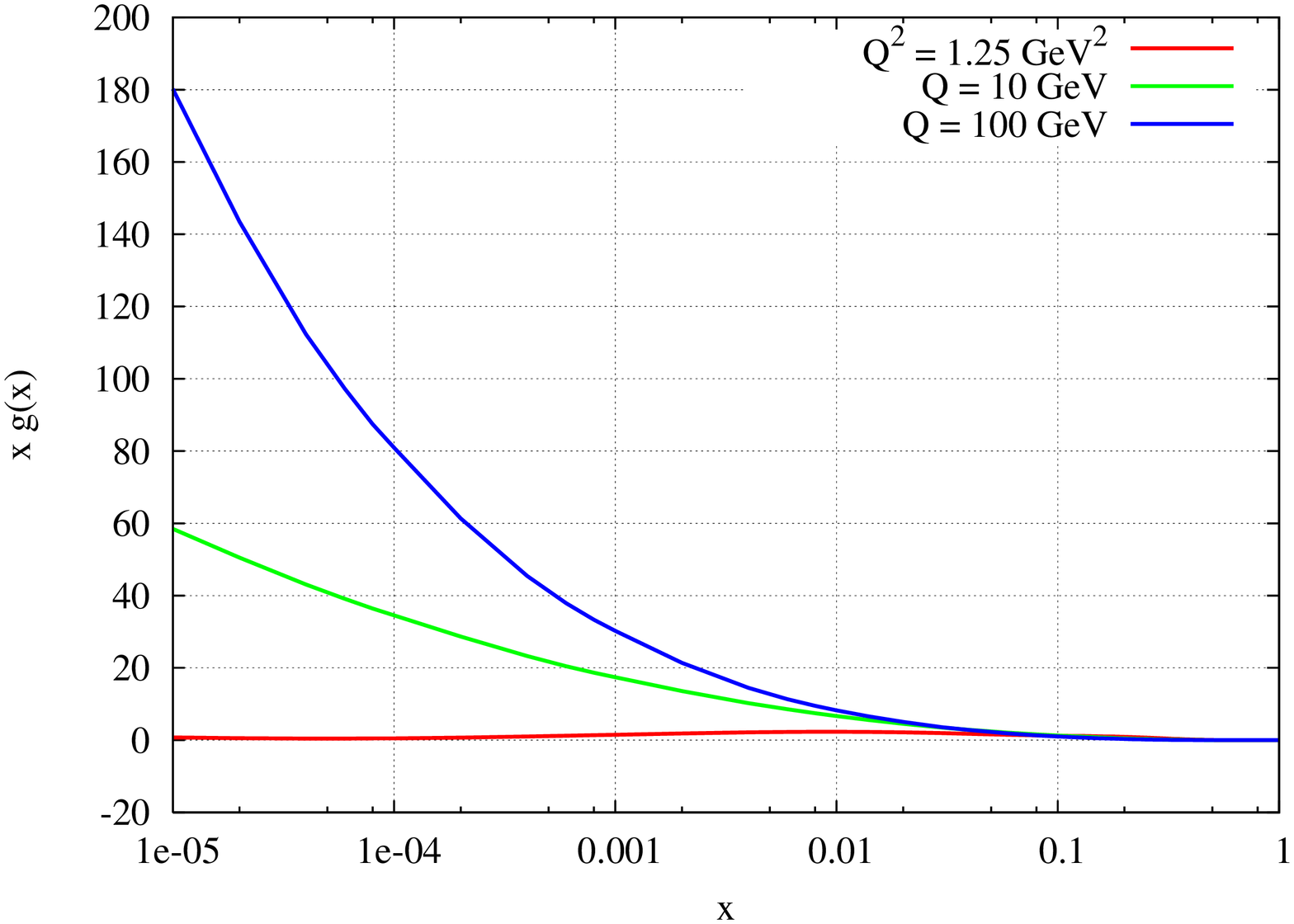}}\end{center}

\caption{Evolution of the gluon distribution $xg$ versus $x$ at different
$Q$ values. All perturbative orders and both input models are shown.\label{fig:Unp_gluon}}
\end{figure}

In Figg.~(\ref{fig:Unp_up} -- \ref{fig:Unp_gluon}), the evolution
of up, down and gluon distributions at $Q=10\,\textrm{GeV}$ and $Q=100\,\textrm{GeV}$
in the unpolarized case is plotted at LO, NLO and NNLO. As initial
distributions we have used both MRST and Alekhin distributions at
the scale $Q_{0}^{2}=1.25\,\textrm{GeV}^{2}$. Some details about
the inputs are shown in the Table (\ref{tab:initialconditions}).

\begin{table}[!tbh]
\begin{center}\begin{sideways}
\begin{tabular}{|c|ccc|ccc|}
\hline 
&
\multicolumn{3}{c|}{MRST}&
\multicolumn{3}{c|}{Alekhin}\tabularnewline
&
 LO&
 NLO&
 NNLO&
 LO&
 NLO&
 NNLO\tabularnewline
\hline
$m_{c}$&
\multicolumn{3}{c|}{$1.43$}&
\multicolumn{3}{c|}{$1.5$}\tabularnewline
$m_{b}$&
\multicolumn{3}{c|}{$4.3$}&
\multicolumn{3}{c|}{$4.5$}\tabularnewline
$m_{t}$&
\multicolumn{3}{c|}{$175$}&
\multicolumn{3}{c|}{$180$}\tabularnewline
\hline
$\Lambda^{(3)}$&
 $0.253$&
 $0.374$&
 $0.277$&
 $0.253$&
 $0.364$&
 $0.282$\tabularnewline
$\Lambda^{(4)}$&
 $0.220$&
 $0.323$&
 $0.235$&
 $0.219$&
 $0.311$&
 $0.238$\tabularnewline
$\Lambda^{(5)}$&
 $0.170$&
 $0.225$&
 $0.166$&
 $0.168$&
 $0.215$&
 $0.167$\tabularnewline
$\Lambda^{(6)}$&
 $0.088$&
 $0.091$&
 $0.068$&
 $0.086$&
 $0.086$&
 $0.069$\tabularnewline
\hline
file&
 \texttt{lo2002.dat}&
 \texttt{alf119.dat}&
 \texttt{vnvalf1155.dat}&
 \texttt{a02m.pdfs\_1\_vfn}&
 \texttt{a02m.pdfs\_2\_vfn}&
 \texttt{a02m.pdfs\_3\_vfn}\tabularnewline
Ref.&
 \cite{MRST2}&
 \cite{MRST1}&
 \cite{MRST2}&
 \cite{Alekhin}&
 \cite{Alekhin}&
 \cite{Alekhin} \tabularnewline
\hline
\end{tabular}
\end{sideways}\end{center}

\caption{Various parton density sets used in the plots shown in this section.
Quark masses and the $\Lambda^{(n_{f})}$ parameter are in GeV.\label{tab:initialconditions}}
\end{table}


\chapter{NNLO precision studies of the Higgs total cross section\label{chap:Higgs}}
\fancyhead[LO]{\nouppercase\leftmark}

\section{Introduction}
One of the main objective at the LHC will be the search of the Higgs particle, which is responsible for the 
mechanism of mass generations for the fermions and the gauge bosons within the Standard Model. 
At the moment one of the most widely accepted ways to generate a mass for these particles relies 
on a scalar sector with a single Higgs field and a symmetry breaking potential. While the Lagrangian 
is invariant under a local symmetry of $SU(2)\times U(1)$ and this, of course, 
requires the same symmetry for the scalar sector as well, the vacuum of the theory, which is a stable state 
of minimum energy, doesn't share the same symmetry. The presence of a 
minimal Higgs sector is also specific of the Standard Model, though many extensions of the Standard 
Model, supersymmetric and not, require a far more complex Higgs structure to give mass to both 
up-type and down-type quarks and to the leptons by some trilinear interactions termed Yukawa couplings.   
Our objective, in this chapter, is to present very accurate predictions for the total Higgs cross section 
by combining results available in the literature, specifically the computation of NNLO hard 
scatterings for the production of a  scalar and a pseudoscalar Higgs in pp collisions. 
The analytical expressions are available through the work of Ravindran, Smith and van Neerven \cite{Smith}. These 
have been combined with the NNLO evolution of the parton distributions in order to generate very accurate 
cross sections for the Higgs search at the LHC.

After the submission of this thesis, the preliminar work presented in this chapter has been completed and published in \cite{CCGS}.

\section{Higgs detection at hadron colliders} 
We summarize some generalities on the Standard Model and the mechanism of mass generation, 
with an emphasis on their phenomenological implications to hadron colliders. 

The electroweak sector of the standard model 
is a chiral theory  $SU(2)_L \times U(1)_Y$ gauge theory containing
$3$ $SU(2)$ gauge bosons, $W_\mu^i$, and one $U(1)$
gauge boson, $B_\mu$. Left-handed and right-handed quarks couple differently 
to the 3 gauge bosons of the weak isospin and to $B_\mu$. 
The Lagrangian is given by
\beq
{\cal L}_{\rm KE} =-{1\over 4}W_{\mu\nu}^i W^{\mu\nu i}
-{1\over 4} B_{\mu\nu} B^{\mu\nu}
\eeq
where
\beqn
W_{\mu\nu}^i&=& \partial_\nu W_\mu^i-\partial _\mu W_\nu^i
+g \epsilon^{ijk}W_\mu^j W_\nu^k
\nonumber \\
B_{\mu\nu}&=&\partial_\nu B_\mu-\partial_\mu B_\nu\quad .
\eeqn
A complex scalar field, doublet of $SU(2)$ is introduced and defined as 
\beq
\Phi={1\over \sqrt{2}}
\left(\begin{array}{c}
\phi_1+i\phi_2  \\
H+i\phi_0   \end{array}\right) \quad ,
\eeq
and with a  scalar potential given by
\beq
 V(\Phi)=\mu^2 \mid \Phi^\dagger\Phi\mid +\lambda
\biggl(\mid \Phi^\dagger \Phi\mid\biggr)^2\quad ,
\label{wspot}
\eeq
($\lambda>0$).
This turns out to be the most general renormalizable and $SU(2)$ invariant
potential allowed. The minimum of the potential is now a valley of equivalent 
vacua, among which we choose a vacuum by a suitable parameterization. One possibility 
is given by 

\beq
\langle \Phi\rangle
= {1\over\sqrt{2}} \left(\begin{array}{c}
 0   \\
 v   \end{array}\right)\quad 
\label{vevdef}
\eeq
with a $U(1)_Y$ hypercharge $Y_\Phi=1$. The electromagnetic charge is
$Q=T_3 +{Y\over 2}$ and once can easily check that this state has not electric charge since 
\beq
Q \langle \Phi\rangle
= 0
\eeq
The vacuum expectation value of Eq. \ref{vevdef} allows to leave the $U(1)_{em}$ intact 
by making one linear combination of the third component of $SU(2)$ and $U(1)_Y$ gauge symmetric, 
with $Q$ identified with the generator of $U(1)_{em}$, according to the braking pattern
 
$SU(2)_L\times U(1)_Y\rightarrow U(1)_{EM}$.
The masses of the gauge bosons are generated by the kinetic part of the Higgs Lagrangian

\beq
{\cal L}_s=(D^\mu \Phi)^\dagger (D_\mu \Phi)-V(\Phi)
\eeq
where $\tau_i$ are the Pauli matrices. The explicit form of the covariant derivatives is

\beq
D_\mu=\partial_\mu +i {g\over 2}\tau\cdot W_\mu+i{g^\prime\over 2}
B_\mu.
\eeq
In unitary gauge the scalar field can be written as
\beq
\Phi={1\over \sqrt{2}}\left(\begin{array}{c}  0 \\
 v+H\end{array}\right)
\eeq
and the contribution to the gauge boson masses
from the scalar kinetic energy term are
\beq
{1\over 2} (0 \quad v )
\biggl({1\over 2}g \tau\cdot W_\mu
+{1\over 2} g^\prime B_\mu
\biggr)^2 \left(\begin{array}{c}  0 \\  v \end{array}
\right).
\eeq
To summarize, the masses of the massive gauge bosons
\beqn
W^{\pm}_\mu&=&
{1\over \sqrt{2}}(W_\mu^1 \mp i W_\mu^2)\nonumber \\
Z^\mu&=& {-g^\prime B_\mu+ g W_\mu^3\over \sqrt{g^2+g^{\prime~2}}}
\nonumber \\
A^\mu&=& {g B_\mu+ g^{\prime} W_\mu^3\over \sqrt{g^2+g^{\prime~2}}}.
\eeqn
are given by
\beqn
M_W^2 &=& {1\over 4} g^2 v^2\nonumber \\
M_Z^2 &=& {1\over 4} (g^2 + g^{\prime~2})v^2\nonumber \\
M_\gamma& = & 0.
\eeqn

and the coupling constants satisfy relations
\beqn
e&=& g \sin\theta_W \nonumber \\
e&=& g^\prime \cos\theta_W
\eeqn
with $theta_W$ being the Weinberg angle. 
Notice that in the unitary gauge a counting of the degrees of freedom for the Higgs scalar and the 
gauge bosons remains the same, as expected, without the appearance of unphysical Goldstone bosons. 
Alternative parameterizations, a la Kibble 

\beq
\Phi={ e^{i{\omega\cdot\tau\over v}}\over \sqrt{2}}
\left(\begin{array}{c}  0 \\
 v+H\end{array}\right).
\eeq
let the Goldstone modes appear from the scalar sector. The appearance of this modes 
is a gauge variant effect. Non unitary gauges are important especially in the context of testing the unitarity of the model.  
In the Standard Model, there are three Goldstone bosons,
${\vec \omega}=(\omega^\pm,z)$, with masses $M_W$ and $M_Z$ in
the Feynman gauge.  

In addition to giving the $W$ and $Z$ bosons their masses, the Higgs
boson is also responsible for giving the mass to the fermions. 
Thegauge invariant  Yukawa coupling of the
Higgs boson to fermions is given by trilinear terms of the form 
\beq
{\cal L}_f=-\lambda_d {\overline Q}_L \Phi d_R + h.c.\quad ,
\eeq
where the left handed $SU(2)$ fermion doublet  is constructed by acting with the chiral 
Lorenz projectors on both components 
\beq
Q_L=
\left(\begin{array}{c}
u\\ d \end{array}\right)_L
{}.
\eeq
This gives the effective coupling
\beq
\lambda_d {1\over\sqrt{2}}
({\overline u}_L,~ {\overline d}_L)\left(
\begin{array}{c}  0 \\
v+ H \end{array} \right) d_R + h.c.
\eeq
which can be seen to yield a mass term for the down quark if
we make the identification
\beq
\lambda_d = {m_d \sqrt{2}\over v}.
\eeq
The mass term for the up quark is obtained by giving a value expectation value to the field 
 $\Phi^c \equiv - i \tau_2 \Phi^*$ which is an $SU(2)$
doublet and  allows to write down the $SU(2)$ invariant coupling
\beq
\lambda_u {\overline Q}_L \Phi^c u_R + h.c.
\eeq
which is the mass term for the up quark. For the upper components of the lepton $SU(2)$  doublets 
the procedure is the same, though the generation of a mass term for the neutrino is still a problem within 
the Standard Model. The neutrino appears only in its left-component and therefore has no Yukawa coupling.  

For the multi-family case, the Yukawa couplings, $
\lambda_d$ and $\lambda_u$, become matrices which are valued in family space  ($N_F \times N_F$) 
with $N_F$ being the number of families. If we introduce a family mixing matrix ${\cal U}$ and define $g_2$ 
and $g_Y$ to be the $SU(2)$ and $U(1)_Y$ coupling constant the final Lagrangian describing the interactions of the leptons with the massive 
gauge bosons is 

\beqa
{\cal L}_l &=& e\overline{e}_i\gamma^\m e_i 
-\frac{g_2}{2 \sqrt{2}} W^+ \overline{e}_j\gamma^\m(1 - \gamma_5) \nu_i\, {\cal U}^\nu_{ji} 
-\frac{g_2}{2 \sqrt{2}} W^- \overline{\nu}_j\gamma^\m(1 - \gamma_5) e_i\, 
{\cal U}^{\nu\dag}_{ji} \nonumber \\
&& - \frac{g_2}{2\cos\theta_W} Z^\mu \overline{\nu}_i\left( g^{\nu-Z}_V\gamma^\mu - g^{\nu-Z}_A\gamma^\mu \gamma^5\right)\nu_i
- \frac{g_2}{2\cos\theta_W} Z^\mu \overline{e}_i\left( g^{e-Z}_V\gamma^\mu - g^{e-Z}_A\gamma^\mu \gamma^5\right)e_i \nonumber \\
\eeqa

where 
\beq
g_V^{f-Z^0}=T_{w3}^{(f)}- 2 \sin^2\theta_W Q_{el}^{(f)}
\qquad g_A^{f-Z^0}=T_{w3}^{(f)}
\eeq

while for the quarks one obtains
\beqa
{\cal L}_q &=& -e \left( \frac{2}{3} \overline{u}_i\gamma^\m u_i -
\frac{1}{3} \overline{d}_i\gamma^\m d_i \right)A_\mu \nonumber \\
&& - \frac{g_2}{2 \sqrt{2}} W^+ \overline{u}_j\gamma^\m(1 - \gamma_5) d_i\, {\cal U}^q_{ji} 
-\frac{g_2}{2 \sqrt{2}} W^- \overline{d}_j\gamma^\m(1 - \gamma_5) u_i\, {\cal U}^{q\dag}_{ji} \nonumber \\
&& - \frac{g_2}{2\cos\theta_W} Z^\mu \overline{u}_i\left( g^{u-Z}_V\gamma^\mu - g^{u-Z}_A\gamma^\mu \gamma^5\right)u_i
-  \frac{g_2}{2\cos\theta_W} Z^\mu \overline{d}_i\left( g^{d-Z}_V\gamma^\mu - g^{d-Z}_A\gamma^\mu \gamma^5\right)d_i. \nonumber \\
\eeqa

\begin{figure}
{\centering \resizebox*{8cm}{!}{\rotatebox{0}
{\includegraphics{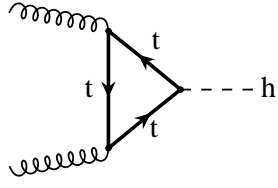}}}\par}  
\caption{The leading order diagram for Higgs production by gluon fusion}
\label{g_fusion}
\end{figure}

\begin{figure}
{\centering \resizebox*{8cm}{!}{\rotatebox{0}
{\includegraphics{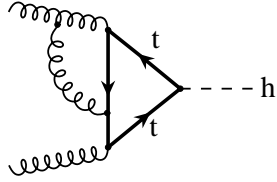}}}\par}  
\caption{A typical NLO diagram for Higgs production by gluon fusion}
\label{g_fusion1}
\end{figure}

A complete set of Feynman rules describing the interaction of the Higgs with the massive states of the 
Standard Model in both unitary and non unitary gauges, which involve the goldstones, can be found in 
\cite{CORE}. The Higgs field, being responsible for the mechanism of mass generation, can be radiated off 
by any massive state and its coupling is proportional to the mass of the same state. At the LHC one of the 
golden plated modes to search for the Higgs is its production via the mechanism of gluon fusion. 
The leading order contribution is shown in fig.~(\ref{g_fusion}) which shows the dependence 
of the amplitude through the quark loop. Most of the contribution comes from the top quark, since this 
is the heaviest and has a larger coupling to the Higgs field. NLO and NNLO corrections 
have been computed in the last few years by various groups. A typical NLO correction is shown in 
fig.~(\ref{g_fusion1}). From the computation of the LO diagram one gets 
\beq
{\cal A}(g g \rightarrow H)
=- {\alpha_s m^2 \over  \pi v}
\delta_{ab}\biggl(g^{\mu\nu}{M_H^2\over 2}-p^\nu q^\mu
\biggr)\int dx dy \biggl({1-4 x y
\over m^2-M_H^2 x y}\biggr)
\epsilon_\mu(p)\epsilon_\nu(q).
\label{fusion}
\eeq
where $a$ and $b$ are color indices and p and q are the momenta of the 
two gluons. When the fermion in the loop is much heavier than the
Higgs boson, $m>>M_H$ the expression above becomes
\beq
{\cal A}(gg\rightarrow H)\longrightarrow_{m >> M_H}
-{\alpha_s\over 3 \pi v}\delta_{a b}
\biggl(g^{\mu\nu}{M_H^2\over 2}- p^\nu q^\mu
\biggr)\epsilon_\mu(p)\epsilon_\nu(q).
\label{heavym}
\eeq
which apparently does not depend on the quark mass. This implies that if there are additional generators, 
then the contribution of this process is proportional to the number of additional generations, 
providing an important window into new physics beyond the Standard Model. 
The resonant production of a heavy Higgs is instead given by the formula 
\beq
{\hat\sigma}(g g\rightarrow H)={\pi^2\over 8  M_H^3}
\Gamma(H\rightarrow g g)\delta(1-{M_H^2\over {\hat s}})  .
\eeq
relating the cross section for Higgs production via gluon fusion to the corresponding 
decay rate $\Gamma(H\to g g)$.
At parton level we obtain 
\beq {\hat \sigma}(g g \rightarrow H)=
{\alpha_s^2\over 64 \pi v^2}M_H^2 \mid I\biggl({M_H^2\over m^2}
\biggr)\mid^2 \delta({\hat s}-M_H^2)
\eeq
where $\sqrt{{\hat s}}$ is the energy in the gluon -gluon
center of mass and the integral $I$ is defined by
\beq
 I(a) \equiv \int_0^1 dx \int^{1-x}_0  dy{1-4xy\over 1-a x y}.
\label{intdef}
\eeq
\subsection{The effective Lagrangian}
 An important feature of the result for Higgs boson production
from gluon fusion is that it is independent of the
heavy quark mass for a light Higgs boson.  
Eq.~\ref{heavym} can be derived
from an effective vertex 
\beqn
{\cal L}_{\rm eff}&=&{\alpha_s \over 12 \pi} G_{\mu\nu}^A G^{A~\mu\nu}
\biggl({H\over v}\biggr)\nonumber \\
&=&{\beta_F \over g_s} G_{\mu\nu}^A G^{A~\mu\nu}
\biggl({H\over 2v}\biggr)(1-2 \alpha_s/\pi),\nonumber
\label{effth}
\eeqn
with
\beq
\beta_F={g_s^3 N_H\over 24 \pi^2}
\eeq
being the contribution of heavy fermion loops
to the QCD beta function. 
The effective Lagrangian can be used to compute the
radiative corrections in the gluon sector. 
The correction in principle involve
2-loop diagrams.  However, using the effective vertices from
Eq. \ref{effth}, the ${\cal O}(\alpha_s^3)$ corrections can be found
from a 1-loop calculation. In this way the contribution from the top quark loop is 
shrunk to a point. 
A discussion of the NNLO approach to the computation of the gluon fusion contributions to 
Higgs production has been presented in \cite{Smith}, work to which we 
refer for more details. We recall that in this work the authors present a study 
for both scalar and pseudoscalar Higgs 
production, the pseudoscalar appearing in 2-Higgs doublets models. In these extended models 
there is a mixing between the two scalars which is usually parameterized in terms of the ratio 
between the two vacuum expectation values at the minimum of the potential $v_1/v_2=\tan\beta$. Diagonalization of the mass matrix for the Higgs at the minimum introduces scalars 
and pseudoscalars interactions between the various Higgs and the quarks, as shown from the structure of the operator $O_2$. Here we will briefly summarize their results. 
 
In the large top-quark mass limit the Feynman rules 
for scalar Higgs  production (${\rm H}$) can be derived from the
effective Lagrangian density
\begin{eqnarray}
\label{eqn2.1}
{\cal L}^{\rm H}_{eff}=G_{\rm H}\,\Phi^{\rm H}(x)\,O(x) \quad 
\mbox{with} \quad O(x)=-\frac{1}{4}\,G_{\mu\nu}^a(x)\,G^{a,\mu\nu}(x)\,,
\end{eqnarray}
whereas the production of a pseudo-scalar Higgs (${\rm A}$) is obtained from
\begin{eqnarray}
\label{eqn2.2}
&&{\cal L}_{eff}^{\rm A}=\Phi^{\rm A}(x)\Bigg [G_{\rm A}\,O_1(x)+
\tilde G_{\rm A}\,O_2(x)\Bigg ] \quad \mbox{with} \quad
\nonumber\\[2ex]
&&O_1(x)=-\frac{1}{8}\,\epsilon_{\mu\nu\lambda\sigma}\,G_a^{\mu\nu}\,
G_a^{\lambda\sigma}(x) \,,
\nonumber\\[2ex]
&&O_2(x) =-\frac{1}{2}\,\partial^{\mu}\,\sum_{i=1}^{n_f}
\bar q_i(x)\,\gamma_{\mu}\,\gamma_5\,q_i(x)\,,
\end{eqnarray}
where $\Phi^{\rm H}(x)$ and  $\Phi^{\rm A}(x)$ represent the scalar and 
pseudo-scalar fields respectively and $n_f$ denotes the number of light 
flavours.
$G_a^{\mu\nu}$ is the field strength of QCD 
and the quark field are denoted by $q_i$.
With an appropriate normalization the constants turn out to be given by
\begin{eqnarray}
\label{eqn2.3}
G_{\rm B}&=&-2^{5/4}\,a_s(\mu_r^2)\,G_F^{1/2}\,
\tau_{\rm B}\,F_{\rm B}(\tau_{\rm B})\,{\cal C}_{\rm B}
\left (a_s(\mu_r^2),\frac{\mu_r^2}{m_t^2}\right )\,,
\nonumber\\[2ex]
\tilde G_{\rm A}&=&-\Bigg [a_s(\mu_r^2)\,C_F\,\left (\frac{3}{2}-3\,
\ln \frac{\mu_r^2}{m_t^2}\right )+\cdots \Bigg ]\,G_{\rm A}\,,
\end{eqnarray}
and $a_s(\mu_r^2)$ is defined by
\begin{eqnarray}
\label{eqn2.4}
a_s(\mu_r^2)=\frac{\alpha_s(\mu_r^2)}{4\pi}\,,
\end{eqnarray}
where $\alpha_s(\mu_r^2)$ is the running coupling constant and $\mu_r$ denotes 
the renormalization scale. $G_F$ is the Fermi constant and the
functions $F_{\rm B}$ are given by
\begin{eqnarray}
\label{eqn2.5}
&& F_{\rm H}(\tau)=1+(1-\tau)\,f(\tau)\,, \qquad  F_{\rm A}(\tau)=f(\tau)\,\cot
\beta\,,
\nonumber\\[2ex]
&& \tau=\frac{4\,m_t^2}{m^2} \,, 
\nonumber\\[2ex]
&&f(\tau)=\arcsin^2 \frac{1}{\sqrt\tau}\,, \quad \mbox{for} \quad \tau \ge 1\,,
\nonumber\\[2ex]
&& f(\tau)=-\frac{1}{4}\left ( \ln \frac{1-\sqrt{1-\tau}}{1+\sqrt{1-\tau}}
+\pi\,i\right )^2 \quad \mbox{for} \quad \tau < 1\,,
\end{eqnarray}
where $\cot \beta$ denotes the mixing angle in the Two-Higgs-Doublet Model.
Further $m$ and $m_t$ denote the masses of the (pseudo-) scalar Higgs
boson and the top quark respectively. 
The coefficient functions ${\cal C}_{\rm B}$ originate from the corrections 
to the top-quark triangular graph provided one takes the 
limit $m_t\rightarrow \infty$. The coefficient functions are known
up to order $\alpha_s^2$ and are given by 
\begin{eqnarray}
\label{eqn2.7}
&&{\cal C}_{\rm H}\left (a_s(\mu_r^2),\frac{\mu_r^2}{m_t^2}\right )=
1+a_s^{(5)}(\mu_r^2)\,\Bigg [5\,C_A-3\,C_F\Bigg ] +
\left (a_s^{(5)}(\mu_r^2)\right )^2\,\Bigg [
\frac{27}{2}\,C_F^2
\nonumber\\[2ex]
&&-\frac{100}{3}\,C_A\,C_F+\frac{1063}{36}\,C_A^2-\frac{4}{3}\,
C_F\,T_f-\frac{5}{6}\,C_A\,T_f +\Big (7\,C_A^2
\nonumber\\[2ex]
&& -11\,C_A\,C_F\Big )\,\ln \frac{\mu_r^2}{m_t^2}
+ n_f\,T_f\,\left (-5\,C_F-\frac{47}{9}\,C_A
+8\,C_F\,\ln \frac{\mu_r^2}{m_t^2} \right ) \Bigg ]\,,
\\[2ex]
\label{eqn2.8}
&&{\cal C}_{\rm A}\left (a_s(\mu_r^2),\frac{\mu_r^2}{m_t^2}\right )=1\,,
\end{eqnarray}
where $a_s^{(5)}$ is given in the five-flavour number scheme. 

Using the effective Lagrangian approach one can calculate the 
total cross section of the reaction
\begin{eqnarray}
\label{eqn2.10}
H_1(P_1)+H_2(P_2)\rightarrow {\rm B}+'X'\,,
\end{eqnarray}
where $H_1$ and $H_2$ denote the incoming hadrons and $X$ represents an 
inclusive hadronic state. The total cross section is given by
\begin{eqnarray}
\label{eqn2.11}
&&\sigma_{\rm tot}=\frac{\pi\,G_{\rm B}^2}{8\,(N^2-1)}\,\sum_{a,b=q,\bar q,g}\,
\int_x^1 dx_1\, \int_{x/x_1}^1dx_2\,f_a(x_1,\mu^2)\,f_b(x_2,\mu^2)\,
\nonumber\\[2ex] && \qquad\qquad\times
\Delta_{ab,{\rm B}}\left ( \frac{x}{x_1\,x_2},\frac{m^2}{\mu^2} \right ) \,,
\quad {\rm B}={\rm H},{\rm A}\,,
\nonumber\\[2ex]
&&\mbox{with}\quad x=\frac{m^2}{S} \quad\,,\quad S=(P_1+P_2)^2\quad
\,,\quad p_5^2=m^2\,,
\end{eqnarray}
where the factor $1/(N^2-1)$ is due to the average over colour
and the parton distributions $f_a(y,\mu^2)$ ($a,b=q,\bar q,g$)
depend on the mass factorization/renormalization scale $\mu$. 
$\Delta_{ab,{\rm B}}$ denotes the partonic hard scattering coefficient computed with NNLO 
accuracy.
\section{Results}
\begin{figure}[!tbh]
\begin{center}\subfigure[MRST]{\includegraphics[%
  width=5.5cm,
  angle=-90]{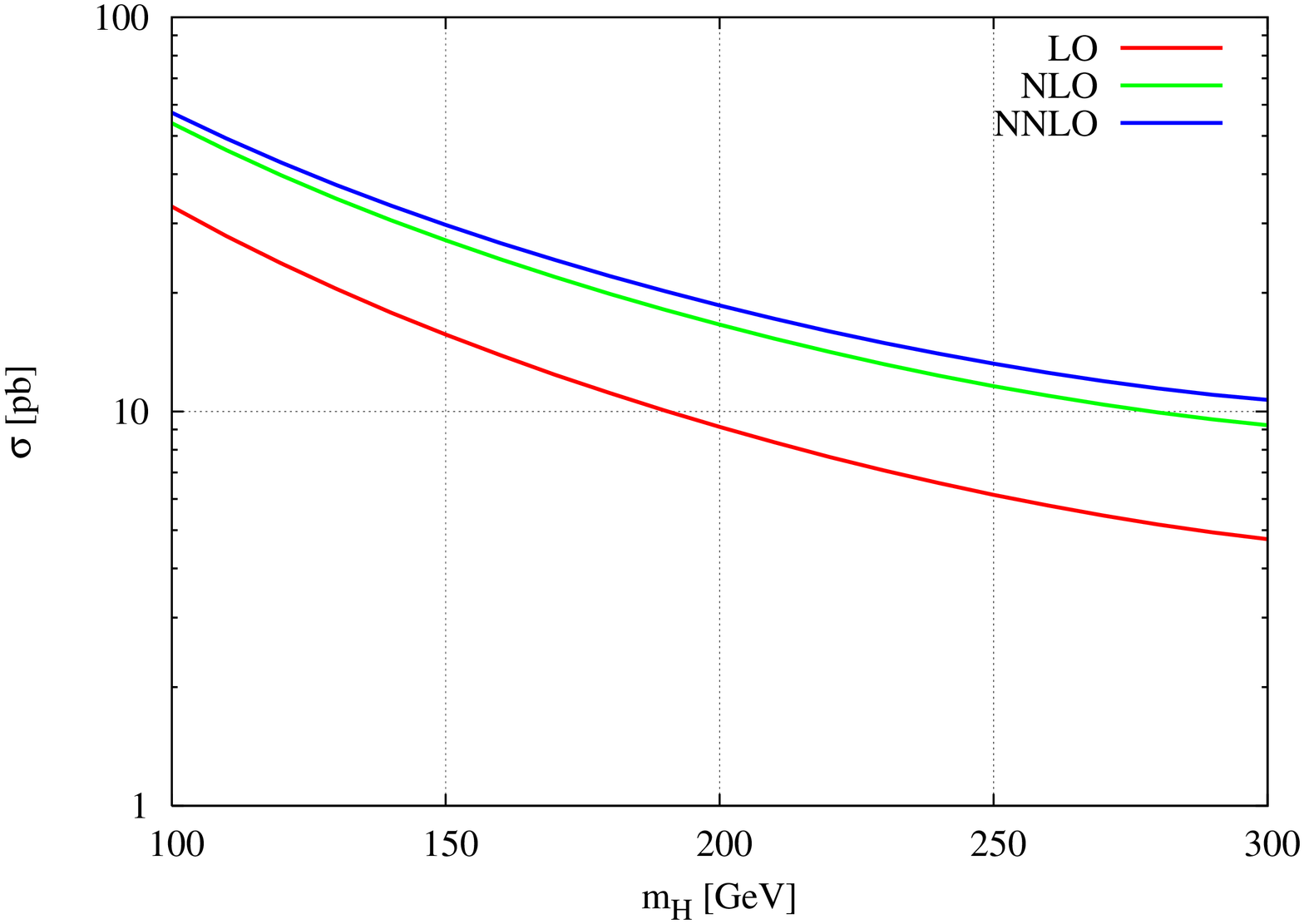}}\subfigure[Alekhin]{\includegraphics[%
  width=5.5cm,
  angle=-90]{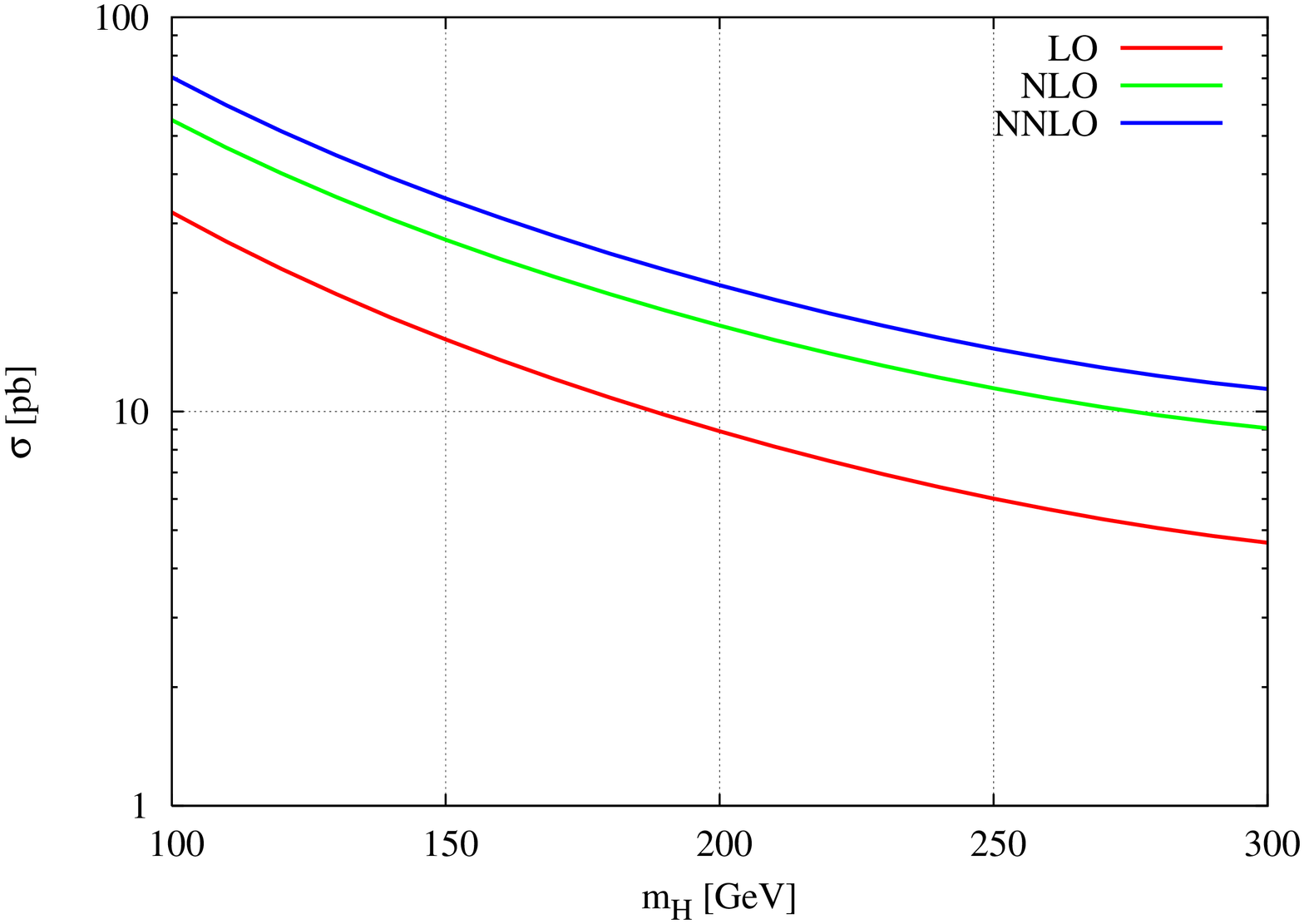}}\end{center}

\caption{Total cross section as a function of Higgs mass at $\sqrt{S}=14\,\textrm{TeV}$
in $pp$ collision, for a scalar Higgs boson.}
\label{3one}
\end{figure}

\begin{figure}[!tbh]
\begin{center}\subfigure[MRST]{\includegraphics[%
  width=5.5cm,
  angle=-90]{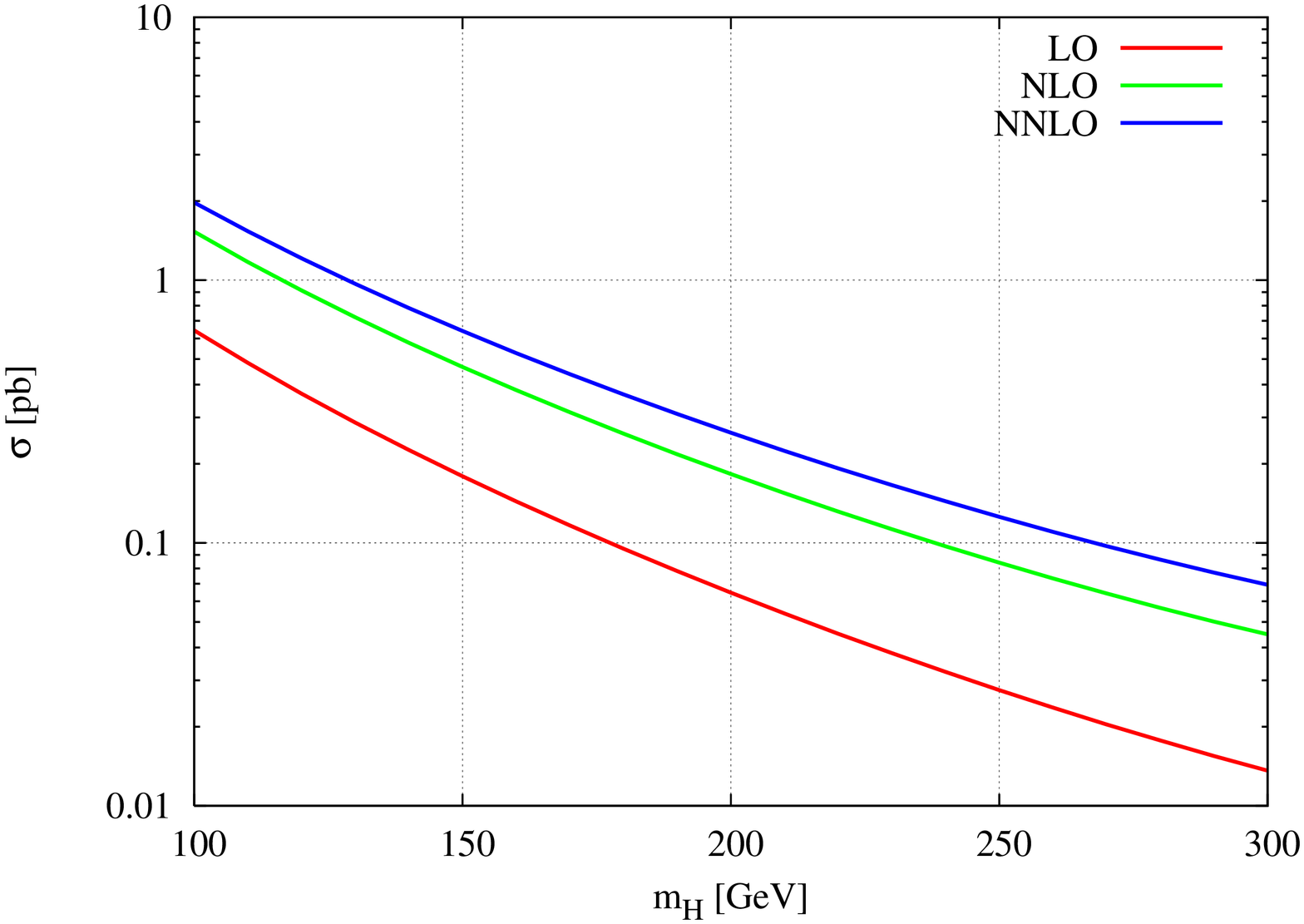}}\subfigure[Alekhin]{\includegraphics[%
  width=5.5cm,
  angle=-90]{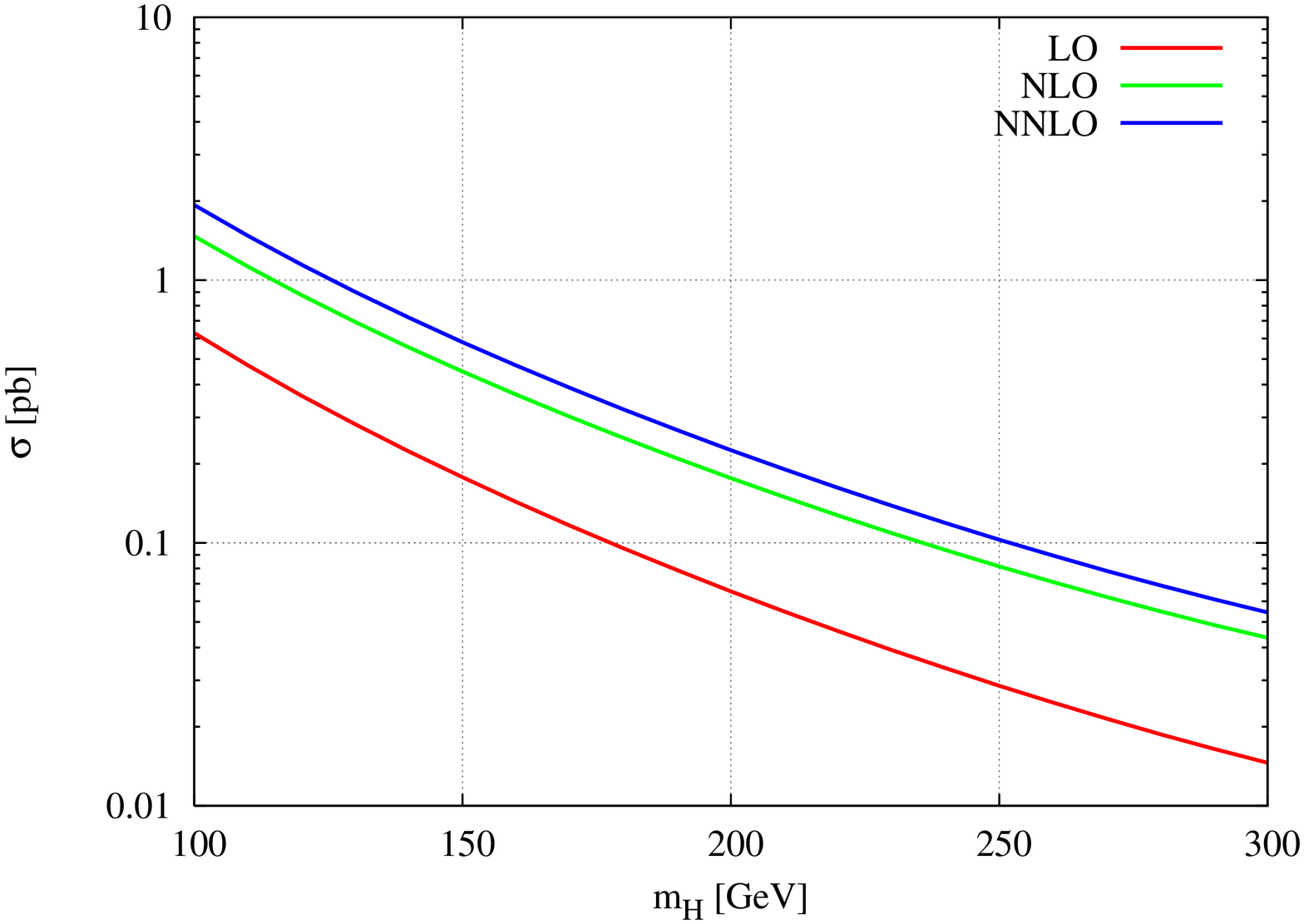}}\end{center}

\caption{Total cross section as a function of Higgs mass at $\sqrt{S}=2\,\textrm{TeV}$
in $p\bar{p}$ collision, for a scalar Higgs boson.}
\label{3two}
\end{figure}

\begin{figure}[!tbh]
\begin{center}\subfigure[MRST]{\includegraphics[%
  width=5.5cm,
  angle=-90]{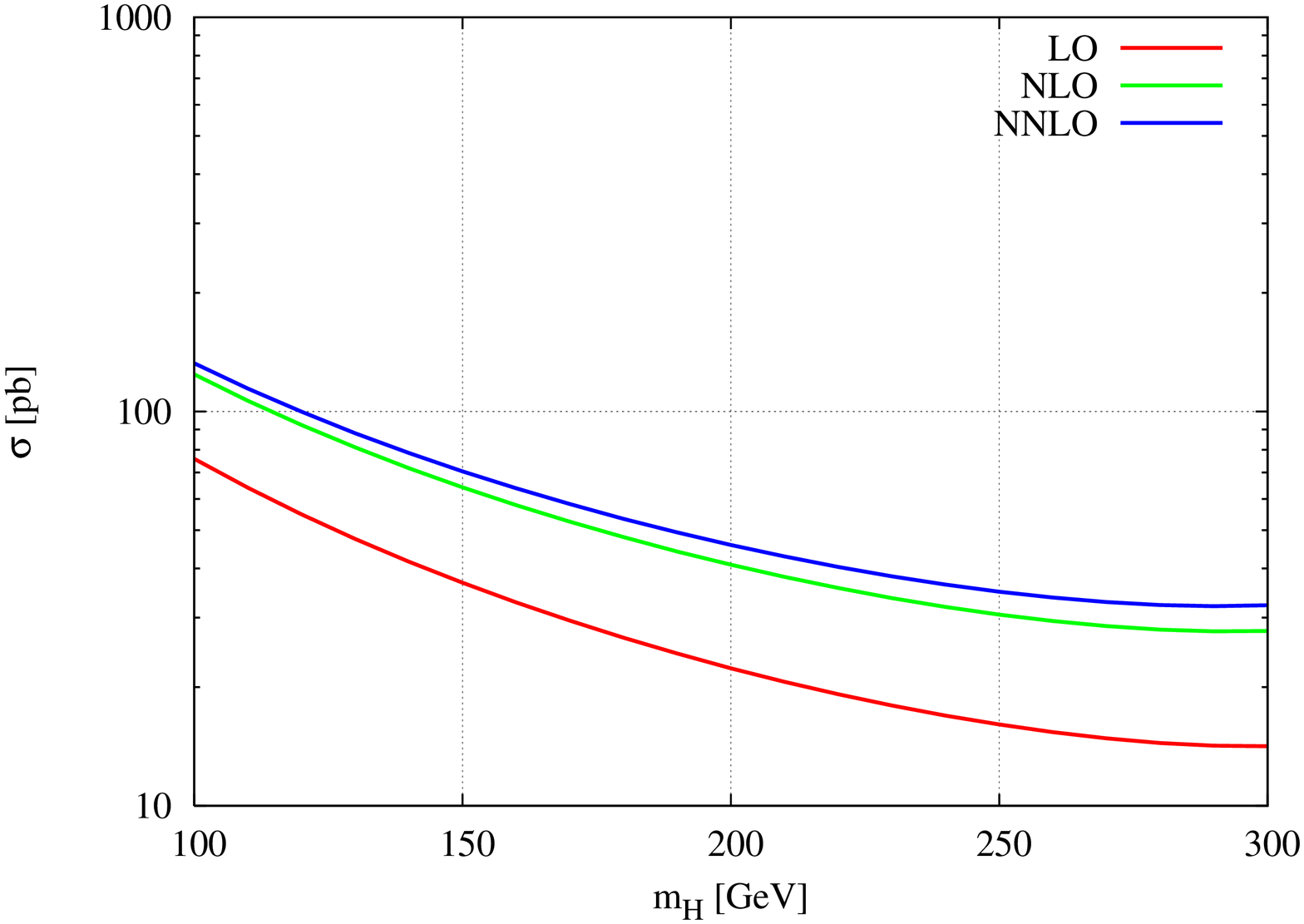}}\subfigure[Alekhin]{\includegraphics[%
  width=5.5cm,
  angle=-90]{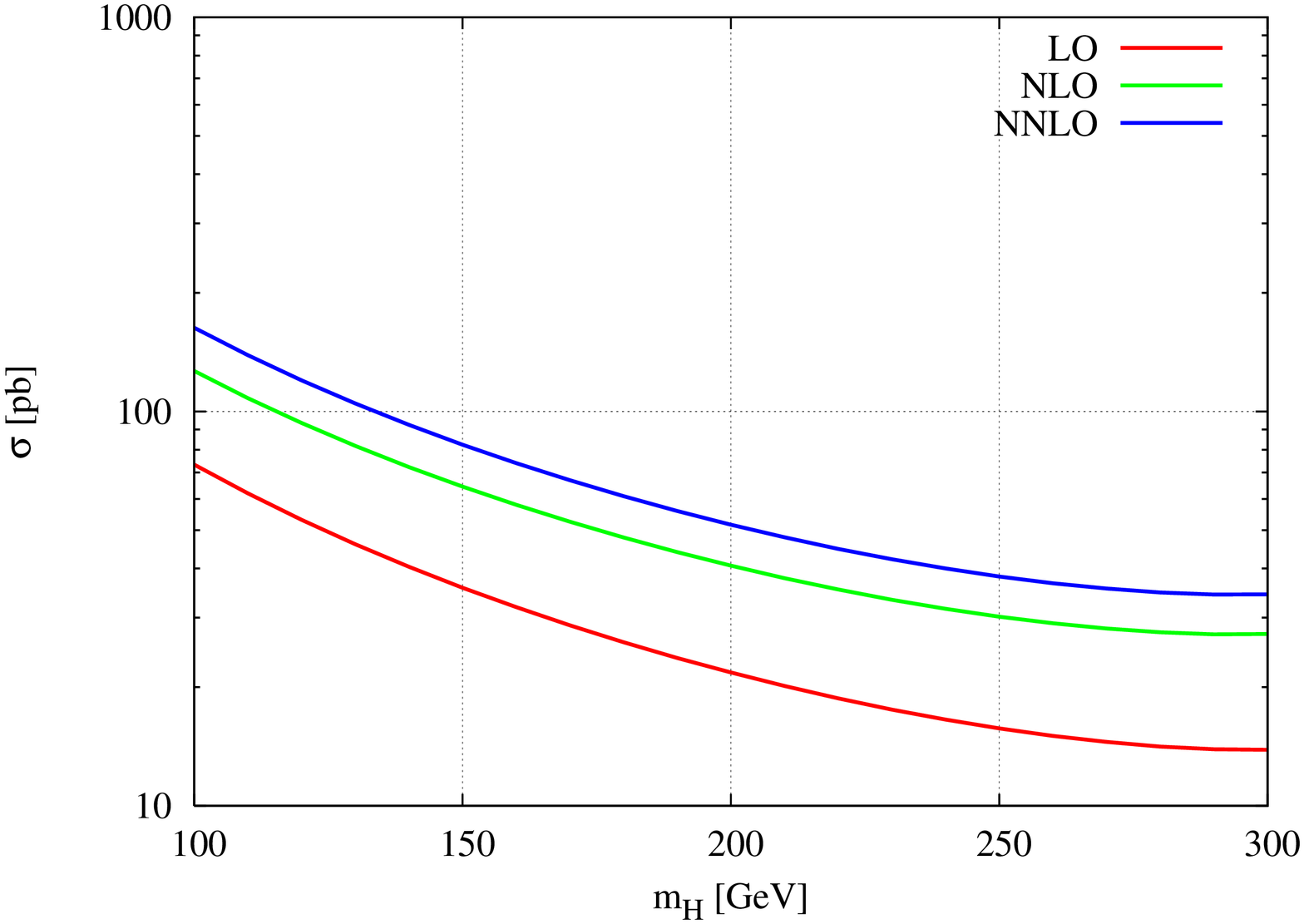}}\end{center}

\caption{Total cross section as a function of Higgs mass at $\sqrt{S}=14\,\textrm{TeV}$
in $pp$ collision, for a pseudoscalar Higgs boson.}
\label{3three}
\end{figure}

\begin{figure}[!tbh]
\begin{center}\subfigure[MRST]{\includegraphics[%
  width=5.5cm,
  angle=-90]{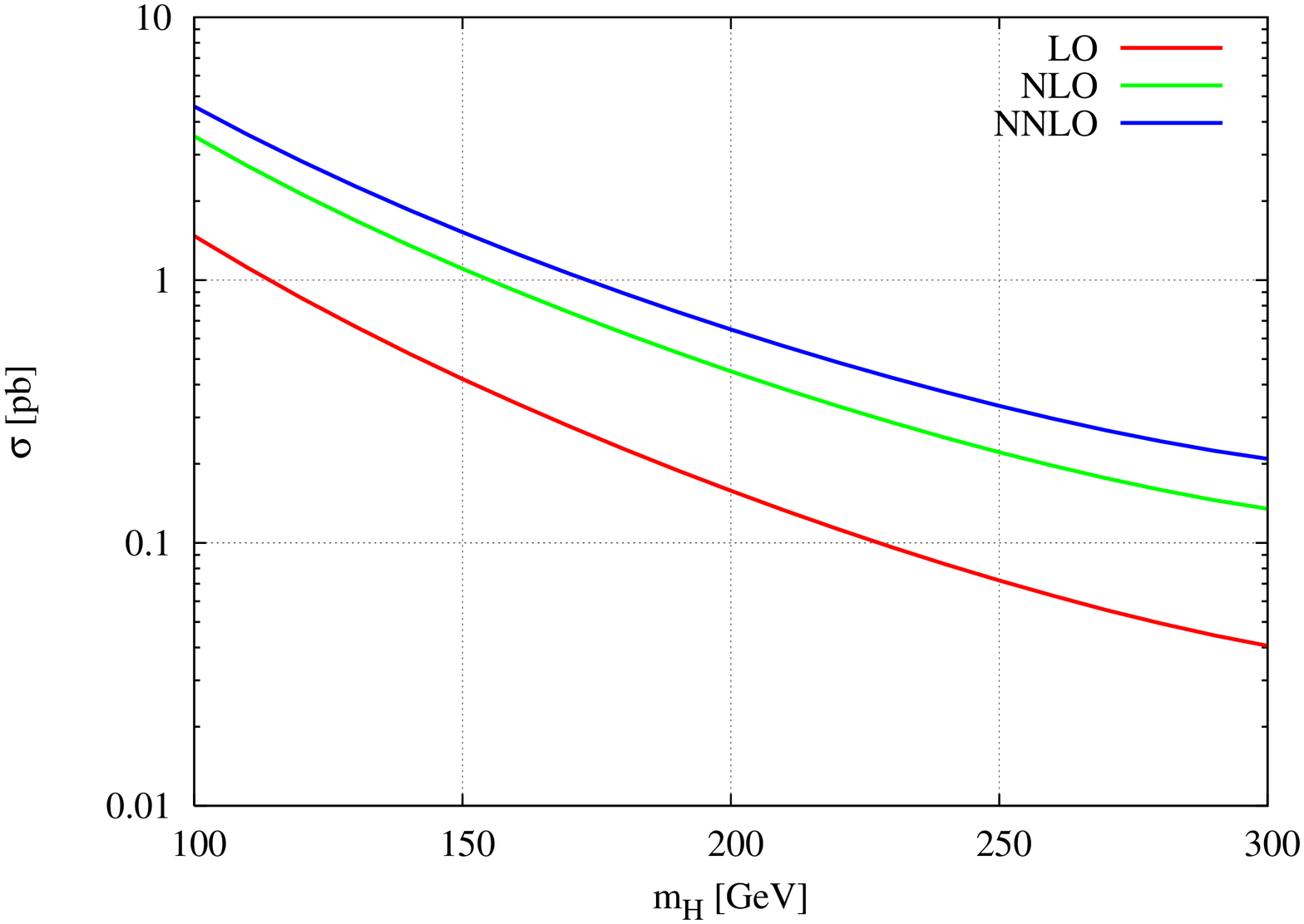}}\subfigure[Alekhin]{\includegraphics[%
  width=5.5cm,
  angle=-90]{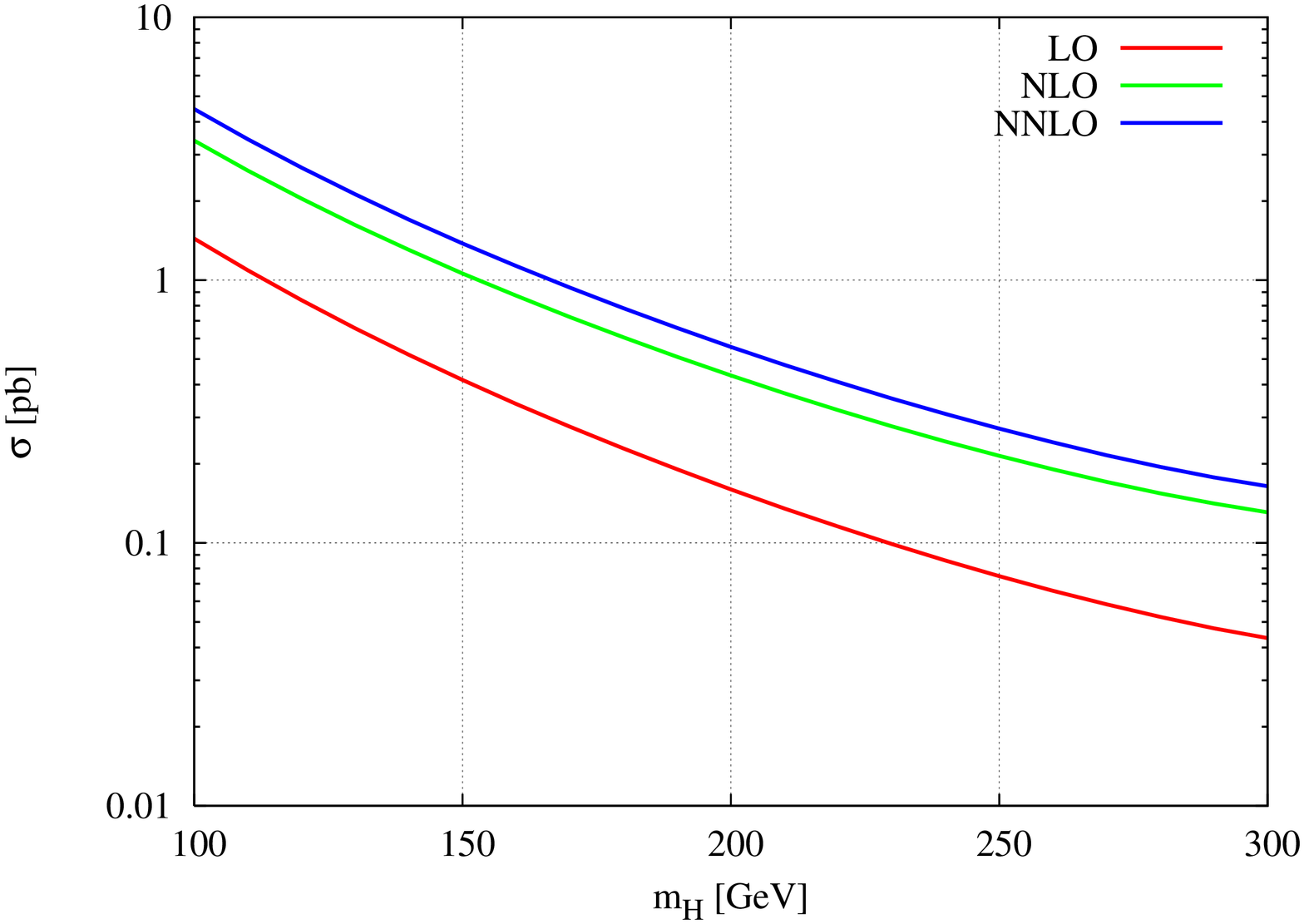}}\end{center}

\caption{Total cross section as a function of Higgs mass at $\sqrt{S}=2\,\textrm{TeV}$
in $p\bar{p}$ collision, for a pseudoscalar Higgs boson.}
\label{3four}
\end{figure}

The use of our results on the NNLO evolution of the parton distributions together 
with the results of \cite{Smith} allows us to provide accurate 
predictions for the total cross section for Higgs production. Here we summarize our numerical 
results which are illustrated in Figs.~(\ref{3one} -- \ref{3four}) and in Tables (\ref{tableone} -- 
\ref{table4}). 
The sets of distributions compared are those of MRST and Alekhin, currently the only 
ones available at NNLO. The plots refer to center of mass energies which are typical at the LHC, 
14 TeV being the largest achievable in a not so distant future. 
The pseudoscalar cross section is systematically larger than the scalar one and both drop as the mass of the Higgs particles increases. The interval of variability which has been considered 
for this parameter is light-to-heavy (100 GeV to 300 GeV). The factorization scale and the renormalization scales have been chosen to coincide. For a given value of the Higgs mass, the cross section 
raises considerably with energy by a factor approximately 50-70 as we move from the lower toward 
the higher value of $S$, the center of mass energy. Other distinct features of this study 
is the dependence of the result on the initial model chosen for the evolution. Finally the 
inclusion of the NNLO corrections, in all models, causes an increase in the actual values of the cross sections compared to the NLO results. A possible extension of this study is the inclusion 
of renormalization effects in the evolution and in the hard scatterings, which can be studied
along the lines discussed in chapter \ref{chap:NNLOcode}. 


\begin{table}[!tbh]
\begin{center}\begin{tabular}{|c|rr|rr|rr|}
\hline 
&
\multicolumn{2}{c|}{LO}&
\multicolumn{2}{c|}{NLO}&
\multicolumn{2}{c|}{NNLO}\tabularnewline
$m_{H}$&
 MRST&
 Alekhin&
 MRST&
 Alekhin&
 MRST&
 Alekhin\tabularnewline
\hline
$100$&
 $33.12$&
$32.02$&
$53.87$&
$54.89$&
$57.31$&
$70.46$\tabularnewline
$110$&
 $27.86$&
$26.97$&
$45.99$&
$46.69$&
$49.24$&
$59.81$\tabularnewline
$120$&
 $23.75$&
$23.01$&
$39.74$&
$40.21$&
$42.81$&
$51.42$\tabularnewline
$130$&
 $20.48$&
$19.86$&
$34.69$&
$35$&
$37.58$&
$44.69$\tabularnewline
$140$&
 $17.83$&
$17.31$&
$30.57$&
$30.77$&
$33.29$&
$39.22$\tabularnewline
$150$&
 $15.67$&
$15.23$&
$27.17$&
$27.28$&
$29.74$&
$34.73$\tabularnewline
$160$&
 $13.89$&
$13.5$&
$24.33$&
$24.37$&
$26.75$&
$30.99$\tabularnewline
$170$&
 $12.39$&
$12.06$&
$21.92$&
$21.92$&
$24.22$&
$27.84$\tabularnewline
$180$&
 $11.13$&
$10.84$&
$19.87$&
$19.84$&
$22.05$&
$25.16$\tabularnewline
$190$&
 $10.06$&
$9.8$&
$18.12$&
$18.05$&
$20.19$&
$22.88$\tabularnewline
$200$&
 $9.138$&
$8.912$&
$16.61$&
$16.52$&
$18.58$&
$20.91$\tabularnewline
$210$&
 $8.35$&
$8.149$&
$15.3$&
$15.19$&
$17.18$&
$19.22$\tabularnewline
$220$&
 $7.673$&
$7.492$&
$14.17$&
$14.05$&
$15.97$&
$17.75$\tabularnewline
$230$&
 $7.089$&
$6.925$&
$13.18$&
$13.06$&
$14.92$&
$16.49$\tabularnewline
$240$&
 $6.584$&
$6.435$&
$12.33$&
$12.2$&
$14.01$&
$15.39$\tabularnewline
$250$&
 $6.148$&
$6.012$&
$11.6$&
$11.46$&
$13.22$&
$14.44$\tabularnewline
$260$&
 $5.771$&
$5.646$&
$10.96$&
$10.81$&
$12.53$&
$13.62$\tabularnewline
$270$&
 $5.446$&
$5.33$&
$10.41$&
$10.26$&
$11.94$&
$12.91$\tabularnewline
$280$&
 $5.169$&
$5.061$&
$9.94$&
$9.786$&
$11.44$&
$12.31$\tabularnewline
$290$&
 $4.935$&
$4.834$&
$9.548$&
$9.39$&
$11.03$&
$11.81$\tabularnewline
$300$&
 $4.743$&
$4.648$&
$9.23$&
$9.069$&
$10.7$&
$11.4$ \tabularnewline
\hline
\end{tabular}\end{center}

\caption{Total cross section as a function of Higgs mass at $\sqrt{S}=14\,\textrm{TeV}$
in $pp$ collision, for a scalar Higgs boson. The mass of the Higgs
boson is in GeV and the cross sections are in pb.\label{tableone}}
\end{table}

\begin{table}[!tbh]
\begin{center}\begin{tabular}{|c|rr|rr|rr|}
\hline 
&
\multicolumn{2}{c|}{LO}&
\multicolumn{2}{c|}{NLO}&
\multicolumn{2}{c|}{NNLO}\tabularnewline
$m_{H}$&
 MRST&
 Alekhin&
 MRST&
 Alekhin&
 MRST&
 Alekhin\tabularnewline
\hline
$100$&
 $0.6434$&
$0.6285$&
$1.528$&
$1.47$&
$1.974$&
$1.93$\tabularnewline
$110$&
 $0.4838$&
$0.4737$&
$1.172$&
$1.126$&
$1.534$&
$1.474$\tabularnewline
$120$&
 $0.37$&
$0.3632$&
$0.9136$&
$0.8773$&
$1.211$&
$1.145$\tabularnewline
$130$&
 $0.2871$&
$0.2826$&
$0.7219$&
$0.6931$&
$0.9682$&
$0.9021$\tabularnewline
$140$&
 $0.2256$&
$0.2228$&
$0.5775$&
$0.5545$&
$0.783$&
$0.7196$\tabularnewline
$150$&
 $0.1793$&
$0.1777$&
$0.467$&
$0.4486$&
$0.6399$&
$0.5806$\tabularnewline
$160$&
 $0.1439$&
$0.1432$&
$0.3812$&
$0.3665$&
$0.5277$&
$0.4731$\tabularnewline
$170$&
 $0.1165$&
$0.1164$&
$0.3139$&
$0.302$&
$0.4387$&
$0.3888$\tabularnewline
$180$&
 $0.09506$&
$0.09537$&
$0.2603$&
$0.2507$&
$0.3673$&
$0.3219$\tabularnewline
$190$&
 $0.07814$&
$0.07874$&
$0.2174$&
$0.2095$&
$0.3095$&
$0.2685$\tabularnewline
$200$&
 $0.06466$&
$0.06546$&
$0.1828$&
$0.1763$&
$0.2625$&
$0.2254$\tabularnewline
$210$&
 $0.05385$&
$0.05478$&
$0.1546$&
$0.1493$&
$0.2239$&
$0.1904$\tabularnewline
$220$&
 $0.04514$&
$0.04615$&
$0.1316$&
$0.1272$&
$0.1922$&
$0.1618$\tabularnewline
$230$&
 $0.03806$&
$0.03912$&
$0.1127$&
$0.109$&
$0.1659$&
$0.1384$\tabularnewline
$240$&
 $0.03228$&
$0.03336$&
$0.09704$&
$0.09395$&
$0.144$&
$0.119$\tabularnewline
$250$&
 $0.02754$&
$0.02862$&
$0.08408$&
$0.08146$&
$0.1256$&
$0.1029$\tabularnewline
$260$&
 $0.02363$&
$0.0247$&
$0.07326$&
$0.07102$&
$0.1102$&
$0.08954$\tabularnewline
$270$&
 $0.0204$&
$0.02144$&
$0.0642$&
$0.06227$&
$0.09728$&
$0.07833$\tabularnewline
$280$&
 $0.01771$&
$0.01872$&
$0.0566$&
$0.05492$&
$0.08633$&
$0.06893$\tabularnewline
$290$&
 $0.01547$&
$0.01646$&
$0.05022$&
$0.04874$&
$0.07709$&
$0.06103$\tabularnewline
$300$&
 $0.01361$&
$0.01457$&
$0.04486$&
$0.04355$&
$0.0693$&
$0.05441$ \tabularnewline
\hline
\end{tabular}\end{center}

\caption{Total cross section as a function of Higgs mass at $\sqrt{S}=2\,\textrm{TeV}$
in $p\bar{p}$ collision, for a scalar Higgs boson. The mass of the
Higgs boson is in GeV and the cross sections are in pb.\label{tabletwo}}
\end{table}

\begin{table}[!tbh]
\begin{center}\begin{tabular}{|c|rr|rr|rr|}
\hline 
&
\multicolumn{2}{c|}{LO}&
\multicolumn{2}{c|}{NLO}&
\multicolumn{2}{c|}{NNLO}\tabularnewline
$m_{H}$&
 MRST&
 Alekhin&
 MRST&
 Alekhin&
 MRST&
 Alekhin\tabularnewline
\hline
$100$&
 $75.83$&
$73.31$&
$124.3$&
$126.7$&
$132.6$&
$163.1$\tabularnewline
$110$&
 $64.04$&
$61.98$&
$106.5$&
$108.1$&
$114.3$&
$139$\tabularnewline
$120$&
 $54.82$&
$53.12$&
$92.44$&
$93.54$&
$99.84$&
$120$\tabularnewline
$130$&
 $47.5$&
$46.06$&
$81.09$&
$81.81$&
$88.08$&
$104.8$\tabularnewline
$140$&
 $41.59$&
$40.37$&
$71.85$&
$72.3$&
$78.46$&
$92.47$\tabularnewline
$150$&
 $36.77$&
$35.72$&
$64.23$&
$64.49$&
$70.5$&
$82.37$\tabularnewline
$160$&
 $32.79$&
$31.89$&
$57.88$&
$58$&
$63.84$&
$73.98$\tabularnewline
$170$&
 $29.47$&
$28.68$&
$52.54$&
$52.54$&
$58.22$&
$66.95$\tabularnewline
$180$&
 $26.68$&
$25.99$&
$48.01$&
$47.93$&
$53.43$&
$60.99$\tabularnewline
$190$&
 $24.32$&
$23.7$&
$44.15$&
$44$&
$49.35$&
$55.94$\tabularnewline
$200$&
 $22.32$&
$21.77$&
$40.87$&
$40.65$&
$45.86$&
$51.64$\tabularnewline
$210$&
 $20.61$&
$20.12$&
$38.05$&
$37.79$&
$42.87$&
$47.96$\tabularnewline
$220$&
 $19.17$&
$18.72$&
$35.66$&
$35.36$&
$40.33$&
$44.84$\tabularnewline
$230$&
 $17.95$&
$17.53$&
$33.63$&
$33.31$&
$38.18$&
$42.2$\tabularnewline
$240$&
 $16.92$&
$16.54$&
$31.94$&
$31.59$&
$36.39$&
$39.99$\tabularnewline
$250$&
 $16.07$&
$15.71$&
$30.54$&
$30.16$&
$34.91$&
$38.15$\tabularnewline
$260$&
 $15.37$&
$15.04$&
$29.41$&
$29.02$&
$33.74$&
$36.68$\tabularnewline
$270$&
 $14.83$&
$14.51$&
$28.55$&
$28.14$&
$32.87$&
$35.55$\tabularnewline
$280$&
 $14.43$&
$14.13$&
$27.97$&
$27.53$&
$32.3$&
$34.76$\tabularnewline
$290$&
 $14.2$&
$13.91$&
$27.68$&
$27.22$&
$32.07$&
$34.35$\tabularnewline
$300$&
 $14.15$&
$13.87$&
$27.74$&
$27.26$&
$32.25$&
$34.38$ \tabularnewline
\hline
\end{tabular}\end{center}

\caption{Total cross section as a function of Higgs mass at $\sqrt{S}=14\,\textrm{TeV}$
in $pp$ collision, for a pseudoscalar Higgs boson. The mass of the
Higgs boson is in GeV and the cross sections are in pb.\label{tablethree}}
\end{table}

\begin{table}[!tbh]
\begin{center}\begin{tabular}{|c|rr|rr|rr|}
\hline 
&
\multicolumn{2}{c|}{LO}&
\multicolumn{2}{c|}{NLO}&
\multicolumn{2}{c|}{NNLO}\tabularnewline
$m_{H}$&
 MRST&
 Alekhin&
 MRST&
 Alekhin&
 MRST&
 Alekhin\tabularnewline
\hline
$100$&
 $1.473$&
$1.439$&
$3.526$&
$3.393$&
$4.579$&
$4.477$\tabularnewline
$110$&
 $1.112$&
$1.089$&
$2.715$&
$2.609$&
$3.572$&
$3.432$\tabularnewline
$120$&
 $0.8541$&
$0.8385$&
$2.125$&
$2.041$&
$2.831$&
$2.677$\tabularnewline
$130$&
 $0.6659$&
$0.6556$&
$1.688$&
$1.62$&
$2.274$&
$2.119$\tabularnewline
$140$&
 $0.5261$&
$0.5196$&
$1.357$&
$1.303$&
$1.849$&
$1.699$\tabularnewline
$150$&
 $0.4205$&
$0.4168$&
$1.104$&
$1.06$&
$1.52$&
$1.379$\tabularnewline
$160$&
 $0.3397$&
$0.338$&
$0.9071$&
$0.8719$&
$1.262$&
$1.131$\tabularnewline
$170$&
 $0.2771$&
$0.2768$&
$0.7522$&
$0.7236$&
$1.056$&
$0.9361$\tabularnewline
$180$&
 $0.2279$&
$0.2287$&
$0.6288$&
$0.6055$&
$0.8914$&
$0.7813$\tabularnewline
$190$&
 $0.189$&
$0.1904$&
$0.5298$&
$0.5106$&
$0.7578$&
$0.6572$\tabularnewline
$200$&
 $0.1579$&
$0.1599$&
$0.4497$&
$0.4338$&
$0.6489$&
$0.557$\tabularnewline
$210$&
 $0.1329$&
$0.1352$&
$0.3844$&
$0.3712$&
$0.5595$&
$0.4756$\tabularnewline
$220$&
 $0.1128$&
$0.1153$&
$0.3311$&
$0.32$&
$0.4859$&
$0.4091$\tabularnewline
$230$&
 $0.09636$&
$0.09904$&
$0.2873$&
$0.2779$&
$0.4249$&
$0.3544$\tabularnewline
$240$&
 $0.08297$&
$0.08573$&
$0.2512$&
$0.2432$&
$0.3744$&
$0.3094$\tabularnewline
$250$&
 $0.07198$&
$0.07479$&
$0.2213$&
$0.2144$&
$0.3322$&
$0.2722$\tabularnewline
$260$&
 $0.06295$&
$0.06578$&
$0.1965$&
$0.1905$&
$0.2971$&
$0.2413$\tabularnewline
$270$&
 $0.05553$&
$0.05837$&
$0.176$&
$0.1708$&
$0.268$&
$0.2157$\tabularnewline
$280$&
 $0.04945$&
$0.05228$&
$0.1592$&
$0.1545$&
$0.2439$&
$0.1947$\tabularnewline
$290$&
 $0.04452$&
$0.04735$&
$0.1455$&
$0.1412$&
$0.2244$&
$0.1776$\tabularnewline
$300$&
 $0.0406$&
$0.04346$&
$0.1348$&
$0.1308$&
$0.2091$&
$0.1642$ \tabularnewline
\hline
\end{tabular}\end{center}

\caption{Total cross section as a function of Higgs mass at $\sqrt{S}=2\,\textrm{TeV}$
in $p\bar{p}$ collision, for a pseudoscalar Higgs boson. The mass
of the Higgs boson is in GeV and the cross sections are in pb.\label{table4}}
\end{table}

\chapter{The kinetic interpretation of the DGLAP Equation, its Kramers-Moyal\\expansion and positivity of helicity distributions\label{chap:kinetic}}

\fancyhead[LO]{\nouppercase{Chapter 4. The kinetic interpretation of the DGLAP Equation}}

\section{Background on the topic}
In this chapter and in the chapter that follows we elaborate on what has come to be known 
as the ``kinetic description'' of the dynamics of partons under the action of the renormalization 
group. In the DGLAP evolution the parton densities tend to be supported by smaller and smaller 
values of the Bjorken variable $x$ as the energy variable $\log(Q)$ increases. Building on previous observations by Teryaev and by Collins and Qiu we clarify that the DGLAP evolution 
can be interpreted as a kinetic master equation which can be expanded formally to generate 
a hierarchy of new equations using the Kramers-Moyal approach. Arrested to the second order, 
this expansion generates a Fokker-Planck approximation to the DGLAP evolution. The issue 
of positivity of the kernels is important in order for this picture to hold. The LO kernels 
have this property, while at NLO a formal proof of positivity is more intricate and requires a 
numerical analysis to be supported. The hierarchy of equations generated by the Kramers-Moyal 
expansion could be studied independently, possibly numerically in order to gain more 
insight into the property of this expansion. One of the consequence of the kinetic interpretation, as 
shown in this chapter is the proof is that parton dynamics is reminscent of a ``downward biased 
random walk''. These issues are analyzed in detail. 

\section{Introduction}
According to a rederivation - due to Collins and Qiu - 
the DGLAP equation 
can be reinterpreted (in leading order) in a probabilistic way.
This form of the equation has been used indirectly to prove the bound 
$|\Delta f(x,Q)| < f(x,Q)$ between polarized and unpolarized distributions, or positivity 
of the helicity distributions, for any $Q$. 
We reanalyze this issue by performing a detailed numerical study of the positivity bounds of the helicity distributions. 
We also elaborate on some of the formal properties of the Collins-Qiu form 
and comment on the underlying regularization, 
introduce a Kramers-Moyal expansion of the equation and
briefly analyze its Fokker-Planck approximation. 
These follow quite naturally
once the master version is given. We illustrate this expansion both for the valence 
quark distribution $q_V$ and for the transverse spin distribution $h_1$ and analyze 
the role of one well known inequality from this perspective.
In fact, an interesting constraint relating longitudinally polarized, unpolarized and transversely polarized distributions is Soffer's inequality, which deserves a special attention, since has to be
respected by the evolution to any order in $\alpha_s$. Some tests of the
inequality have been performed in the near past, bringing support to it. 
However, other inequalities are supposed to hold as well. 

In this work we perform a NLO analysis of an inequality which relates longitudinally polarized distributions and unpolarized ones. The inequality can be summarized in the statement 
that helicity distributions (positive and negative) for quarks and gluons have to be 
positive. The inequality states that

\beq
|\Delta f(x,Q^2)| < f(x,Q^2) 
\eeq
or
\beq
f^\pm (x,Q^2) > 0  
\label{helicities}
\eeq
where the $\pm$ refers to the the possible values of the helicities 
of quarks and gluons. The statement is supposed to hold, at least in 
leading order, for any $Q$. To analyze the renormalization group evolution of this relation,
especially to next-to-leading order, requires some effort 
since this study involves a combined study of the (longitudinally) 
polarized and unpolarized evolutions. In this work we present 
a complete NLO study of the evolution equations starting directly from the 
helicity basis. Helicities are in fact the basic parton distributions 
from which other distributions can be built. 

Compared to other implementations, 
in our work we perform a NLO test of the positivity of the
helicity distributions using an ansatz due to Rossi \cite{Rossi} which reduces the
evolution equations to an infinite set of recursion relations for some scale invariant coefficients.

Various arguments to validate eq. (\ref{helicities}) 
have been presented in the literature. From our perspective, 
an interesting one has been formulated by Teryaev 
and Collaborators who have tried to establish a link, to leading order, 
between evolution equations and their probabilistic interpretation in 
order to prove Soffer's inequality. Similar arguments hold also in the 
analysis of eq. (\ref{helicities}).

We should remark that a complete probabilistic 
picture exists only for the leading order unpolarized evolution 
\cite{CollinsQiu} and the arguments of \cite{Teryaev} are inspired 
by the fact that the subtraction terms (the $x=1$ contributions 
in the expressions of the kernels, where $x$ is Bjorken's variable), 
being positive, once they are combined with the bulk ($x<1$) 
contributions give a form of the evolution equations which are 
diagonal in parton type and resemble ``kinetic'' equations. 
Our arguments, on this issue, are just a 
refinement of this previous and influential analysis. 
 
In the recent literature there has been some attention to this feature of the 
DGLAP evolution, limited to the nonsinglet sector, in connection 
with kinetic theory and the "dynamical renormalization group'', in the words of ref.~\cite{devega}. 

All the arguments, so far, go back to some important older work of Collins 
and Qiu who provided an interesting derivation of the (unpolarized) 
DGLAP equation using 
Mueller's formalism of cut diagrams. In their paper \cite{CollinsQiu} the authors reinterpreted 
the DGLAP equation as a kinetic probabilistic equation of Boltzmann type. 
The authors gave no detail on some of the issues concerning the regularization of their 
diagrammatic expansion, on which we will elaborate since we need it for our 
accurate numerical analysis. In our work the Collins-Qiu form of the DGLAP 
equation is interpreted simply as a {\em master equation} rather than a Boltzmann 
equation, given the absence of a 2-to-2 scattering cross section in the 
probabilistic partonic interpretation. A master equation is governed by transition 
probabilities and various formal approximations find their way once this conceptual
step is made. We illustrate, in the spirit of a stochastic approach to the DGLAP dynamics,
how to extract standard differential equations of Kramers-Moyal type for the simplest nonsinglet evolutions, those involving valence distributions 
and transverse spin distributions. 

We show that the DGLAP dynamics can be described, 
at least in a formal way, by a differential equation of arbitrarily high order. 
Truncations of this expansion to the first few orders provide the usual 
link with the Fokker-Planck approximation, the Langevin equation 
and its path integral version \footnote{For an example of this interplay between differential and 
stochastic descriptions we refer the reader to \cite{BCM}}. 
The picture one should have in mind, at least in this approximation, 
is that of a stochastic (Brownian) dynamics of Bjorken's variable $x$ in a 
fictitious time $\log(Q)$, describing the evolution under 
the renormalization group (RG). In this interpretation 
the probability function is the parton distribution itself. 

This chapter is based on the paper \cite{kinetic}.

\section{Master Equations and Positivity}
Let's start considering  a generic 1-D master equation for transition
probabilities $w(x|x')$ which we interpret as the probability of making
a transition to a point $x$ given a starting point $x'$ for a given physical system.
The picture we have
in mind is that of a gas of particles making collisions in 1-D and entering the
interval $(x,x + dx)$ with a probability $w(x|x')$ per single transition,
or leaving it with a transition probability $w(x'|x)$.
In general one writes down a master equation
\beq
\frac{\partial }{\partial \tau}f(x,\tau)=\int dx'\left(
w(x|x') f(x',\tau) -w(x'|x) f(x,\tau)\right) dx'.
\eeq
describing the time $\tau$ evolution of the density of the gas
undergoing collisions or the motion of a many replicas of walkers of density
$f(x,\tau)$ jumping with a pre-assigned probability,
according to taste.

The result of Collins a Qiu, who were after a
derivation of the DGLAP equation that could include automatically also the ``edge point''
contributions (or x=1 terms of the DGLAP kernels) is in pointing out the existence of a
probabilistic picture of the DGLAP dynamics.
These edge point terms had been always introduced in the past only by hand and serve to enforce the baryon number sum rule and the momentum sum rule as $Q$, the momentum scale, varies.

The kinetic interpretation was used in \cite{Teryaev} to provide an alternative proof
of Soffer's inequality.
We recall that this inequality
\beq
|h_1(x)| < q^+(x)
\eeq
famous by now, sets a bound on the transverse spin distribution $h_1(x)$ in terms of the
components of the positive helicity component of the quarks, for a given flavour.
The inequality has to be respected by the evolution.
We recall that $h_1$, also denoted by the symbol 
\begin{equation}
\Delta _{T}q(x,Q^{2})\equiv q^{\uparrow }(x,Q^{2})-q^{\downarrow }(x,Q^{2}),
\end{equation} 
has the property
of being purely nonsinglet and of appearing at leading twist. It is
identifiable in transversely polarized 
hadron-hadron collisions and not in Deep Inelastic Scattering (DIS), where can
appear only through an insertion of the electron mass in the unitarity 
graph of DIS. 

The connection between the Collins-Qiu form of the DGLAP equation and the master equation
is established as follows.
The DGLAP equation, in its original formulation is generically written as

\beq
\frac{d q(x,Q^2)}{d \log( Q^2)} = \int_x^1 \frac{dy}{y} P(x/y)q(y,Q^2),
\eeq
where we are assuming a scalar form of the equation, such as in the nonsinglet sector. The generalization to the singlet sector of the arguments given below is, of course, quite straightforward.
To arrive at a probabilistic picture of the equation we start reinterpreting
$\tau=\log (Q^2)$ as a time variable, while the parton density $q(x,\tau)$
lives in a one dimensional (Bjorken) $x$ space.

We recall that the kernels are defined as ``plus'' distributions.
Conservation of baryon number, for instance, is enforced by the addition of
edge-point contributions proportional to $\delta(1-x)$.

We start with the following form of the kernel
\beq
P(z) = \hat{P}(z) - \delta(1-z) \int_0^1 \hat{P}(z)\, dz,
\label{form}
 \eeq
where we have separated  the edge point contributions from the rest
of the kernel, here called $\hat{P}(z)$. This manipulation is understood in all the
equations that follow.
The equation is rewritten in the following form

\beq
\frac{d}{d \tau}q(x,\tau) = \int_x^1 dy \hat{P}\left(\frac{x}{y}\right)\frac{q(y,\tau)}{y}
-\int_0^x \frac{dy}{y}\hat{ P}\left(\frac{y}{x}\right)\frac{q(x,\tau)}{x}
\label{bolz}
\eeq

\begin{figure}[t]
{\centering \resizebox*{6cm}{!}{\rotatebox{0}{\includegraphics{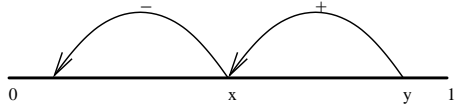}}} \par}
\caption{ The constrained random walk of the parton densities}
\label{walk}
\end{figure}

Now, if we define
\beq
w(x|y)= \frac{\alpha_s}{2 \pi} \hat{P}(x/y)\frac{\theta(y > x)}{y}
\eeq
(\ref{bolz})
becomes a master equation for the probability function $q(x,\tau)$
\beq
\frac{\partial }{\partial \tau}q(x,\tau)=\int dx'\left(
w(x|x') q(x',\tau) -w(x'|x) q(x,\tau)\right) dx'.
\label{masterf}
\eeq
There are some interesting features of this special master equation. Differently from
other master equations, where transitions are allowed from  a given x both toward $y>x$
and $y< x$, in this case, transitions toward x take place only from values $y>x$ and
leave the momentum cell $(x, x+ dx)$ only toward smaller y values (see Fig.(\ref{walk}).

Clearly, this sets a direction of the kinetic evolution of the densities from large
x values toward smaller-x values as $\tau$, the fictitious ``time'' variable, increases.

Probably this is the simplest illustration of the fact that parton densities, at large final evolution scales, are dominated by their
small-x behaviour. As the ``randomly moving partons'' reach the $x\approx 0$ region
of momentum space, they find no space where to go, while other partons tend to pile up
toward the same region from above. This is the picture of a 
random walk biased to move downward (toward the small-x region) and is illustrated in Fig.~(\ref{walk}). 

\section{Probabilistic Kernels}

 We briefly discuss some salient features of the structure of the kernels in this approach and comment on the type 
of regularization involved in order to define them appropriately.  

We recall that unpolarized and polarized kernels, in leading 
order, are given by
\beqa
P^{(0)}_{NS} &=&P_{qq}^{(0)}=C_F\left( \frac{2}{(1-x)_+} -1 - x + 
\frac{3}{2} \delta(1-x)\right) \nonumber\\
P_{qg}^{(0)}&=& 2 T_f\left(x^2 + (1-x)^2)\right) \nonumber \\
P_{gq}^{(0)}&=& C_F \frac{1 + (1 -x)^2}{x}\nonumber \\
P_{gg}^{(0)}&=& 2 N_c\left(\frac{1}{(1-x)_+} +\frac{1}{x} -2 + x(1-x)\right)
 +\frac{\beta_0}{2}\delta(1-x)
\eeqa
where
\beq
C_F=\frac{N_C^2 -1}{2 N_C}, \,T_f=T_R n_f= \frac{1}{2} n_f,
\,\,\beta_0=\frac{11}{3} N_C  - \frac{4}{3} T_f \nonumber\\
\eeq
and
\beqa
\Delta P^{(0)}_{NS} &=&\Delta P_{qq}^{(0)}\nonumber\\
\Delta P_{qq}^{(0)}&=&C_F\left( \frac{2}{(1-x)_+} - 1 - x +
 \frac{3}{2}\delta(1-x)\right)\nonumber \\
\Delta P_{qg}^{(0)}&=& 2 T_f( 2 x - 1) \nonumber \\
\Delta P_{gq}^{(0)}&=& C_F (2 -x)\nonumber \\
\Delta P_{gg}^{(0)}&=& 2 N_c\left(\frac{1}{(1-x)_+} - 2 x + 1\right)
+\delta(1-x)\frac{\beta_0}{2},
\label{stand2}
\eeqa

while the LO transverse kernels are given by
\beq
\Delta_T P^{(0)}_{qq}=C_F\left( \frac{2 }{(1-x)_+} - 2  
+ \frac{3}{2}\delta(1-x)\right).
\eeq

The unpolarized kernels should be compared with the Collins-Qiu form 
\beqa
P_{qq}&=&\gamma_{qq} -\delta(1-x) \int_0^1 dz \gamma_{qq}
\nonumber \\
P_{gg}&=&\gamma_{gg} - 
\left(n_f \int_0^1 {dz} \gamma_{qg} +\frac{1}{2}\int_0^1 dz 
\gamma_{gg}\right) \delta(1-x) \nonumber \\
P_{qg}&=& \gamma_{qg}\nonumber \\
P_{gq}&=& \gamma_{gq}\nonumber \\
\label{cq1}
\eeqa
where
\beqa
\gamma_{qq}&=& C_F \left(\frac{2}{1-x} -1 - x 
\right)\nonumber \\
\gamma_{qg} &=& (2 x -1)\nonumber \\
\gamma_{gq} &=& C_F(2 - x) \nonumber \\
\gamma_{gg} &=& 2 N_c\left( \frac{1}{1-x} +\frac{1}{x} -2 + x(1-x) \right). \nonumber \\
\label{cq2}
\eeqa
These kernels need a suitable regularization to be well defined. 
Below we will analyze the implicit regularization 
underlying eq.~(\ref{cq1}). One observation is however almost immediate: 
the component $P_{gg}$ is not of the form given by eq.~(\ref{form}). 
In general, therefore, in the singlet case, the generalization of 
eq.~(\ref{form}) is given by 
\beq
P(x)=\hat{P}_1(x) - \delta(1-x)\int_0^1 \hat{P}_2(z) dz
\eeq
and a probabilistic interpretation is more complex compared to the nonsinglet 
case and has been discussed in the original literature \cite{CollinsQiu}. 

\section{Convolutions and Master Form of the Singlet}
Distributions are folded with the kernels and the result rearranged
in order to simplify the structure of the equations. Since
in the previous literature this is done in a rather involute way 
\cite{Hinchliffe} 
we provide here a simplification, from which the equivalence of the 
various forms of the kernel, in the various regularizations adopted, will be 
apparent.     
All we need is the simple relation
\beq
\int_x^1 \frac{dy}{y (1-y)_+}f(x/y)=\int_x^1\frac{dy}{y}
\frac{ yf(y) - x f(x)}{y-x} +\log(1-x) f(x)
\label{simplerel}
\eeq
in which, on the right hand side, regularity of both the first 
and the second term is explicit. For instance, the evolution equations become
\beqa
\frac{d q}{d \log(Q^2)} &=& 2 C_F 
\int\frac{dy}{y}\frac{ y q(y) - x q(x)}{y-x} +2 C_F \log(1-x)\, q(x) -
\int_x^1\frac{dy}{y}\left( 1 + z\right)q(y) + 
\frac{3}{2} C_F q(x) \nonumber \\
&& + n_f\int_x^1\frac{dy}{y}
\left( z^2 +(1-z)^2\right)g(y)\nonumber \\
\frac{d g}{d \log(Q^2)} &=& 
C_F \int_x^1\frac{dy}{y}\frac{1 +(1-z)^2}{z}q(y)
+ 2 N_c \int_x^1\frac{dy}{y}
\frac{ y f(y) - x f(x)}{y-x}g(y) 
\nonumber \\
&& + 2 N_c \log(1-x) g(x) 
+2 N_c\int_x^1 \frac{dy}{y}\left( \frac{1}{z} -2 + z(1-z)\right)g(y) + 
\frac{\beta_0}{2}g(x) \nonumber \\
\label{standard}
\eeqa
with $z\equiv x/y$. The same simplified form is obtained 
from the probabilistic version, having defined a suitable regularization 
of the edge point singularities in the integrals over the components 
$\gamma_{f f'}$ in eq. (\ref{cq2}). The canonical expressions of the 
kernels (\ref{stand2}), expressed in terms of ``+'' distributions, can also be rearranged to look like their equivalent probabilistic 
form by isolating the edge-point contributions hidden in their ``+''
distributions. We get the expressions

\beqa
{P_{qq}^{(0)}}_{NS} &=&P_{qq}^{(0)}=C_F\left( \frac{2}{(1-x)} -1 - x\right) - 
\left(C_F\int_0^1 \frac{dz}{1-z} -\frac{3}{2}\right) \delta(1-x) \nonumber\\
P_{gg}^{(0)}&=& 2 N_c\left(\frac{1}{(1-x)} +\frac{1}{x} -2 + x(1-x)\right)
-\left(2 N_c \int_0^1 \frac{dz}{1-z}-\frac{\beta_0}{2}\right)\delta(1-x)
\eeqa

and
\beqa
\Delta P_{qq}^{(0)}&=&C_F\left( \frac{2}{(1-x)} - 1 - x\right) -
C_F\left( \int_0^1 \frac{dz}{1-z} - \frac{3}{2}\right)\delta(1-x)\nonumber \\
\Delta P_{gg}^{(0)}&=& 2 N_c\left(\frac{1}{1-x} - 2 x + 1\right)
-\left(2 N_c  \int_0^1 \frac{dz}{1-z}  -\frac{\beta_0}{2}\right) \delta(1-x),
\eeqa
the other expressions remaining invariant. In appendix A we provide some 
technical details on the equivalence between the convolutions obtained 
using these kernels with the standard ones.

A master form of the singlet (unpolarized) equation 
is obtained by a straightforward change of variable in the decreasing 
terms. We obtain 
\beqa
\frac{d q}{d \tau}&=&\int_x^{1 - \Lambda}\frac{dy}{y}\gamma_{qq}(x/y)q(y) 
-\int_0^{x - \Lambda}\frac{dy}{y}\gamma_{qq}(y/x) q(x)\nonumber \\
\frac{d g}{d \tau}&=&\int_x^{1 - \Lambda}\frac{dy}{y}\gamma_{gg}(x/y)
- n_f\int_0^x\gamma_{qg}(y/x) g(x)  \nonumber \\
&& -\frac{1}{2} \int_\Lambda^{x - \Lambda}\gamma_{gg}(y/x) g(x) + 
\int_x^1\frac{dy}{y}\gamma_{gq}(x/y)q(y)
\eeqa 
with a suitable (unique) cutoff $\Lambda$ needed to cast the 
equation in the form (\ref{standard}). A discussion of this aspect is 
left in appendix B.  
The (regulated) transition probabilities are then given by 
\beqa
w_{qq}(x|y)&=&\gamma_{qq}(x/y)\frac{\theta(y>x)\theta(y< 1 -\Lambda)}{y}\nonumber \\  
w_{qq}(y|x)&=&\gamma_{qq}(y/x)\frac{\theta(y< x- \Lambda) \theta(y>0)}{x} \nonumber \\ 
w_{gg}(x|y) &=&\gamma_{gg}(x/y)\frac{\theta(y>x)\theta(y< 1 -\Lambda)}{y} \nonumber \\ 
w_{qq}(y|x)&=&\left(n_f\gamma_{qg}(y/x)
-\frac{1}{2}\gamma_{gg}(y/x)\right)\frac{\theta(y< x- \Lambda) \theta(y>0)}{x} \nonumber \\
w_{gq}(y|x)&=&\gamma_{gq}(x/y)\frac{\theta(y>x)\theta(y< 1 -\Lambda)}{y}\nonumber \\  
w_{gq}(x|y)&=& 0, \nonumber \\
\eeqa
as one can easily deduct from the form of eq. (\ref{masterf}).

\section{A Kramers-Moyal Expansion for the DGLAP Equation}
Kramers-Moyal (KM) expansions of the master equations (backward or forward)
are sometimes useful in order to gain insight into the master equation itself,
since they may provide a complementary view of the underlying dynamics.

The expansion allows to get rid of the integral which characterizes the master
equation, at the cost of introducing a differential operator of arbitrary
order. For the approximation to be useful, one has to stop the expansion after
the first few orders. In many cases this turns out to be possible.
Examples of processes of this type are special Langevin processes and
processes described by a Fokker-Planck operator. In these cases
the probabilistic interpretation allows us to write down a fictitious
Lagrangian, a corresponding path integral and solve for the
propagators using the Feynman-Kac formula. 
 For definitess we take the integral to cover all
the real axis in the variable $x'$ 

\beq
\frac{\partial }{\partial \tau}q(x,\tau)=\int_{-\infty}^{\infty} dx'\left(
w(x|x') q(x',\tau) -w(x'|x) q(x,\tau)\right) dx'.
\eeq

As we will see below, in the DGLAP case some
modifications to the usual form of the KM expansion will appear. 
At this point we perform a KM expansion of the equation in the usual way.
We make the substitutions in the master equation $y\to x-y$ in the first term and
$y\to x + y$ in the second term

\beq
\frac{\partial }{\partial \tau}q(x,\tau)=\int_{-\infty}^{\infty} dy\left(
w(x|x -y) q(x - y,\tau) -w(x + y|x) q(x,\tau)\right) ,
\eeq
identically equal to
\beq
\frac{\partial }{\partial \tau}q(x,\tau)=\int_{-\infty}^{\infty} dy\left(
w(x + y - y'|x -y') q(x - y',\tau) - w(x + y'|x) q(x,\tau)\right),
\label{shift}
\eeq
with $y=y'$. First and second term in the equation above differ by a shift (in $-y'$) and can be related using a Taylor (or KM) expansion of the first term

\beqa
\frac{\partial }{\partial \tau}q(x,\tau)&=&\int_{-\infty}^{\infty} dy \sum_{n=1}^\infty
\frac{(-y)^n}{n!}\frac{\partial^n}{\partial x^n}\left( w(x + y|x)
q(x,\tau)\right)
\eeqa
where the $n=0$ term has canceled between the first and the second
contribution coming from (\ref{shift}). The result can be
written in the form
\beq
\frac{\partial }{\partial \tau}q(x,\tau)= \sum_{n=1}^\infty
\frac{(-y)^n}{n!}\frac{\partial^n}{\partial x^n}\left( a_n(x)q(x,\tau)\right)
\eeq
where
\beq
a_n(x)=\int_{-\infty}^{\infty} dy(y-x)^n w(y|x).
\eeq

In the DGLAP case we need to amend the former derivation, due to the
presence of boundaries $( 0 < x < 1)$ in the Bjorken variable $x$. For simplicity 
we will focus on the nonsinglet case. 
We rewrite the master equation using the same change of variables used above 

\beqa
\frac{\partial}{\partial\tau}q(x,\tau) &=& \int_x^1 dy w(x|y)q(y,\tau) - 
\int_0^x dy w(y|x) q(x,\tau)
\nonumber \\
&& -\int_0^{\alpha(x)} dy w(x+y|x)* q(x,\tau)+ 
\int_0^{-x} dy w(x+y|x)q(x,\tau),
\eeqa

where we have introduced the simplest form of the Moyal product 
\footnote{A note for noncommutative geometers: 
this simplified form is obtained for a dissipative dynamics when the 
{\bf p}'s of phase space are replaced by constants. Here we have only one variable: $x$} 
 
\beq
w(x+y|x)*q(x)\equiv w(x+y|x) e^{-y \left(\overleftarrow{\partial}_x + 
\overrightarrow{\partial}_x\right)} q(x,\tau)
\eeq
and $\alpha(x) =x-1$. 
The expansion is of the form 
\beqa
\frac{\partial}{\partial \tau}q(x,\tau)=\int_{\alpha(x)}^{-x}dy\,  w(x+y|x)q(x,\tau) - 
\sum_{n=1}^{\infty}\int_0^{\alpha(x)}dy \frac{(-y)^n}{n!}{\partial_x}^n
\left(w(x+y|x)q(x,\tau)\right)
\label{expans}
\eeqa
which can be reduced to a differential equation of arbitrary order using simple manipulations. 
We recall that the Fokker-Planck approximation is obtained stopping the expansion at 
the second order 
\beq
\frac{\partial}{\partial \tau}q(x,\tau)= a_0(x) -\partial_x\left(a_1(x) q(x)\right) + 
\frac{1}{2}\partial_x^2\left( a_2(x) q(x,\tau)\right)
\label{fpe}
\eeq
with 
\beq
a_n(x)=\int dy \,y^n\, w(x+y,x)
\eeq
being moments of the transition probability function $w$.
Given the boundary conditions on the Bjorken variable x, even in the 
Fokker-Planck approximation, the Fokker-Planck version of the DGLAP equation 
is slightly more involved than Eq. (\ref{fpe}) and the coefficients $a_n(x)$ need to be redefined.

\section{The Fokker-Planck Approximation}
The probabilistic interpretation of the DGLAP equation motivates us to investigate the role of the Fokker-Planck (FP) approximation to the equation and its possible practical use. 
We should start by saying a word of caution regarding this expansion. 

In the context of a random walk, an all-order derivative expansion of the master equation 
can be arrested to the first few terms either if the conditions of Pawula's 
theorem are satisfied -in which case the FP approximation turns out to be exact- 
or if the transition probabilities show an exponential decay above 
a certain distance allowed to the random walk. Since the DGLAP kernels 
show only an algebraic decay in x, 
and there isn't any explicit scale in the kernel themselves, 
the expansion is questionable. However, from a formal viewpoint, it is still allowed. 
With these caveats in mind we proceed to investigate the features of this expansion. 
  
We redefine 

\beqa
\tilde{a}_0(x) &=& \int_{\alpha(x)}^{-x} dy w(x+y|x)q(x,\tau) \nonumber \\
a_n(x) &=&\int_0^{\alpha(x)} dy y^n w(x+y|x) q(x,\tau) \nonumber \\
\tilde{a}_n(x)&=&\int_0^{\alpha(x)}
dy y^n {\partial_x}^n \left(w(x+y|x)q(x,\tau)\right) \,\,\,n=1,2,...
\eeqa
For the first two terms $(n=1,2)$ one can easily work out the relations
\beqa
\tilde{a}_1(x) &=&\partial_x a_1(x) - \alpha(x) \partial_x \alpha(x)
w(x + \alpha(x)|x)q(x,\tau) \nonumber \\
\tilde{a}_2(x) &=&\partial_x^2 a_2(x) - 
2 \alpha(x) (\partial_x \alpha(x))^2 w(x+ \alpha(x)|x) q(x,\tau) - 
\alpha(x)^2 \partial_x\alpha(x)
\partial_x\left( w(x+ \alpha(x)|x)q(x,\tau)\right)\nonumber \\ 
&& - \alpha^2(x)\partial_x \alpha(x) \partial_x\left( w(x + y|x)q(x,\tau)\right)|_{y=\alpha(x)}
\eeqa

Let's see what happens when we arrest the expansion (\ref{expans}) to the first 3 terms.
The Fokker-Planck version of the equation is obtained by including in the approximation
only $\tilde{a}_n$ with $n=0,1,2$.

The Fokker-Planck limit of the (nonsinglet) equation is then given by
\beq
\frac{\partial}{\partial \tau}q(x,\tau) = \tilde{a}_0(x)
+\tilde{a}_1(x) - \frac{1}{2} \tilde{a}_2(x)
\eeq
which we rewrite explicitly as 
\beqa
\frac{\partial}{\partial \tau}q(x,\tau) &=& 
C_F\left(\frac{85}{12} +\frac{3}{4 x^4} - \frac{13}{3 x^3} +
\frac{10}{x^2} -\frac{12}{x}
+ 2\log\left(\frac{1-x}{x}\right)\right)q(x)\nonumber \\
&& +C_F\left( 9 - \frac{1}{2 x^3} +\frac{3}{x^2} -\frac{7}{x} 
-\frac{9}{2}\right) \partial_x q(x,\tau) \nonumber \\
&& + C_F\left(\frac{9}{4} +\frac{1}{8 x^2} -\frac{5}{6 x} -
\frac{5 x}{2}  +\frac{23 x^2}{24}\right) \partial_x^2 q(x,\tau).
\eeqa

A similar approach can be followed also for other cases, for which a 
probabilistic picture (a derivation of Collins-Qiu type) has not been established yet, 
such as for $h_1$. 
We describe briefly how to proceed in this case. 

First of all, we rewrite the evolution equation for the transversity 
in a suitable master form. This is possible since the subtraction terms 
can be written as integrals of a positive function. A possibility is 
to choose the transition probabilities
\beqa
w_1[x|y] &=& \frac{C_F}{y}\left(\frac{2}{1- x/y} - 2 \right)
\theta(y>x) \theta(y<1)\nonumber \\
w_2[y|x] &=& \frac{C_F}{x} \left(\frac{2}{1- y/x} - \frac{3}{2}\right)
\theta(y > -x)\theta(y<0)
\nonumber \\
\eeqa
which reproduce the evolution equation for $h_1$ in master form

\beq
\frac{d h_1}{d \tau}= \int_0^1 dy w_1(x|y)h_1(y,\tau) 
-\int_0^1 dy w_2(y|x) h_1(x,\tau).
\eeq
The Kramers-Moyal expansion is derived as before, with some slight
modifications. The result is obtained introducing an intermediate cutoff which is 
removed at the end. In this case we get 
\beqa
\frac{d h_1}{d\tau} &=& C_F\left( \frac{17}{3}
-\frac{2}{3 x^3} + \frac{3}{x^2} - \frac{6}{x}
+ 2 \log\left(\frac{1-x}{x}\right)\right) h_1(x,\tau) \nonumber \\
&& + C_F\left( 6 + \frac{2}{3 x^2} -\frac{3}{x} 
- \frac{11 x }{3}\right)\partial_x h_1(x,\tau) \nonumber \\
&& + C_F\left( \frac{3}{2} - \frac{1}{3 x } 
- 2 x + \frac{5 x^2}{6}\right)\partial_x^2 h_1(x,\tau). \nonumber 
\eeqa

Notice that compared to the standard Fokker-Planck approximation, the boundary now 
generates a term on the left-hand-side of the equation proportional to $q(x)$ 
which is absent in eq. (\ref{fpe}). This and higher order approximations to the 
DGLAP equation can be studied systematically both analytically and numerically 
and it is possible to assess the validity of the approximation \cite{CPC}.

\section{Helicities to LO}
As we have mentioned above, an interesting version of the usual DGLAP equation involves
the helicity distributions.

We start introducing \cite{Teryaev} the DGLAP kernels for fixed
helicities
$P_{++}(z)=(P(z)+\Delta P(z))/2$ and $ P_{+-}(z)=(P(z)-\Delta P(z))/2$
which will be used below. $P(z)$
denotes (generically) the unpolarized kernels, while the $\Delta P(z)$ are the longitudinally polarized ones. These definitions, throughout the chapter, are meant to be
expanded up to NLO, the order at which our numerical analysis holds.

\begin{figure}
{\centering \resizebox*{12cm}{!}{\rotatebox{-90}{\includegraphics{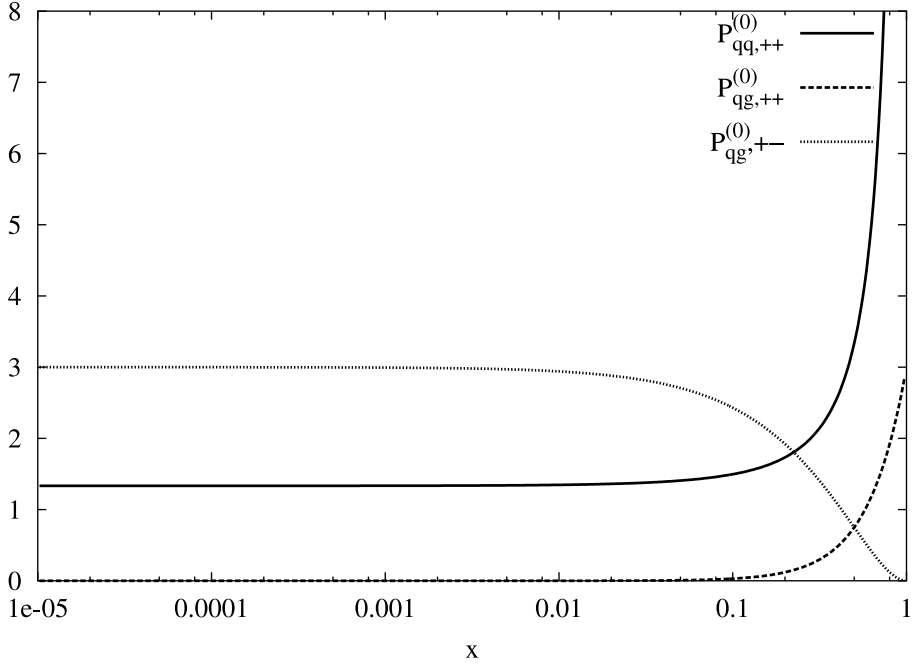}}} \par}
\caption{LO kernels (\protect\( qq\protect \) and \protect\( qg\protect \))
in the helicity basis.}
\label{p0qx}
\end{figure}

 The equations, in the helicity basis, are 

\begin{eqnarray}
{dq_+(x) \over{dt}}=
{\alpha_s \over {2 \pi}} (P_{++}^{qq} ({x \over y}) \otimes q_+(y)+
P_{+-}^{qq} ({x \over y}) \otimes q_-(y)  \nonumber \\
+P_{++}^{qg} ({x \over y}) \otimes g_+(y)+
P_{+-}^{qg} ({x \over y}) \otimes g_-(y)),
\nonumber \\
{dq_-(x) \over{dt}}=
{\alpha_s \over {2 \pi}} (P_{+-} ({x \over y}) \otimes q_+(y)+
P_{++} ({x \over y}) \otimes q_-(y) \nonumber \\
+P_{+-}^{qg} ({x \over y}) \otimes g_+(y)+
P_{++}^{qg} ({x \over y}) \otimes g_-(y)),  \nonumber \\
{dg_+(x) \over{dt}}=
{\alpha_s \over {2 \pi}} (P_{++}^{gq} ({x \over y}) \otimes q_+(y)+
P_{+-}^{gq} ({x \over y}) \otimes q_-(y) \nonumber \\
+P_{++}^{gg} ({x \over y}) \otimes g_+(y)+
P_{+-}^{gg} ({x \over y}) \otimes g_-(y)),  \nonumber \\
{dg_-(x) \over{dt}}=
{\alpha_s \over {2 \pi}} (P_{+-}^{gq} ({x \over y}) \otimes q_+(y)+
P_{++}^{gq} ({x \over y}) \otimes q_-(y) \nonumber \\
+P_{+-}^{gg} ({x \over y}) \otimes g_+(y)+
P_{++}^{gg} ({x \over y}) \otimes g_-(y)).
\end{eqnarray}
The nonsinglet (valence) analogue of this equation is also easy to
write down
\begin{eqnarray}
{dq_{+, V}(x) \over{dt}}=
{\alpha_s \over {2 \pi}} (P_{++} ({x \over y}) \otimes q_{+,V}(y)+
P_{+-} ({x \over y}) \otimes q_{-,V}(y)), \nonumber \\
{dq_{-,V}(x) \over{dt}}=
{\alpha_s \over {2 \pi}} (P_{+-} ({x \over y}) \otimes q_{+,V}(y)+
P_{++} ({x \over y}) \otimes q_{-,V}(y)).
\end{eqnarray}
where the $q_{\pm,V}=q_\pm - \bar{q}_\pm$ are the valence components of fixed helicities.
The kernels in this basis are given by 
\beqa
P_{NS\pm,++}^{(0)} &=&P_{qq, ++}^{(0)}=P_{qq}^{(0)}\nonumber \\
P_{qq,+-}^{(0)}&=&P_{qq,-+}^{(0)}= 0\nonumber \\
P_{qg,++}^{(0)}&=& n_f x^2\nonumber \\
P_{qg,+-}&=& P_{qg,-+}= n_f(x-1)^2 \nonumber \\
P_{gq,++}&=& P_{gq,--}=C_F\frac{1}{x}\nonumber \\ 
P_{gg,++}^{(0)}&=&P_{gg,++}^{(0)}= N_c
\left(\frac{2}{(1-x)_+} +\frac{1}{x} -1 -x - x^2 \right) +{\beta_0}\delta(1-x) \nonumber \\
P_{gg,+-}^{(0)}&=& N_c
\left( 3 x +\frac{1}{x} -3 - x^2 \right) 
\eeqa

Taking linear combinations of these equations (adding and subtracting),
one recovers the usual evolutions for unpolarized $q(x)$ and longitudinally
polarized $\Delta q(x)$ distributions.
We recall that the unpolarized distributions, the polarized and the transversely
polarized $q_T(x)$ are related by

\begin{figure}
{\centering \resizebox*{12cm}{!}{\rotatebox{-90}{\includegraphics{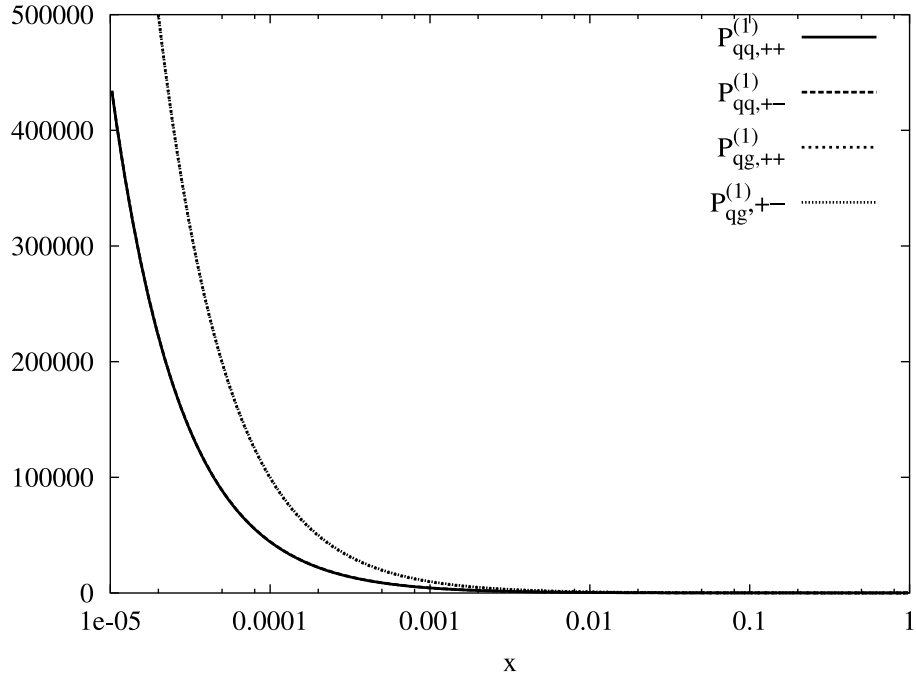}}} \par}

\caption{NLO kernels (\protect\( qq\protect \) and \protect\( qg\protect \))
in the helicity basis.}
\label{p1qx}
\end{figure}

\beqa
q(x)&=&q_+(x) + q_-(x)=q_{+T}(x) + q_{-T}(x)\nonumber \\
\Delta q(x)&=& q_+(x) - q_-(x)
\eeqa
at any $Q$ of the evolution and, in particular, at the boundary of the
evolution. 

\begin{figure}
{\centering \resizebox*{12cm}{!}{\rotatebox{-90}{\includegraphics{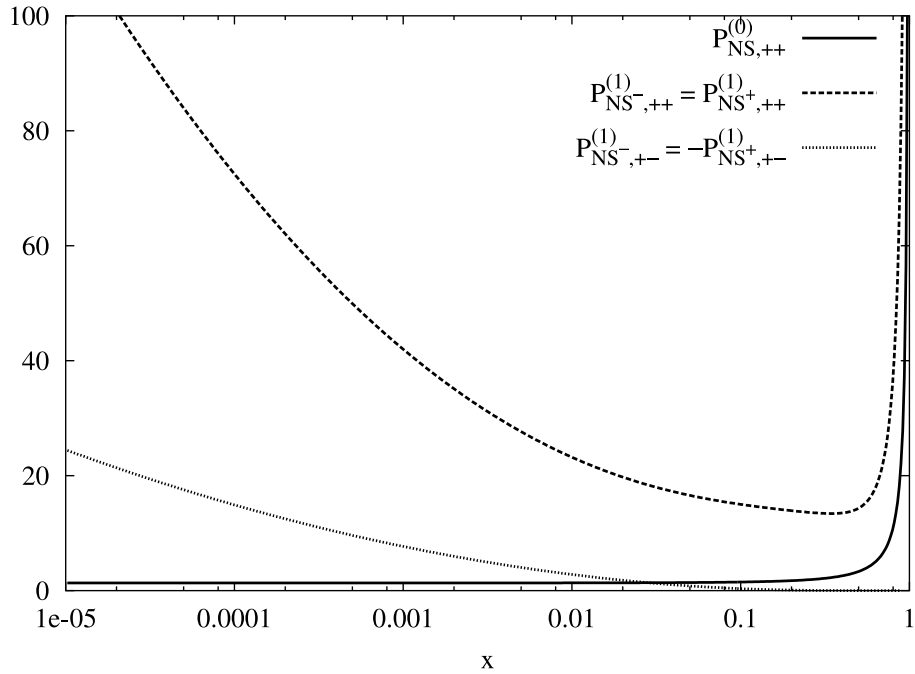}}} \par}
\caption{Nonsinglet kernels in the helicity basis.}
\label{pns}
\end{figure}

Similar definition have been introduced for the gluon sector
with $G_\pm(x)$ denoting the fixed helicities of the gluon distributions with
$\Delta g(x) = g_+(x) - g_-(x)$ and  $g(x)= g_+(x) + g_-(x)$ being the
corresponding longitudinal asymmetry and the unpolarized density respectively.

\begin{figure}
{\centering \resizebox*{12cm}{!}{\rotatebox{-90}{\includegraphics{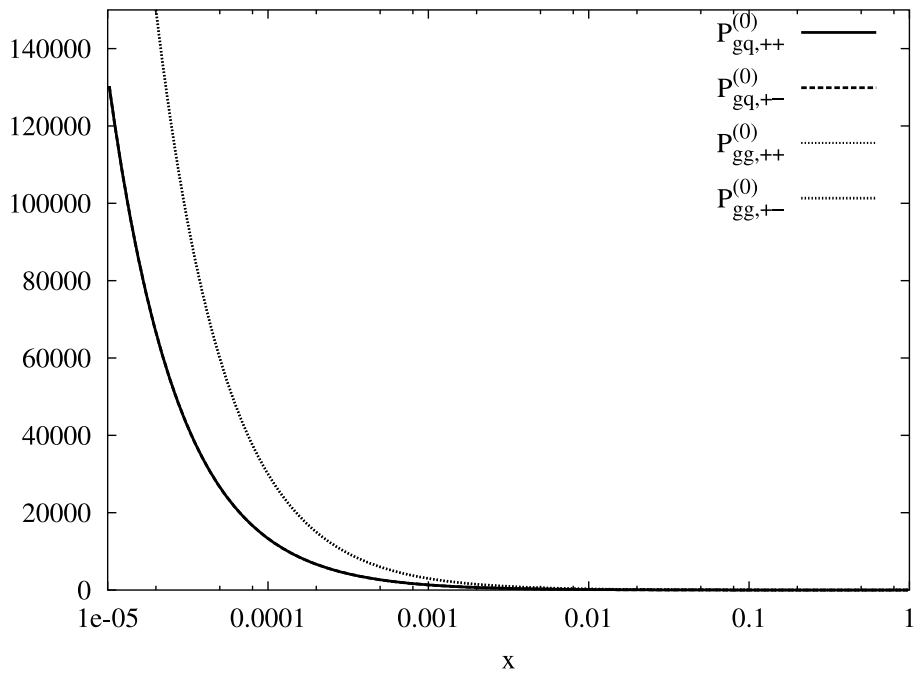}}} \par}

\caption{LO kernels (\protect\( gq\protect \) and \protect\( gg\protect \))
in the helicity basis.}
\label{p0gx}
\end{figure}

\section{Summary of Positivity Arguments}

Let us recapitulate here the basic arguments \cite{Teryaev} that are
brought forward in order to prove the positivity of the evolution to NLO.

If
\beq
|\Delta P(z)| \leq P(z), z < 1
\label{condi}
\eeq
then both kernels $P_{++}(z)$ and $P_{+-}(z)$ are positive as far as $z < 1$.

The singular contributions at $z=1$, which appear as subtraction terms
in the evolution and which could, in principle, alter positivity, appear
only in diagonal form, which means that they are only contained
in $P_{++}$, multiplied by the single functions $q_+(x)$ or $q_-(x)$

\beq
{dq_+(x) \over{dt}}=
{\alpha_s \over {2 \pi}} P_{++}^{qq} ({x \over y}) \otimes q_+(y) + . . .
\label{qplus}
\eeq
\beq
{dq_-(x) \over{dt}}=
{\alpha_s \over {2 \pi}} P_{++}^{qq} ({x \over y}) \otimes q_+(y) + . . .
\eeq
\beq
{d g_+(x) \over{dt}}=
{\alpha_s \over {2 \pi}} P_{++}^{gg} ({x \over y}) \otimes g_+(y) + . . .
\eeq

Let's focus just on the equation for $q_+$ (\ref{qplus}). Rewriting the diagonal contribution as a master equation
\beq
{dq_+(x) \over{dt}}=\int dx'\left(
w_{++}(x|x') q_+(x',\tau) -w_{++}(x'|x) q_+(x,\tau)\right) dx' + . . .
\label{formx}
\eeq
in terms of a transition probability
\beq
w_{++}(x|y)= \frac{\alpha_s}{2 \pi} \hat{P}_{++}(x/y)\theta(y > x)
\eeq
which can be easily established to be positive, as we are going to show rigorously below, 
as far as all the remaining
terms (the ellipses) are positive. We have performed a detailed 
numerical analysis to show the positivity of the contributions at $x=1$. 
 
This last condition is also clearly satisfied, since the
$\delta(1-z)$ contributions appear only in $P_{++}$ and are diagonal in
the helicity of the various flavours (q, g). 
For a rigorous proof of the positivity of the solutions of master equations 
we proceed as follows. 

\begin{figure}
{\centering \resizebox*{12cm}{!}{\rotatebox{-90}{\includegraphics{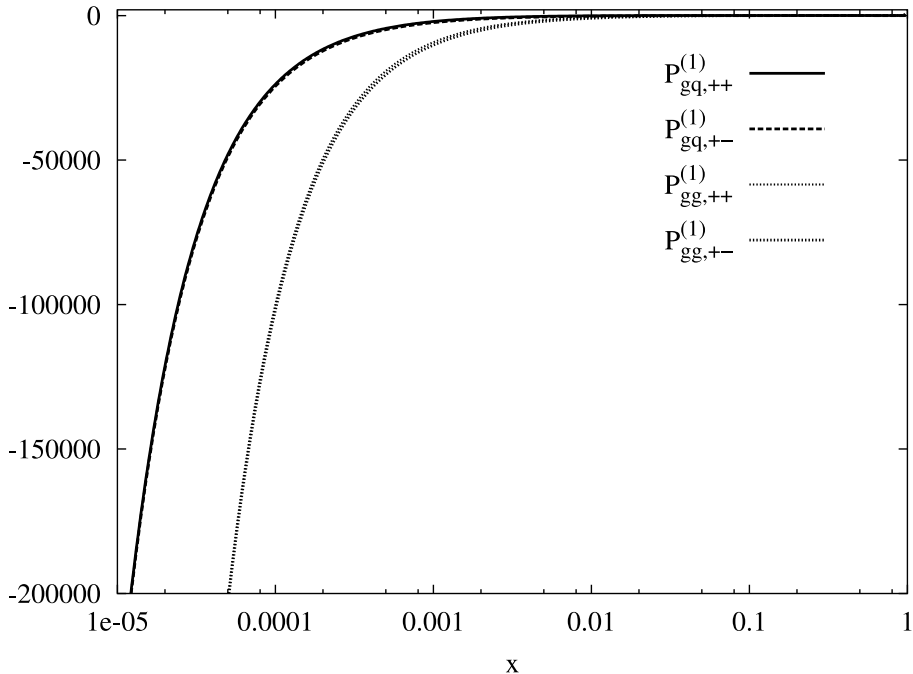}}} \par}

\caption{NLO kernels (\protect\( gq\protect \) and \protect\( gg\protect \))
in the helicity basis.}
\label{p1gx}
\end{figure}

Let $q(x,\tau)$  be a positive distribution for 
$\tau < \tau_c$ and let us assume that it vanishes at $\tau=\tau_c$, after which 
it turns negative. We also assume that the evolution of 
$q(x,\tau)$ is of the form (\ref{form}) with positive 
transition probabilities $w(x|y)$ and $w(y|x)$. Notice that since 
the function is continuous together with its first derivative and decreasing, 
continuity of its first derivative will require $\frac{d q(x,\tau)}{d \tau} < 0$ at 
$\tau=\tau_c$ and in its neighbor. However, eq. (\ref{form}) requires 
that at $\tau=\tau_c$   
\beq
{dq(x) \over{dt}}=\int dx'
w (x|x') q(x',\tau) 
\label{form1}
\eeq

which is positive, and we have a contradiction. 
We can picture the evolution in $\tau$ of these functions as a family 
of curves getting support to smaller and smaller x-values as $\tau$ grows 
and being almost vanishing at intermediate and large x values. 
We should mention that this proof does not require a complete 
probabilistic picture of the evolution, but just the positivity of the 
bulk part of the kernels, the positivity 
of the edge point subtractions and their diagonality in flavour. 
From Figs. (\ref{p0qx})  and (\ref{p0gx}) it is also evident that the leading order 
kernels are positive, together with the $qg$ and $qq$ (Fig. (\ref{p1qx})) sectors. 

The edge point contributions, generating the ``subtraction terms'' in the master 
equations for the ``++'' components of the kernels are positive, 
as is illustrated in 3 Tables included in Appendix A. 
There we have organized these terms in the form $\sim C\delta(1-x)$ with
\beq
C=-\log(1- \Lambda) A + B 
\label{sub}
\eeq
with A and B being numerical coefficients depending on the number 
of flavours included in the kernels. Notice that the subtraction terms are 
always of the form (\ref{sub}), with the (diverging) logarithmic contribution 
($\sim \int_0^\Lambda dz/(1-z)$) regulated by a cutoff. This divergence 
in the convolution cancels when these terms are combined with the divergence at 
$x=1$ of the first term of the master equation for all the relevant components 
containing ``+'' distributions. 
It is crucial, however, to establish positivity of the evolution of the 
helicities that the boundary conditions on the evolution
$|\Delta q(x,Q_0^2)| \leq q(x,Q_0^2)$ be satisfied. 
Initial conditions have this special property, in most of models, and the proof 
of positivity of all the distributions therefore holds at any $Q$. 

\begin{figure}
{\centering \resizebox*{12cm}{!}{\rotatebox{-90}{\includegraphics{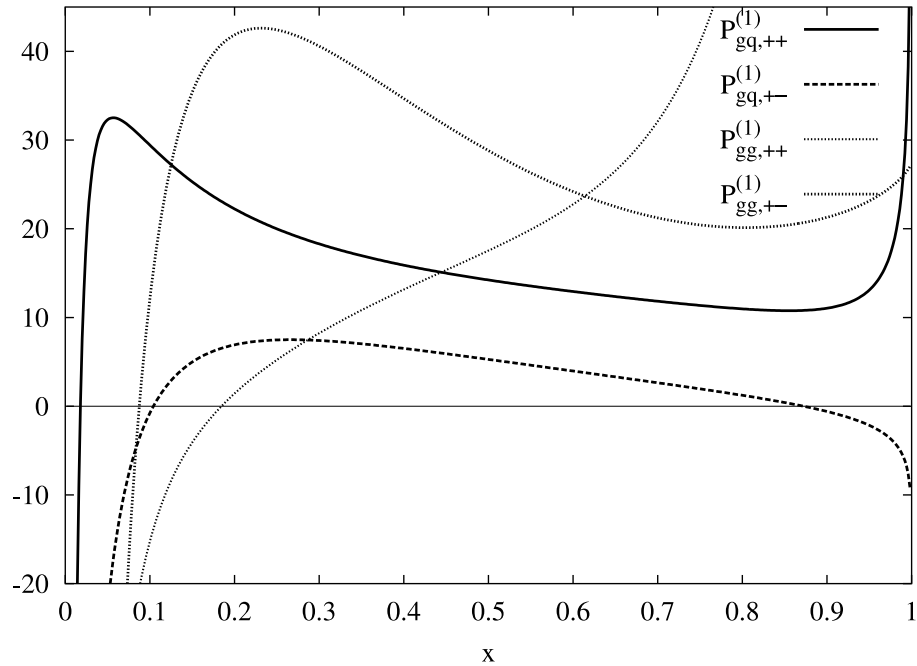}}} \par}
\caption{The crossing of NLO kernels (\protect\( gq\protect \) and \protect\( gg\protect \))
in the helicity basis.}
\label{p1gx_nonlog}
\end{figure}

As we move to NLO, the pattern gets more complicated.  
In fact, from a numerical check, one can see that some NLO kernels 
turn to be negative, including the unpolarized kernels and the 
helicity kernels, while others (Fig. (\ref{pns})) are positive.  
One can also notice 
the presence of a crossing of several helicity components in the $gq$ and $gg$ 
sectors (see Fig. (\ref{p1gx_nonlog})) at larger x values, while in the 
small-x region some components turn negative (Fig. (\ref{p1gx})). 
There is no compelling proof of positivity, in this case, either 
than that coming from a direct numerical analysis.  

We have seen that 
master forms of evolution equations, for evolutions of all kinds, when found, 
can be used to establish positivity of the evolution itself. 

The requirements  have been spelled out above and can be
summarized in the following points: 1) diagonality of the decreasing terms,
2) initial positivity of the distributions, 3) positivity
of the remaining (non diagonal) kernels. 
As we have also seen, some of these conditions are not satisfied by the NLO evolution.

\section{Results}

In regard to the input distributions we used, we refer the reader to Sec.~\ref{sec:initial}. 

\begin{figure}[tbh]
{\centering \resizebox*{12cm}{!}{\rotatebox{-90}{\includegraphics{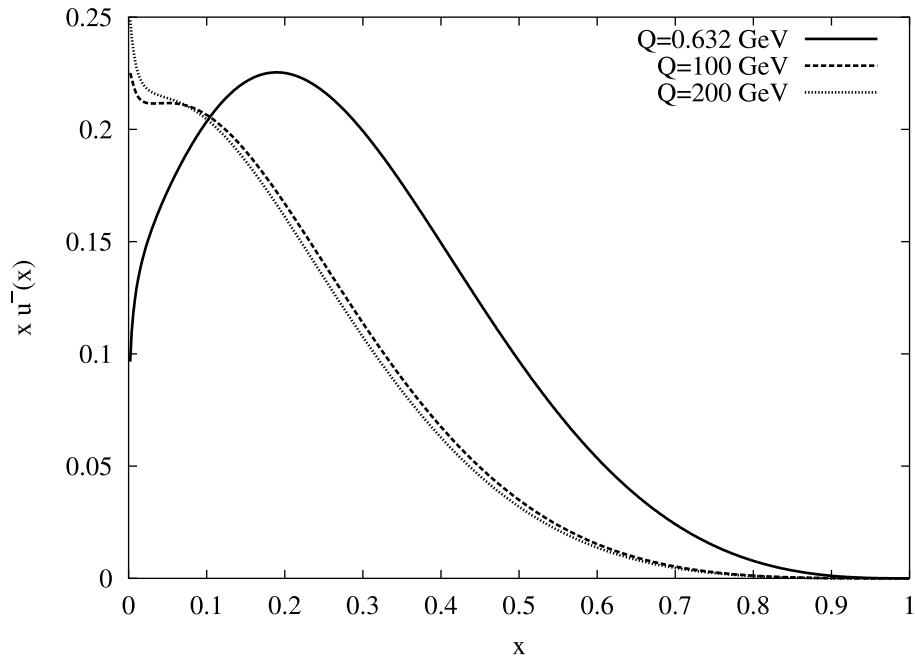}}} \par}

\caption{Evolution of \protect\( u^{+}\protect \) versus \protect\( x\protect \)
at various \protect\( Q\protect \) values.}
\label{uplus}
\end{figure}

\begin{figure}[tbh]
{\centering \resizebox*{12cm}{!}{\rotatebox{-90}{\includegraphics{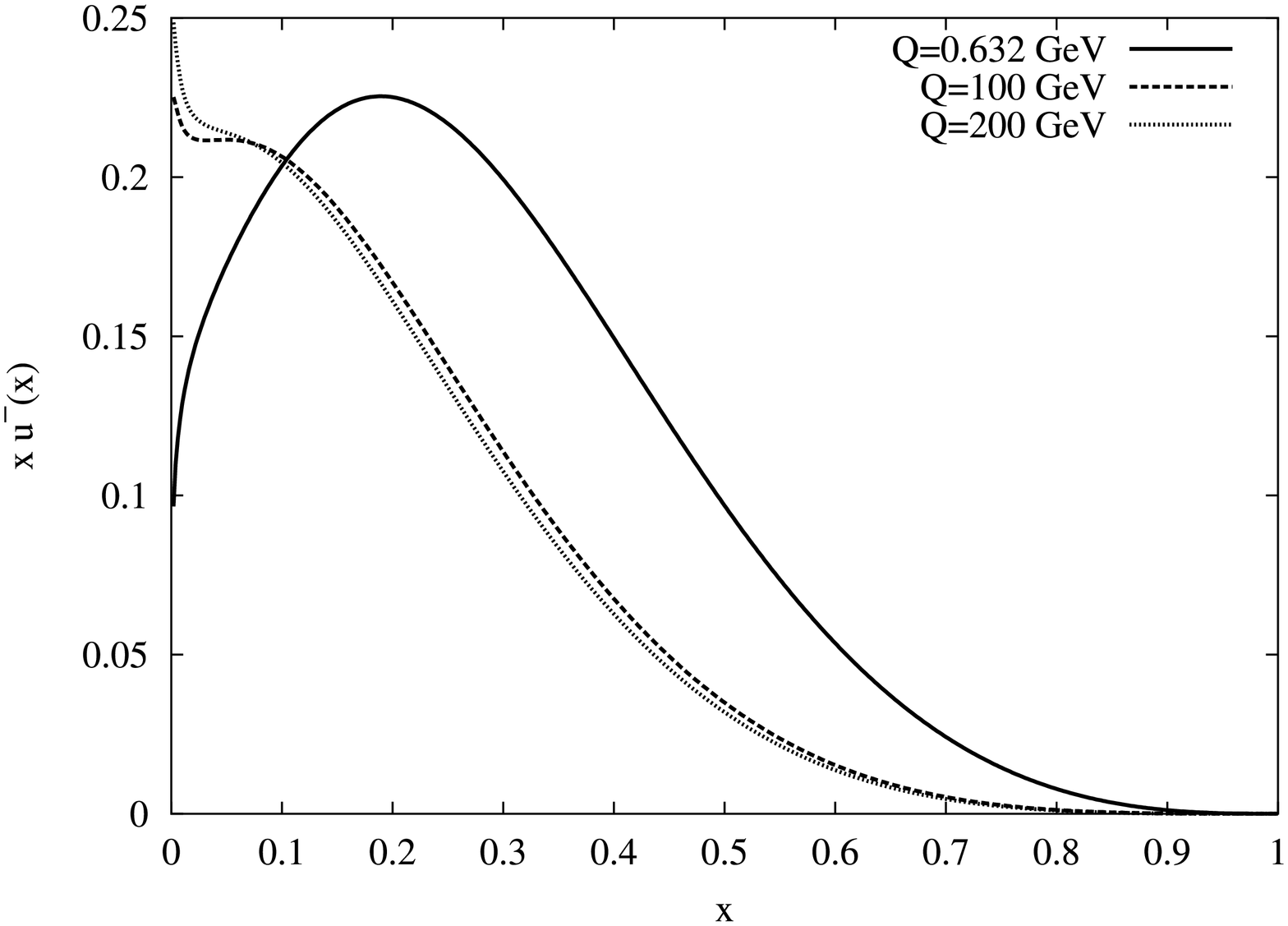}}} \par}

\caption{Evolution of \protect\( u^{-}\protect \) versus \protect\( x\protect \)
at various \protect\( Q\protect \) values.}
\label{uminus}
\end{figure}

\begin{figure}[tbh]
{\centering \resizebox*{12cm}{!}{\rotatebox{-90}{\includegraphics{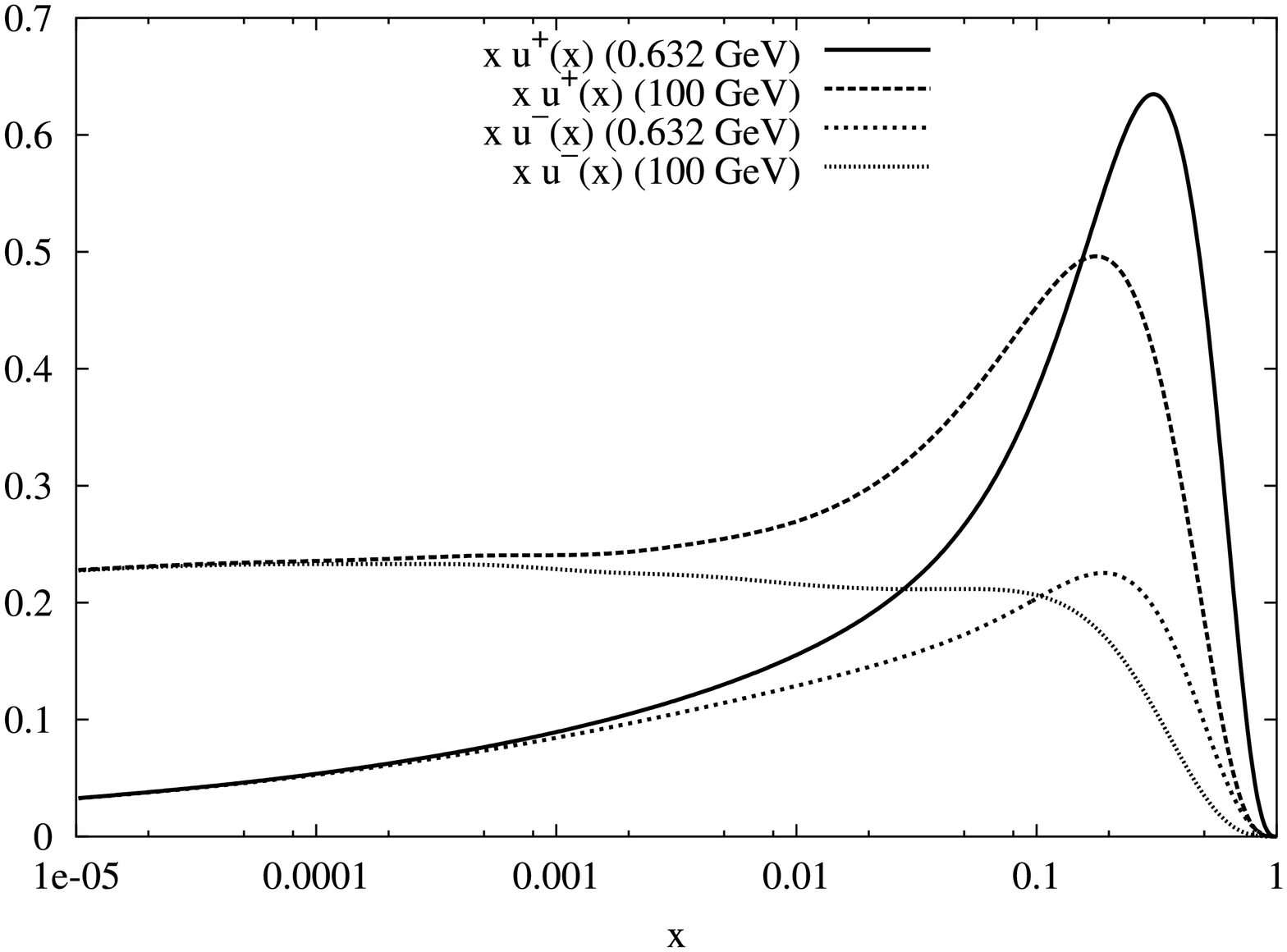}}} \par}
\caption{Small-$x$ behaviour of $u^{\pm}$ at 100 GeV.}
\label{ulog}
\end{figure}

\begin{figure}[tbh]
{\centering \resizebox*{12cm}{!}{\rotatebox{-90}{\includegraphics{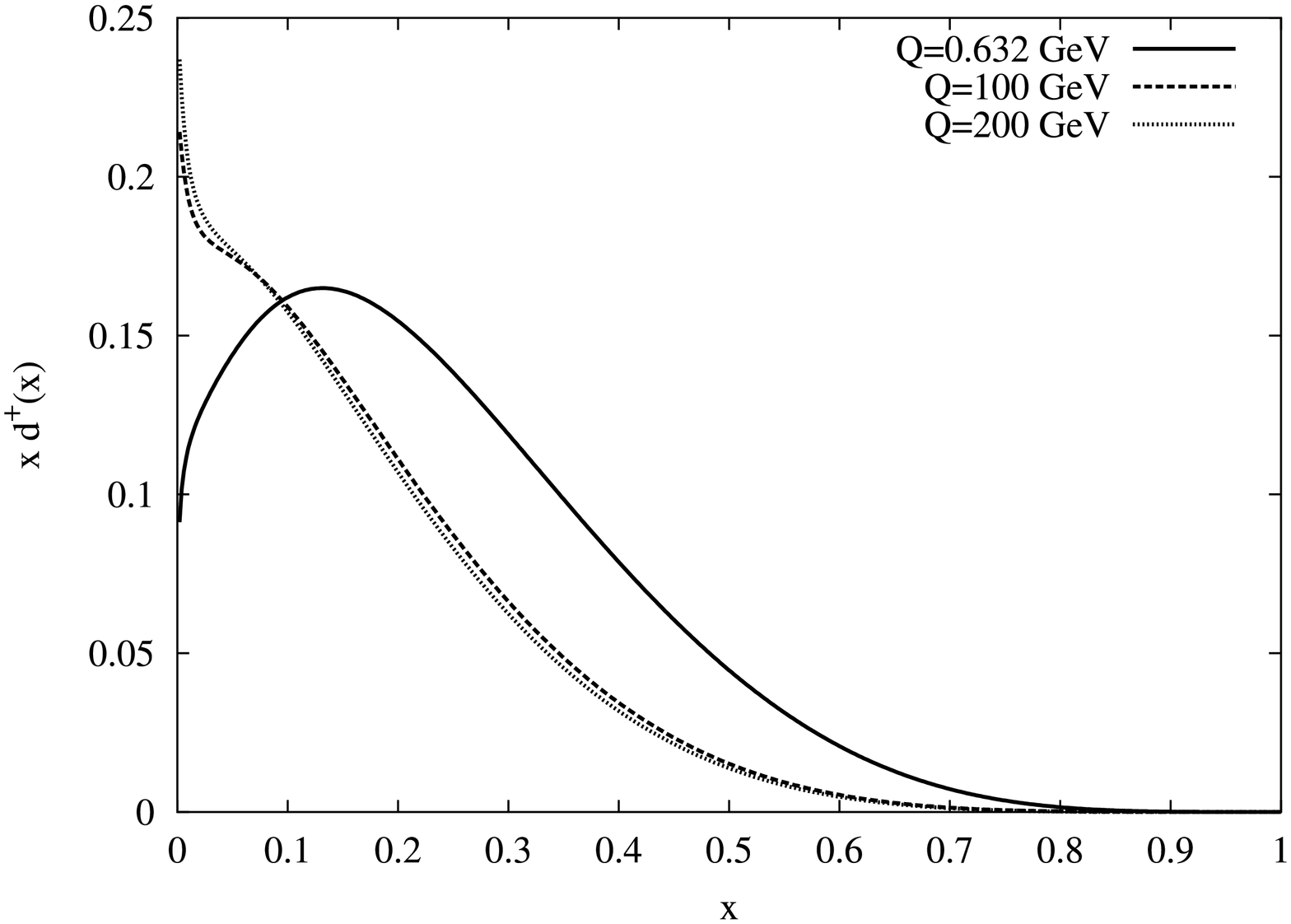}}} \par}
\caption{Evolution of \protect\( d^{+}\protect \) versus \protect\( x\protect \)
at various \protect\( Q\protect \) values.}
\label{dplus}
\end{figure}

\begin{figure}[tbh]
{\centering \resizebox*{12cm}{!}{\rotatebox{-90}{\includegraphics{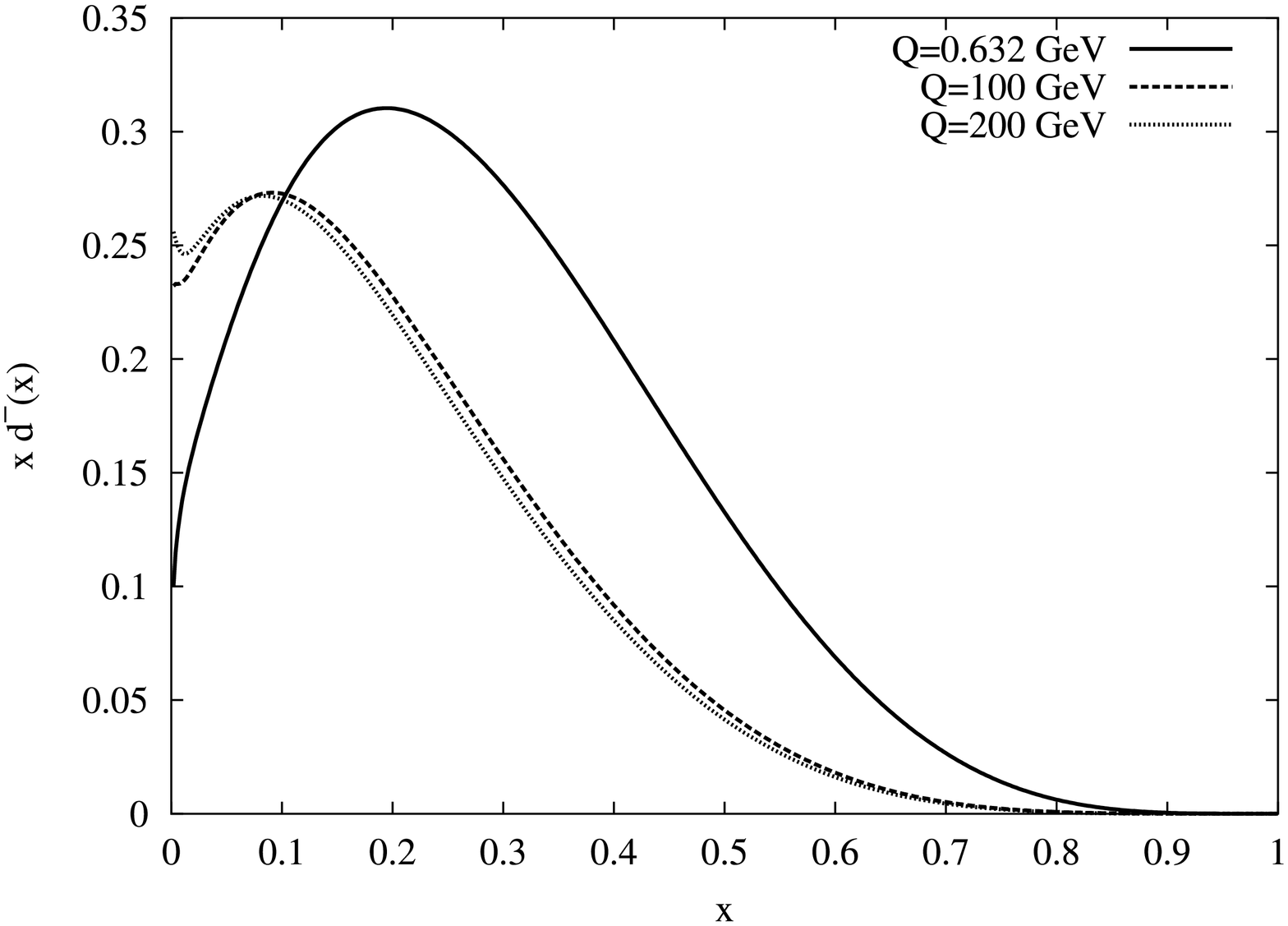}}} \par}
\caption{Evolution of \protect\( d^{-}\protect \) versus \protect\( x\protect \)
at various \protect\( Q\protect \) values.}
\label{dminus}
\end{figure}

\begin{figure}[tbh]
{\centering \resizebox*{12cm}{!}{\rotatebox{-90}{\includegraphics{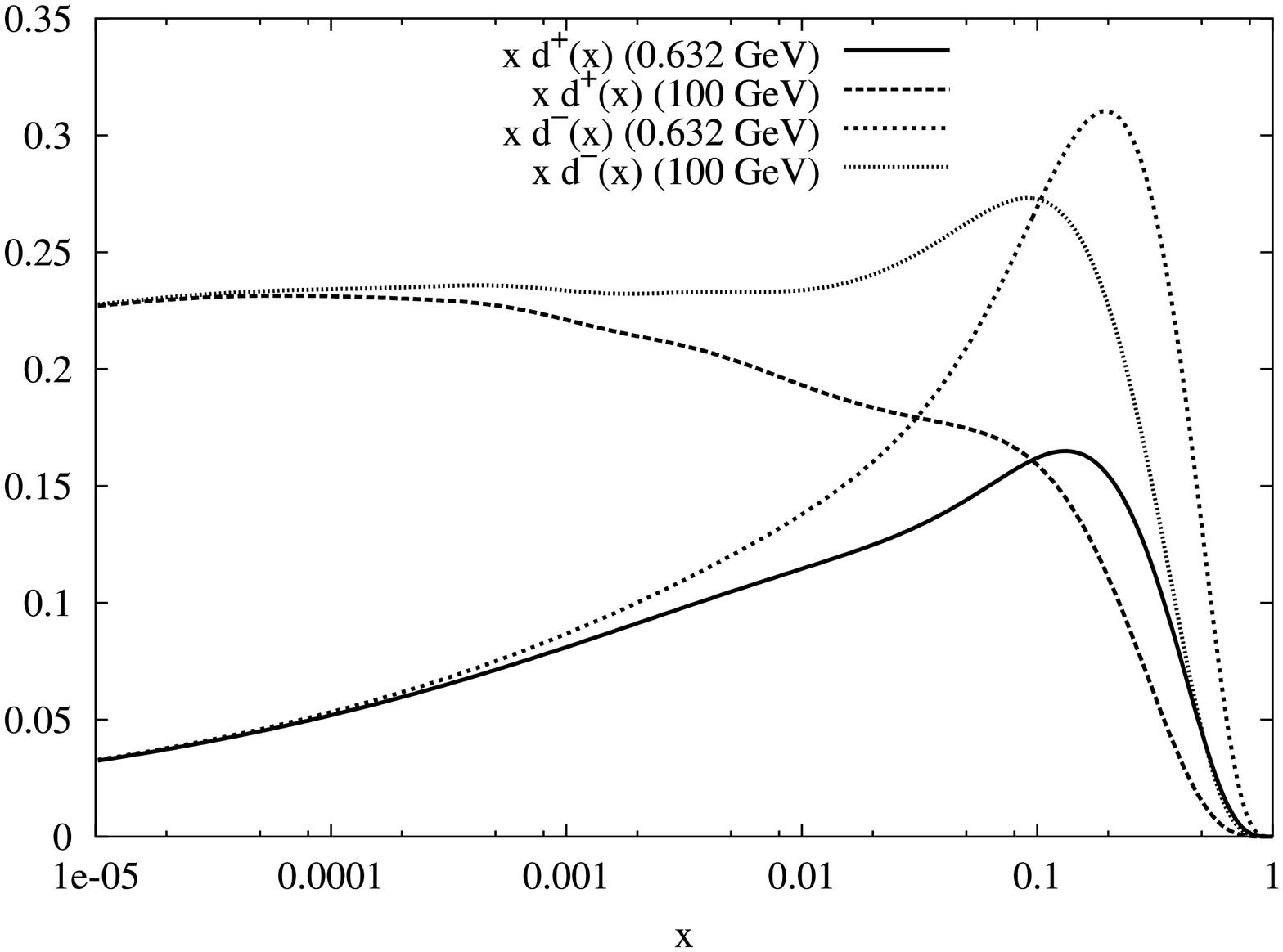}}} \par}
\caption{Small-$x$ behaviour of $d^{\pm}$ at 100 GeV.}
\label{dlog}
\end{figure}

We show in Fig. \ref{uplus} results for the evolution of the $u^+$ 
distribution at the initial scale ($0.632$ GeV) and at two final scales, 
100 GeV and 200 GeV respectively. The peaks at the various scales 
get lowered and become more pronounced toward the smaller x region as Q increases. 
In Fig. \ref{uminus} $u^-$ show an apparent steeper growth at small-x compared to 
$u^+$. For the $d$ distributions the situation is inverted, with $d^-$ 
growing steeper compared to $d^+$ (Figs. \ref{dplus} and \ref{dminus} respectively). 
This apparent behaviour is resolved in Figs. (\ref{ulog}) and (\ref{dlog}) 
from which it is evident that both plus and minus components converge, at very small-x 
values, toward the same limit.   
 
\begin{figure}[tbh]
{\centering \resizebox*{12cm}{!}{\rotatebox{-90}{\includegraphics{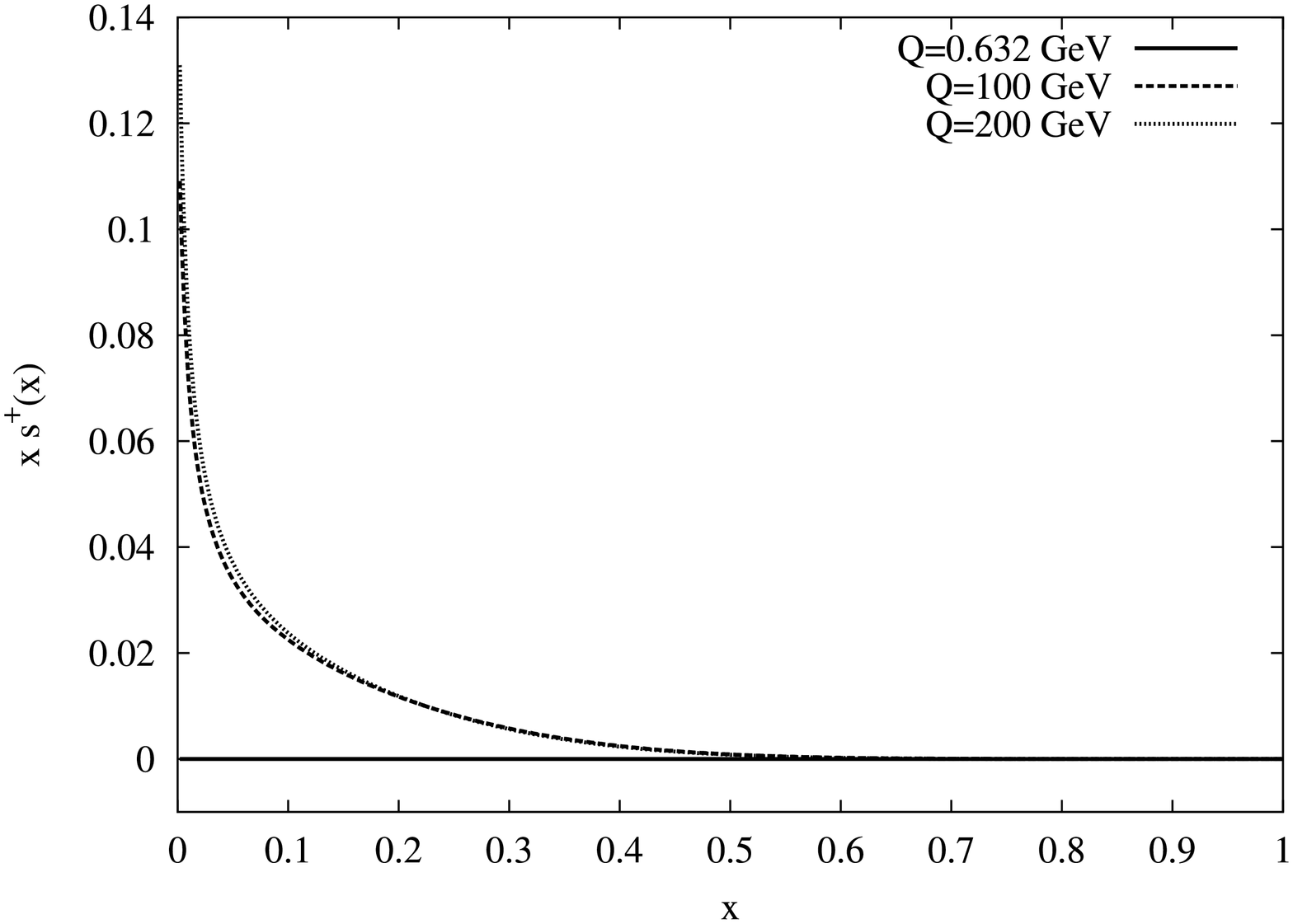}}} \par}

\caption{Evolution of \protect\( s^{+}\protect \) versus \protect\( x\protect \)
at various \protect\( Q\protect \) values.}
\label{splus}
\end{figure}

\begin{figure}[tbh]
{\centering \resizebox*{12cm}{!}{\rotatebox{-90}{\includegraphics{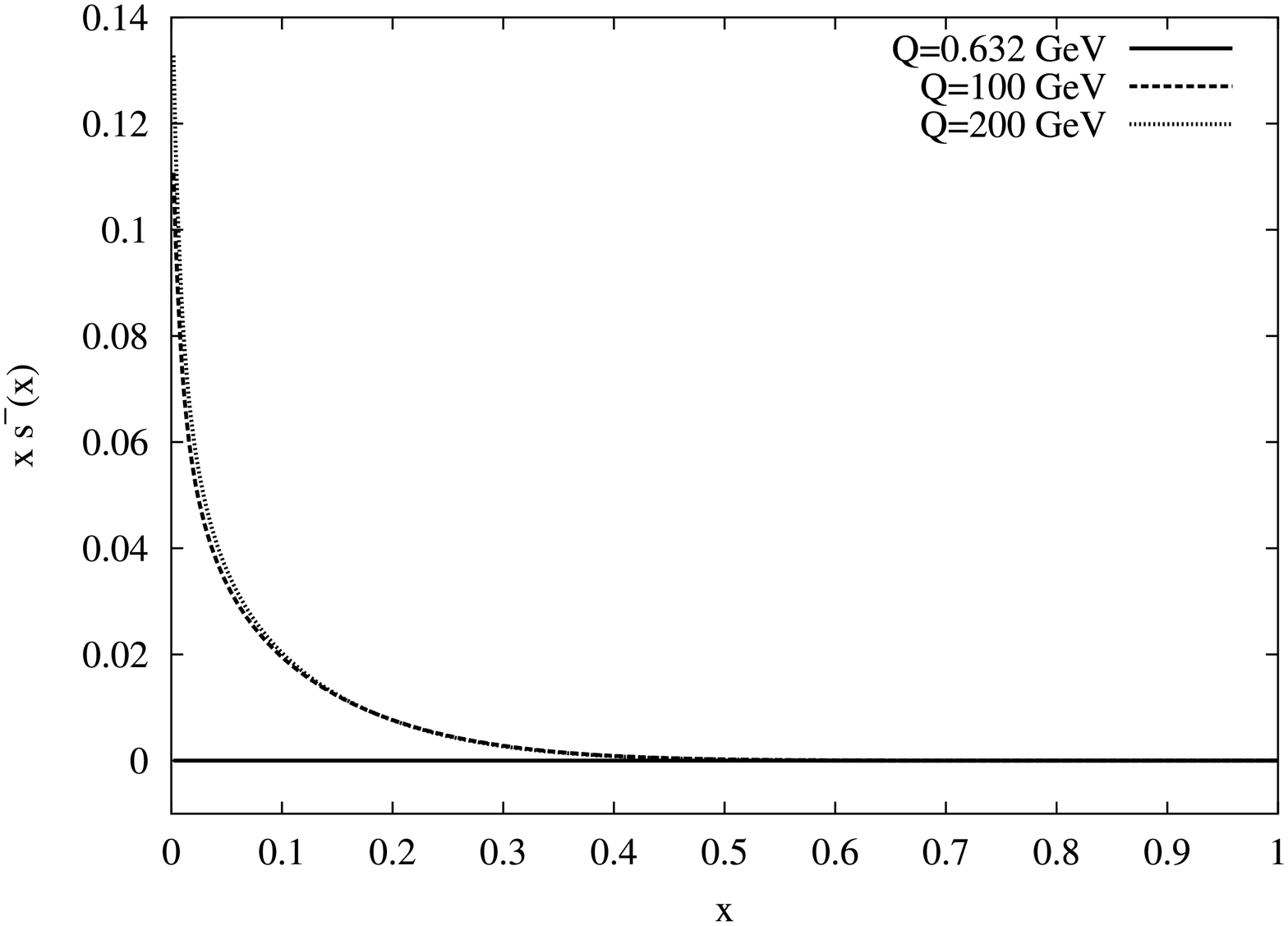}}} \par}

\caption{Evolution of \protect\( s^{-}\protect \) versus \protect\( x\protect \)
at various \protect\( Q\protect \) values.}
\label{sminus}
\end{figure}

\begin{figure}[tbh]
{\centering \resizebox*{12cm}{!}{\rotatebox{-90}{\includegraphics{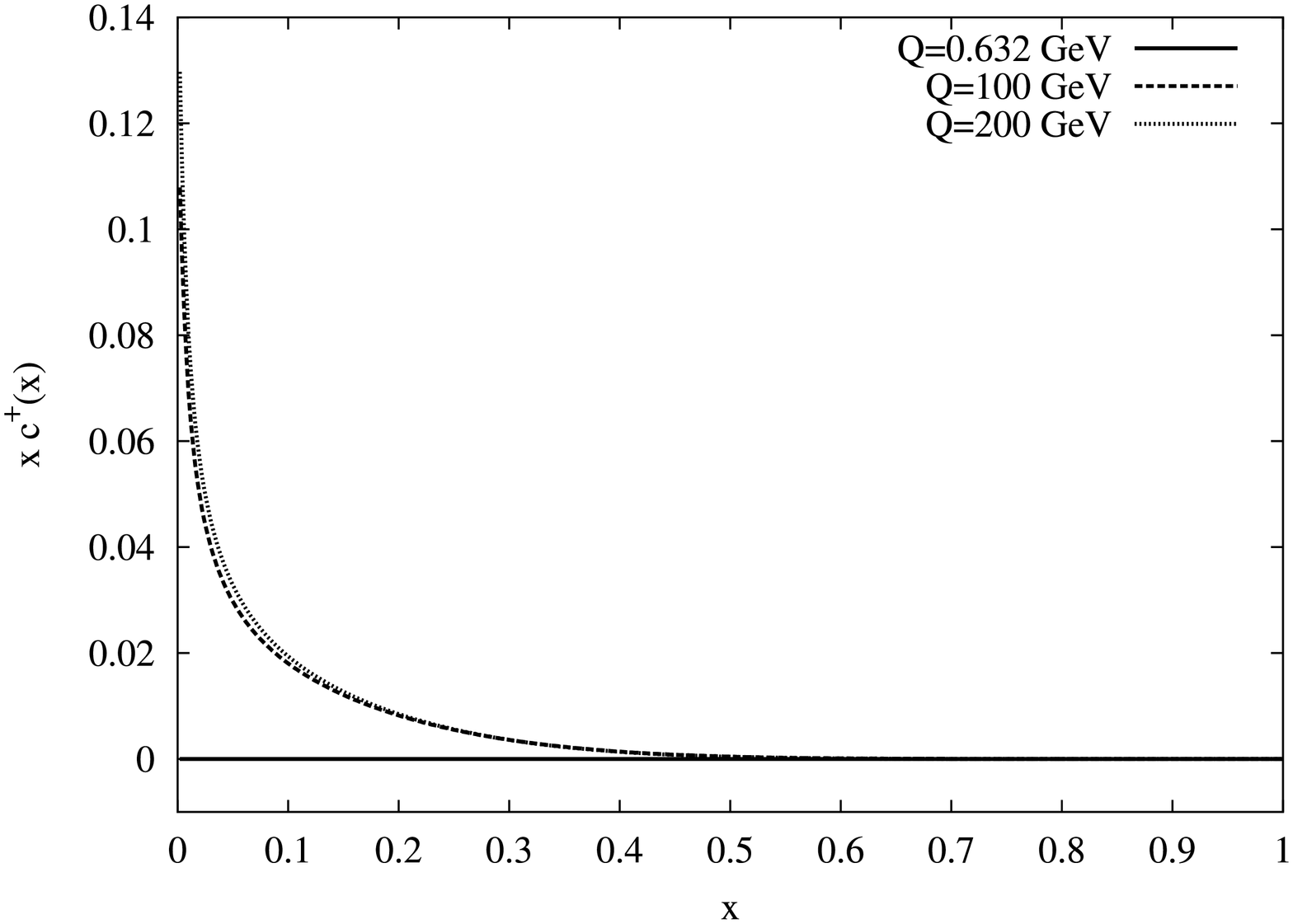}}} \par}

\caption{Evolution of \protect\( c^{+}\protect \) versus \protect\( x\protect \)
at various \protect\( Q\protect \) values.}
\label{cplus}
\end{figure}

\begin{figure}[tbh]
{\centering \resizebox*{12cm}{!}{\rotatebox{-90}{\includegraphics{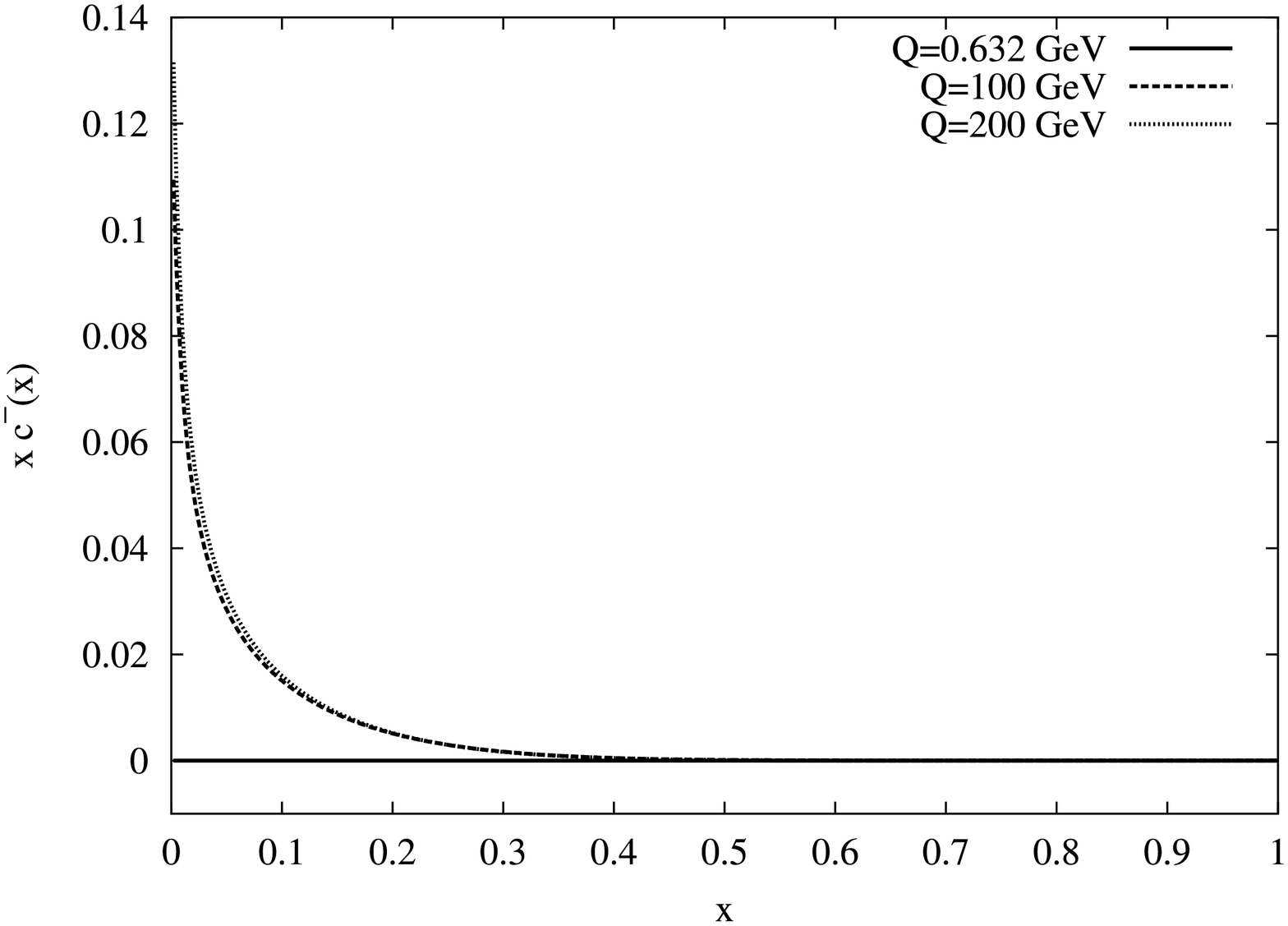}}} \par}

\caption{Evolution of \protect\( c^{-}\protect \) versus \protect\( x\protect \)
at various \protect\( Q\protect \) values.}
\label{cminus}
\end{figure}

\begin{figure}[tbh]
{\centering \resizebox*{12cm}{!}{\rotatebox{-90}{\includegraphics{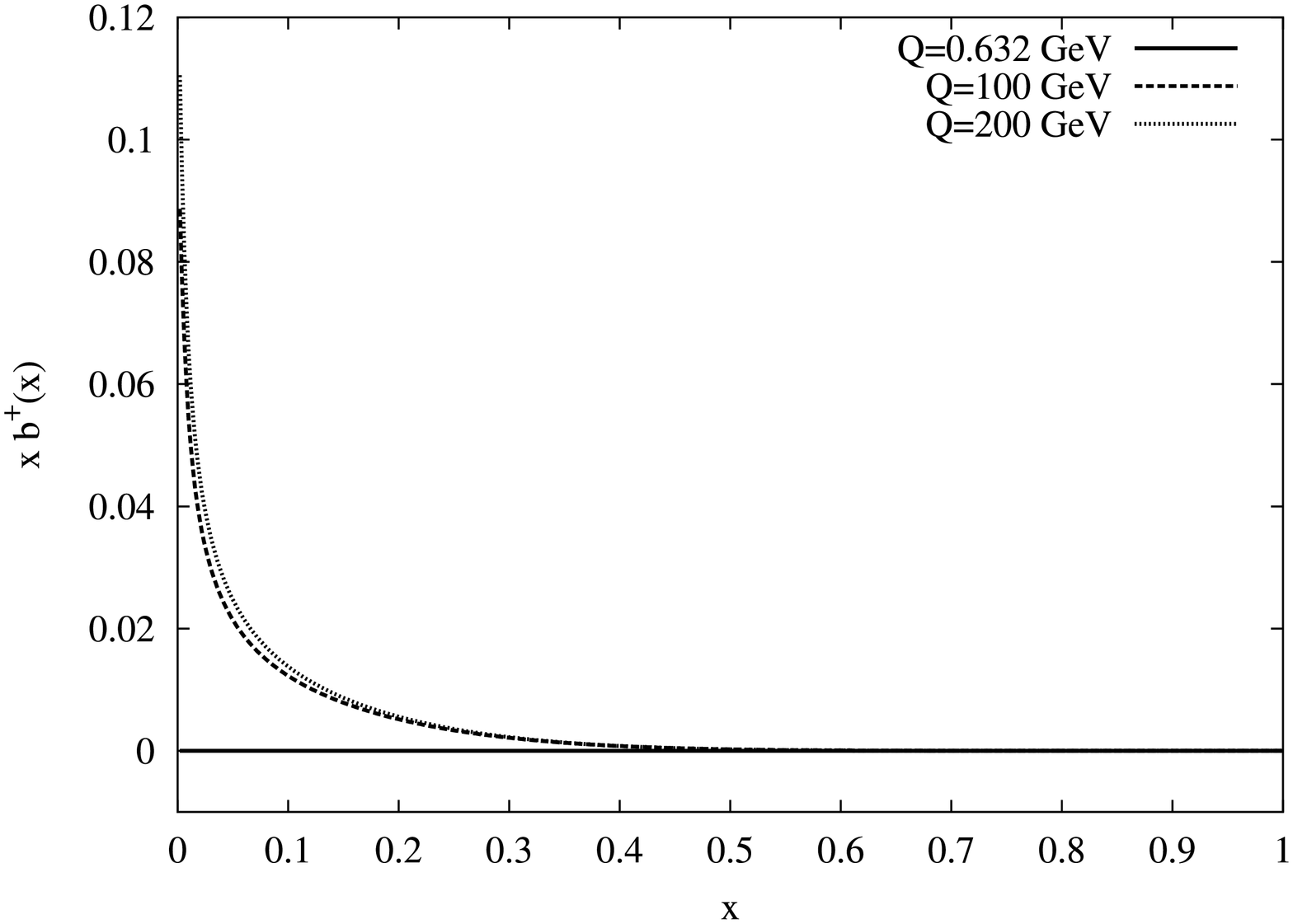}}} \par}

\caption{Evolution of \protect\( b^{+}\protect \) versus \protect\( x\protect \)
at various \protect\( Q\protect \) values.}
\label{bplus}
\end{figure}

\begin{figure}[tbh]
{\centering \resizebox*{12cm}{!}{\rotatebox{-90}{\includegraphics{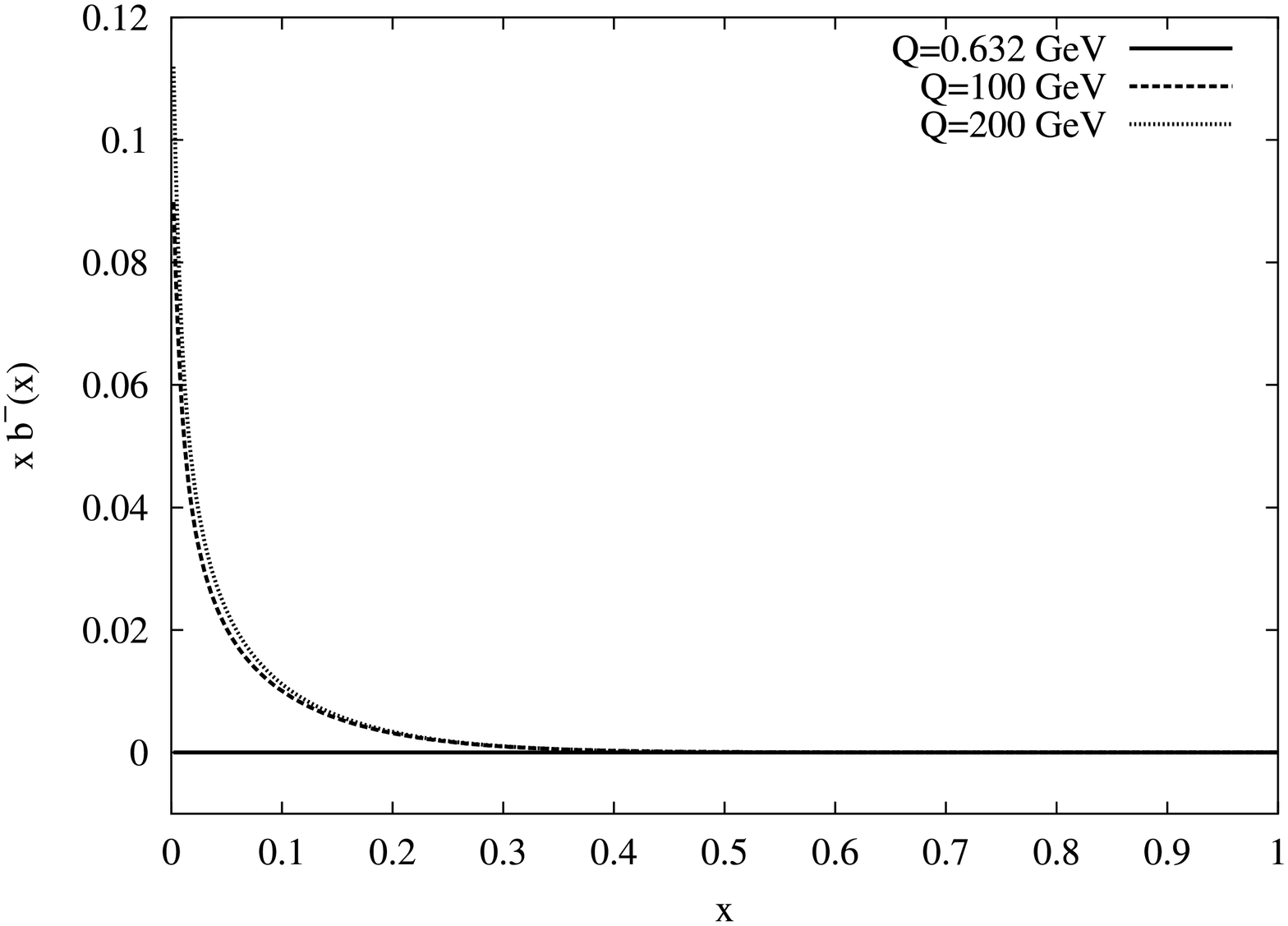}}} \par}

\caption{Evolution of \protect\( b^{-}\protect \) versus \protect\( x\protect \)
at various \protect\( Q\protect \) values.}
\label{bminus}
\end{figure}

\begin{figure}[tbh]
{\centering \resizebox*{12cm}{!}{\rotatebox{-90}{\includegraphics{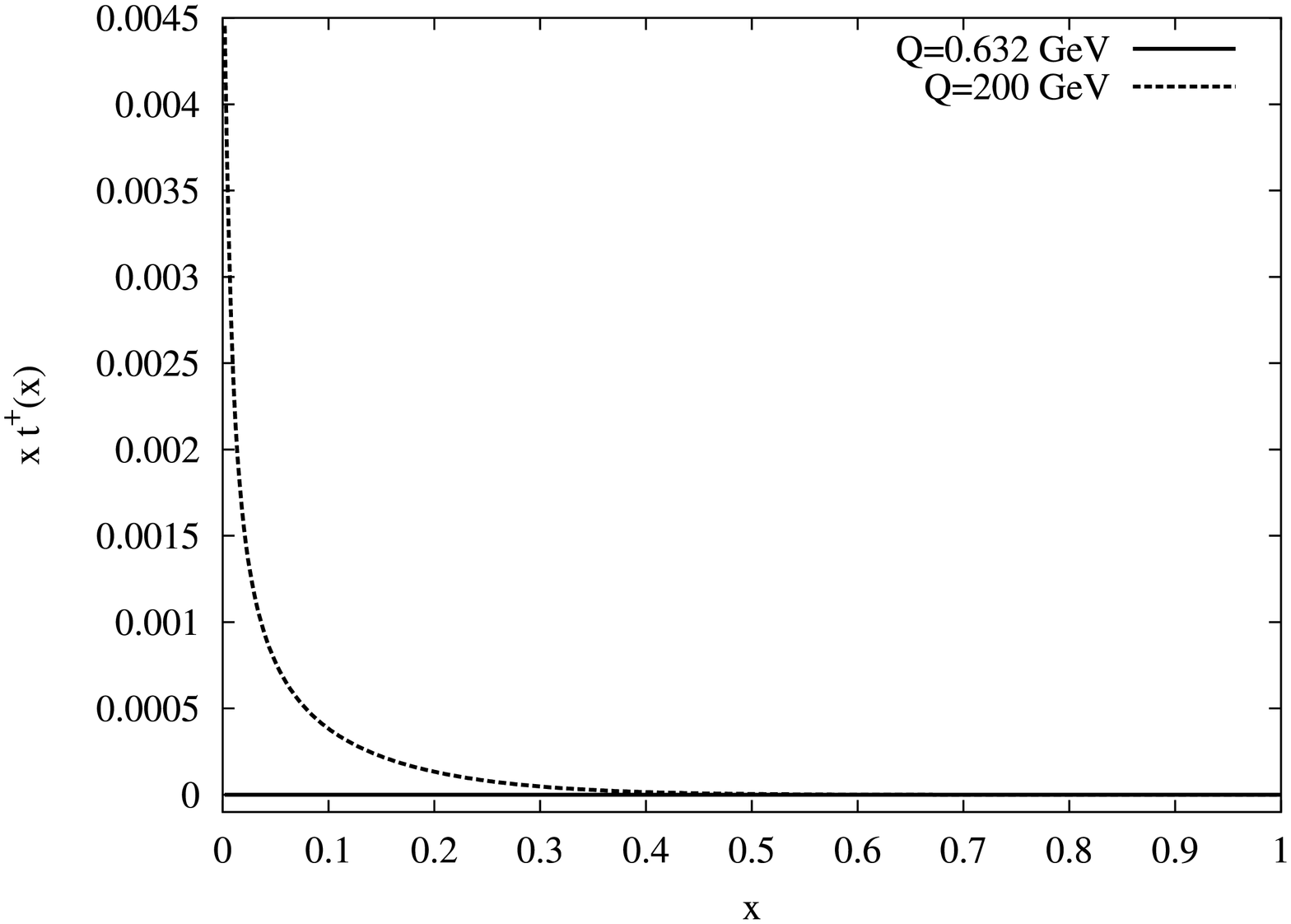}}} \par}

\caption{Evolution of \protect\( t^{+}\protect \) versus \protect\( x\protect \)
at various \protect\( Q\protect \) values.}
\label{tplus}
\end{figure}

\begin{figure}[tbh]
{\centering \resizebox*{12cm}{!}{\rotatebox{-90}{\includegraphics{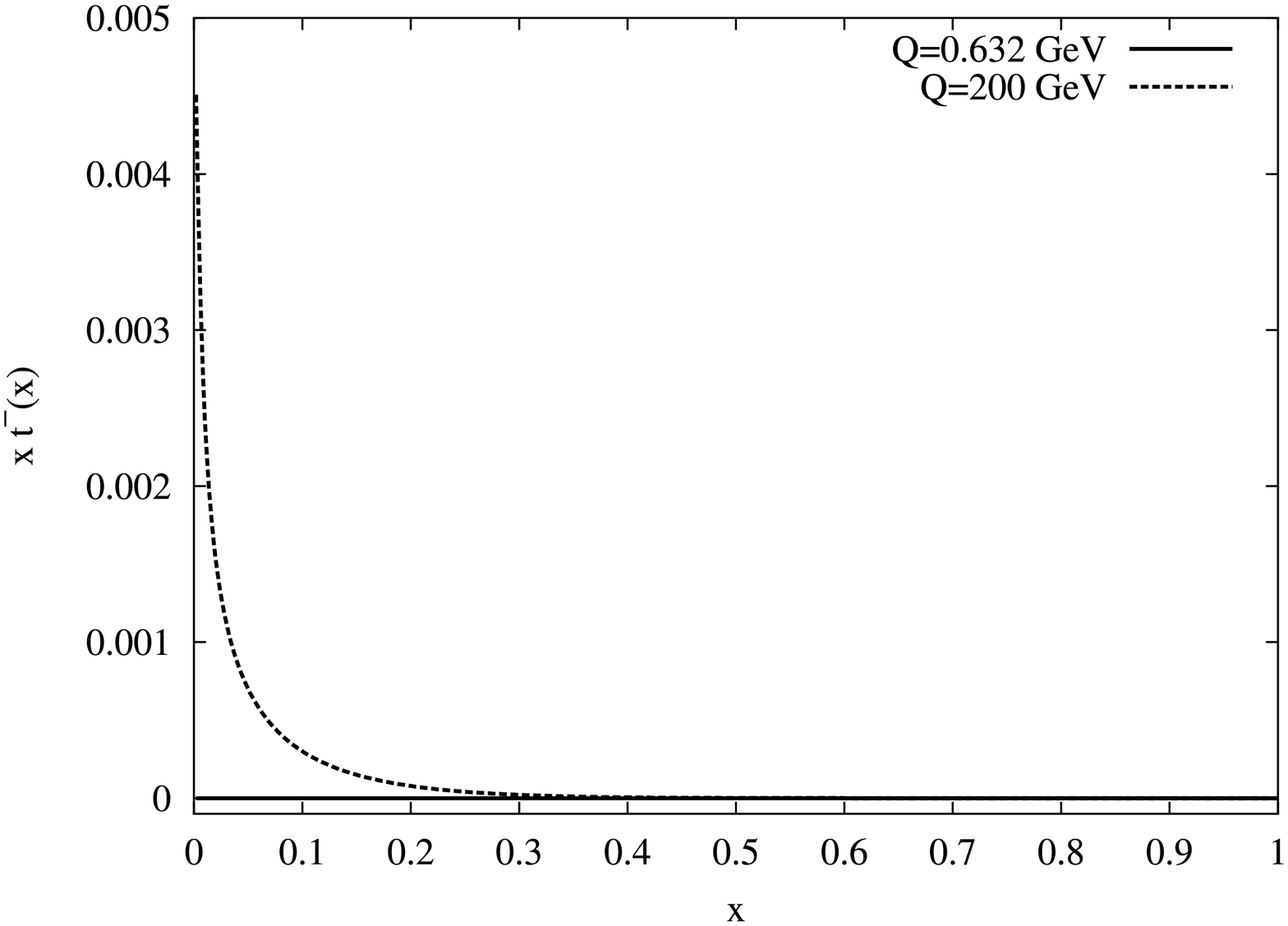}}} \par}

\caption{Evolution of \protect\( t^{-}\protect \) versus \protect\( x\protect \)
at various \protect\( Q\protect \) values.}
\label{tminus}
\end{figure}

The components s,c,t and b (Figs.~\ref{splus}-\ref{tminus})  
have been generated radiatively from vanishing 
initial conditions for final evolution scales of 100 and 200 GeV $(s,c,b)$ 
and 200 GeV (t). 

\begin{figure}[tbh]

{\centering \resizebox*{12cm}{!}{\rotatebox{-90}{\includegraphics{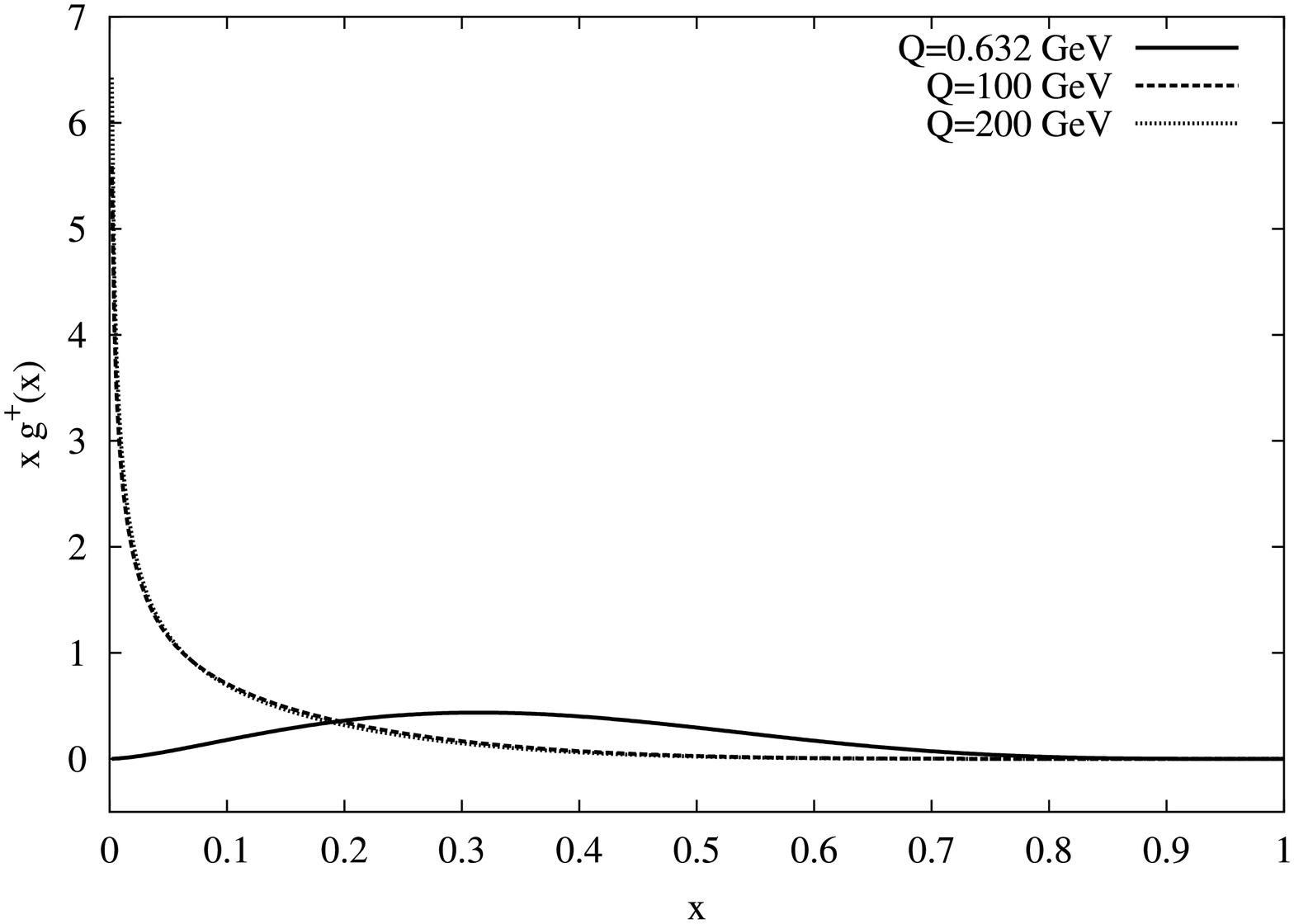}}} \par}

\caption{Evolution of \protect\( g^{+}\protect \) versus \protect\( x\protect \)
at various \protect\( Q\protect \) values.}
\label{gplus}
\end{figure}

\begin{figure}[tbh]
{\centering \resizebox*{12cm}{!}{\rotatebox{-90}{\includegraphics{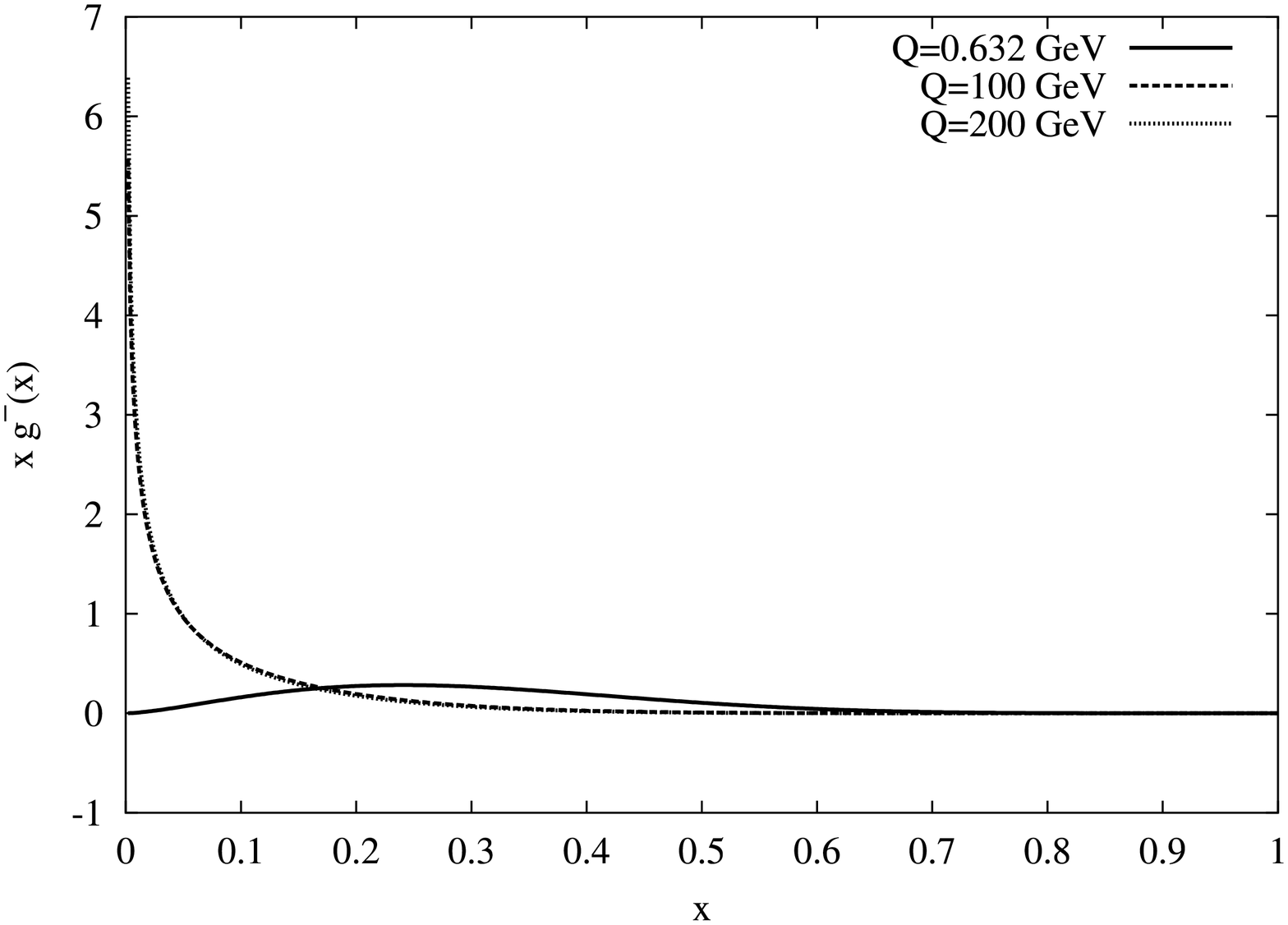}}} \par}

\caption{Evolution of \protect\( g^{-}\protect \) versus \protect\( x\protect \)
at various \protect\( Q\protect \) values.}
\label{gminus}
\end{figure}

Both positive and negative components grow steadily at small-x and are negligible at larger x values. The distribution for the top quark $(t)$ has been included for
completeness. Given the smaller evolution interval the helicity 
distributions for heavier generations 
are suppressed compared to those of lighter flavours. Gluon helicities 
(Figs. \ref{gplus}, \ref{gminus}) are also enhanced at small-x, and show a 
similar growth. The fine difference between the quark u and d 
distributions are shown in Figs. \ref{ulog} and \ref{dlog}.

\begin{figure}[tbh]
{\centering \resizebox*{12cm}{!}{\rotatebox{-90}{\includegraphics{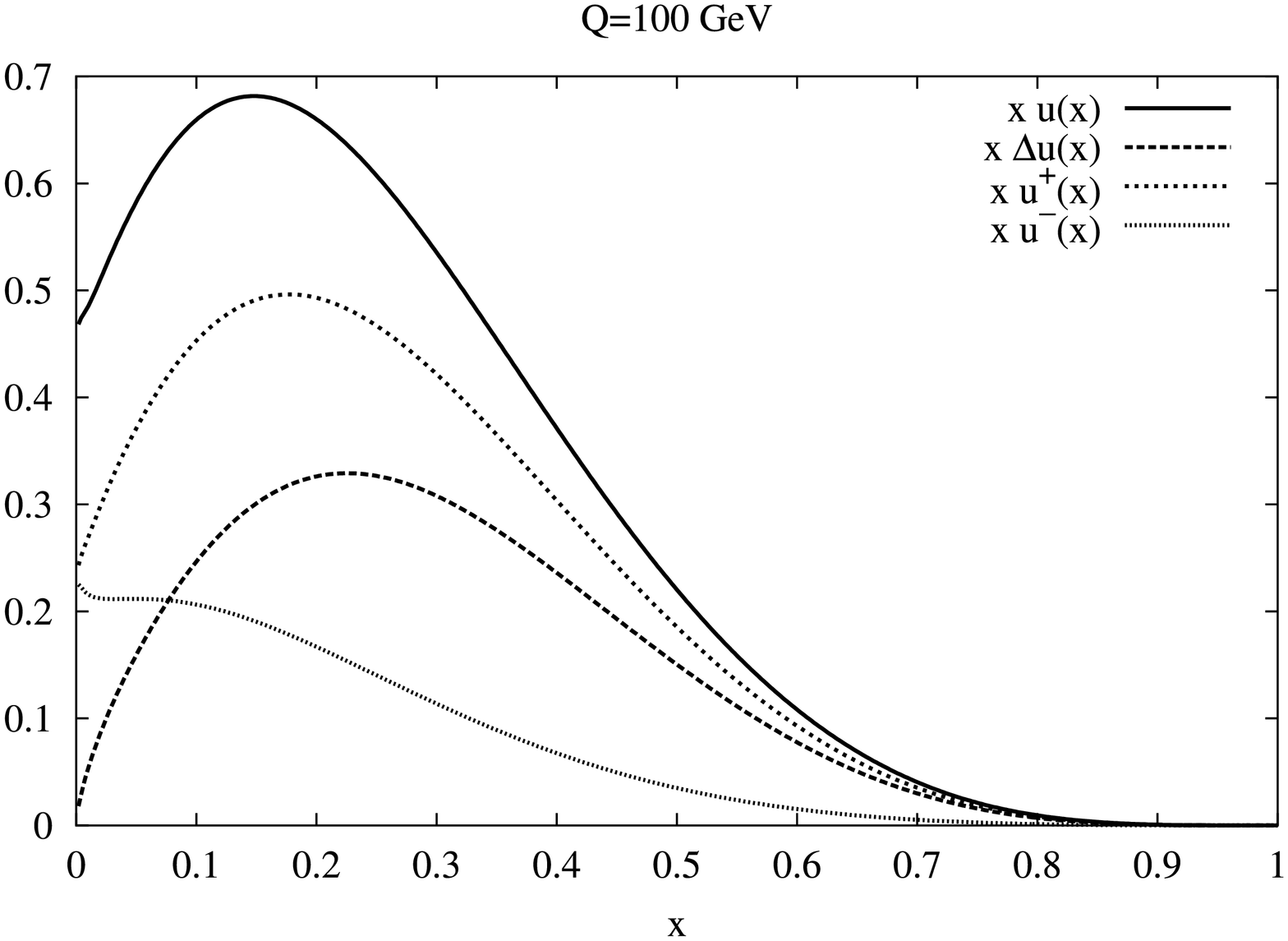}}} \par}
\caption{Various kinds of distributions of quark up at \protect\( Q=100\, \textrm{GeV}\protect \).}
\label{ups}
\end{figure}

\begin{figure}[tbh]
{\centering \resizebox*{12cm}{!}{\rotatebox{-90}{\includegraphics{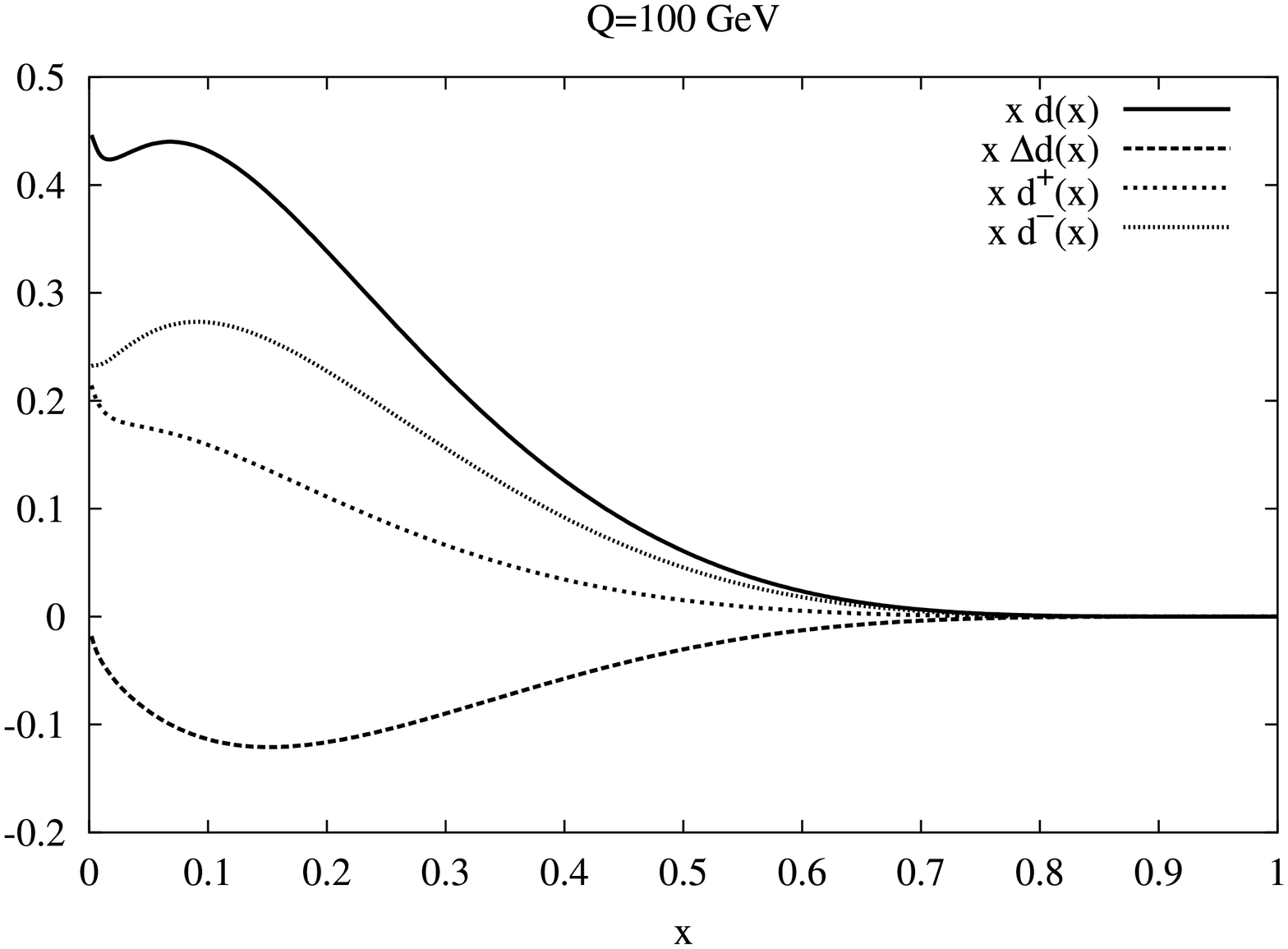}}} \par}
\caption{Various kinds of distributions of quark down at \protect\( Q=100\, \textrm{GeV}\protect \).}
\label{downs}
\end{figure}

\begin{figure}[tbh]
{\centering \resizebox*{12cm}{!}{\rotatebox{-90}{\includegraphics{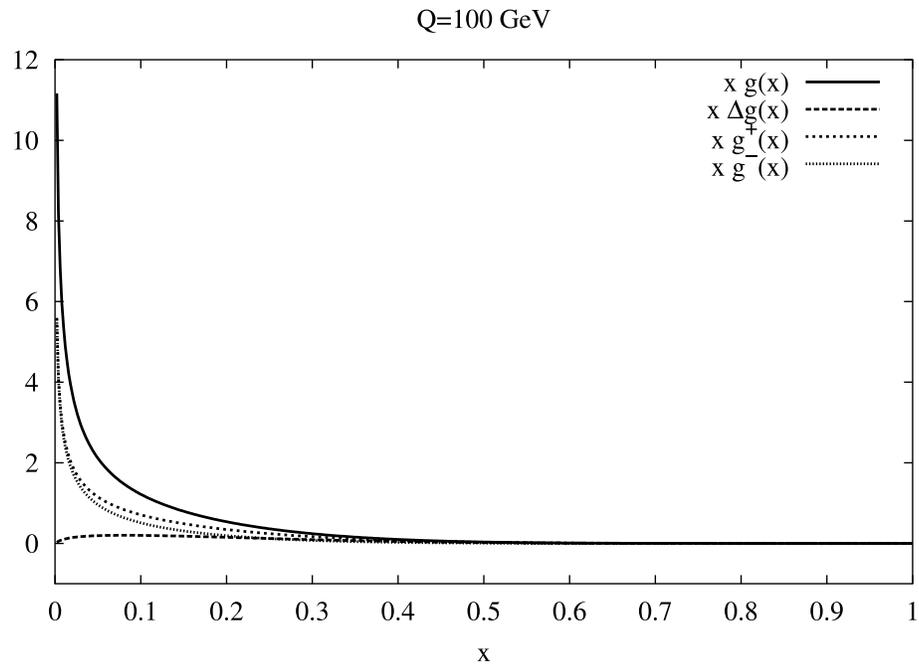}}} \par}

\caption{Various kinds of gluon distributions at \protect\( Q=100\, \textrm{GeV}\protect \).}
\label{gluons}
\end{figure}

\begin{figure}[tbh]
{\centering \resizebox*{12cm}{!}{\rotatebox{-90}{\includegraphics{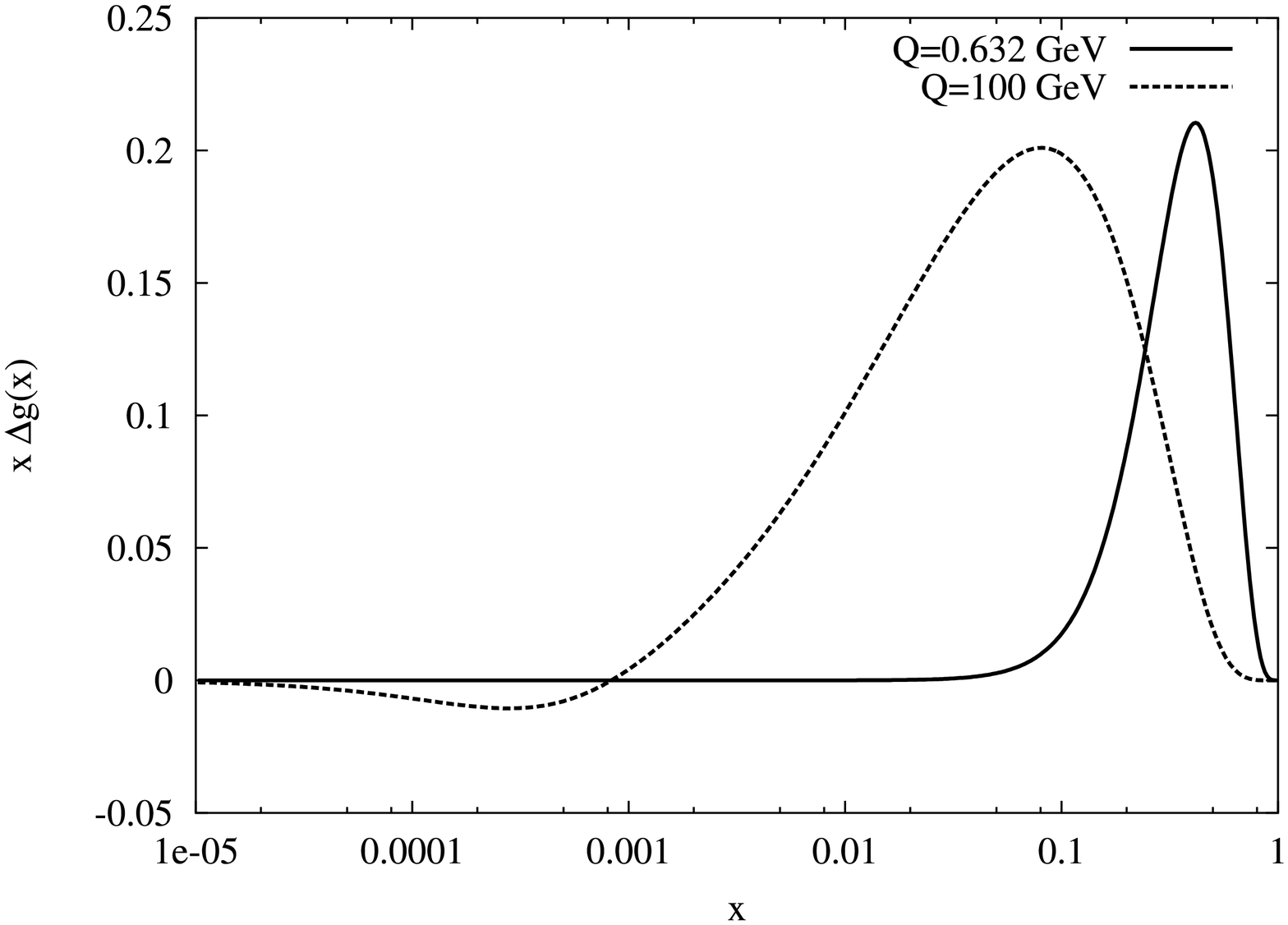}}} \par}

\caption{Small-x behaviour of $\Delta g$ at 100 GeV.}
\label{Dgluon}
\end{figure}

Finally in Figs. \ref{ups}, \ref{downs} and \ref{gluons} 
we plot simultaneously longitudinally polarized, unpolarized and 
helicity distributions for up quarks, down quarks and gluons 
at an intermediate factorization scale of 100 GeV, relevant
for experiments at RHIC. Notice that while $u^+$ and $u^-$ are positive 
and their difference $(\Delta u)$ is also positive, for down quarks the two helicity components are positive while their difference $(\Delta d)$ is negative. Gluons, in the model studied 
here, have a positive longitudinal polarization, and their helicity components are also positive. The positive and negative gluon helicities are plotted 
in two separate figures, Figs. \ref{gplus} and \ref{gminus}, while their 
difference, $\Delta g(x)$ is shown in Fig. \ref{Dgluon}. One can observe, at 
least in this model, a crossing at small-x in this distribution. 

We conclude that, at least for this set of boundary conditions, positivity 
of all the components holds to NLO, as expected.  

We also report three tables (\ref{edge1} -- \ref{edge3}) illustrating the (positive) numerical values of the 
contributions coming from the subtraction terms in the NLO kernels. Coefficients $A$ and $B$
refer to the subtraction terms $-\log(1- \Lambda ) A + B$ as explained in the section above.

\begin{table}
\begin{tabular}{|c|c|c|}
\hline
\( N_{f} \)&
\( A \)&
\( B \)\\
\hline
\hline
3&
12.5302&
12.1739\\
\hline
4&
10.9569&
10.6924\\
\hline
5&
9.3836&
9.2109\\
\hline
6&
7.8103&
7.7249\\
\hline
\end{tabular}
\caption{Coefficients $A$ and $B$  for $P^{(1)}_{NS,++}$\label{edge1}}

\end{table}

\begin{table}
\begin{tabular}{|c|c|c|}
\hline
\( N_{f} \)&
\( A \)&
\( B \)\\
\hline
\hline
3&
12.5302&
12.1739\\
\hline
4&
10.9569&
10.6924\\
\hline
5&
9.3836&
9.2109\\
\hline
6&
7.8103&
7.7249\\
\hline
\end{tabular}
\caption{Coefficients $A$ and $B$ as in Table \ref{edge1} for $P^{(1)}_{qq,++}$}
\end{table}

\begin{table}
\begin{tabular}{|c|c|c|}
\hline
\( N_{f} \)&
\( A \)&
\( B \)\\
\hline
\hline
3&
48.4555&
27.3912\\
\hline
4&
45.7889&
24.0579\\
\hline
5&
43.1222&
20.7245\\
\hline
6&
40.4555&
17.3912\\
\hline
\end{tabular}
\caption{Coefficients $A$ and $B$ as in Table \ref{edge1} for $P^{(1)}_{gg,++}$\label{edge3}}

\end{table}

\section{Regularizations}

The ``+'' plus form of the kernels and all the other forms introduced before, 
obtained by separating the 
contributions from the edge-point $(x=1)$ from those coming from the bulk $(0< x < 1)$
  are all equivalent, as we are going to show, 
with the understanding that a linear (unique) cutoff is used to regulate the divergences 
both at x=0 and at x=1. We focus here on the two possible sources of singularity, i.e. on 
$P_{qq}$ and on the $P_{gg}$ contributions, which require some attention. 
Let's start from the $P_{qq}$ case.   
We recall that ``+'' plus distributions are defined as  

\beq
\frac{1}{(1-x)_+}=\frac{\theta(1- x- \Lambda)}{1-x} -
\delta(1-x) \int_0^{1-\Lambda}\frac{dz}{1-z}
\eeq
with $\Lambda$ being a cutoff for the edge-point contribution. 

We will be using the relations 
\beqa
\int_x^1\frac{dy}{y}\delta(1-y) &=& 1 \nonumber \\
\int_0^1\frac{dz}{1-z}&=&\int_x^1\frac{dz}{1-z} - log(1-x)\nonumber \\
\int_x^1\frac{dy}{y} f(y)g(x/y)&=& \int_x^1\frac{dy}{y} f(x/y)g(y).
\label{refs}
\eeqa

Using the expressions above it is easy to obtain 
\beqa
\frac{1}{(1-x)_+}\otimes f(x)&=&\int_x^1\frac{dy}{y}\frac{1}{(1- x/y)_+} q(y) 
\nonumber \\
&=&\int_x^1\frac{dy}{y}\frac{1}{1- x/y}f(y) -\int_x^1\frac{dy}{y}
\delta(1-y)\int_0^1 \frac{dz}{z} \nonumber \\
&=&\int_x^1\frac{dy}{y}
\frac{ yf(y) - x f(x)}{y-x} +\log(1-x) f(x)\nonumber \\
\eeqa
which is eq. (\ref{simplerel}). If we remove the ``+'' distributions 
and adopt (implicitly) a cutoff regularization, we need special care. 
In the probabilistic version of the kernel, the handling of 
$P_{qq}\otimes q$ is rather straightforward 

\beqa
P_{qq}\otimes q(x) &=& C_F\int_x^1\frac{dy}{y}
\left( \frac{2}{1- x/y} -1- x/y\right)q(y)\nonumber \\
&& - C_F\int_x^1\frac{dy}{y}\delta(1-y)\int_0^1 dz'
\left( \frac{2}{1- z'} -1- z'\right) \nonumber \\
\eeqa
and using eqs. (\ref{refs}) we easily obtain 
\beqa
P_{qq}\otimes q(x) &=& 2 C_F \int_x^1\frac{dy}{y}
\frac{y q(y) - x q(x)}{y-x} + 2 C_F \log(1-x) q(x) \nonumber \\
 && - C_F\int_x^1\frac{dy}{y}\left( 1 + x/y\right)q(y) + \frac{3}{2}C_F q(x).
\eeqa
Now consider the convolution $P_{gg}\otimes g(x)$ in the Collins-Qiu form. 
We get 
\beqa 
P_{gg}\otimes g(x)&=& 2 C_A \int_x^1\frac{dy}{y}\frac{1}{1- x/y}g(y) 
\nonumber \\
&& + 2 C_A \int_x^1\frac{dy}{y}\left( x/y (1 - x/y) + \frac{1}{x/y} -2 
\right)g(y)
- g(x)\int_0^1\frac{dz} C_A \left( \frac{1}{z} 
+ \frac{1}{1-z}\right)
\nonumber \\
&& -\frac{1}{2}g(x)\int_0^1 dz \,\,2 C_A\left( z(1-z) - 2\right)
- n_f g(x)\int_0^1 dz \frac{1}{2}\left( z^2 + (1-z)^2\right).
\eeqa

There are some terms in the expression above that require some care. 
The appropriate regularization is 
\beq
\int_0^1\frac{dz}{z} + \int_0^1 \frac{dz}{1-z}\rightarrow 
I(\Lambda)= \int_\Lambda^1\frac{dz}{z} + \int_0^{1- \Lambda} \frac{dz}{1-z}.
\eeq

Observe also that 
\beq
 \int_\Lambda^1\frac{dz}{z} = \int_0^{1- \Lambda} \frac{dz}{1-z}= -\log \Lambda.
\eeq
Notice that in this regularization the singularity of $1/z$ at $z=0$ 
is traded for a singularity at z=1 in $1/(1-z)$. 
 It is then rather straightforward to show that 
\beqa
P_{gg}\otimes g(x) &=& 2 C_A 
\int_x^1\frac{dy}{y}
\frac{ yg(y) - x g(x)}{y-x} + 2 C_A\log(1-x) g(x) \nonumber \\
&& + 2 C_A \int_x^1\frac{dy}{y}\left( x/y ( 1 - x/y) 
+ \frac{1}{x/y} - 2 \right)
+ \frac{\beta_0}{2} g(x).
\eeqa

A final comment is due for the form of the sum rule 
\beq
\int_0^1 dz (z - \frac{1}{2})\gamma_{gg}=0
\label{gluonsr}
\eeq
that we need to check with the regularization given above.

The strategy to handle this expression is the same as before. 
We extract  all the $1/z$ and $1/(1-z)$ integration terms and use 
\beqa
\int_0^1 \frac{dz}{z} - \int_0^1\frac{dz}{1-z}
&\rightarrow & \int_\Lambda^1 \frac{dz}{z} - \int_0^{1- \Lambda}\frac{dz}{1-z}
\nonumber \\
&=& -\log \Lambda + \log \Lambda = 0 \nonumber \\
\eeqa
to eliminate the singularities at the boundaries 
$x=0, 1$ and verify eq. (\ref{gluonsr}).

\clearpage

\section{Conclusions}
We have discussed in detail some of the main features of the 
probabilistic approach to the DGLAP evolution in the helicity basis. 
Numerical results for the evolution of 
all the helicities have been provided, using a 
special algorithm, based in $x$-space. We have also illustrated some of the 
essential differences between the standard distributional 
form of the kernels and their probabilistic version, clarifying 
some issues connected to their regularization. Then we have turned to 
the probabilistic picture, stressing on the connection between 
the random walk approach to parton diffusion in $x$-space and the master 
form of the DGLAP equation. The link between the two descriptions 
has been discussed especially in the context of the Kramers-Moyal 
expansion. A Fokker-Planck approximation to the expansion 
has also been presented which may turn useful for the study 
of formal properties of the probabilistic evolution. We have also seen 
that positivity of the helicity distributions, to NLO, 
requires a numerical analysis, as already hinted in \cite{Teryaev}.  
Our study also validates the use of a very fast evolution algorithm, 
alternative to other standard algorithms based on Mellin algorithms, 
whose advantage is especially in the analysis of the evolution of nonforward 
parton distributions, as we will show elsewhere.

\chapter{An $x$-space analysis of evolution equations: Soffer's inequality and the\\nonforward evolution\label{chap:Soffer}}
\fancyhead[LO]{\nouppercase{Chapter 5. An $x$-space analysis of evolution equations: Soffer's inequality}}
We analyze the use of algorithms based in $x$-space for the 
solution of renormalization group equations of DGLAP-type 
and test their consistency by studying bounds 
among partons distributions - in our specific case Soffer's inequality and the perturbative behaviour of the nucleon tensor charge -  
to next-to-leading order in QCD.
 
We also comment on the (kinetic) proof of positivity 
of the evolution of $h_1$, using a kinetic analogy, along the lines of the previous chapter, and 
illustrate the extension of the algorithm to the evolution of generalized parton 
distributions. We prove positivity of the nonforward evolution in a special case 
and illustrate a Fokker-Planck approximation to it.

\section{Introduction} 
One of the most fascinating aspects of the structure of the nucleon
is the study of the distribution of spin among its constituents, a topic of 
remarkable conceptual complexity which has gained a lot of attention 
in recent years. This study is entirely based on the classification and on the 
phenomenological modeling of all the leading-twist parton distributions, used 
as building blocks for further investigations in hadronic physics. 

There are various theoretical 
ways to gather information on these non-local matrix elements. One among the various 
possibilities is to discover sum rules connecting moments of these distributions to other 
fundamental observables. Another possibility is to discover 
bounds - or inequalities - among them and use these results in the process of their modeling. 
There are various bounds that can be studied, particularly in the context of the 
new generalized parton dynamics typical of the skewed distributions 
\cite{Ji,Radyushkin}. 
All these relations can be analyzed in perturbation theory and studied using 
the Renormalization Group (RG), although a complete description of their perturbative dynamics 
is still missing. This study, we believe, 
may require considerable theoretical effort since it involves a global understanding both of the (older) forward (DGLAP) dynamics and of the generalized 
new dynamics encoded in the skewed distributions. 

In this context, a program aimed at the study of various bounds in perturbation theory using primarily a parton dynamics in $x$-space has been outlined \cite{CPC}. 
This requires accurate algorithms to solve the equations up to 
next-to-leading order (NLO). Also, underlying this type of description is, in many cases, a 
probabilistic approach \cite{Teryaev1} which has some interesting consequences worth 
of a closer look . In fact, the DGLAP equation, viewed as a probabilistic process, 
can be rewritten in a {\em master form} which is at the root of some interesting 
formal developments. In particular, a wide set of results, available from the theory 
of stochastic processes, find their way in the study of the evolution. 
We have elaborated on this issue in previous work \cite{CPC}
and proposed a Kramers-Moyal expansion of the DGLAP equation as an alternative 
way to describe its dynamics. Here, this analysis will be extended 
to the case of the nonforward evolution. 

With these objectives in mind, in this study we
test  $x$-space algorithms up to NLO and verify their accuracy using a stringent test: 
Soffer's inequality. As usual, we are bound to work with specific models of initial conditions.  
The implementations on which our analysis are based are general, 
with a varying flavour number $n_f$ at any threshold of intermediate quark mass 
in the evolution.  
Here, we address Soffer's inequality using an approach 
based on the notion of ``superdistributions'' 
\cite{Teryaev}, 
which are constructs designed to have a simple (positive) 
evolution thanks to the existence of an underlying master form \cite{Teryaev1,CPC}. 
The original motivation for using such a master form 
(also termed {\em kinetic} or {\em probabilistic}) 
to prove positivity has been presented in \cite{Teryaev}, 
while further extensions of these arguments have been presented in \cite{CPC}.  
In a final section we propose the extension of the evolution algorithm 
to the case of the skewed distributions, and illustrate its implementation 
in the nonsinglet case. As for the forward case, 
numerical tests of the inequality are performed for 
two different models. We show that even starting from a saturated inequality at the 
lowest evolution scale, the various models differ significantly even for a moderate final 
factorization scale of $Q=100$ GeV. Finally, we illustrate in another application the 
evolution of the tensor charge and show that, in the models considered, differences 
in the prediction of the tensor charge are large.

\section{Prelude to $x$-space: A Simple Proof of Positivity of $h_1$ to NLO} 
There are some nice features of the parton dynamics, at least in the leading logarithmic approximation (LO), when viewed in $x$-space, once a suitable ``master form'' of the 
parton evolution equations is identified.    

The existence of such a master form, as firstly shown by Teryaev, 
is a special feature of the evolution equation 
itself. The topic has been addressed before 
in LO \cite{Teryaev} and reanalyzed in more detail in
\cite{CPC} where, starting from a kinetic interpretation of the 
evolution, a differential equation obtained 
from the Kramers-Moyal expansion of the DGLAP equation 
has also been proposed. 

The arguments of refs.~\cite{Teryaev,CPC} 
are built around a form of the evolution equation 
which has a simple kinetic interpretation and is written 
in terms of transition probabilities constructed from the kernels. 
  
The strategy used, at least in leading order,  
to demonstrate the positivity of the LO evolution for 
special combinations of parton distributions 
$\Q_\pm$ \cite{Teryaev}, to be defined below, or the NLO evolution for $h_1$, 
which we are going to address, is based on some results of
ref.\cite{Teryaev}, briefly reviewed here, in order 
to be self-contained. 

 A master equation is typically given by 
\beq
\frac{\partial }{\partial \tau}f(x,\tau)=\int dx'\left(
w(x|x') f(x',\tau) -w(x'|x) f(x,\tau)\right) dx'
\label{masterforms}
\eeq

and if through some manipulations, a DGLAP equation 

\beq
\frac{d q(x,Q^2)}{d \log( Q^2)} = \int_x^1 \frac{dy}{y} P(x/y)q(y,Q^2),
\eeq
with kernels $P(x)$, is rewritten in such a way to resemble  
eq. (\ref{masterforms})  

\beq
\frac{d}{d \tau}q(x,\tau) = \int_x^1 dy \hat{P}\left(\frac{x}{y}\right)\frac{q(y,\tau)}{y}
-\int_0^x \frac{dy}{y}\hat{ P}\left(\frac{y}{x}\right)\frac{q(x,\tau)}{x},
\eeq
with a (positive) transition probability 
\beq
w(x|y)= \frac{\alpha_s}{2 \pi} \hat{P}(x/y)\frac{\theta(y > x)}{y}
\eeq
then positivity of the evolution is established. 

For equations of nonsinglet type, such 
as those evolving $q^{(-)}=q - \bar{q}$, the valence quark distribution, 
or $h_1$, the transverse spin distribution, 
this rewriting of the equation is possible, at least in LO. 
NLO proofs are, in general, impossible to construct by this method, 
since the kernels turn out, in many cases, to be negative. The only possible proof, in these cases, is just a numerical one, for suitable (positive) 
boundary conditions observed by the initial form of the parton distributions. 
Positivity of the evolution is then a result of a non obvious interplay between 
the various contributions to the kernels in various regions in $x$-space.  

In order to discuss the probabilistic version of the DGLAP equation it 
is convenient to separate the bulk contributions of the kernels $(x<1)$ from the 
edge point contributions at $x=1$. For this purpose 
we recall that the structure of the kernels is, in general, given by 
\beq
P(z) = \hat{P}(z) - \delta(1-z) \int_0^1 \hat{P}(z)\, dz,
 \eeq
where the bulk contributions $(z<1)$ and the edge point contributions 
$(\sim \delta(z-1))$ have been explicitly separated.
We focus on the transverse spin distributions as an example. 
With these prerequisites, 
proving the LO and NLO positivity of the transverse spin distributions 
is quite straightforward, but requires a numerical inspection of the transverse 
kernels. Since the evolutions for $\Delta_T q^{(\pm)}\equiv h^q_1$ are purely nonsinglet, 
diagonality in flavour of the subtraction terms $(\sim \int_0^x w(y|x)q(x,\tau))$ 
is satisfied, while the edge-point subtractions can be tested 
to be positive numerically. 
We illustrate the explicit construction of the master equation for $h_1$ in LO, since extensions to NLO of this construction are rather straightforward. 

In this case the LO kernel is given by 

\beqn
\Delta_{T}P^{(0)}_{qq}(x)= C_{F}\left[\frac{2}{(1-x)_{+}}-2 +\frac{3}{2}\delta(1-x)\right] 
\eeqn
and by some simple manipulations we 
can rewrite the corresponding evolution equation 
in a suitable master form. That this is possible is an elementary fact 
since the subtraction terms 
can be written as integrals of a positive function. For instance, 
a possibility is 
to choose the transition probabilities
\beqa
w_1[x|y] &=& \frac{C_F}{y}\left(\frac{2}{1- x/y} - 2 \right)
\theta(y>x) \theta(y<1)\nonumber \\
w_2[y|x] &=& \frac{C_F}{x} \left(\frac{2}{1- y/x} - \frac{3}{2}\right)
\theta(y > -x)\theta(y<0)
\nonumber \\
\eeqa
which reproduce the evolution equation for $h_1$ in master form

\beq
\frac{d h_1}{d \tau}= \int_0^1 dy w_1(x|y)h_1(y,\tau) 
-\int_0^1 dy w_2(y|x) h_1(x,\tau).
\label{masterix}
\eeq

\begin{figure}
{\centering \resizebox*{12cm}{!}{\rotatebox{-90}{\includegraphics{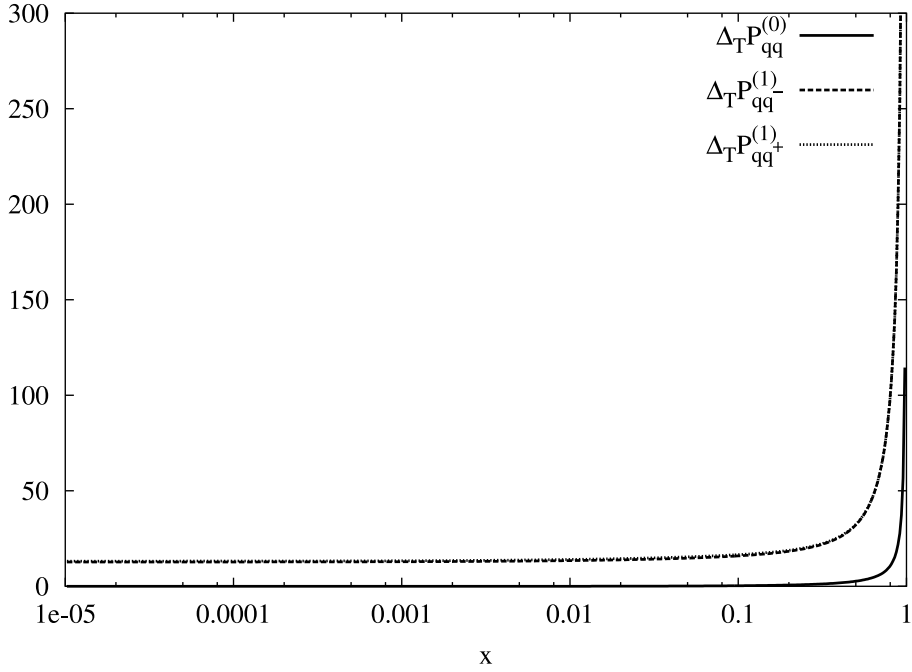}}} \par}
\caption{Plot of the transverse kernels.}
\label{transversekernels}
\end{figure}

The NLO proof of positivity is also rather straightforward. 
For this purpose we have analyzed numerically the behaviour of the NLO kernels both 
in their bulk region and at the edge-point. 
We show in Table 1 of Appendix B results 
for the edge point contributions to NLO for both of 
the $\Delta_T P^{(1)}_\pm$ components, 
which are numerically the same.
There we have organized these terms in the form $\sim C\delta(1-x)$ with
\beq
C=-\log(1- \Lambda) A + B 
, 
\eeq
with A and B being numerical coefficients depending on the number 
of flavours included in the kernels. 
The (diverging) logarithmic contribution ($\sim \int_0^\Lambda dz/(1-z)$) 
have been regulated by a cutoff. This divergence 
in the convolution cancels when these terms are combined with the divergence at 
$x=1$ of the first term of the master equation (\ref{masterix}) 
for all the relevant components 
containing ``+'' distributions. As for the 
bulk contributions $(x<1)$, positivity up to NLO of the transverse kernels 
is shown numerically in Fig. (\ref{transversekernels}). 
All the conditions of positivity are therefore satisfied and therefore 
the $\Delta_{T\pm}q$ distributions evolve positively up to NLO. 
The existence of a master form of the equation is then guaranteed.

Notice that the NLO positivity of $\Delta_{T\pm}q$ implies positivity of the
nucleon tensor charge \cite{JJ}
\beq
\delta q\equiv\int_0^1 dx \left( h_1^q(x) - h_1^{\bar{q}(x)}\right)
\eeq
for each separate flavour for positive initial conditions. 
As we have just shown, this proof of positivity is very short, as far as one 
can check numerically that both components of eq.(\ref{masterix}) 
are positive. 

\begin{table}
\begin{tabular}{|c|c|c|}
\hline
\( n_{f} \)&
\( A \)&
\( B \)\\
\hline
\hline
3&
12.5302&
12.1739\\
\hline
4&
10.9569&
10.6924\\
\hline
5& 
9.3836&
9.2109\\
\hline
6&
7.8103&
7.7249\\
\hline
\end{tabular}

\caption{Coefficients $A$ and $B$ for various flavor numbers, to NLO 
for $\Delta_T P_{qq, \pm}$} 
\end{table}

\section{Soffer's inequality}
Numerical tests of Soffer's inequality can be performed 
either in moment space or, as we are going to illustrate 
in the next section, directly in $x$-space, using suitable 
algorithms to capture the perturbative nature of the evolution.  
We recall that Soffer's inequality
\beq
|h_1(x)| < q^+(x)
\eeq
sets a bound on the transverse spin distribution $h_1(x)$ in terms of the
components of the positive helicity component of the quarks, for a given flavour.
An original proof of Soffer's inequality 
in LO has been discussed in ref.\cite{Barone}, while 
in \cite{Teryaev} an alternative proof was presented, based 
on a kinetic interpretation of the evolution equations. 

We recall that $h_1$, also denoted by the symbol 
\begin{equation}
\Delta _{T}q(x,Q^{2})\equiv q^{\uparrow }(x,Q^{2})-q^{\downarrow }(x,Q^{2}),
\end{equation} 
has the property
of being purely nonsinglet and of appearing at leading twist. It is
identifiable in transversely polarized 
hadron-hadron collisions and not in Deep Inelastic Scattering (from now on we will 
omit sometimes the $x$-dependence in the kernels and in the distributions when obvious).
In the following we will use interchangeably the notations $h_1\equiv h_1^q$ 
and $\Delta_T q$ to 
denote the transverse asymmetries. We introduce also the combinations 
\beqa
\Delta_T(q + \bar{q}) &=& h_1^q + h_1^{\bar{q}} \nonumber \\
\Delta_T q^{(-)}=\Delta_T(q - \bar{q}) &=& h_1^q - h_1^{\bar{q}} \nonumber \\
\Delta_T q^{(+)} &=& \sum_i \Delta_T(q_i + \bar{q}_i) \nonumber \\
\eeqa
where we sum over the flavor index $(i)$, and we have introduced singlet and nonsinglet 
contributions for distributions of fixed helicities 
\beqa
q_+^{(+)}&=&\sum_i\left( q_{+ i} + \bar{q}_{+ i}\right)\nonumber \\
q_+^{(-)}&=& q_{+ i} -\bar{q}_{+ i}\equiv \Sigma. \nonumber \\
\eeqa
In our analysis we solve all the equations in the helicity basis and reconstruct 
the various helicities after separating singlet and nonsinglet sectors. 
We mention that 
the nonsinglet sector is now given by a set of 2 equations, each involving 
$\pm$ helicities and the singlet sector is given by a 4-by-4 matrix.   

In the singlet sector we have 

\begin{eqnarray}
{dq_+^{(+)} \over{dt}}=
{\alpha_s \over {2 \pi}} (P_{++}^{qq}\otimes q_+^{(+)}+
P_{+-}^{qq} \otimes q_-^{(-)}  \nonumber \\
+P_{++}^{qG} \otimes G_++
P_{+-}^{qG} \otimes G_-),
\nonumber \\
{dq_-^{(+)}(x) \over{dt}}=
{\alpha_s \over {2 \pi}} (P_{+-} \otimes q_+^{(+)} +
P_{++}  \otimes q_-^{(+)} \nonumber \\
+P_{+-}^{qG} \otimes G_+ +
P_{++}^{qG} \otimes G_-),  \nonumber \\
{dG_+(x) \over{dt}}=
{\alpha_s \over {2 \pi}} (P_{++}^{Gq} \otimes q_+^{(+)}+
P_{+-}^{Gq} \otimes q_-^{(+)} \nonumber \\
+P_{++}^{GG}\otimes G_+ +
P_{+-}^{GG} \otimes G_-),  \nonumber \\
{dG_-(x) \over{dt}}=
{\alpha_s \over {2 \pi}} (P_{+-}^{Gq} \otimes q_+^{(+)} +
P_{++}^{Gq} \otimes q_-^{(+)} \nonumber \\
+P_{+-}^{GG} \otimes G_+ +
P_{++}^{GG} \otimes G_-).
\end{eqnarray}

while the nonsinglet (valence) analogue of this equation is also easy to
write down
\begin{eqnarray}
{dq_{+ i}^{(-)}(x) \over{dt}}=
{\alpha_s \over {2 \pi}} (P^{NS}_{++} \otimes q_{+ i}^{(-)}+
P^{NS}_{+-} \otimes q_{-}^{(-)}(y)), \nonumber \\
{dq_{- i}^{(-)}(x) \over{dt}}=
{\alpha_s \over {2 \pi}} (P^{NS}_{+-} \otimes q_{+}^{(-)}+
P^{NS}_{++} \otimes q_{- i}^{(-)}).
\end{eqnarray}
Above, $i$ is the flavor index, $(\pm)$ indicate $q\pm \bar{q}$ components and the lower subscript $\pm$ stands for the helicity.

Similarly to the unpolarized case the flavour reconstruction is done by adding 
two additional equations for each flavour in the helicity $\pm$
\beq
\chi_{\pm,i}= q_{\pm i}^{(+)}- \frac{1}{n_f}q^{(+)}_\pm
\eeq
whose evolution is given by 
\beqa
{d \chi_{+ i}^{(-)}(x) \over{dt}} &=&
{\alpha_s \over {2 \pi}} (P^{NS}_{++} \otimes \chi_{+ i} +
P^{NS}_{+-} \otimes \chi_{- i}), \nonumber \\
{d \chi_{- i} (x) \over{dt}} &=&
{\alpha_s \over {2 \pi}} (P^{NS}_{+-} \otimes \chi_{+ i} +
P^{NS}_{++} \otimes \chi_{- i}). \nonumber \\
\label{h11}
\end{eqnarray}

The reconstruction of the various contributions in flavour space 
for the two helicities is finally done 
using the linear combinations 
\beq
q_{\pm i}=\frac{1}{2}\left( q_{\pm i}^{(-)} + \chi_{\pm i} +\frac{1}{n_f}q_{\pm}^{(+)}\right).
\eeq

We will be needing these equations below when we present
a proof of positivity up to LO, and we will thereafter proceed with a NLO implementation of these and other evolution equations. For this we will be needing some more notations. 

We recall that the following relations are also true to all orders 
\beqa
P(x) &=&\frac{1}{2}\left( P_{++}(x) + P_{+-}(x)\right)\nonumber \\
&=&\frac{1}{2}\left( P_{--}(x) + P_{-+}(x)\right)\nonumber 
\eeqa
between polarized and unpolarized $(P)$ kernels 
and 
\beq
P_{++}(x) =  P_{--}(x),\,\,\,P_{-+}(x)=P_{+-}(x)
\eeq
relating unpolarized kernels to longitudinally polarized ones. 
Generically, the kernels of various type are expanded up to NLO as 
\beq
P(x)= \frac{\alpha_s}{2 \pi} P^{(0)}(x) + \left(\frac{\alpha_s}{2 \pi}\right)^2 P^{(1)}(x),
\eeq
and specifically, in the transverse case we have

\begin{eqnarray} \label{pm}
\Delta_T P_{qq,\pm}^{(1)} &\equiv& \Delta_T P_{qq}^{(1)} \pm \Delta_T 
P_{q\bar{q}}^{(1)} \; , \\
\end{eqnarray}
with the corresponding evolution equations 

\begin{equation} \label{evol3}
\frac{d}{d\ln Q^2} \Delta_T q_{\pm} (Q^2) = \Delta_T P_{qq,\pm} 
(\alpha_s (Q^2))\otimes \Delta_T q_\pm (Q^2) \; .
\end{equation}

We also recall that the kernels in the helicity basis in LO are given by 
\beqa
P_{NS\pm,++}^{(0)} &=&P_{qq, ++}^{(0)}=P_{qq}^{(0)}\nonumber \\
P_{qq,+-}^{(0)}&=&P_{qq,-+}^{(0)}= 0\nonumber \\
P_{qg,++}^{(0)}&=& n_f x^2\nonumber \\
P_{qg,+-}&=& P_{qg,-+}= n_f(x-1)^2 \nonumber \\
P_{gq,++}&=& P_{gq,--}=C_F\frac{1}{x}\nonumber \\ 
P_{gg,++}^{(0)}&=&P_{gg,++}^{(0)}= N_c
\left(\frac{2}{(1-x)_+} +\frac{1}{x} -1 -x - x^2 \right) +{\beta_0}\delta(1-x) \nonumber \\
P_{gg,+-}^{(0)}&=& N_c
\left( 3 x +\frac{1}{x} -3 - x^2 \right). 
\eeqa

An inequality, such as Soffer's inequality, can be stated as positivity condition 
for suitable linear combinations of parton distributions \cite{Teryaev} 
and this condition can be analyzed - as we have just shown 
for the $h_1$ case -  in a most direct way using the master form.

For this purpose consider the linear valence combinations
\beqa
\Q_+ &=& q_+ + h_1 \nonumber \\
\Q_-  &=& q_+ - h_1 \nonumber \\
\eeqa
which are termed ``superdistributions'' in ref.\cite{Teryaev}. Notice that a proof 
of positivity of the $\Q$ distributions is equivalent to verify Soffer's inequality. 
However, given the mixing of singlet and nonsinglet sectors, the analysis of 
the master form is, in this case, more complex. As we have just mentioned, what can spoil the proof of 
positivity, in general, is the negativity of the kernels to higher order. 
We anticipate here the result that we will illustrate below where we show 
that a LO proof of the positivity of the evolution for $\Q$ can be established using 
kinetic arguments, being the kernels are positive at this order. However 
we find 
that the NLO kernels do not satisfy 
this condition. 
In any case, let's see how the identification of such master form proceeds in general. 
We find useful to illustrate the result using the separation between singlet and nonsinglet 
sectors. In this case we introduce the combinations

\beqa
\Q_\pm^{(-)} &=& q_+^{(-)} \pm h_1^{(-)} \nonumber \\
\Q_\pm^{(+)} &=& q_+^{(+)} \pm h_1^{(+)} \nonumber \\
\label{separation}
\eeqa
with $h_1^{(\pm)}\equiv \Delta_T q^{(\pm)}$.  

Differentiating these two linear combinations  (\ref{separation}) we get
\beqa
\frac{d \Q_\pm^{(-)}}{d\log(Q^2)}= P^{NS}_{++} q_+^{(-)} 
+ P^{NS}_{+ -} q_-^{(-)} \pm P_T h_1^{(-)} \nonumber \\
\eeqa
which can be rewritten as
\beqa
\frac{d \Q_+^{(-)}}{d\log(Q^2)} &=& \frac{1}{2}\left(P_{++}^{(-)} + P_T^{(-)}\right)\Q_+^{(-)}
 + \frac{1}{2}\left(P_{++}^{(-)} - P_T^{(-)}\right)\Q_-^{(-)} + P_{+ -}^{(-)}q_-^{(-)} \nonumber \\
\frac{d \Q_+^{(-)}}{d\log(Q^2)} &=& \frac{1}{2}\left(P_{++}^{(-)} - P_T^{(-)}\right)\Q_+^{(-)}
+ \frac{1}{2}\left(P_{++}^{(-)} + P_T\right)^{(-)}\Q_-^{(-)} + P_{+ -}^{(-)}q_-^{(-)} \nonumber \\
\eeqa
with $P^{(-)}\equiv P^{NS}$ being the nonsinglet (NS) kernel. 

At this point we define the linear combinations
\beqa
{\bar{P}^Q}_{+\pm}= \frac{1}{2}\left(P_{++} \pm P_T\right)
\eeqa
and rewrite the equations above as
\beqa
\frac{d \Q_+ i}{d\log(Q^2)} &=& \bar{P}^Q_{ ++}\Q_{i+}
 + \bar{P}^Q_{+-}\Q_{i-} + P^{qq}_{+ -}q_{i-} \nonumber \\
\frac{d \Q_{i+}}{d\log(Q^2)} &=& \bar{P}^Q_{+-}\Q_{i+}
 + \bar{P}^Q_{++}\Q_{i-} + P_{+ -}^{qq}q_{i-} \nonumber \\
\label{pos1}
\eeqa
where we have reintroduced $i$ as a flavour index.
From this form of the equations it is easy to establish the leading order positivity of the evolution, after checking the positivity of the kernel and the existence of a master form.

The second nonsinglet sector is defined via the variables 

\beq
\chi_{i\pm}=q_{i \pm}^{(+)} - \frac{1}{n_f}q_{i\pm}^{(+)}
\eeq
which evolve as nonsinglets 
and the two additional distributions 
\beqa
\Q_{\chi i,\pm}= \chi_{i +} \pm h_1^{i(+)}.
\eeqa
Also in this case we introduce  the kernels 
\beqa
{\bar{P}^{Q_\chi}}_{+\pm} &=& \frac{1}{2}\left(P_{++} \pm 
\Delta_T P^{(+)}\right)
\eeqa
to obtain the evolutions 
\beqa
\frac{d \Q_{\chi i+}}{d\log(Q^2)} &=& \bar{P}^{Q_\chi}_{ ++}\Q_{\chi i+}
 + \bar{P}^{Q_\chi}_{+-}\Q_{\chi i-} + P^{qq}_{+ -}\chi_{i-} \nonumber \\
\frac{d \Q_{i+}}{d\log(Q^2)} &=& \bar{P}^Q_{\chi +-}\Q_{\chi i+}
 + \bar{P}^{Q_\chi}_{++}\Q_{\chi i-} + P_{+ -}^{qq}\chi_{i-}. \nonumber \\
\label{pos12}
\eeqa

For the singlet sector, we simply define $Q_+^{(+)}=q^{(+)}$, and the 
corresponding evolution is similar to the singlet equation of the helicity basis. 
Using the equations above, 
the distributions $\Q_{i\pm}$ are then reconstructed as 
\beq
\Q_{i\pm} = \frac{1}{2}\left(\Q_{i \pm}^{(-)} + 
\Q_{\chi i \pm}^{(-)} + \frac{1}{n_f}Q_+^{(+)}\right)
\eeq
and result positive for any flavour if the addends are positive as well. 
However, as we have just mentioned, positivity of all the kernels introduced above is easy to check numerically to LO, together with their diagonality in flavour which guarantees the existence 
of a master form.  

As an example, consider the LO evolution of $\Q\pm$. 
The proof of positivity is a simple consequence of the structure of eq.~(\ref{pos1}).
In fact the edge-point
contributions appear only in $P^Q_{++}$, i.e. they are diagonal in the evolution of
$\Q_{\pm}$. The inhomogeneous terms on the right hand side of (\ref{pos1}), proportional to
$q_-$ are are harmless, since the $P_{+-}$ kernel has no edge-point contributions. Therefore under 1) diagonality in flavour of the subtraction terms and 
2) positivity of first and second term 
(transition probabilities) we can have positivity of the evolution. A 
refined arguments to support this claim has been presented in \cite{CPC}.    

\begin{figure}
{\centering \resizebox*{12cm}{!}{\rotatebox{-90}{\includegraphics{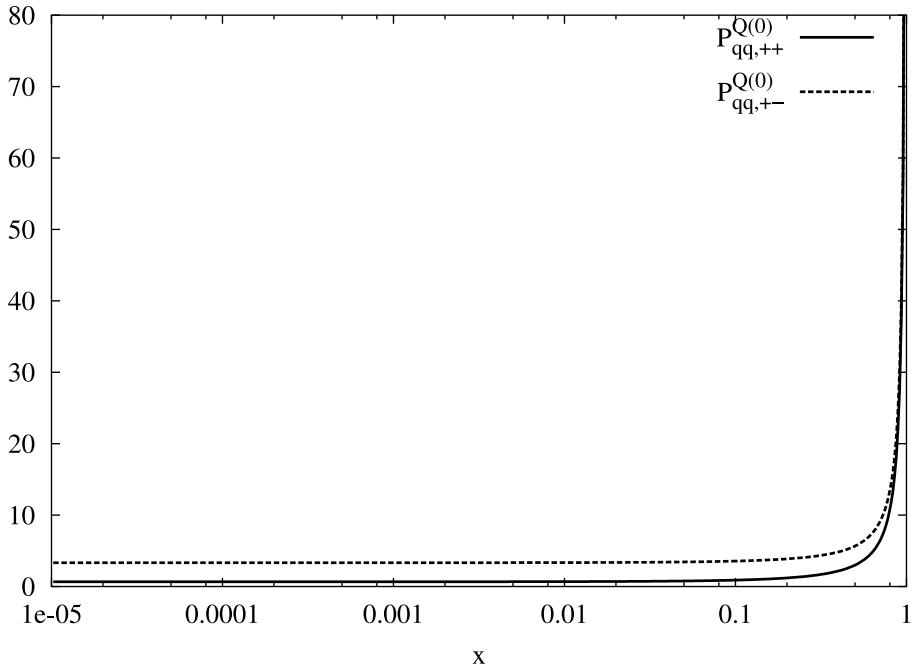}}} \par}
\caption{Plot of the LO kernels for the $\Q$ distributions}
\label{Qkernels}
\end{figure}

\begin{figure}
{\centering \resizebox*{12cm}{!}{\rotatebox{-90}{\includegraphics{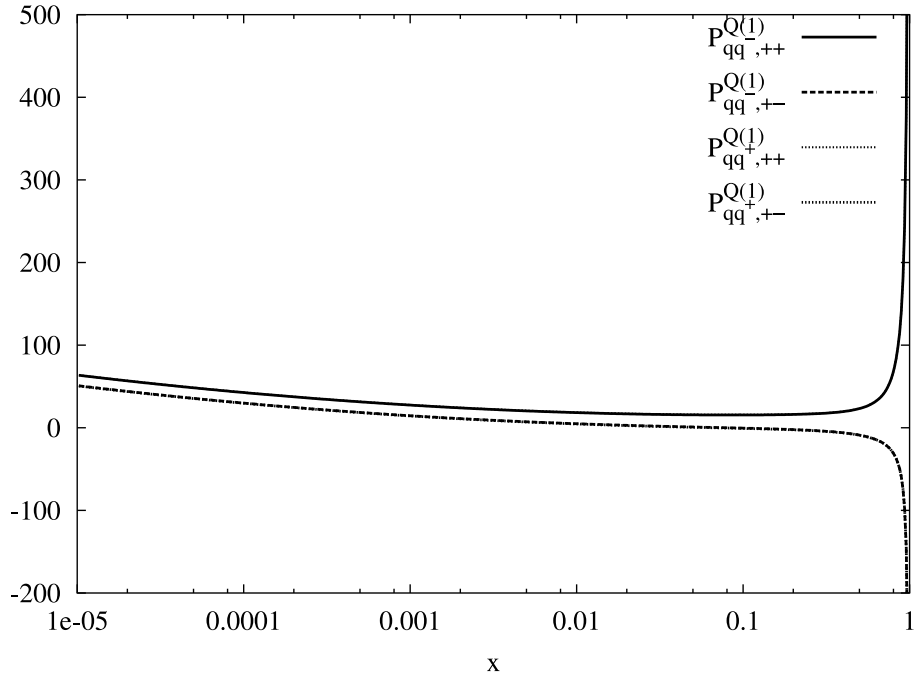}}} \par}
\caption{Plot of the NLO kernels for the $\Q$ distributions, 
showing a negative behaviour at large $x$}
\label{QNLOkernels}
\end{figure}

This construction is not valid to NLO. In fact, while the features of flavour 
diagonality of the 
master equation  are satisfied, the transition probabilities $w(x,y)$ 
are not positive in the whole $x,y$ range. The existence 
of a crossing from positive to negative values in $P^{\Q}_{++}$ 
can, in fact, be established quite easily using a numerical analysis. We illustrate in Figs. 
(\ref{Qkernels}) and (\ref{QNLOkernels}) plots of the $\Q$ kernels at LO and NLO, 
showing that, at NLO, the requirement of positivity of some components is violated.  
The limitations of this sort of proofs -based on kinetic arguments- are strictly 
linked to the positivity of the transition probabilities once a master form of the 
equation is identified. 
\section{ An $x$-space Expansion}
We have seen that NLO proofs of positivity, can be -at least partially- obtained 
only for suitable sets of boundary conditions. To this purpose, we choose to 
investigate the numerical behaviour of the solution using $x$-space 
based algorithms which need to be tested up to NLO.   

Our study validates a method which can be used to solve evolution equations 
with accuracy in leading and in next-to-leading order. The method is entirely 
based on an expansion \cite{Rossi} used in the context of spin physics \cite{Gordon}
and in supersymmetry \cite{Coriano}. An interesting feature of the
expansion, once combined with Soffer's inequality, is to generate an infinite set of 
relations among the scale invariant coefficients $(A_n, B_n)$ which characterize it. 

In this approach, the NLO expansion of the distributions in the DGLAP equation is generically given by
\beq
f(x,Q^2)=\sum_{n=0}^{\infty} \frac{A_n(x)}{n!}\log^n
\left(\frac{\alpha(Q^2)}{\alpha(Q_0^2)}\right) +
\alpha(Q^2)\sum_{n=0}^\infty \frac{B_n(x)}{n!}\log^n
\left(\frac{\alpha(Q^2)}{\alpha(Q_0^2)}\right)
\label{expansion}
 \eeq
where, to simplify the notation, we assume a short-hand matrix notation for all the convolution products.
Therefore $f(x,Q^2)$ stands for a vector having as components 
any of the helicities of the various flavours  $(\Q_{\pm},q_\pm,G_\pm)$.

If we introduce Rossi's expansion for $h_1$, $q_+$, and the linear combinations $\Q_\pm$ (in short form)

\beqa
h_1 &\sim&\left(A^{h}_n,B^{h +}_n\right)\nonumber \\
q_\pm &\sim&\left(A^{q_\pm}_n,B^{q_\pm}_n\right)\nonumber \\
\Q_\pm &\sim&\left(A^{Q +}_n,B^{Q +}_n\right) \nonumber \\
\eeqa

we easily get the inequalities
\beq
(-1)^n\left(A^{q_+}_n + A^{h}_n\right) >0
\eeq
and
\beq
(-1)^n\left(A^{q_+}_n - A^{h}_n\right) >0
\eeq
valid to leading order,which we can check numerically. 
Notice that the signature factor has to be included due to the alternation
in sign of the expansion.
To next to leading order we obtain 

\begin{eqnarray}
(-1)^{n+1} \left(A^{q_+}_n(x) +\alpha(Q^2) B^{q_+}_n (x)\right)\ & < & (-1)^n \left(A^{h}_n(x)\,+\alpha(Q^2) B^{h}_n (x)\right)\nonumber \\
 & < & (-1)^n \left(A^{q_+}_n(x) +\alpha(Q^2) 
B^{q_+}_n (x)\right)\end{eqnarray}

valid for $n\geq 1$, obtained after identification of the corresponding 
logarithmic powers $\log\left(\alpha(Q^2)\right)$ at any $Q$. 
In general, one can assume a saturation of the inequality at the initial evolution scale 
\beq
\Q_-(x,Q_0^2)=h_1(x,Q_0^2) -\frac{1}{2}q_+(x,Q_0^2)=0. 
\eeq
This initial condition has been evolved in $Q$ solving the equations 
for the $\Q_\pm$ distributions  to NLO.

\section{Nonforward Extensions}
In this section we finally discuss the nonforward extension of the evolution 
algorithm. In the case of nonforward distributions a second scaling parameter $\zeta$ 
controls the asymmetry between the initial and the final nucleon momentum in the 
deeply virtual limit of nucleon Compton scattering. The solution of the evolution 
equations, in this case, are known in operatorial form. Single and double parton distributions are obtained sandwiching the operatorial solution with 4 possible types of initial/final states $<p|...|p>, <p|...|0>, <p'|...|p>$, corresponding, respectively, 
to the case of diagonal parton distributions, distribution amplitudes and, in the latter 
case, skewed and double parton distributions \cite{Radyushkin}. Here we will simply analyze the nonsinglet case and discuss the extension of the forward algorithm to this more general case. 
Therefore, given the off-forward distributions $H_q(x,\xi)$, in Ji's notation, 
we set up the expansion 
\beq 
H_q(x,\xi)=\sum_{k=0}^{\infty} \frac{A_k(x,\xi)}{k!}\log^k
\left(\frac{\alpha(Q^2)}{\alpha(Q_0^2)}\right) +
\alpha(Q^2)\sum_{k=0}^\infty \frac{B_k(x,\xi)}{k!}\log^k
\left(\frac{\alpha(Q^2)}{\alpha(Q_0^2)}\right),
\label{expansionx}
 \eeq
which is the natural extension of the forward algorithm discussed 
in the previous sections. We recall that in the light-cone gauge $H(x,\xi)$ 
is defined as 
\beq
H_q(x,\xi,\Delta^2))= \frac{1}{2}\int \frac{dy^-}{2 \pi}e^{-i x \bar{P}^+y^-}
\langle P'| \bar{\psi}_q(0,\frac{y^-}{2},{\bf 0_\perp})
\frac{1}{2}\gamma^+ \psi_q (0,\frac{y^-}{2},{\bf 0_\perp})|P\rangle
\eeq
with $\Delta= P' - P$, $\bar{P}^+=1/2(P + \bar{P})$ \cite{Ji} (symmetric choice) and 
$\xi \bar{P}=1/2\,\,\Delta^+$.

This distribution describes for $x>\xi$ and $x < -\xi$ the DGLAP-type region for the quark and the antiquark distribution respectively, and the ERBL 
\cite{ERBL} (see also \cite{CL} for an overview) distribution amplitude 
for $-\xi <x < \xi$. In the following we will omit the $\Delta$ dependence from $H_q$.  

Again, once we insert the ansatz (\ref{expansionx}) into the evolution equations we obtain an infinite set of recursion relations which we can solve numerically. In LO, it is rather simple to relate the Gegenbauer moments of the skewed distributions and those of the generalized 
scaling coefficients $A_n$.
We recall that in the 
nonforward evolution, the multiplicatively renormalizable operators appearing 
in the light cone expansion are given in terms of Gegenbauer polynomials 
\cite{Radyushkin}. 
The Gegenbauer moments of the coefficients $A_n$ of our expansion (\ref{expansionx}) 
can be easily related to those of the off-forward distribution 

\beq
C_n(\xi, Q^2) =\zeta^n\int_{-1}^{1} C_n^{3/2}(z/\xi)H(z,\xi, Q^2) dz. 
\eeq  
The evolution of these moments is rather simple  
\beq
C_n(\zeta, Q^2)=C_n(\zeta, Q_0^2) \left(\frac{\alpha(Q^2)}{\alpha(Q_0^2)}\right)^{\gamma_n/\beta_0} 
\eeq
with
\beq
\gamma_n= C_F \left(\frac{1}{2} - \frac{1}{(n+1)(n+2)} + 2 \sum_{j=2}^{n+1}\frac{1}{j}\right)
\eeq
being the nonsinglet anomalous dimensions. If we define the Gegenbauer moments of 
our expansion 
\beq
A_k^{(n)}(\xi, Q^2) =\xi^n\int_{-1}^{1} C_n^{3/2}(z/\xi)H(z\,\xi, Q^2) dz 
\eeq  
we can relate the moments of the two expansions as 
\beq
A_k^{(n)}(\xi)=C_n(\zeta,Q_0^2)\left( \frac{\gamma_n}{\beta_0}\right)^k. 
\eeq
Notice that expansions similar to (\ref{expansionx}) hold also for other choices 
of kinematical variables, such as those defining the nonforward distributions 
\cite{Radyushkin}, where the t-channel longitudinal momentum exchange $\Delta^+$ is related to the longitudinal momentum of the incoming nucleon as $\Delta=\zeta P$. We recall 
that $H_q(x.\xi)$ as defined in \cite{Ji} can be mapped 
into two independent distributions $\hat{\mathcal{F}}_q(X,\zeta)$ and 
$\hat{\mathcal{F}}_{\bar{q}}(X,\zeta)$ through the mappings \cite{Golec}
\beqa
X_1 &=& \frac{(x_1+ \xi)}{(1 +\xi)} \nonumber \\
X_2 &=& \frac{\xi - x_2}{(1 +\xi)} \nonumber \\
\xi &=&\zeta/(2 - \zeta) \nonumber \\
\mathcal{F}_q(X_1,\zeta) &=& \frac{1}{1 - \zeta/2}H_q(x_1,\xi) \nonumber \\
\mathcal{F}_{\bar{q}}(X_2,\zeta) &=& \frac{-1}{1 - \zeta/2}H_q(x_2,\xi),\nonumber \\
\eeqa
in which the interval  $-1 \leq x \leq 1$ is split 
into two coverings, partially overlapping (for $-\xi\leq x \leq \xi$, or ERBL region) 
in terms of the two variables $-\xi \leq x_1 \leq 1$ ($0\leq X_1 \leq 1$) and 
$-1 \leq x_2 \leq \xi$ ($0\leq X_2 \leq 1$). In this new parameterization, the 
momentum fraction carried by the emitted quark is $X$, 
as in the case of ordinary distributions, where it is parametrized by Bjorken $x$. 
For definitess, we focus here on the DGLAP-like $(X> \zeta)$ region of 
the nonsinglet evolution. The nonsinglet kernel is given in this case by 
$(x\equiv X)$
\beqa
P_\zeta(x,\zeta)=\frac{\alpha}{\pi}C_F\left( \frac{1}{y - x}\left[1 + \frac{x x'}{y y'}\right] -
\delta(x - y)\int_0^1 dz \frac{1 + z^2}{1 - z}\right),
\label{nfkernel}
\eeqa
 we introduce a LO ansatz 
\beq 
\mathcal{F}_q(x,\zeta)=\sum_{k=0}^{\infty} \frac{\mathcal{A}_k(x,\zeta)}{k!}\log^k
\left(\frac{\alpha(Q^2)}{\alpha(Q_0^2)}\right)
\eeq
and insert it into the evolution of this region to obtain the very simple recursion relations 
\beqa
\A_{n+1}(X,\zeta) &=& -\frac{2}{\beta_0} C_F 
\int_X^1 \frac{dy}{y} \frac{y \A_n(y,\zeta) -
x \A_n(X,\zeta)}{y-X}  -\frac{2}{\beta_0} C_F 
\int_X^1 \frac{dy (X-\zeta)}{y(y-\zeta)} \frac{\left(y \A_n(X,\zeta) -X \A_n(y,\zeta)\right)}{y-X} 
\nonumber \\
&& -\frac{2}{\beta_0}C_F\hat{\A}_n(X,\zeta)\left[\frac{3}{2}+\ln\frac{(1 - X)^2(1 - x/\zeta)}{1 - \zeta}\right].
\eeqa
The recursion relations can be easily reduced to a weighted sum of contributions in which $\zeta$ is a spectator parameter. Here we will not make a complete implementation, but we will illustrate 
in an appendix the general strategy to be followed. There we show a very accurate analytical 
method to evaluate the logarithms generated by the expansion without having 
to rely on brute-force computations.  
 
\section{Positivity of the nonsinglet Evolution}
Positivity of the nonsinglet evolution is a simple consequence of the master-form associated to the 
nonforward kernel (\ref{nfkernel}). As we have already emphasized above, 
positivity of the initial conditions are sufficient to guarantee a positivity of the 
solution at any scale $Q$. The master-form of the equation allows to reinterpret the parton dynamics as a random walk biased toward small-x values as $\tau=\log(Q^2)$ 
increases.

In the nonforward case the identification of a transition probability 
for the random walk \cite{CPC} associated with the evolution of the parton distribution is obtained 
via the nonforward transition probability 
\beqa
w_\zeta(x|y) &=&\frac{\alpha}{\pi}C_F \frac{1}{y- x}\left[1 + \frac{x}{y}\frac{(x-\zeta)}{y - \zeta}\right]
\theta(y>x)\nonumber \\
w'_\zeta(y|x)&=&\frac{\alpha}{\pi}C_F \frac{x^2 + y^2}{x^2(x - y)}\theta(y<x)
\eeqa
and the corresponding master equation is given by 
\beq
\frac{d \mathcal{F}_q}{d\tau}=\int_x^1 dy\, w_\zeta(x|y)\mathcal{F}_q(y,\zeta,\tau)-
\int_0^x dy\, w'_\zeta(y|x)\mathcal{F}_q(x,\zeta,\tau),
\eeq
that can be re-expressed in a form which is a simple generalization of the formula for the 
forward evolution \cite{CPC}.

\beqa
\frac{d \mathcal{F}_q}{d\log Q^2} &=& \int_x^1 dy\, w_\zeta(x|y)\mathcal{F}_q(y,\zeta,\tau) - 
\int_0^x dy \,w'_\zeta(y|x) \mathcal{F}_q(x,\zeta,\tau)
\nonumber \\
&=& -\int_0^{\alpha(x)} dy w_\zeta(x+y|x)* \mathcal{F}_q(x,\zeta,\tau)+ 
\int_0^{-x} dy\, w'_\zeta(x+y|x)\mathcal{F}_q(x,\zeta,\tau),
\eeqa
where a Moyal-like product appears 
\beq
w_\zeta(x+y|x)*\mathcal{F}_q(x,\zeta,\tau)\equiv w_\zeta(x+y|x) e^{-y \left(\overleftarrow{\partial}_x + 
\overrightarrow{\partial}_x\right)} \mathcal{F}_q(x,\zeta,\tau)
\eeq
and $\alpha(x) =x-1$. 
A Kramers-Moyal expansion of the equation allows to generate a differential equation 
of infinite order with a parametric dependence on $\zeta$ 

\beqa
\frac{d \mathcal{F}_q}{d\log Q^2} &=&\int_{\alpha(x)}^{0}dy\,  
w_\zeta(x+y|x)\mathcal{F}_q(x,\zeta,\tau) + 
\int_{0}^{-x}dy\,  
w'_\zeta(x+y|x)\mathcal{F}_q(x,\zeta,\tau) \nonumber \\
&& - \sum_{n=1}^{\infty}\int_0^{\alpha(x)}dy \frac{(-y)^n}{n!}{\partial_x}^n
\left(w_\zeta(x+y|x)\mathcal{F}_q(x,\zeta,\tau)\right).
\eeqa
We define
\beqa
\tilde{a}_0(x,\zeta) &=& \int_{\alpha(x)}^{0} dy w_\zeta(x+y|x)\mathcal{F}_q(x,\zeta,\tau)
+ \int_{0}^{-x}dy\,  
w'_\zeta(x+y|x)\mathcal{F}_q(x,\zeta,\tau)
 \nonumber \\
a_n(x,\zeta) &=&\int_0^{\alpha(x)} dy \,y^n w_\zeta (x+y|x) \mathcal{F}_q(x,\zeta,\tau) \nonumber \\
\tilde{a}_n(x,\zeta)&=&\int_0^{\alpha(x)}
dy y^n {\partial_x}^n \left(w_\zeta(x+y|x)\mathcal{F}_q(x,\zeta,\tau)\right) \,\,\,n=1,2,...
\eeqa
If we arrest the expansion at the first two terms $(n=1,2)$ we are able to derive an 
approximate equation describing the dynamics of partons for non-diagonal transitions. 
The procedure is a slight generalization of the method presented in \cite{CPC}, 
to which we refer for further details. For this purpose we use the identities

\beqa
\tilde{a}_1(x,\zeta) &=&\partial_x a_1(x,\zeta) - \alpha(x) \partial_x \alpha(x)
w_\zeta(x + \alpha(x)|x)\mathcal{F}_q(x,\zeta,\tau) \nonumber \\
\tilde{a}_2(x,\zeta) &=&\partial_x^2 a_2(x,\zeta) - 
2 \alpha(x) (\partial_x \alpha(x))^2 w_\zeta(x+ \alpha(x)|x)\mathcal{F}_q(x,\zeta,\tau)
\nonumber \\
&&  - 
\alpha(x)^2 \partial_x\alpha(x)
\partial_x\left( w_\zeta(x+ \alpha(x)|x)\mathcal{F}_q(x,\zeta,\tau)\right)\nonumber \\ 
&& - \alpha^2(x)\partial_x \alpha(x) \partial_x\left( w_\zeta(x + y|x)
\mathcal{F}_q(x,\zeta,\tau)\right)|_{y=\alpha(x)}.
\eeqa
which allow to compute the first few coefficients of the expansion. 
Using these relations, the Fokker-Planck approximation to this equation 
can be worked out explicitly. We omit details on the derivation which is non obvious 
since particular care is needed to regulate the (canceling) divergences and just quote 
the result. 

A lengthy computation gives 
\beqa
\frac{d \mathcal{F}_q}{d\tau}&=& \frac{\alpha}{\pi}C_F\left(\frac{x_{0,-3}}{(x-\z)^3}
+ \frac{x_{0,-1}}{(x-\z)} + x_{0,0}\right)\mathcal{F}_q(x,\z,\tau) 
\nonumber \\
&& + \frac{\alpha}{\pi}C_F\left(\frac{x_{1,-3}}{(x-\z)^3}
+ \frac{x_{1,-1}}{(x-\z)}\right)\partial_x\mathcal{F}_q(x,\z,\tau) 
+ \frac{\alpha}{\pi}C_F\frac{x_{0,-3}}{(x-\z)^3}\partial_x^2\mathcal{F}_q(x,\z,\tau) 
\nonumber \\	
\eeqa
where we have defined  
\beqa
x_{0,-3}&=&\frac{-\left( {\left( -1 + x \right) }^3\,
      \left( 17 x^3 - \z^2 \left( 3 + 4\z \right)  + 
        3 x \z\left( 3 + 5 \z \right)  - 3 x^2\left( 3 + 7 \z \right) 
        \right)  \right) }{12\,x^3}\nonumber \\
x_{0,-1} &=&
\frac{-29 x^4 - 3 + x^2\,\left( -1 + \z \right)  + 2 \z - 
    2 x \left( 1 + 3 \z \right)  + x^3 \left( 12 + 23 \z \right) }{3 x^3}
\nonumber \\
x_{0,0} &=& 4 + \frac{1}{2 x^2} - \frac{3}{x} + 2 \log \frac{(1 - x)}{x}
\nonumber \\
x_{1,-1}&=& 
\frac{-\left( \left( -1 + 6 x - 15 x^2 + 14 x^3 \right) \,
      \left( x - \z \right)  \right) }{3 x^2}\nonumber \\
x_{1,-3}&=&
\frac{1}{2} - \frac{5 x}{3} + 5 x^3 - \frac{23 x^4}{6} + \frac{7 \z}{3} - 
  \frac{3 \z}{4 x} + \frac{5 x \z}{2} \nonumber \\
&& - 15 x^2 \z + 
  \frac{131 x^3 \z}{12} - \frac{5 \z^2}{2} + \frac{\z^2}{4 x^2} - 
  \frac{\z^2}{x} + 13 x \z^2 - \frac{39 x^2 \z^2}{4} - 3 \z^3 + 
  \frac{\z^3}{3 x^2} + \frac{8 x \z^3}{3} \nonumber \\
x_{2,-3}&=&\frac{-\left( {\left( -1 + x \right) }^2\,{\left( x - \z \right) }^2\,
      \left( 3 + 23 x^2 + 4 \z - 2 x\left( 7 + 8 \z \right)  \right) 
      \right) }{24 x}.
\eeqa

This equation and all the equations obtained by arresting the 
Kramers-Moyal expansion to higher order provide a complementary 
description of the nonforward dynamics in the DGLAP region, at least 
in the nonsinglet case. Moving to higher order is straightforward 
although the results are slightly lengthier. A full-fledged 
study of these equations is under way and we expect that the DGLAP dynamics 
is reobtained - directly from these equations - as the order of the approximation increases.

\section{Model comparisons, saturation and the tensor charge}

In this last section we discuss some implementations of our methods to the 
standard (forward) evolution by doing a NLO model comparisons both in the 
analysis of Soffer's inequality and for the evolution of the tensor charge. 
We have selected two models, motivated quite independently 
and we have compared the predicted evolution of the Soffer bound at an 
accessable final evolution scale around $100$ GeV for the light quarks and 
around $200$ GeV for the heavier generations. At this point we recall that 
in order to generate suitable initial 
conditions for the analysis of Soffer's inequality, one needs an ansatz 
in order to quantify the difference between its left-hand side and right-hand side 
at its initial value.

The well known strategy to build reasonable initial conditions for the transverse 
spin distribution consists in generating polarized distributions (starting 
from the unpolarized ones) and then saturate the inequality at some lowest scale, 
which is the approach we have followed for all the models that we have implemented. 

The first model we have used is the GRV-GRSV, that we have described in Sec.~\ref{sec:initial}.

In the implementation of the second model (GGR model)
we have used as input distributions in the unpolarized case the 
CTEQ4 parametrization \cite{CTEQ}, calculated to NLO in the
\( \overline{\textrm{MS}} \) scheme at a scale \( Q_{0}=1.0\, \textrm{GeV} \)

\begin{eqnarray}
x(u-\overline{u})(x,Q_{0}^{2}) & = & 1.344x^{0.501}(1-x)^{3.689}(1+6.402x^{0.873})\nonumber \\
x(d-\overline{d})(x,Q_{0}^{2}) & = & 0.64x^{0.501}(1-x)^{4.247}(1+2.69x^{0.333})\nonumber \\
xs(x,Q_{0}^{2})=x\overline{s}(x,Q_{0}^{2}) & = & 0.064x^{-0.143}(1-x)^{8.041}(1+6.112x)\nonumber \\
x(\overline{d}-\overline{u})(x,Q_{0}^{2}) & = & 0.071x^{0.501}(1-x)^{8.041}(1+30.0x)\nonumber \\
x(\overline{u}+\overline{d})(x,Q_{0}^{2}) & = & 0.255x^{-0.143}(1-x)^{8.041}(1+6.112x)\nonumber \\
xg(x,Q_{0}^{2}) & = & 1.123x^{-0.206}(1-x)^{4.673}(1+4.269x^{1.508})
\end{eqnarray}
and \( xq_{i}(x,Q_{0}^{2})=x\overline{q_{i}}(x,Q_{0}^{2})=0 \) for
\( q_{i}=c,b,t \)
and we have related the unpolarized input distribution to the longitudinally
polarized ones by the relations \cite{GGR}

\begin{eqnarray}
x\Delta \overline{u}(x,Q_{0}^{2}) & = & x\eta _{u}(x)xu(x,Q_{0}^{2})\nonumber \\
x\Delta u(x,Q_{0}^{2}) & = & \cos \theta _{D}(x,Q_{0}^{2})\left[ x(u-\overline{u})-\frac{2}{3}x(d-\overline{d})\right] (x,Q_{0}^{2})+x\Delta \overline{u}(x,Q_{0}^{2})\nonumber \\
x\Delta \overline{d}(x,Q_{0}^{2}) & = & x\eta _{d}(x)xd(x,Q_{0}^{2})\nonumber \\
x\Delta d(x,Q_{0}^{2}) & = & \cos \theta _{D}(x,Q_{0}^{2})\left[ -\frac{1}{3}x(d-\overline{d})(x,Q_{0}^{2})\right] +x\Delta \overline{d}(x,Q_{0}^{2})\nonumber \\
x\Delta s(x,Q_{0}^{2})=x\Delta \overline{s}(x,Q_{0}^{2}) & = & x\eta _{s}(x)xs(x,Q_{0}^{2})
\end{eqnarray}
and \( x\Delta q_{i}(x,Q_{0}^{2})=x\Delta \overline{q_{i}}(x,Q_{0}^{2})=0 \)
for \( q_{i}=c,b,t \).

\begin{figure}[tbh]
{\centering \resizebox*{12cm}{!}{\rotatebox{-90}{\includegraphics{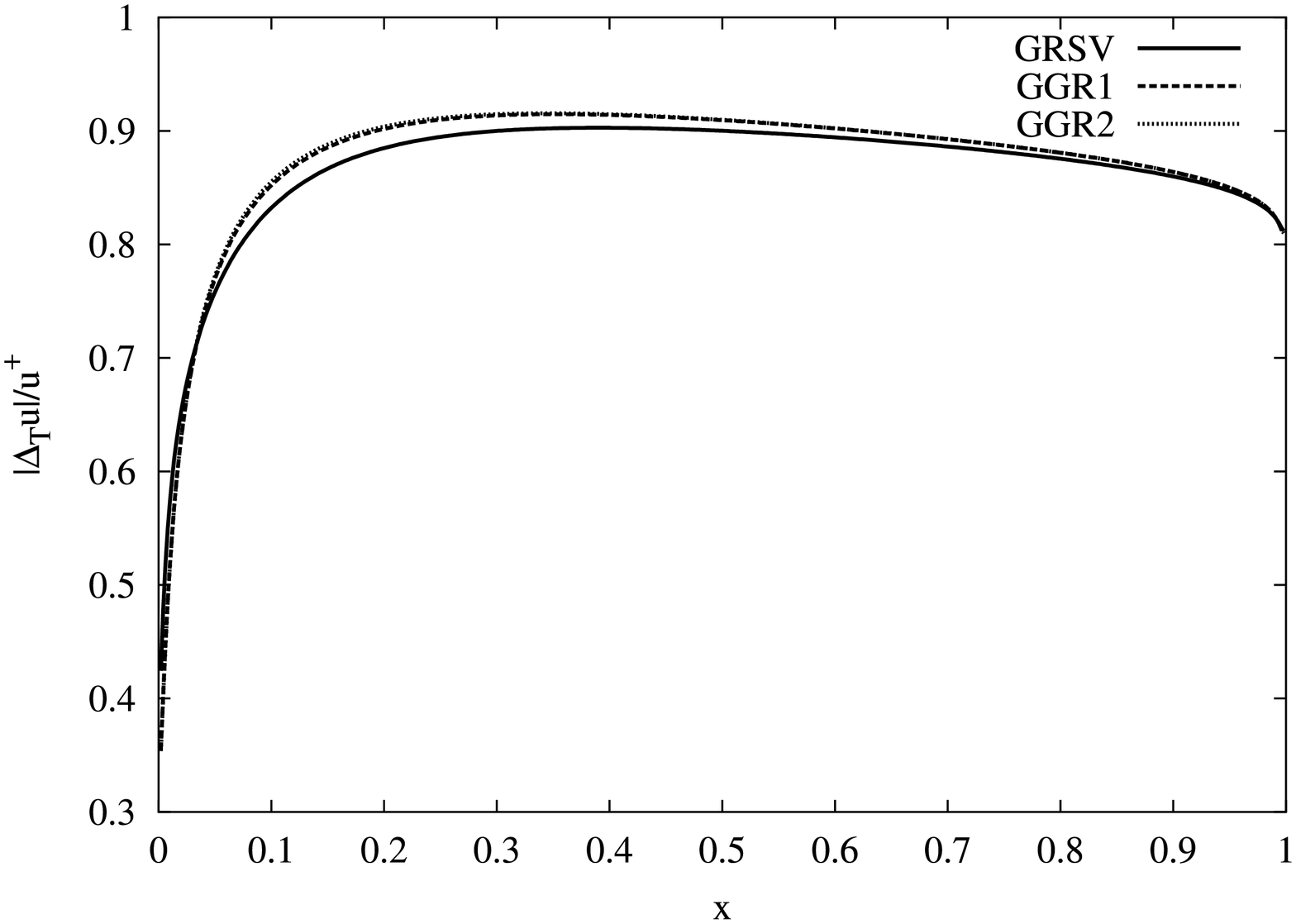}}} \par}
\caption{ Test of Soffer's inequality for quark up at \protect\( Q=100\protect \)
GeV for different models.}
\label{upsof}
\end{figure}

\begin{figure}[tbh]
{\centering \resizebox*{12cm}{!}{\rotatebox{-90}{\includegraphics{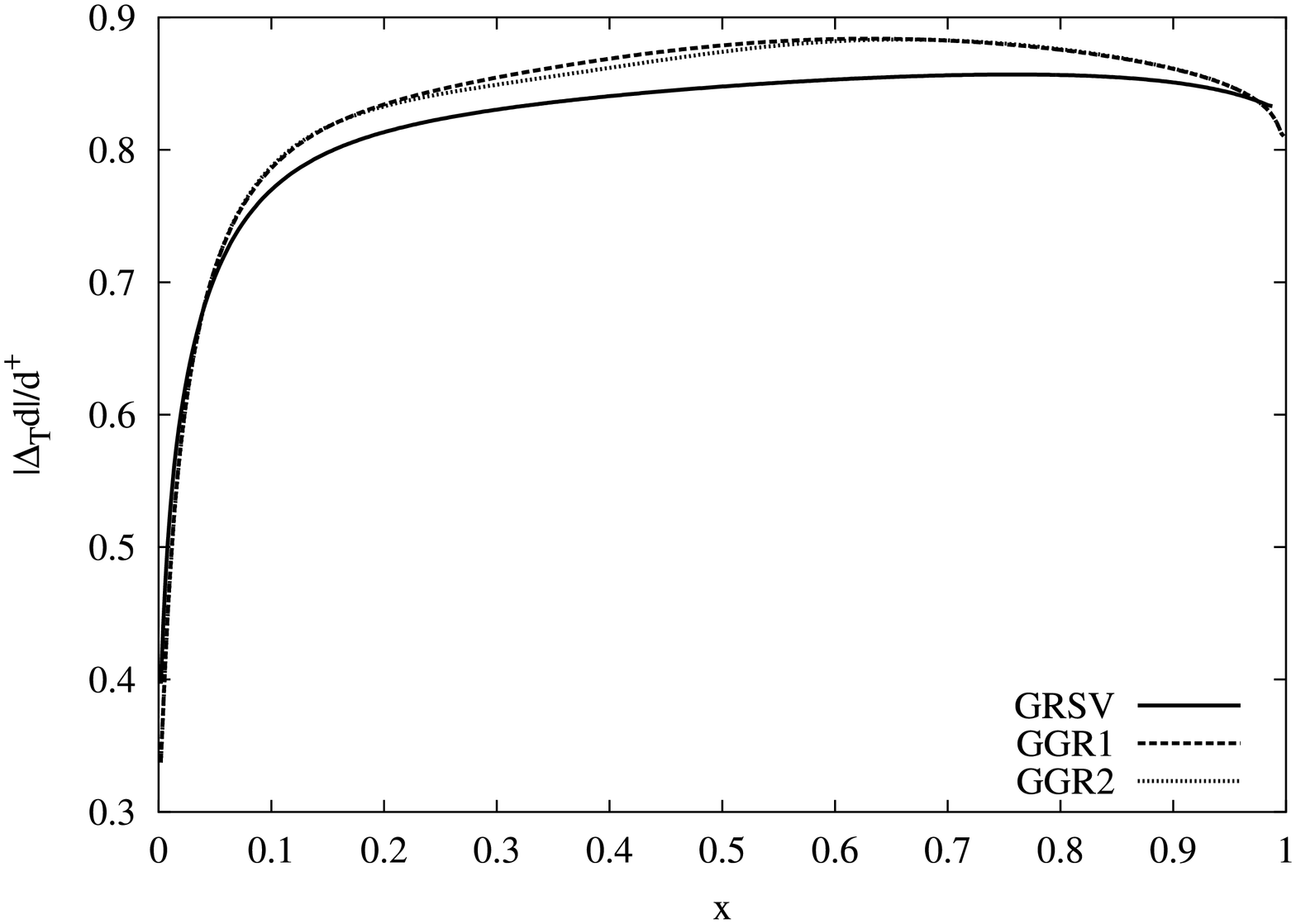}}} \par}
\caption{ Test of Soffer's inequality for quark down at \protect\( Q=100\protect \)
GeV for different models}
\label{downsof}
\end{figure}

\begin{figure}[tbh]
{\centering \resizebox*{12cm}{!}{\rotatebox{-90}{\includegraphics{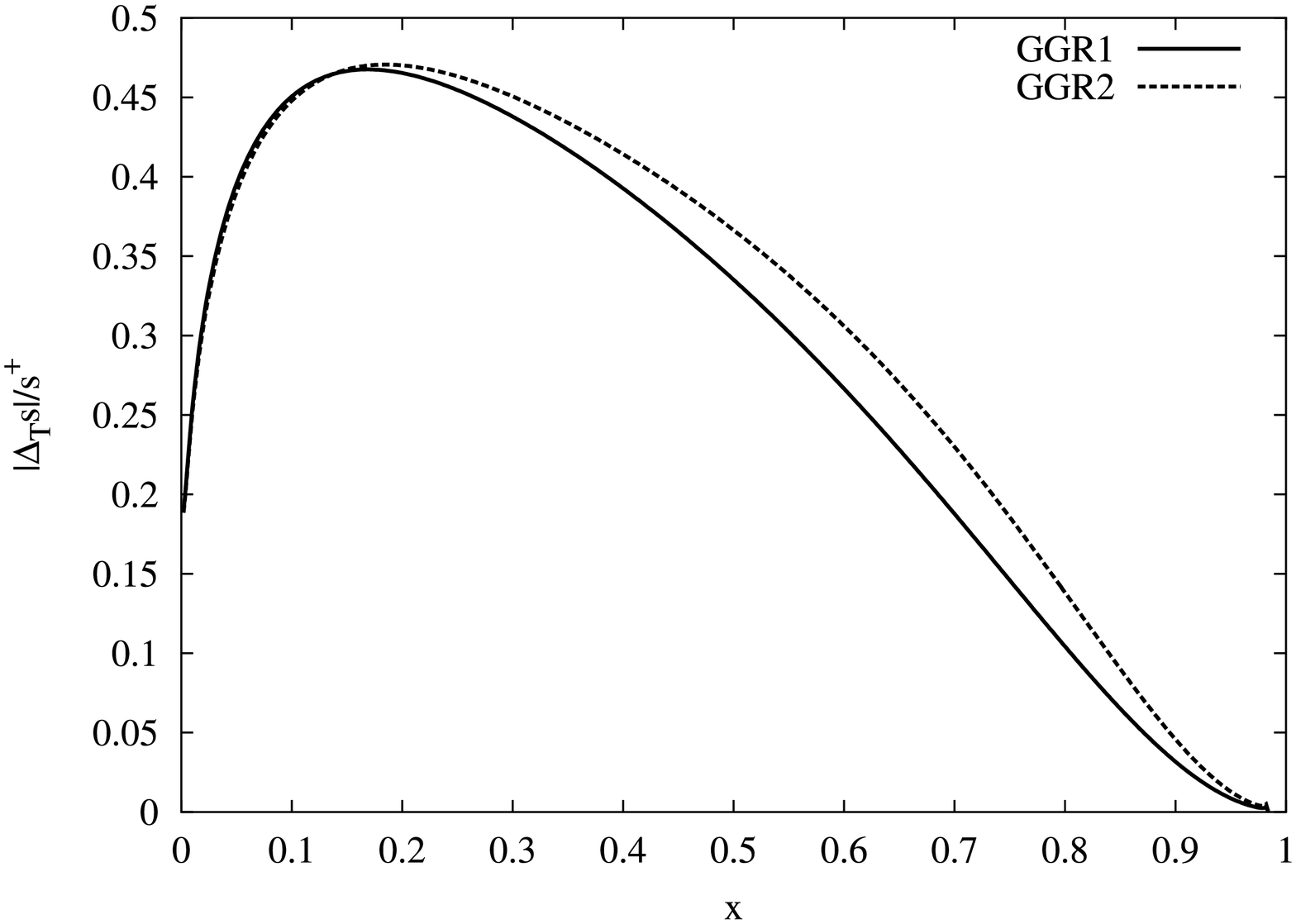}}} \par}
\caption{ Test od Soffer's inequality for quark strange at \protect\( Q=100\protect \)
GeV for different models}
\label{strangesof}
\end{figure}

\begin{figure}[!tbh]
\begin{center}\subfigure[strange]{\includegraphics[%
  width=5.5cm,
  angle=-90]{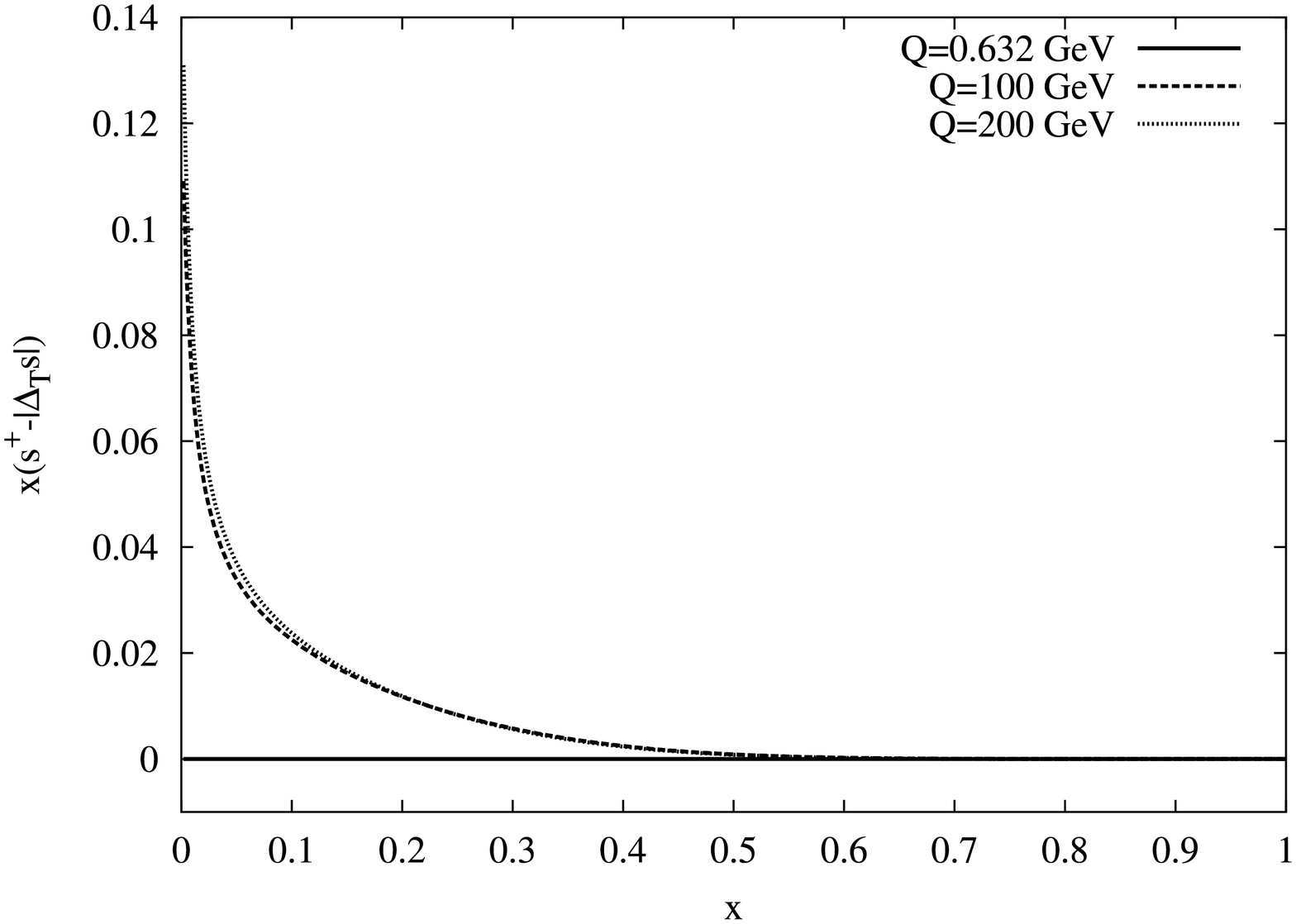}}\subfigure[charm]{\includegraphics[%
  width=5.5cm,
  angle=-90]{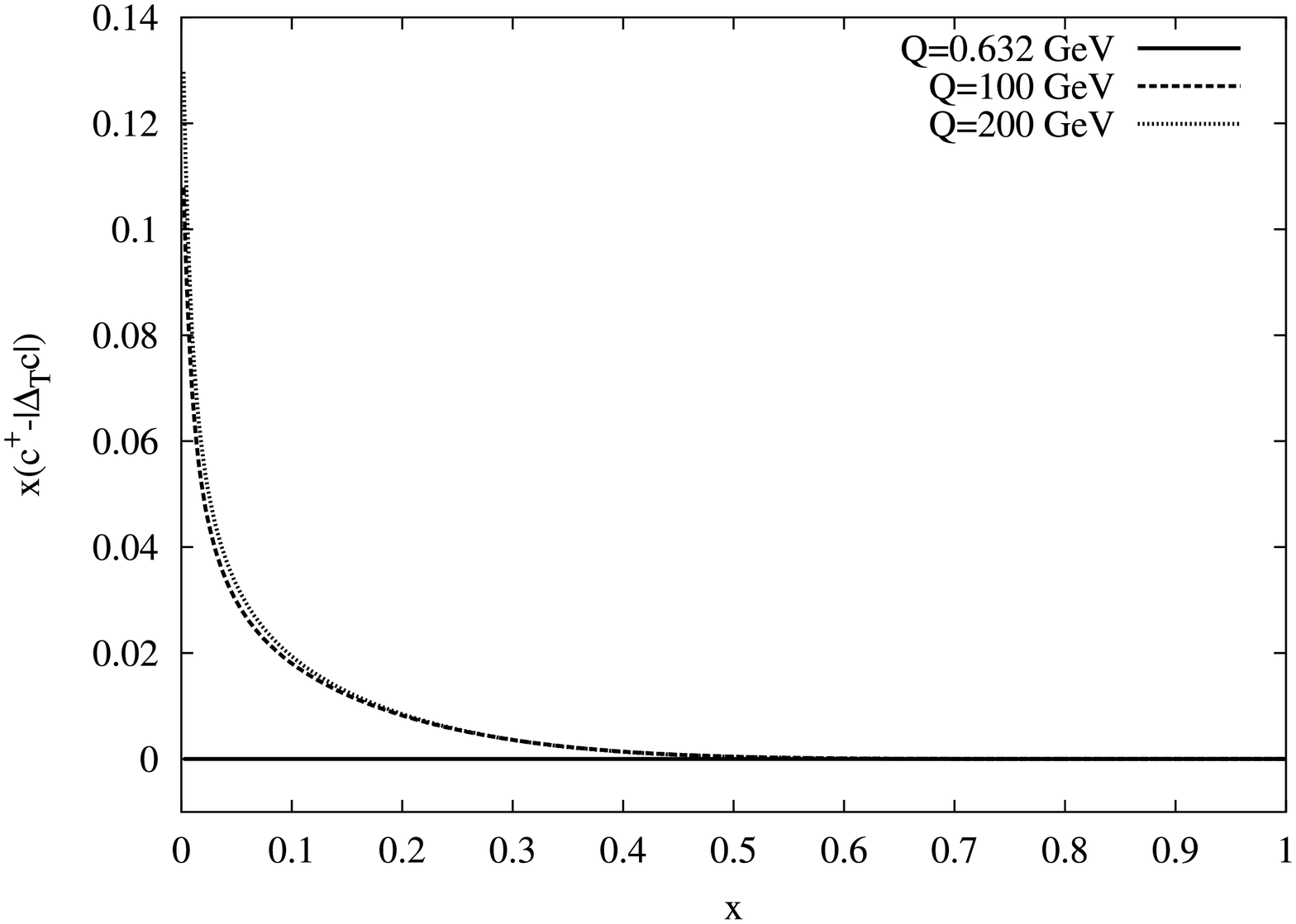}}\end{center}

\caption{Soffer's inequality for strange and charm in the GRSV model.}
\label{scsof}
\end{figure}

\begin{figure}[!tbh]
\begin{center}\subfigure[bottom]{\includegraphics[%
  width=5.5cm,
  angle=-90]{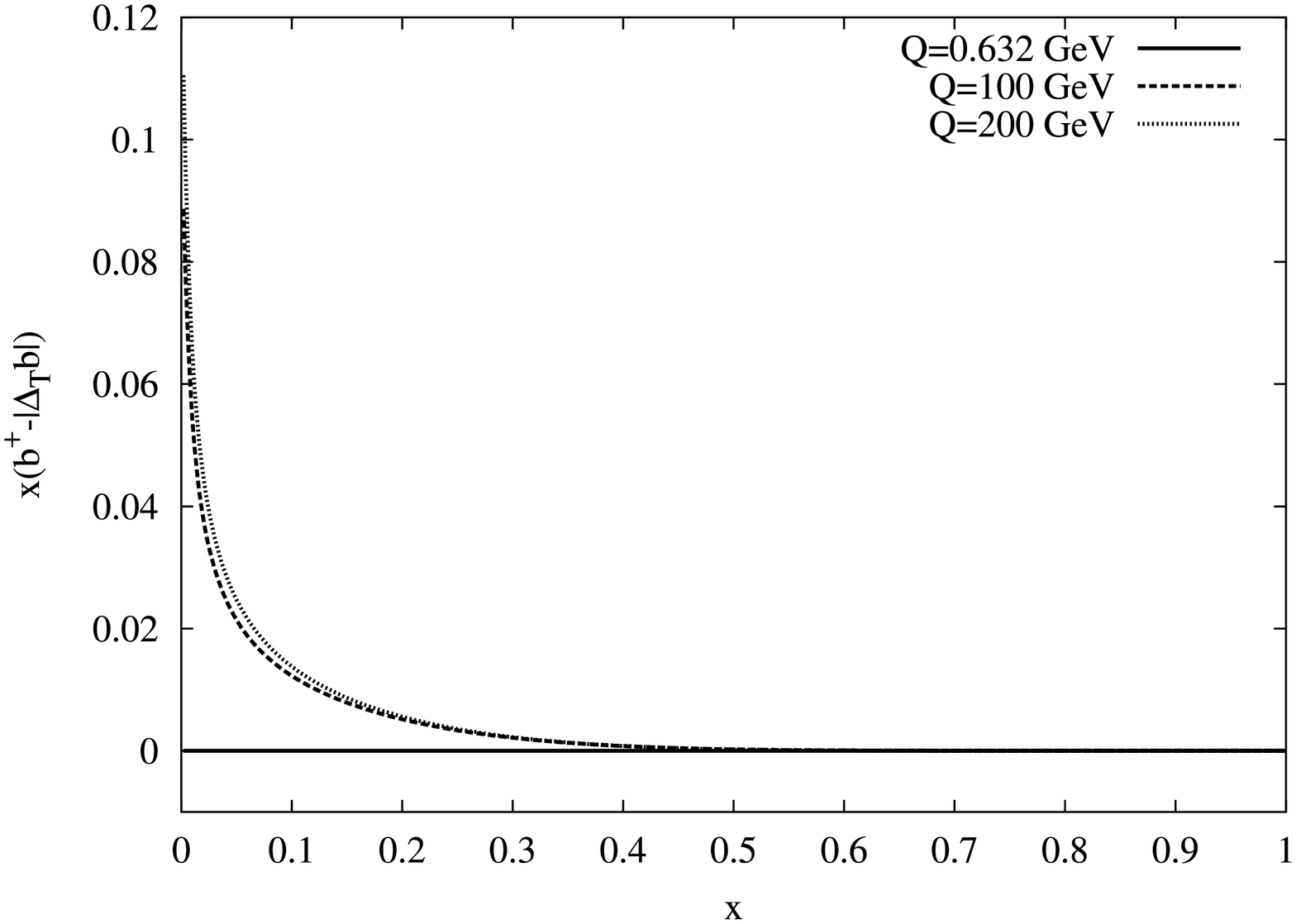}}\subfigure[top]{\includegraphics[%
  width=5.5cm,
  angle=-90]{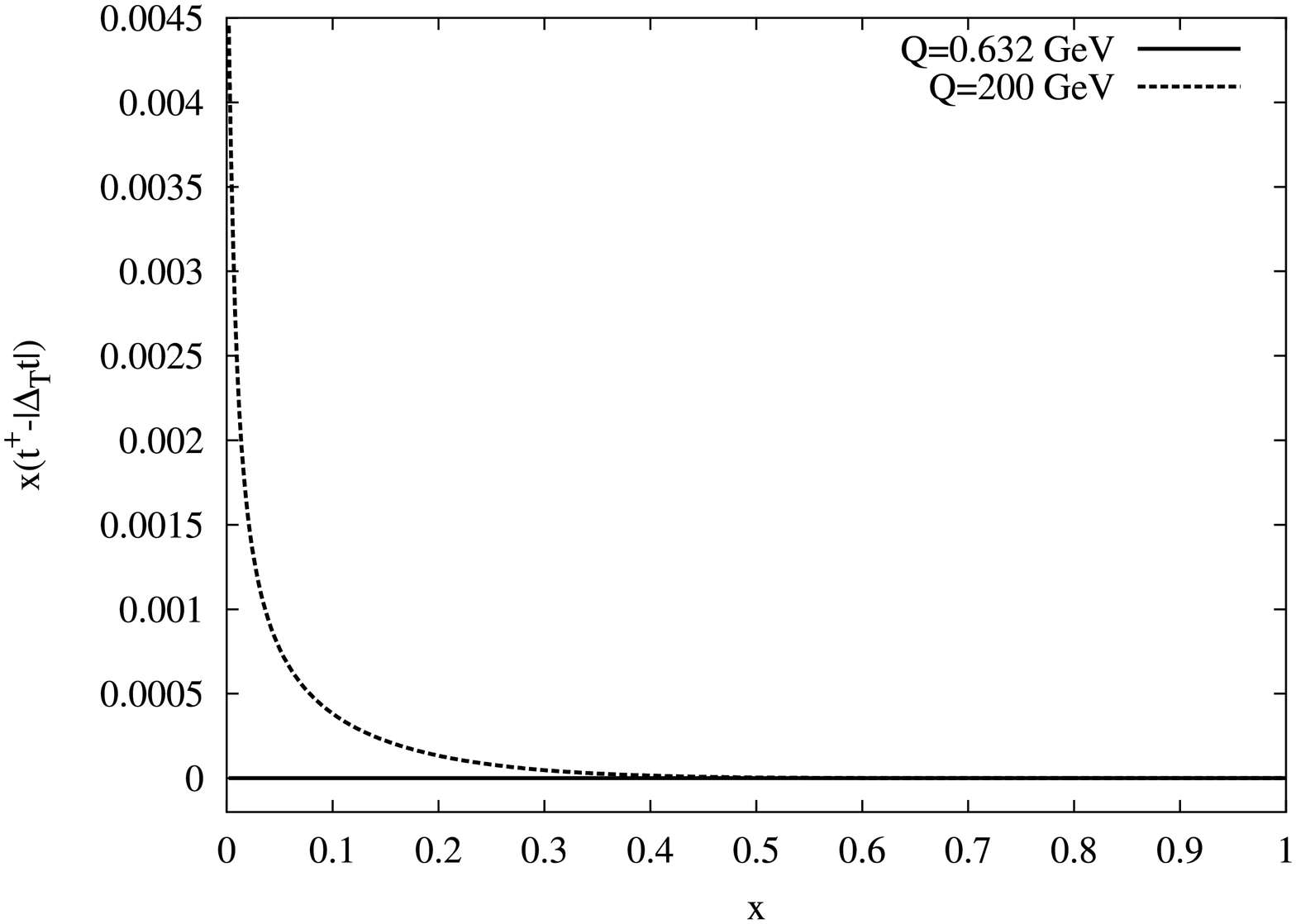}}\end{center}

\caption{Soffer's inequality for bottom and top in the GRSV model.}
\label{btsof}
\end{figure}

A so-called ``spin dilution factor'' as defined in \cite{GGR}, which appears 
in the equations above is given by
\begin{equation}
\cos \theta _{D}(x,Q_{0}^{2})=\left[ 1+\frac{2\alpha _{s}(Q^{2})}{3}\frac{(1-x)^{2}}{\sqrt{x}}\right] ^{-1}.
\end{equation}
In this second (GGR) model, in regard to the initial conditions for the gluons, 
we have made use of two different options, characterized by a parameter 
\( \eta  \) dependent on the corresponding option. 
The first option, that
we will denote by GGR1, assumes that gluons are moderately polarized 
\begin{eqnarray}
x\Delta g(x,Q_{0}^{2}) & = & x\cdot xg(x,Q_{0}^{2})\nonumber \\
\eta _{u}(x)=\eta _{d}(x) & = & -2.49+2.8\sqrt{x}\nonumber \\
\eta _{s}(x) & = & -1.67+2.1\sqrt{x},
\end{eqnarray}
while the second option (GGR2) assumes that gluons are not polarized 
\begin{eqnarray}
x\Delta g(x,Q_{0}^{2}) & = & 0\nonumber \\
\eta _{u}(x)=\eta _{d}(x) & = & -3.03+3.0\sqrt{x}\nonumber \\
\eta _{s}(x) & = & -2.71+2.9\sqrt{x}.
\end{eqnarray}
We have plotted both ratios $\Delta_T/f^+$ and differences 
$(x f^+ - x\Delta_T f)$ for 
various flavours as a function of $x$. For the up quark, while the two models GGR1 and GGR2 
are practically overlapping, the difference between the GGR 
and the GRSV models in the the ratio $\Delta_T u/u^+$ 
is only slightly remarked in the intermediate x region $(0.1-0.5)$. In any case, it is just at the few percent level (Fig. (\ref{upsof})), while the inequality is satisfied 
with a ratio between the plus helicity distribution and transverse around 10 percent from the saturation value, and above. There is a wider gap in the inequality at small x, region characterized by larger transverse distribution, with values up to 40 percent from saturation. A similar trend is noticed for the x-behaviour of the inequality in the case 
of the down quark (Fig. \ref{downsof}). In this latter case the GGR 
and the GRSV model show a more remarked difference, especially for 
intermediate x-values. An interesting features appears in the corresponding 
plot for the strange quark (Fig.(\ref{strangesof})), showing a much 
wider gap in the inequality (50 percent and higher) compared 
to the other quarks. Here we have plotted results for the two GGR models (GGR1 and GGR2). 
Differently from the case of the other quarks, in this case we observe 
a wider gap between lhs and rhs at larger x values, increasing as $x\rightarrow 1$. 
In figs. (\ref{scsof})and (\ref{btsof}) we plot the differences $(x f^+ - x\Delta_T f)$ 
for strange and charm and for bottom and top quarks respectively, which 
show a much more reduced evolution from the saturation value up to the final corresponding 
evolving scales (100 and 200 GeV). 
As a final application we finally discuss the behaviour of the tensor charge 
of the up quark for the two models as a function of the final 
evolution scale $Q$. We recall that like the isoscalar and the isovector 
axial vector charges defined from the forward matrix element 
of the nucleon, the nucleon tensor charge is defined from the matrix 
element of the tensor current 
\beq
\langle P S_T|\bar{\psi}\sigma^{\mu\nu}\gamma_5 \lambda^a \psi|P,S_T\rangle 
=2 \delta q^a(Q_0^2)\left( P^\mu S_T^\nu - P^\nu S_T^\mu\right)
\eeq
where $\delta^a q(Q_0^2)$ denotes the flavour (a) contribution to the nucleon tensor charge at a scale $Q_0$ and $S_T$ is the transverse spin. 

\begin{figure}[tbh]
{\centering \resizebox*{12cm}{!}{\rotatebox{-90}{\includegraphics{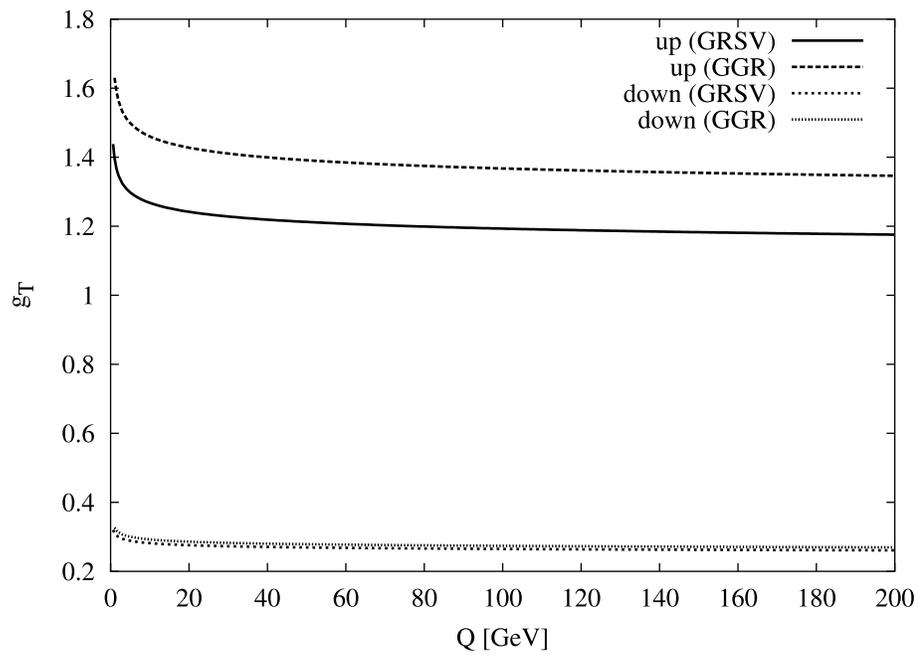}}} \par}
\caption{Tensor charge \protect\( g_{T}\protect \) as a function of \protect\( Q\protect \)
for up and down quark for the GRSV and GGR models.}
\label{figure}
\end{figure}

In fig. (\ref{figure}) we plot the evolution of the tensor charge for the models 
we have taken in exam. At the lowest evolution scales the charge is, in these models, 
above 1 and decreases slightly as the factorization scale $Q$ increases. 
We have performed an evolution up to 200 GeV as an illustration of this behaviour. 
There are substantial differences between these models, as one can easily observe, which are around 20 percent. From the analysis of these differences at various factorization 
scales we can connect low energy dynamics to observables at higher energy, thereby 
distinguishing between the various models. Inclusion of the correct evolution, 
up to subleading order is, in general, essential.

\section{Conclusions}
We have illustrated the use of $x$-space based algorithms for the solution 
of evolution equations in the leading and in the next-to-leading 
approximation and we have 
provided some applications of the method both in the analysis 
of Soffer's inequality and in the investigation 
of other relations, such as the evolution of 
the proton tensor charge, for various models. 
The evolution has been implemented using a suitable base, 
relevant for an analysis of positivity in LO, using kinetic arguments. 
The same kinetic argument 
has been used to prove the positivity of the evolution of 
$h_1$ and of the tensor charge up to NLO. 
In our implementations we have completely relied on recursion relations without 
any reference to Mellin moments. We have provided several illustrations of the 
recursive algorithm and extended it to the nonforward evolution up to NLO. 
Building on previous work 
for the forward evolution, we have presented a master-form of the nonsinglet 
evolution of the skewed distributions, a simple proof of positivity
and a related Kramers Moyal expansion, 
valid in the DGLAP region of the skewed evolution for any value 
of the asymmetry parameter $\zeta$. We hope to return with a complete study of the 
nonforward evolution and related issues not discussed here in the near future.

\begin{figure}[tbh]
{\centering \resizebox*{12cm}{!}{\rotatebox{-90}{\includegraphics{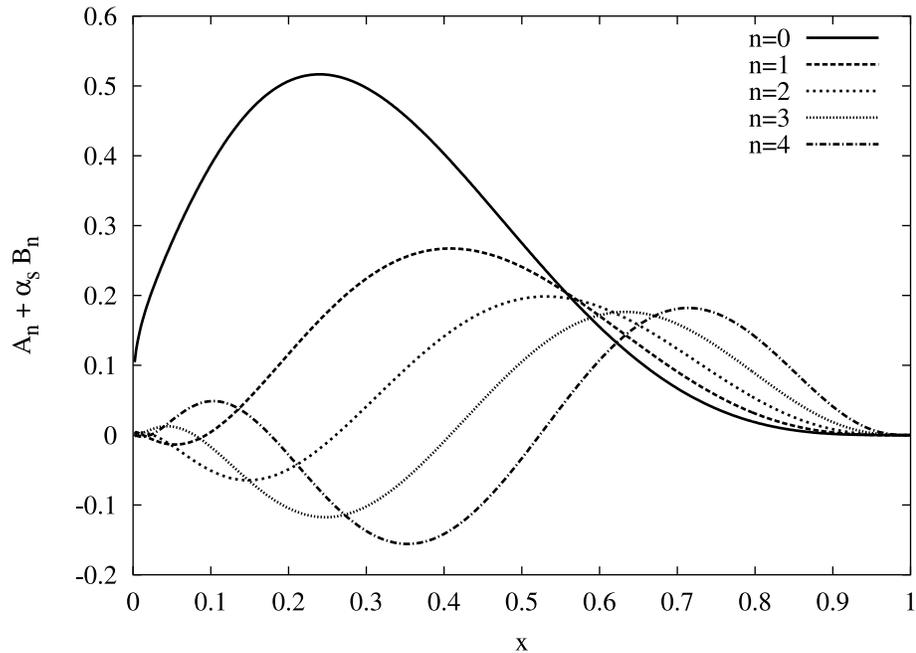}}} \par}
\caption{Coefficients \protect\( A_{n}(x)+\alpha _{s}(Q^{2})B_{n}\protect \),
with \protect\( n=0,\ldots ,4\protect \) for a final scale \protect\( Q=100\protect \)
GeV for the quark up.}
\label{an}
\end{figure}

\begin{figure}[tbh]
{\centering \resizebox*{12cm}{!}{\rotatebox{-90}{\includegraphics{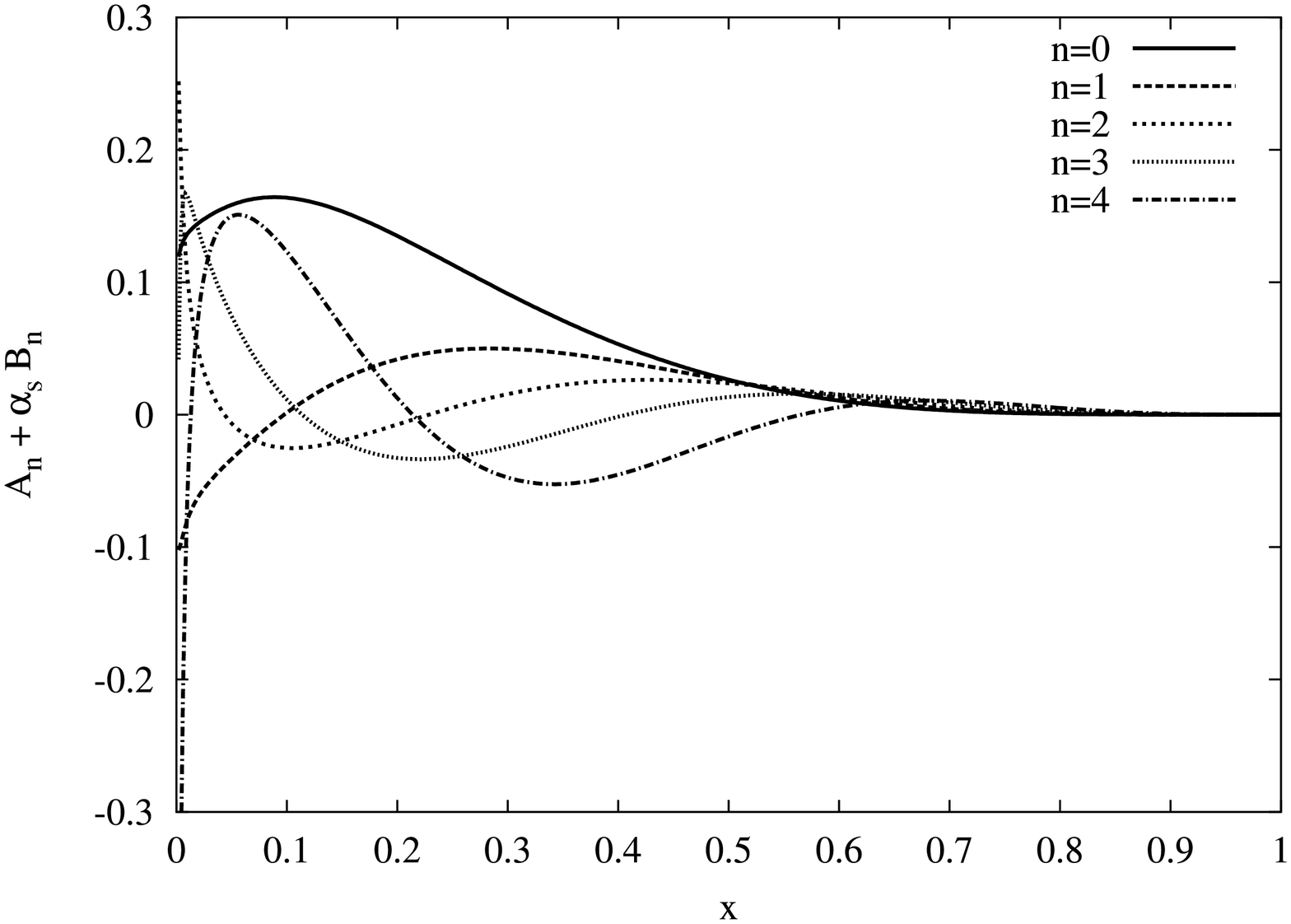}}} \par}

\caption{Coefficients \protect\( A_{n}(x)+\alpha _{s}(Q^{2})B_{n}\protect \),
with \protect\( n=0,\ldots ,4\protect \) for a final scale \protect\( Q=100\protect \)
GeV for the quark down.}
\label{anprime}
\end{figure}

\chapter{Simulation of air showers: lateral distributions, multiplicities 
 and the\\search for new physics at Auger\label{chap:airshowers}}
\fancyhead[LO]{\nouppercase{Chapter 6. Simulation of air showers}}
\section{A brief overview} 
The inclusive high energy cosmic rays spectrum has been the focus point of a lot of 
attention in the last few years. The spectrum, at a first look, appears to be deprived of any 
structure and, on the other hand, drops dramatically as we move upward in energy. The region 
which we are interested in, in our study, is the high energy region, where a correct description 
of the underlying quark-gluon dynamics is important in order to gain some insight into the 
physics of the extensive air showers that form when a high energy primary 
(say a proton or a neutrino) collides with the nuclei in the atmosphere. Here 
we would like to put things into place in order to set up the stage for our analysis, 
which will be developed in this chapter and will be specialized in the next chapter to the case 
of mini black hole production in cosmic rays in the context of  brane models. 
Our journey has been guided by our interest in applying standard QCD tools, such as the RG 
evolution of fragmentation functions, to this challenging environment. It turns out, 
as expected, that the issues to be analyzed, using the parton model, 
are manifold and do not have a clear-cut answer at the moment, although there 
is a general agreement that models based on reggeon theory, popular in the sixties, can capture 
the complex dynamics of a proton-air collision at those energies. The upper end of the cosmic 
rays spectrum (these cosmic rays are classified as ultra high energy cosmic rays or UHECR), 
in particular, has been affected by a controversy which has spanned several decades over the existence of a cutoff which 
would more or less sharply limit it to stay below $10^{19}$ eV.      
 
In this chapter we perform large scale air shower simulations around the GZK cutoff are
performed. An extensive analysis of the behaviour
of the various subcomponents of the cascade is presented.
We focus our investigation both
on the study of total and partial multiplicities along the entire
atmosphere and on the geometrical structure of the various
cascades, in particular on the lateral distributions.
The possibility of detecting new physics in Ultra High Energy
Cosmic Rays (UHECR) at Auger is
also investigated. We try to disentangle
effects due to standard statistical fluctuations in the first proton
impact in the shower formation from the underlying interaction and comment on
these points.
We argue that theoretical models predicting large missing energy may have a
chance
to be identified, once the calibration errors in the energy measurements
are resolved by the experimental collaborations, in measurements
of inclusive multiplicities.

\section{Introduction}

One of the most intriguing experimental observations of recent years
is the detection of Ultra--High--Energy--Cosmic--Rays (UHECR),
with energy in excess of the Greisen--Zatsepin--Kuzmin (GZK)
cutoff \cite{uhecr} (for a review see \cite{la}).
While its validity is still under some dispute,
it is anticipated that the forthcoming Auger \cite{auger}
and EUSO \cite{euso} experiments will provide enough statistics to 
resolve the debate. From a theoretical perspective, the Standard Model of
particle physics and its Grand Unified extensions indicate
that many physical structures may lie far beyond the reach
of terrestrial collider experiments. If this eventuality
materializes it may well be that the only means of unlocking the
secrets of the observed world will be mathematical rigor and peeks
into the cosmos in its most extreme conditions. In this
context the observation of UHECR is especially puzzling
because of the difficulty in explaining the events without
invoking some new physics.
There are apparently no astrophysical sources
in the local neighborhood that can account for the
events. The shower profile of the
highest energy events is consistent with identification of the
primary particle as a hadron but not as a photon or a neutrino.
The ultrahigh energy events observed in the air shower arrays
have muonic composition indicative of hadrons.
The problem, however, is that the propagation of hadrons
over astrophysical distances is affected by the
existence of the cosmic background radiation, resulting
in the GZK cutoff on the maximum energy of cosmic ray
nucleons $E_{\rm GZK}\le10^{20}\;{\rm eV}$ \cite{gzk}.
Similarly, photons of such high energies have a mean free path of less
than 10 Mpc due to scattering {}from the cosmic background radiation and
radio photons. Thus, unless the primary is a neutrino,
the sources must be nearby. On the other hand, the primary
cannot be a neutrino because the neutrino interacts very weakly
in the atmosphere. A neutrino primary would imply that the
depths of first scattering would be uniformly distributed
in column density, which is contrary to the observations.

The most exciting aspect of the UHECR
is the fact that the Auger and EUSO
experiments will explore the physics associated with these events,
and provide a wealth of observational data.
Clearly, the first task of these experiments is to establish
whether the GZK cutoff is violated, and to settle the
controversy in regard to the air shower
measurement.

\section{Probing new physics with UHECR}
We may, however, entertain the
possibility that these experiments can probe
various physics scenarios. In the first place,
the center of mass energy in the collision
of the primary with the atmosphere is of the order
of 100 TeV and exceed the contemporary, and forthcoming,
collider reach by two orders of magnitude.
Thus, in principle the air shower analysis should
be sensitive to any new physics that is assumed
to exist between the electroweak scale and the
collision scale due to the interaction of the
primaries with the atmospheric nuclei. Other
exciting possibilities include the various
explanations that have been put forward to
explain the existence of UHECR events \cite{la, explanations,berez, ben, sb, cfp},
and typically assume some form of new physics.
One of the most intriguing possible solutions
is that the UHECR primaries originate {}from the decay of long--lived
super--heavy relics, with mass of the order of $10^{12-15}\;{\rm GeV}$
\cite{berez, ben, sb, cfp}.
In this case the primaries for the observed UHECR would originate
from decays in our galactic halo, and the GZK bound would not apply.
This scenario is particularly interesting due to the possible
connection with superstring theory.
{}From the particle physics perspective the meta--stable super--heavy
candidates should possess several properties.
First, there should exist a stabilization mechanism which produces
the super--heavy state with a lifetime of the order of
$
10^{17}s\le \tau_X \le 10^{28}s,~
$
and still allows it to decay and account for the observed UHECR events.
Second, the required mass scale of the meta--stable state
should be of order, $M_X~\sim~10^{12-13}{\rm GeV}.$
Finally, the abundance of the super--heavy relic
should satisfy the relation
$
({\Omega_X/\Omega_{0}})({t_0/\tau_X})\sim5\times10^{-11},~
$
to account for the observed flux of UHECR events.
Here $t_0$ is the age of the universe, $\tau_X$ the lifetime
of the meta--stable state, $\Omega_{0}$ is the critical mass density
and $\Omega_{X}$ is the relic mass density of the meta--stable state.
It is evident that the parameters of the super--heavy meta--stable states
are sufficiently flexible to accommodate the observed flux of UHECR,
while evading other constraints \cite{cfp}.

Superstring theory inherently possesses the ingredients
that naturally give rise to super--heavy meta--stable states.
Such states arise in string theory due to the
breaking of the non--Abelian gauge symmetries by Wilson lines.
The massless spectrum then contains states with fractional electric
charge or ``fractional'' $U(1)_{Z^\prime}$ charge \cite{ww,eln,ccf}.
The lightest states are meta--stable due to a local gauge, or discrete,
symmetry \cite{ccf,lds}.
This phenomenon is of primary importance for superstring phenomenology.
The main consequence is that it generically results in
super--massive states that are meta--stable.
The super--heavy
states can then decay via the nonrenormalizable operators,
which are produced from exchange of heavy string modes,
with lifetime $\tau_x>10^{7-17}\;{\rm years}$ \cite{eln,ben,cfp}.
The typical mass scale of the exotic states will exceed the
energy range accessible to future collider experiments by several
orders of magnitude. The exotic states are rendered super--massive
by unsuppressed mass terms \cite{fcp}, or are confined
by a hidden sector gauge group \cite{eln}.
String models may naturally produce mass scales of the
required order, $M_X\approx10^{12-13}{\rm GeV}$,
due to the existence of an hidden sector that typically contains
non--Abelian $SU(n)$ or $SO(2n)$ group factors.
The hidden sector dynamics are set by the initial
conditions at the Planck scale,
and by the hidden sector gauge and matter content,
$M_X~\sim~\Lambda_{\rm hidden}^{\alpha_s,M_S}(N,n_f).$
Finally, the fact that $M_X\sim10^{12-13}{\rm GeV}$ implies
that the super--heavy relic is not produced in thermal
equilibrium and some other production mechanism is responsible
for generating the abundance of super--heavy relic \cite{ccf}.

The forthcoming cosmic rays observatories can therefore provide
fascinating experimental probes, both to the physics
above the electroweak scale as well as to more exotic
possibilities at a much higher scale.
It is therefore
imperative to develop the theoretical tools to decipher
the data from these experiments.
Moreover, improved information on the colliding
primaries may reveal important clues on the properties
of the decaying meta--stable state, which further motivates
the development of such techniques.
In this chapter we make a
modest step in this direction, by studying possible modifications
of air shower simulations, that incorporate the possible
effects of new physics above the electroweak scale. This
is done by varying the cross section in the air shower
codes that are used by the experimentalists.
In this respect we assume here for concreteness
that the new physics above the electroweak scale remains
perturbative and preserve unitarity, as in the
case of supersymmetric extensions of the Standard Model.
This in turn is motivated by the success of supersymmetric
gauge coupling unification \cite{mssmunification} and their
natural incorporation in string theories. In the case of
supersymmetry the deviations from the Standard Model are
typically in the range of a few percent,
a quantitative indication which we take as our reference point for
study.

\subsection{Possible Developments}
Even if the forthcoming experiments
will confirm the existence of UHECR
events, it remains to be seen whether
any new physics can be inferred from the results. We
will argue that this is a very difficult question.

A possible way, in the top-down models of the UHECR
interaction is to optimize the analysis of any new high energy
primary interaction.
One should keep in mind that the information carried by the
primaries in these collisions
is strongly ``diluted'' by their interaction with the atmosphere and that
large statistical fluctuations are immediately generated both by the
randomness of the first impact, the variability
in the zenith angle of the impact, and the natural fluctuations in the
- extremely large - phase space available
at those energies. We are indeed dealing with {\em extreme} events.
These uncertainties are clearly
mirrored even in the existing Monte Carlo codes for the simulation of air
showers, and, of course,
in the real physical process that these complex Monte Carlo
implementations try, at their best,
to model (see also \cite{Sciutto} for the discussion of simulation issues) .
Part of our work will be concerned primarily  with trying to assess, by
going through extensive
air-shower simulations using existing interaction models - at the
GZK and comparable energies - the main features
of the showers, such as the multiplicities at various heights and
on the detector plane. We will
illustrate the geometry of a typical experimental
setup to clarify our method of analysis and
investigate in detail some geometrical observables.

A second part of our analysis will be centered
around the implications of a modified first impact on the multiplicities
of the subcomponents. Our analysis here is just
a first step in trying to see whether a modified first impact
cross section has any implication on the multiplicity structure of the
shower. The analysis is computationally very expensive
and has been carried out using a rather simple strategy
to render it possible. We critically comment on
our results, and suggest some possible improvements
for future studies.

\section{Simulation of air showers}
The quantification of the variability and parametric dependence
of the first impact in the formation of extensive air showers can
be discussed, at the moment,
only using Monte Carlo event generators. Although various attempts
have been made
in the previous literature to model the spectrum of a
generic ``X-particle'' decay  in various
approximations, all of them include - at some
level and with variants - some new physics
in the generation of the original spectrum.
In practice what is seen at experimental level is
just a single event, initiated by a single hadron (a proton)
colliding with an air nucleus
(mostly of oxygen or nitrogen) within the 130 km depth of the
Earth atmosphere.
Our studies will show that
the typical strength of the interaction
of the primary at the beginning of the showers - at
least using the existing Monte Carlo codes -
has to increase fairly dramatically in order
to be able to see - at the experimental level -
any new physics.

Our objective here is to assess the actual possibility, if any, to
detect new physics from the
high energy impact of the primary cosmic ray assuming that other channels
open up at those energies.
Our investigation here is focused on the case of supersymmetry,
which is the more widely accepted extension of the Standard Model.
Other scenarios are left for future studies.

We recall that at the order of the GZK cutoff,
the center of mass energy of the first collision
reaches several hundreds of TeVs and is, therefore,
above any supersymmetric scale, according to current MSSM models.
It is therefore reasonable to ask weather supersymmetric interactions
are going to have any impact on some of the observables that are
going to be measured.

We will provide enough evidence that supersymmetric effects in
total hadronic cross sections cannot
raise the hadronic nucleon nucleon cross section
above a (nominal) 100\% upper limit. We will then
show that up to such limit the fluctuations in 1)
the multiplicity distributions of the most important
components of the (ground) detected air showers and 2)
the geometric distributions of particles
on the detector are overwhelmingly affected by natural (statistical)
fluctuations in the formation of the
air showers and insignificantly by any interaction whose strength
lays below such 100\% nominal limit.

In order to proceed with our analysis we need to define a set
of basic observables which can be used in the characterization of
the shower at various heights in the atmosphere.

There are some basic features of the shower that are important in
order to understand its structure and can be summarized in: 1)
measurements of its multiplicities in the main
components; 2) measurements of the geometry of the shower.
Of course there are obvious limitations in the study of the
development of the shower at the various levels,
since the main observations are carried out on the ground. However,
using both Cerenkov telescopes and fluorescence measurements by satellites
one hopes to reconstruct the actual shape of the shower as it develops
in the atmosphere.

To illustrate the procedure that we have
implemented in order to characterize the shower,
we have assumed that the first (random) impact of the incoming
primary (proton)
cosmic ray takes place at zero zenith angle, for simplicity. We have not
carried out simulations at variable zenith, since our objective is to
describe the
main features of the shower in a rather simple, but realistic, setting.
We have chosen a
flat model for the atmosphere and variable first impacts, at energies mainly
around the GZK cutoff region. Our analysis has been based on CORSIKA
\cite{CORSIKA}
and the hadronization model chosen has been QGSJET \cite{QGSJET}.

Measurements at any level are performed taking the arrival axis (z-axis)
of the primary as center of the detector. The geometry of the shower on
the ground and
at the various selected observation levels has been always measured
with respect to this
axis. The ``center'' of the detector is, in our simulations, assumed to be the
point at which the z-axis intersects the detector plane. Below,
the word ``center'' refers to this particular geometrical setting.

The shower develops according to an obvious cylindrical symmetry around the
vertical z-axis, near the center. The various components of the showers
are characterized at any observation level by this cylindrical symmetry.
Multiplicities are plotted after integration over the azimuthal angle and shown
as a function of the distance from the core (center), in the
sense specified above.

The showers show for each subcomponent specific locations of
the maxima and widths
of the associated distributions. We will plot the positions
of the maxima along the entire spatial extent of the shower in the
atmosphere. These plots are useful in order to have an idea of
what is the geometry of the
shower in the 130 km along which it develops.

\section{Features of the Simulation}
Most of our simulations are carried out at two main energies,
$10^{19}$ and $10^{20}$ eV. Simulations have been performed on a
small cluster running a communication protocol (openmosix)
which distributes automatically the computational load. The simulation
program follows each secondary from beginning to end and is extremely time
and memory intensive. Therefore,
in order to render our computation manageable we have
implemented in CORSIKA the thinning option \cite{thinning},
which allows to select only a
fraction of the entire shower and followed its development from
start to end.
We recall that CORSIKA is, currently, the main program used by the
experimental collaborations for the analysis of cosmic rays.
The results have been corrected statistically in order to reproduce
the result of the actual (complete) shower. The CORSIKA output has been
tokenized
and then analyzed using various intermediate software written by us.
The number of events generated, even with the thinning algorithm, is huge
at the GZK energy and requires an appropriate handling of the final data.
We have performed sets of run and binned the data using bins of 80 events,
where an event is a single impact with its given parametric dependence.
The memory cost of a statistically significant set of
simulations is approximately 700 Gigabytes, having selected in our
simulations a maximal number of
observations levels (9) along the entire height of the atmosphere.

\subsection{Multiplicities on the Ground}
We show in Fig.~\ref{first} results for the
multiplicities of the photon component and of the $e^\pm$ components at
$10^{19}$ eV plotted against the distance from the core (center) of the
detector. For photons,
the maximum of the shower is around 90 meters from the center, as measured on
the plane of the detector. As evident from the plot,
the statistics is lower as we get closer to the central
axis (within the first 10
meters from the vertical axis), a feature which is typical of all
these distributions, given the
low multiplicities measured at small distances from the center.
For electrons and positrons
the maxima also lie within the first 100 meters, but slightly
closer to the center and are down
by a factor of 10 in multiplicities with
respect to the photons.
Positron distributions are
suppressed compared to
electron distributions. It is also easily noticed that the lower
tail of the photon distribution
is larger compared to the muon distribution, but all the
distributions show overall similar widths, about 1 km wide.

Multiplicities for muons
(see Fig.~\ref{second}) are a factor 1000 down with respect
to photons and 100 down with respect to electrons. The maxima of the muon
distributions are also at comparable distance as for the photons,
and both muon and antimuons
show the same multiplicity.

At the GZK energy (Figs \ref{third}, \ref{fourth})
the characteristics of the distributions of
the three main components (photons, electrons, muons)
do not seem to vary appreciably, except for the values of the multiplicities,
all increased by a factor of 10 respect to the previous plots.
The maxima of the
photons multiplicities are pushed away from the center,
together with their tails.
There appears also to be an increased
separation in the size of the multiplicities of electrons and positrons and a
slightly smaller width for the photon distribution compare to the lower energy
result ($10^{19}$ eV). We should mention
that all these gross features of the showers can possibly be tested after
a long run time of the experiment.
Our distributions have been obtained averaging over sets of 80
events with independent first impacts.

\section{Missing multiplicities?}
Other inclusive observables which are worth studying are the total
multiplicities, as measured
at ground level, versus the total energy of the primary. We show in
Fig.~\ref{fifth} a double logarithmic plot of the total multiplicities
of the various components versus the primary energy in the range
$10^{15}-10^{20}$ eV, which appears to be strikingly linear.
{}From our result it appears that the multiplicities can be
fitted by a relation of the form  y=m*x+q,
where  $y={\rm Log(N)}$ and $x= {\rm Log(E)}$
or $N(E)= 10^q  x^m$. The values of $m$ and  $q$ are given by

\beqa
\gamma : && m = 1.117 \pm 0.011;\,\,\, q = -11.02 \pm 0.19\nonumber \\
e^+: && m = 1.129 \pm 0.011; \,\,\,    q = -12.36 \pm 0.18 \nonumber \\
e^-: && m = 1.129 \pm 0.012;  \,\,\,   q = -12.17 \pm 0.20 \nonumber\\
\mu^+: && m = 0.922 \pm 0.006; \,\,\,  q = -10.12 \pm 0.10 \nonumber \\
\mu^- && m = 0.923 \pm 0.006;  \,\,\,  q = -10.15 \pm 0.09 \nonumber \\
\eeqa
where $m$ is the slope. $m$ appears to be almost universal for all the
components,
while the intercept $(q)$ depends on the component.
The photon component is clearly dominant, followed by the electron, positron
and the
two muon components which appear to be superimposed.
It would be interesting to see whether missing energy effects, due, for
instance, to
an increased multiplicity rate toward the production of weakly
interacting particles can modify this type of inclusive measurements,
thereby predicting variations in the slopes of the multiplicities  respect
to the Monte Carlo predictions reported here. One could entertain
the possibility that a failure
to reproduce this linear behaviour could be a serious problem for the theory
and a possible signal of new physics. Given the large sets
of simulations that we have performed, the statistical
errors on the Monte Carlo results are quite small, and the Monte Carlo
prediction
appear to be rather robust. The difficulty of these measurements however,
lay mainly in the energy reconstruction of the primary, with the possibility of
a
systematic error. However, once the reconstruction of the energy of the primary
is under control among the various UHECR detectors, with a global calibration,
these measurements could be a possible test for new physics. At the moment,
however, we still do not have a quantification of the deviation
from this behaviour such as that induced by supersymmetry or other
competing theoretical models.

\subsection{Directionality of the Bulk of the Shower}
\begin{figure}[tbh]
{\centering
\resizebox*{12cm}{!}{\rotatebox{-90}{\includegraphics{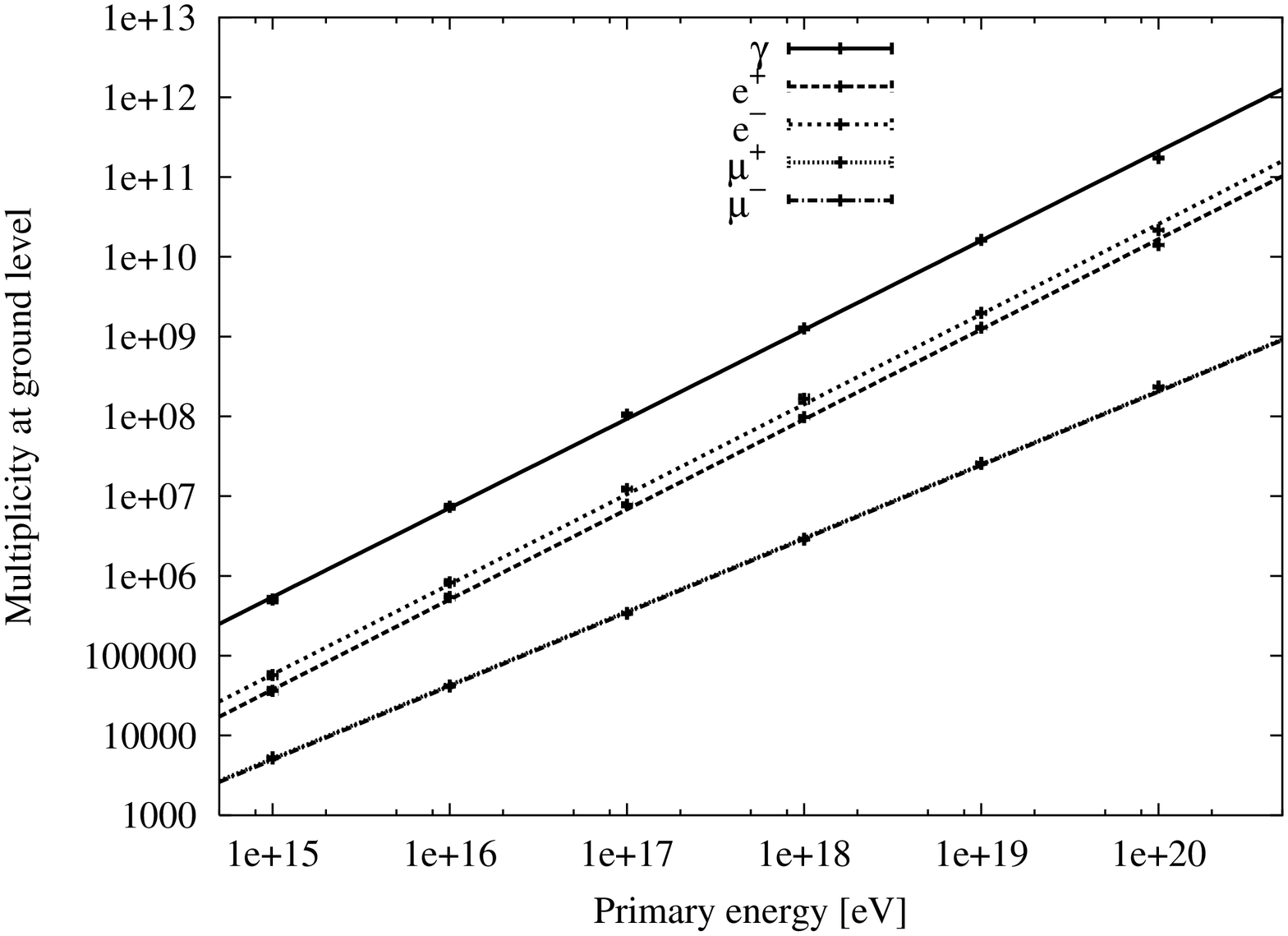}}} \par}
\caption{Multiplicities of photons, \protect\( e^{\pm }\protect \), \protect\(
\mu ^{\pm }\protect \)
at the ground level as a function of the primary energy. Due to the
logarithmic scale, \protect\( \mu ^{+}\protect \) and \protect\( \mu
^{-}\protect \)
look superimposed.}
\label{fifth}
{\centering
\resizebox*{12cm}{!}{\rotatebox{-90}{\includegraphics{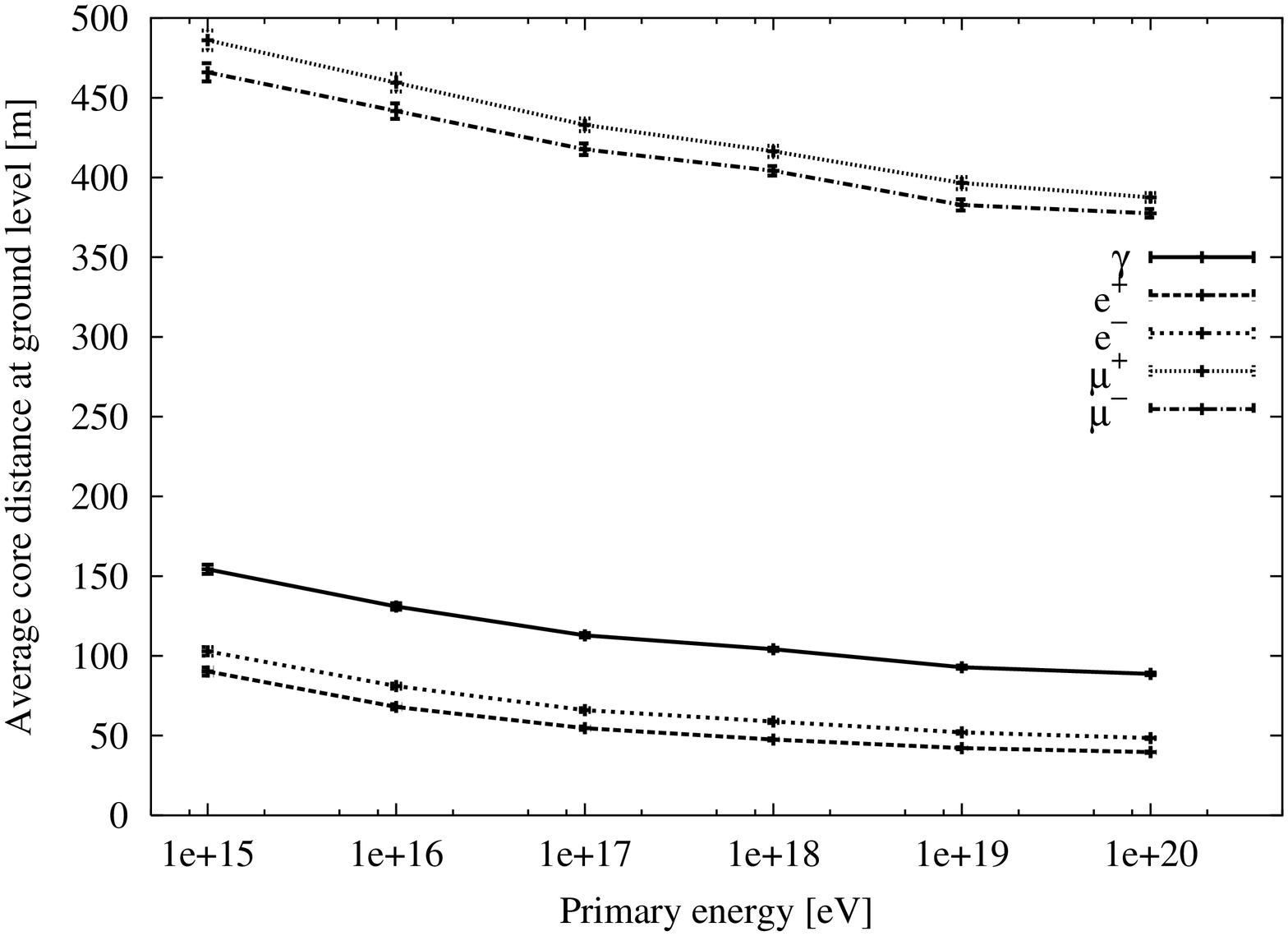}}} \par}
\caption{Average distance from the core of the shower of photons, \protect\(
e^{\pm }\protect \),
\protect\( \mu ^{\pm }\protect \) at the ground level as a function
of the primary energy.}
\label{sixth}
\end{figure}

Another geometrical feature of the shower is the position
of its bulk
(measured as the opening of the radial cone at the radial distance
where the maximum is achieved)
as a function of the energy. This feature is illustrated in Fig.~\ref{sixth}.
Here we have plotted the averaged location
of the multiplicity distribution as a function of the energy of the incoming
primary. The geometrical center of the distributions
tend to move slightly toward the vertical axis (higher collimation) as the
energy
increases. From the same plot it appears that
the distributions of electrons, positrons and photons
are closer to the center of the detector compared to the muon-antimuon
distributions. As shown in the figure, the
statistical errors on these results appear to be rather small.

\subsection{The Overall Geometry of the Shower}
As the shower develops in the atmosphere, we can monitor
both the multiplicities of the various components and the average
location of the bulk of the distributions at various observation levels.
As we have mentioned above, we choose up to 9 observation
levels spaces at about 13 km from one another. The lowest observation
level is, according to our conventions, taken to coincide
with the plane of the detector.

\begin{figure}[tbh]
{\centering \subfigure[
]{\resizebox*{12cm}{!}{\rotatebox{-90}{\includegraphics{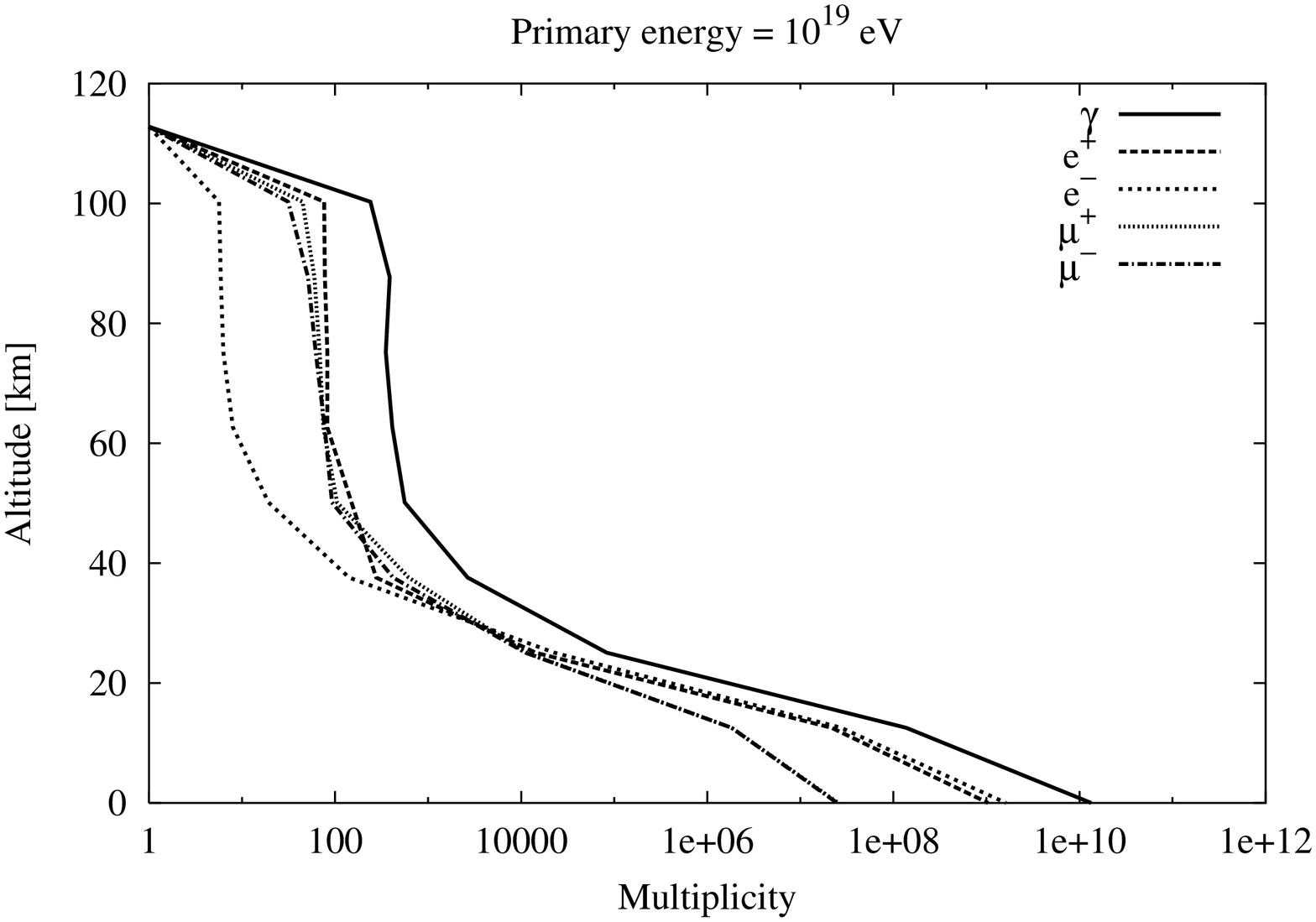}}}}
\par}
{\centering \subfigure[
]{\resizebox*{12cm}{!}{\rotatebox{-90}{\includegraphics{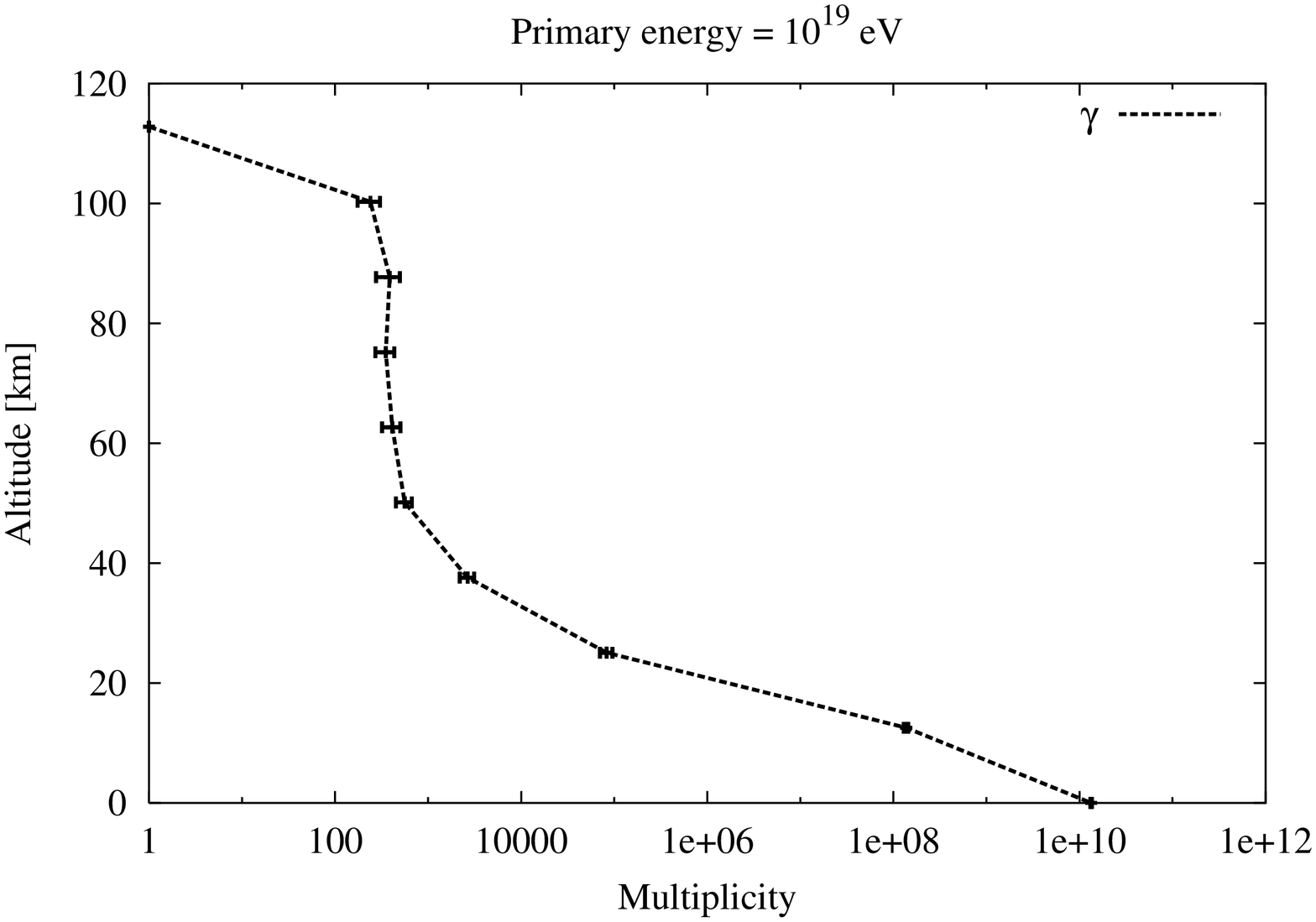}}}}
\par}
\caption{Multiplicities of photons, \protect\( e^{\pm }\protect \), \protect\(
\mu ^{\pm }\protect \)
at various levels of observations for a primary energy of \protect\(
10^{19}\protect \)
eV. The first impact is forced to occur at the top of the atmosphere.
In the subfigure (b) we show the uncertainties just in the case of
the photons.}
\label{seventh}
\end{figure}

We show in Fig.~\ref{seventh}a a complete simulation of the shower using a set
of 9
observation levels, as explained above, at an energy of $10^{19}$ eV.
In the simulation we assume that a first impact occurs near the top of the
atmosphere
at an eight of 113 km and we have kept this first impact fixed.
The multiplicities
show for all the components a rather fast growth within the first 10 km of
crossing of the atmosphere after the impact,
with the photon components growing faster compared to the others.
The electron component also grows rather fast, and a similar behaviour is
noticed for the muon/antimuon components, which show a linear growth
in a logarithmic scale (power growth). In the following 40 km downward,
from a height of 100 km down to 60 km, all the components largely conserve
their
multiplicities. Processes of regeneration of the various components and their
absorption seem to balance. For the next 20 km, from a height of 60 km down to
40 km, all the components starts to grow,
with the photon component showing a faster (power growth) with the traversed
altitude. Slightly below 40 km of altitude the multiplicities of three
components seem to merge
(muons and electrons), while the photon component is still dominant
by a factor of 10 compared to the others. The final development of the
air shower is characterized by a drastic growth
of all the components, with a final reduced muon component, a larger electron
component and a dominant photon component. The growth
in this last region (20 km wide) and in the first 20 km after
the impact of the primary -in the upper part of the atmosphere- appear to be
comparable.
The fluctuations in the multiplicities of the components
are rather small at all levels, as shown (for the photon case) in
Fig.\ref{seventh}b.

\begin{figure}[tbh]
{\centering \subfigure[
]{\resizebox*{12cm}{!}{\rotatebox{-90}{\includegraphics{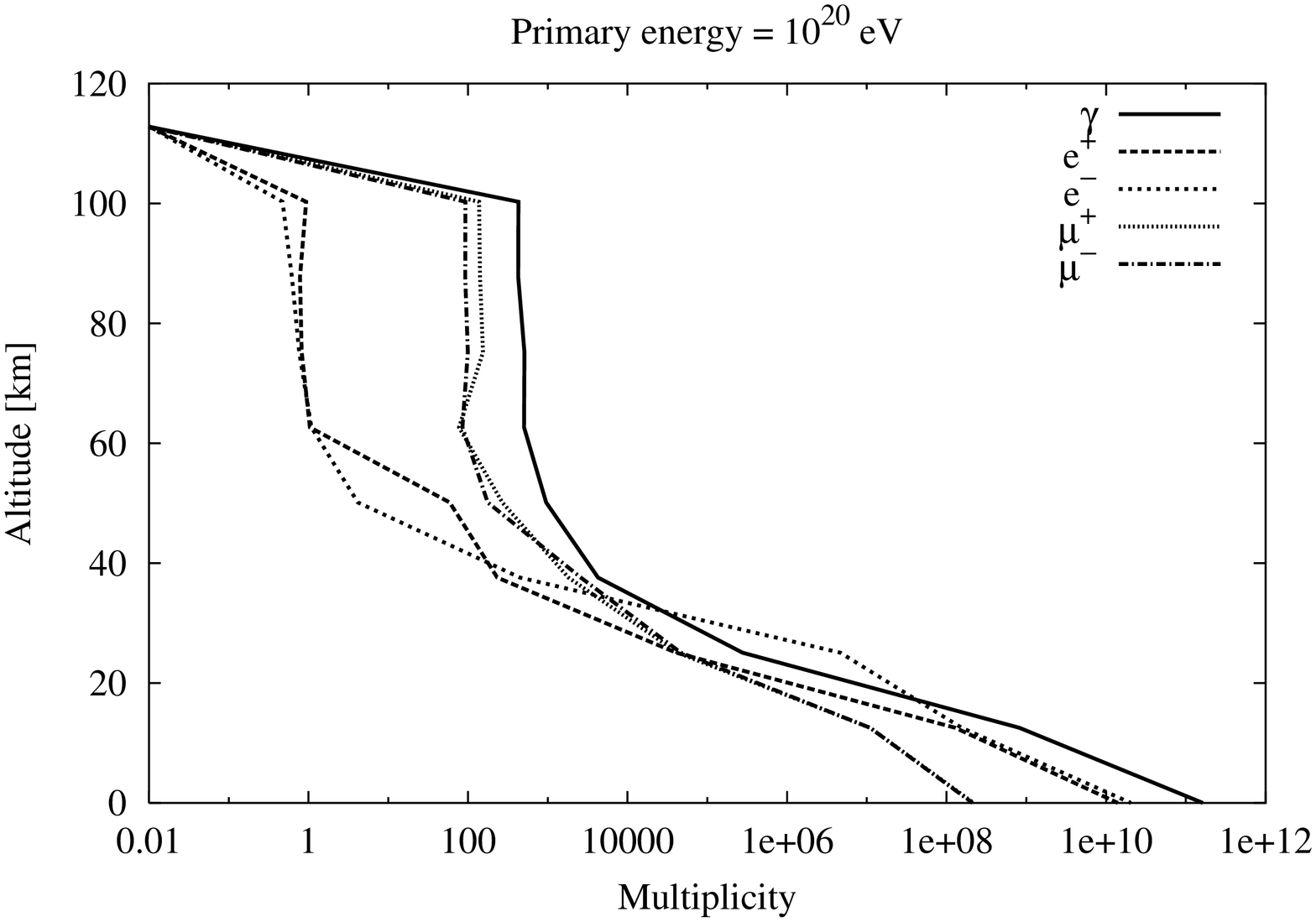}}}}
\par}

{\centering \subfigure[
]{\resizebox*{12cm}{!}{\rotatebox{-90}{\includegraphics{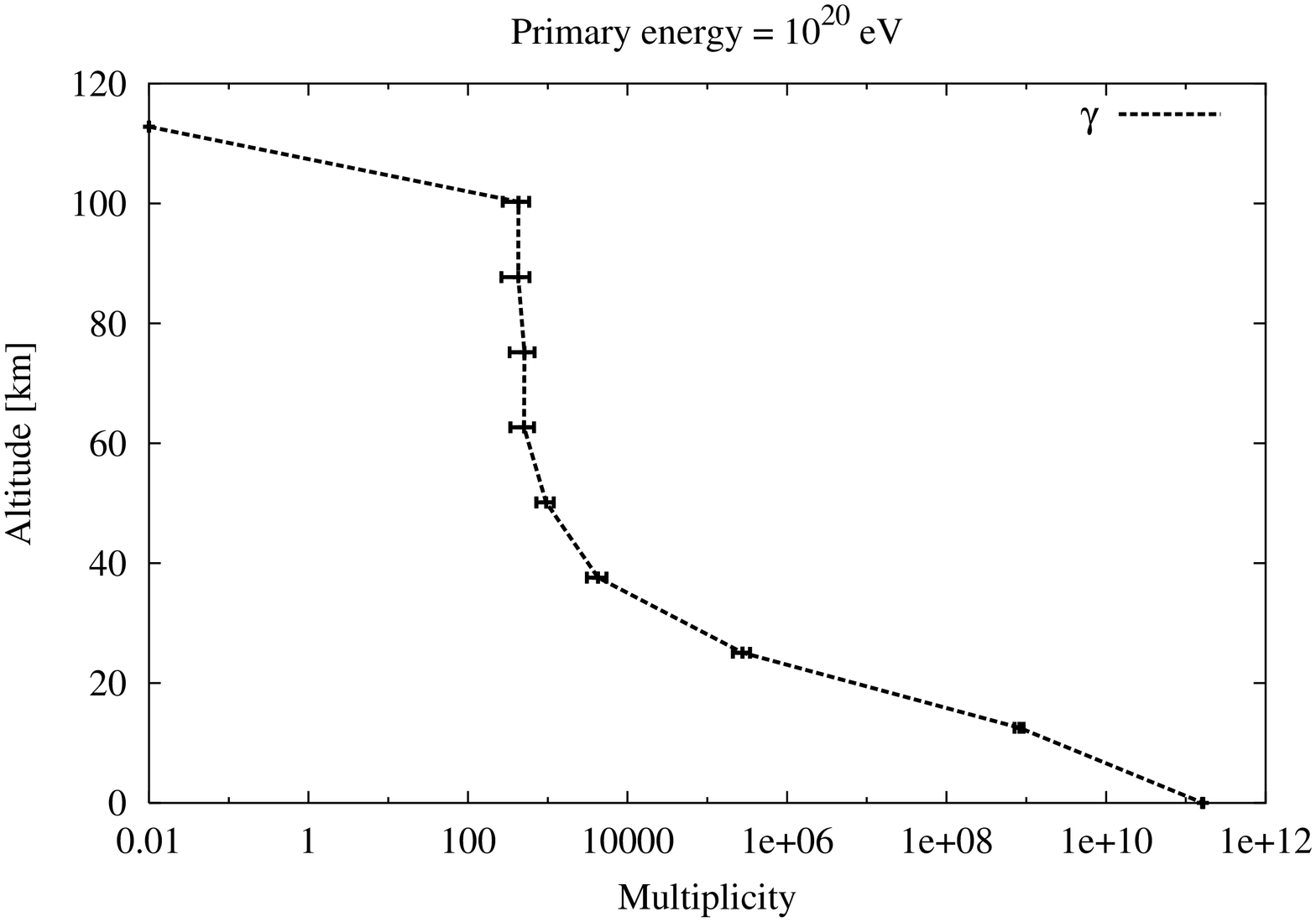}}}}
\par}
\caption{Multiplicities of photons, \protect\( e^{\pm }\protect \), \protect\(
\mu ^{\pm }\protect \)
at various levels of observations for a primary energy of \protect\(
10^{20}\protect \)
eV. The first impact is forced to occur at the top of the atmosphere.
In the subfigure (b) we show the uncertainties just in the case of
the photons.}
\label{eigth}
\end{figure}

As we reach the GZK cutoff, increasing the energy of the primary
by a factor of 10, the pattern just discussed in Fig.~\ref{seventh}a is
reproposed in Fig.~\ref{eigth}a, though -in this case- the growth of the
multiplicities
of the subcomponents in the first 20 km from the impact and in the last 20 km
is much stronger. The electron and the muon components appear
to be widely separated, while the electron and positron components
tend to be more overlapped. To the region of the first impact
-and subsequent growth- follows an intermediate region, exactly
as in the previous plots, where the two phenomena of production and absorption
approximately balance one another and the multiplicities undergo minor
variations. The final
growth of all the components is, a this energy, slightly anticipated compared
to
Fig. \ref{seventh}a, and starts to take place at a height of 40 km and above
and continues
steadily until the first observation level. The photon remains the
dominant component, followed by the electron and the muon component. Also
in this case the fluctuations (pictured for the photon component only, Fig.
\ref{eigth}b)
are rather small.

\subsection{The Opening of the Shower}
\begin{figure}[tbh]
{\centering \subfigure[
]{\resizebox*{12cm}{!}{\rotatebox{-90}{\includegraphics{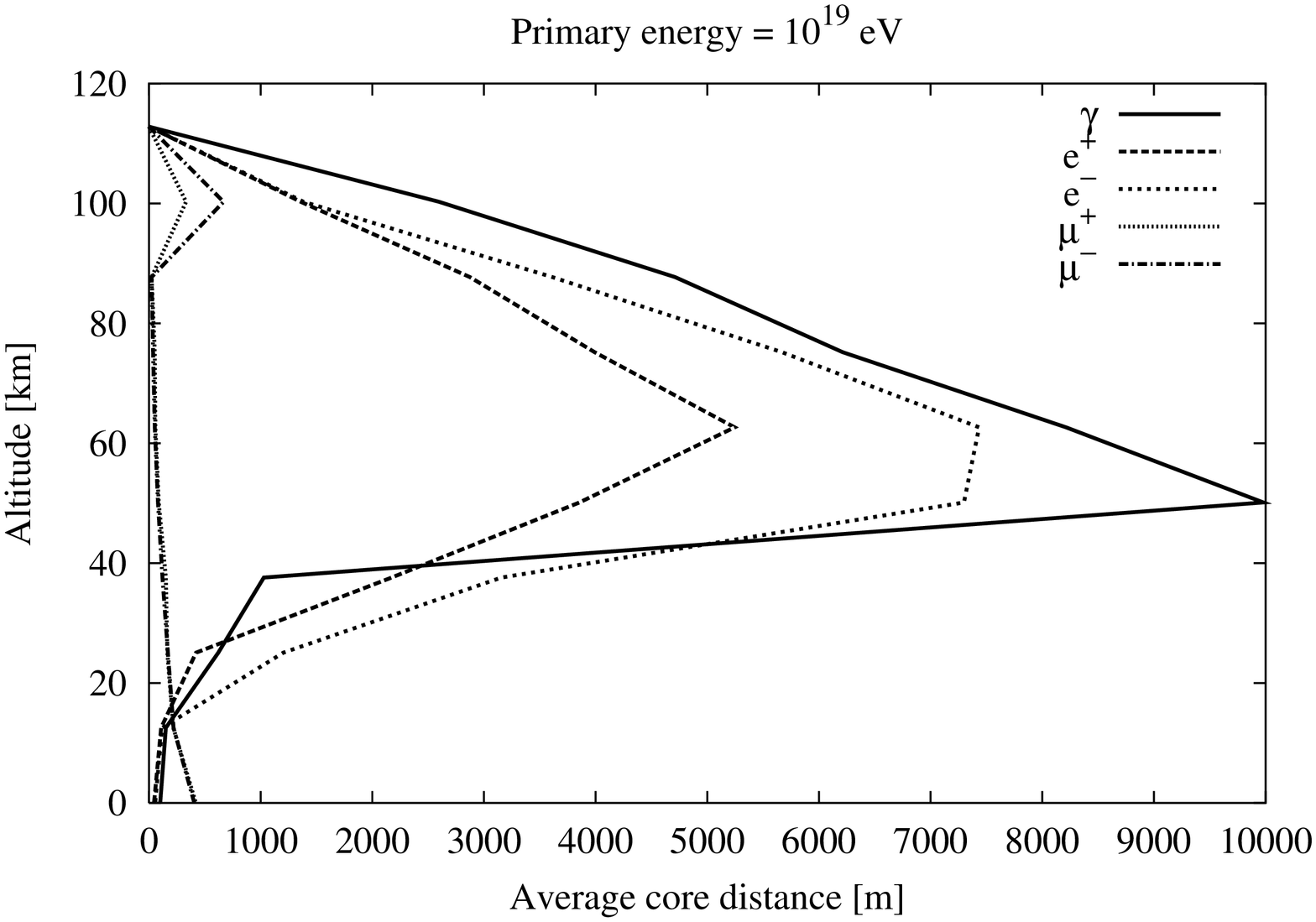}}}} \par}
{\centering \subfigure[
]{\resizebox*{12cm}{!}{\rotatebox{-90}{\includegraphics{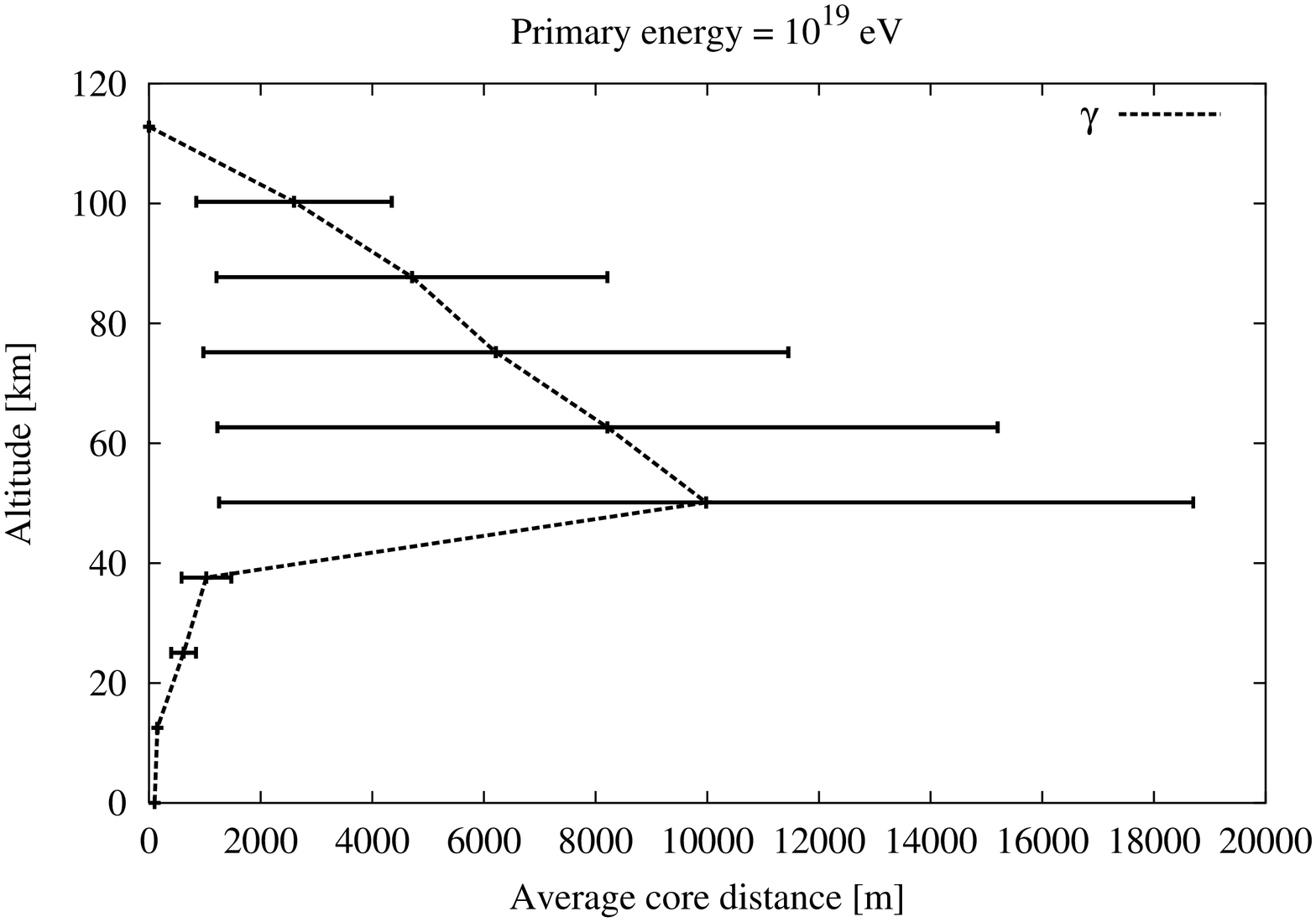}}}}
\par}
\caption{Average core distances of photons, \protect\( e^{\pm }\protect \),
\protect\( \mu ^{\pm }\protect \) at various levels of observations
for a primary energy of \protect\( 10^{19}\protect \) eV. The first
impact is forced to occur at the top of the atmosphere. In the subfigure
(b) we show the uncertainties just in the case of the photons.}
\label{ninth}
\end{figure}

\begin{figure}[tbh]
{\centering \subfigure[
]{\resizebox*{12cm}{!}{\rotatebox{-90}{\includegraphics{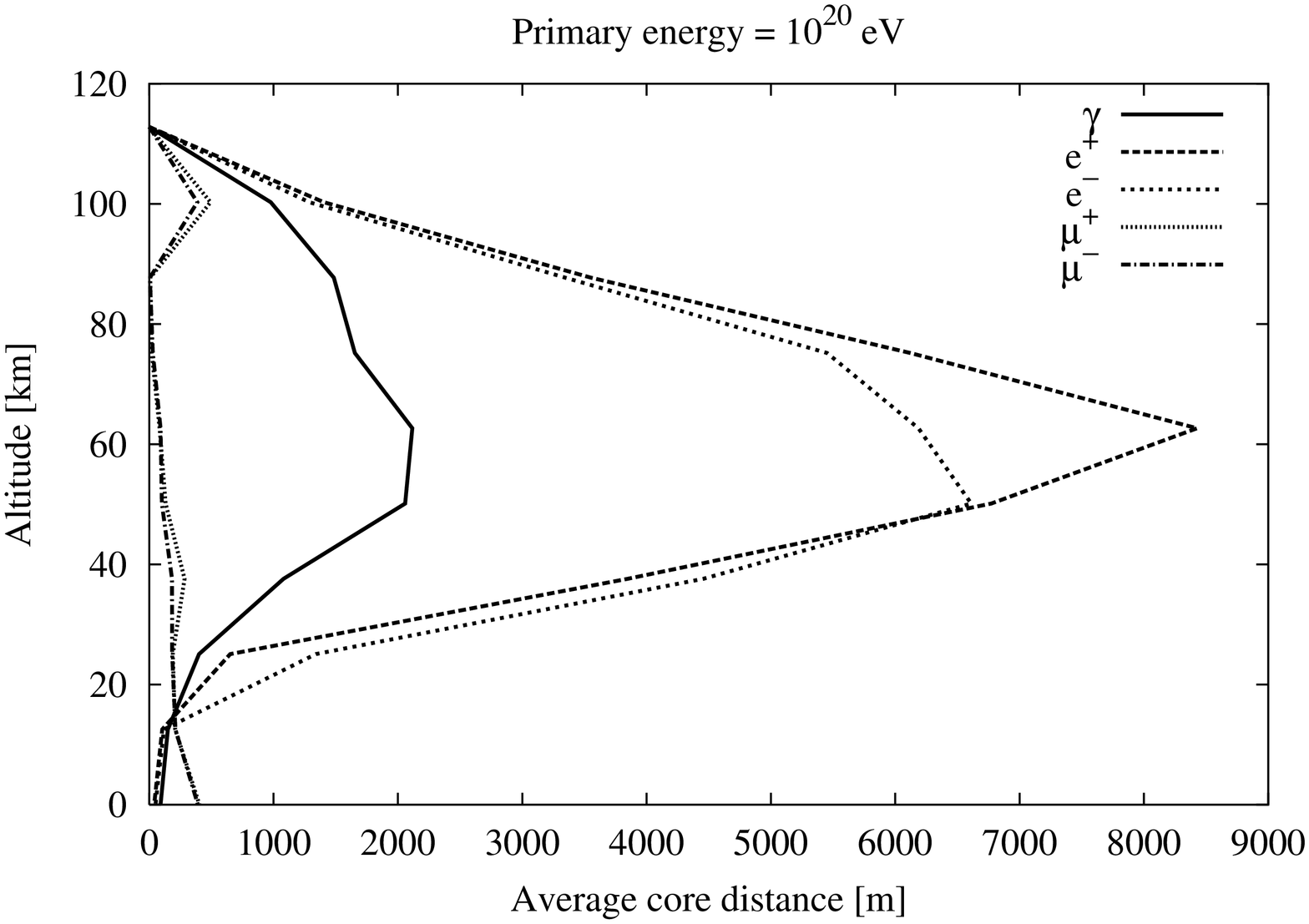}}}} \par}
{\centering \subfigure[
]{\resizebox*{12cm}{!}{\rotatebox{-90}{\includegraphics{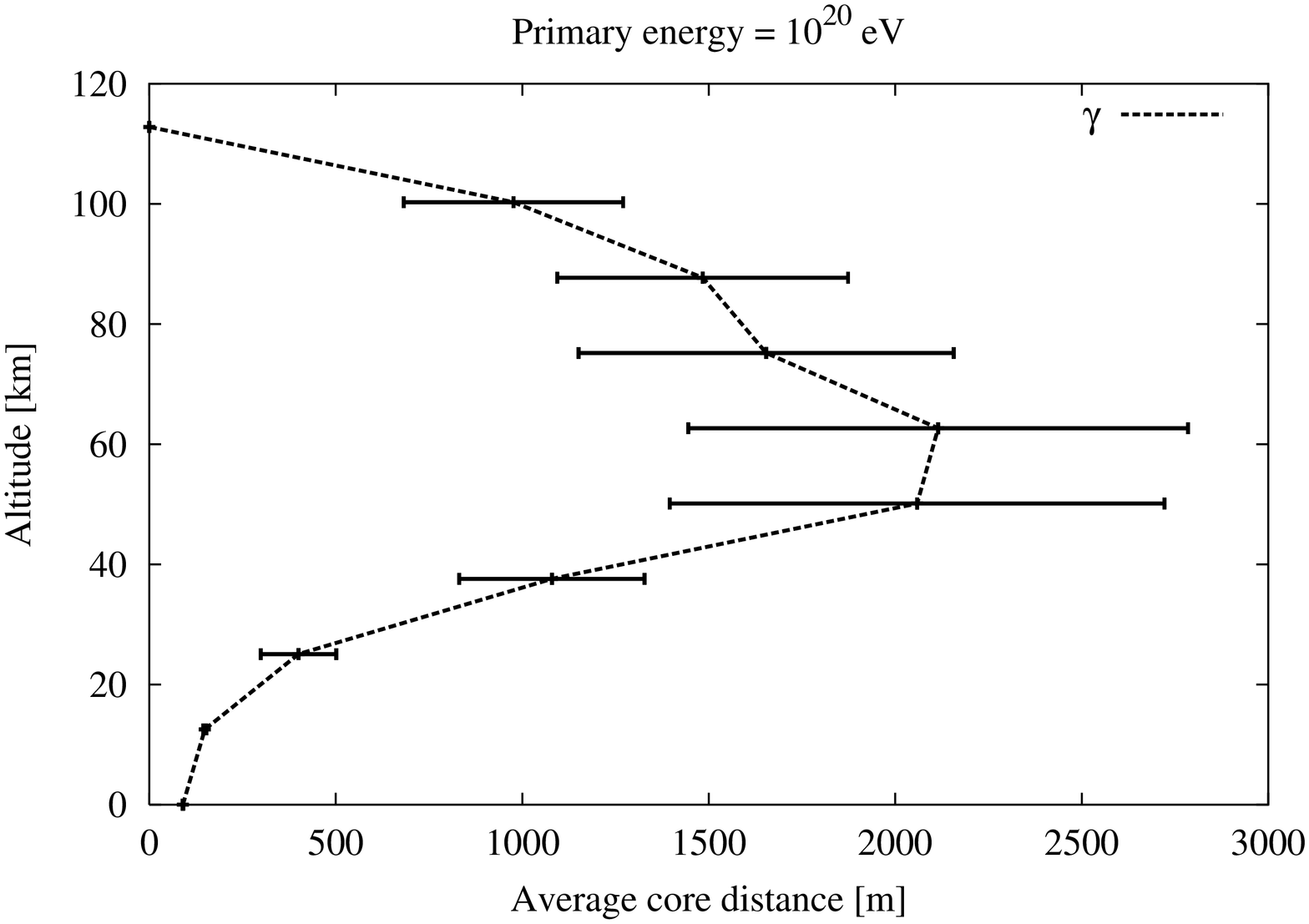}}}}
\par}
\caption{Average core distances of photons, \protect\( e^{\pm }\protect \),
\protect\( \mu ^{\pm }\protect \) at various levels of observations
for a primary energy of \protect\( 10^{20}\protect \) eV. The first
impact is forced to occur at the top of the atmosphere. In the subfigure
(b) we show the uncertainties just in the case of the photons.}
\label{tenth}
\end{figure}

In our numerical study the geometrical center of each component of the shower
is identified through
a simple average with respect to all the distances from the core
\beq
R_M\equiv=\frac{1}{N}\sum_i R_i
\eeq
where $N$ is the total multiplicity at each selected observation level, $R_i$
is the position
of the produced particle along the shower and $i$ runs over the single events.
This analysis has been carried out for 9 equally
spaced levels and the result of this study are shown in
Figs.~\ref{ninth},\ref{tenth} and
\ref{eigth} at two different values of energy ($10^{19}$ and $10^{20}$ eV). The
opening of the various components are clearly identified by these plots. We
start from Fig.~\ref{ninth}.
We have taken in this figure an original point of impact at a height of 113 Km,
as in the previous simulations.
It is evident that the photon component of the shower tends to spread rather
far and within
the first 20 km of depth into the atmosphere
has already reached an extension of about 2 km;
reaching a lateral extension of
10 km within the first 60 km of crossing of the atmosphere.

Starting from a height of 50 km down to 10 km,
the shower gets reabsorbed (turns toward the center) and is characterized
by a final impact which lays rather close to the vertical axis. Electrons
and photons follow a similar behaviour, except that for electrons whose
lateral distribution in the first section of the development of the shower is
more reduced.
The muonic
(antimuons) subcomponents appear to have a rather small opening and develop
mostly along the vertical axis of impact. In the last section of the shower
all the components get aligned near the vertical axis and hit the detector
within 1 km.

Few words should be said about the fluctuations.
At $10^{19}$ eV, as shown in Fig.~\ref{ninth}b, the fluctuations are rather
large,
especially in the first part of the development of the shower. These turn
out to be more pronounced for photons, whose multiplicity growth is large
and very broad. As we increase the energy of the primary to $10^{20}$ eV the
fluctuations in the lateral distributions (see Fig.~\ref{tenth}b) are overall
reduced, while the lateral distributions of the photons appear to be
drastically reduced (Fig.~\ref{tenth}a).

\section{Can we detect new physics at Auger?}
There are various issues that can be addressed,
both at theoretical and at experimental level, on this point,
one of them being an eventual confirmation
of the real existence of events above the cutoff.
However, even if these measurement will confirm their existence, it
remains yet to be seen whether any additional new physics can be inferred
just from an analysis of the air shower.
A possibility might be supersymmetry or any new underlying interaction,
given the large energy available in the first impact.
We recall that the spectrum of the decaying X-particle
(whatever its origin may be),
prior to the atmospheric impact of the
UHECR is of secondary relevance, since
the impact is always due to a single proton.
Unless correlations are found among different events - and by
this we mean that
a large number of events should be initiated by special types of primaries -
we tend to believe that effects due to new interactions are likely to
play a minor role.

In previous works we have analyzed in great detail the
effects of supersymmetry in the formation
of the hadronic showers. These studies, from our previous experience
\cite{CC,CC1,fragfun},
appear to be rather complex since they involve several possible
intermediate and large
final scales and cannot possibly be conclusive.
There are some obvious doubts
that can be raised over these analysis, especially when the DGLAP
equations come
to be extrapolated to such large evolution scales,
even with a partial resummation
of the small-x logarithms. In many cases results obtained in this
area of research by extrapolating results from collider phenomenology
to extremely
high energies should be taken with extreme caution in order to reduce
the chances of inappropriate hasty conclusions.
What is generally true in a first approximation is that supersymmetric effects
do appear to be mild \cite{CC,CC1,fragfun}. Rearrangements in the fragmentation
spectra or supersymmetric effects in initial state scaling violations are down
at the few percent level. We should mention that the generation of
supersymmetric
scaling violations in parton distributions, here considered to be the bulk of
the supersymmetric
contributions,
are rather mild if the entrance into the SUSY region takes
place ``radiatively''
as first proposed in \cite{CC,CC1}. This last picture might change in
favour of a more
substantial signal if threshold enhancements are also included in the
evolution,
however this and other related points have not yet been analyzed
in the current literature.

\subsection{The Primary Impact and a Simple Test}
Our objective, at this point, is to describe the structural properties
of the shower with an emphasis on the dynamics of the first impact of
the primary with the
atmosphere, and at this stage one may decide to look for the emergence of
possible new interactions, the most popular one being supersymmetry.

One important point to keep into consideration is
that the new physical signal carried by the primaries in these collisions
is strongly ``diluted'' by their interaction with the atmosphere and that
large
statistical fluctuations are immediately generated both by the
randomness of the first impact, the variability
in the zenith angle of the impact, and the -extremely
large- phase space available at those energies in terms of
fragmentation channels. We can't possibly
underestimate these aspects of the dynamics,
which are at variance with previous
analysis, where the search for supersymmetric effects
(in the vacuum) seemed
to ignore the fact that our detectors are on the ground and not in space.

For this reason we have resorted to a simple and realistic analysis of
the structure of
air showers as can be obtained from the current Monte Carlo.

The simplest way to test whether a new interaction at the first proton-proton
impact can have any effect on the shower is to modify the cross section
at the first atmospheric impact using CORSIKA in combination with
some
of the current hadronization models which are supposed to work at and around
the GZK cutoff. There are obvious limitations in this approach, since none of
the
existing codes incorporates any new physics beyond the standard model,
but this is possibly one of the simplest ways to proceed.
For this purpose we have used SYBILL \cite{SIBYLL},
with the appropriate modifications discussed below.

To begin let's start by recalling one feature of the behaviour
of the hadronic
cross section ($p\,p$ or $p\,\bar{p}$) at asymptotically large energies.
There is evidence (see \cite{Ferreiro}) demonstrating
a saturation of the Froissart bound of the total cross section
with rising total energy, $s$. This ${\rm log}^2(s)$ growth
of the total cross section is usually embodied into many of the hadronization
models used in the analysis of first impact and leaves, therefore,
little room for other substantial growth with the opening
of new channels, supersymmetry being one of them.
We should also mention that various significant elaborations \cite{Alan1}
on the growth of the total cross section and the soft pomeron dominance
have been discussed in the last few years and the relation
of this matter \cite{Alan2} with the UHECR events is of utmost relevance.

With this input in mind we can safely
``correct'' the total cross section by at most a (nominal)
factor of 2 and study whether these nominal changes can have any impact on the
structural properties of the showers.

We run simulations on the showers generated by this modification and try
to see whether there is any signal in the multiplicities which points toward a
structural (multiplicity, geometrical) modification of the air showers in all
or some of its subcomponents. For this purpose we have performed runs at two
different
energies, at the GZK cutoff and 1 decade below, and analyzed the effects
due to these changes.

We show in two figures results on the multiplicities,
obtained at zero zenith angle, of some selected particles
(electrons and positrons, in our case, but similar results hold
for all the dominant components of the final shower) obtained from a large
scale simulation of air showers at and around the
GZK cutoff. We have used the simulation code CORSIKA for this purpose.

\begin{figure}[t]
{\centering
\resizebox*{12cm}{!}{\rotatebox{-90}{\includegraphics{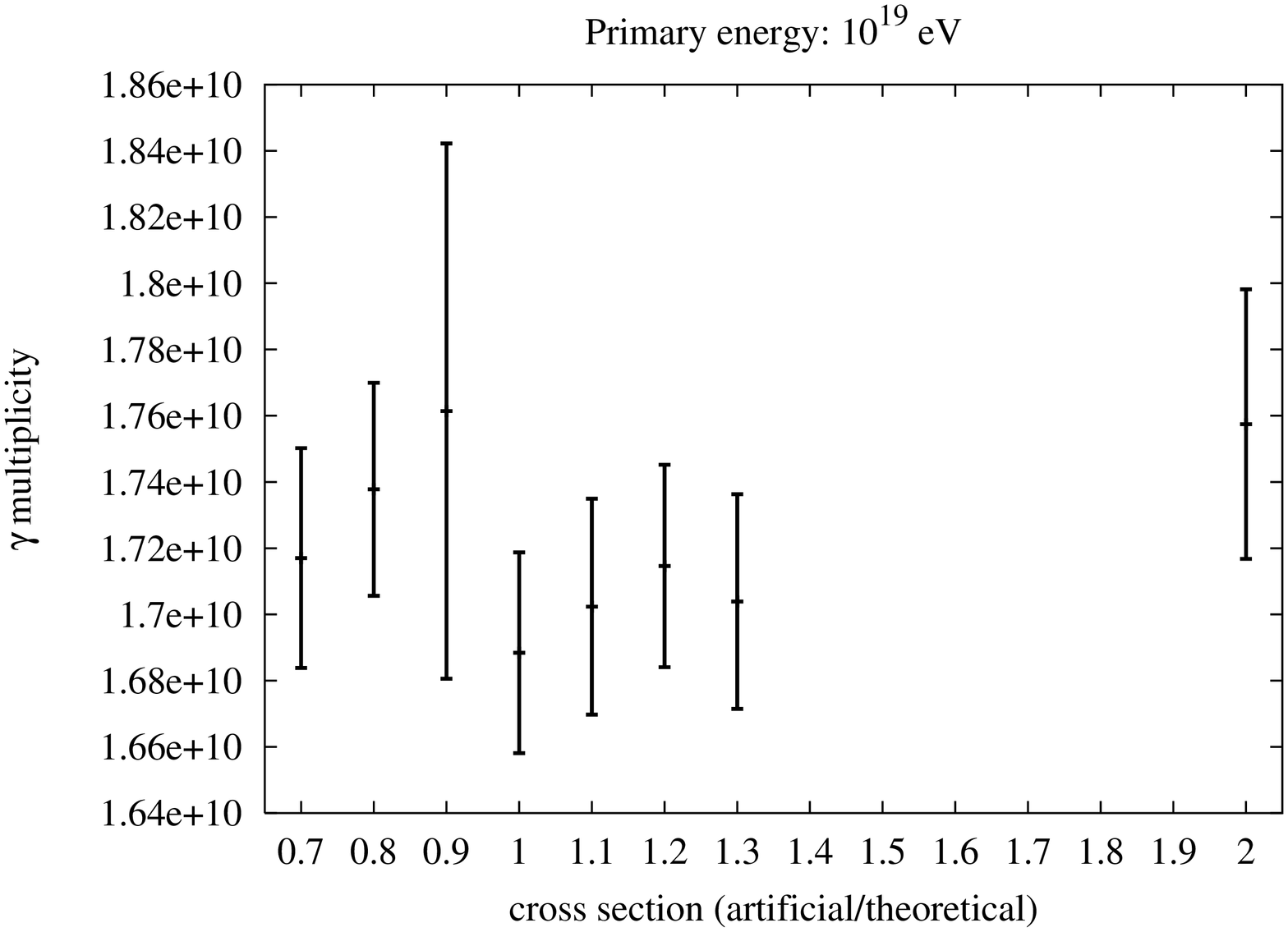}}} \par}
{\centering
\resizebox*{12cm}{!}{\rotatebox{-90}{\includegraphics{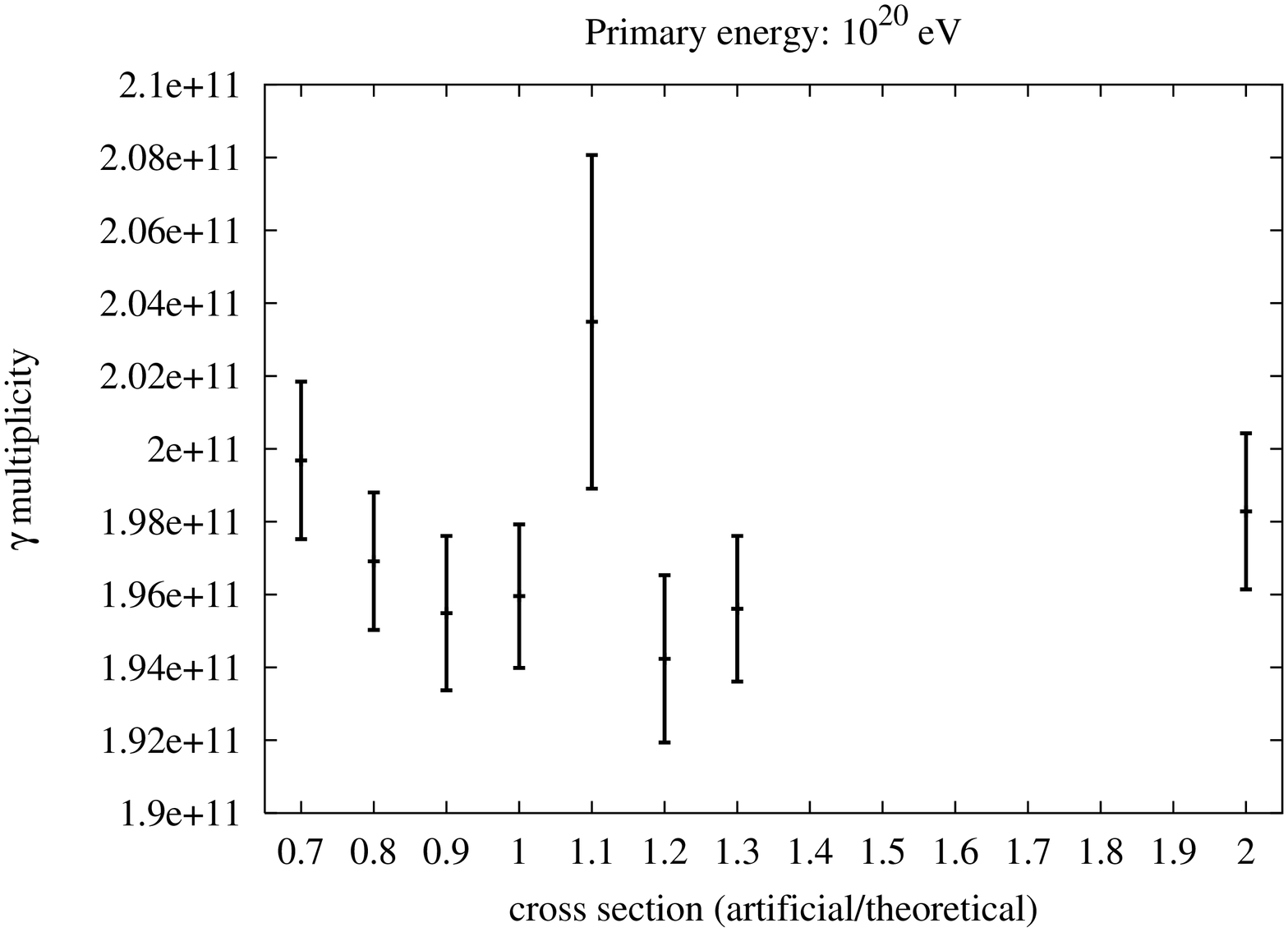}}} \par}
\caption{variation of the photon multiplicity as a function
of the first impact cross section at $10^{19}$ and at $10^{20}$ eV}
\label{multiphoton}
\end{figure}

\begin{figure}[t]
{\centering \resizebox*{12cm}{!}{\rotatebox{-90}{\includegraphics{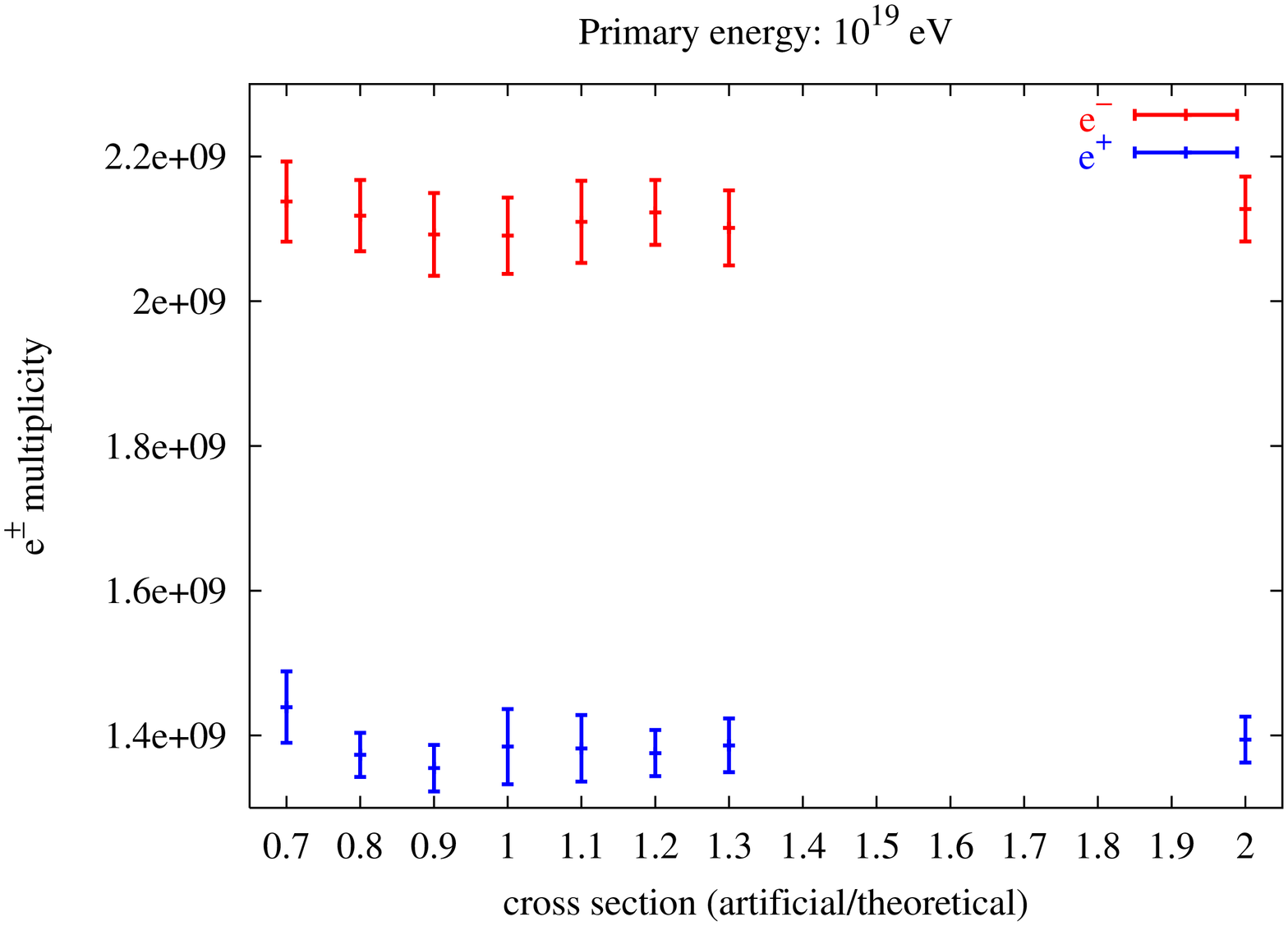}}}
\par}
{\centering \resizebox*{12cm}{!}{\rotatebox{-90}{\includegraphics{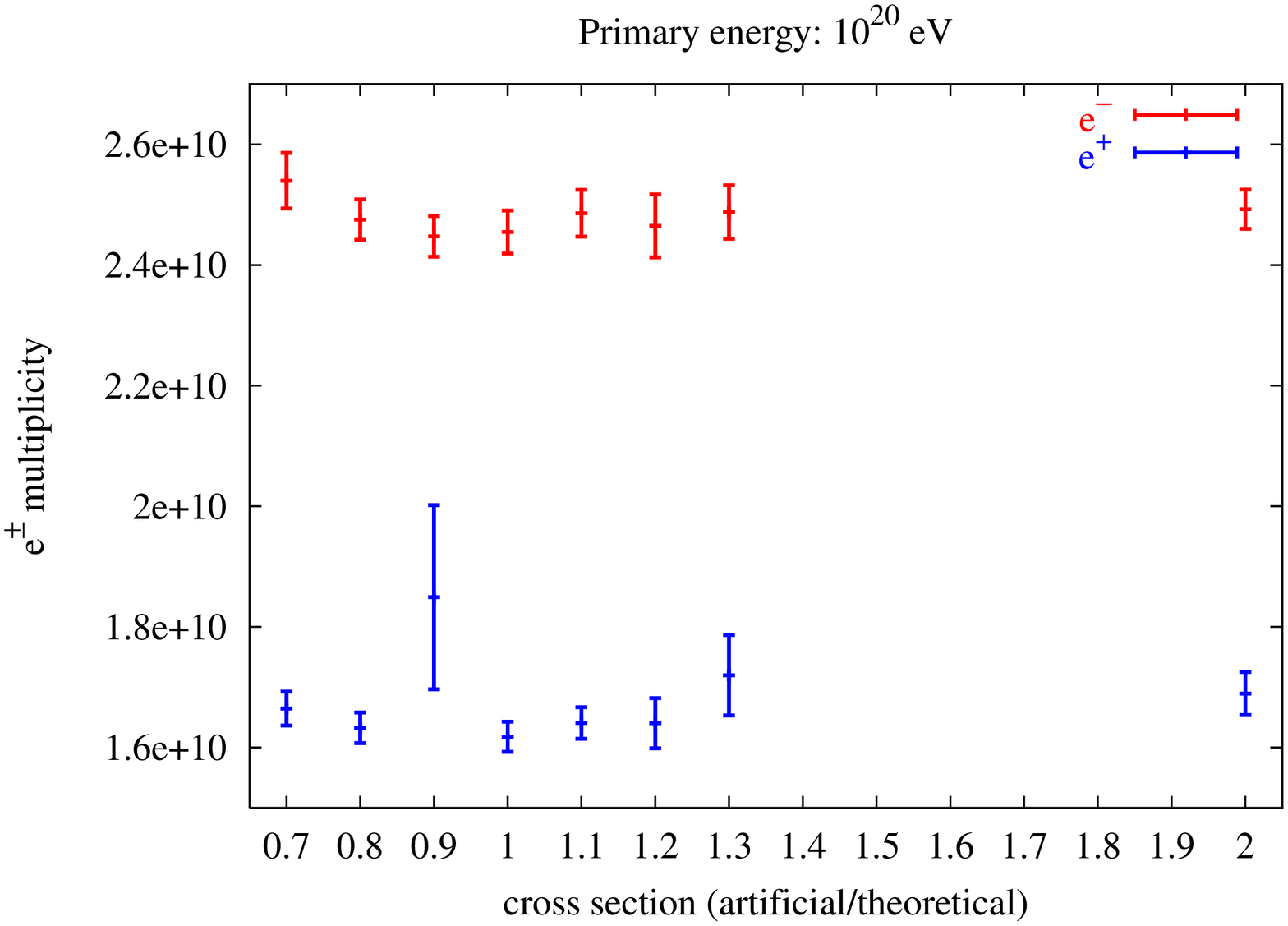}}}
\par}
\caption{variation of the $e^\pm$ multiplicity as a function
of the first impact cross section at $10^{19}$ and at $10^{20}$ eV}
\label{multielectron}
\end{figure}

\begin{figure}[t]
{\centering \resizebox*{12cm}{!}{\rotatebox{-90}{\includegraphics{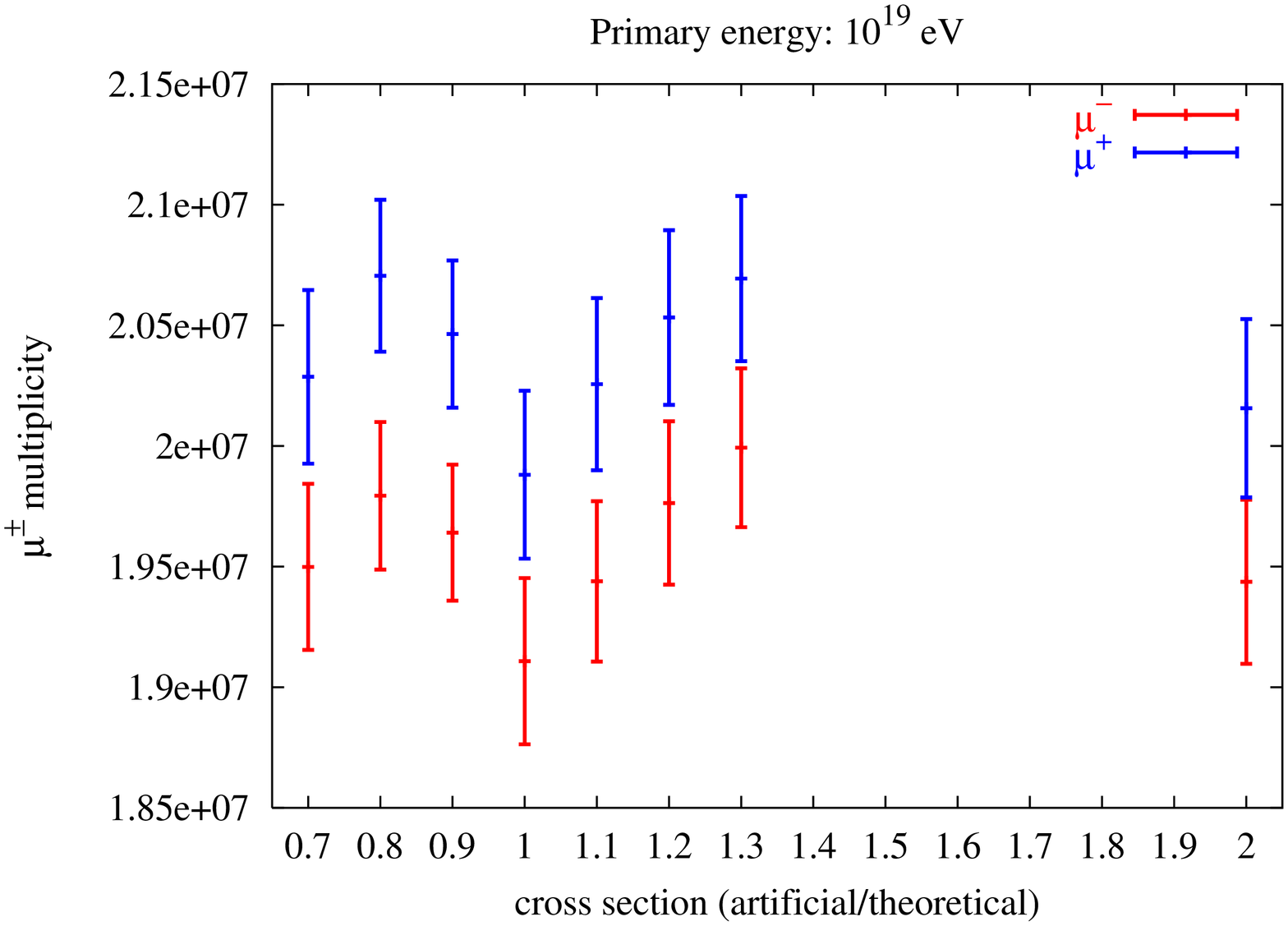}}}
\par}
{\centering \resizebox*{12cm}{!}{\rotatebox{-90}{\includegraphics{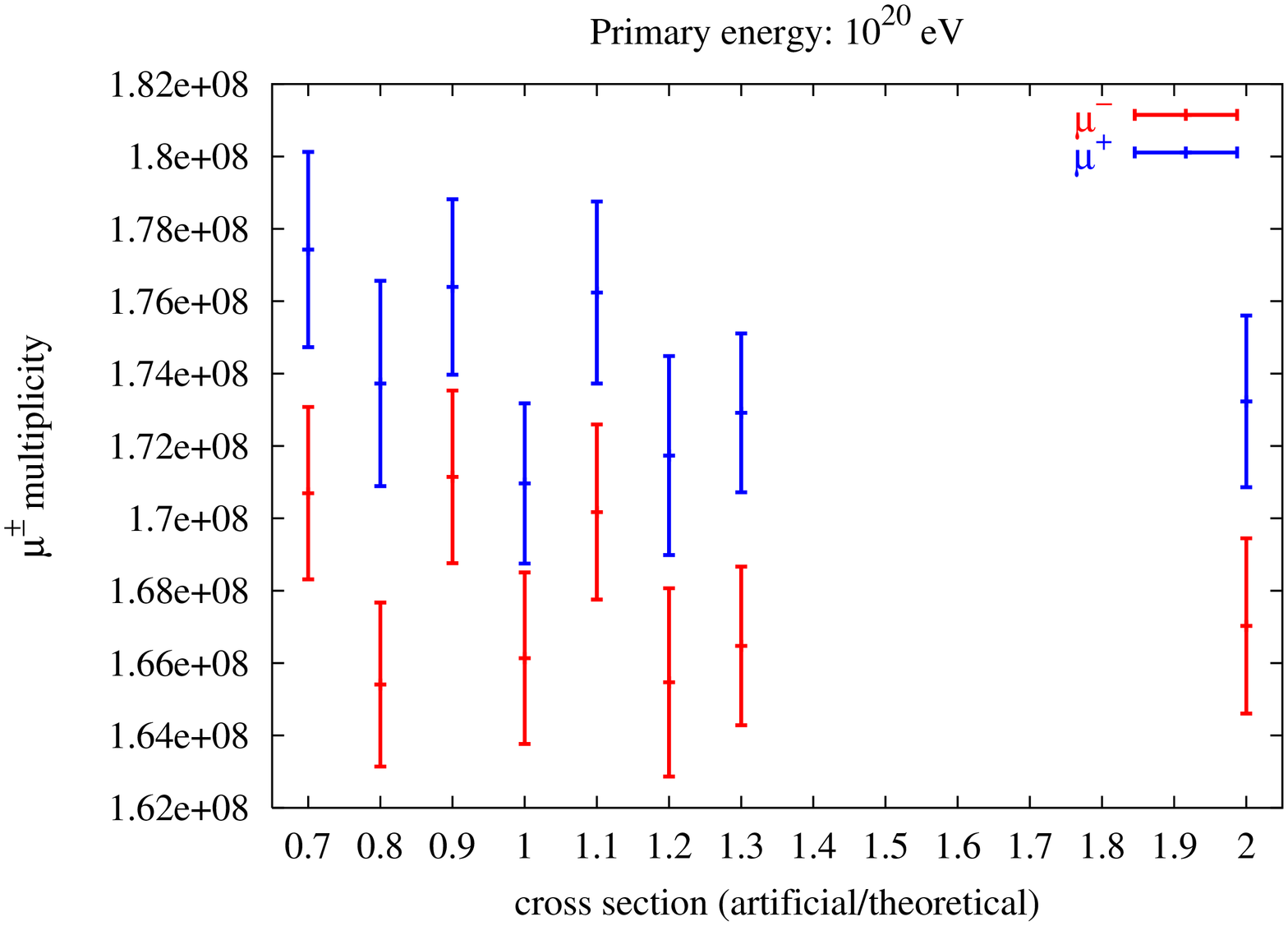}}}
\par}
\caption{variation of the $\mu^\pm$ multiplicity as a function
of the first impact cross section at $10^{19}$ and at $10^{20}$ eV}
\label{multimuon}
\end{figure}

\begin{figure}[t]
{\centering \resizebox*{12cm}{!}{\rotatebox{-90}{\includegraphics{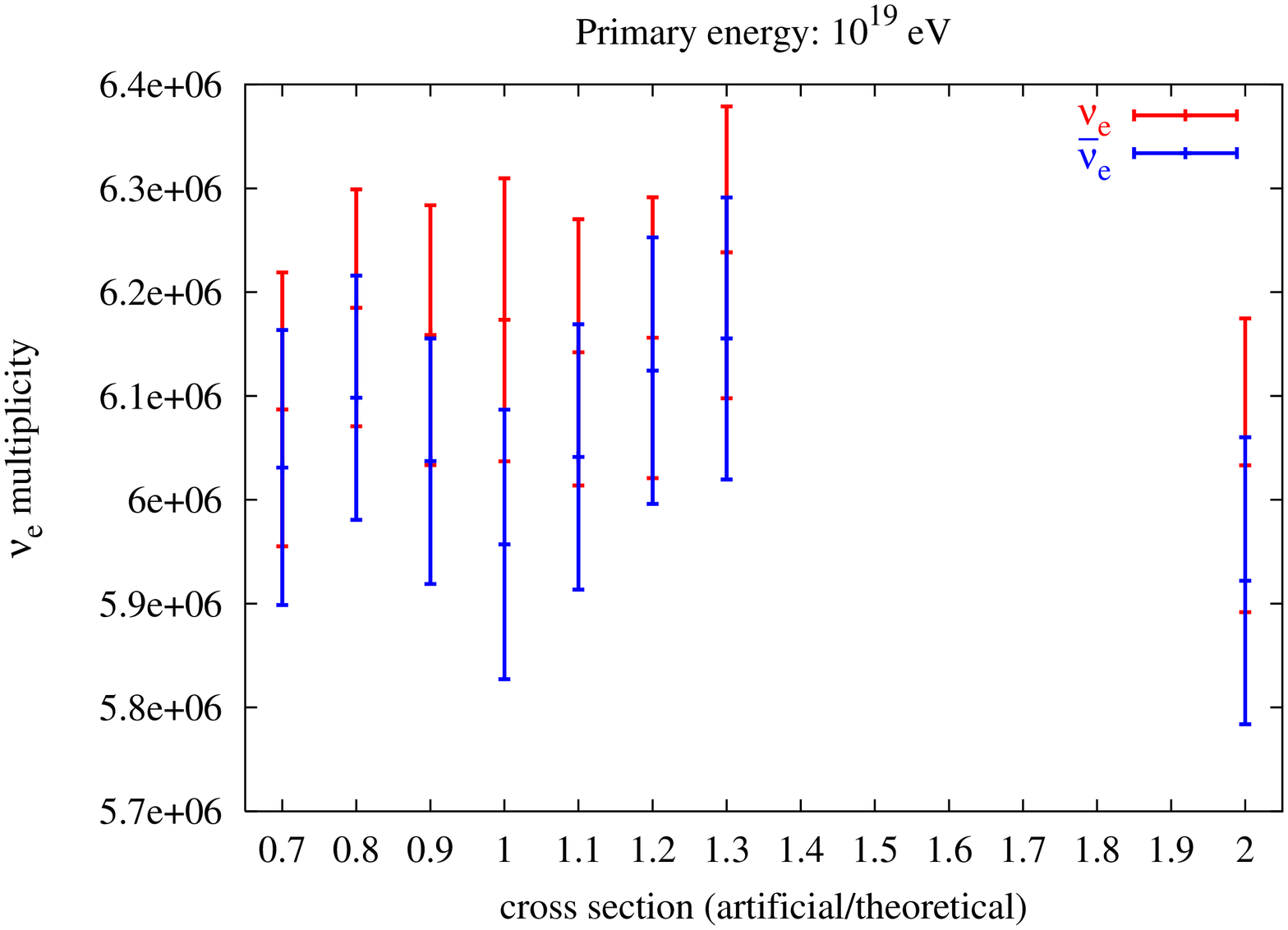}}}
\par}
{\centering \resizebox*{12cm}{!}{\rotatebox{-90}{\includegraphics{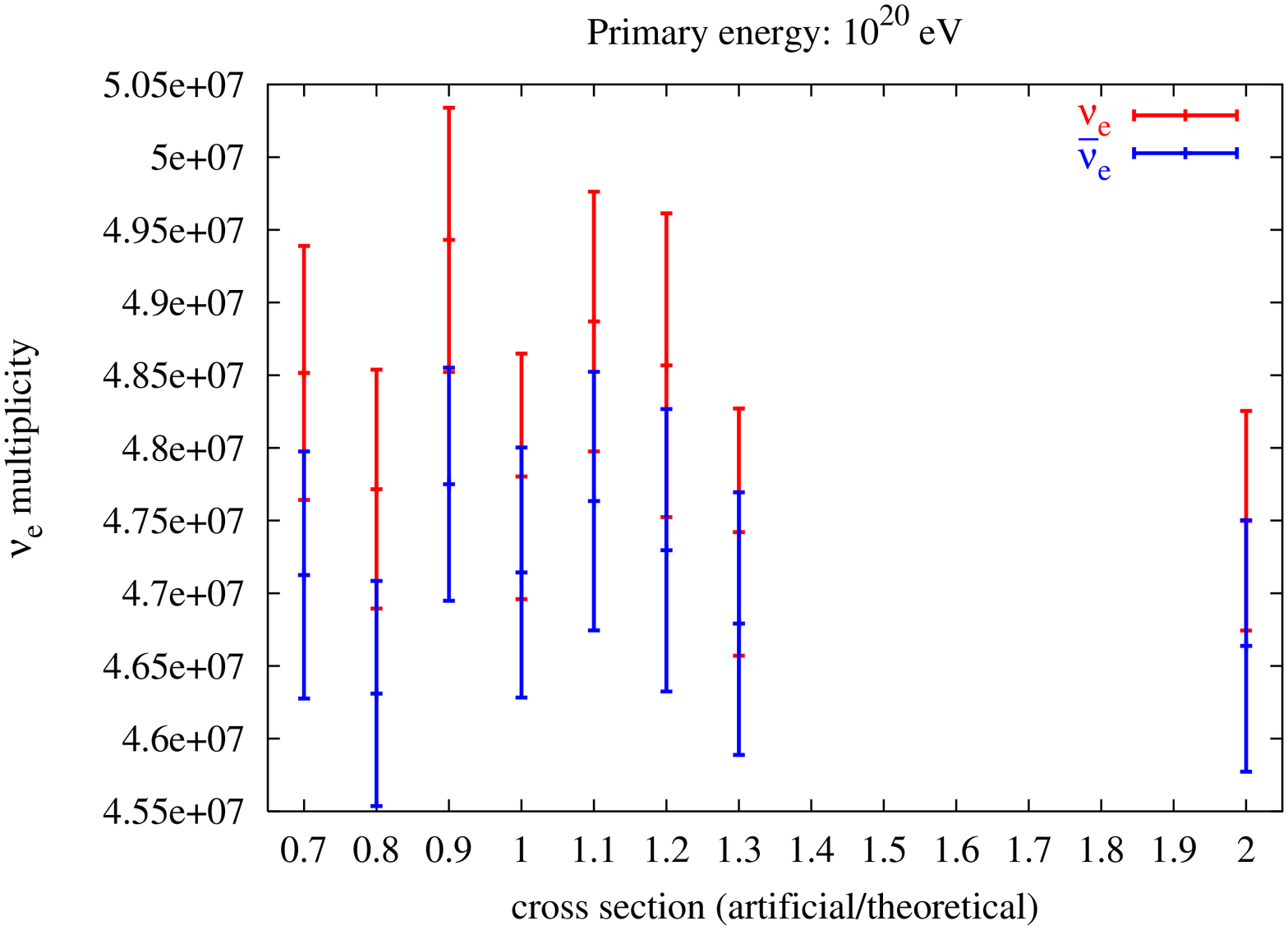}}}
\par}
\caption{variation of the $\nu_e$ multiplicity as a function
of the first impact cross section at $10^{19}$ and at $10^{20}$ eV}
\label{multielectronneutrino}
\end{figure}

\begin{figure}[t]
{\centering
\resizebox*{12cm}{!}{\rotatebox{-90}{\includegraphics{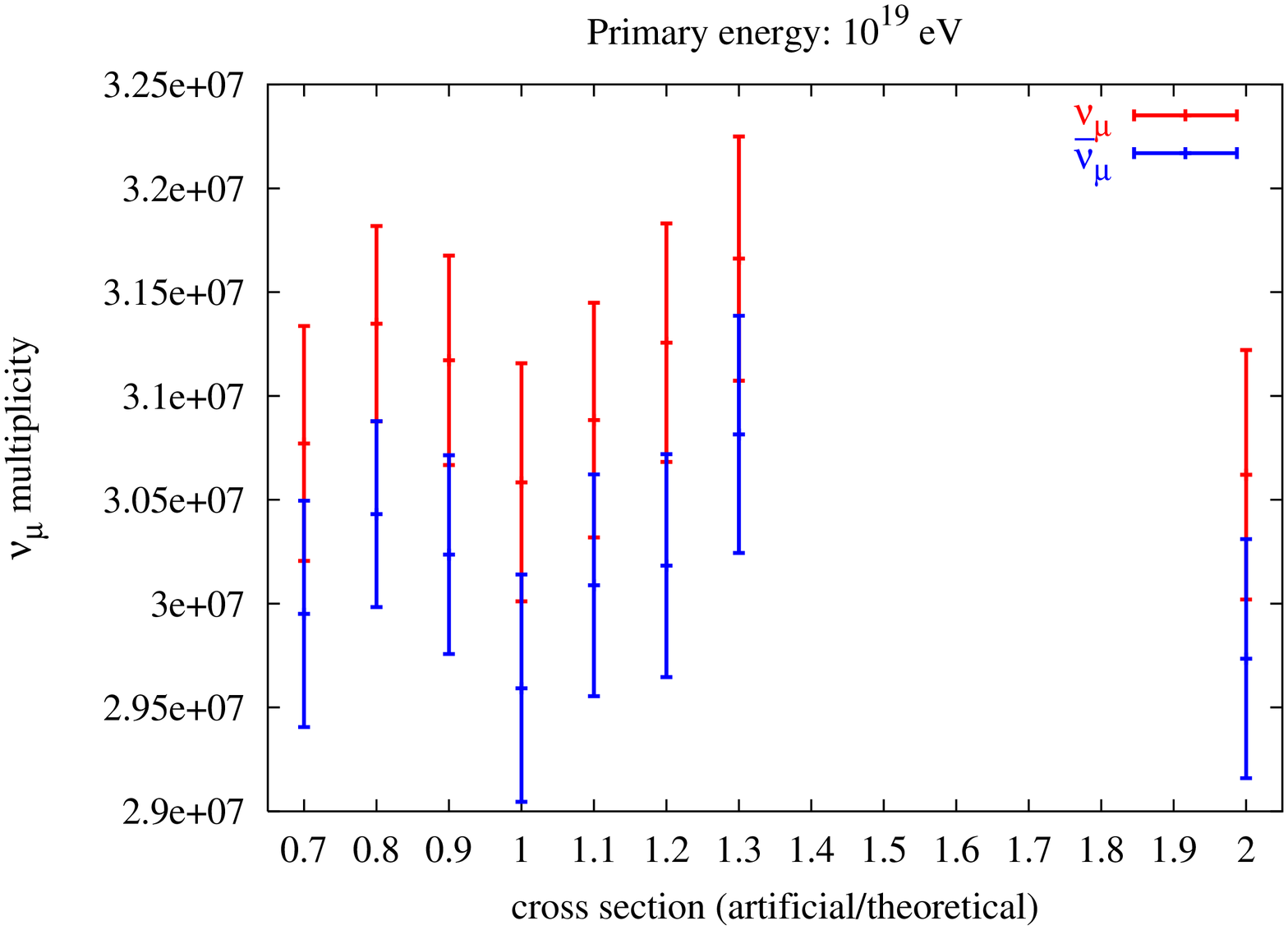}}} \par}
{\centering
\resizebox*{12cm}{!}{\rotatebox{-90}{\includegraphics{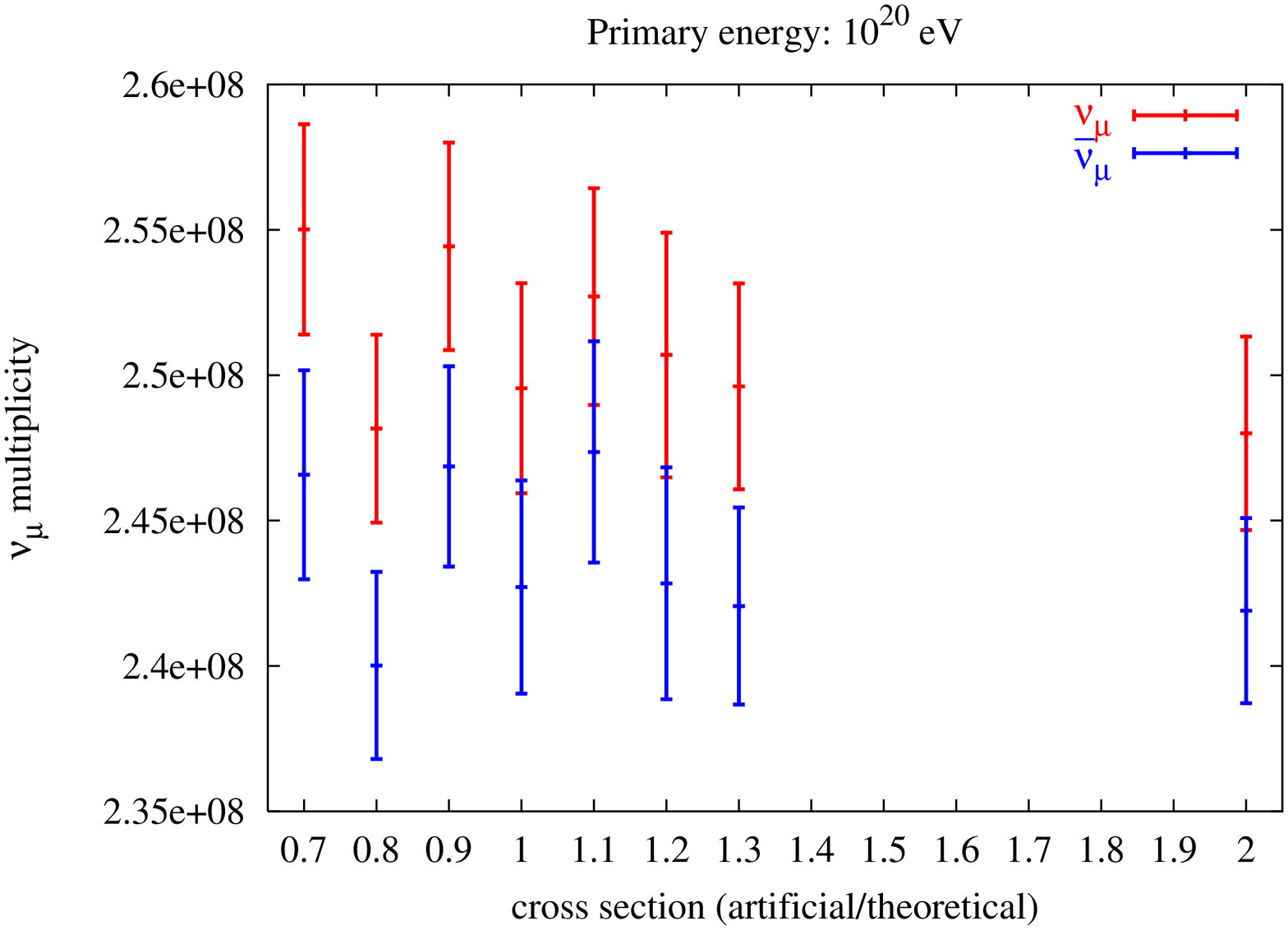}}} \par}
\caption{variation of the $\nu_\mu$ multiplicity as a function
of the first impact cross section at $10^{19}$ and at $10^{20}$ eV}
\label{multimuonneutrino}
\end{figure}

In Figs.~\ref{multiphoton}-\ref{multimuonneutrino}
we show plots obtained simulating an artificial first proton
impact in which we have modified the first interaction cross section by a
nominal factor ranging from 0.7 to 2. We plot on the y-axis the corresponding
fluctuations in the multiplicities both for electrons and positrons.
Statistical fluctuations \footnote{we keep the height of the first proton
impact with the atmosphere arbitrary for each selected correction factor
(x-axis)}
have been estimated using bins of 80 runs. The so-developed showers have been
thinned using the Hillas
algorithm, as usually done in order to make the results of these simulations
manageable,
given the size of the showers at those energies. As one can immediately see,
the artificial
corrections on the cross section are compatible with ordinary fluctuations of
the air-shower. We have analyzed all the major subcomponents of the air shower,
photons and leptons,
together with the corresponding neutrino components.
We can summarize these findings by assessing that a modified first impact,
at least for such correction factors in the 0.7-2 range
in the cross section, are unlikely to modify the multiplicities in any
appreciable way.

\begin{figure}[t]
{\centering
\resizebox*{12cm}{!}{\rotatebox{-90}{\includegraphics{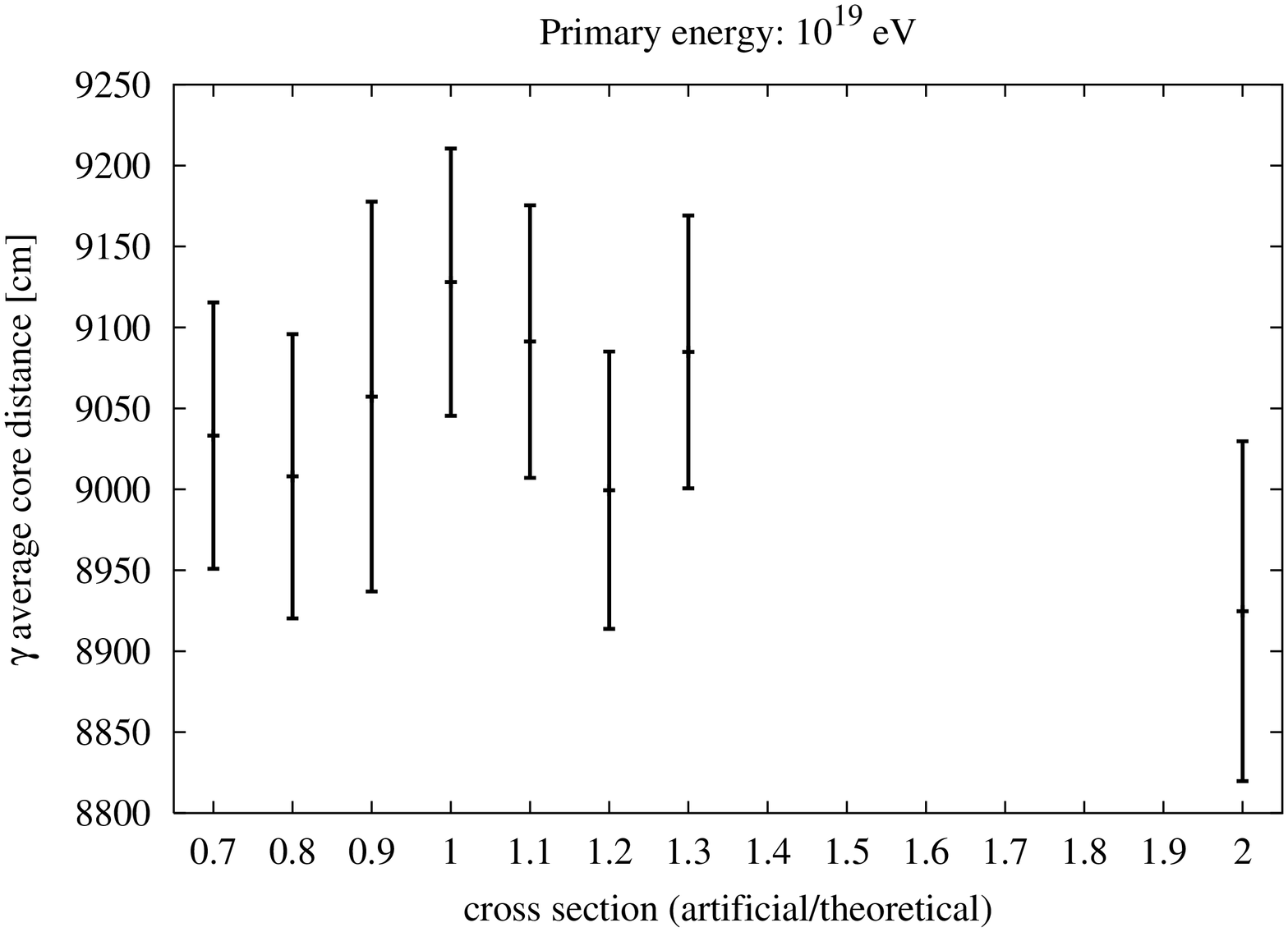}}} \par}
{\centering
\resizebox*{12cm}{!}{\rotatebox{-90}{\includegraphics{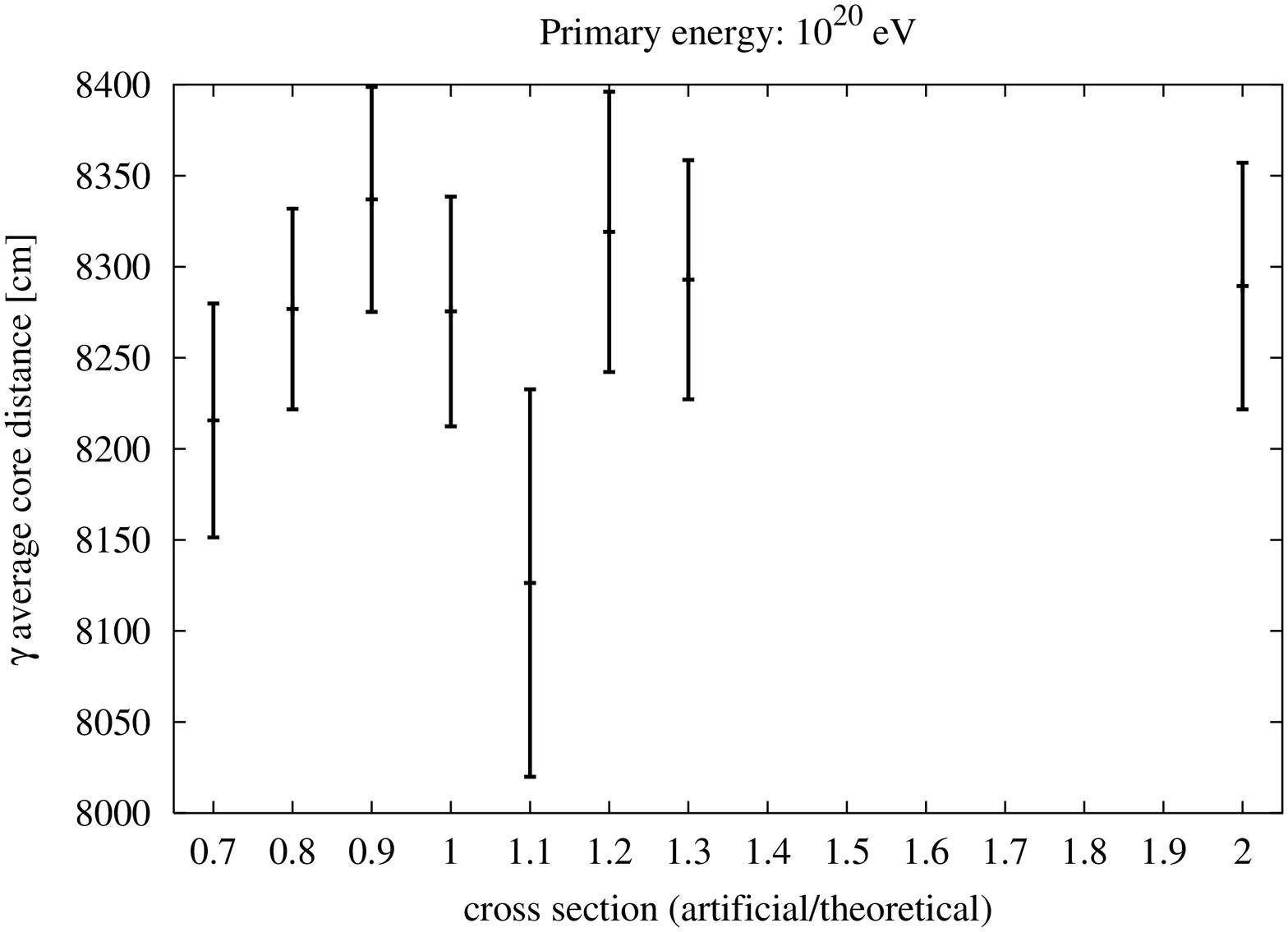}}} \par}
\caption{Lateral distributions of photons as a function of the first impact
cross section at $10^{19}$ and at $10^{20}$ eV}
\label{lateralphoton}
\end{figure}

\begin{figure}[t]
{\centering \resizebox*{12cm}{!}{\rotatebox{-90}{\includegraphics{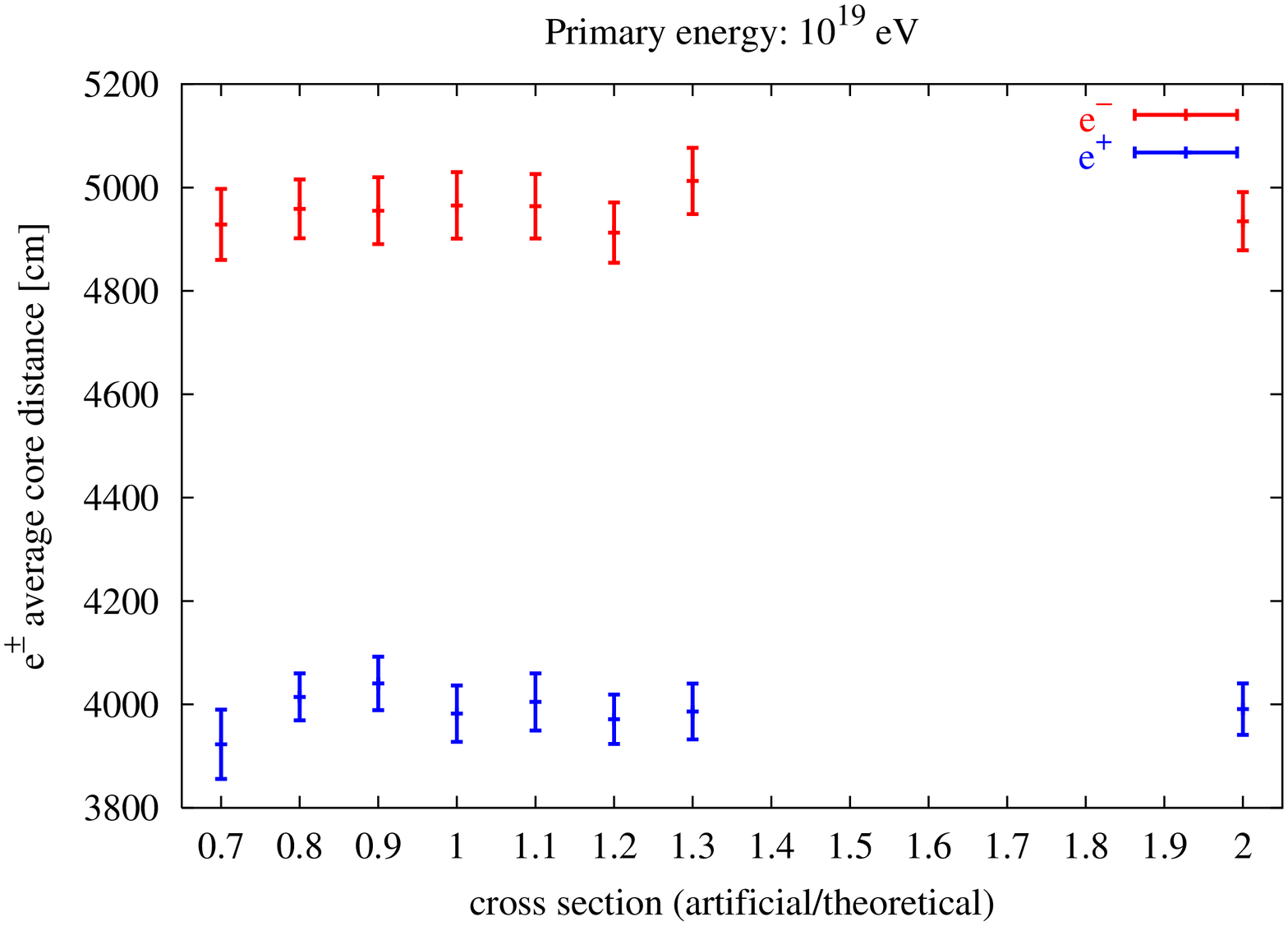}}}
\par}
{\centering \resizebox*{12cm}{!}{\rotatebox{-90}{\includegraphics{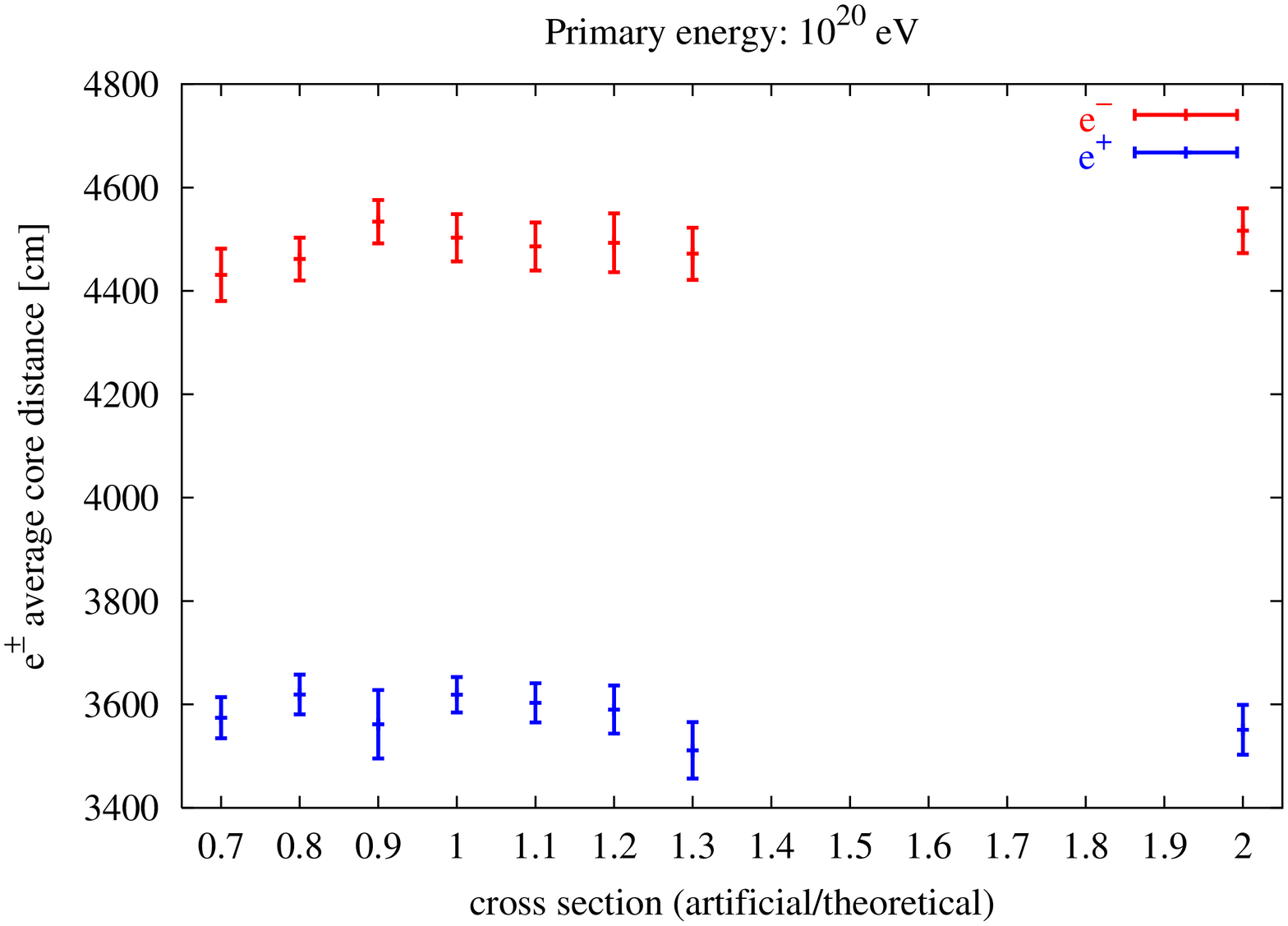}}}
\par}
\caption{Lateral distributions of $e^\pm$
as a function of the first impact cross section at $10^{19}$ and at $10^{20}$
eV}
\label{lateralepm}
\end{figure}

\begin{figure}[t]
{\centering \resizebox*{12cm}{!}{\rotatebox{-90}{\includegraphics{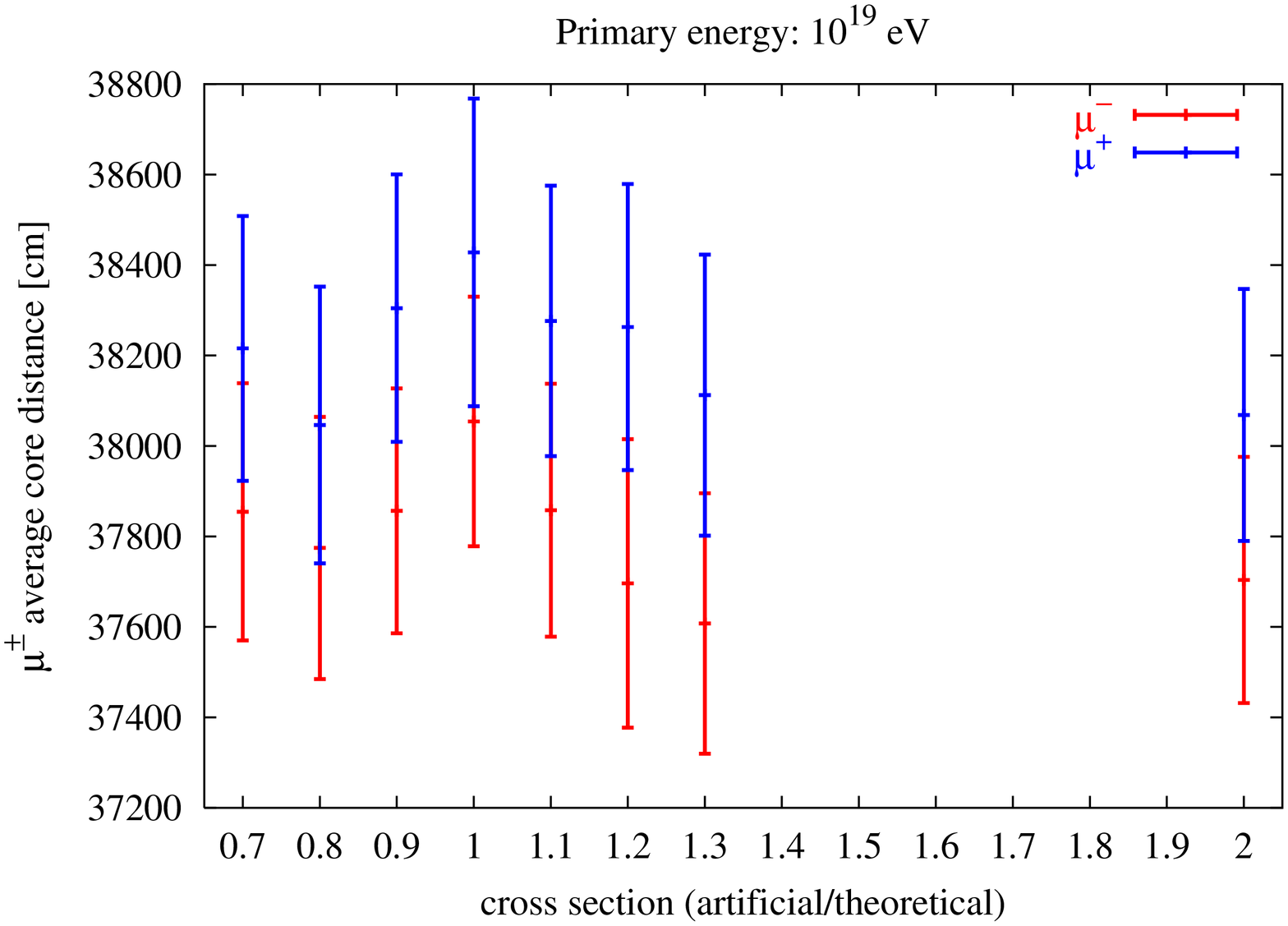}}}
\par}
{\centering \resizebox*{12cm}{!}{\rotatebox{-90}{\includegraphics{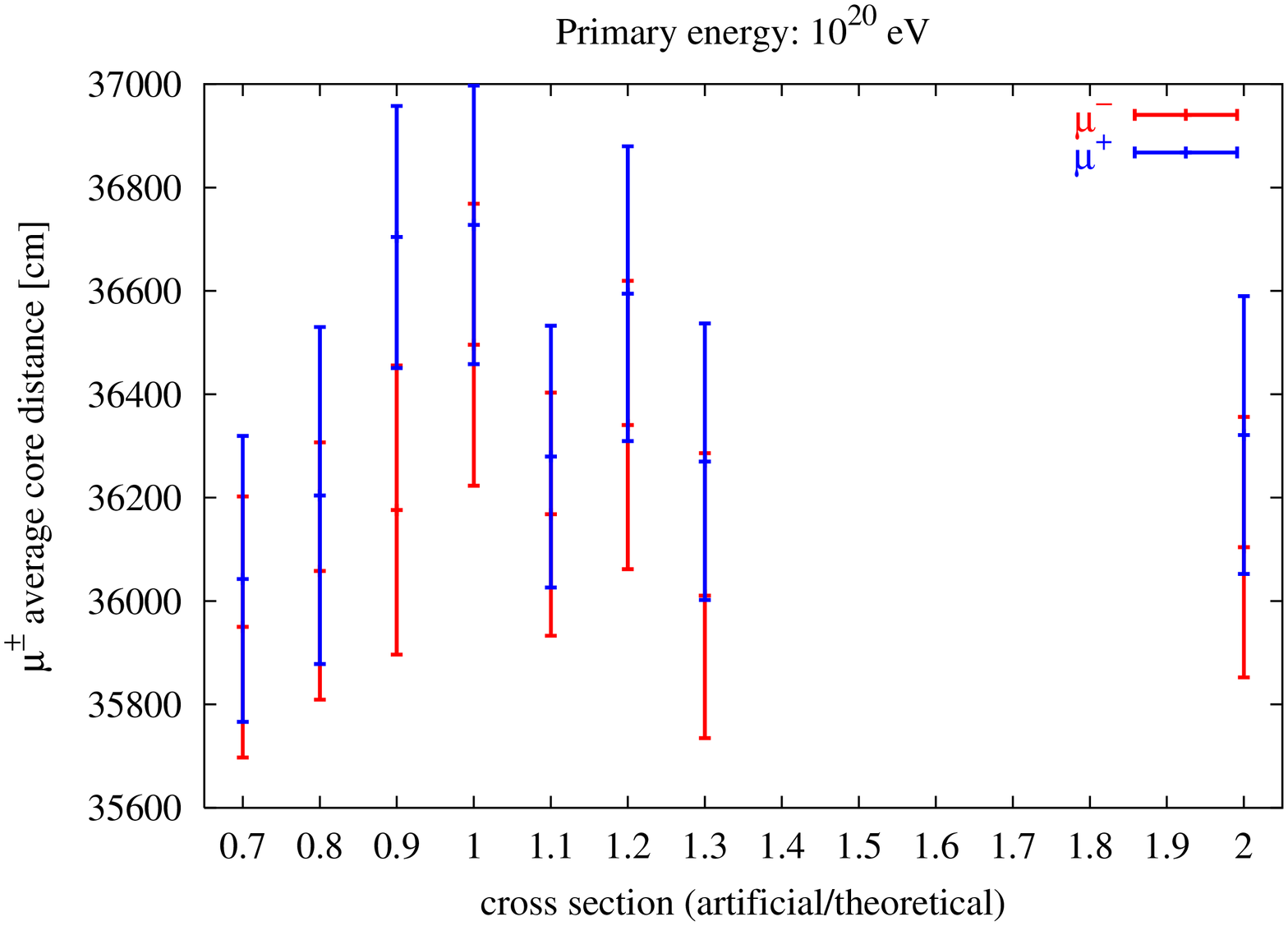}}}
\par}
\caption{Lateral distributions of $\mu^\pm$
as a function of the first impact cross section at $10^{19}$ and at $10^{20}$
eV}
\label{lateralmupm}
\end{figure}

\begin{figure}[t]
{\centering \resizebox*{12cm}{!}{\rotatebox{-90}{\includegraphics{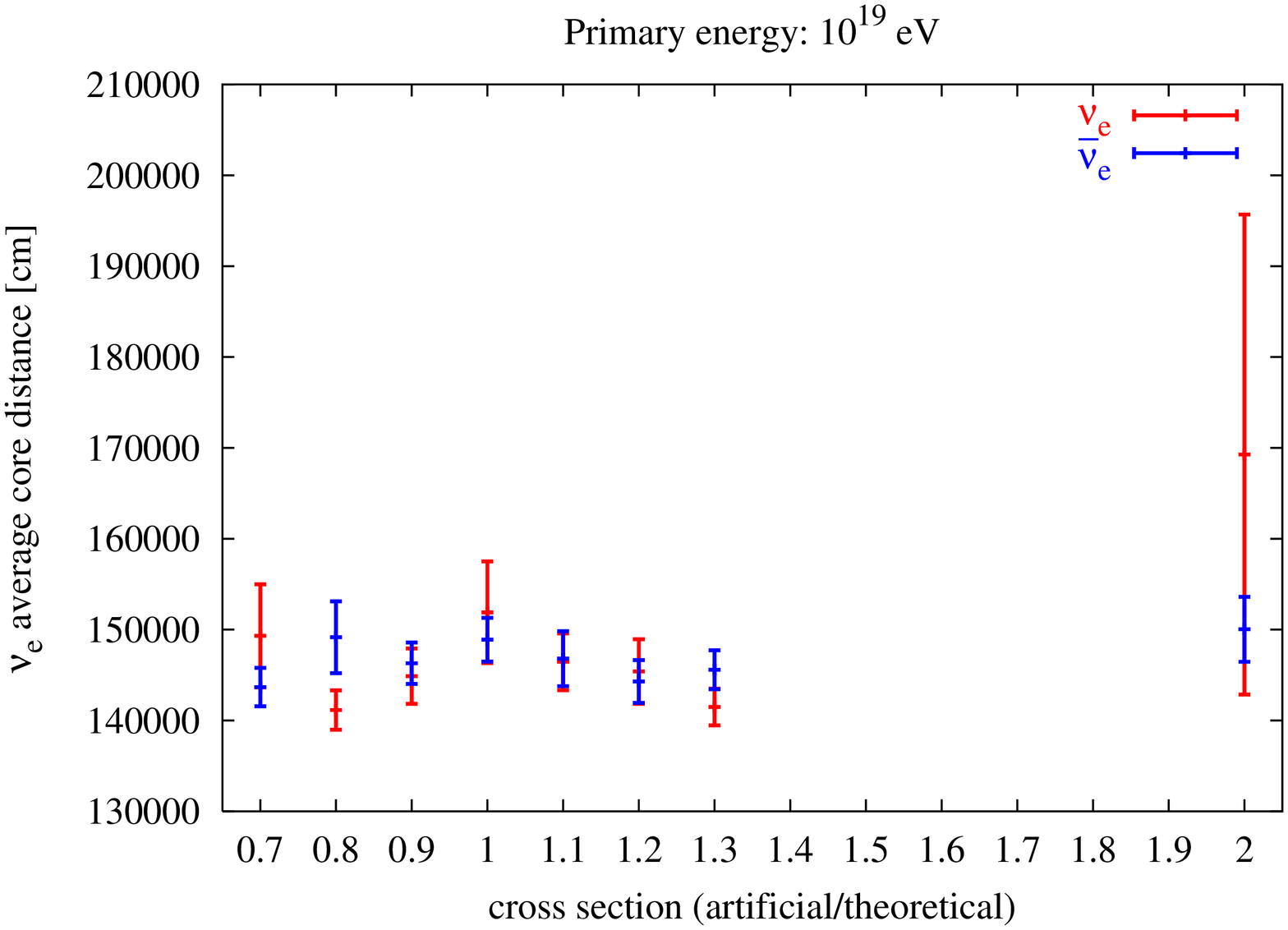}}}
\par}
{\centering \resizebox*{12cm}{!}{\rotatebox{-90}{\includegraphics{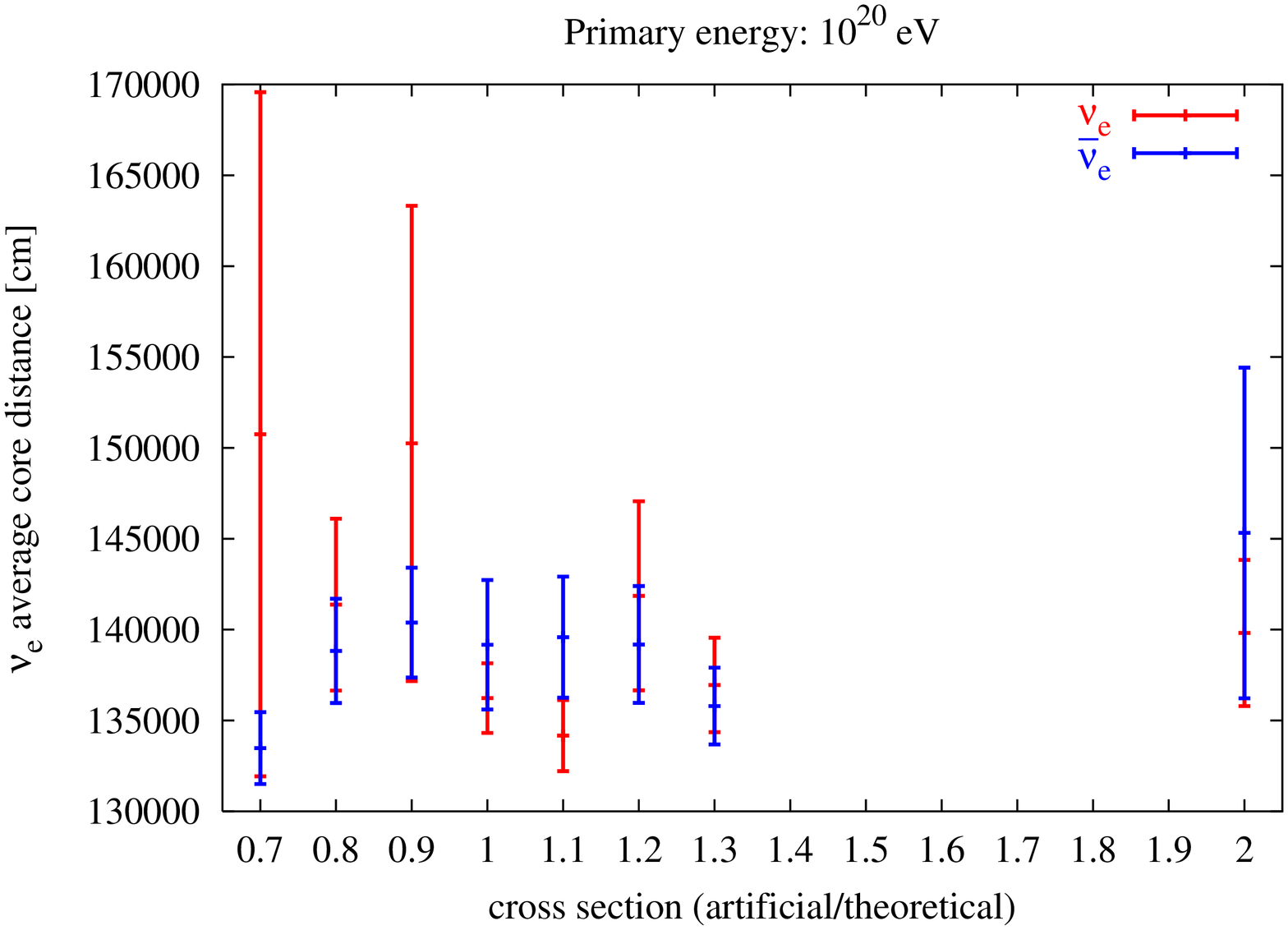}}}
\par}
\caption{Lateral distributions of $\nu_e$
as a function of the first impact cross section at $10^{19}$ and at $10^{20}$
eV}
\label{lateralnue}
\end{figure}

\begin{figure}[t]
{\centering
\resizebox*{12cm}{!}{\rotatebox{-90}{\includegraphics{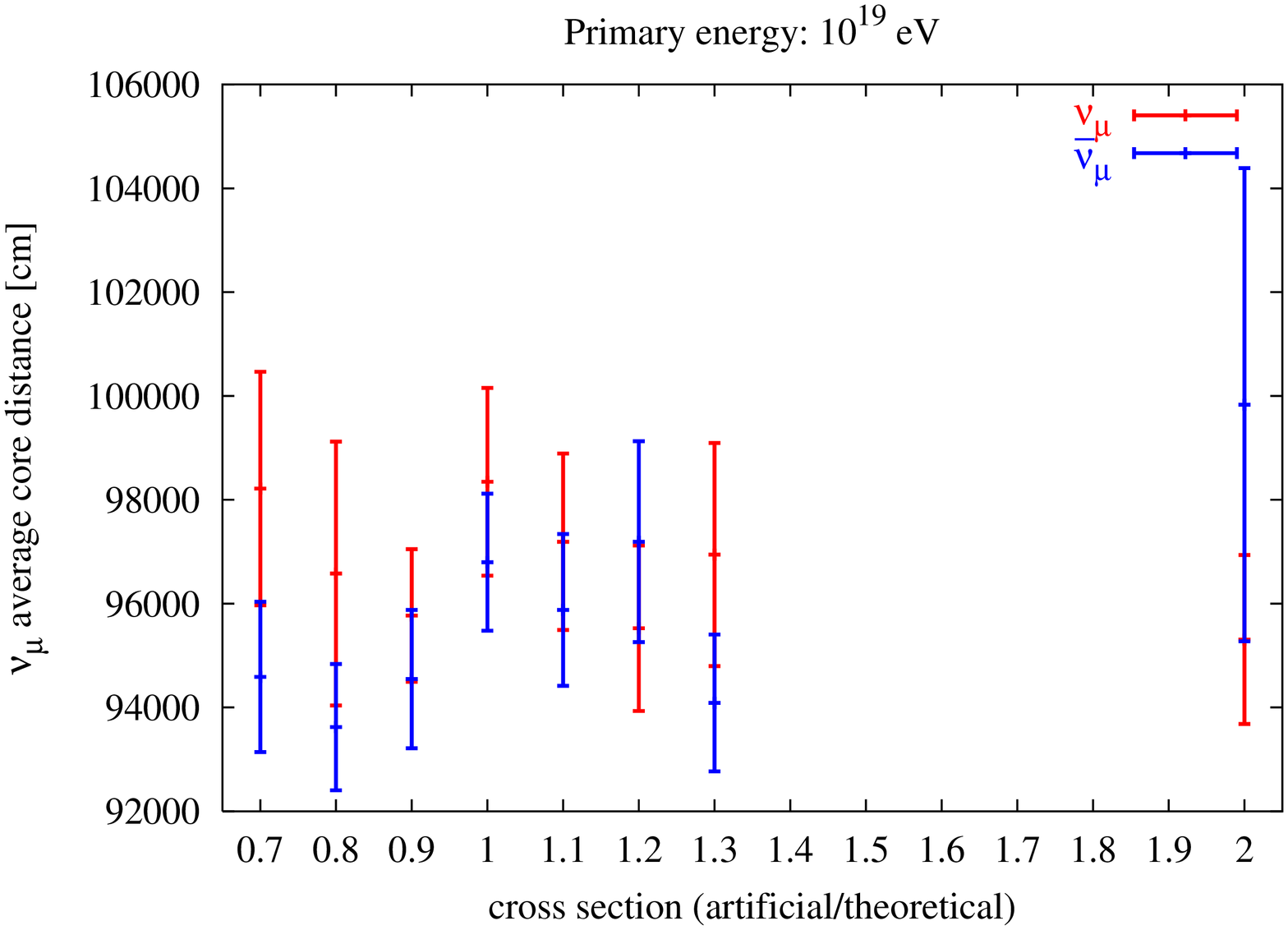}}} \par}
{\centering
\resizebox*{12cm}{!}{\rotatebox{-90}{\includegraphics{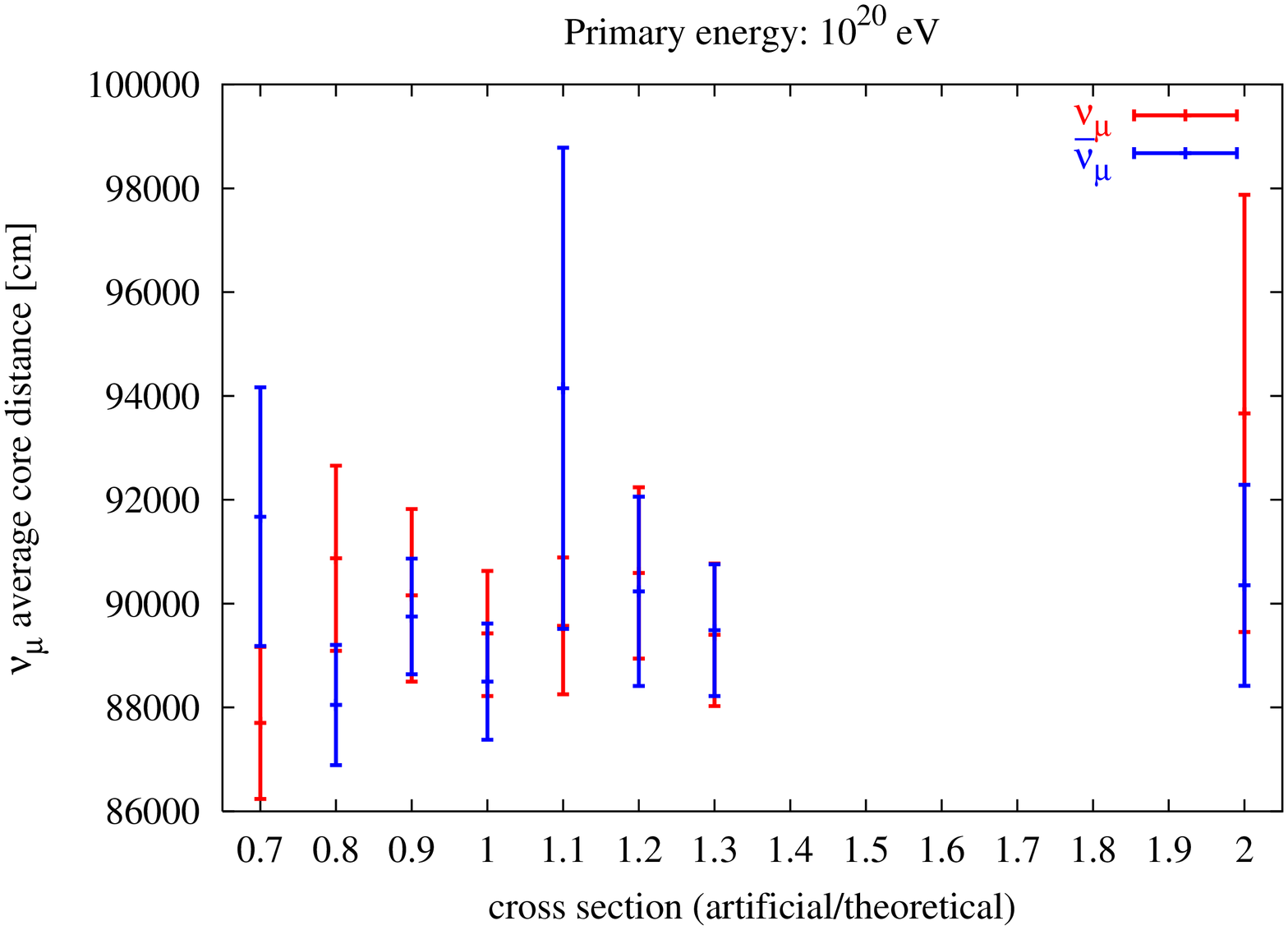}}} \par}
\caption{Lateral distributions of $\nu_\mu$
as a function of the first impact cross section at $10^{19}$ and at $10^{20}$
eV}
\label{lateralnumu}
\end{figure}

A second test is illustrated in Figs.~\ref{lateralphoton}-\ref{lateralnumu}.
Here we plot the same correction factors on the
x-axis as in the previous plots (\ref{multiphoton}-\ref{multimuonneutrino})
but we show on the y-axis (for the same particles) the average point
of impact on the detector and its corresponding statistical fluctuations.
As we increase the correction
factor statistical fluctuations in the formation of the air shower
seem to be compatible with the modifications induced by the ``new physics''
of the first impact and no special new effect is observed.

Fluctuations of these type, generated by a minimal modification of the existing
codes
only at the first impact may look simplistic, and can possibly
be equivalent to ordinary simulations with a simple rescaling
of the atmospheric height at which the first collision occurs, since the
remaining
interactions are, in our approach, unmodified.
The effects we have been looking for, therefore, appear subleading compared to
other standard fluctuations which take place in the formation of the cascade.
On the other hand, drastic changes in the
structure of the air shower should possibly depend mostly on the physics
of the first impact and only in a less relevant way on the modifications
affecting the
cascade that follows up.
We have chosen to work at an energy of $10^{20}$ eV but we do not
observe any substantial modifications of our results at lower energies
($10^{19}$ eV), except for the multiplicities which are down by a factor
of 10. Our brief analysis, though simple, has the purpose to
illustrate one of the many issues which we believe
should be analyzed with great care in the near future: the physics of the first
impact and substantial additional modifications to the existing codes
in order to see whether any new physics can be extracted from
these measurements.

\section{Conclusions}
We have tried to analyze with a searching criticism the possibility of
detecting new physics at Auger using current ideas about supersymmetry,
the QCD evolution and such.
While the physics possibilities of these experiments
are far reaching and may point toward a validation
or refusal of the existence of a GZK cutoff, we have argued that
considerable progress still needs to be done
in order to understand better the hadronization models at very large
energy scales. Our rather conservative viewpoint stems from the fact
that the knowledge of the
structure of the hadronic showers at large energies is still
under debate and cannot be conclusive.
We have illustrated by an extended and updated simulation some of the
characteristics of the showers, the intrinsic fluctuations in the lateral
distributions, the multiplicities of the various subcomponents and of the
total spectrum, under some realistic conditions. We have also tried
to see whether nominal and realistic changes in the cross section of the
first impact may affect the multiplicities, with a negative outcome.
We have however pointed out, in a positive way, that new physics models
predicting large missing
energy may have a chance to be identified, since the trend followed by the
total multiplicities
(in a log-log scale) appear to be strikingly linear.

We do believe, however, that other and even more extensive simulation studies
should be done, in combination with our improved understanding of small-x
effects
in QCD at large parton densities, in order to further enhance the physics
capabilities
of these experiments. Another improvement in the extraction of new physics
signals from the
UHECR experiments will come from the incorporation of new physics in the
hadronization
models.

\begin{figure}[tbh]
{\centering \subfigure[
]{\resizebox*{12cm}{!}{\rotatebox{-90}{\includegraphics{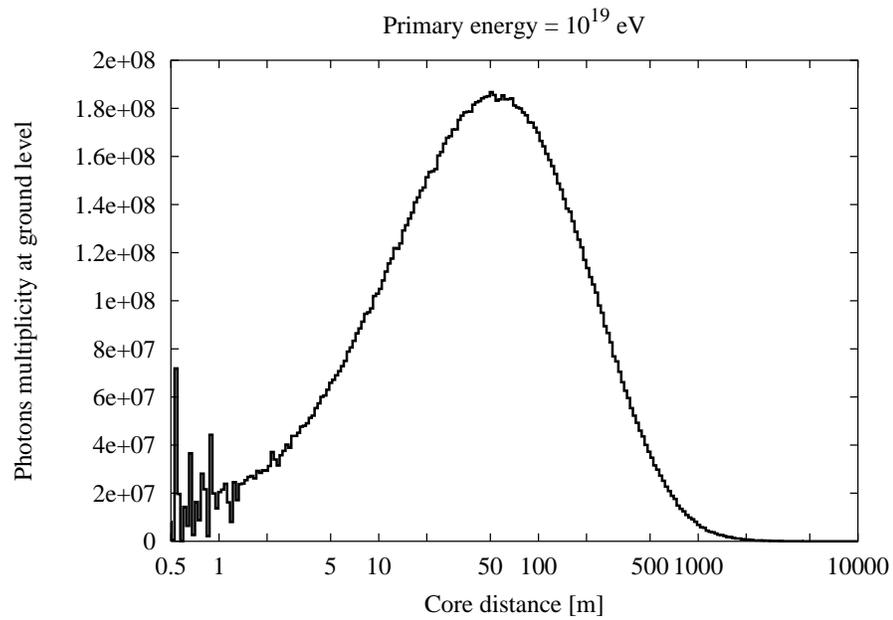}}}} \par}
{\centering \subfigure[
]{\resizebox*{12cm}{!}{\rotatebox{-90}{\includegraphics{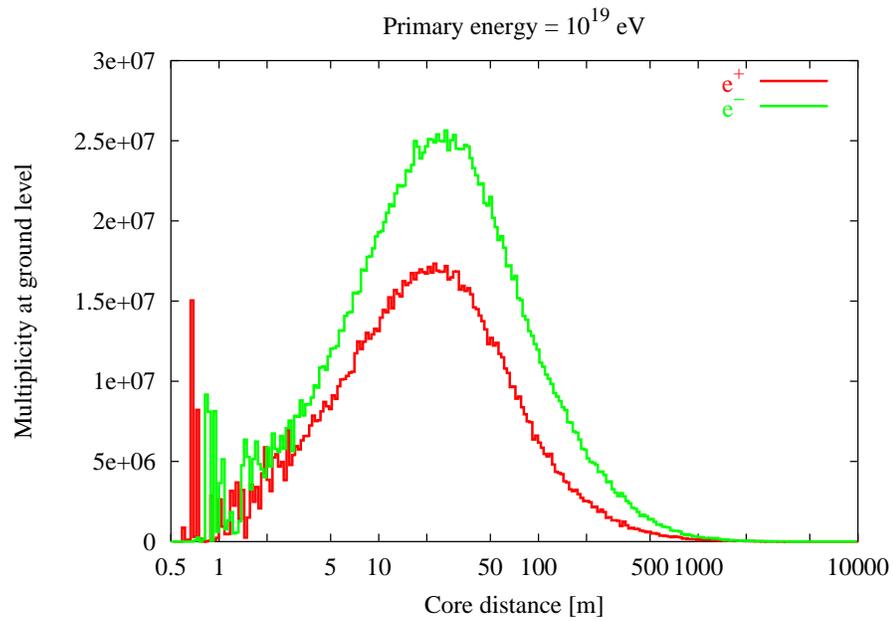}}}} \par}
\caption{Multiplicities of photons and \protect\( e^{\pm }\protect \)
at the ground level with a proton primary of \protect\( 10^{19}\protect \)
eV as a function of the distance from the core of the shower.}
\end{figure}
\begin{figure}[tbh]
{\centering \subfigure[
]{\resizebox*{12cm}{!}{\rotatebox{-90}{\includegraphics{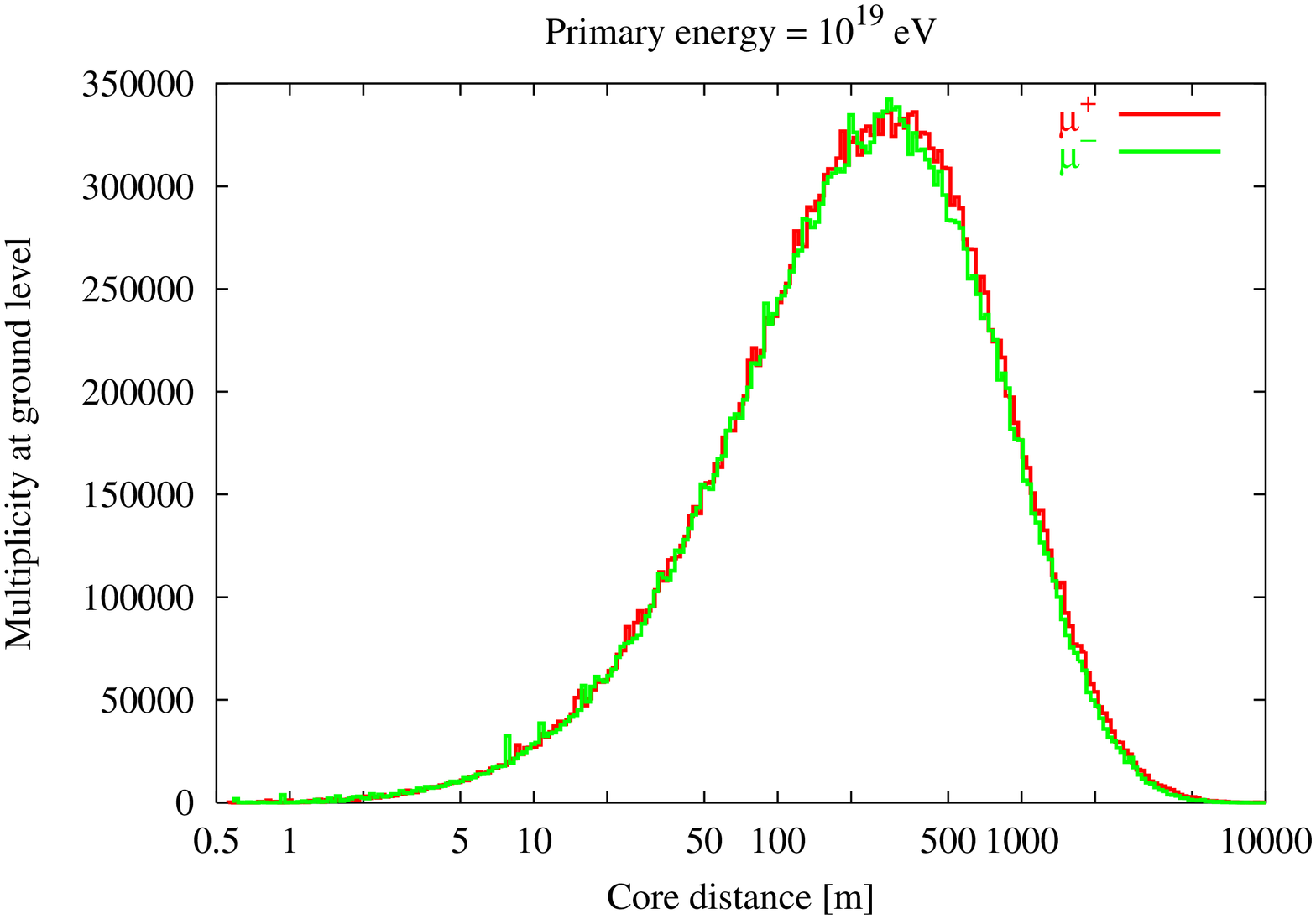}}}} \par}
\caption{Multiplicities of \protect\( \mu ^{\pm }\protect \)
at the ground level with a proton primary of \protect\( 10^{19}\protect \)
eV as a function of the distance from the core of the shower.}
\end{figure}

\begin{figure}[tbh]
{\centering \subfigure[
]{\resizebox*{12cm}{!}{\rotatebox{-90}{\includegraphics{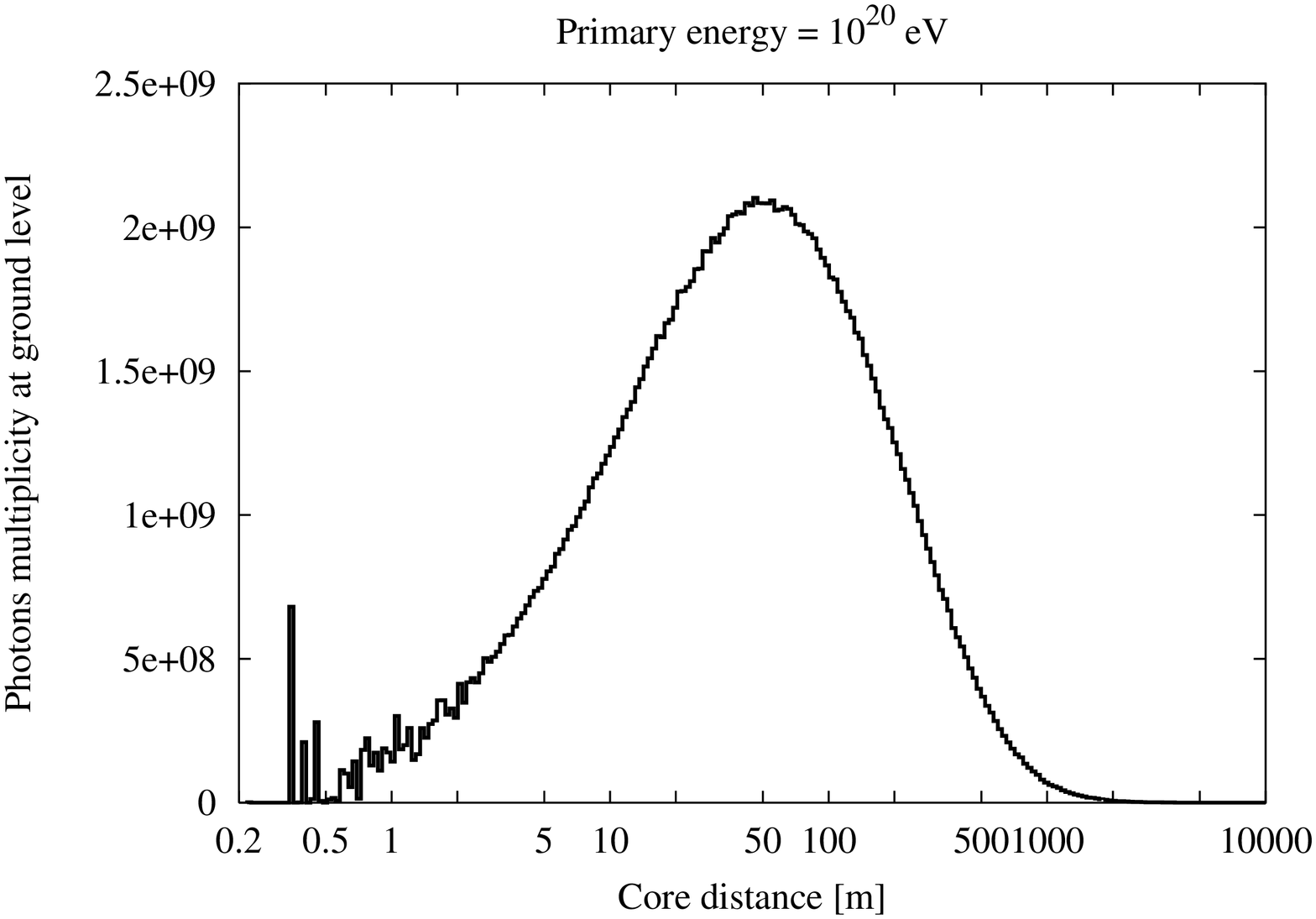}}}} \par}
{\centering \subfigure[
]{\resizebox*{12cm}{!}{\rotatebox{-90}{\includegraphics{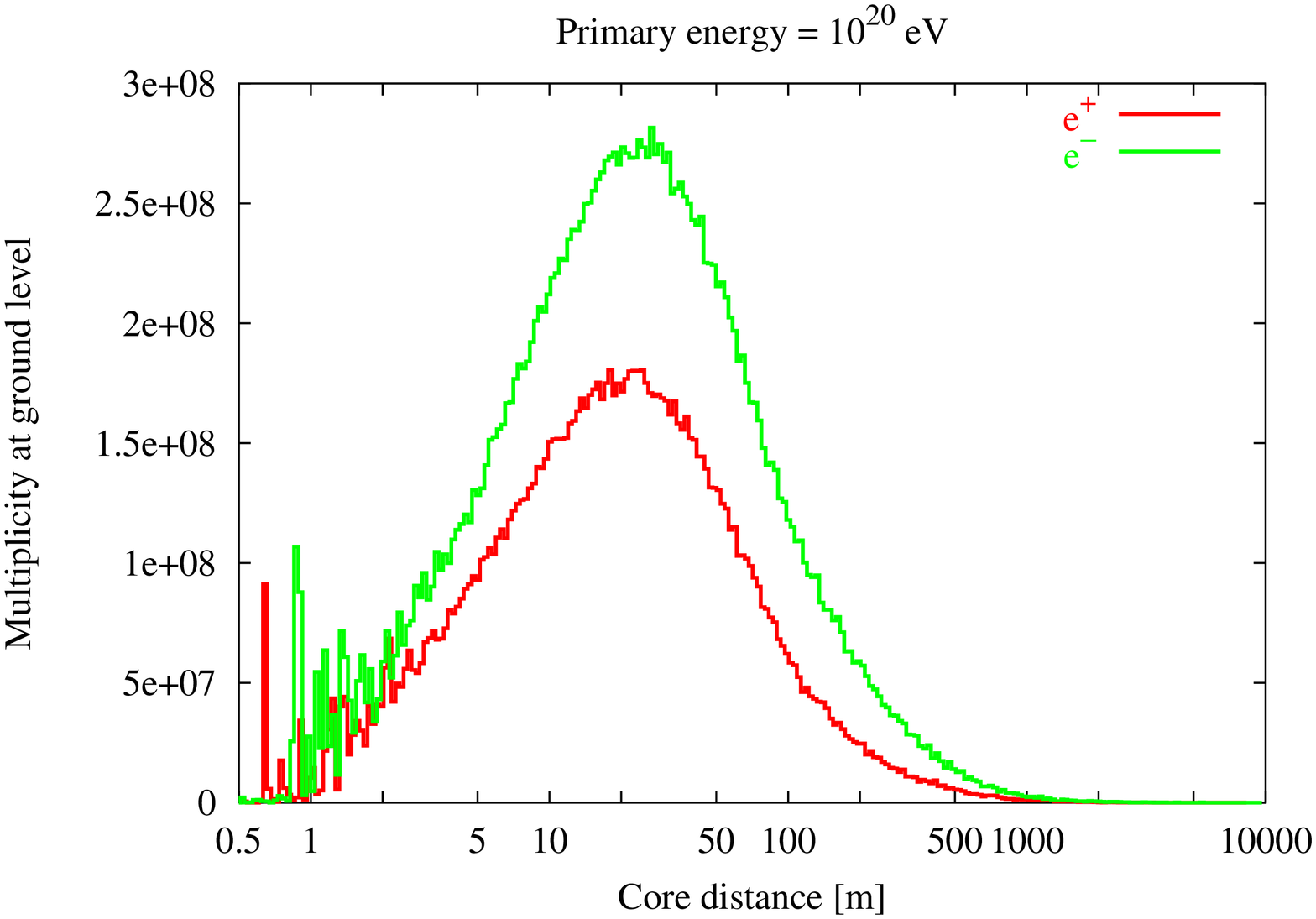}}}} \par}
\caption{Multiplicities of photons and \protect\( e^{\pm }\protect \)
at the ground level with a proton primary of \protect\( 10^{20}\protect \)
eV as a function of the distance from the core of the shower.}
\end{figure}

\begin{figure}[tbh]
{\centering \subfigure[
]{\resizebox*{12cm}{!}{\rotatebox{-90}{\includegraphics{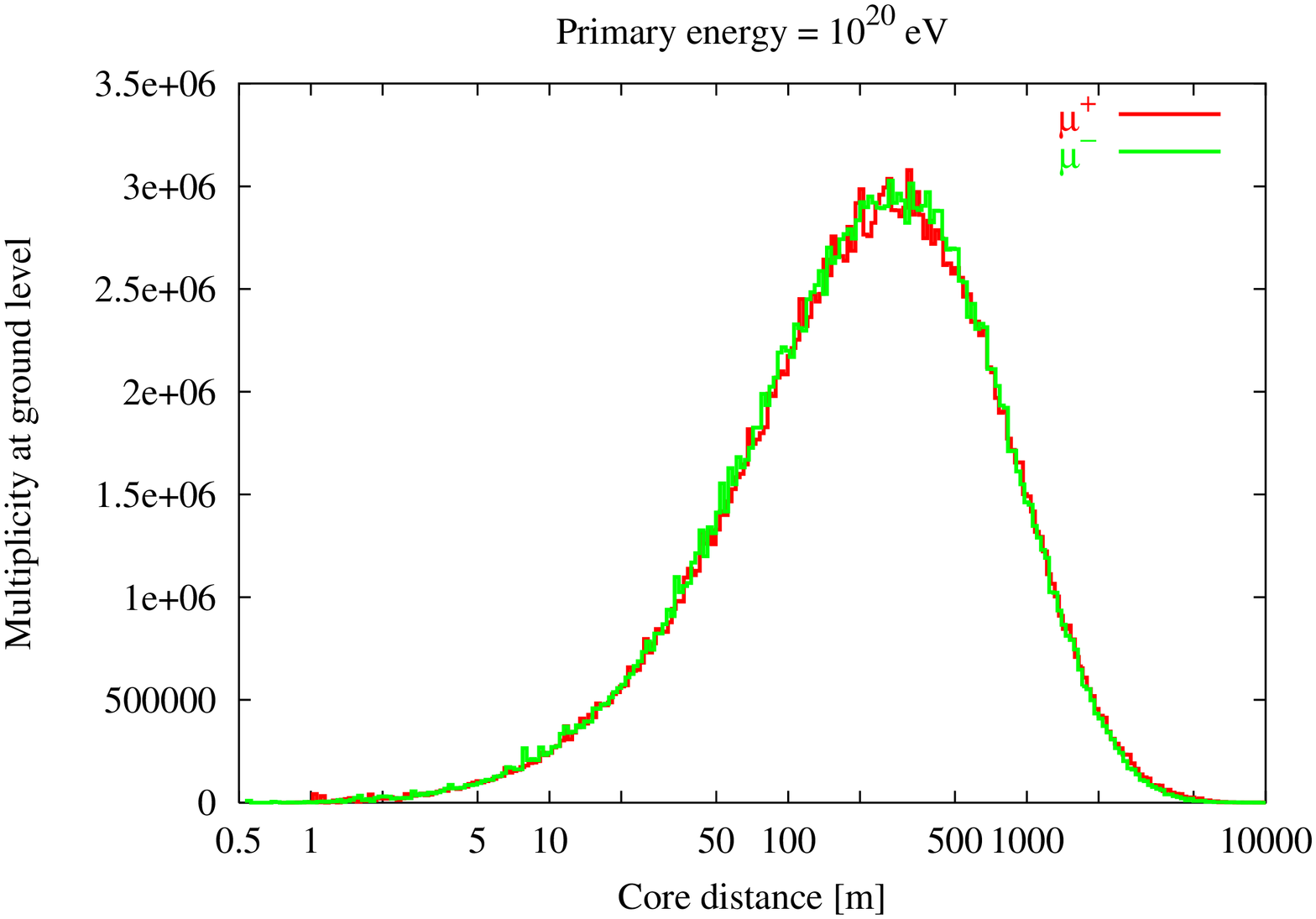}}}} \par}
\caption{Multiplicities of \protect\( \mu ^{\pm }\protect \)
at the ground level with a proton primary of \protect\( 10^{20}\protect \)
eV as a function of the distance from the core of the shower.}
\end{figure}

\clearpage

\chapter{Searching for extra dimensions in high energy cosmic rays:\\the fragmentation of mini black holes\label{chap:MBH}}  
\fancyhead[LO]{\nouppercase{Chapter 7. Searching for extra dimensions}}
\section{Background on the topic} 
Extra Dimensions have been around the theory landscape for some time, starting from the 
20's with the work of Kaluza and Klein. Their original idea was to use a five dimensional 
manifold, the fifth dimension being a circle, and identify 
the $g_{\mu 5}$ component of the metric in this 5-dimensional 
space as a 4-dimensional vector field. The field, after a Fourier expansion in the 
extra variable, has a massless component which can be identified with the photon, plus an infinite tower of massive states. The masses of these additional states 
are inversely proportional to the size of the extra dimensions and their disappearance from the low energy spectrum can be simply obtained by tuning the extra dimensions appropriately. 
Then, the natural question to ask is: how large can the extra dimensions be without 
getting into conflict with the experiment? The answer to this question requires the 
Standard Model fields to be localized on the brane with gravity free to propagate 
in the extra space, called ``the bulk''. This scenario is characterized by a 
rich phenomenology, as we are going to discuss next, with implications which may 
appear also in the structure of cosmic ray showers. One possibility is the 
formation, due to a lower gravity scale, 
of mini black holes which decay into an ``s-wave'' or isotropically into all the 
particles of the Standard Model. For this reason 
we present a study of the main observables of the air showers initiated by 
an incoming primary which collides with the atmosphere and characterized by the 
formation of an intermediate mini black hole. 
We study particle multiplicities, lateral distributions 
and the ratio of the electromagnetic to the hadronic
components of these special air showers.
In this chapter  we illustrate a simulation study of the resulting
cascades over the entire range ($10^{15}-10^{19}$eV)
of ultra high initial energies, for several values of the number of large extra 
dimensions, for a variety of altitudes of the initial interaction and 
with the energy losses in the bulk taken into account.
The results are compared with a representative of the standard 
events, namely the shower due to the collision of a primary proton with 
a nucleon in the atmosphere.
Both the multiplicities and the lateral distribution of the showers
show important differences between the two cases and, consequently, may be
useful for the observational characterization of the events. The
electromagnetic/hadronic ratio is strongly fluctuating and, thus,  
less decisive for the altitudes considered.

This chapter is based on the paper \cite{CafarellaCorianoTomaras}.

\section{Introduction}
In the almost structureless fast falling with energy inclusive 
cosmic ray spectrum, two kinematic regions have drawn considerable 
attention 
for a long time \cite{la, explanations}. These regions are the only ones 
in which the spectral index of the cosmic ray flux shows
a sharper variation as a function of energy, probably signaling 
some ``new physics'', according to many. These two regions, 
termed the {\em knee} and the {\em ankle} \cite{uhecr} have been 
puzzling theorists and experimentalists alike and
no clear and widely accepted explanation of this 
unusual behaviour in the propagation of the primaries
- prior to their impact with the earth atmosphere - exists yet. 
A large experimental effort \cite{auger,euso} in the next several 
years will hopefully clarify several of the issues related 
to this behaviour. 

While the {\em ankle} is mentioned in the debate regarding the 
possible existence of the so called Greisen, Zatsepin and Kuzmin 
(GZK) cutoff \cite{gzk}, due to the interaction of the
primaries with the cosmic background radiation, the proposed 
resolutions of this puzzle are several, ranging from a resonant 
Z-burst mechanism \cite{zburst} to string relics and other exotic 
particle decays \cite{berez, ben, sb, cfp}. 
The existence of data beyond
the cutoff has also been critically discussed \cite{BW}.

Given the large energy involved in the first stage of the 
formation of the air showers, the study of the properties of the 
cascade should
be sensitive to any new physics between the electroweak scale and the
original collision scale. Especially in the highest energy region 
of the spectrum, the energy available in the interaction of the 
primaries with the atmospheric nuclei is far above 
any conceivable energy scale attainable at future ground-based accelerators. 
Therefore, the possibility of detecting supersymmetry, for instance, 
in cosmic ray showers has also been contemplated \cite{fragfun}.
Thus, it is not surprising, that most of the attempts to explain these 
features of the cosmic ray spectrum typically 
assume some form of new physics at those energies.
 
With the advent of theories with a low fundamental scale of 
gravity \cite{ED} and large compact or non-compact extra dimensions,
the possibility of copiously producing mini black holes (based on 
Thorne's hoop conjecture \cite{Thorne}) in collisions involving hadronic 
factorization scales above 1 TeV has received considerable 
attention \cite{DL} \cite{Sarcevic} and these ideas, naturally, 
have found their way also in the literature of high energy cosmic 
rays \cite{Anchordoqui,pp4} and astrophysics \cite{DR}. For instance,
it was recently suggested that the long known {\em Centauro} events 
might be understood as evaporating mini black holes, produced 
by the collision of a very energetic primary (maybe a neutrino) with 
a nucleon (quark) in the atmosphere \cite{Theodore}. Other proposals
\cite{Ewa} also either involve new forms of matter (for example 
strangelets) or 
speculate about major changes in the strong interaction dynamics 
\cite{White}.  

While estimates for the frequencies of these types of processes 
both in cosmic rays \cite{Cavaglia,Anchordoqui,Theodore} and at colliders 
\cite{Sarcevic} have been presented, detailed studies of the multiplicities of
the particles collected at the detectors, generated by the extensive 
atmospheric air showers following the first impact of the primary rays,
are far from covering all the main features of the cascade \cite{refhtml}. 
These studies 
will be useful in order to eventually disentangle new physics 
starting from an analysis of the geometry of the shower, of the multiplicity 
distributions of its main sub-components \cite{CCF}
and of its directionality from deep space. 
For instance, the study of the location of the 
maxima of the showers at positions which can be
detected by fluorescence mirrors \cite{review_Anchordoqui},
generated as they go across the atmosphere,
and their variations 
as a function of the parameters of the underlying physical theory, 
may help in this effort \cite{Cavaglia}; other observables which also 
contain potential new information are the multiplicities of the 
various particle sub-components and the opening of the showers as 
they are detected on the ground \cite{CCF}. 
We will focus on this last type of observables.

To summarize: in the context of the TeV scale gravity with large extra dimensions
it is reasonable to assume that mini black holes, black holes with mass
of a few TeV, can form at the first impact of ultra high energy
primary cosmic rays with nucleons in the atmosphere. The black hole 
will evaporate into all types of particles of 
the Standard Model and gravity. The initial partons will hadronize and
all resulting particles as they propagate in the atmosphere will develop
into a shower(s), which eventually will reach the detectors. 
The nature and basic characteristics of these showers is the 
question that is the main subject of the present work. What is the signature 
on the detector of the showers arising from the decay of such mini
black holes and how it compares with a normal (not black hole mediated)
cosmic ray event, due, for instance, to a primary proton with the 
same energy colliding with an
atmospheric nucleon (the "benchmark" event used here). The comparison will be 
based on appropriate observables of the type mentioned above.

Our incomplete control of the quantum gravity/string theory effects, of the physics
of low energy non-perturbative QCD and of the nature of the quark-gluon
plasma phase in QCD, makes a fully general analysis of the above phenomena 
impossible at this stage. To proceed, we made the following 
simplifying assumptions and approximations. (1) The brane 
tension was assumed much smaller than the fundamental gravity scale, so it does
not modify the flat background metric. It is not clear at this point
how severe this assumption is, since it is related to the 
``cosmological constant problem'' and to the concrete realization 
of the Brane-World scenario. (2) The black hole 
was assumed to evaporate instantly, leading to initial ``partons'', whose 
number and distributions are obtained semiclassically. No virtual holes
were discussed and no back reaction was taken into account. 
(3) The initial decay products were assumed
to fly away and hadronize, with no intermediate formation of a quark-gluon plasma 
or of a disoriented chiral condensate (DCC). (4) We used standard
simulation programs for the investigation of the extensive air showers produced
in the cases of interest. To this purpose, we have decided
to use the Monte Carlo program CORSIKA \cite{CORSIKA} with the 
hadronic interaction implemented in SIBYLL \cite{SIBYLL} 
in order to perform this comparison, selecting a benchmark process 
which can be realistically simulated by this Monte Carlo, though other 
hadronization models are also available \cite{QGSJET}.
Finally, (5) a comment is in order about our selection of benchmark process
and choice of interesting events. 
In contrast to the case of a hadronic primary, the mini black hole 
production cross section due to the collision of a $\geq 10^3$ TeV neutrino
with a parton is of the order of the weak interaction neutrino-parton
cross section \cite{Theodore}. It would, thus, be interesting to compare the atmospheric 
showers of a normal neutrino-induced cosmic ray event to one with a 
black hole intermediate state.  
Unfortunately, at present neutrinos are not available as primaries 
in CORSIKA, a fact which sets a limitation on our benchmark study. 
However, it has to be mentioned that neutrino scattering off 
protons is not treated coherently at very high energy, since effects of 
parton saturation have not yet been implemented in the existing
codes \cite{CCF}. As shown in \cite{KK} these effects tend
to lower the cross section in the neutrino case.  For 
a proton-proton impact, the distribution of momenta among the partons 
and the presence of a lower factorization scale should render this 
effect less pronounced. For these reasons we have selected 
as benchmark process a proton-to-air collision at the same depth ($X_0$) 
and with the same energy as the corresponding ``signal event''. 
In order to reduce the large statistical fluctuations in the
formation of the extensive air showers after the collisions,
we have chosen at a first stage, in the bulk of our work, to simulate collisions taking place in the lower part of the atmosphere,
up to 1 km above the detector, in order to see whether 
any deviation from a standard scattering scenario can be identified. 
Another motivation for the analysis of such deeply penetrating 
events is their relevance in the study of the possibility to interpret the 
Centauro events as evaporating mini black holes \cite{Theodore}.
A second group of simulations have been performed at a higher altitude, for comparison. 

The present chapter consists of seven sections, of which this Introduction
is the first. In Section \ref{sec:mbh2} we briefly describe the D-brane world scenario,
in order to make clear the fundamental theoretical assumptions in our
study. A brief review of the properties of black holes
and black hole evaporation is offered here, together with all basic 
semiclassical formulas used in the analysis, with the dependence on the 
large extra dimensions shown explicitly.
In Section \ref{sec:mbh3} a detailed
phenomenological description of the modeling of the decay of the 
black hole is presented, which is 
complementary to the previous literature and provides an independent 
characterization of the structure of the decay. Incidentally, a Monte 
Carlo code for black hole decay 
has also been presented recently \cite{Webber}.
We recall that this description 
-as done in all the previous works on the subject-
is limited to the {\em Schwarzschild phase} of the lifetime of the mini 
black hole. 
The modeling of the radiation emission from the black hole - as obtained in
the semiclassical picture - (see \cite{kanti2} for an overview) is
performed here independently, using semi-analytical methods, and has been
included in the computer code that we have written and used, and which is 
interfaced with CORSIKA. 
Recent computations of the greybody factors
for bulk/brane emissions \cite{kanti2}, which match
well with the analytical approach of \cite{kanti1} valid in the low energy
limit of particle emission by the black hole, have also 
been taken into account.
Section \ref{sec:mbh4} contains our modeling of the 
hadronization process.
The hadronization of the partons emitted by the black hole
is treated analytically in the black hole rest frame, by solving
the evolution equations for the parton fragmentation functions, making use of a
special algorithm \cite{CPC} and of a specific set of initial conditions
for these functions \cite{kkp}.
After a brief discussion in Section \ref{sec:mbh5} of the transformation of the 
kinematics of the black hole decay event from the black hole frame 
to the laboratory frame, we proceed in Section \ref{sec:mbh6} with a Monte Carlo  
simulation of the extensive air showers of the particles produced by taking 
these particles as primaries. 
The simulations are quite intensive and have been 
performed on a small computer cluster. 
As we have already mentioned, in this work we focus on the multiplicities, on the 
lateral distributions of the events and on the ratio of electromagnetic 
to hadronic energies and multiplicities and scan the entire ultra high energy 
part of the cosmic ray spectrum. 
Our results are summarized in a series of plots and are commented upon in
the final discussion in Section \ref{sec:mbh7}.

\section{TeV Scale Gravity, Large Extra Dimensions and Mini Black Holes \label{sec:mbh2}}

The theoretical framework of the present study is the D-brane 
world scenario \cite{ED}. 
The World, in this scenario, 
is 10 dimensional, but all the Standard Model matter and forces
are confined on a $4+n_L$ dimensional hypersurface (the $D_{3+n_L}$-brane).
Only gravity with a characteristic scale $M_*$ can propagate in the bulk.
The $n_L$ longitudinal dimensions are constrained experimentally to be
smaller than ${\cal O}(TeV^{-1})$. However, for our purposes these dimensions 
may be neglected, since the Kaluza-Klein excitations related to these have 
masses at least of ${\cal O}(TeV^{-1})$, too large to affect our discussion 
below.
Consistency with the observed Newton's law, on the other hand, leads to the 
relation $M_{Pl}^2=M_*^{n+2}V_n$, between $M_{Pl}\simeq 10^{19}$ GeV, the 
fundamental gravity scale $M_*$ and the volume $V_n$ of the $n=6-n_L$ 
dimensional compact or non-compact transverse space.  
A natural choice for $M_*$, dictated a priori by the ``gauge hierarchy'' 
puzzle,
is $M_*={\cal O}(M_W)={\cal O}$(1 TeV), while the simplest choice for
the transverse space is an $n$-dimensional torus with all radii 
equal to $R$. Thus, one obtains a condition
between the number $n$ and the size $R(n)$ of the transverse dimensions.
Notice that under the above assumptions and for all values of $n$, $R$ 
is much larger than $10^{-33}$cm,
the length scale at which one traditionally expects possible deviations
from the 3-dimensional gravity force, and the corresponding dimensions 
are termed ``large extra dimensions'' (LED). 
For $n=2$ one obtains $R(n=2)$ of the order of a fraction of a mm. 
At distances much smaller than $R$ one should observe $3+n$-dimensional
Newton's law, for instance, as in torsion balance experiments \cite{Hoyle}. 
Current bounds on the size of these large extra dimensions and on $M_*$
come from various arguments, mostly of astrophysical
(for instance $M_* > 1500$ TeV for $n=2$) or cosmological ($M_* > 1.5$ TeV 
for $n=4$) origin \cite{kanti2}. A larger number of LED ($n$) translates into 
a reduced lower bound on $M_*$. It should be pointed out, that in general 
it is possible, even if ``unnatural'', that the transverse space has a few 
dimensions large and the others small. 
Here we shall assume a value of $M_*$ of
order 1TeV, neglect the small extra dimensions and treat the number of 
LED ($n$) as a free parameter. 

The implications of the existence of LED are quite direct
in the case of black hole physics. The black hole is effectively
4-dimensional if its horizon ($r_H$) is larger than the size of the extra 
dimensions. In the opposite case ($r_H\ll R$, or equivalently for black hole masses
$M_{BH}\ll 10^{13}$kg for $n=6$ \cite{Theodore}) it is $4+n$ dimensional, it 
spreads over the full space and its properties are those of a genuine 
higher dimensional hole. 
According to some estimates, over which however there is no universal 
consensus \cite{suppress}, black holes should be produced copiously \cite{DL}
\cite{Giddings} in particle collisions, whenever the center of mass energy 
available in the 
collision is considerably larger than the effective scale $M_*$ ($\sqrt{s} >> M_*$). 
With $M_*\sim$1 TeV, one may contemplate the possibility of producing 
black holes with masses of order a few TeV. 

Their characteristic temperature $T_H$ is 
inversely proportional to the radius $r_H$ 
of the horizon, or roughly of order $M_*$ and evaporate by emitting particles, 
whose mass is smaller than $T_H$. The radiation emitted depends both on the 
spin of the emitted particle, on the dimension of the ambient space 
and on the amount of back-scattering outside the horizon,
contributions which are commonly included in the so called 
``greybody factor'', which are particularly relevant 
in the characterization of the spectrum at lower and at intermediate energy. 
A main feature of the decaying mini black hole 
is its large partonic multiplicity, 
with a structure of the event which is approximately spheroidal in the 
black hole rest frame. 

Once produced, these mini black holes evaporate almost instantly.
The phenomenological study of 4-dimensional black holes of large mass 
and, in particular, 
of their Hawking radiation \cite{Hawking} 
\cite{Page} \cite{MacGibbon}, as well as the study 
of the scattering of states of various spins ($s=0,1/2,1$) on a 
black hole background, all performed in the semiclassical 
approximation, have a long history.
For rotating black holes one identifies four phases characterizing its decay, 
which are
(1) the balding phase (during which the hole gets rid of its hair); 
(2) the spin-down phase (during which the hole slows down its rotational
motion);
(3) the Schwarzschild phase (the usual semiclassically approximated 
evaporation phase) and, finally, 
(4) the Planck phase (the final explosive part of the evaporation 
process, with important quantum gravitational contributions). 
Undoubtedly, the best understood among these phases 
is the Schwarzschild phase, which is characterized by the emission of 
a (black body) energy spectrum which is approximately thermal, 
with a superimposed energy-dependent modulation, especially at larger values 
of the energy. The modulation is a function of the spin and is calculable 
analytically only at small energies. 
Extensions of these results to 4+n dimensions are now available,
especially in the Schwarzschild phase, where no rotation and no charge 
parameter characterize the background black hole solutions. 
Partial results exist for the spin down phase, where the behaviour 
of the greybody factors have been studied (at least for 1 additional 
extra dimension) both analytically and numerically.
The Planck phase, not so relevant for a hole of large mass (say of the mass 
of the sun ($M\sim 2\times 10^{33}$ gr) which emits in the nano-Kelvin region,
is instead very relevant for the case of mini-black holes, for which the 
separation between the mass of the hole and the corresponding (effective) 
Planck mass $M_*$ gets drastically reduced as the temperature 
of the hole raises and the back-reaction of the metric has to be taken into account.

In the discussion below we shall use the semiclassical formulas derived
for large black holes in the Schwarzschild phase and naively extrapolate them 
to the mini black holes as well. 
This is not, we believe, a severe approximation for the phenomena we shall discuss.
As the hole evaporates, it looses energy, its mass decreases, its temperature
increases and the rate of evaporation becomes faster. Thus, the lifetime of
the hole is actually shorter than the one derived ignoring the back reaction.
As we shall see below, the naive lifetime is already many orders of magnitude
smaller than the hadronization time. This justifies the use
of the ``sudden approximation'' we are making of the decay process
and explains why the neglect of the back reaction is not severe.

We recall that the metric of the $4+n$
dimensional hole in the Schwarzschild phase is given by 
\cite{Myers}
\begin{equation}
ds^2 = \left[1-\left(\frac{r_H}{r}\right)^{n+1}\right]\,dt^2 -
\left[1-\left(\frac{r_H}{r}\right)^{n+1}\right]^{-1}\,dr^2 - 
r^2 d\Omega_{2+n}^2\,,
\label{metric-n}
\end{equation}
where $n$ denotes the number of extra spacelike dimensions, 
and $d\Omega_{2+n}^2$ is the area of the 
($2+n$)-dimensional unit sphere which, using coordinates
 $0 <\varphi < 2 \pi$ and $0< \theta_i < \pi$, with $i=1, ..., n+1$
takes the form 
\begin{equation}
d\Omega_{2+n}^2=d\theta^2_{n+1} + \sin^2\theta_{n+1} \,\biggl(d\theta_n^2 +
\sin^2\theta_n\,\Bigl(\,... + \sin^2\theta_2\,(d\theta_1^2 + \sin^2 \theta_1
\,d\varphi^2)\,\Bigr)\biggr)\,.
\label{unit}
\end{equation}
The temperature $T_H$ of the black hole is related to the size 
of its horizon by \cite{Myers}
\begin{equation}
T_H=\frac{n+1}{4\pi\,r_H}\,
\label{temp}
\end{equation}
and the formula for the horizon $r_H$ can be expressed in general in terms 
of the mass of the black hole $M_{BH}$ and the gravity scale $M_*$ \cite{Myers}
\begin{equation}
r_H= \frac{1}{\sqrt{\pi}M_*}\left(\frac{M_{BH}}{M_*}\right)^
{\frac{1}{n+1}}\left(\frac{8\Gamma\left(\frac{n+3}{2}\right)}{n+2}\right)
^{\frac{1}{n+1}}\,.
\label{horizon}
\end{equation}
For $n=0$ and $M_*=M_{Pl}\simeq 10^{19}$ GeV it reproduces the usual formula 
for the horizon ($r_H=2 G M_{BH}$) of a 4 dimensional black hole. 
For $n>0$ the relation 
between $r_H$ and $M_{BH}$ becomes nonlinear and the presence of $M_*$ in 
the denominator of 
Eq.~(\ref{horizon}) in place of $M_{Pl}$ increases the horizon size 
for a given $M_{BH}$. For $M_{BH}/M_*\sim 5$ and $M_*=1$ TeV the size of the 
horizon is around $10^{-4}$ fm and decreases with increasing $n$.

In the Schwarzschild/spin-down phase, the number of particles emitted per unit 
time by the black hole as a function of energy 
is expressed in terms of the absorption/emission cross sections 
$\sigma^{(s)}_{j,n}(\omega)$ (or equivalently of the greybody factors 
$\Gamma(\omega)$), which, apart from $n$, depend on the spin $(s)$ of the 
emitted particle, the angular momentum $(j)$ of the partial wave and the 
corresponding energy $(\omega)$, 
\begin{equation}
\label{flux}
\frac{dN^{(s)}(\omega)}{dt d\omega} = 
\sum_{j} \frac{\sigma^{(s)}_{j,n}(\omega)}{2 \pi^2}
{\omega^2 \over \exp\left(\omega/T_{H}\right) \pm 1}d\,\omega  \,.
\label{rate}
\end{equation}
Multiplying the rate of emitted particles per energy interval 
$dN^{(s)}(\omega)/dt d\omega$ by the particle energy $\omega$ one obtains for
the power emission density 
\begin{equation}
\frac{d E^{(s)}(\omega)}{dt d\omega} = \sum_{j} 
\frac{\sigma^{(s)}_{j,n}(\omega)}{2 \pi^2}\,
{\omega^3  \over \exp\left(\omega/T_{H}\right) \pm 1}d\,\omega\,
\label{power}
\end{equation}
where the sum is over all Standard Model particles and the +($-$) in the denominator 
correspond to fermions (bosons), respectively.
$\sigma^{(s)}_{j,n}$ are the cross sections for the various partial waves and 
depend on the spin $s$ of each particle. 
We recall, that in the geometric optics approximation a black hole acts as a perfect 
absorber of slightly larger radius $r_c$ than $r_H$ \cite{Sanchez}, 
which can be identified as the critical radius for null geodesics 
\beq
r_c=\left(\frac{n+3}{2}\right)^{1/(n+1)}\sqrt{{n+3}\over {n+1}}\, r_H \,.
\label{radius}
\eeq

The optical cross section is then defined in function of  
$r_c$ (or equivalently $r_H$ via Eq.~(\ref{radius})), 
such that $A_k$, the effective surface area of the black hole hole 
projected over a k-dimensional sub-manifold 
becomes \cite{ehm}:
\begin{equation}
 A_k = \Omega_{k-2}\left( {d-1 \over 2} \right)^{{2\over d-3}}
 \left( {d-1\over d-3} \right)^{{k-2\over 2}} r_{H}^{k-2}
\label{ak}
\end{equation}
and 
\begin{equation}
 \Omega_k= {2\pi^{k+1\over 2} \over \Gamma({k+1\over 2})}.
\end{equation}
is the volume of a k-sphere. 

It is convenient to rewrite the greybody factors as a 
dimensionless constant $\Gamma_s=\sigma_s /A_4$ normalized to the 
effective area of the horizon $A_4$, obtained from (\ref{ak}) 
setting $k=4$ and $d= 4 + n$
\beq
A_4=4 \pi \left(\frac{n+3}{2}\right)^{2/(n+1)} \frac{n+3}{n +1}\, r_H^2,
\eeq
and replacing the particle cross section $\sigma$ in terms of a thermal 
averaged graybody factor $\Gamma_i$($\Gamma_{1/2}=2/3,\Gamma_{1}=1/4, 
\Gamma_0=1$, $i$ denoting the spin or species \cite{page}). 
Eqs.~(\ref{rate}) integrated over the frequency give (for particle $i$)

\beq
\frac{d N_i}{d t}=\alpha(n,r_H) \, T_H^3
\eeq
with
\beq
\alpha(n,r_H)=\frac{f_i}{2\pi^2} \,\Gamma_i \,\Gamma(3)\,\zeta(3)\, 
c_i\, A_4\, T_H^3, 
\eeq
where $c_i$ is the number of degrees of freedom of particle $i$ and $f_i$ 
is defined by the integral ($s_i$ is the spin)
\beq
\int_0^\infty d\, \omega\frac{\omega^2}{e^{\omega/T_H} - (-1)^{2 s_i}}= 
f_i\, \Gamma(3)\,\zeta(3)\, T_H^3
\eeq
from which $f_i=1$ $(f_i=3/4)$ for bosons (fermions). 
These numbers depend on the dimension of the brane, which in our case is 3.
$\Gamma(x)$ and $\zeta(x)$ are the Gamma and the Riemann function respectively.
Since $A_4$ depends on the temperature (via $r_H$), after some manipulations 
one obtains 
\beq
A_4 T_H^3= \frac{1}{4\pi} \left(\frac{n+3}{2}\right)^{2/(n+1)}(n+3)(n+1)\, T_H
\eeq
and 
\beq
\frac{d N_i}{d t}=\frac{f_i}{8\pi^3}\frac{(n+3)^{(n+3)/(n+1)}}{2^{2/(n+1)}} 
(n+1)\Gamma(3)\zeta(3) \Gamma_i c_i T_H.
\eeq
Summing over all the particles $i$ we obtain the compact expression \
\beq
\frac{d N }{d t}=\frac{1}{2\pi}\left(\sum_i{f_i}\,\overline{\Gamma}_i\, 
c_i\right)\,\Gamma(3)\,\zeta(3)\,T_H
\label{enne}
\eeq
with 
\beq
\overline{\Gamma}_i= 
\frac{\Gamma_i (n+1)\,(n+3)^{(n+3)/(n+1)}}{4 \pi^2 \, 2^{2/(n+1)}}.
\eeq

The emission rates are given by
\begin{equation}
\dot{N}_i \approx 4\times 3.7 \times 10^{21}\,
\frac{(n+3)^{(n+3)/(n+1)}(n+1)}{2^{2/(n+1)} \,}\,
\left(\frac{T_H}{{\rm GeV}}\right)\,\, {\rm s}^{-1} \,\,,
\end{equation}
\begin{equation}
\dot{N}_i \approx 4\times 3.7 \times 10^{21}\,
\frac{(n+3)^{(n+3)/(n+1)}(n+1)}{2^{2/(n+1)}}\,
\left(\frac{T_H}{{\rm GeV}}\right)\,\, {\rm s}^{-1} \,\,,
\end{equation}
\begin{equation}
\dot{N}_i \approx 4\times 1.85 \times 10^{20}\,
\frac{(n+3)^{(n+3)/(n+1)}(n+1)}{2^{2/(n+1)}
\,}\,
\left(\frac{T_H}{{\rm GeV}}\right)\,\, {\rm s}^{-1} \,\,,
\end{equation}
for particles with $s = 0,\, 1/2, \,1,$ respectively. Furthermore,
integration of Eq.~(\ref{power}) gives for the black hole mass evolution
\beqa
\frac{d M}{d t} &\equiv& -\frac{dE}{dt} = \beta(n, r_H)\, T_H^4 \nonumber \\
&=& \frac{1}{2\pi}\left(\sum_i{f_i}\,\overline{\Gamma}_i\, c_i\right)\,
\Gamma(4)\,\zeta(4)\,T_H^2, \nonumber \\
\label{emme}
\eeqa
with
\beq
\beta=\frac{1}{2 \pi^2}\sum_i \left(c_i\, \Gamma_i\, {f'}_i \right)\, A_4\, \Gamma(4)\,\zeta(4)
\eeq
where now ${f'_i}=1$ $(7/8)$ for bosons (fermions). 
Taking the ratio of the two equations (\ref{emme}) and (\ref{enne}) 
we obtain 
\beqa
\frac{d N}{d M}&=&\left(\frac{\alpha}{\beta}\right)\,\frac{1}{T_H}\nonumber \\
 &=& \rho \frac{4 \pi\theta(n)}{n+1}\frac{1}{M_*}
\left(\frac{M}{M_*}\right)^{\frac{1}{(n+1)}},
\label{better}
\eeqa
where we have defined 

\beq
\theta(n)= \left(\frac{8\Gamma\left(\frac{n+3}{2}\right)}{n+2}\right)^{\frac{1}{n+1}}\frac{1}{\sqrt{\pi}},
\eeq
and 
\beq
\rho=\frac{ \sum_i c_i\, f_i\, {\Gamma}_i\, \Gamma(3)\, \zeta(3)}
{ \sum_i c_i \,{f'}_i\, {\Gamma}_i\, \Gamma(4)\, \zeta(4)}.
\eeq
This formula does not include corrections from emission in the bulk. 

\begin{figure}[tbh]
{\centering \resizebox*{12cm}{!}{\rotatebox{-90}{\includegraphics{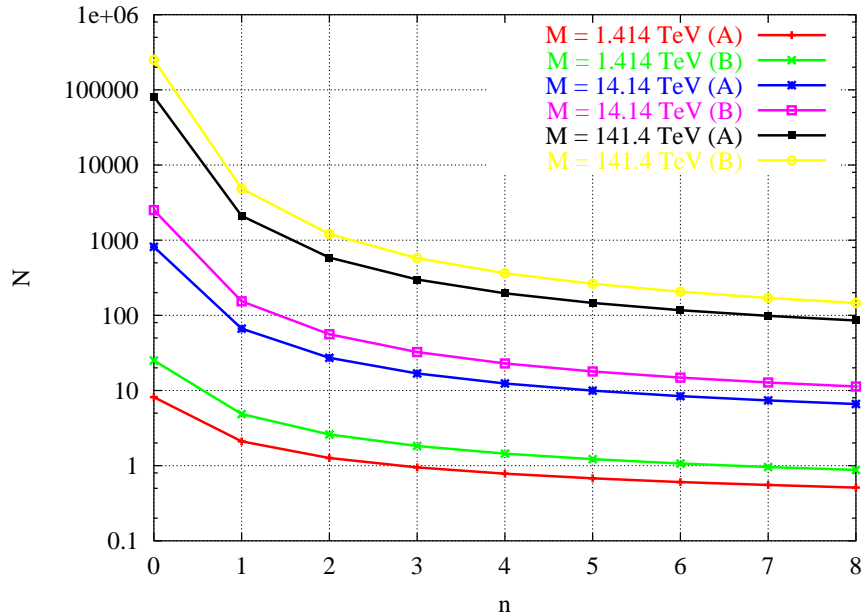}}} \par}
\caption{Multiplicities computed with Eq.~(\ref{improved}) (A) 
and Eq.~(\ref{nav}) (B) for a varying number of extra dimensions $n$.}
\label{DLL}
\end{figure}

In the ``sudden approximation'' in which the black hole decays at its original 
formation temperature one easily finds $ N = \left\langle
\frac{M_{BH}}{E} \right\rangle$, where $E$ is the energy spectrum
of the decay products, and using Boltzmann statistics 
$ N  \approx \frac{M_{BH}}{2T_H}$ one obtains the expression \cite{DL}

\begin{equation}
    N  = \frac{2\pi}{n+1}
    \left(\frac{M_{BH}}{M_*}\right)^\frac{n+2}{n+1}\,\theta(n).
\label{nav}
\end{equation}
This formula is approximate as are all the formulas for the multiplicities. 
A more accurate expression is obtained integrating Eq.~(\ref{better}) to obtain 

\beq
N=\rho \frac{4 \pi}{n+2}\left(\frac{M_{BH}}{M_*}\right)^\frac{n+2}{n+1}\theta(n),
\eeq
and noticing that the entropy of a black hole is given semiclassically by 
the expression 
\beqa
S_0 &=&\frac{n+1}{n+2} \frac{M}{T} \nonumber \\
&=& \frac{4 \pi}{n+2}\left(\frac{M_{BH}}{M_*}\right)^\frac{n+2}{n+1}\theta(n),
\nonumber 
\eeqa
one finds that 
\beq
N= \rho\, S_0, \nonumber \\
\label{improved}
\eeq
which can be computed numerically for a varying $n$. As one can see from Fig.~\ref{DLL}, 
the two formulas for the multiplicities are quite close, as expected, 
but Eq.~(\ref{nav}) gives larger values for the multiplicities 
compared to (\ref{improved}) as noted by \cite{Cavaglia1}. 
Other expressions for the multiplicities can be found in \cite{Cavaglia-Das-Casadio}.

Since the number of elementary states becomes quite large as we raise 
the black hole mass compared to the (fixed) gravity scale, and given 
the (large) statistical fluctuations induced by the formation of the air shower, which reduce the dependence on the multiplicity formula used, we will adopt 
Eq.~(\ref{nav}) in our simulations. Overall, in the massless approximation,
the emission of the various species for a 3-brane is characterized by 
approximately $2\%$ into spin zero, $85\%$ into spin half and $13\%$ 
into spin one particles, with similar contributions 
also for the power emissivities. These numbers change as we vary the 
dimension of the brane (d) and so does the formula for the emissivities, 
since the number of brane degrees of freedom $(c_i(d))$ 
has to be recomputed, together with the integrals on the emission spectra 
$(f_i(d))$ \cite{Cavaglia}.     

The integration of the equation for the power spectrum, 
in the massless approximation, can be used to compute the total time of decay 
(assuming no mass evolution during the decay)
\begin{equation}
\tau \sim \frac{1}{M_*}\,\biggl(\frac{M_{BH}}{M_*}\biggr)^{(n+3)/(n+1)}\,
\label{lifetime}
\end{equation}
which implies that at an energy of approximately 1 TeV the decay time 
is of the order of $10^{-27}$ seconds. Therefore 
strong interaction effects and gravity effects appear to be widely separated 
and hadronization of the partons takes place after their crossing of the horizon. 
The black hole is assumed to decay isotropically (s-wave) to a set of 
$N$ elementary states, selected with equal probability from 
all the possible states available in the Standard Model. 
We mention that in most of the analysis presented so far \cite{pp4,Anchordoqui, Cavaglia} 
the (semiclassical) energy loss due to bulk emission has not been thoroughly analyzed. 
We will therefore correct our numerical studies by keeping into account some estimates 
of the bulk emission.

\section{Modeling of the Black Hole Decay \label{sec:mbh3}}

The amount of radiation emitted by the black hole in the ED is viewed, 
by an observer living on the brane, as missing energy compared to 
the energy available at the time when the black hole forms. 
From the point of view of cosmic ray physics 
missing energy channels imply 
reduced multiplicities in the final air shower and modified lateral distributions, 
these two features being among the main observables of the cosmic ray event. 
However, since the initial energy of the original cosmic ray 
is reconstructed by a measurement of the multiplicities, 
an event of reduced multiplicity will simply 
be recorded as an event of lower energy. It is then 
obvious that an additional and independent reconstruction of 
the energy of the primary cosmic ray is needed in order to correctly identify the energy of 
these events. 

In our study we will compute all the observables of the induced 
air shower using both the lab frame (LF) 
and the black hole frame (BHF) to describe the impact and the formation of the 
intermediate black hole resonance. 
Also, in the simulations that we will perform, the observation level at which  
we measure the properties of the air showers 
will be selected to take properly into account the actual position of a 
hypothetical experimental detector. The target of the first impact of mass $M$ 
is assumed to be a nucleon (or a quark) at rest in the atmosphere and the 
center of mass energy, corrected by emission loss in the bulk, is made 
promptly available for an instantaneous black hole formation and decay.
We will also assume that the energy $E_1$ of the incoming primary 
varies over all the highest part of the cosmic ray spectrum, from $10^{15}$ eV up to
$10^{20}$ eV. 

We denote by 
$\beta$ the speed of the black hole in the lab frame. In our notations, $E^*$
is the typical energy of 
each elementary state in the decay products (parton, lepton) 
in the BH frame and $P^*$ is its corresponding momentum. 

We will assume that a black hole decays ``democratically'' into all the 
possible partonic states, proportionally to the number of Standard Model 
states which are available to it at a given energy. 

The energy per partonic channel will be appropriately weighted   
and we will assume that each parton $(f)$ will decay into a final state hadron 
$h$ (carrying a fraction $x$ of the original momentum), 
with a probability distribution given by the corresponding fragmentation 
function $D_f^h(x,Q^2)$, which is evolved from some low energy 
input scale $Q_0$ up to the relevant scale characterizing the decay. This is given by 
the available energy per fundamental state, equally distributed among all the 
states.

The quantification of the injection spectrum 
involves a computation of the relevant probabilities 
for the formation of all the possible hadronic/leptonic states prior to the simulation 
of the air shower. Let's briefly elaborate on this.

To move from the parton level to hadron level, 
we let $D_q^h(x,Q^2)$, $D_{\overline{q}}^h(x,Q^2)$, 
and $D_g^h(x,Q^2)$ be the fragmentation functions 
of $N_F$ quarks $q$, antiquarks $\overline{q}$, and of the gluon $g$, 
respectively, into
some hadron $h$ with momentum fraction $x$ at the scale $Q$. 
From the fragmentation 
functions we obtain, for each hadron $h$, the mean multiplicity of the corresponding s-wave  and the corresponding average energy and momentum. Specifically 
we obtain
\beqa
  <D_h> &=&\sum_f\int_{z_{min}}^1 dz\, D_f^h(z,Q^2)
\eeqa
for the probability of producing a hadron $h$, and 
\beqa
E^*_h &=& \sum_f \int_{z_{min}}^1 z\, dz\, D_f^h(z,Q^2)
\eeqa
for the average energy of the same hadron. We recall that $z_{min}$
is the minimal fraction of energy a hadron $(h)$, of mass $m_h$, can take at a scale $Q$, 
and can be defined as $z_{min}=m_h/(Q/2)$. In practical applications 
one can take the nominal value $z_{min}=0.05$ for every hadron, without affecting
much the mean multiplicities and the related probabilities.  
This implies that 

\beq
<D_r^h> + \sum_f <D_f^h> + <D_g^h> + <D_\gamma^h>\equiv \textrm{Pr}_h \nonumber \\
\eeq
together with the condition $\sum_h \textrm{Pr}_h =1$, where the sum runs over all the 
types of hadrons allowed by the fragmentation. In all the equations above, 
the fragmentation takes place at the typical scale $Q=E/N$, scale at 
which the moments are computed numerically. 
For the identification of the probabilities it is convenient 
to organize the 123 fundamental states of the Standard Model 
into a set of flavour states 
$(q_f)$, with $f$ running over all the flavours except for the top quark, 
where in $(q_f)$ we lump antiquark states and color states, 
plus some additional states. The weight of the $(q_f)$ set is 
$p_f=2\times 2\times 3/123$, where the factors 2 and 3 refer to spin, quark-antiquark 
degeneracy and color. It is worth to
recall that quark and antiquark states of the same flavour 
have equal fragmentation functions in all the hadrons, and this 
justifies the $q/\bar{q}$ degeneracy of the set.
The additional states are the gluon (g) with a weight $p_g=2\times 8/123$, the photon 
$(\gamma)$, with a weight $p_\gamma=2/123$ and the remaining states $(r)$ 
in which we lump all the states which have been unaccounted for, 
whose probabilities 
$p_r$ are computed by difference. These include the top and the antitop 
$(12/123)$, the W's and Z $(9/123)$ and the leptons $(24/123)$.
The fragmentation functions into hadrons, corresponding to these states, 
$<D_r^h>$ are computed by difference from the remaining ones
$<D_g^h>$, $<D_f^h>$ and $<D_\gamma^h>$, which are known at any scale 
$Q$ from the literature. 
Beside the favour index $f=u,d,c,s,b$, introduced above, we introduce a second index 
$i$ running over the $(r)$ states, the photon and the gluons $(i=g,\gamma,r)$. 

The probability of generating a specific sequence of $N$ states in the course 
of the evaporation of the black hole is then given by a multinomial distribution of the form
\beq
f(n_f,n_i,p_f,p_i)= {N!\over \prod_f n_f! \prod_i n_i!}\prod_f 
p_f^{n_f}\prod_i {p_i}^{n_i}
\eeq
which describes a typical multi-poissonian process with $N$ trials. 
Notice that, to ensure proper normalization, we need to require that 
\beqa
\prod_i n_i! &=& n_g! n_\gamma! n_r! \nonumber \\
&=& n_g! n_\gamma! (N - \sum_f n_f - n_g - n_\gamma)! 
\eeqa

The computation of the cumulative probabilities to produce any number of hadrons of type 
$h$ by the decay of the black hole are obtained from the multinomial 
distribution multiplied by the fragmentation probabilities of each elementary 
state into $(h)$ and summing over all the possible sequences
\beq
\textrm{Pr}_{\textrm{cum h}}\equiv \sum_{n_f, n_i} {N!\over \prod_f n_f! \prod_i n_i!}\prod_f 
\left(p_f <D_f^h>\right)^{n_f}\prod_i \left(p_i <D_i^h>\right)^{n_i}.
\label{probab}
\eeq
A possible way to compute 
$\textrm{Pr}_{\textrm{cum h}}$ when
$N$ is large is to multiply the multinomial distribution by a suppression 
factor  $\textrm{Exp}[{-\Lambda( \sum_i n_i + \sum_f n_f -N)}]$, with $\Lambda$
a very large number, and interpret this factor as a Boltzmann weight, as in 
standard Monte Carlo computations of the partition function 
for a statistical system. Simulations can be easily done by a 
Metropolis algorithm and the configurations of integers selected are those for 
which the normalization condition $N= \sum_i n_i + \sum_f n_f$ is satisfied. 
In our case, since we are interested only in the mean number of hadrons produced 
in the decay and in their thermal spectrum, 
the computation simplifies if we average over all the relevant configurations.

\section{Fragmentation and the Photon Component \label{sec:mbh4}}

\begin{figure}
{\centering \resizebox*{9cm}{!}{\rotatebox{-90}{\includegraphics{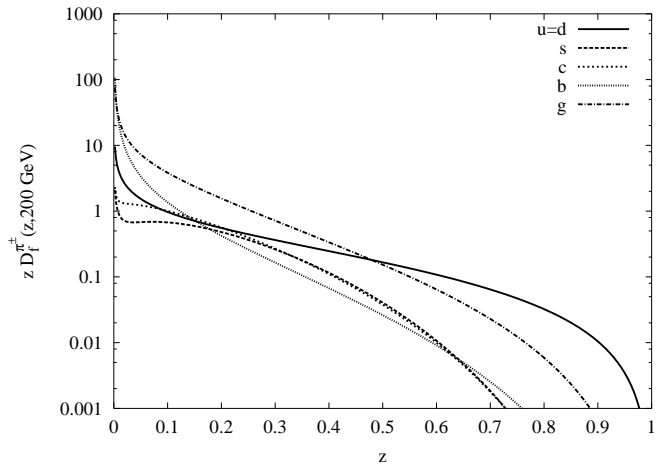}}} \par}

\caption{Fragmentation functions into \protect\( \pi ^{\pm }\protect \) at
\protect\( 200\, \textrm{GeV}\protect \).}
\label{frag1}
\end{figure}

\begin{figure}
{\centering \resizebox*{9cm}{!}{\rotatebox{-90}{\includegraphics{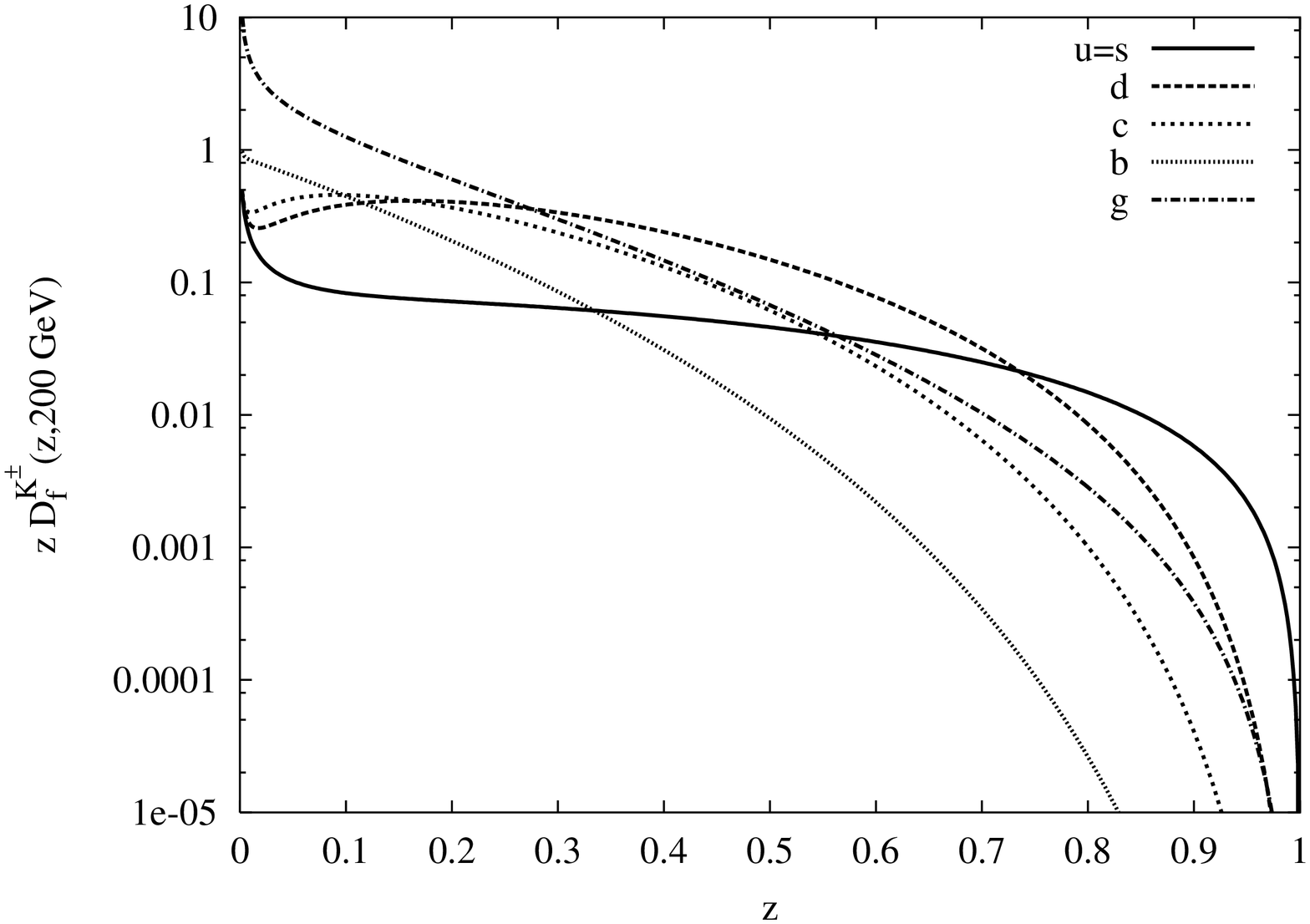}}} \par}

\caption{Fragmentation functions into \protect\( K^{\pm }\protect \) 
at \protect\( 200\, \textrm{GeV}\protect \).}
\label{frag2}
\end{figure}

\begin{figure}
{\centering \resizebox*{9cm}{!}{\rotatebox{-90}{\includegraphics{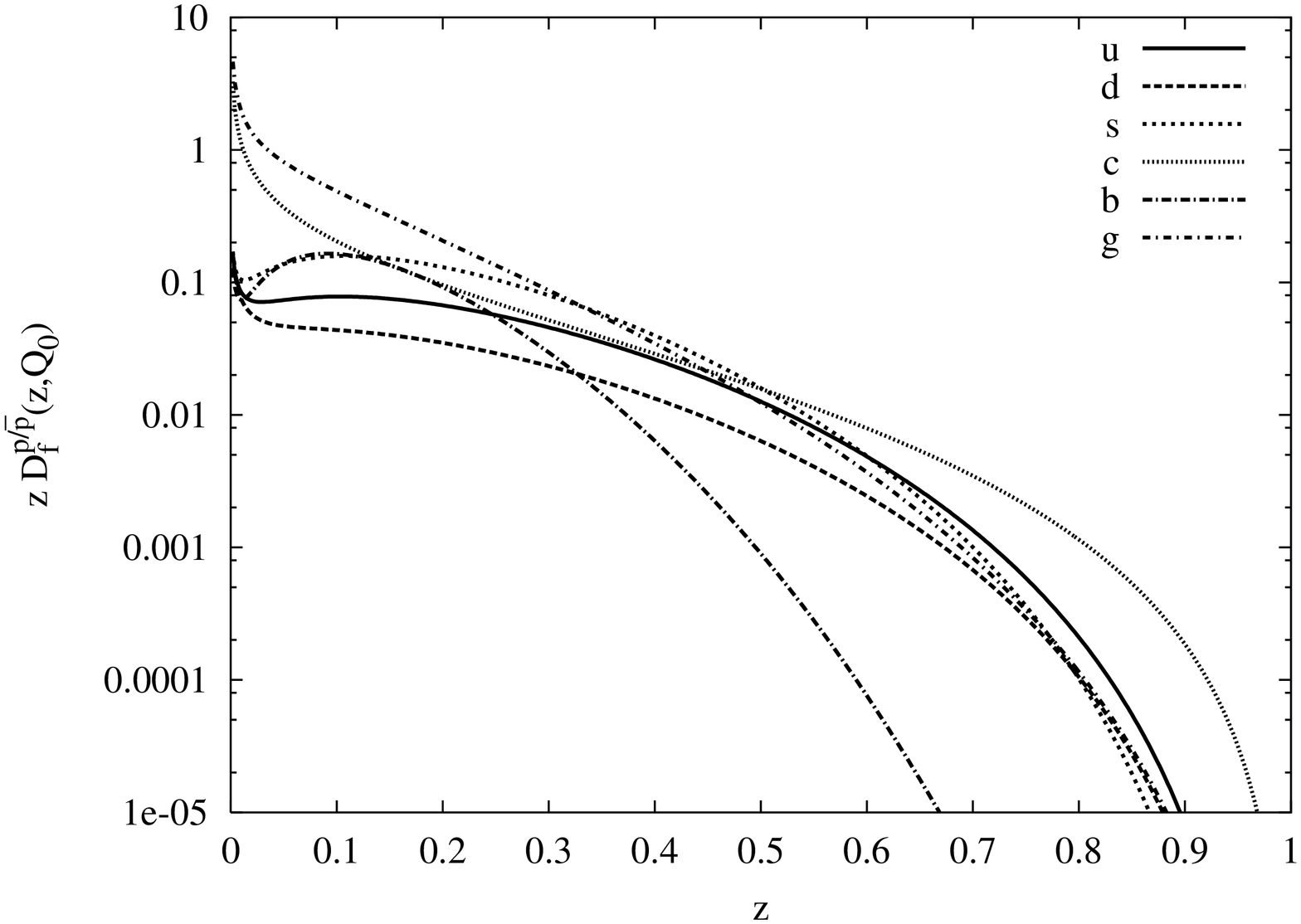}}} \par}

\caption{Fragmentation functions into \protect\( p/\overline{p}\protect \)
at \protect\( 200\, \textrm{GeV}\protect \).}
\label{frag3}
\end{figure}

The evolution with $Q^2$ of the fragmentation functions is 
conveniently formulated in terms of the linear combinations
\begin{eqnarray}
D_{\Sigma}^h(x,Q^2)&\n=\n&\sum_{i=1}^{N_F}\left( 
      D_{q_i}^h(x,Q^2)+D_{\overline{q}_i}^h(x,Q^2) \right)
\COMMA
\\
\nonumber\\
D_{(+),i}^h(x,Q^2)&\n=\n&D_{q_i}^h(x,Q^2) 
     + D_{{\overline{q}},i}^h(x,Q^2)
   -\frac{1}{N_F} D_{\Sigma}^h(x, Q^2)
\COMMA
\\
\nonumber\\
D_{(-),i}^h(x,Q^2)&\n=\n&D_{q_i}^h(x,Q^2)
     - D_{{\overline{q}},i}^h(x,Q^2)
\COMMA
\end{eqnarray}
as for these the gluon decouples from the 
non--singlet $\scriptstyle (+)$ and the
asymmetric $\scriptstyle (-)$
combinations, leaving only the singlet and the gluon 
fragmentation functions coupled;
\begin{eqnarray}
Q^2\frac{d}{d Q^2}
D_{(+),i}^h(x,Q^2)
&\n=\n&\left[P_{(+)}\left(\alpha_s(Q^2)\right)\otimes
D_{(+),i}^h(Q^2)\right](x) 
\COMMA \label{APplus}
\\
\nonumber\\
Q^2\frac{d}{d Q^2}D_{(-),i}^h(x,Q^2)
&\n=\n&\left[P_{(-)}\left(\alpha_s(Q^2)\right)\otimes
D_{(-),i}^h(Q^2)\right](x)
\COMMA \label{APminus}
\\
\nonumber\\
Q^2\frac{d}{d Q^2}D_{\Sigma}^h(x,Q^2)
&\n=\n&\left[
P_{\Sigma}\left(\alpha_s(Q^2)\right)\otimes D_{\Sigma}^h(Q^2)
\right](x)
\nonumber\\
& & \hspace{1cm}
+2N_F\left[P_{q\to G}\left(\alpha_s(Q^2)\right)\otimes D_G^h(Q^2)
\right](x)
\COMMA 
\\
\nonumber\\
Q^2\frac{d}{d Q^2}D_G^h(x,Q^2)
&\n=\n&\frac{1}{2N_F}\left[
P_{G\to q}\left(\alpha_s(Q^2)\right)\otimes D_{\Sigma}^h(Q^2)
\right](x)
\nonumber\\
& & \hspace{1cm}
+\left[
P_{g\to g}\left(\alpha_s(Q^2)\right)D_g^h(Q^2)\right](x)
\STOP 
\end{eqnarray}
The kernels 
that appear in the equations above are defined by
\begin{eqnarray}
P_{(+)}\left(x,\alpha_s(Q^2)\right)&\n=\n&
P_{q\to q}^{V}\left(x,\alpha_s(Q^2)\right) +
P_{q\to \overline{q}}^{V}\left(x,\alpha_s(Q^2)\right)
\COMMA
\\
\nonumber\\
P_{\Sigma}\left(x,\alpha_s(Q^2)\right)&\n=\n&
P_{(+)}\left(x,\alpha_s(Q^2)\right)+
2N_FP_{q\to q}^{S}\left(x,\alpha_s(Q^2)\right)
\COMMA
\\
\nonumber\\
P_{(-)}\left(x,\alpha_s(Q^2)\right)&\n=\n&
P_{q\to q}^{V}\left(x,\alpha_s(Q^2)\right) -
P_{q\to \overline{q}}^{V}\left(x,\alpha_s(Q^2)\right), 
\end{eqnarray}
with $\alpha_s(Q^2)$ being the QCD coupling constant.
In the perturbative expansion of the splitting functions,
\begin{equation}
P(x,\alpha_s(Q^2))=\frac{\alpha_s(Q^2)}{2\pi}P^{(0)}(x)
+\left(\frac{\alpha_s(Q^2)}{2\pi}\right)^2 P^{(1)}(x)
+{\cal O}\left(\left(\frac{\alpha_s(Q^2)}{2\pi}\right)^3\right).
\end{equation}
The timelike kernels that we use are given by
\begin{eqnarray}
P_{q\to q}^{V,(0)}(x) &\n=\n& C_F\left[\frac{3}{2}\delta(1-x)
   +2\left(\frac{1}{1-x}\right)_+ -1 -x\right] \COMMA
\label{P0Vqq}
\\
P_{q\to \overline{q}}^{V,(0)}(x) &\n=\n& P_{q\to q}^{S,(0)}(x) = 0
\COMMA
\label{P0Vqqbar}
\\
P_{q\to G}^{(0)}(x) &\n=\n& C_F\left[\frac{1+(1-x)^2}{x}\right]
\COMMA
\label{P0qG}
\\
P_{q\to q}^{(0)}(x) &\n=\n& 2N_F T_R\left[x^2+(1-x)^2\right]
\COMMA
\label{P0Gq}
\\
P_{G\to G}^{(0)}(x) &\n=\n& \left(\frac{11}{6}N_C-\frac{2}{3}N_F T_R
  \right)\delta(1-x) 
\nonumber\\
& & \hspace{1cm}
+2N_C\left[\left(\frac{1}{1-x}\right)_+ 
  +\frac{1}{x}-2+x-x^2\right]
\STOP
\label{P0GG}
\end{eqnarray}
\smallskip
The formal solution of the equations is given by
\begin{equation}
D_a^h(x,Q^2)=D_a^h(x,Q_0^2)+\int_0^{\log(Q^2/Q_0^2)} d \log(Q^2/Q_0^2)
\frac{\alpha_s(Q^2)}{2\pi}
   \sum_b \left[P_{a\to b}(\alpha_s(Q^2)) 
    \otimes D_b^h(Q^2)\right]
\label{ap}
\end{equation}
 where $Q_0$ is the starting scale of the initial conditions, given by 
$D(x,Q_0^2)$.
At leading order in $\alpha_s$, we solve this equation using a special ansatz
\beq
D_f^{h}(x,Q^2)=\sum_n {A_n(x)\over n!} \log\left( {\alpha_s(Q^2)\over 
\alpha_s (Q_0^2)}\right)^n
\eeq
and generating recurrence relations at the $n+1-th$ order for the 
$A_{n+1}$ coefficients in terms of the $A_n$ \cite{CPC}. It is easy to see 
that this corresponds to the numerical implementation 
of the formal solution 
\beq
D_f^{h}(x,Q^2)={\it{Exp}}\left(t P\otimes\right) D_f^{h}(x,Q_0^2)
\eeq
with $t=\left(\alpha_s(Q^2)/ \alpha_s (Q_0^2)\right)$, where the 
exponential is a formal expression for an infinite iteration of convolution products. 
We show in Figures~\ref{frag1}-\ref{frag3} results for some of the 
fragmentation functions into pions, kaons and protons 
computed for a typical parton scale of 200 GeV.  

The photon contributions to the decay of the black hole is treated separately. 
The evolution equation for the fragmentation functions of photons and parton 
fragmentation into photon 
$D_\gamma^\gamma(x,Q^2), \,\, D_q^\gamma(x,Q^2)$ satisfy at leading order in 
$\alpha_{\rm{em}}$ (the QED fine structure constant) and $\alpha_s$ 
(the QCD coupling), the evolution 
equations \cite{Klasen}
\beq
\frac{\textrm{d} D_{\gamma}^\gamma(x,Q^2)}{\textrm{d}\ln Q^2} =
 \frac{\alpha}{2\pi} P_{\gamma\rightarrow \gamma}(x)\otimes D_{\gamma}^{\gamma}(x,Q^2)
\eeq
which can be integrated with the initial conditions $D_{\gamma}^\gamma(x,Q^2)=\delta (1-x)$,  
and 
\beq
 \frac{\textrm{d} D_q^{\gamma}(x,Q^2)}{\textrm{d}\ln Q^2} =
 \frac{\alpha}{2\pi} P_{\gamma\rightarrow \gamma}(x)\otimes D_{\gamma}^\gamma(x, Q^2)
\eeq
which can also be integrated with the result \cite{gehrmann}
\beq
D_{\gamma/q}(x,Q^2)=\frac{\alpha}{2\pi}P_{q\rightarrow \gamma}(x)\ln\frac{Q^2} 
{Q_0^2}+D_{q}^\gamma (x,Q_0^2).
\label{frag}
\eeq
In Eq.~(\ref{frag}) the second term is termed the hadronic boundary conditions, 
which come from an experimental fit, 
while the first term is the pointlike contributions, which can be obtained perturbatively.  
In \cite{ALEPH} a leading order fit was given 
\beq
D_q^\gamma(x,Q_0^2)=\frac{\alpha}{2\pi}\left[-\, P_{q\rightarrow \gamma }(x)
\ln(1-x)^2-13.26\right]
\eeq
at the starting scale $Q_0=0.14$ GeV. 
The kernels in these cases are given by simple modifications of the ordinary 
QCD kernels, for instance  
$P_{q\rightarrow \gamma}(x)= e_q^2/C_F P_{q\rightarrow g}(x)$ \cite{Klasen}.

For the fragmentation function of quarks
to photons with virtuality $M_\gamma$, the perturbative result is given in \cite{qiu}
\beq
D_{\gamma/q}(x,Q^2,M_{\gamma}^2) = e_q^2\frac{\alpha}{2\pi}\left[\frac{1+(1-x)^2}{x}
\ln\frac{xQ^2}{M_{\gamma}^2}- x\left( 1-\frac{P^2}{xQ^2}\right)\right],
\eeq
where $P^2$ is the virtuality of the photon. 
The gluon to photon transitions are neglected, since $P_{g\rightarrow \gamma}$ vanishes 
in leading order. 

We recall that each elementary
state emitted is characterized by an average energy 
given by \( \left\langle \varepsilon \right\rangle =M_{BH}/\left\langle N\right\rangle  \).

The leptonic component \( e^{\pm } \), \( \mu ^{\pm } \), produced by the decay
is left unaltered and provides an input for the air shower simulator as soon as 
these particles 
cross the horizon. The $\tau^\pm$s are left to decay into their main channels,
while the hadronization of the \( u,\, d,\, s,\, c \) quarks and the gluons 
is treated with our code, that evolves the
fragmentation functions to the energy scale \( \left\langle \varepsilon \right\rangle  \).
The top (\( t \)) quark is treated consistently 
with all its fundamental decays included; hadronization of the $b$ quark is treated 
with suitable fragmentation functions and also involves a suitable evolution. 
As we vary $M_{BH}$ and we scan over the spectrum of the incoming cosmic rays 
the procedure is repeated and rendered automatic by combining in a single algorithm 
all the intermediate steps. 
Tables 1 and 2 contain the results of a renormalization group analysis of the 
fragmentation functions for all the 
partons (except the top quark), where we show 
both the initial conditions at the input scale, whose lowest value is $Q=1.414$ GeV, 
and the results of the evolution, at a final scale of $Q=200$ GeV,  
the initial set being taken from ref.~\cite{kkp}.

\smallskip

\begin{table}
\begin{tabular}{|c||c|c|c|c|c|c|}
\hline
&
\( \pi ^{\pm } \)&
\( \pi ^{0} \)&
\( K^{\pm } \)&
\( K^{0}/\overline{K}^{0} \)&
\( p/\overline{p} \)&
\( n/\overline{n} \)\\
\hline
\hline
\emph{u}&
\( 0.451 \)&
\( 0.226 \)&
\( 0.048 \)&
\( 0.174 \)&
\( 0.067 \)&
\( 0.034 \)\\
&
\( 0.463 \)&
\( 0.231 \)&
\( 0.084 \)&
\( 0.252 \)&
\( 0.070 \)&
\( 0.035 \)\\
\hline
\emph{d}&
\( 0.451 \)&
\( 0.226 \)&
\( 0.174 \)&
\( 0.048 \)&
\( 0.034 \)&
\( 0.067 \)\\
&
\( 0.463 \)&
\( 0.231 \)&
\( 0.252 \)&
\( 0.084 \)&
\( 0.035 \)&
\( 0.070 \)\\
\hline
\emph{s}&
\( 0.391 \)&
\( 0.195 \)&
\( 0.068 \)&
\( 0.068 \)&
\( 0.139 \)&
\( 0.139 \)\\
&
\( 0.295 \)&
\( 0.147 \)&
\( 0.084 \)&
\( 0.084 \)&
\( 0.108 \)&
\( 0.108 \)\\
\hline
\emph{c}&
\( 0.329 \)&
\( 0.165 \)&
\( 0.167 \)&
\( 0.167 \)&
\( 0.085 \)&
\( 0.085 \)\\
&
\( 0.309 \)&
\( 0.155 \)&
\( 0.194 \)&
\( 0.194 \)&
\( 0.071 \)&
\( 0.071 \)\\
\hline
\emph{b}&
\( 0.438 \)&
\( 0.219 \)&
\( 0.129 \)&
\( 0.129 \)&
\( 0.042 \)&
\( 0.042 \)\\
&
\( 0.324 \)&
\( 0.162 \)&
\( 0.115 \)&
\( 0.115 \)&
\( 0.041 \)&
\( 0.041 \)\\
\hline
\emph{g}&
\( 0.303 \)&
\( 0.152 \)&
\( 0.253 \)&
\( 0.253 \)&
\( 0.020 \)&
\( 0.020 \)\\
&
\( 0.807 \)&
\( 0.404 \)&
\( 0.317 \)&
\( 0.317 \)&
\( 0.034 \)&
\( 0.034 \)\\
\hline
\end{tabular}

\caption{Initial conditions. For each couple of parton and hadron, 
the upper number in the box is the probability for 
the parton \protect\( f\protect \) to hadronize
into the hadron \protect\( h\protect \), \protect\( \left( \int _{z_{min}}^{1}D_{f}^{h}(z,Q)dz\right) /\sum _{h'}\left( \int _{z_{min}}^{1}D_{f}^{h'}(z,Q)dz\right) \protect \),
while the lower number is the average energy fraction of \protect\( h\protect \),
\protect\( \int _{z_{min}}^{1}zD_{f}^{h}(z,Q)dz\protect \). In this table 
the energy of the parton is \protect\( Q=1.414\, \textrm{GeV}\protect \) for
\protect\( u\protect \), \protect\( d\protect \), \protect\( s\protect \)
and \protect\( g\protect \), \protect\( Q=2m_{c}=2.9968\, \textrm{GeV}\protect \)
for \protect\( c\protect \) and \protect\( Q=2m_{b}=9.46036\, \textrm{GeV}\protect \)
for the \protect\( b\protect \) quark, generated via the set of ref.~\cite{kkp}}
\end{table}

\begin{table}
\begin{tabular}{|c||c|c|c|c|c|c|}
\hline
\( Q=200\, \textrm{GeV} \) &
\( \pi ^{\pm } \)&
\( \pi ^{0} \)&
\( K^{\pm } \)&
\( K^{0}/\overline{K}^{0} \)&
\( p/\overline{p} \)&
\( n/\overline{n} \)\\
\hline
\hline
\emph{u}&
\( 0.446 \)&
\( 0.223 \)&
\( 0.079 \)&
\( 0.166 \)&
\( 0.053 \)&
\( 0.033 \)\\
&
\( 0.385 \)&
\( 0.193 \)&
\( 0.077 \)&
\( 0.178 \)&
\( 0.047 \)&
\( 0.027 \)\\
\hline
\emph{d}&
\( 0.446 \)&
\( 0.223 \)&
\( 0.166 \)&
\( 0.079 \)&
\( 0.033 \)&
\( 0.053 \)\\
&
\( 0.385 \)&
\( 0.193 \)&
\( 0.178 \)&
\( 0.077 \)&
\( 0.027 \)&
\( 0.047 \)\\
\hline
\emph{s}&
\( 0.425 \)&
\( 0.213 \)&
\( 0.093 \)&
\( 0.093 \)&
\( 0.088 \)&
\( 0.088 \)\\
&
\( 0.295 \)&
\( 0.147 \)&
\( 0.077 \)&
\( 0.077 \)&
\( 0.070 \)&
\( 0.070 \)\\
\hline
\emph{c}&
\( 0.371 \)&
\( 0.185 \)&
\( 0.158 \)&
\( 0.158 \)&
\( 0.064 \)&
\( 0.064 \)\\
&
\( 0.295 \)&
\( 0.147 \)&
\( 0.150 \)&
\( 0.150 \)&
\( 0.051 \)&
\( 0.051 \)\\
\hline
\emph{b}&
\( 0.431 \)&
\( 0.216 \)&
\( 0.132 \)&
\( 0.132 \)&
\( 0.045 \)&
\( 0.045 \)\\
&
\( 0.292 \)&
\( 0.146 \)&
\( 0.101 \)&
\( 0.101 \)&
\( 0.036 \)&
\( 0.036 \)\\
\hline
\emph{g}&
\( 0.428 \)&
\( 0.214 \)&
\( 0.135 \)&
\( 0.135 \)&
\( 0.044 \)&
\( 0.044 \)\\
&
\( 0.577 \)&
\( 0.289 \)&
\( 0.175 \)&
\( 0.175 \)&
\( 0.057 \)&
\( 0.057 \)\\
\hline
\end{tabular}
\caption{For each couple of parton/hadron, the first number is the probability of fragmentation of the parton $f$ into the hadron
 $\left( \int _{z_{min}}^{1}D_{f}^{h}(z,Q)dz\right) /\sum _{h'}\left( \int _{z_{min}}^{1}D_{f}^{h'}(z,Q)dz\right)$,
while the second is the average energy fraction of $h$,
$ \int _{z_{min}}^{1}zD_{f}^{h}(z,Q)dz $. The
energy of the parton is $Q=200$ GeV.\label{incon}}

\end{table}

This concludes the computation of the probabilities for each hadron/lepton 
present in the decay products of the mini black hole. It is 
reasonable to assume that these particles will be produced
spherically, since higher angular momenta are suppressed by the corresponding
centrifugal barrier. However, the analysis of the shower profile has to be 
performed in the lab frame. This requires the transformation of the 
initial configurations above to the laboratory frame, which is exactly what is 
discussed next.

\section{Sphericity and Boost \label{sec:mbh5}}

\begin{figure}
{\centering \resizebox*{12cm}{!}{\includegraphics{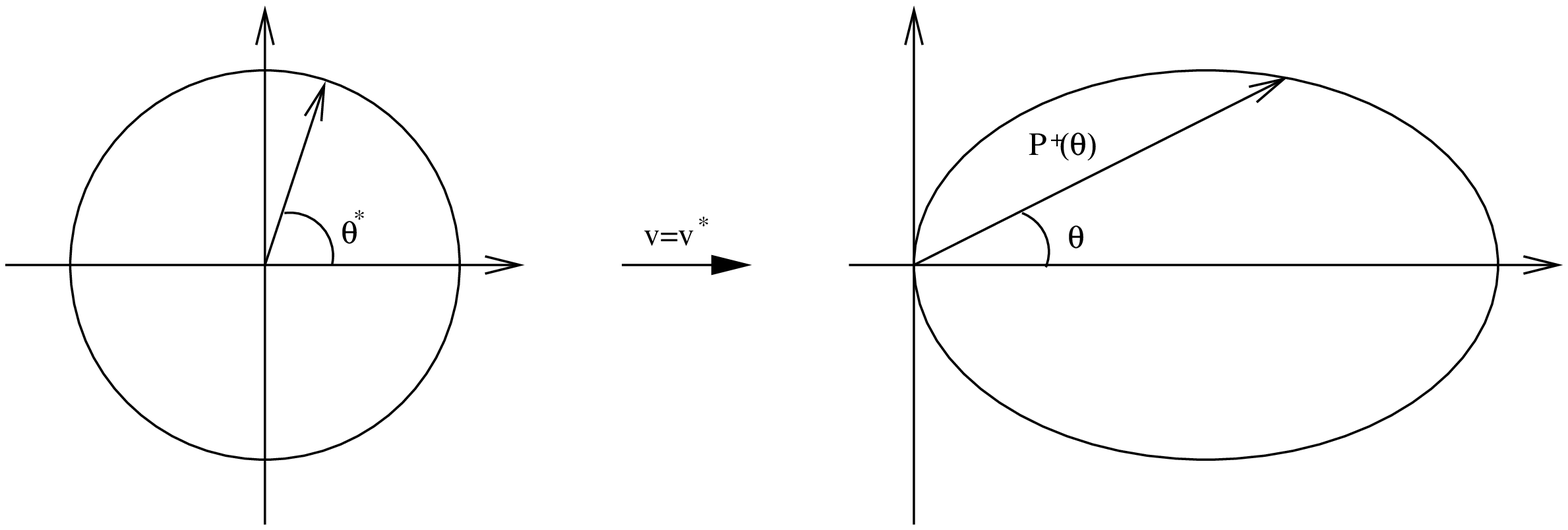}} 
\resizebox*{12cm}{!}{\includegraphics{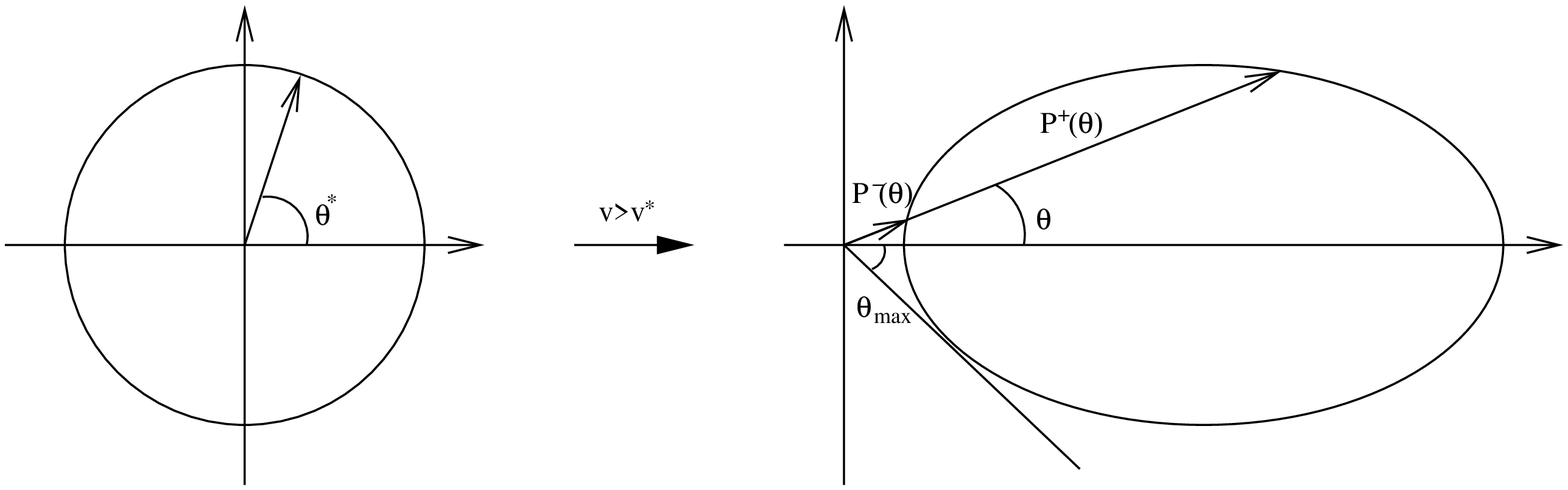}} \par}
\caption{An s-wave distribution in the BHF is transformed in the detector
frame to an elliptical one, whose detailed shape depends on the value 
of $g^*$}
\label{shapes}
\end{figure}

The transformation from the black hole frame (BHF) to the laboratory frame (LF)
is performed by a Lorentz boost with speed $\beta$, the speed of the black
hole in the LF. 
Assuming that the black hole is produced in the collision of a primary of 
energy $E_1$ in the LF and negligible mass compared to $E_1$, 
with a parton of mass $M$ in the atmosphere, one 
obtains $\beta=E_1/(E_1+M)$. 
A spherical distribution of a particular particle of mass $m$ 
among the decay products 
in the BHF is transformed to an elliptical one,
whose detailed form is conveniently parametrized by 
\beq
g^*={\beta \over \beta^*}=
\frac{1-\frac{M}{E_1+M}}{\left(1 - {{m^2}\over {E^{*2}}}\right)^{1/2}}
\eeq
where  
$\beta^*=P^*/E^*$ is the speed of this particle in the BHF, 
the ratio of its BHF momentum and energy. Figure \ref{shapes}
depicts the relevant kinematics.
A particle emitted in the direction $\theta^*$ in the BHF, is seen in the 
direction $\theta$ in the LF, with
\beq
\tan\theta=\sqrt{1-\beta^2}\frac{\sin\theta^*}{g^*+\cos\theta^*}.
\label{theta}
\eeq
For values of $g^*\geq 1$ the shape of the 
1-particle distribution in the LF is characterized by a maximum 
angle of emission
\beq
|\tan\theta_{max}|=\sqrt{\frac{1-\beta^2}{g^{*2}-1}}\,, 
\label{thetamax}
\eeq
which for $g^*=1$ is equal to $90^o$. 
Only for $g^*<1$ there is backward emission in the LF. 
As a relevant example, let us consider the case of a hadron of mass
$m=1$ GeV and energy $E^*=100$ GeV emitted by
a black hole, formed by an initial primary 
of energy $E_1=1000$ TeV, which hit a quark of mass $M\sim 10$MeV to form 
a black hole. The gives $g^*=1.00005$ and the corresponding maximum angle in
the LF is
$\theta_{max}\simeq \tan\theta_{max}=1.4\times 10^{-2}$, giving an angular
opening of the decay products of about 2 degrees. 

As shown in Figure \ref{shapes}, for $g^*>1$, which is relevant for our 
purposes, the mapping from $\theta^*$ to $\theta$ is not one to one. In
a given direction $\theta$ in the LF, one receives particles emitted in
two different directions $\theta^{*\pm}$ in the BHF. They satisfy \cite{KB}
\beq
\frac{ d\cos\theta^{*\pm}}{d\cos\theta}=\left(\frac{P^\pm}{P^*}\right)^2 
\frac{1}{\left(\pm \cos\theta\sqrt{K}\right)}
\label{wh}
\eeq
where
\beq
K=1 + \gamma^2(1 - {g^*}^2)\tan^2 \theta,
\eeq
and with the momenta $P^{\pm}$ and energies $E^{\pm}$ 
of the two branches given by
\beq
P^{\pm}={P^* \cos\theta (g^* \pm \sqrt{K}) \over \gamma (1 - v^2 \cos^2\theta)}.
\eeq
and 
\beq
E^{\pm}=\frac{m\left(\gamma^* \pm v\cos\theta \left(v^{* 2}\gamma^{* 2} - 
v^2\gamma^2\sin^2\theta\right)^{1/2}\right)}{\gamma (1- v^2\cos^2\theta)}
\eeq
respectively. In the above formulas $\gamma^{-1}=\sqrt{1-\beta^2}$, $v(v^*)$
is the speed of the hadron in the LF (BHF), and $(\gamma^*)^{-1}=
m/E^*_h=\sqrt{1-v^{*2}}$.
For massless final state particles, in particular, these relations become 
\beq
P=E={P^*\over \gamma (1 - \cos\theta)}
\eeq
and reduce to the familiar Doppler formula when $\theta=0$.

The probability distribution $W_h^*(\cos\theta^*,\phi)$ of a hadron (h) 
as a function of the direction $\Omega=(\cos\theta^*,\phi)$ in the BHF, 
assumed spherically symmetric and normalized to the 
total probability $Pr_h$ of detecting this hadron among the decay products
with $N$ elementary states, is
\beq
W_h^*(\cos\theta^*)={ \textrm{Pr}_h \over 2}.
\eeq
The corresponding one in the LF is
\beq
W_h(\cos\theta)=\sum_{\pm}\frac{d\cos\theta^{*\pm}}{d\cos\theta}\,
W^*_h(\cos\theta^{*\pm})\,.
\eeq

In the special case $g^*=1$, the probability distribution simplifies to  
\beq
W_h(\cos\theta)=2 \textrm{Pr}_h{ \cos\theta \over 
\gamma^2(1- \beta^2\cos\theta^2)^2}\,,
\eeq
peaked in the forward direction, symmetric around the maximum value, 
obtained for $\theta=0$ and equal to $2 \gamma^2$, while the 
momentum distribution is
\beq
P(\theta)=m { \beta\gamma^*\cos\theta\over \gamma
(1- \beta^2\cos\theta^2)}\,.
\eeq   

\begin{figure}
{\centering
\resizebox*{10cm}{!}{\rotatebox{-90}{\includegraphics{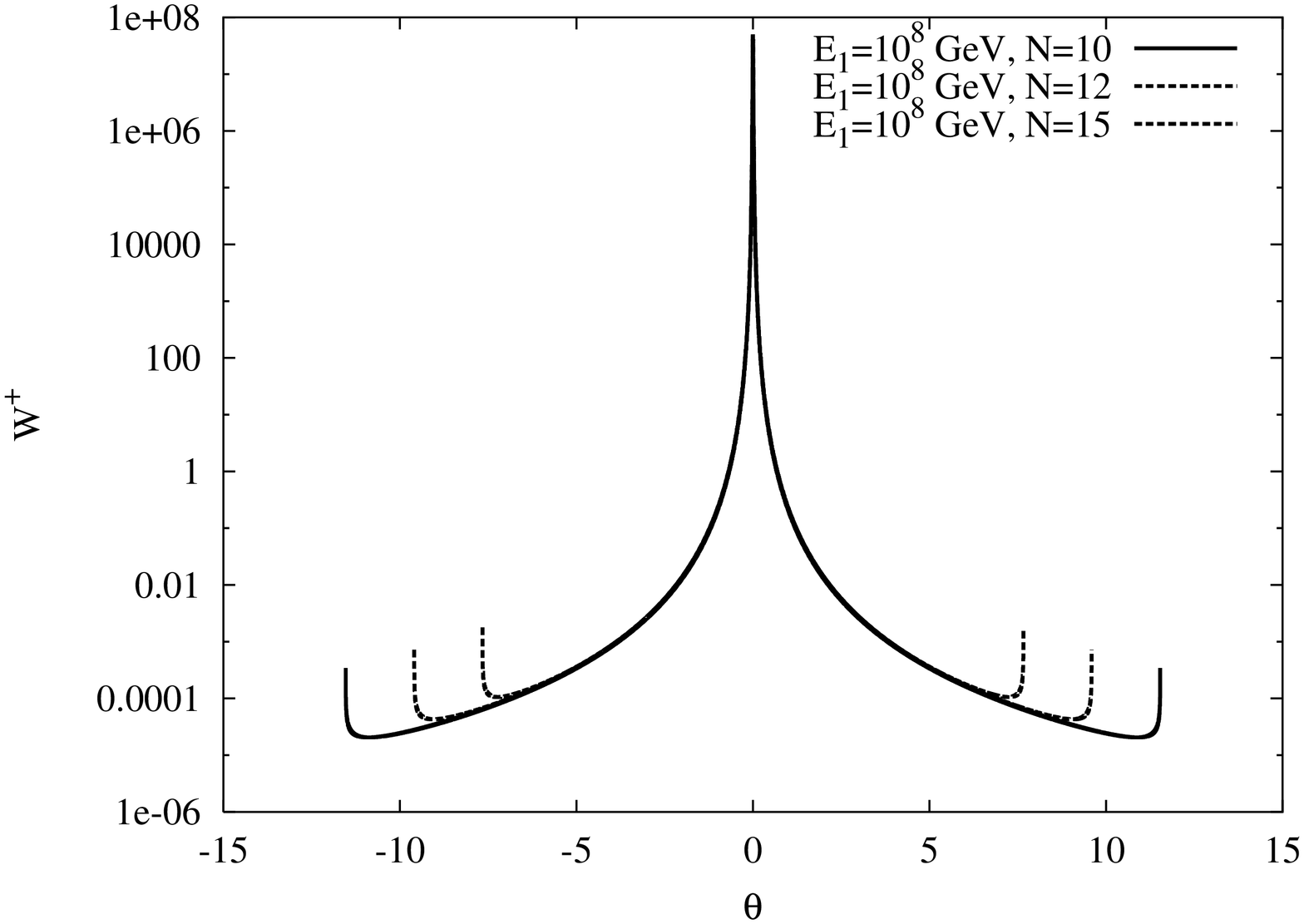}}} \par}
\caption{Plot of the $W^+$ branch of the probability distribution
for an incoming energy of the cosmic ray $E_1= 10^8$ GeV and for various 
values of $N$ of the elementary partonic states emitted during the decay
of the black hole}
\label{wplus}
\vspace{3cm}
{\centering
\resizebox*{10cm}{!}{\rotatebox{-90}{\includegraphics{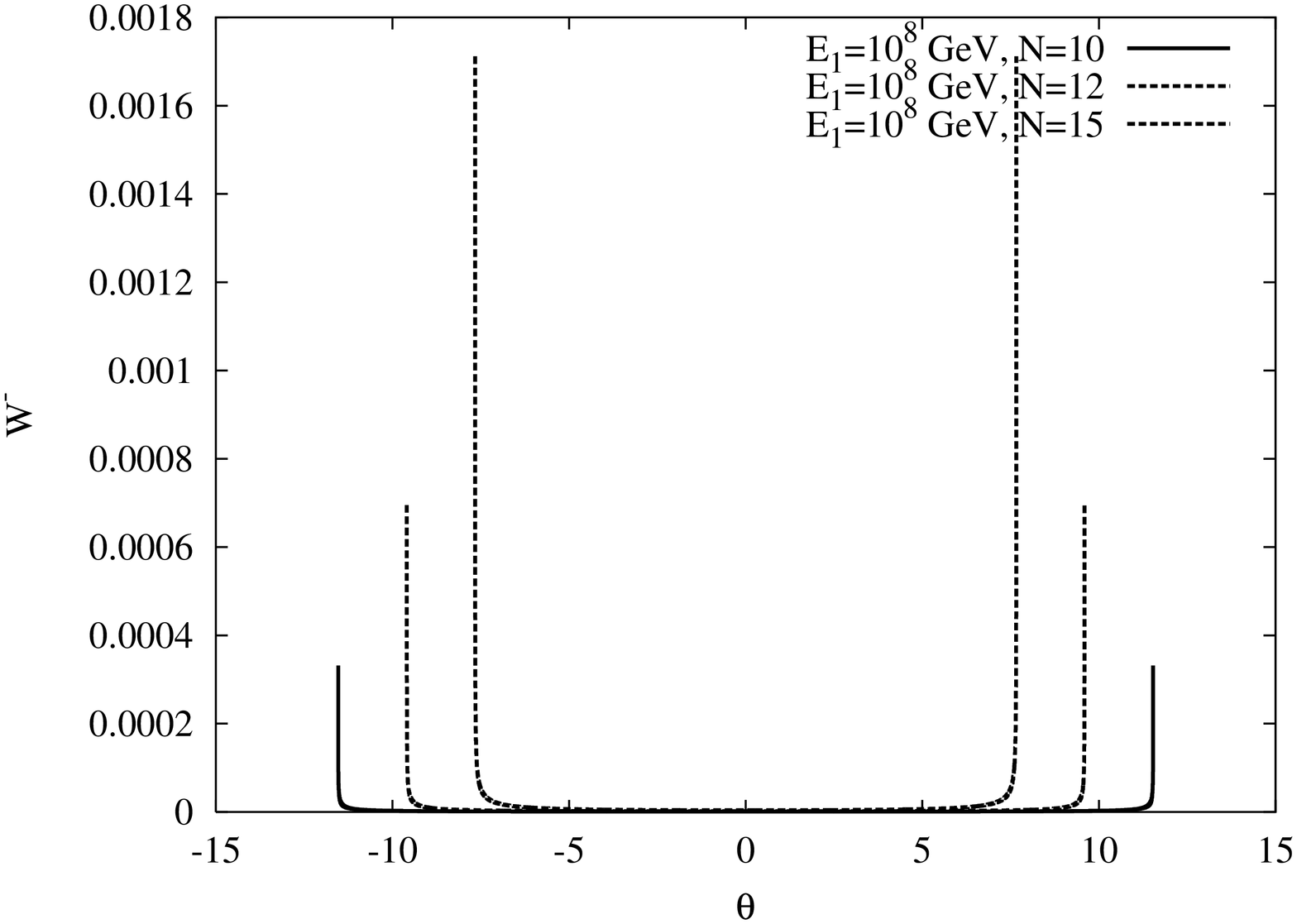}}} \par}
\caption{Same plot as above but for the $W^-$ branch of the distribution}
\label{wminus}
\end{figure}

\begin{figure}
{\centering
\resizebox*{8cm}{!}{\rotatebox{-90}{\includegraphics{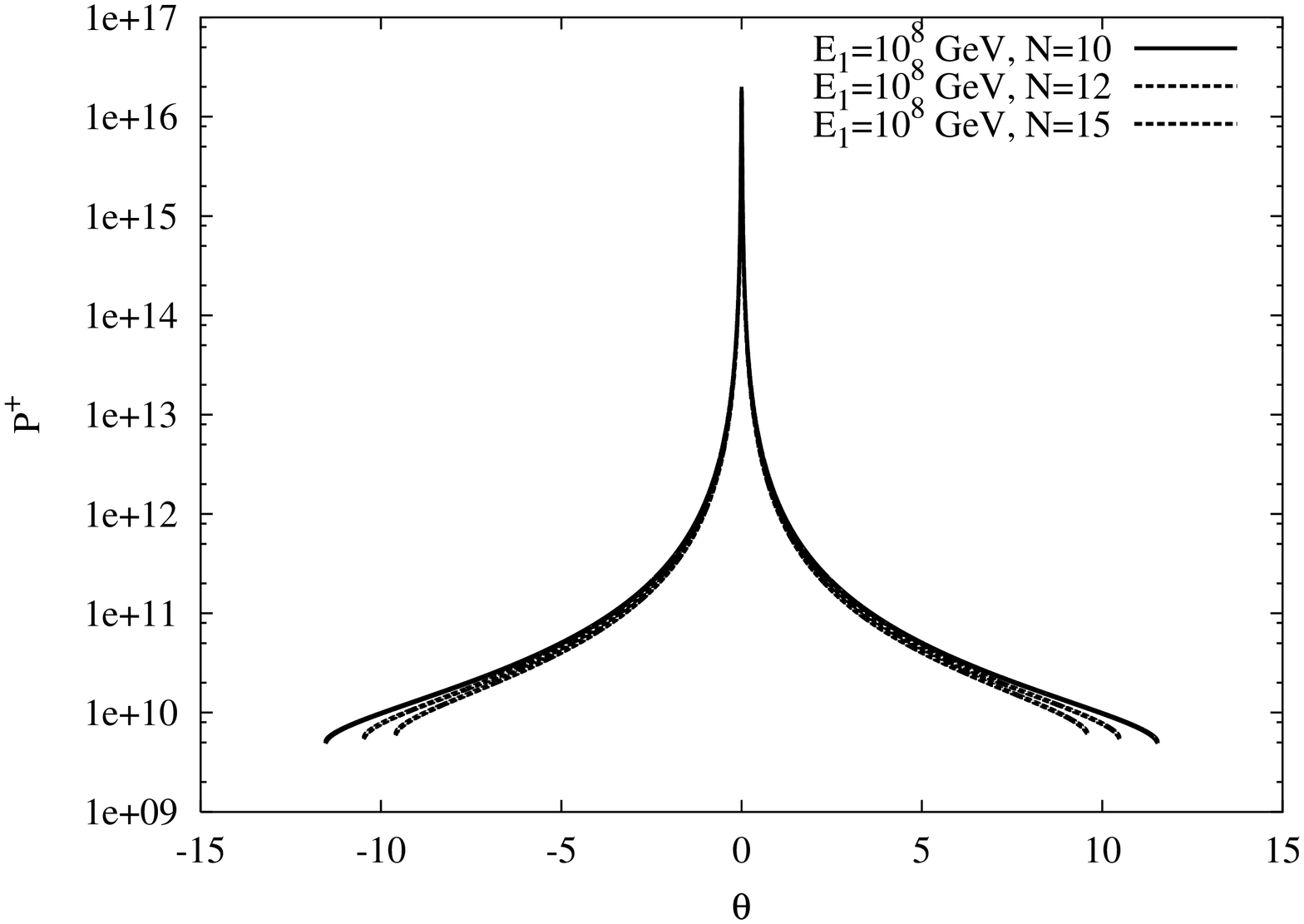}}} \par}
\caption{Plot of the $P^+$ branch of the momentum distribution in eV
for an incoming energy of the cosmic ray $E_1= 10^8$ GeV and for various 
values of $N$ of the elementary partonic states emitted during the decay
of the black hole}
\label{ppbranch}
\vspace{3cm}
{\centering
\resizebox*{8cm}{!}{\rotatebox{-90}{\includegraphics{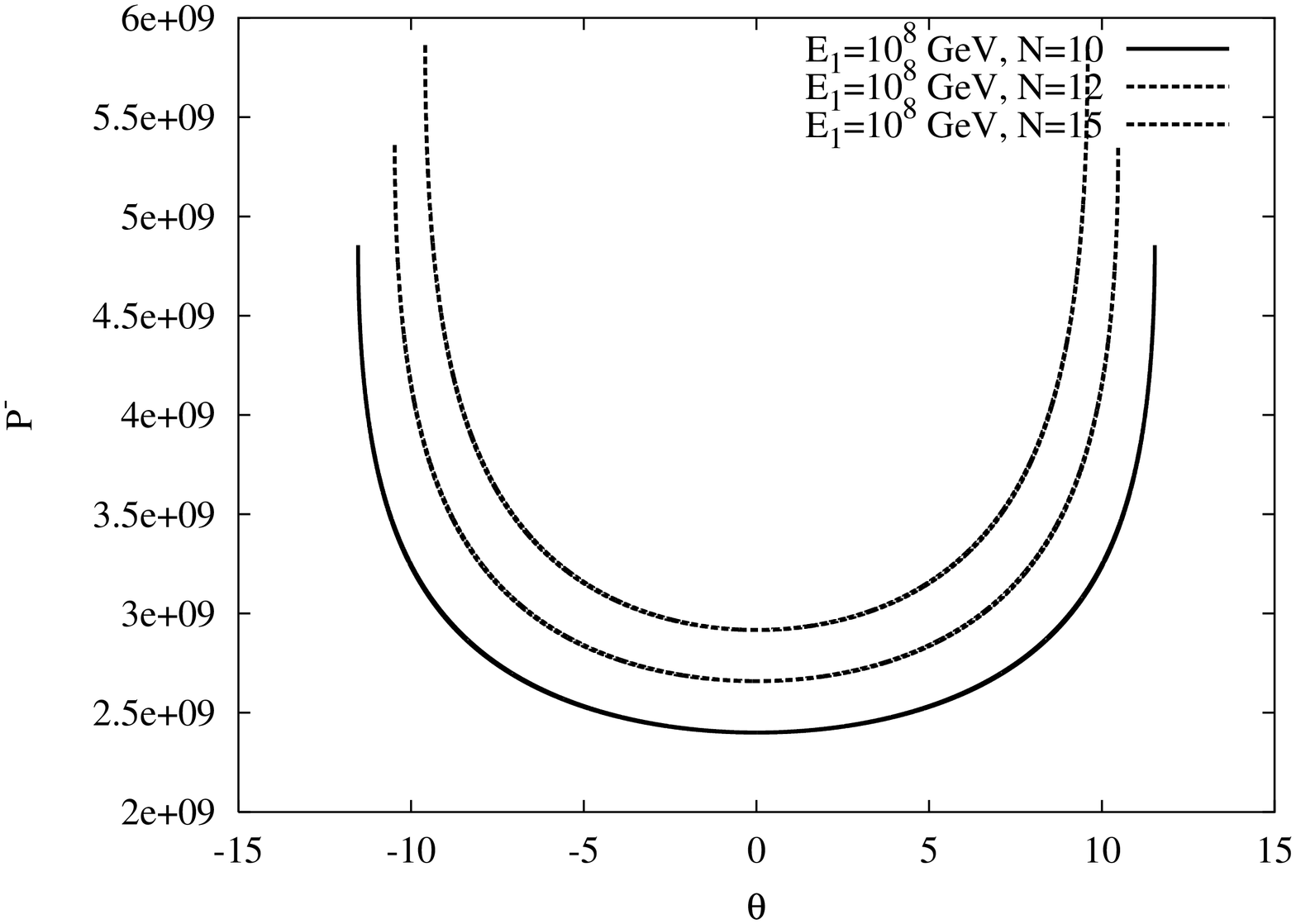}}} \par}
\caption{Same plot as above but for the $P^-$ branch of the momentum 
distribution}
\label{pmbranch}
\end{figure}

As we have already mentioned, 
the structure of the partonic event (and, similarly, of the hadronic 
event after fragmentation) is characterized by the formation of an 
elliptical distribution of partons, strongly boosted toward the detector 
along the vertical direction. Each uniform (s-wave) distribution 
is strongly elongated along the arrival direction (due to the large 
speed of the black hole along this direction) and is characterized by 
two sub-components ($W^\pm$), identified by a $\pm$ superscript. Their sum
is the total probability distribution given in (\ref{wh}).
The ``+'' momentum component is largely dominant and strongly peaked around 
the vertical direction with rather small opening angles and this 
behaviour can be analyzed numerically with its $N$ dependence. 
In the explicit identification of the two independent distributions 
$\pm$ in terms of the opening angle $\theta$, as measured in the LF, we 
use the relations 
\beq
W^\pm(\theta) = \frac{1}{2} W^\pm(\cos \theta^*) 
\mid {d\cos(\theta^*)\over d\cos \theta}^\pm\mid \sin\theta,
\eeq
where we have introduced a factor of 1/2 for a correct normalization of the 
new distribution in the $\theta$ variable. 
In Figures~\ref{wplus} and \ref{wminus} we show the structure of these 
distributions in the LF. Both are characterized by a very small 
opening angle ($\theta$) with respect to the azimuthal direction of the 
incoming cosmic ray, $W^+$ being the dominant one. Two similar 
plots (Figures~\ref{ppbranch} and \ref{pmbranch}) illustrate the two 
components $P^\pm$ as functions of the same angle.

One can easily check that we can 
now integrate symmetrically on both distributions to obtain 
the correct normalization (to $\textrm{Pr}_h$) for a given hadron (or parton)
\beq
\int_{-\theta_{max}}^{\theta_{max}}  W^\pm(\theta) d\theta = 
\int_{-{\theta_1}^*}^{{\theta_1}^*} W^+(\theta) d\theta +
\int_{{\theta_1}^*}^{\pi} W^-(\theta) d\theta +
\int_{-\pi}^{{-\theta_1}^*} W^-(\theta) d\theta 
={\textrm{Pr}_h}
\eeq
or, equivalently, using $\cos \theta$ as a distribution variable
\beq
W^\pm(cos \theta)=W^\pm(\cos \theta^*) \mid 
{d\cos \theta^*\over d\cos \theta}^\pm\mid
\sin\theta
\eeq
with
\beq
\int_{cos\theta_{max}}^{1}  W^\pm(\cos\theta) d \cos\theta =
\int_{\cos{\theta_1}^*}^{1} W^+(\cos\theta^*) d\ cos\theta^* +
\int_{-1}^{\pi} W^-(\cos\theta^*) d\cos\theta^*
={\textrm{Pr}_h}
\eeq
and ${\theta^*}_1$ obtained from (\ref{thetamax}).

\clearpage
\section{Air Shower Simulations \label{sec:mbh6}}

The simulation of the events is performed at the last stage, using 
an air shower simulator. We have used CORSIKA \cite{CORSIKA} with 
appropriate initial conditions on the spectrum of the incoming particles 
in order to generate the full event measured at detector level.
In most of the simulations we have assumed that the first impact takes place in the 
lower part of the atmosphere, not far from the level of the detector, 
at a varying altitude. The reason is that one of our interests is the investigation
of the possibility that the Centauro events may be related to evaporating
mini black holes, formed by the collision of weakly interacting particles (e.g.
neutrinos) which penetrate the atmosphere. Of course, we have simulated 
events happening at higher altitudes as well.  
We have performed two separate sets of simulations, the first set being 
benchmark events with an ``equivalent'' proton replacing the neutrino-nucleon 
event, colliding at the same height, the second being the signal event, 
i.e. the black hole resonance. The difference between the first and the 
second set is attributed to the different components of the final state 
prior to the development of the air shower.

We compute the average number of particles produced in the process
of BH evaporation using the formula
\beq
M_{BH}=E_{CM}f_{n}
\eeq
where \( E_{CM} \) is the energy in the center of mass frame in the
neutrino-nucleon collision and \( f_{n} \) is the fraction of \( E_{CM} \)
that is bound into the black hole as a function of the number \( n \)
of extra-dimensions. Numerical values for \( f_{n} \) in head-on
collisions are taken from Ref. \cite{EardleyGiddings} and 
reported in Table \ref{tab:eardley}.
\begin{table}
\begin{tabular}{|c|c|}
\hline
\( n \)&
\( f_{n} \)\\
\hline
\hline
0&
\( 0.70711 \)\\
\hline
1&
\( 0.66533 \)\\
\hline
2&
\( 0.63894 \)\\
\hline
3&
\( 0.62057 \)\\
\hline
4&
\( 0.60696 \)\\
\hline
\end{tabular}

\caption{Fraction \protect\( f_{n}\protect \) of \protect\( E_{CM}\protect \)
that is bound into the black hole as a function of the number \protect\( n\protect \)
of extra-dimension in head-on collisions.\label{tab:eardley}}

\end{table}

The overall shower, defined as the superposition of the various
sub-components, develops according to an obvious cylindrical symmetry around 
the vertical z-axis near the center. We assume in all the studies that the 
incoming primary undergoes a collision with a nucleon in the atoms of the 
atmosphere at zero zenith with respect to the plane of the detector.

The model of the atmosphere that we have adopted consists 
of \( N_{2} \), \( O_{2} \) and
\( Ar \) with the volume fractions of \( 78.1\% \), \( 21.0\% \)
and \( 0.9\% \) \cite{HCP}. The density variation of the atmosphere
with altitude is modeled by 5 layers. The pressure \( p \) as a function
of the altitude \( h \) is given by\begin{equation}
p(h)=a_{i}+b_{i}\exp (-h/c_{i}),\quad i=1,\ldots ,4
\end{equation}
in the lower four layers and by\begin{equation}
p(h)=a_{5}-h\frac{b_{5}}{c_{5}}
\end{equation}
in the fifth layer.

\begin{figure}
{\centering \resizebox*{10cm}{!}{\rotatebox{-90}
{\includegraphics{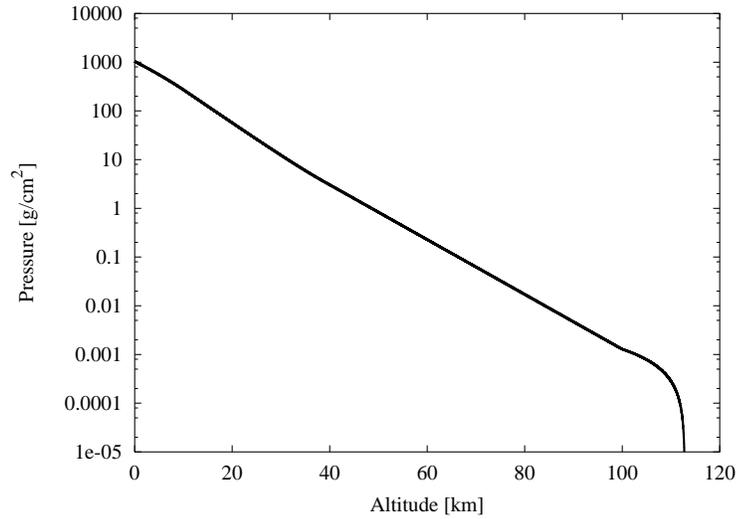}}} \par}
\caption{Pressure versus altitude for the U.S.~standard 
atmosphere model.\label{fig:USstandard}}
\label{pressure}
\end{figure}

The \( a_{i} \), \( b_{i} \), \( c_{i} \) parameters, that we report
in Table~\ref{tab:USstandard}, are those of the U.S.~standard atmospheric model
\cite{CORSIKA}. The boundary of the atmosphere in this model is
defined at the height \( 112.8\, \textrm{km} \), where the pressure vanishes.
In Figure~\ref{fig:USstandard} we show a plot of the pressure ($p=X_v$, also 
called vertical depth) as a function of the height. 
\begin{table}
\hspace{2cm}
\begin{tabular}{|c|c|c|c|c|}
\hline
Layer \( i \)&
Altitude \( h \) {[}km{]}&
\( a_{i}\, [\textrm{g}/\textrm{cm}^{2}] \)&
\( b_{i}\, [\textrm{g}/\textrm{cm}^{2}] \)&
\( c_{i}\, [\textrm{cm}] \)\\
\hline
\hline
1&
\( 0\ldots 4 \)&
\( -186.5562 \)&
\( 1222.6562 \)&
\( 994186.38 \)\\
2&
\( 4\ldots 10 \)&
\( -94.919 \)&
\( 1144.9069 \)&
\( 878153.55 \)\\
3&
1\( 0\ldots 40 \)&
\( 0.61289 \)&
\( 1305.5948 \)&
\( 636143.04 \)\\
4&
\( 40\ldots 100 \)&
\( 0.0 \)&
\( 540.1778 \)&
\( 772170.16 \)\\
\hline
5&
\( >100 \)&
\( 0.01128292 \)&
\( 1 \)&
\( 10^{9} \)\\
\hline
\end{tabular}
\caption{Parameters of the U.S.~standard atmosphere.\label{tab:USstandard}}

\end{table}

\vspace{2cm}

This is defined via the integral  
\beq
X_v=\int_h^\infty \rho(h')d h'
\eeq
of the atmospheric density $\rho(h)$ for zero zenith angle, 
while the corresponding slant depth is given by 
\beq
X=\int_l^\infty \rho\left(l \cos\theta + \frac{1}{2} \frac{l^2}{R_T} 
\sin^2\theta\right) 
\eeq
for a zenith angle $\theta$ and $R_T$ is the radius of the earth. 

To put into perspective our Monte Carlo study it is convenient to briefly 
summarize the basic features of the theory of cascades on an analytical ground. 
The theory consists of the system of transport equations 
\cite{Gaisser,Farrar} for the numbers $N_n(E,X)$ of particles
of type $n$ with energy $E$ at height $X$
\begin{equation}
\frac{d N_{n}(E,X)}{d X}=-\frac{N_{n}(E,X)}{\lambda _{n}(E)}-\frac{1}{c\tau_n 
\gamma \rho _{\mathrm{Air}}}N_{n}(E,X)~,
\end{equation}
where \( \lambda _{n}(E) \) is their interaction length, $\tau_n$ is their 
lifetime and
\( \gamma  \) the Lorentz-factor corresponding to their given energy. 
In the simple case of an isothermal 
atmosphere \( \rho _{\mathrm{Air}}=\rho _{0}\exp (-h/h_{0})=X/h_{0} \), 
at a scale height $h_0$
\begin{equation}
\label{for:reacprob}
\frac{d N_{n}(E,X)}{d X}=-\frac{N_{n}(E,X)}{\lambda _{n}(E)}-
\frac{1}{d_{n}}N_n(E,X)
\end{equation}
where $d_n$ is their decay length, defined by
\beq
\frac{1}{d_n}=\frac{m c^2 h_0}{E c \tau_n X}.
\eeq
Particles produced at higher energies are also accounted for by an additional 
term in the cascade 
\begin{eqnarray}
\frac{\partial N_n(E,X)}{\partial X} & = & -N_{n}(E,X)\left[ \frac{1}
{\lambda _{n}(E)}+\frac{1}{d_n(E)} \right]\label{cascade} \\
&  & +\sum _{m}\int _{E}^{E^{max}} N_{m}(E',X)\left[ \frac{W_{mn}(E',E)}
{\lambda_{m}(E')}\right. \nonumber \\
&  & \, \, \, \, \, \, \, \, \, \, \, \, \, \, \, \left. +
\frac{1}{d_n(E')}D_{mn}(E',E)\right] dE'\nonumber ~~,
\end{eqnarray}
describing the change in the number of particles of type $n$ due to particles 
of type $m$ by interaction or decay, integrated over an allowed 
interval of energy. 
The functions \( W_{mn}(E',E) \) are the
energy-spectra of secondary particles of
type \( n \) in a collision of particle \( m \) with an air-molecule, while 
\( D_{mn}(E',E) \) are the corresponding decay-functions. The advantage of 
a transport equation compared to a Monte Carlo 
is, that it provides a rather simple analytical view of the development of the 
cascade across the atmosphere. 
Most common in the study of these equations is to use a factorized ansatz for 
the solution $N(E,X)=A(E)B(X)$, which 
assumes a scaling in energy of the transition functions \cite{Gaisser}. 
In our case an analytical treatment of the cascade corresponds to the 
boundary condition 
\beq
N_n(E,X_0)=\textrm{Pr}_n(E) \,\delta(E - f M_{BH}/\langle N\rangle)
\label{index_n}
\eeq
with $\textrm{Pr}_n(E)$ being the probability that the black hole decays 
into a specific state $n$. As we have already discussed 
above, these decays are uncorrelated and Eq.~(\ref{index_n}) is replicated 
for all the elementary states 
after hadronization. The emission probabilities $\textrm{Pr}_n(E)$ 
have been computed by us for a varying initial energy 
$E=f M_{BH}$ using renormalization group equations as described before, 
having corrected for energy loss in the bulk. The interactions 
in the injection spectrum of the original primaries at our $X_0$ 
($X_0=517$ g/cm$^2$) has been neglected since this is not implemented in CORSIKA. 
The showers have been performed independently and the results of the 
simulations have been statistically superimposed at the end 
with multiplicities computed at detector level ($X_1=553$ g/cm$^2$).
We have kept the gravity scale $M_*$ constant at 1 TeV and varied the mass 
of the black hole according to the available center of mass 
energy $E$. As we have already discussed in the previous sections, as a 
benchmark process we have selected a proton-air impact at the same 
$X_0$ with the boundary condition 
\beq
N_{p}(E,X_0)= \textrm{Pr}_p \delta(E - f M_{BH})
\eeq
which occurs with probability 1 ($Pr_p=1$).

We are interested both in the behaviour of the multiplicities and in the 
lateral distributions of the cascades developed at detector level. 
For this purpose we have
defined the opening of the conical shower after integration over the azimuthal 
angle, as in \cite{CCF}, 
and given the symmetry of the event, we plot only the distance from the center 
as a relevant parameter of the conical shower.

A varying number of extra dimensions \( n=0,\, 1,\ldots ,4 \) implies a 
different ratio for bulk-to-brane energy emission, 
a different average number of elementary decaying states and different energy 
distributions among these. 
We have varied the energy $E_1$ of the primary, thereby 
varying the mass of the black hole 
resonance, in the interval $10^{15}-10^{20}$\,
\textrm{eV}. The hadronization part of our code has been done by 
changing the $\textrm{Pr}_h(E)$ obtained from a numerical solution of the 
fragmentation functions separately for each value of the energy shared.

\begin{itemize}

\vspace{0.6cm}

\item{\em Preliminary studies}

We start with the numerical study of the partial and total multiplicities 
of the various sub-components and of their lateral distributions in the benchmark 
event. The results for photons and leptons are shown in 
Figures~\ref{multiphoton} to \ref{multiphoton3} as functions of the 
altitude of the impact and for two different observation points 
at 4,500 m and 5,000 m, approximately the altitudes 
of the detectors at Pamir and Chacaltaya, respectively.  
These preliminary plots, based on actual simulations of 
a proton-to-air nucleus impact at $10^{15}$ eV, 
show a steady growth of the 
multiplicities of the secondaries as we raise the point of first impact above the detector.
\begin{figure}
{\centering \resizebox*{10cm}{!}{\rotatebox{-90}
{\includegraphics{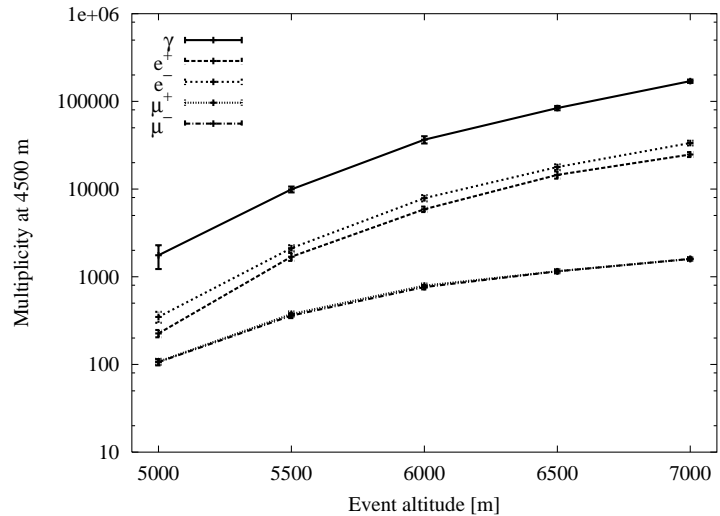}}} \par}
\caption{Benchmark event: Multiplicities of photons, 
\protect\( e^{\pm }\protect \), \protect\( \mu ^{\pm }\protect \)
at an observation level of 4500 m as a function of the altitude of the first impact.
\label{fig:moltVh_4500}}
\end{figure}
\begin{figure}
{\centering \resizebox*{10cm}{!}{\rotatebox{-90}
{\includegraphics{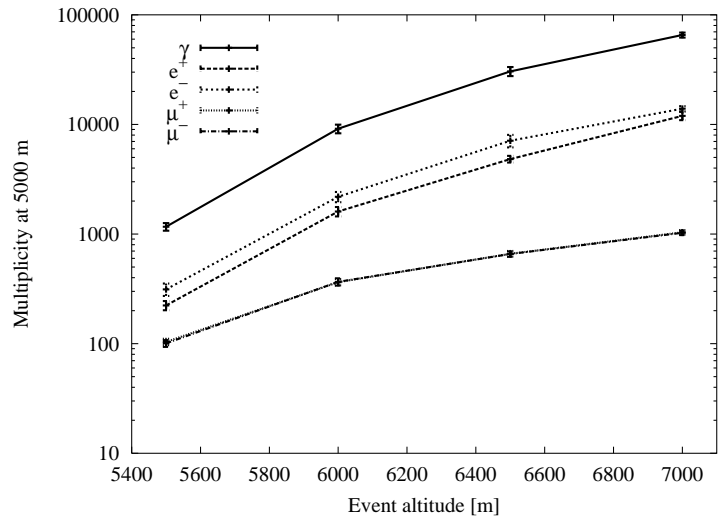}}} \par}
\caption{As in Figure \ref{fig:moltVh_4500}, but at an observation level of
5000 m.}
\label{multiphoton1}
\end{figure}
\begin{figure}
{\centering \resizebox*{10cm}{!}{\rotatebox{-90}
{\includegraphics{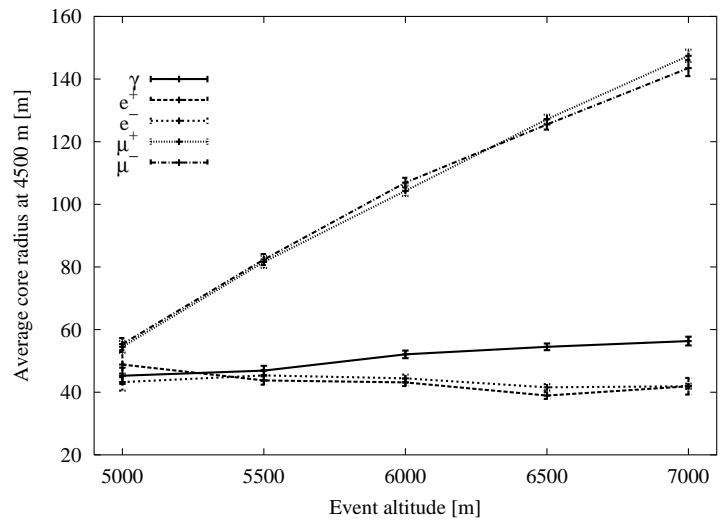}}} \par}
\caption{Average radial opening of the shower of photons, 
\protect\( e^{\pm }\protect \),
\protect\( \mu ^{\pm }\protect \) at an observation level of 4500
m as a function of the altitude of proton's first interaction.\label{distVh_4500}}
\label{multiphoton2}
\end{figure}
\begin{figure}
{\centering \resizebox*{10cm}{!}{\rotatebox{-90}
{\includegraphics{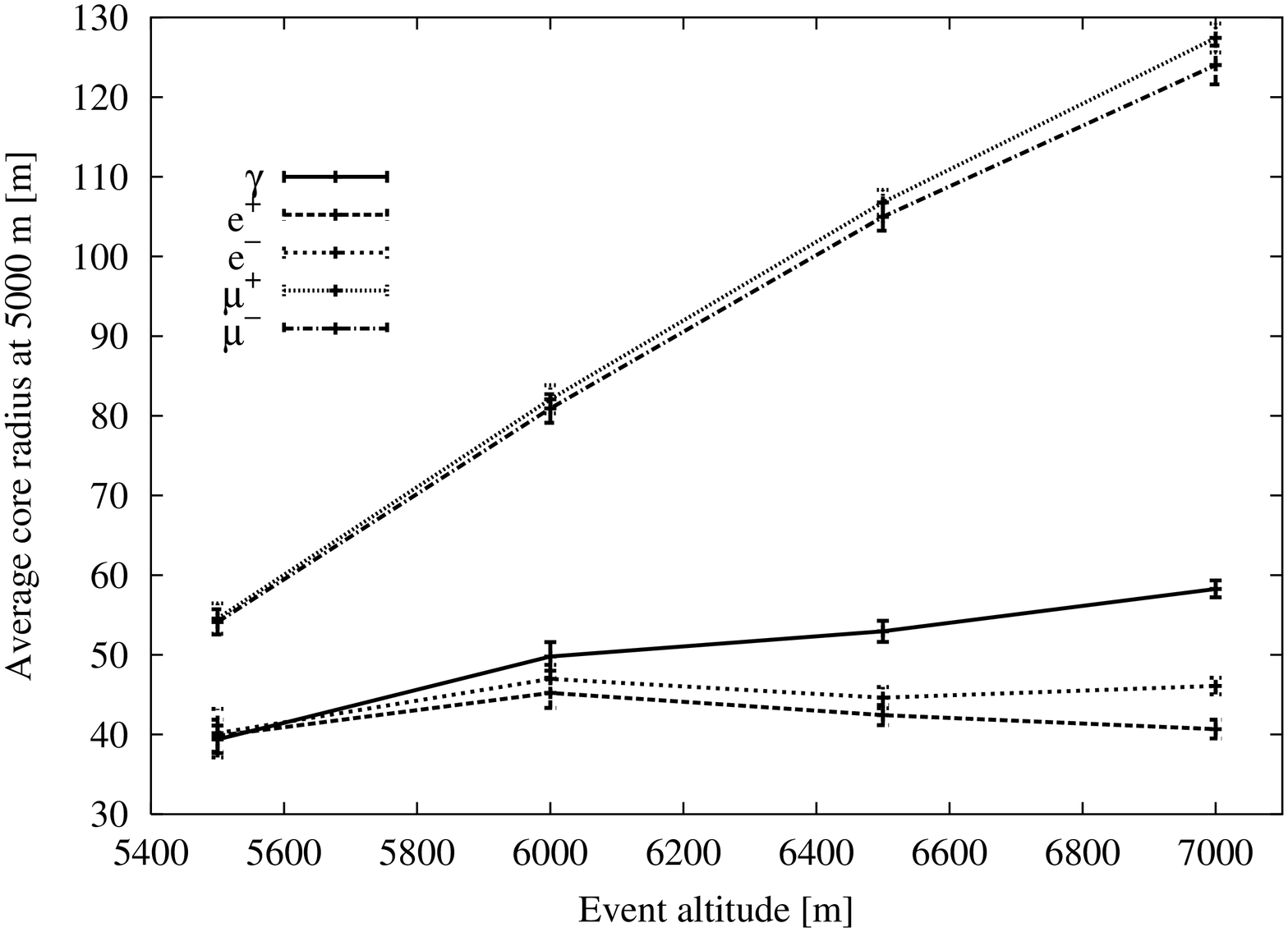}}} \par}
\caption{As in Fig. \ref{distVh_4500}, but at an observation level of 5000m.}
\label{multiphoton3}
\end{figure}
The statistical errors in the simulations 
(80 uncorrelated events have been collected per point) are 
rather small, quite uniformly over all the altitudes of the impact, 
and indicate a satisfactory stability of the result. 
The positions of the detectors 
do not seem to have an appreciable impact on the characteristics of the 
secondaries. As for the lateral distributions we observe an increase in 
the opening of the showers with the event altitude, 
which is more enhanced for the muonic component and for the photons and 
less for the electrons and positrons. Also in this case 
the statistical fluctuations are rather small. On the basis of these 
results we have selected for the remaining simulations 
a first impact at 5,500 \textrm{m} and the observation (detector) level at 
5,000 m. However, for comparison, we will also show later the results
of a second set of simulations that have been 
performed with the first impact at 15,000 m.

\vspace{0.6cm}

\item{\em Choice of scales and corrections}

To compare standard and black holes events, 
we have selected a gravity scale of $M_*=1$TeV and varied 
the black hole mass, here taken to be equal to the available 
center of mass energy during the collision. Therefore, a varying $E_1$ 
is directly related to a varying $M_{\rm{BH}}$ 
and we have corrected, as explained above, for the energy 
loss into gravitational emission. Unfortunately, this can be 
estimated only heuristically, with bounds largely 
dependent on the impact parameter of the primary collision. 
A reasonable estimate may be of the order of $10-15 \%$ \cite{kanti1}.   
Corrections related to emission in the bulk have also been included, 
in the way discussed in previous sections.

\vspace{0.6cm}

\item{\em Energy ratios: electromagnetic versus hadronic} 

Not all observables are statistically insensitive to the natural 
fluctuations of the air showers. In the study of black hole versus 
standard (benchmark) events, the study of the ratios 
$N_{\rm{em}}/N_{\rm{hadron}}$ and 
$E_{\rm{em}}/E_{\rm{hadron}}$ have been proposed 
as a way to distinguish between ordinary showers and other extra-ordinary 
ones. {\em Centauro} events, for instance, have been claimed to be 
characterized by a rather small ratio of electromagnetic over hadronic 
energy deposited in the detectors, contrary to normal showers, in which
this ratio is believed to be $E_{\rm{em}}/E_{\rm{hadron}}\sim 2$. 
Instead, as one can easily recognize from the results presented in 
Figures \ref{ratio1} and \ref{ratio2}, the multiplicity 
ratio takes values in two different 
regimes. In the ``band'' of values $1-5$ for the case of 
the lower first impact and $100-160$ for the higher impact. 
The larger values of the band in this latter case are justified by the 
fact that the shower is far more developed, given the altitude of the 
impact, and therefore is characterized by an even more dominant
electromagnetic component. The energy ratio, on the other hand
can take small values, in agreement with the values observed in Centauros.
However, notice that both the black hole and standard simulations
show a complex pattern for these ratios and in addition they are 
characterized by large 
fluctuations for varying energy and number of extra dimensions. 
\begin{figure}
{\centering \subfigure[ ]{\resizebox*{10cm}{!}{\rotatebox{-90}
{\includegraphics{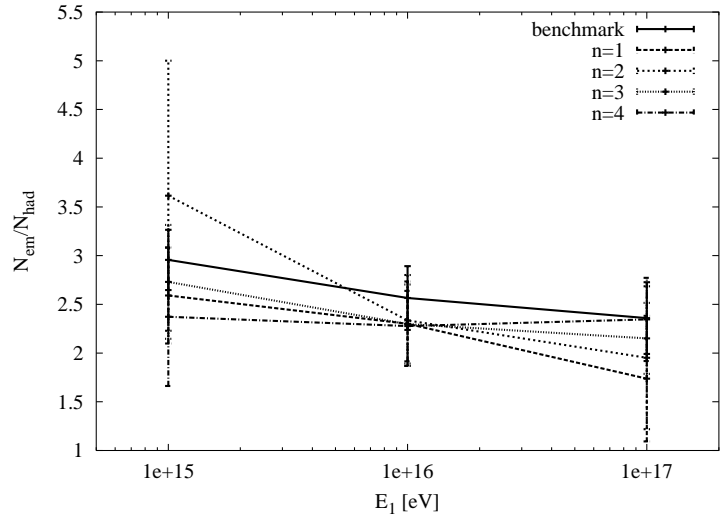}}}} \par}
{\centering \subfigure[ ]{\resizebox*{10cm}{!}{\rotatebox{-90}
{\includegraphics{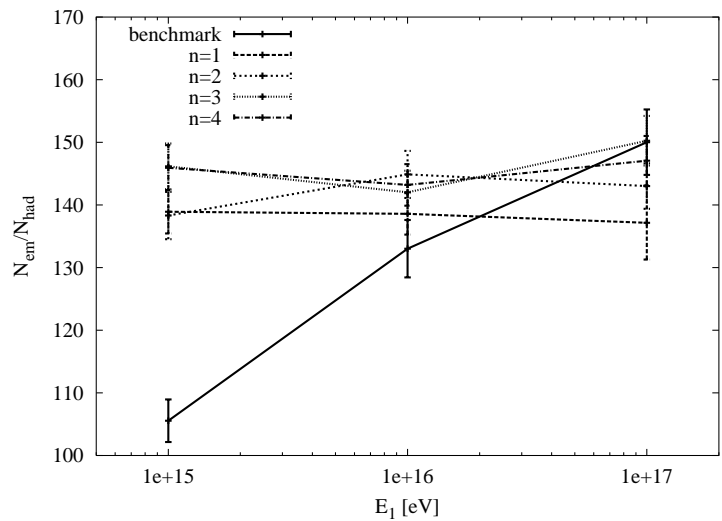}}}} \par}
\caption{Ratio between \protect\( N_{em}\protect \) (total multiplicity of
photons and \protect\( e^{\pm }\protect \)) and \protect\( N_{had}\protect \)
(total multiplicity of everything else) as a function of \protect\( E_{1}\protect \).
The first interaction is kept fixed at \protect\( 5500\, \textrm{m}\protect \)
(517 $g/cm^2$) (a), or at \protect\( 15000\, \textrm{m}\protect \)
(124 $g/cm^2$) (b),
and the observation level is \protect\( 5000\, \textrm{m}\protect \)
(553 $g/cm^2$).
We show in the same plot the benchmark (where the primary is a proton)
and mini black holes with different numbers of extra-dimensions 
\protect\( n\protect \).\label{fig:bhwl_NrappVE}}
\label{ratio1}
\end{figure}
\begin{figure}
{\centering \subfigure[ ]{\resizebox*{10cm}{!}{\rotatebox{-90}
{\includegraphics{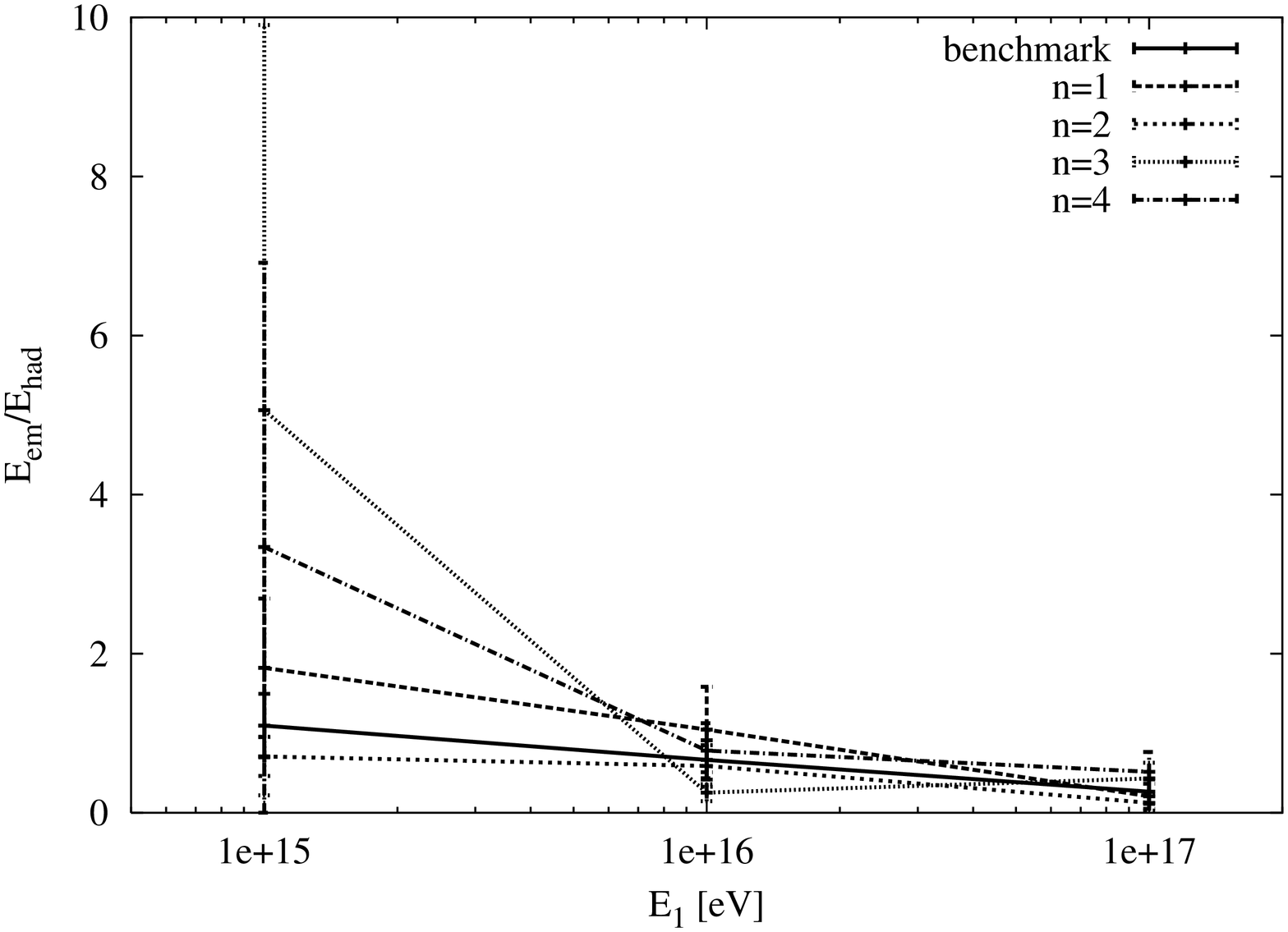}}}} \par}
{\centering \subfigure[ ]{\resizebox*{10cm}{!}{\rotatebox{-90}
{\includegraphics{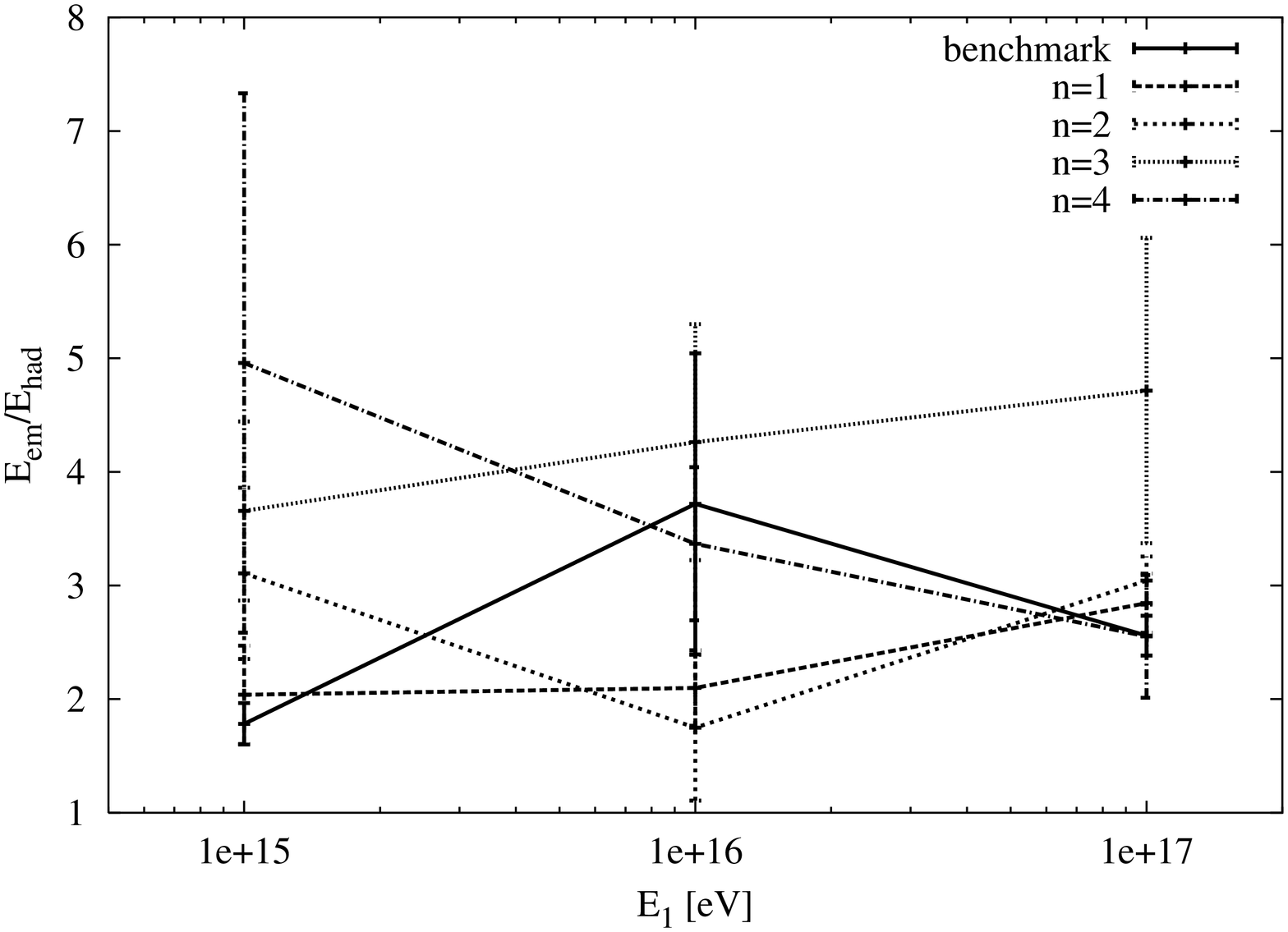}}}} \par}
\caption{As in Figure~\ref{fig:bhwl_NrappVE}, but this time we show the ratio
between \protect\( E_{em}\protect \) (total energy of photons and
\protect\( e^{\pm }\protect \)) and \protect\( E_{had}\protect \)
(total energy of everything else) as a function of \protect\( E_{1}\protect \).}
\label{ratio2}
\end{figure}
Furthermore, there 
does not seem to be a statistically significant difference in these observables 
between the benchmark event and the black hole mediated ones, at least
for event heights greater than 500 m above the detector. 
We conclude that (a) either these observables may not be suitable, 
especially given the limited 
statistics of the existing and future experiments, 
to discriminate between black hole mediated events versus standard ones, or (b) that
the differences in these ratios appear in events with initial impact
in the range $0-500$ m from the detector.



\vspace{0.6cm}

\item{\em Multiplicities}

For the multiplicities themselves the situation is much cleaner.
In Figures~\ref{prima}-\ref{multipro} we show the behaviour of the 
total as well as of some partial multiplicities in black hole mediated 
events and in standard 
events as a function of the energy and for a varying number of extra 
dimensions. 

\begin{figure}
{\centering \subfigure[ ]{\resizebox*{10cm}{!}{\rotatebox{-90}
{\includegraphics{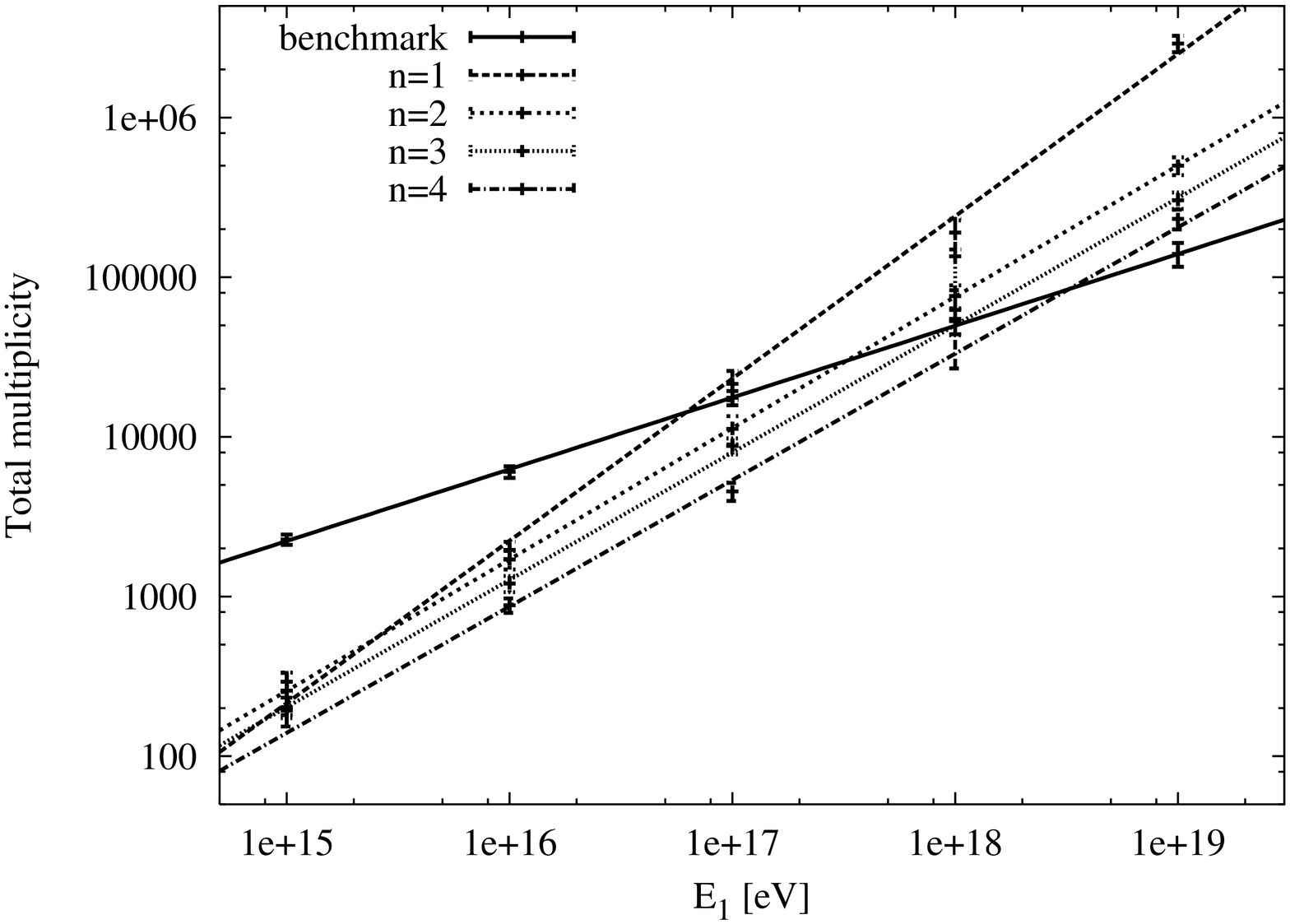}}}} \par}
{\centering \subfigure[ ]{\resizebox*{10cm}{!}{\rotatebox{-90}
{\includegraphics{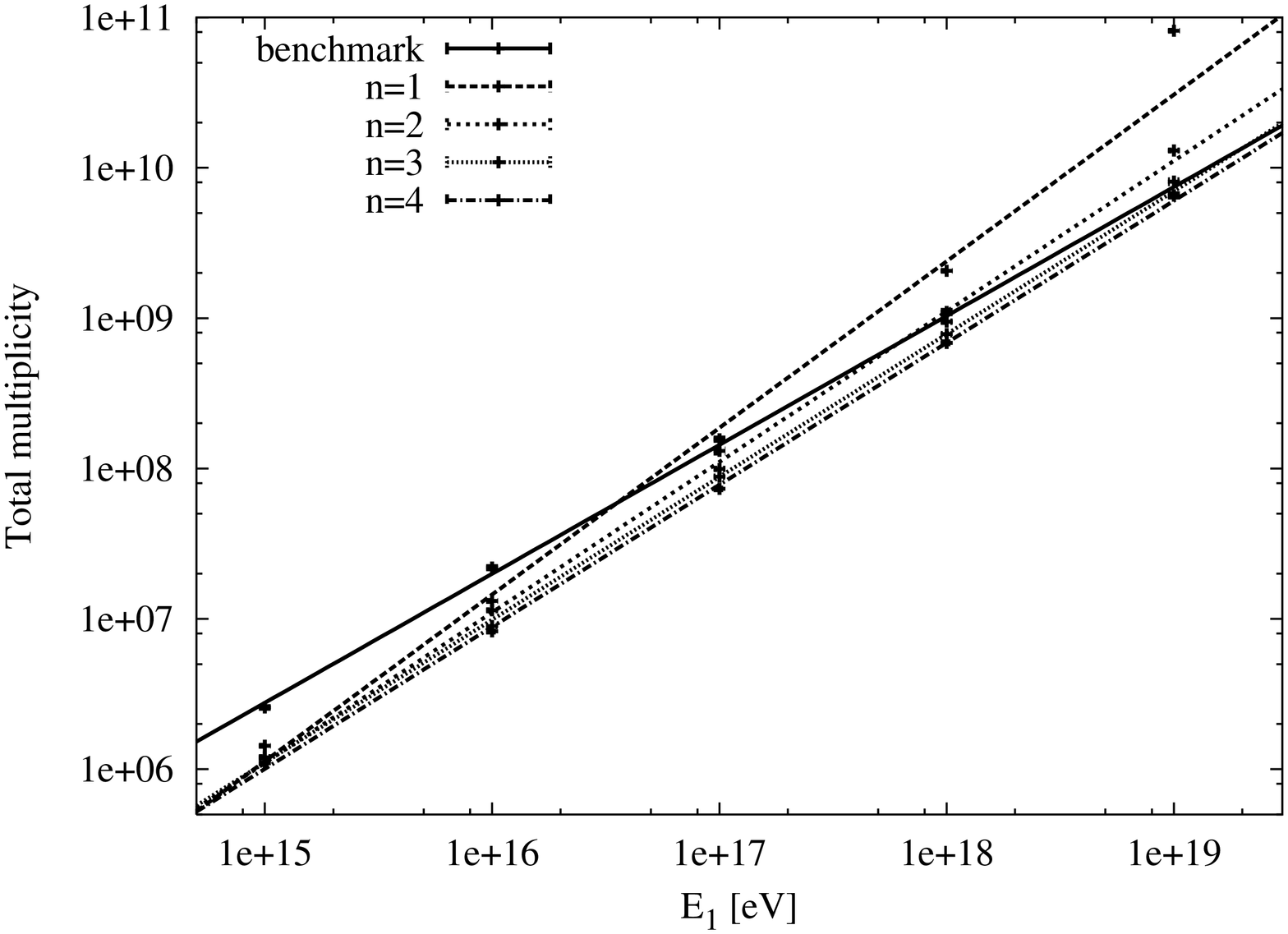}}}} \par}
\caption{Plot of the total particle multiplicity
as a function of \protect\( E_{1}\protect \). Case (a) is for an impact point 
of 5,500 m and case (b) for 15,000 m.}
\label{prima}
\end{figure}

\begin{figure}
{\centering \subfigure[ ]{\resizebox*{10cm}{!}{\rotatebox{-90}
{\includegraphics{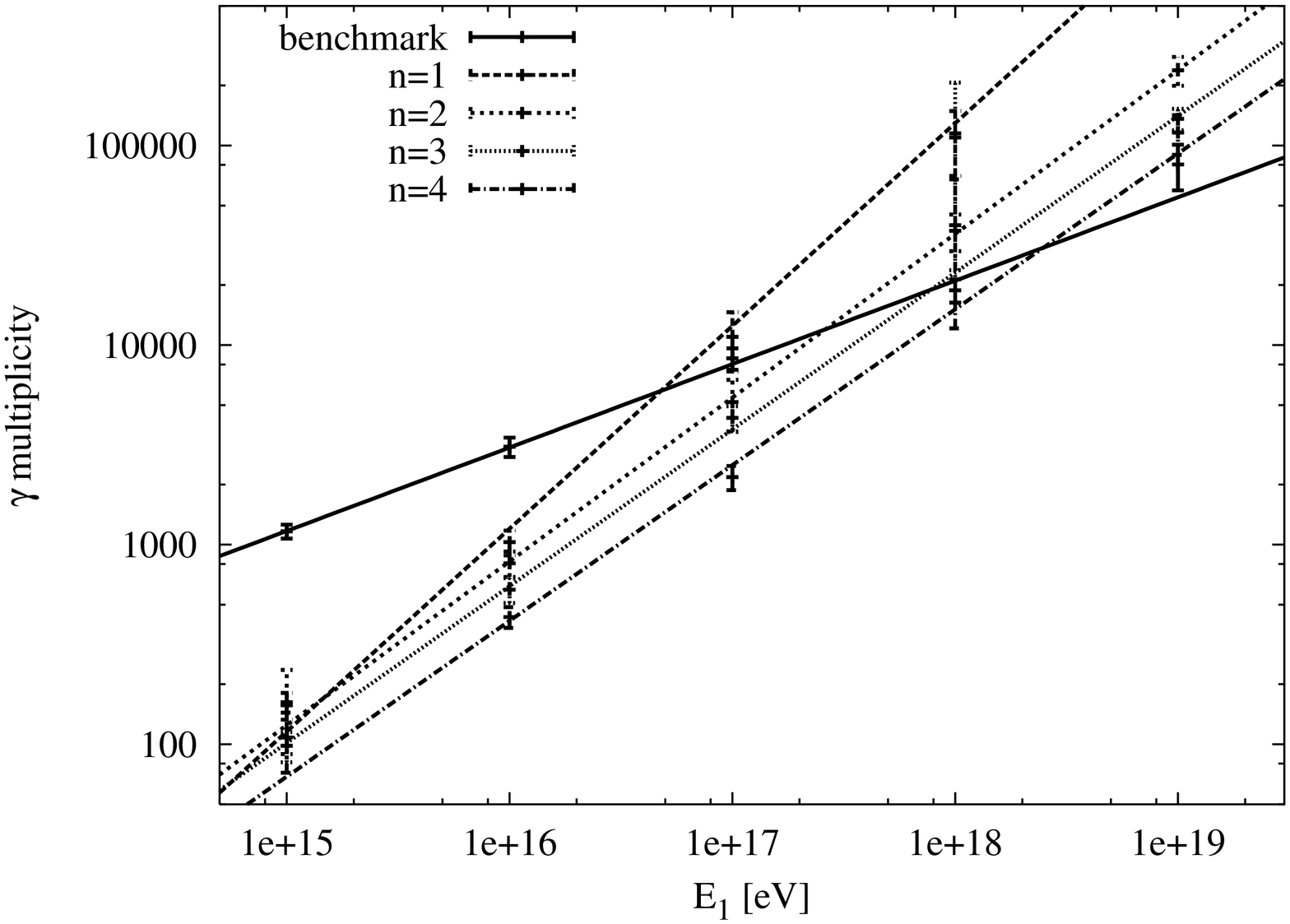}}}} \par}
{\centering \subfigure[ ]{\resizebox*{10cm}{!}{\rotatebox{-90}
{\includegraphics{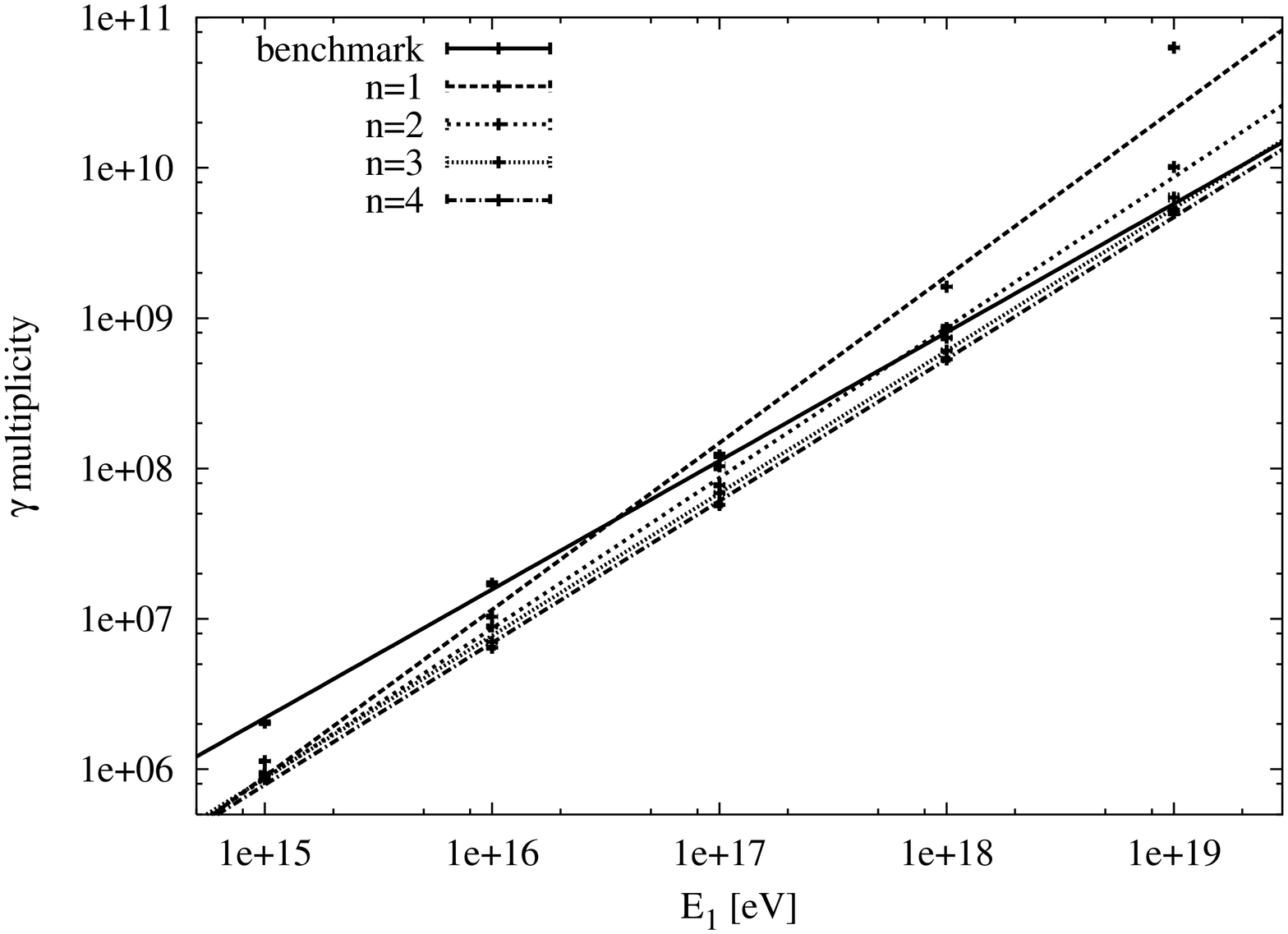}}}} \par}
\caption{Plot of the multiplicity of photons as a function of 
\protect\( E_{1}\protect \), (a) 5,500 m, (b) 15,000 m }
\label{seconda}
\end{figure}

\begin{figure}
{\centering \subfigure[ ]{\resizebox*{10cm}{!}{\rotatebox{-90}
{\includegraphics{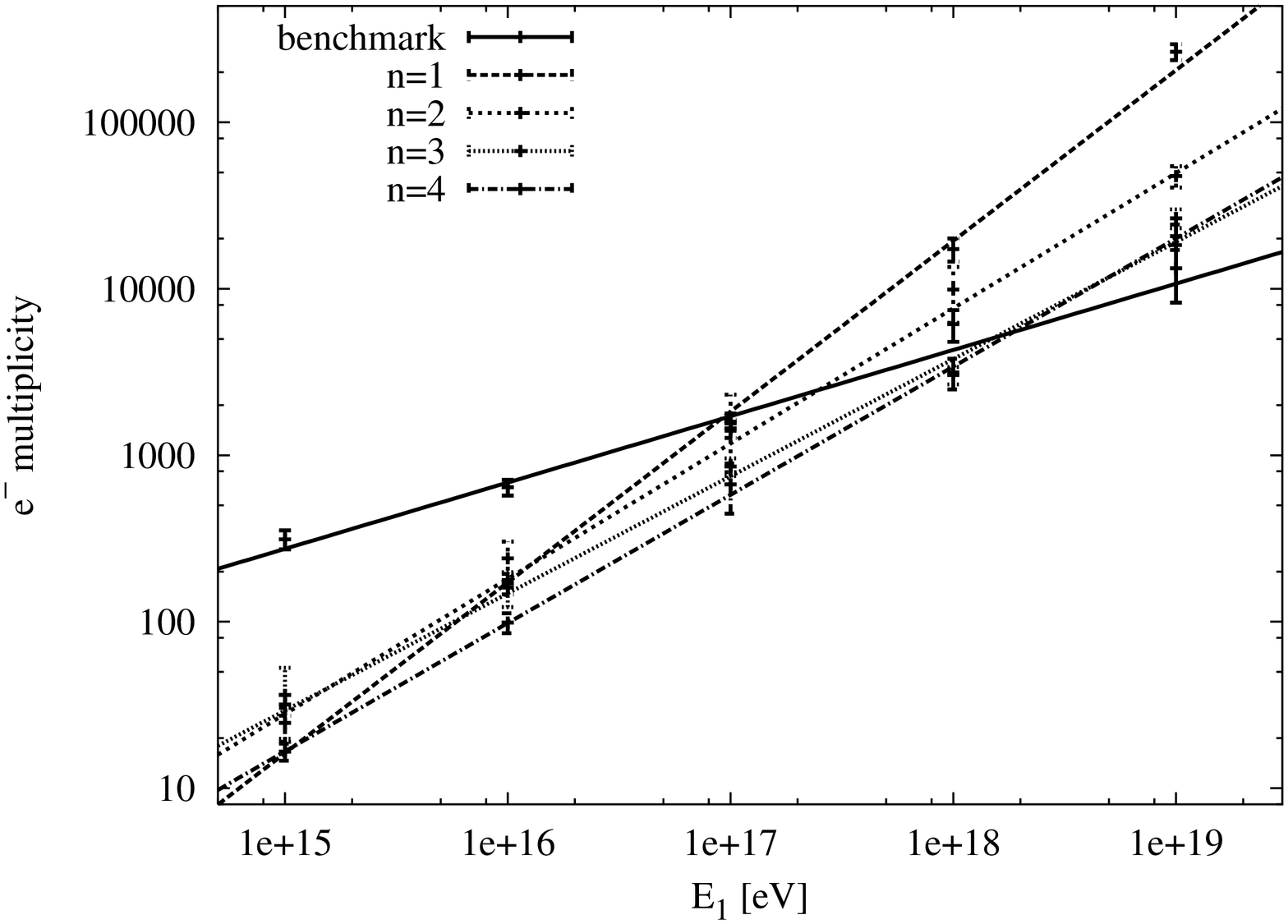}}}} \par}
{\centering \subfigure[ ]{\resizebox*{10cm}{!}{\rotatebox{-90}
{\includegraphics{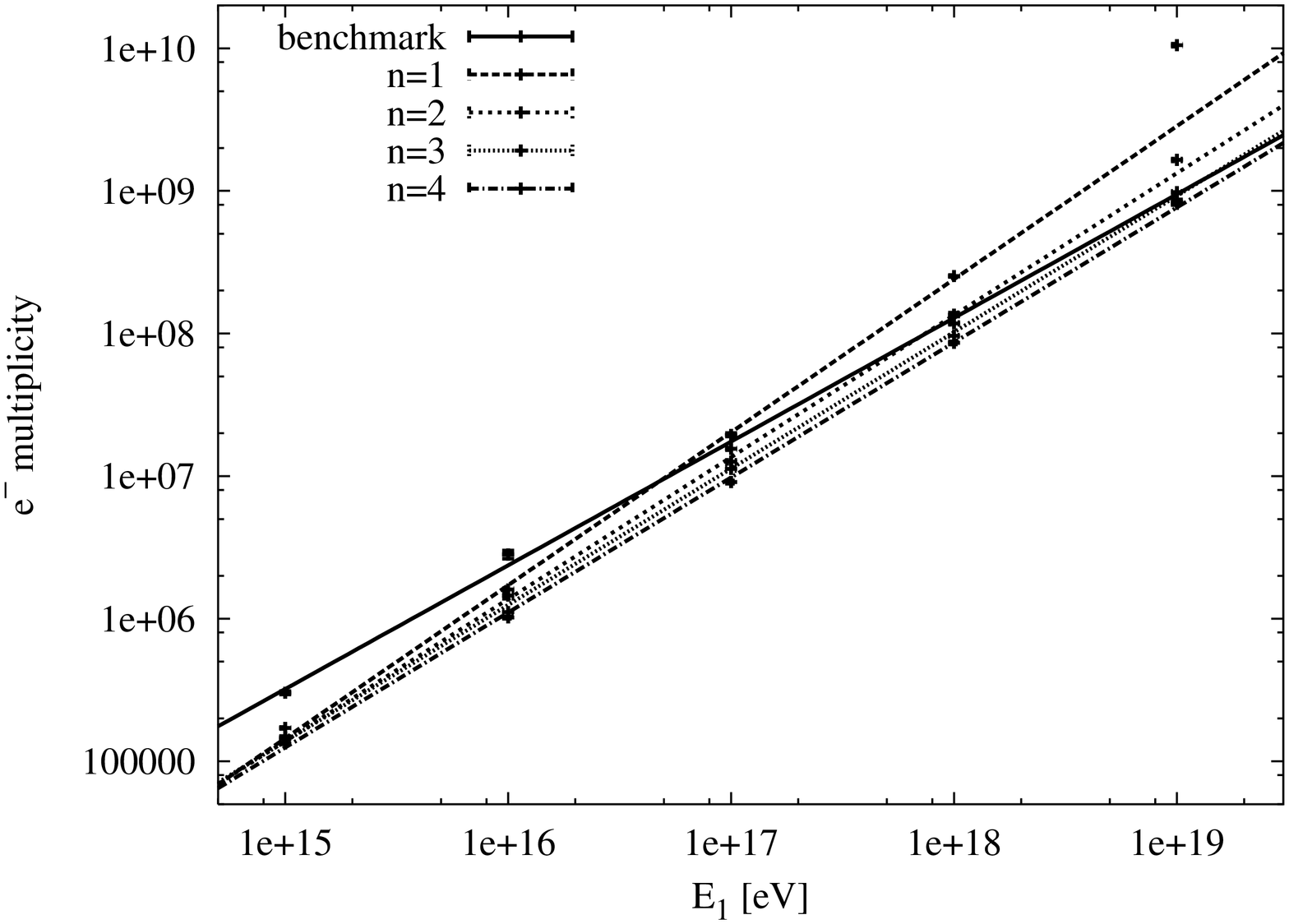}}}} \par}
\caption{Plot of the multiplicity of \protect\( e^{-}\protect \) 
as a function of \protect\( E_{1}\protect \), 
(a) 5,500 m, (b) 15,000 m}
\label{terza}
\end{figure}

\begin{figure}
{\centering \subfigure[ ]{\resizebox*{10cm}{!}{\rotatebox{-90}
{\includegraphics{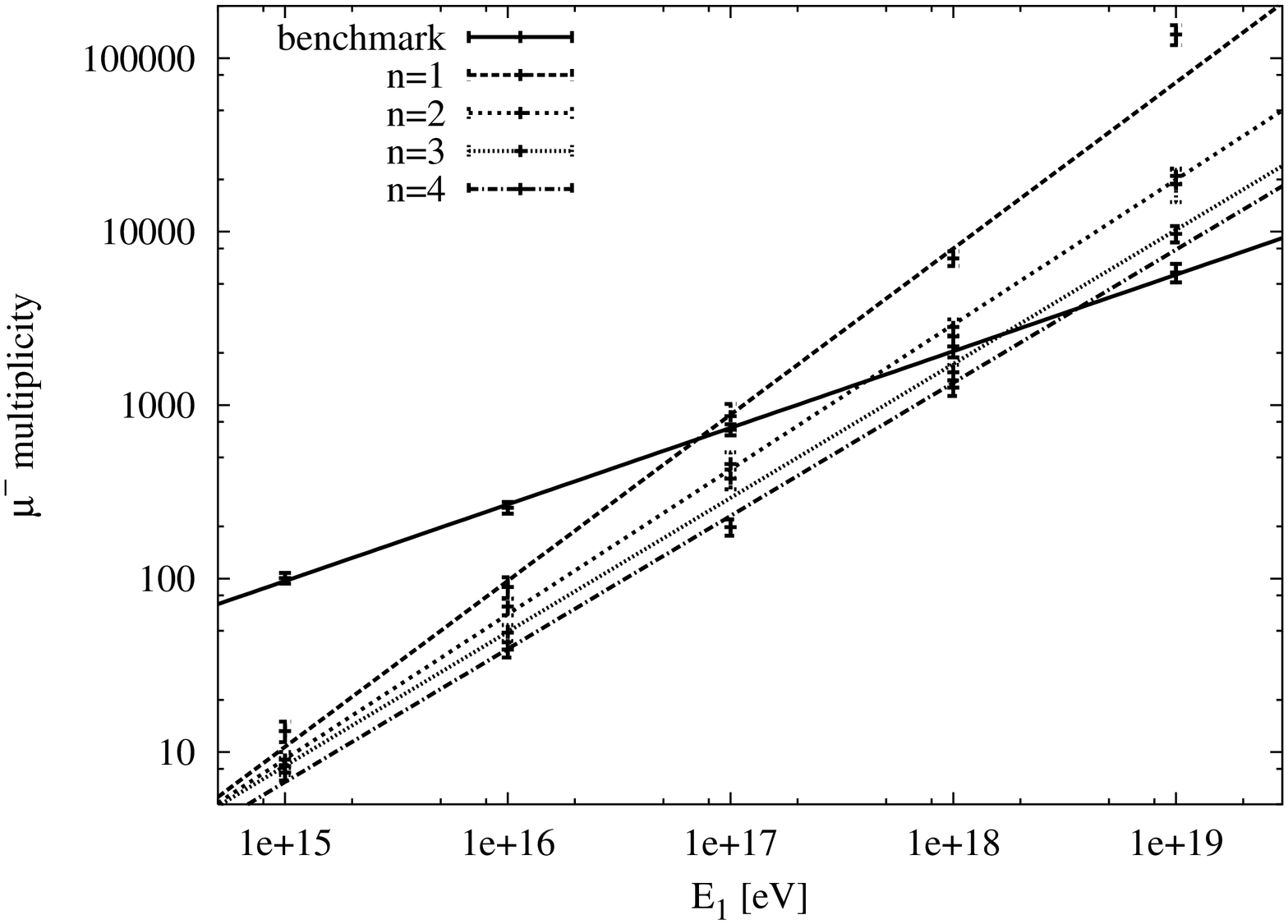}}}} \par}
{\centering \subfigure[ ]{\resizebox*{10cm}{!}{\rotatebox{-90}
{\includegraphics{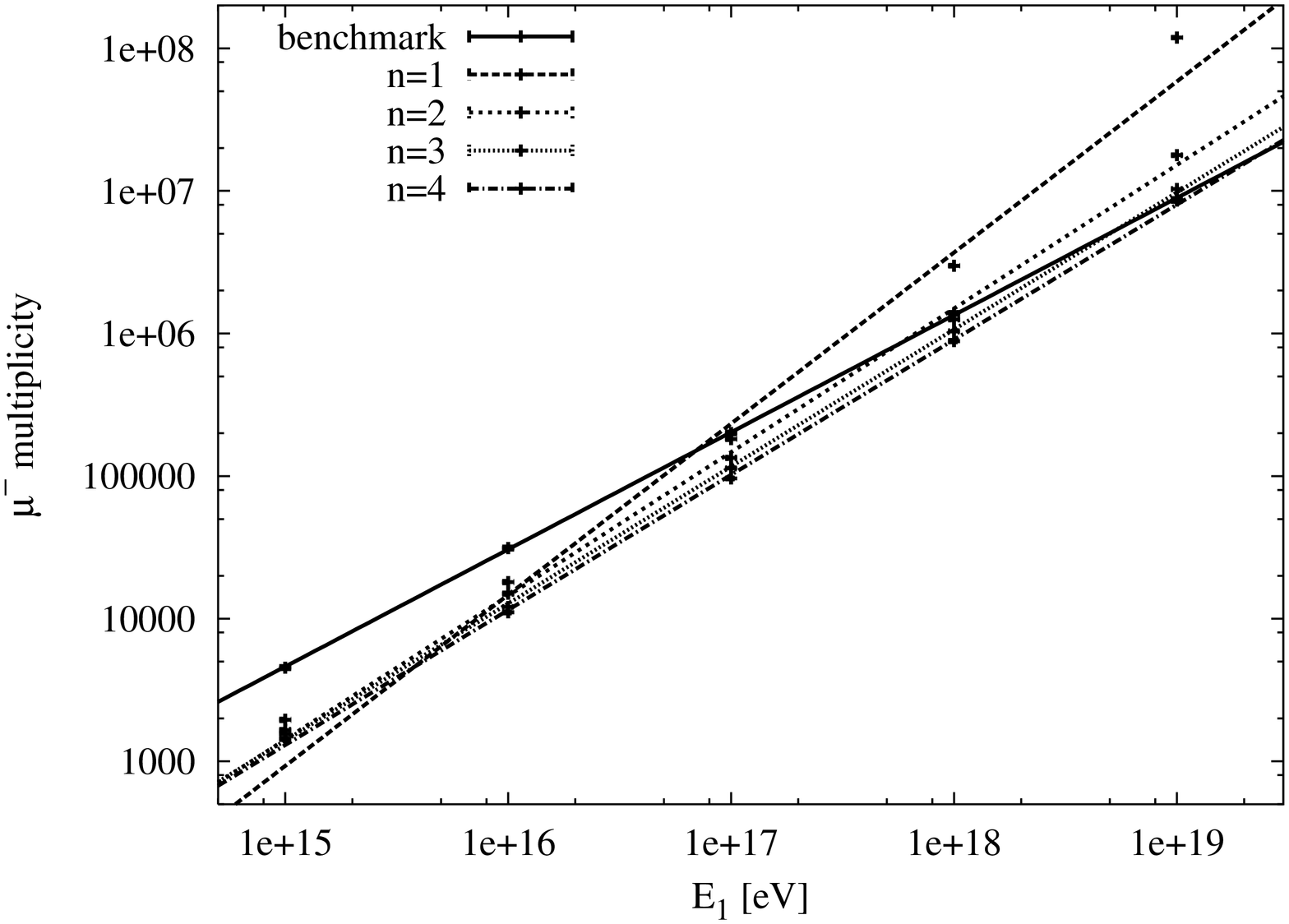}}}} \par}
\caption{Plot of the multiplicity
of \protect\( \mu ^{-}\protect \) as a function of \protect\( E_{1}\protect \), 
(a) 5,500 m, (b) 15,000 m}
\label{quarta}
\end{figure}

\begin{figure}
{\centering \subfigure[ ]{\resizebox*{10cm}{!}{\rotatebox{-90}
{\includegraphics{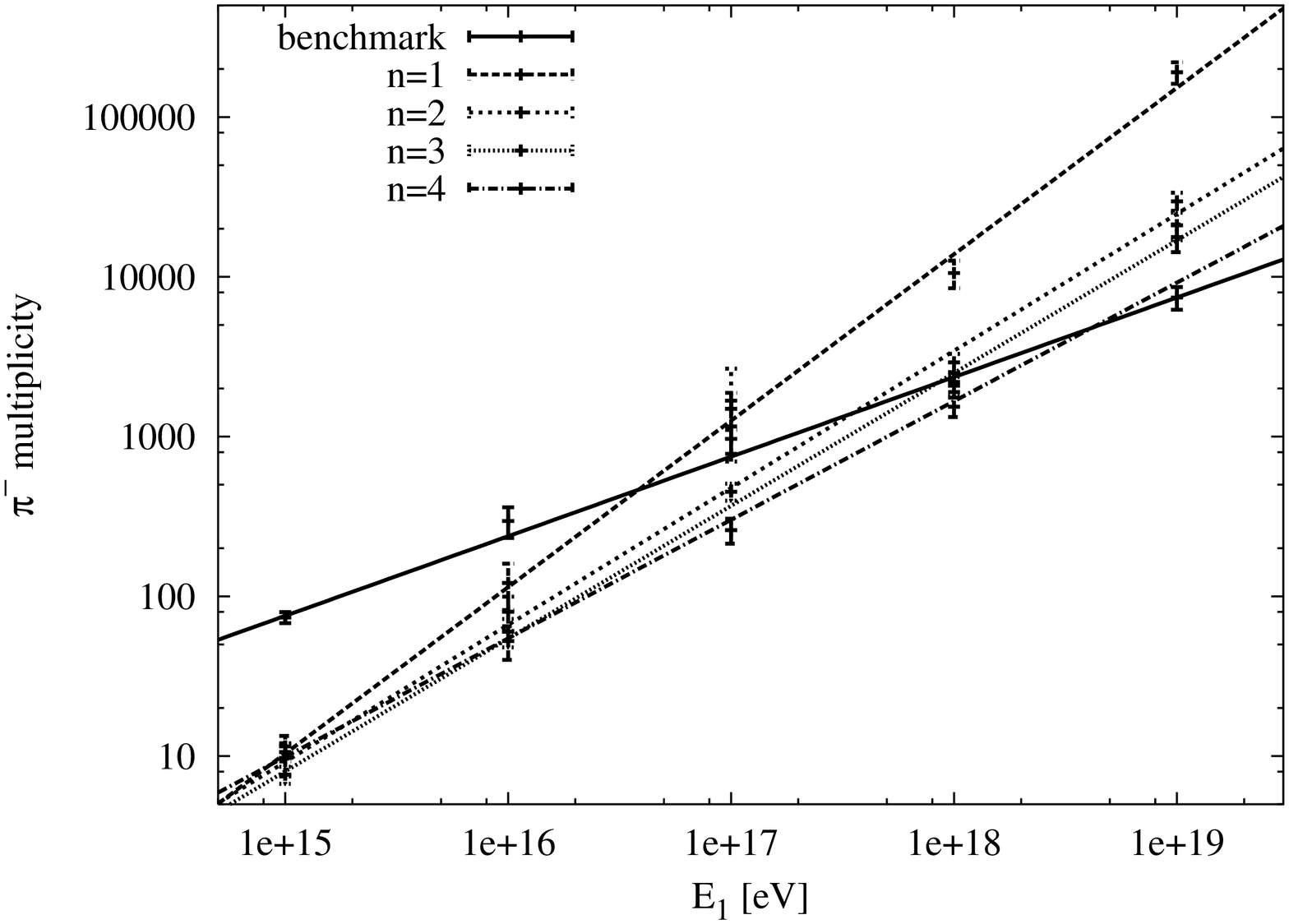}}}} \par}
{\centering \subfigure[ ]{\resizebox*{10cm}{!}{\rotatebox{-90}
{\includegraphics{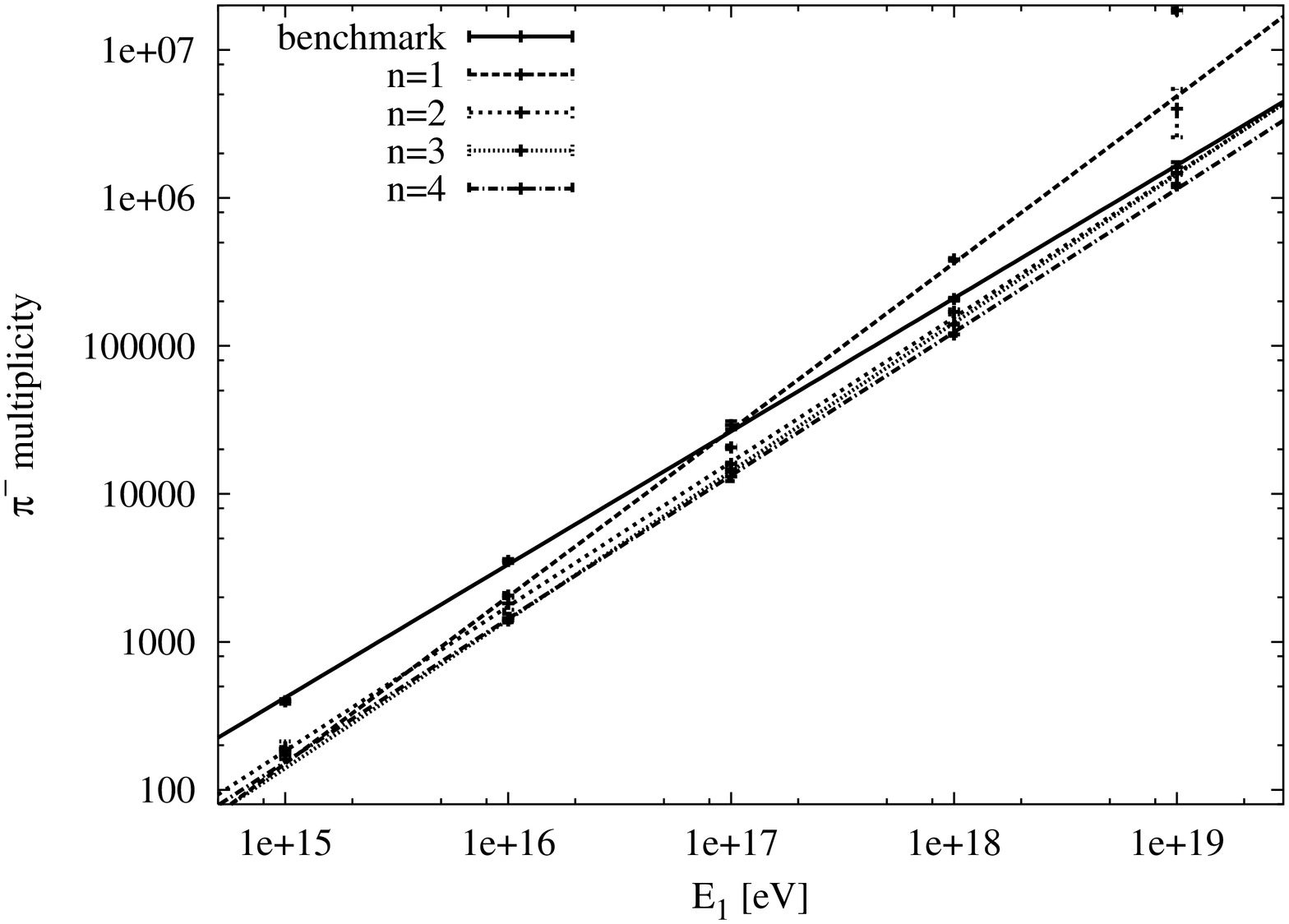}}}} \par}
\caption{Plot of the multiplicity
of \protect\( \pi ^{-}\protect \) as a function of \protect\( E_{1}\protect \), 
(a) 5,500 m, (b) 15,000 m.}
\label{multipi}
\end{figure}

\begin{figure}
{\centering \subfigure[ ]{\resizebox*{10cm}{!}{\rotatebox{-90}
{\includegraphics{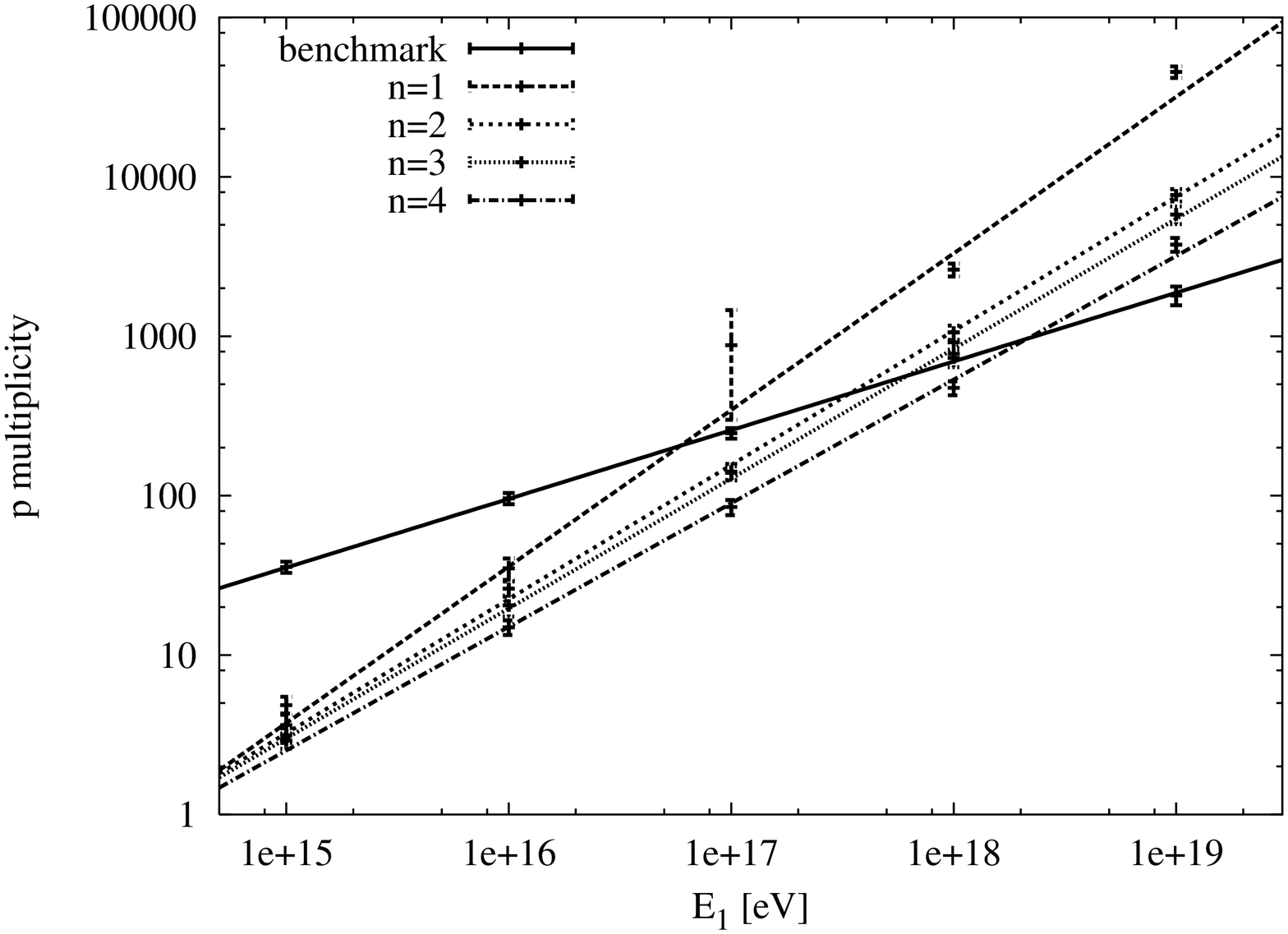}}}} \par}
{\centering \subfigure[ ]{\resizebox*{10cm}{!}{\rotatebox{-90}
{\includegraphics{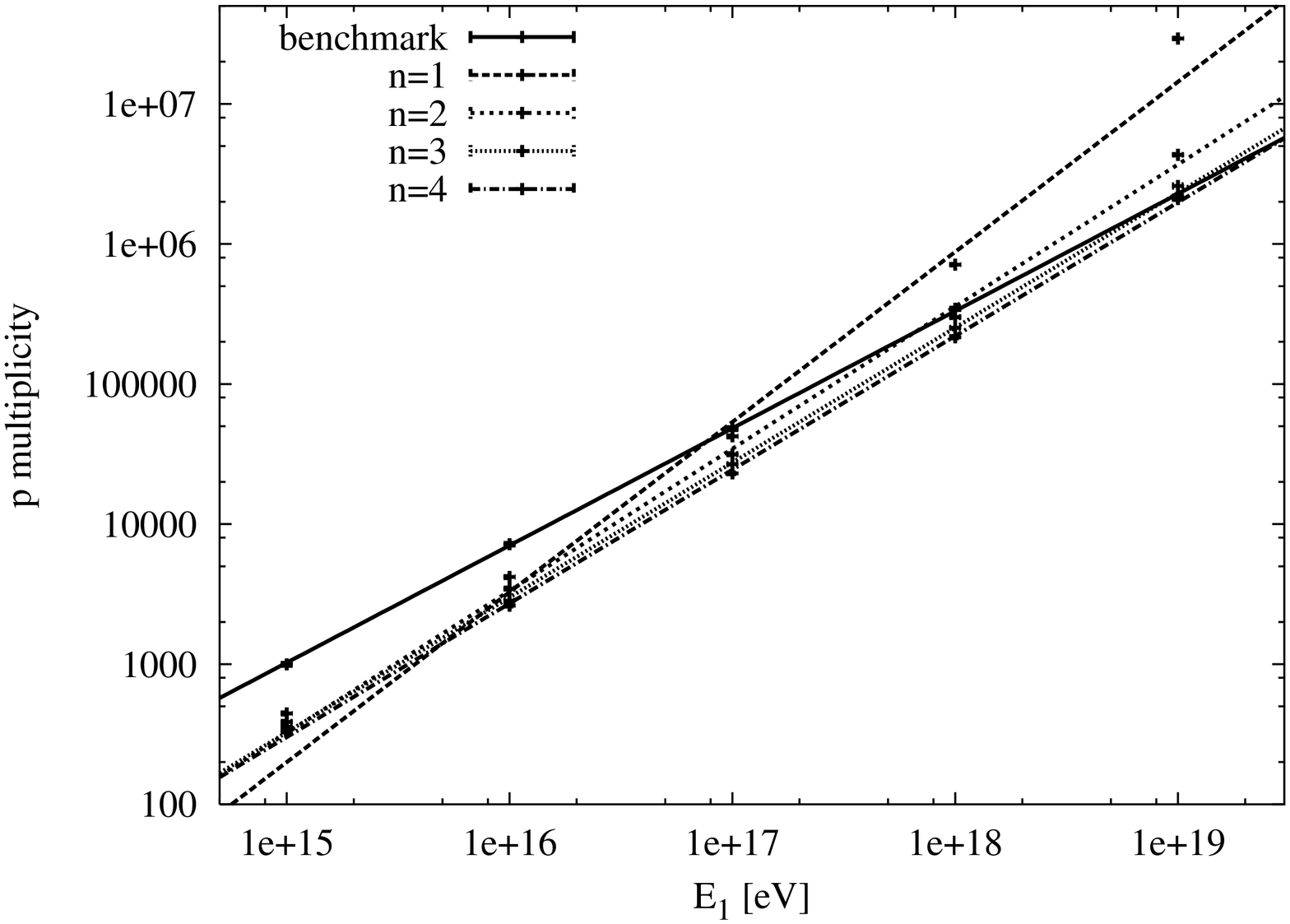}}}} \par}
\caption{Plot of the multiplicity
of protons as a function of \protect\( E_{1}\protect \), (a) 5,500 m, (b) 15,000 m.}
\label{multipro}
\end{figure}
The curves are very well fitted in a log-log plot by 
a linear relation of the form 
\beq
N=10^{q(n)} E_1^{\sigma(n)}  
\label{ntotal}
\eeq
with intercepts $q(n)$ and slopes $\sigma(n)$, 
that increase with the number of extra dimensions $n$. 
We present in Figures (a) results of the simulations performed with a first 
impact taken at 5,500 m, while Figures (b) refer to a first collision at 
15,000 m. In Figures of type (a) the slopes of the benchmark events are 
smaller than those of the black hole events and show a larger intercept. 
This feature is common to all the  
sub-components of the air showers. A simple explanation of this fact is 
that at lower value of the impact energy, the 
number of states available for the decay of the black hole is smaller 
than the number of partonic degrees of freedom available in 
a proton-proton collision. We recall, as we have already discussed in 
the previous sections, that our benchmark results define in this case 
an upper bound for the total multiplicities expected in a 
neutrino-proton collision. Therefore, in a more realistic comparison, 
we would discover that the black hole and the standard results should 
differ more noticeably. 
The large multiplicity of the states available for the decay of the black 
hole dominates over that of a standard hadronic interaction, and this 
justifies the larger multiplicities produced at detector level.  
As we increase the altitude of the impact, in plots of type (b) we find 
a similar trend but the differences in the total and partial multiplicities 
are much harder to discern for black holes and benchmark events. In fact, 
for collisions starting at higher altitudes the showers are all 
fully developed and the differences between the two underlying events are 
less pronounced. 

Another feature of the black hole events is that the slopes and the 
intercepts of the various plots, for a given choice of altitude of 
the impact, are linearly correlated. To illustrate this 
point we refer to Figures~\ref{fitcurve1}-\ref{fitcurve2} from which 
this behaviour is 
immediately evident. To generate each of these figures we have plotted 
the parameters 
$(\sigma,q)$ of a corresponding plot - for the total or for the partial 
multiplicities - 
independently of the specific number of extra dimensions. The results 
shown in these 
figures clearly indicate that the relation between the intercept q and 
the slope $\sigma$ 
appearing in Eq.~(\ref{ntotal}) is linear and independent of $n$ 
\beq
q=\alpha\,\sigma+\beta
\eeq
with $\alpha$ and $\beta$ typical of a given setup (photons, total multiplicities,
etc.) 
but insensitive to the parameter $n$. Therefore, black hole events are characterized 
by particle multiplicities on the ground of the form  
\beq
N_{ground}=10^{\alpha\,\sigma + \beta} E_1^{\sigma}.  
\label{ntotalfinal}
\eeq

\begin{figure}
{\centering \subfigure[ ]{\resizebox*{10cm}{!}{\rotatebox{-90}
{\includegraphics{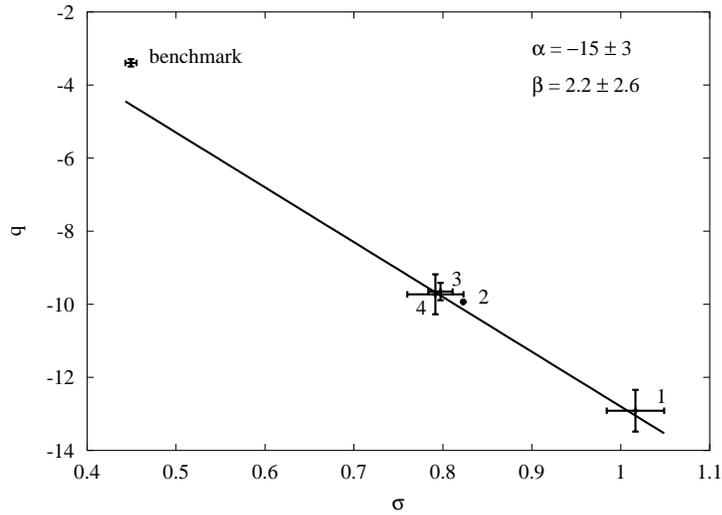}}}} \par}
{\centering \subfigure[ ]{\resizebox*{10cm}{!}{\rotatebox{-90}
{\includegraphics{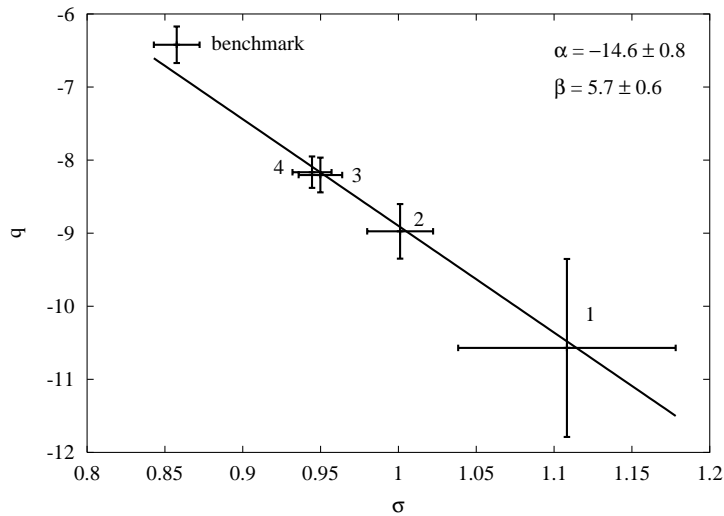}}}} \par}
\caption{Parameter fit for the intercepts and the
slopes of the curves in Fig.~\ref{prima} for the total
multiplicities. The numbers over each point in this plot indicate the
value of the extra-dimensions. The \protect\( (\sigma,q)\protect \) parameters
are fitted to a straight line \protect\( q= \alpha\,\sigma+\beta \protect \)
independently of the
numbers of extra dimensions. (a) is the fit for 5,500 m, (b) for 15,000 m.
The benchmark is also shown in the plot, but has not been used in the fit.}
\label{fitcurve1}
\end{figure}
\begin{figure}
{\centering \subfigure[ ]{\resizebox*{10cm}{!}{\rotatebox{-90}
{\includegraphics{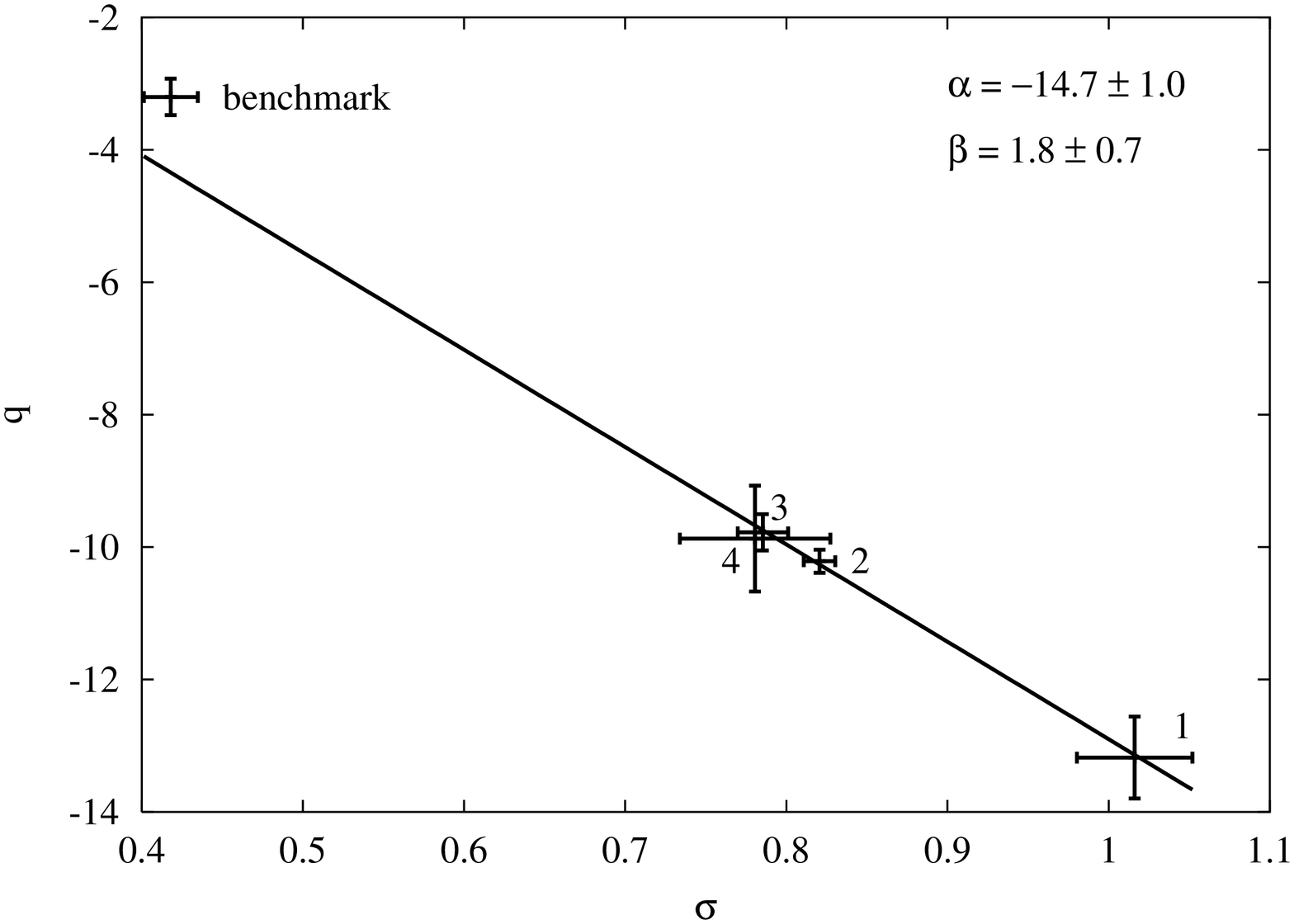}}}} \par}
{\centering \subfigure[ ]{\resizebox*{10cm}{!}{\rotatebox{-90}
{\includegraphics{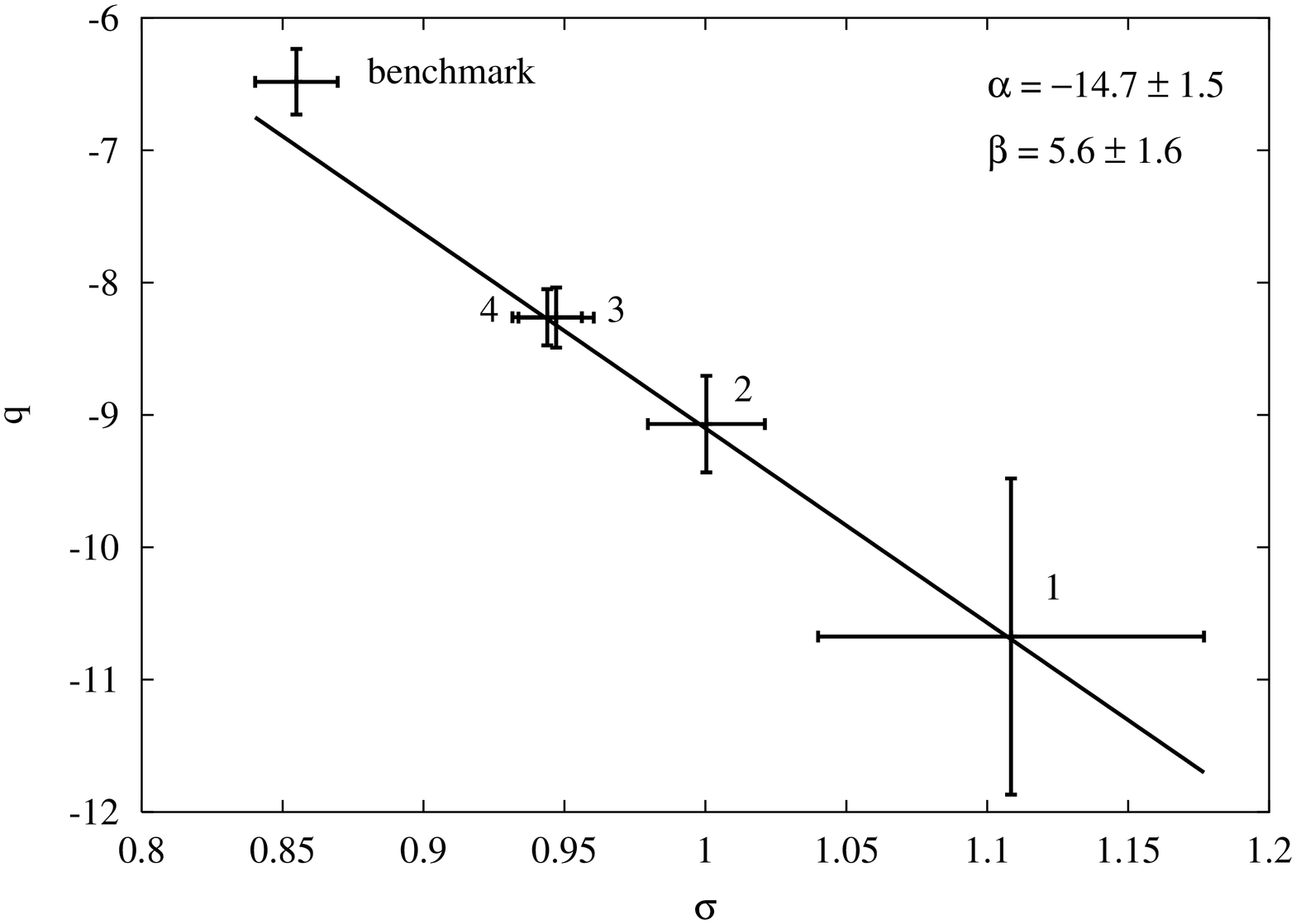}}}} \par}
\caption{Paramater fit for the intercepts and the slopes for the
curves in Fig.~\ref{seconda}, now for the multiplicity of photons. 
(a) is the fit for 5,500 m, (b) for 15,000 m.}
\label{fitcurve2}
\end{figure}


\vspace{0.6cm}

\item{\em Lateral distributions}

In Figures~\ref{core1}-\ref{core5} we illustrate 
the results of our study of the lateral 
distributions for the total inclusive shower and the various sub-components 
as a function of the incoming energy $E_1$. The average opening 
of the shower as measured at detector level is plotted versus energy in a log-log 
scale. Notice a growing opening of the shower as 
we raise the energy of the black hole resonance, which is more 
remarked for a lower number of extra dimensions. In contrast, 
the benchmark simulation 
shows a small decrease (negative slope) with energy. The larger opening 
of the shower in black hole mediated events - compared to standard air 
showers - is due to the s-wave emission typical of a black hole decay, 
which is very different from an ordinary collision. Contrary to the 
case of multiplicities, here simulations of type (a) and 
(b) show a similar trend, with very distinct features between standard and 
black hole events. 
Notice that in this case the difference in the partonic content of the 
two different events (benchmark versus black hole mediated) is less relevant, 
since it is the geometrical fireball emission in the black hole case 
which is responsible for the generation of larger lateral distributions.

\begin{figure}
{\centering \subfigure[ ]{\resizebox*{10cm}{!}{\rotatebox{-90}
{\includegraphics{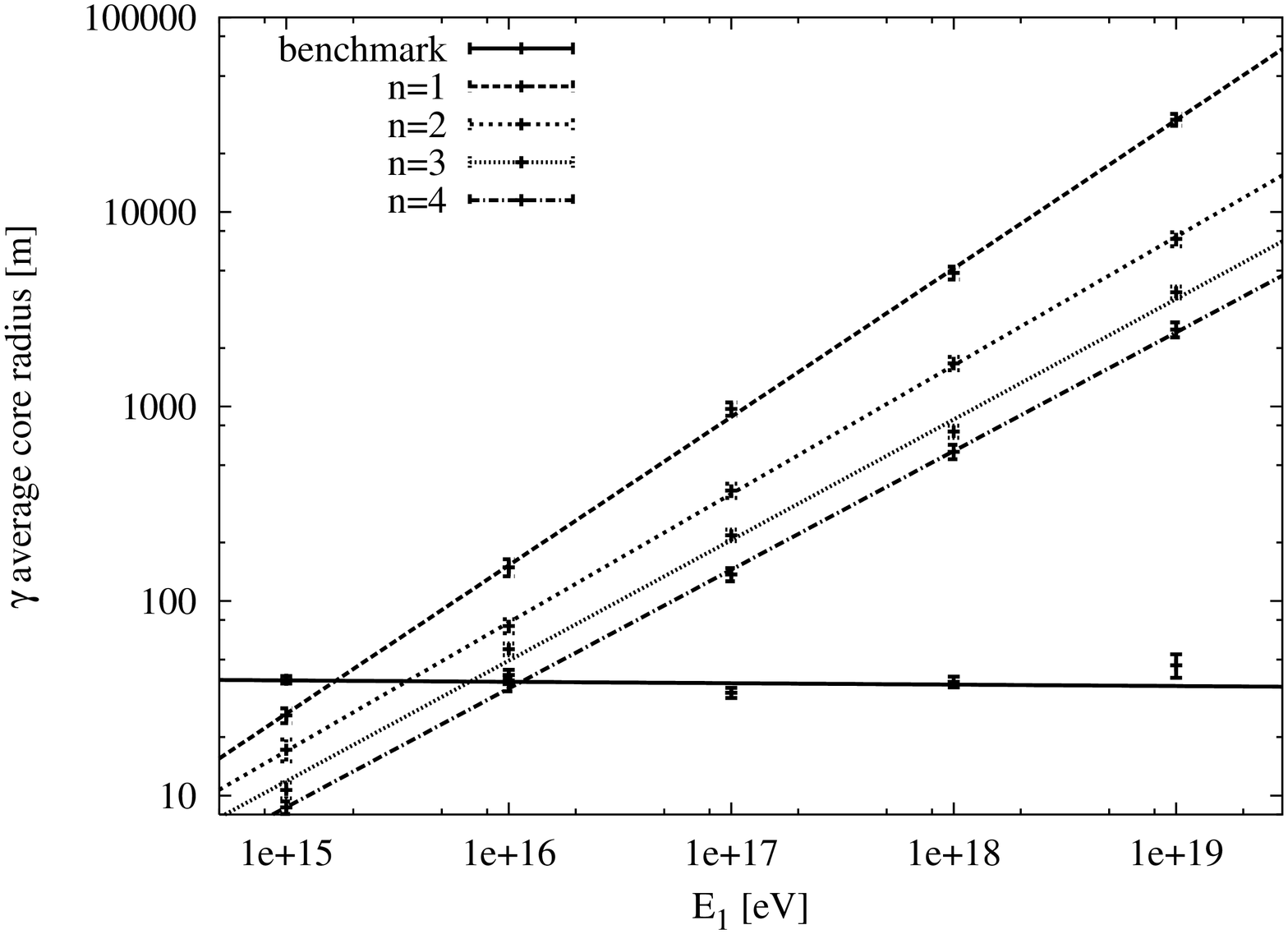}}}} \par}
{\centering \subfigure[ ]{\resizebox*{10cm}{!}{\rotatebox{-90}
{\includegraphics{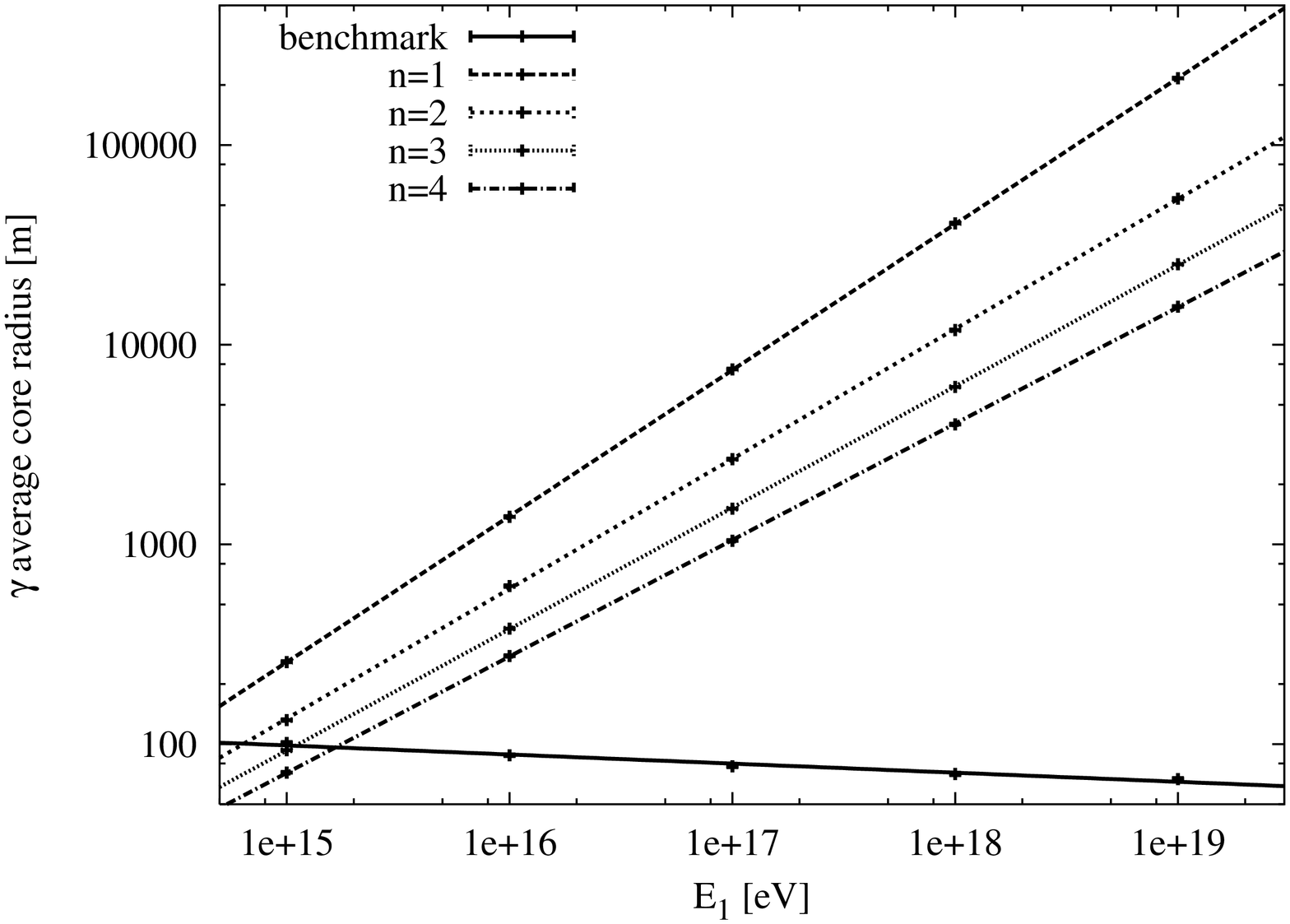}}}} \par}
\caption{Plot of the average
radius $R$ of the core of the shower of photons as a function 
of \protect\( E_{1}\protect \) for a black hole with a varying number 
of extra dimensions.
The benchmark result is also shown for comparison. (a) 5,500 m, (b) 15,000 m.}
\label{core1}
\end{figure}

\begin{figure}
{\centering \subfigure[ ]{\resizebox*{10cm}{!}{\rotatebox{-90}
{\includegraphics{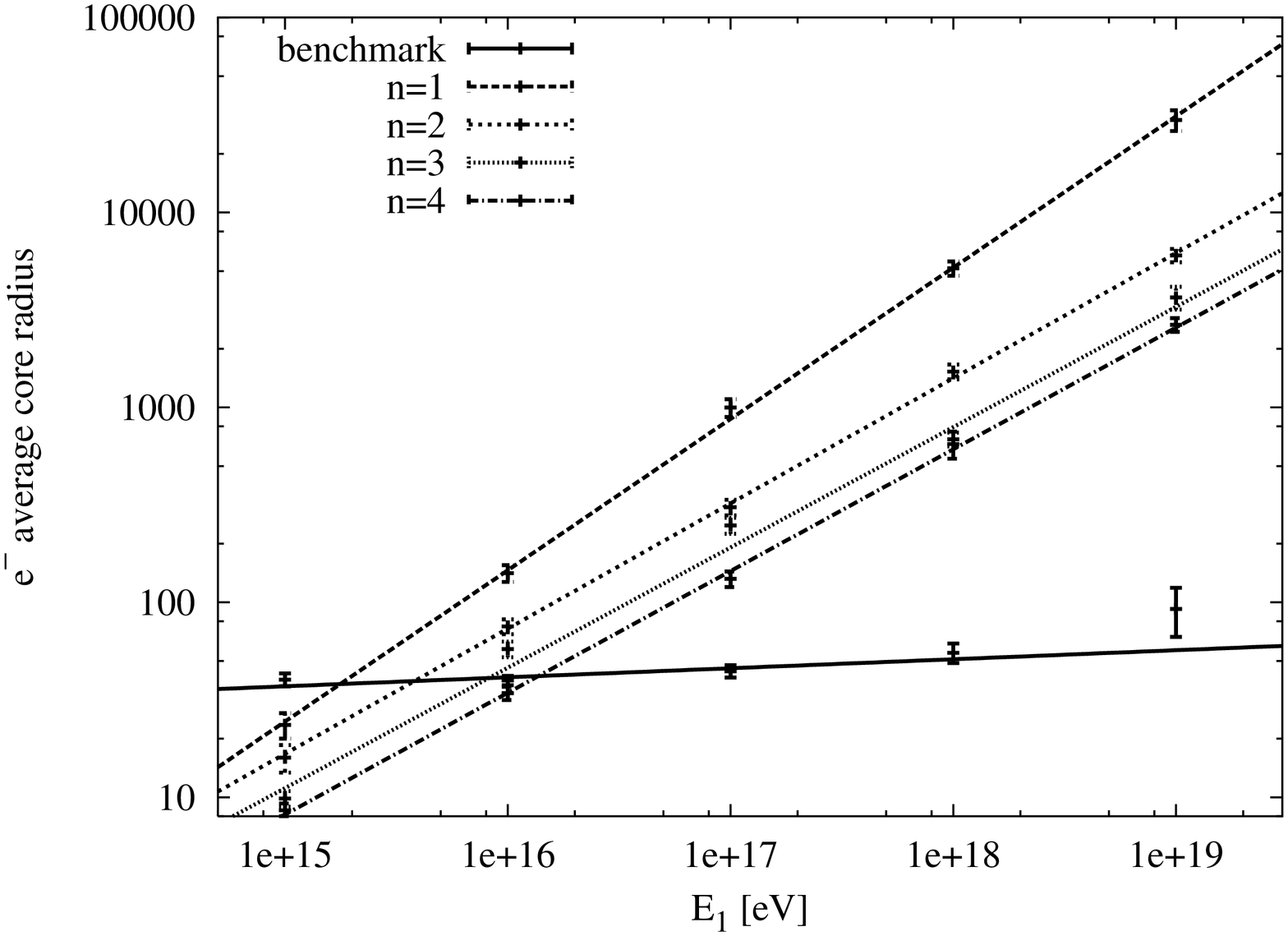}}}} \par}
{\centering \subfigure[ ]{\resizebox*{10cm}{!}{\rotatebox{-90}
{\includegraphics{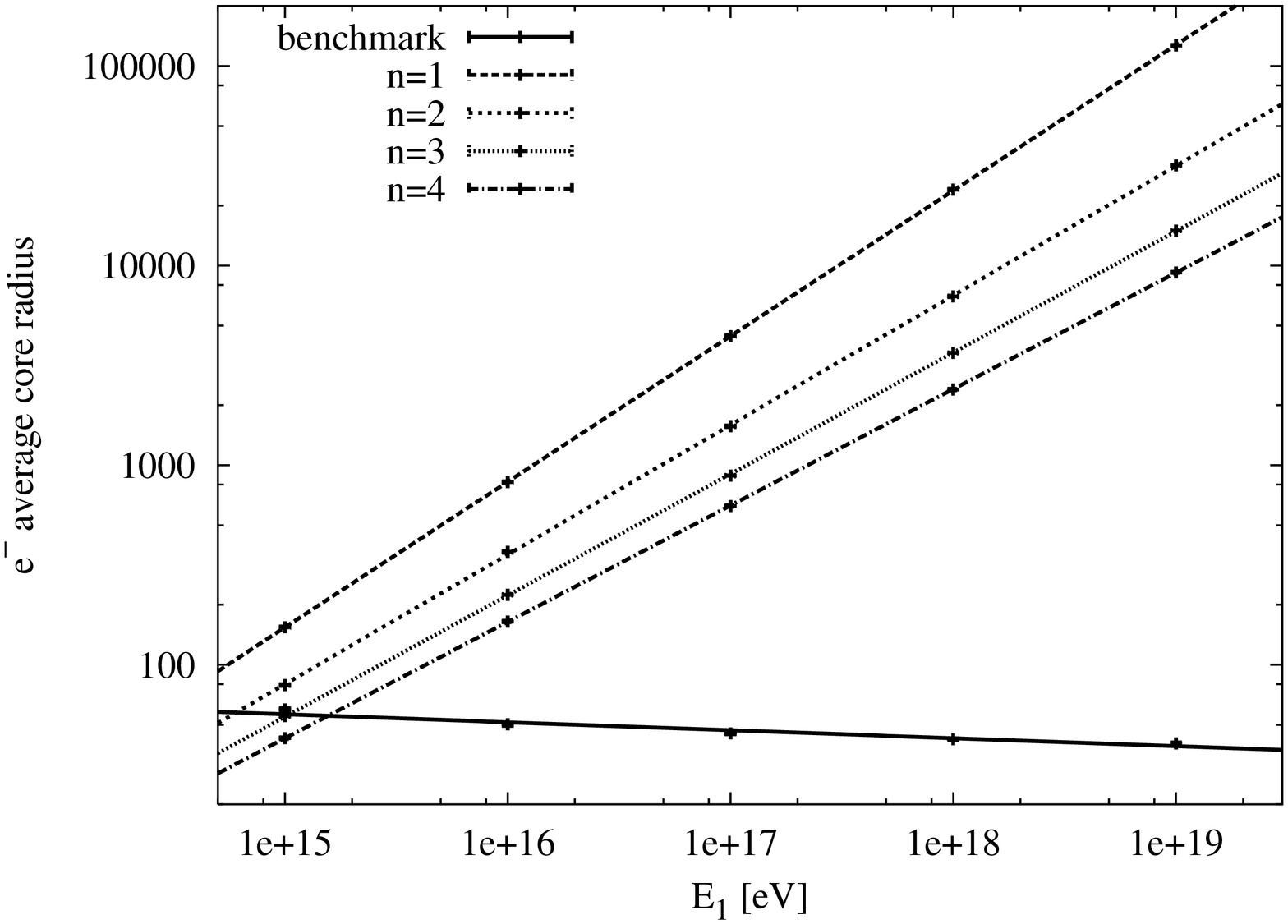}}}} \par}
\caption{Plot of the average
size of the core $R$ of the shower of \protect\( e^{-}\protect \)
as a function of \protect\( E_{1}\protect \). (a) 5,500 m, (b) 15,000 m.}
\label{core2}
\end{figure}

\begin{figure}
{\centering \subfigure[ ]{\resizebox*{10cm}{!}{\rotatebox{-90}
{\includegraphics{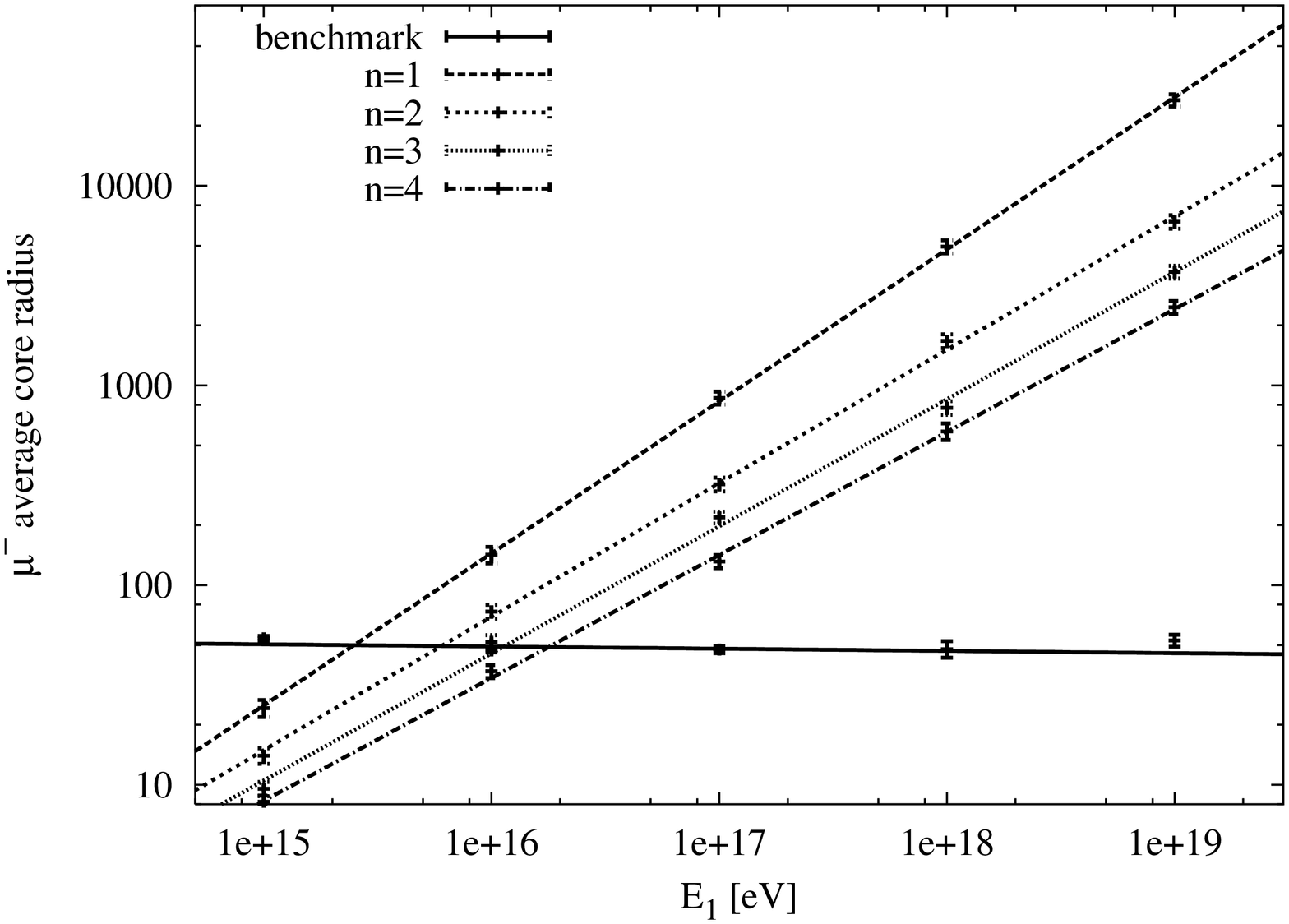}}}} \par}
{\centering \subfigure[ ]{\resizebox*{10cm}{!}{\rotatebox{-90}
{\includegraphics{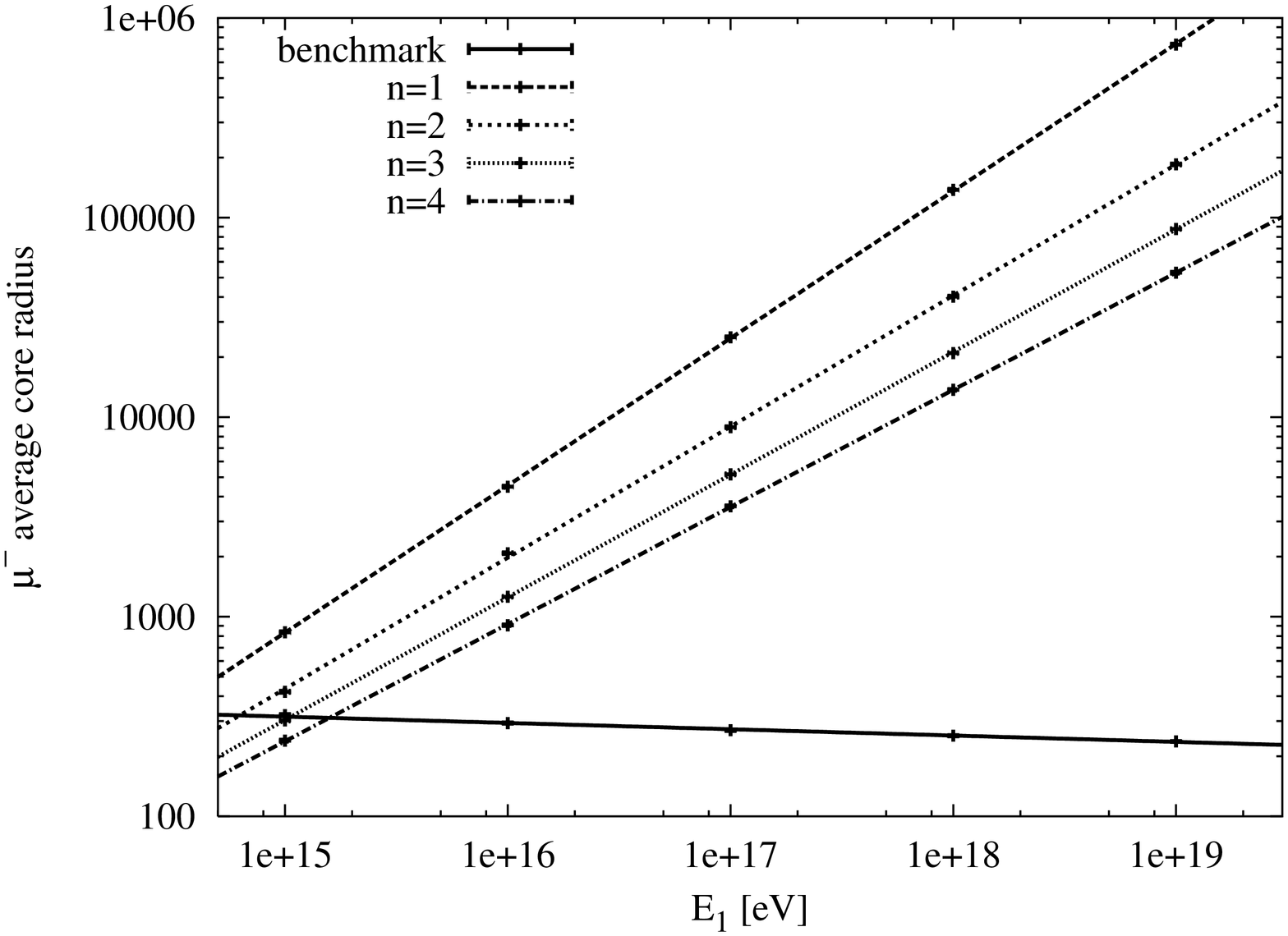}}}} \par}
\caption{Plot of the average
size $R$ of the core of the shower of \protect\( \mu ^{-}\protect \)
as a function of \protect\( E_{1}\protect \), (a) 5,500 m, (b) 15,000 m.}
\label{core3}
\end{figure}

\begin{figure}
{\centering \subfigure[ ]{\resizebox*{10cm}{!}{\rotatebox{-90}
{\includegraphics{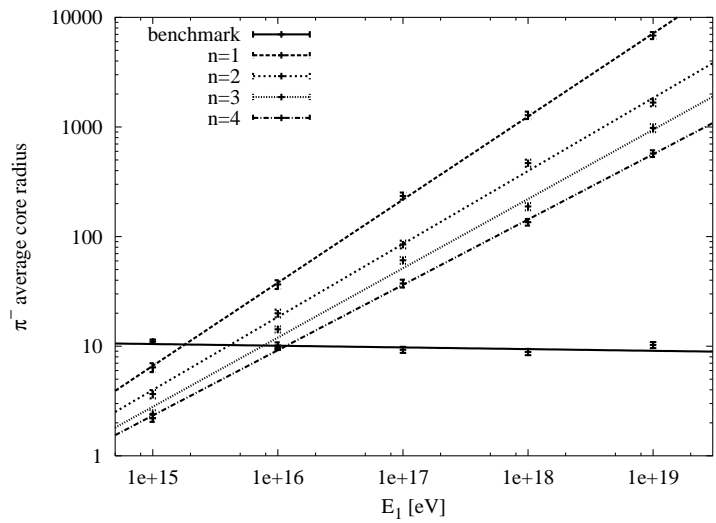}}}} \par}
{\centering \subfigure[ ]{\resizebox*{10cm}{!}{\rotatebox{-90}
{\includegraphics{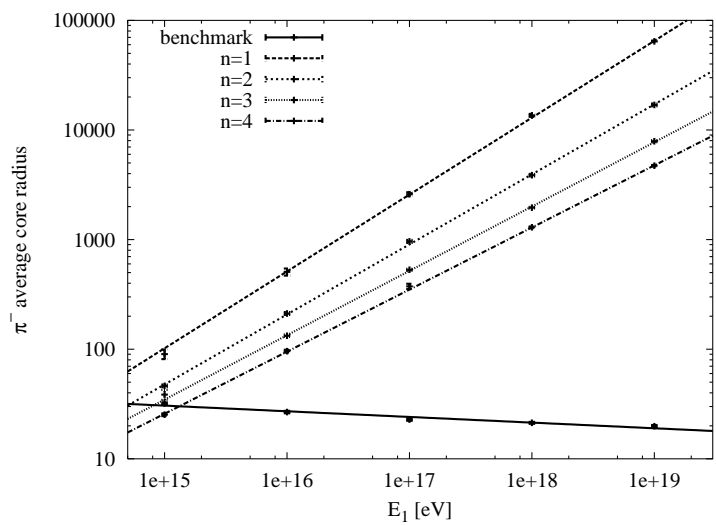}}}} \par}
\caption{As above for \protect\( \pi ^{-}\protect \)}
\label{core4}
\end{figure}

\begin{figure}
{\centering \subfigure[ ]{\resizebox*{10cm}{!}{\rotatebox{-90}
{\includegraphics{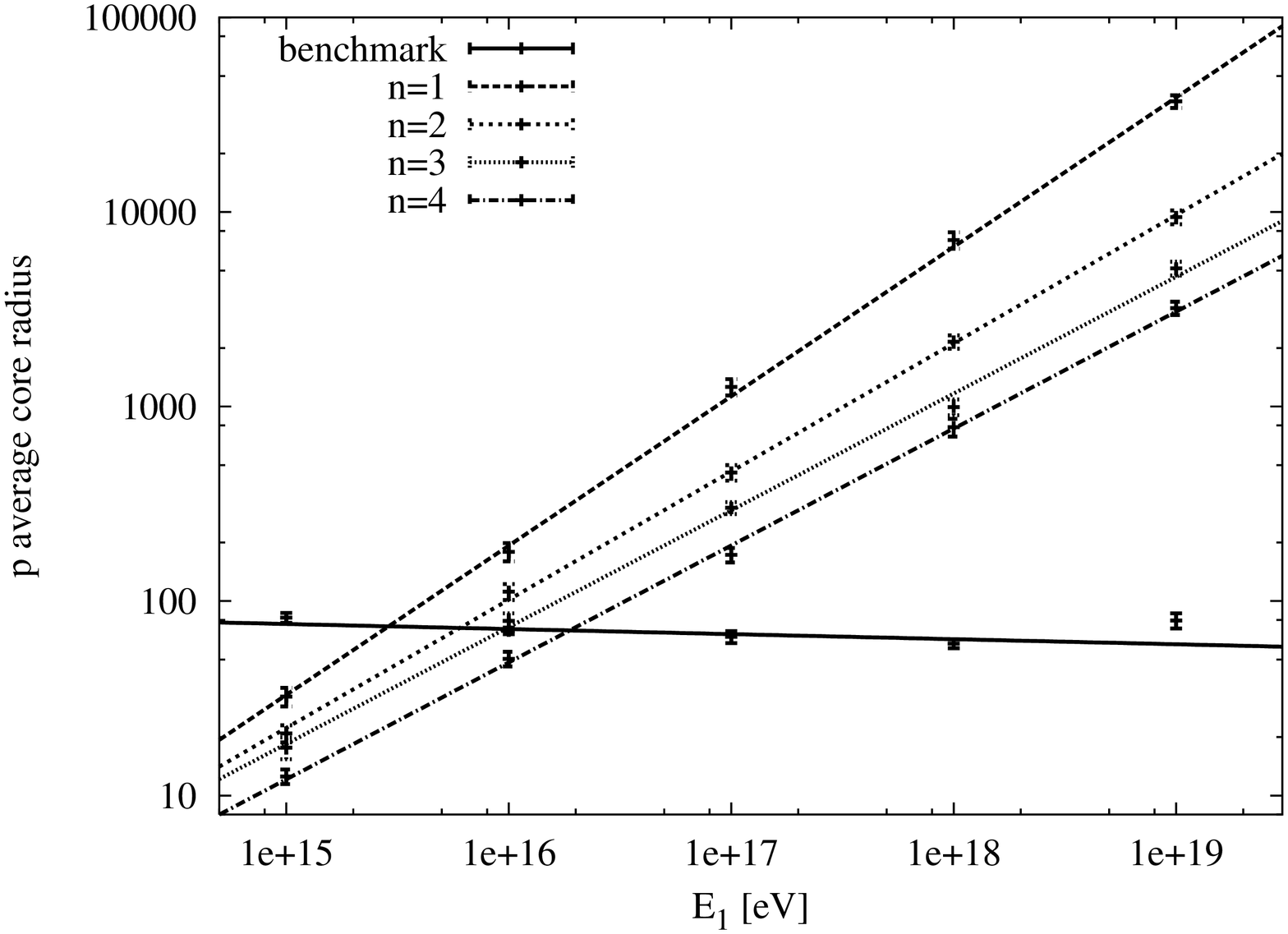}}}} \par}
{\centering \subfigure[ ]{\resizebox*{10cm}{!}{\rotatebox{-90}
{\includegraphics{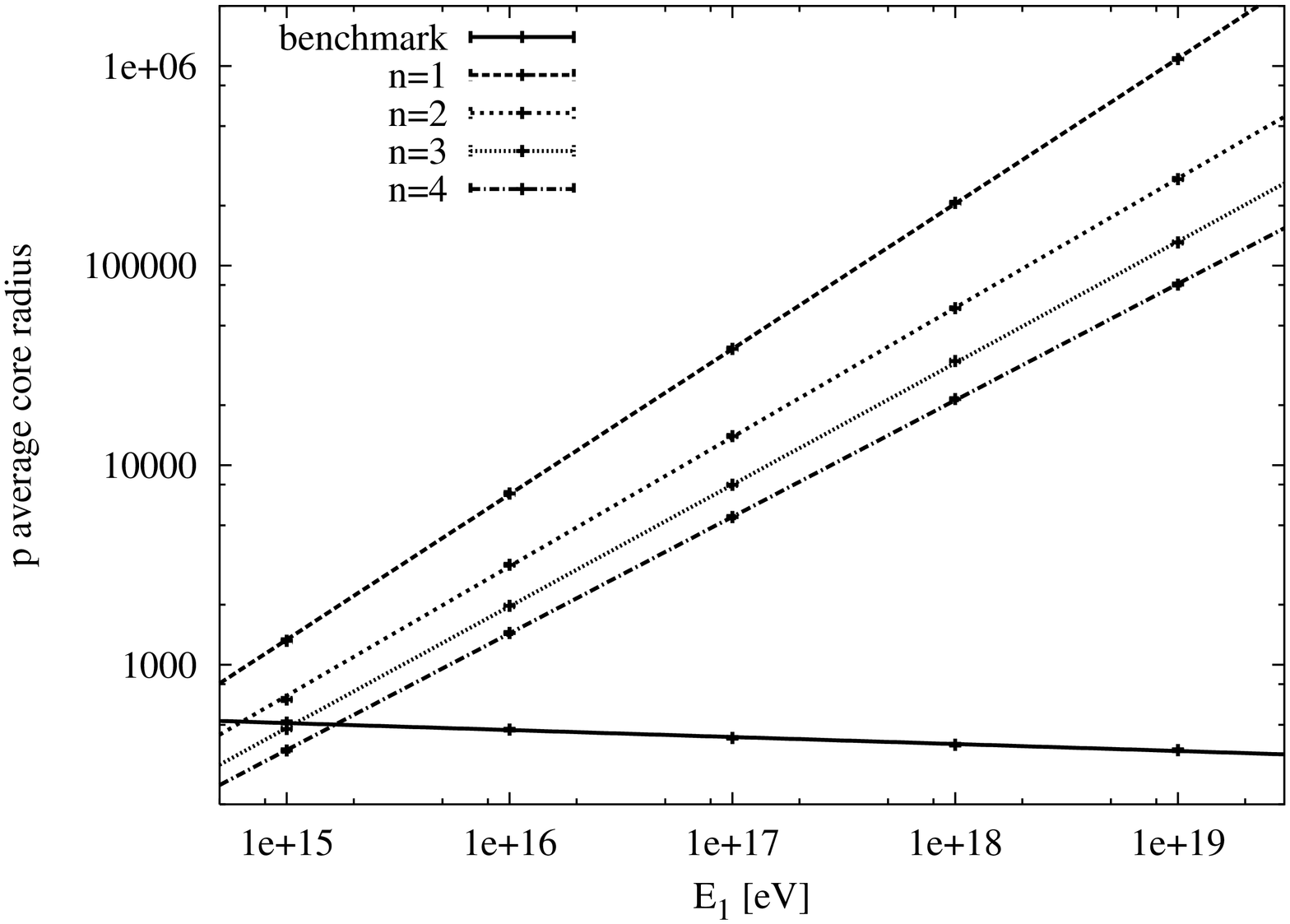}}}} \par}
\caption{The proton core size as a function of \protect\( E_{1}\protect \), 
(a) 5,500 m, (b) 15,000 m.}
\label{core5}
\end{figure}

Also in this case we discover a linear relation between 
average radius $R$ of the conical openings and energy, relation that can be fitted 
to a simple power law
\beq
R=10^{q'(n)} E_1^{\sigma'(n)}.  
\label{ntotal1}
\eeq
In analogy to Figures~\ref{fitcurve1} and \ref{fitcurve2}, we show in 
Figure~\ref{fitcurve3} that for a given setup there is a linear relation between 
slopes and intercepts of Eq.~(\ref{ntotal1})
\beq
q'=\alpha'\,\sigma' +\beta'
\eeq
with $\alpha'$ and $\beta'$ typical for a given setup, 
but again independent of $n$.

\end{itemize}

\begin{figure}
{\centering \subfigure[ ]{\resizebox*{10cm}{!}{\rotatebox{-90}
{\includegraphics{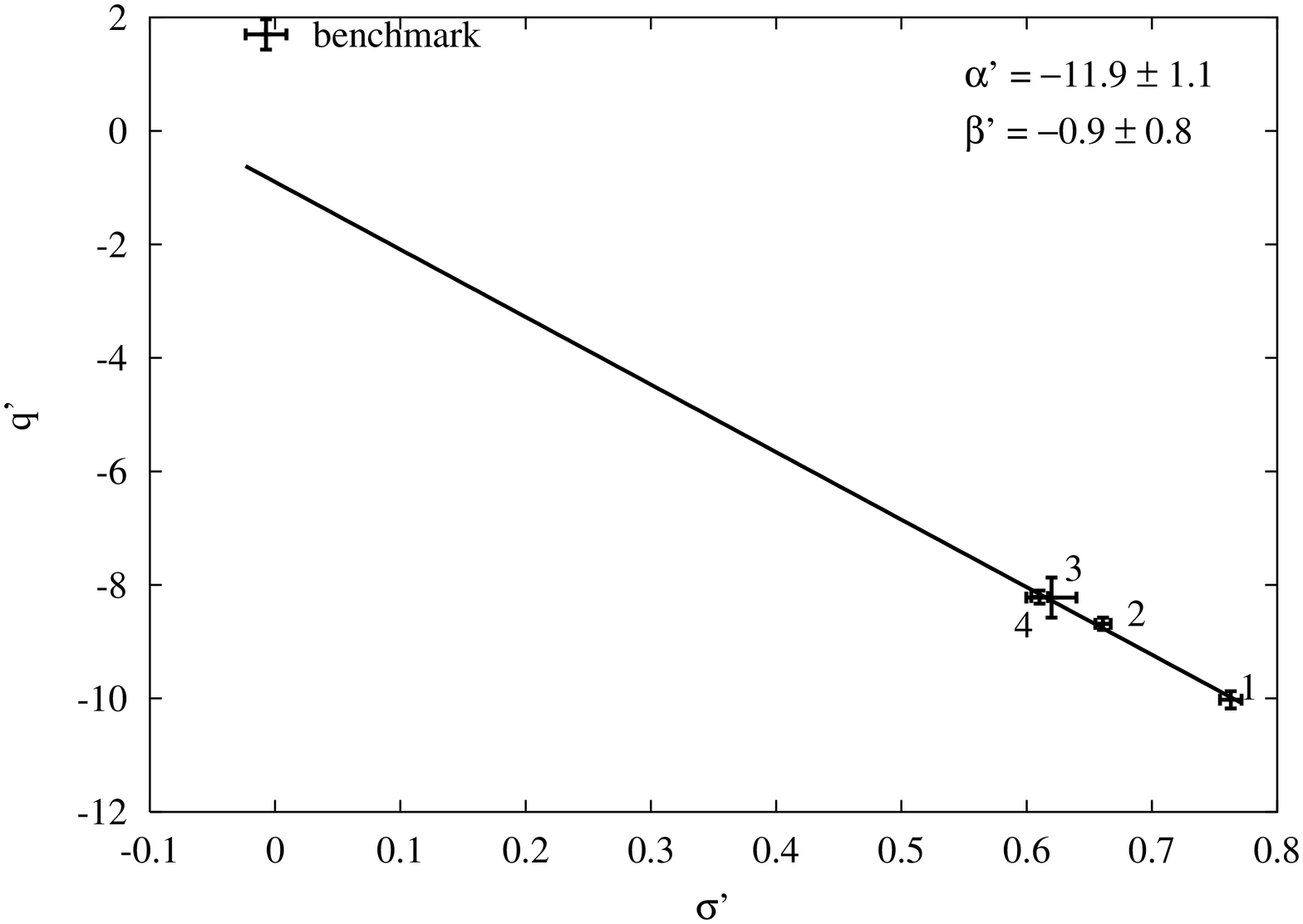}}}} \par}
{\centering \subfigure[ ]{\resizebox*{10cm}{!}{\rotatebox{-90}
{\includegraphics{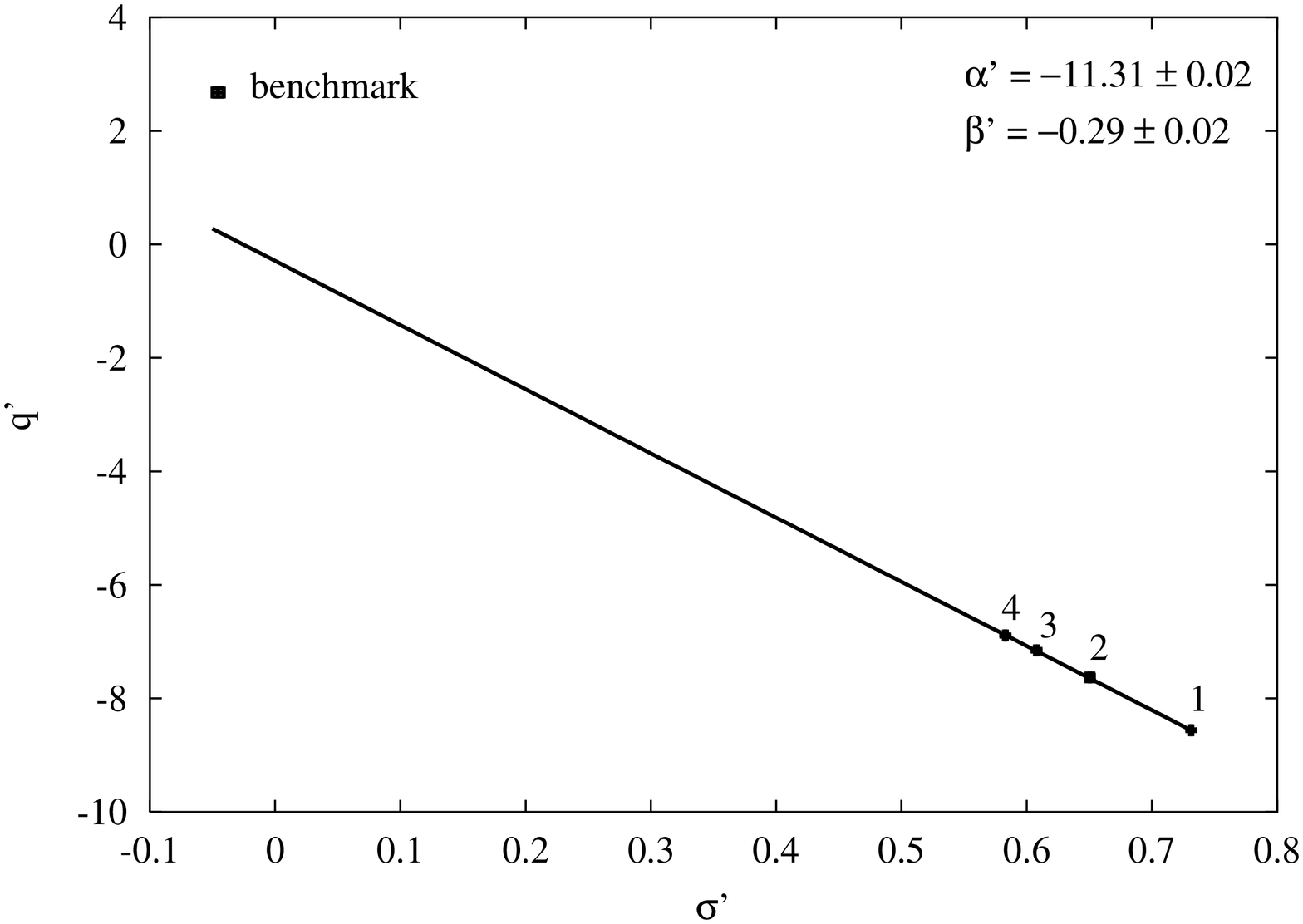}}}} \par}
\caption{Parameter fit for the curves
in Fig.~\ref{core1}, describing the openings of the showers of photons. 
(a) is the fit for 5,500 m, (b) for 15,000 m.}
\label{fitcurve3}
\end{figure}



\section{Discussion \label{sec:mbh7}}

A rather detailed analysis was presented of 
some of the main observables which characterize the air showers
formed, when a high energy collision in the atmosphere
leads to the formation of a mini black hole. We have decided
to focus our attention on the particle
multiplicities of these events, on the geometrical opening
of the showers produced, and on the ratio of their electromagnetic to
hadronic components, as functions of the entire ultra high energy
spectrum of the incoming primary source.
We have shown that in a double logarithmic scale
the energy vs multiplicity as well as the energy vs shower-size 
plots are linear, characterized
by slopes which depend on the number of extra dimensions.
We have compared these predictions with standard (benchmark) simulations 
and corrected for the energy which escaped in the bulk, or emitted 
by the black holes at stages prior to the Schwarzschild phase.
Black hole events are characterized by faster growing 
multiplicities for impacts taking place close to the detector; impacts 
at higher altitudes share a similar trend, but less pronounced. The
multiplicities from the black hole are larger in the lower part of the energy
range, while they become bigger for higher energies.
We should also mention that, given the choice made for our benchmark 
simulations, here we have been considering the worst scenario: in a 
simulation with an impacting neutrino 
it should be possible to discern between the two underlying events, 
whether they are standard or black hole mediated.   
The lateral distributions appear to be the most striking signature 
of a black hole event. Due to the higher $p_T$s involved, they are 
much larger than in the benchmark standard simulations. 

Our analysis can be easily generalized 
to more complex geometrical situations, where 
a slanted entry of the original primary can be envisioned and, in particular, 
to the case of near horizontal air showers, which are relevant for the detection of 
neutrino induced showers. These are characterized by a larger cross section compared 
to the vertical ones and therefore are more likely to occur. 
The strategy presented in this analysis, which is limited to 
primaries entering vertically, does not change substantially in this more general case, 
except for the geometry which should give 
an asymmetric opening due to the directionality of the event.
However, the main physical properties of the showers with an intermediate black hole should remain unchanged.
These characteristics, in fact, are not sensitive to the geometry of the direction of 
the event but are due to the larger multiplicities of the decay of the black hole resonance 
that forms after the impact and to the s-wave structure of its instantaneous decay.

\chapter*{Acknowledgements}

\addcontentsline{toc}{chapter}{Acknowledgements}\markboth{}{Acknowledgements}Above
all I wish to thank my advisor Claudio Corian\`{o} and all the people
with whom I collaborated during my doctoral course: Marco Guzzi, Alon
Faraggi, Daniele Martello, John Smith, Theodore Tomaras, Enzo Barone and Phil Ratcliffe. Thanks
are also due to my undergraduate thesis advisors, Mario Leo and Rosario
Antonio Leo, with whom I completed a paper last year.

Special thanks are due to all the friends with whom I shared these
three years: Andrea Bl., Andrea V., Antonio B., Giovanna and Letizia
(who shared the office with me in different times) and Andrea G.,
Andrea M., Antonio V., Barbara, Ciccio, Claudio, Cristian,
Domenico, Eleonora, Francesco, Gianfranco, Karen, Marco, Michele,
Piero, Raffaele, Saori, Simona, Tonio and many others.

Last but not least, thanks to my family for the constant support.

\end{document}